\newlength{\extralineskip}
  \documentstyle[12pt]{article}
\def\e{\, {\rm e}}

\def\IR{{\rm I}\!{\rm R}}

\def\semi{{\times\!\!\!\lhd}}
\def\inbar{\,\vrule height1.5ex width.4pt depth0pt}
\def\IR{\relax{\rm I\kern-.18em R}}
\def\IC{{\relax\hbox{$\inbar\kern-.3em{\rm C}$}}}
\def\IZ{{{\rm Z}\!\!{\rm Z}}}

\def\lie{{\cal L}}
\def\slash{\!\!\!\!/~}

\newcounter{equnum}[section]
\def\theequnum{\rm{\thesection}.\arabic{equnum}}
\newcommand{\beq}{$$ \refstepcounter{equnum}}
\newcommand{\eeq}{\eqno (\theequnum) $$}

\newcommand{\bd}{\begin{displaymath}}
\newcommand{\ed}{\end{displaymath}}

\def\IF{{{\rm I}\!{\rm F}}}
\def\tr{{\rm tr}}
\def\e{~{\rm e}}
\def\ch{{\rm ch}}
\def\td{{\rm td}}
\def\man{{\cal M}}
\def\hess{{\cal H}}
\def\sgn{{\rm sgn}}
\def\pfaff{{\rm Pfaff}}

\def\bea{\begin{eqnarray}}
\def\eea{\end{eqnarray}}
\def\ba{\beq\new\begin{array}{c}}
\def\ea{\end{array}\eeq}

\makeatletter
\newdimen\normalarrayskip              
\newdimen\minarrayskip                 
\normalarrayskip\baselineskip
\minarrayskip\jot
\newif\ifold             \oldtrue            \def\new{\oldfalse}
\def\arraymode{\ifold\relax\else\displaystyle\fi} 
\def\@arrayskip{\ifold\baselineskip\z@\lineskip\z@
     \else
     \baselineskip\minarrayskip\lineskip2\minarrayskip\fi}
\def\@arrayclassz{\ifcase \@lastchclass \@acolampacol \or
\@ampacol \or \or \or \@addamp \or
   \@acolampacol \or \@firstampfalse \@acol \fi
\edef\@preamble{\@preamble
  \ifcase \@chnum
     \hfil$\relax\arraymode\@sharp$\hfil
     \or $\relax\arraymode\@sharp$\hfil
     \or \hfil$\relax\arraymode\@sharp$\fi}}
\def\@array[#1]#2{\setbox\@arstrutbox=\hbox{\vrule
     height\arraystretch \ht\strutbox
     depth\arraystretch \dp\strutbox
     width\z@}\@mkpream{#2}\edef\@preamble{\halign \noexpand\@halignto
\bgroup \tabskip\z@ \@arstrut \@preamble \tabskip\z@ \cr}%
\let\@startpbox\@@startpbox \let\@endpbox\@@endpbox
  \if #1t\vtop \else \if#1b\vbox \else \vcenter \fi\fi
  \bgroup \let\par\relax
  \let\@sharp##\let\protect\relax
  \@arrayskip\@preamble}

\begin{document}

\begin{titlepage}
\setcounter{footnote}0
\begin{center}
\baselineskip=12pt
\hfill OUTP-96-51P\\
\hfill hep-th/9608068
\begin{flushright}{\today}\end{flushright}
\vspace{0.5in}

{\LARGE\bf Equivariant Localization of Path Integrals}
\vspace{.5in}

\baselineskip=12pt

{\large\bf Richard J. Szabo}
\vskip 0.1truein
{\it Department of Theoretical Physics\\ University of Oxford\\ 1 Keble
Road, Oxford OX1 3NP, U.K.}
\bigskip
\bigskip

\vskip 1.0 truein

\end{center}

\begin{abstract}

\baselineskip=12pt

We review equivariant localization techniques for the evaluation of Feynman
path integrals. We develop systematic geometric methods for studying the
semi-classical properties of phase space path integrals for dynamical systems,
emphasizing the relations with integrable and topological quantum field
theories. Beginning with a detailed review of the relevant mathematical
background -- equivariant cohomology and the Duistermaat-Heckman theorem, we
demonstrate how the localization ideas are related to classical integrability
and how they can be formally extended to derive explicit localization formulas
for path integrals in special instances using BRST quantization techniques.
Various loop space localizations are presented and related to notions in
quantum integrability and topological field theory. We emphasize the common
symmetries that such localizable models always possess and use these symmetries
to discuss the range of applicability of the localization formulas. A number of
physical and mathematical applications are presented in connection with
elementary quantum mechanics, Morse theory, index theorems, character formulas
for semi-simple Lie groups, quantization of spin systems, unitary integrations
in matrix models, modular invariants of Riemann surfaces, supersymmetric
quantum field theories, two-dimensional Yang-Mills theory, conformal field
theory, cohomological field theories and the loop expansion in quantum field
theory. Some modern techniques of path integral quantization, such as coherent
state methods, are also discussed. The relations between equivariant
localization and other ideas in topological field theory, such as the
Batalin-Fradkin-Vilkovisky and Mathai-Quillen formalisms, are presented and
used to discuss the general relationship between topological field theories and
more conventional physical models.

\end{abstract}

\end{titlepage}
\clearpage

\newpage

\baselineskip=18pt

\tableofcontents

\newpage
\setcounter{footnote}0

\baselineskip=14pt

\section{Introduction}

In this Paper we shall review the systematic approaches and applications of a
theory which investigates situations where Feynman path integrals of physical
systems can be evaluated exactly leading to a complete understanding of the
quantum physics. These mathematical formalisms are in large part motivated by
the symmetries present in integrable systems and topological quantum field
theories which make these latter examples exactly solvable problems. Besides
providing conceptual understandings of the solvability features of these
special classes of problems, this framework yields geometric approaches to
evaluating the quantum spectrum of generic quantum mechanical and quantum field
theoretical partition functions. The techniques that we shall present here in
fact motivate an approach to studying generic physical problems by relating
their properties to those of integrable and topological field theories. In
doing so, we shall therefore also review some of the more modern quantum field
theoretical and mathematical ideas which have been at the forefront of
theoretical physics over the past two decades.

\subsection{Path Integrals in Quantum Mechanics, Integrable Models and
Topological Field Theory}

The idea of path integration was introduced by Feynman \cite{feynman} in the
1940's as a novel approach to quantum theory. Symbolically, the
fundamental path integral formula is
\beq
{\cal K}(q',q;T)=\sum_{C_{qq'}}\e^{iTL[C_{qq'}]}
\label{feynmanpropsum}\eeq
where the `sum' is over all paths $C_{qq'}$ between the points $q$ and $q'$ on
the configuration space of a physical system, and $L[C_{qq'}]$
is the length of the path. The quantity on the left-hand side of
(\ref{feynmanpropsum}) represents the probability amplitude for the system to
evolve from a state with configuration $q$ to one with configuration $q'$ in a
time span $T$. One of the great advantages of the path integral formulation is
that it gives a global (integral) solution of the quantum problem in question,
in contrast to the standard approach to quantum mechanics based on the
Schr\"odinger equation which gives a local (differential) formulation of the
problem. Of utmost significance at the time was Feynman's generalization of the
path integral to quantum electrodynamics from which a systematic derivation of
the famous Feynman rules, and hence the basis of most perturbative calculations
in quantum field theory, can be carried out \cite{itzykson}.

The problem of quantum integrability, i.e. the possibility of solving
analytically for the spectrum of a quantum Hamiltonian and the corresponding
energy eigenfunctions, is a non-trivial problem. This is even apparent from the
point of view of the path integral, which describes the time evolution of
wavefunctions. Relatively few quantum systems have been solved exactly and even
fewer have had an exactly solvable path integral. At the time that the
functional integration (\ref{feynmanpropsum}) was introduced, the only known
examples where it could be evaluated exactly were the harmonic oscillator and
the free particle. The path integrals for these 2 examples can be evaluated
using the formal functional analog of the classical Gaussian integration
formula
\cite{zinn}
\beq
\int_{-\infty}^\infty\prod_{k=1}^ndx^k\e^{\frac{i}{2}\sum_{i,j}x^iM_{ij}x^j
+i\sum_i\lambda_ix^i}=\frac{(2\pi\e^{i\pi/2})^{\frac{n}{2}}\e^{\frac{i}{2}
\sum_{i,j}\lambda_i(M^{-1})^{ij}\lambda_j}}{\sqrt{\det M}}
\label{gaussintclass}\eeq
where $M=[M_{ij}]$ is a non-singular symmetric $n\times n$ matrix. In this way,
the Feynman propagator (\ref{feynmanpropsum}) can be evaluated formally for any
field theory which is at most quadratic in the field variables. If this is
not the case, then one can expand the argument of the exponential in
(\ref{feynmanpropsum}), approximate it by a quadratic form as in
(\ref{gaussintclass}), and then take the formula (\ref{gaussintclass}) as
an approximation for the integral. For a finite-dimensional integral this is
the well-known stationary phase (otherwise known as the saddle-point or
steepest-descent) approximation \cite{guillstern}. In the framework of path
integration, it is usually referred to as the Wentzel-Kramers-Brillouin (or WKB
for short) approximation \cite{messiah,schulman}. Since the result
(\ref{gaussintclass}) is determined by substituting into the exponential
integrand the global minimum (i.e. classical value) of the quadratic form
and multiplying it by a term involving the second variation of that form
(i.e. the fluctuation determinant), this approach to functional integration is
also called the semi-classical approximation. In this sense,
(\ref{feynmanpropsum}) interprets quantum mechanics as a sum over paths
fluctuating about the classical trajectories (those with minimal length
$L[C_{qq'}]$) of a dynamical system.
When the semi-classical approximation is exact, one can think of the Gaussian
integration formula (\ref{gaussintclass}) as a `localization' of the
complicated looking integral on the left-hand side of (\ref{gaussintclass})
onto the global minimum of the quadratic form there.

For a long time, these were the only examples of exactly solvable path
integrals. In 1968 Schulman \cite{schul1} found that a path integral
describing the precession of a spin vector was given exactly by its
WKB approximation. This was subsequently generalized by Dowker \cite{dowker}
who proved the exactness of the semi-classical approximation for the path
integral describing free geodesic motion on compact group manifolds.
It was not until the late 1970's that more general methods, beyond the
restrictive range of the standard WKB method, were developed. In these
methods, the Feynman path integral is calculated rigorously in discretized
form (i.e. over piecewise-linear paths) by a careful regularization
prescription \cite{langouche}, and then exploiting information provided by
functional analysis, the theory of special functions, and the theory of
differential equations (see \cite{bohm} and references therein). With these
tricks the list of exactly solvable path integrals has significantly increased
over the last 15 years, so that today one is able to essentially evaluate
analytically the path integral for any quantum mechanical problem for which
the Schr\"odinger equation can be solved exactly. We refer to \cite{kleinert}
and \cite{grosche} for an overview of these methods and a complete
classification of the known examples of exactly solved quantum mechanical path
integrals to present date.

The situation is somewhat better in quantum field theory, which represents the
real functional integrals of interest from a physical standpoint. There are
many non-trivial examples of classically integrable models (i.e. ones whose
classical equations of motion are `exactly solvable'), for example the
sine-Gordon model, where the semi-classical approximation describes the exact
spectrum of the quantum field theory \cite{zinn}. Indeed, for any classically
integrable dynamical system one can canonically transform the phase space
variables so that, using Hamilton-Jacobi theory \cite{goldstein}, the path
integral can be formally manipulated to yield a result which if taken naively
would imply the exactness of the WKB approximation for {\it any} classically
integrable system \cite{schulman}. This is not really the case, because the
canonical transformations used in the phase space path integral do not respect
the ordering prescription used for the properly discretized path integral and
consequently the integration measure is not invariant under these
transformations \cite{blaukesk}. However, as these problems stem mainly from
ordering ambiguities in the discretization of the path integral, in quantum
field theory these ordering ambiguities could be removed by
a suitable renormalization, for instance by an operator commutator ordering
prescription. This has led to the conjecture that properly interpreted
results of semi-classical approximations in integrable field theories
reproduce features of the exact quantum spectrum \cite{zinn}. One of the
present motivations for us is to therefore develop a systematic way to
implement realizations of this conjecture.

Another class of field theories where the path integral is exactly solvable
in most cases is supersymmetric theories and topological quantum field
theories (see \cite{birm} for a concise review). Topological field theories
have lately been of much interest in both the mathematics and physics
literature. A field theory is topological if it has only global degrees of
freedom. This means, for example, that its classical equations of
motion eliminate all propagating degrees of freedom from the problem
(so that the effective quantum action vanishes). In particular, the theory
cannot depend on any metric of the space on which the fields are defined. The
observables of these quantum field theories therefore describe geometrical
and topological invariants of the spacetime which are computable by the
conventional techniques of quantum field theory and are of prime interest in
mathematics. Physically, topological quantum field theories bear resemblances
to many systems of longstanding physical interest and it is hoped that this
special
class of field theories might serve to provide insight into the structure
of more complicated physical systems and a testing ground for new approaches
to quantum field theory. There is also a conjecture that topological quantum
field theories represent different (topological) phases of their more
conventional counterparts (e.g. 4-dimensional Yang-Mills theory).
Furthermore, from a mathematical point of view, these field theories provide
novel representations of some global invariants whose properties are
frequently transparent in the path integral approach.

Topological field theory essentially traces back to the work of Schwarz
\cite{schwarz} in 1978 who showed that a particular topological invariant,
the Ray-Singer analytic torsion, could be represented as the partition function
of a certain quantum field theory. The most important historical work for us,
however, is the observation made by Witten \cite{witten1} in 1982 that the
supersymmetry algebra of supersymmetric quantum mechanics describes exactly the
DeRham complex of a manifold, where the supersymmetry charge is the
exterior derivative. This gives a framework for understanding Morse theory in
terms of supersymmetric quantum mechanics in which the quantum partition
function computes exactly the Euler characteristic of the configuration
manifold, i.e. the index of the DeRham complex.

Witten's partition function computed the so-called Witten index \cite{witten2},
the difference between the number of bosonic and fermionic zero energy states.
In order for the supersymmetry to be broken in the ground state of a
supersymmetric model, the Witten index must vanish. As supersymmetry, i.e.
a boson-fermion symmetry, is not observed in nature, it is necessary to have
some criterion for dynamical supersymmetry breaking if supersymmetric theories
are to have any physical meaning. Witten's construction was subsequently
generalized by Alvarez-Gaum\'e \cite{alvgaume}, and Friedan and Windey
\cite{friedan}, to give supersymmetric field theory proofs of the Atiyah-Singer
index theorem \cite{eguchi}. In this way, the partition function is reduced to
an integral over the configuration manifold $\man$. This occurs because the
supersymmetry of the action causes only zero modes of the fields, i.e. points
on $\man$, to contribute to the path integral, and the integrals over the
remaining fluctuation modes are Gaussian. The resulting integral encodes
topological information about the manifold $\man$ and represents a huge
reduction of the original infinite-dimensional path integration.

This field began to draw more attention around 1988 when Witten introduced
topological field theories in a more general setting \cite{wittentop} (see also
\cite{wittencoh}). A particular supersymmetric non-abelian gauge theory was
shown by Witten to describe a theory with only global degrees of freedom whose
observables are the Donaldson invariants, certain differential invariants
which are used for the study of differentiable structures on 4-manifolds.
Subsequent work then put these ideas into a general framework so that today the
formal field theoretic structures of Witten's actions are well-understood
\cite{birm}. Furthermore, because of their topological nature, these field
theories have become the focal point for the description of topological effects
in quantum systems using quantum field theory, for instance for the description
of holonomy effects in physical systems arising from the adiabatic transports
of
particles \cite{semchern} and extended objects such as strings \cite{bergeron}
(i.e. Aharonov-Bohm type effects). In this way the functional integral has
in recent years become a very popular tool lying on the
interface between string theory, conformal field theory and topological quantum
field theory in physics, and between topology and algebraic geometry in
mathematics. Because of the consistent reliability of results that path
integrals of these theories can produce when handled with care, functional
integration has even acquired a certain degree of respectability among
mathematicians.

\subsection{Equivariant Localization Theory}

The common feature of topological field theories is that their path integrals
are described exactly by the semi-classical approximation. It would be nice
to put semi-classically exact features of functional integrals, as well as the
features which reduce them to integrals over finite-dimensional manifolds as
described above, into some sort of general framework. More generally, we would
like to have certain criteria available for when we expect partition functions
of quantum theories to reduce to such simple expressions, or `localize'. This
motivates an approach to quantum integrability in which one can
systematically study the properties of integrable field theories and their
conjectured semi-classical ``exactness" that we mentioned before. In this
approach we focus on the general features and properties that path integrals
appearing in this context have in common. Foremost among these is the
existence of a large number of (super-)symmetries in the underlying dynamical
theory, so that these functional integrals reduce to Gaussian ones and
essentially represent finite-dimensional integrals\footnote{\baselineskip=12pt
The exact solvability features of path integrals in this context is similar to
the solvability features of the Schr\"odinger equation in quantum mechanics
when
there is a large symmetry group of the problem. For instance, the $O(4)$
symmetry of the 3-dimensional Coulomb problem is what makes the hydrogen atom
an exactly solvable quantum system \cite{messiah}.}. The transition between the
functional and finite-dimensional integrals can then be regarded as a rather
drastic localization of the original infinite-dimensional integral, thereby
putting it into a form that is useable for extracting physical and mathematical
information. The mathematical framework for describing these symmetries, which
turn out to be of a topological nature, is equivariant cohomology and the
approach discussed in this paragraph is usually called `equivariant
localization theory'. This approach introduces an equivariant cohomological
framework as a tool for developing geometric techniques for manipulating path
integrals and examining their localization properties.

Historically, this subject originated in the mathematics
literature in 1982 with the Duistermaat-Heckman theorem \cite{dh}, which
established the exactness of the semi-classical approximation for
finite-dimensional oscillatory integrals (i.e. finite-dimensional versions
of (\ref{feynmanpropsum})) over compact symplectic manifolds in certain
instances. The Duistermaat-Heckman theorem applies to classical systems whose
trajectories all have a common period, so that the symmetry responsible for the
localization here is the existence of a global Hamiltonian torus action on the
manifold. Atiyah and Bott \cite{atiyahbott} showed that the Duistermaat-Heckman
localization was a special case of a more general localization property of
equivariant cohomology (with respect to the torus group action in the case of
the Duistermaat-Heckman theorem). This fact was used by Berline and Vergne
\cite{berline1,berline2} at around the same time to derive a quite general
integration formula valid for Killing vectors on general compact Riemannian
manifolds.

The first infinite-dimensional generalization of the Duistermaat-Heckman
theorem
is due to Atiyah and Witten \cite{atiyah}, in the setting of a supersymmetric
path integral for the index (i.e. the dimension of the space of zero modes)
of a Dirac operator. They showed that a formal application of the
Duistermaat-Heckman theorem on the loop space $L\man$ of a manifold $\man$
to the partition function of $N=\frac{1}{2}$ supersymmetric quantum mechanics
(i.e. a supersymmetric spinning particle in a gravitational background)
reproduced the well-known Atiyah-Singer index theorem correctly. The crucial
idea was the interpretation of the fermion bilinear in the supersymmetric
action as a loop space symplectic 2-form. This approach was then generalized
by Bismut \cite{bismut1,bismut2}, within a mathematically rigorous framework,
to twisted Dirac operators (i.e. the path integral for spinning particles
in gauge field backgrounds), and to the computation of the Lefschetz number
of a Killing vector field $V$ (a measure of the number of zeroes of $V$) acting
on the manifold. Another nice infinite-dimensional generalization of the
Duistermaat-Heckman theorem was suggested by Picken \cite{picken} who
formally applied the theorem to the space of loops over a group manifold
to localize the path integral for geodesic motion on the group, thus
establishing the well-known semi-classical properties of these systems.

It was the beautiful paper by Stone \cite{stone} in 1989 that brought the
Duistermaat-Heckman theorem to the attention of a wider physics audience.
Stone presented a supersymmetric derivation of the Weyl character formula
for $SU(2)$ using the path integral for spin and interpreted the result as a
Duistermaat-Heckman localization. This supersymmetric derivation was
extended by Alvarez, Singer and Windey \cite{alvarez} to more general Lie
groups
using fiber bundle theory, and the supersymmetries in both of these approaches
are very closely related to equivariant cohomology. At around this time,
other important papers concerning the quantization of spin appeared. Most
notably, Nielsen and Rohrlich \cite{nielsen} (see also \cite{nima}) viewed the
path integral for spin from a more geometrical point of view, using as action
functional the solid angle swept out by the closed orbit of the spin. This
approach was related more closely to geometric quantization and group
representation theory by Alekseev, Faddeev and Shatashvili \cite{alek1,alek2},
who calculated the coadjoint orbit path integral for unitary and orthogonal
groups, and also for cotangent bundles of compact groups, Kac-Moody groups and
the Virasoro group. The common feature is always that the path integrals are
given exactly by a semi-classical localization formula that resembles the
Duistermaat-Heckman formula.

The connections between supersymmetry and equivariant cohomology in the
quantum mechanics of spin were clarified by Blau in \cite{blau1}, who related
the Weinstein action invariant \cite{weinstein} to Chern-Simons gauge theory
using the Duistermaat-Heckman integration formula. Based on this
interpretation, Blau, Keski-Vakkuri and Niemi \cite{blaukesk} introduced
a general supersymmetric (or equivariant cohomological) framework for
investigating Duistermaat-Heckman (or WKB) localization formulas for generic
(non-supersymmetric) phase space path integrals, leading to the fair amount
of activity in this field which is today the foundation of equivariant
localization theory. They showed formally that the partition function for the
quantum mechanics of circle actions of isometries on symplectic manifolds
localizes. Their method of proof involves formal techniques of
Becchi-Rouet-Stora-Tyupin (or BRST for short) quantization of constrained
systems \cite{baulieu}. BRST-cohomology is the fundamental structure in
topological field theories, and such BRST supersymmetries are always the
symmetries that are responsible for localization in these models.

\subsection{Outline}

In this Review we shall primarily explore the geometric features of the
localization formalism for phase space path integrals. In particular, we shall
focus on how these models can be used to extract information about integrable
and topological quantum field theories. In this sense, the path integrals we
study for the most part can be thought of as ``toy models" serving as a testing
ground for ideas in some more sophisticated field theories. The main idea
behind this reasoning is that the localization in topological field theories is
determined by their kinematical properties. The path integrals we shall focus
on allow us to study their kinematical (i.e. geometrical and topological)
aspects in isolation from their dynamical properties. These models are
typically dynamically linear (i.e. free field theories) in some sense and the
entire non-triviality of their path integrals lie in the large kinematical
non-linearity that these theories possess. The appropriate relationship between
topological field theories and more conventional, physical interacting quantum
field theories (which are kinematically linear but dynamically highly
non-linear) would then, in principle at least, allow one to incorporate the
approaches and techniques which follow to generic physical models. Indeed, one
of the central themes in what follows will be the interplay between physical,
integrable and topological field theories, and we shall see that the
equivariant localization formalism implies connections between these 3 classes
of models and thus a sort of unified description of functional integration
which provides alternative approximation techniques to the usual perturbative
expansion in quantum field theory.

We shall therefore approach the localization formalism for path integrals in
the
following manner. Focusing on the idea of localizing a quantum partition
function by reducing it using the large symmetry of the dynamical system to a
sum or finite-dimensional integral in analogy with the classical Gaussian
integration formula (\ref{gaussintclass}), we shall first analyse the
symmetries responsible for the localization of finite-dimensional integrals
(where the symmetry of the dynamics is represented by an equivariant
cohomology). The main focus of this Review will then be the formal
generalizations of these ideas to phase space path integrals, where the
symmetry becomes a ``hidden" supersymmetry of the dynamics representing the
infinite-dimensional analog of equivariant cohomology. The subsequent
generalization will be then the extension of these notions to both Poincar\'e
supersymmetric quantum field theories (where the symmetry is represented by the
supersymmetry of that model) and topological quantum field theories (where the
symmetry is represented by a gauge symmetry). The hope is that these serve as
testing grounds for the more sophisticated quantum field theories of real
physical interest, such as quantum chromodynamics (QCD). This gives
a geometric framework for studying quantum integrability, as well as insights
into the structure of topological and supersymmetric field theories, and
integrable models. In particular, from this analysis we can hope to uncover
systematically the reasons why some quantum problems are exactly solvable, and
the reasons why others aren't.

Briefly, the structure of this Review is as follows. In Section 2 we go through
the main mathematical background for localization theory, i.e. equivariant
cohomology, with reviews of some other mathematical ideas that will be
important for later Sections as well. In Section 3 we present the
Duistermaat-Heckman theorem and its generalizations and discuss the connections
they imply between equivariant cohomology and the notion of classical
integrability of a dynamical system. Section 4 then goes through the formal
supersymmetry
and loop space equivariant cohomology arguments establishing the localization
of phase space path integrals when there is a Riemannian structure on the phase
space which is invariant under the classical dynamics of the system. Depending
on the choice of localization scheme, different sets of phase space
trajectories are lifted to a preferred status in the integral. Then all
contributions to the functional integral come from these preferred paths
along with a term taking into account the quantum fluctuations about these
selected loops. Sections 5 and 6 contain the main physical and mathematical
applications of equivariant localization. There we use the fundamental isometry
condition to construct numerous examples of localizable path integrals. In each
case we evaluate and discuss the localization formulas from both physical and
mathematical standpoints. Here we shall encounter numerous examples and
gain much insight into the range of applicability of the localization formalism
in general. We will also see here many interesting features of the localizable
partition functions when interpreted as topological field theories, and we
discuss in detail various other issues (e.g. coherent state quantization and
coadjoint orbit character formulas) which are common to all the localizable
examples that we find within this setting and which have been of interest in
the more modern approaches to the quantization of dynamical systems. Section 7
then takes a slightly different approach to analysing localizable systems, this
time by some geometric constructs of the full loop-expansion on the phase
space. Here we shall discuss how the standard localization symmetries could be
extended to more general ones, and we shall also show how the localization
ideas could be applied to the formulation of a geometric approach to obtaining
corrections to the standard WKB approximation for non-localizable partition
functions. The analysis of this Section is a first step towards a systematic,
geometric understanding of the reasons why the localization formulas may not
apply to a given dynamical system. In Section 8 we turn our attention to field
theoretical applications of equivariant localization and discuss the
relationships that are implied between topological field theories, physical
quantum field theories, and the localization formalism for dynamical systems.
For completeness, 2 Appendices at the end of the paper are devoted to an
overview of some ideas in the BRST quantization formalism and some more
mathematical ideas of equivariant cohomology, all of which play an important
role in the development of the ideas in the main body of this Review.

We close this introductory Section with some comments about the style
of this Review. Although we have attempted to keep things self-contained and at
places where topics aren't developed in full detail we have included ample
references for further reading, we do assume that the reader has a relatively
solid background in many of the mathematical techniques of modern theoretical
physics such as topology, differential geometry and group theory. All of the
group theory that is used extensively in this Review can be found in
\cite{varadarajan} (or see \cite{georgi} for a more
elementary introduction), while most of the material discussing differential
geometry, homology and cohomology, and index theorems can be found in the books
\cite{greensw,bottu,nash} and the review articles \cite{birm,eguchi}. For a
more detailed introduction to algebraic topology, see \cite{massey}. The basic
reference for quantum field theory is the classic text \cite{itzykson}.
Finally,
for a discussion of the issues in supersymmetry theory and BRST quantization,
see \cite{baulieu,birm,henneaux,niemi,sohnius} and references therein.

\section{Mathematical Preliminaries: Equivariant Cohomology and the
Localization Principle}

\subsection{Example: The Height Function of the Sphere}

To help motivate some of the abstract and technical formalism which follows, we
start by considering the evaluation of a rather simple integral.
Consider the 2-sphere $S^2$ of unit radius viewed in Euclidean 3-space $\IR^3$
as a sphere standing on end on the $xy$-plane and centered at $z=a$
symmetrically about the $z$-axis. We introduce the usual spherical polar
coordinates $x=\sin\theta\cos\phi$, $y=\sin\theta\sin\phi$ and $z=a-\cos\theta$
for the embedding of the sphere in 3-space as
$S^2=\{(x,y,z)\in\IR^3:x^2+y^2+(z-a)^2=1\}$, where $0\leq\theta\leq\pi$ and
$0\leq\phi\leq2\pi$. The height of the sphere off of the $xy$-plane is given by
the height function $z$ in $\IR^3$ restricted to $S^2$,
\beq
h_0(\theta,\phi)=a-\cos\theta
\label{heights2}\eeq
We want to evaluate the oscillatory integral
\beq
Z_0(T)=\int_0^\pi\!\!\int_0^{2\pi}d\theta~d\phi~\sin\theta\e^{iTh_0(\theta,
\phi)}
\label{zs2}\eeq
which represents a `toy' version of (\ref{feynmanpropsum}). The integration
measure in (\ref{zs2}) is the standard volume form on $S^2$, i.e. that which is
obtained by restriction of the measure $dx~dy~dz$ of $\IR^3$ to the sphere, and
$T$ is some real-valued parameter. It is straightforward to carry out the
integration in (\ref{zs2}) to get
\beq\new{\begin{array}{ll}
Z_0(T)&=2\pi\int_{-1}^{+1}d\cos\theta~\e^{iT(a-\cos\theta)}\\&=\frac{2\pi
i}{T}\left(\e^{-iT(1+a)}-\e^{iT(1-a)}\right)=\frac{4\pi}{T}\e^{-iTa}\sin T
\end{array}}
\label{zs2eval}\eeq

Although this integral is simple to evaluate explicitly, it illustrates 2
important features that will be the common theme throughout our discussion. The
first characteristic is the second equality in (\ref{zs2eval}). This shows that
$Z_0(T)$ can be expressed as a sum of 2 terms which correspond, respectively,
to the 2 extrema of the height function (\ref{heights2}) -- one from the north
pole $\theta=\pi$, which is the maximum of (\ref{heights2}), and the other from
the south pole $\theta=0$, which is its minimum. The relative minus sign
between these 2 terms arises from the fact that the signature of the Hessian
matrix of $h_0$ at its maximum is negative while that at its minimum is
positive, i.e. the maximum of $h_0$ is unstable in the 2 directions along the
sphere, each of which, heuristically, constributes a factor of $i$ to the sum
in (\ref{zs2eval}). Finally, the factor $2\pi i/T$ can be understood as the
contribution from the 1-loop determinant when (\ref{heights2}) is expanded to
quadratic order in $(\theta,\phi)$ and the standard WKB approximation to the
integral is used. In other words, (\ref{zs2eval}) coincides {\it exactly} with
the Gaussian integral formula (\ref{gaussintclass}), except that it sums over
both the minimum and maximum of the argument of the exponential in (\ref{zs2}).

The second noteworthy feature here is that there is a symmetry responsible for
the simple evaluation of (\ref{zs2}). This symmetry is associated with the
interplay between the globally-defined (i.e. single-valued) integration measure
and the integrand function (\ref{heights2}) (see the first equality in
(\ref{zs2eval})). Both the height function (\ref{heights2}) and the integration
measure in (\ref{zs2}) are independent of the polar coordinate $\phi$ of $S^2$.
This is what led to the simple evaluation of (\ref{zs2}), and it means, in
particular, that the quantities integrated in (\ref{zs2}) are invariant under
the translations $\phi\to\phi+\phi_0$, $\phi_0\in[0,2\pi)$, which correspond to
rigid rotations of the sphere about the $z$-axis. These translations generate
the circle group $S^1\sim U(1)$. The existence of a group acting on $S^2$ which
serves as a mechanism for the `localization' of $Z_0(T)$ onto the stationary
points of $h_0$ gives us hope that we could understand this feature by
exploiting non-trivial global features of the quotient space $S^2/S^1$ within a
mathematically rigorous framework. Our hopes are immediately dashed because the
2-sphere with its points identified under this (continuous) circular symmetry
globally has the same properties as the mathematically trivial interval $[0,1]$
(the space where $\theta/\pi$ lives), i.e.
\beq
S^2/U(1)\simeq[0,1]
\eeq
We shall see that the reason we cannot examine this space in this way is
because  the circle action above leaves fixed the north pole, at $\theta=\pi$,
and the south pole, at $\theta=0$, of the sphere. The fixed points in this case
are at the 2 extrema of the height function $h_0$. The correct mathematical
framework that we should use to describe this situation should take into proper
account of this group action on the space -- this is `equivariant cohomology'.

Equivariant cohomology has over the past few years become an increasingly
popular tool in theoretical physics, primarily in studies of topological models
such as topological gauge theories, topological string theory, and topological
gravity. This theory, and its connection with the ideas of this Subsection,
will be the topic of this Section. Beginning with a quick review of the DeRham
theory, which has for quite a while now been at the forefront of many of the
developments of modern theoretical physics (see \cite{eguchi} for a
comprehensive review), we shall then develop the framework which describes the
topology of a space when there is an action of some Lie group on it. This is
reminiscent of how one changes ordinary derivatives to gauge-covariant ones in
a gauge field theory to properly incorporate local gauge invariance of the
model. We shall ultimately end up discussing the important localization
property of integration in equivariant cohomology, and we will see later on
that the localization theorems are then fairly immediate consequences of this
general formalism.

We close this Subsection with a comment about the above example. Although it
may seem to serve merely as a toy model for some ideas that we wish to pursue,
we shall see that this example can be considered as the classical partition
function of a spin system (i.e. a classical rotor). A quantum mechanical
generalization of it will therefore be associated with the quantization of
spin. If we think of the sphere as the Lie group quotient space $SU(2)/U(1)$,
then, as we shall discuss extensively later on, this example has a nice
generalization to the so-called `homogeneous' spaces of the form $G/T$, where
$G$ is a Lie group and $T$ is a maximal torus of $G$. These sets of examples,
known as `coadjoint orbits', will frequently occur as non-trivial verifications
of the localization formalisms.

\subsection{A Brief Review of DeRham Cohomology}

To introduce some notation and to provide a basis for some of the more abstract
concepts that will be used throughout this Review, we begin with an elementary
`lightning' review of DeRham cohomology theory and how it probes the
topological features of a space. Throughout we shall be working on an abstract
topological space (i.e. a set with a collection of open subsets which is closed
under unions and finite intersections), and we always regard 2 topological
spaces as the same if there is an invertible mapping between the 2 spaces which
preserves their open sets, i.e. a bi-continuous function or `homeomorphism'. To
carry out calculus on these spaces, we have to introduce a smooth structure on
them (i.e. one that is infinitely-continuously differentiable -- or $C^\infty$
for short) which is done in the usual way by turning to the notion of a
differentiable manifold.

Let $\man$ be a $C^\infty$ manifold of
dimension $n$, i.e. $\man$ is a paracompact Hausdorff topological space which
can be covered by open sets $U_i$, $\man=\bigcup_iU_i$, each of which is
homeomorphic to $n$-dimensional Euclidean space $\IR^n$ and the local
homeomorphisms so used induce $C^\infty$ coordinate transformations on the
overlaps of patches in $\IR^n$. This means that locally, in a neighbourhood of
any point on $\man$, we can treat the manifold as a copy of the more familiar
$\IR^n$, but globally the space $\man$ may be very different from Euclidean
space. One way to characterize the global properties of $\man$, i.e. its
topology, which make it quite different from $\IR^n$ is through the theory of
homology and its dual theory, cohomology. Of particular importance to us will
be the DeRham theory \cite{bottu}. We shall always assume that $\man$ is
orientable and path-connected (i.e. any 2 points in $\man$ can be joined by a
continuous path in $\man$). We shall usually assume, unless otherwise stated,
that $\man$ is compact. In the non-compact case, we shall assume certain
regularity conditions at infinity so that results for the compact case hold
there as well.

Around each point of the manifold we choose an open set $U$ which is a copy of
$\IR^n$. In $\IR^n$ we have the natural notion of tangent vectors to a point,
and so we can use the locally defined homeomorphisms to define tangent vectors
to a point $x\in\man$. Using the local coordinatization provided by the
homeomorphism onto $\IR^n$, a general linear combination of tangent vectors
is denoted as
\beq
V=V^\mu(x)\frac{\partial}{\partial x^\mu}
\label{vecdef}\eeq
where throughout we use the Einstein summation convention for repeated upper
and lower indices. A linear combination such as (\ref{vecdef}) will be refered
to here as a vector field. Its components $V^\mu(x)$ are
$C^\infty$ functions on $\man$ and are specified by the introduction
of local coordinates from $\IR^n$. Acting on a smooth function $f(x)$, the
quantity $V(f)\equiv V^\mu\partial_\mu f$ is the directional derivative of $f$
in the direction of the vector components $\{V^\mu\}$. The local derivatives
$\{\frac{\partial}{\partial x^\mu}\}_{\mu=1}^n$ span an $n$-dimensional vector
space over $\IR$ which is called the tangent space to $\man$ at $x$ and it is
denoted by $T_x\man$. The disjoint union of all tangent spaces of the manifold,
\beq
T\man=\bigsqcup_{x\in\man}T_x\man
\label{tanbundle}\eeq
is called the tangent bundle of $\man$.

Any vector space $W$ has a dual vector space $W^*$ which is the space of
linear functionals ${\rm Hom}_\IR(W,\IR)$ on $W\to\IR$. The dual of the tangent
space $T_x\man$ is called the cotangent space $T_x^*\man$ and its basis
elements $dx^\mu$ are defined by
\beq
dx^\mu\left(\frac{\partial}{\partial x^\nu}\right)=\delta^\mu_\nu
\label{dualbasis}\eeq
The disjoint union of all the cotangent spaces of $\man$,
\beq
T^*\man=\bigsqcup_{x\in\man}T_x^*\man
\label{cotbundle}\eeq
is called the cotangent bundle of $\man$.

The space $(T^*_x\man)^{\otimes k}$ is the space of $n$-multilinear functionals
on $T_x\man\times\cdots\times T_x\man$ whose elements are the linear
combinations
\beq
T=T_{\mu_1\cdots \mu_k}(x)dx^{\mu_1}\otimes\cdots\otimes dx^{\mu_k}
\label{tensor}\eeq
The object (\ref{tensor}) is called a rank-$(k,0)$ tensor and its components
are $C^\infty$ functions of $x\in\man$. Similarly, the associated dual space
$(T_x\man)^{\otimes \ell}$ consists of the linear combinations
\beq
\tilde T=\tilde T^{\mu_1\cdots \mu_\ell}(x)\frac{\partial}{\partial x^{\mu_1}}
\otimes\cdots\otimes\frac{\partial}{\partial x^{\mu_\ell}}
\label{dualtens}\eeq
which are called $(0,\ell)$ tensors. The elements of $(T^*_x\man)^{\otimes k}
\otimes(T_x\man)^{\otimes\ell}$ are called $(k,\ell)$ tensors and one can
define tensor bundles analogously to the tangent and cotangent bundles
above. Under a local $C^\infty$ change of coordinates on $\man$
represented by the diffeomorphism $x\to x'(x)$, (\ref{tensor}) and
(\ref{dualtens}) along with the usual chain rules
\beq
\frac{\partial}{\partial x^\mu}=\frac{\partial x'^\lambda}{\partial x^\mu}
\frac{\partial}{\partial x'^\lambda}~~~~~,~~~~~dx^\mu=\frac{\partial x^\mu}{
\partial x'^\lambda}dx'^\lambda
\label{chainrule}\eeq
imply that the components of a generic rank $(k,\ell)$ tensor field
$T^{\mu_1\cdots\mu_\ell}_{\nu_1\cdots\nu_k}(x)$ transform as
\beq
T'^{\lambda_1\cdots\lambda_\ell}_{\rho_1\cdots\rho_k}(x')=\frac{\partial x
^{\prime\lambda_1}}{\partial x^{\mu_1}}\cdots\frac{\partial x^{\prime\lambda_
\ell}}{\partial x^{\mu_\ell}}\frac{\partial x^{\nu_1}}{\partial
x^{\prime\rho_1}
}\cdots\frac{\partial x^{\nu_k}}{\partial x^{\prime\rho_k}}T^{\mu_1\cdots\mu_
\ell}_{\nu_1\cdots\nu_k}(x)
\label{tenstransf}\eeq
Such local coordinate transformations can be thought of as changes of bases
(\ref{chainrule}) on the tangent and cotangent spaces.

We are now ready to define the DeRham complex of a manifold $\man$. Given
the tensor product of copies of the cotangent bundle as above, we define
a multi-linear anti-symmetric multiplication of elements of the cotangent
bundle, called the exterior or wedge product, by
\beq
dx^{\mu_1}\wedge\cdots\wedge dx^{\mu_k}=\sum_{P\in S_k}{\rm sgn}(P)dx^{\mu_
{P(1)}}\otimes\cdots\otimes dx^{\mu_{P(k)}}
\label{wedgeprod}\eeq
where the sum is over all permutations $P$ of $1,\dots,k$ and ${\rm sgn}(P)$
is the sign of $P$, defined as $(-1)^{t(P)}$ where $t(P)$ is the number of
transpositions in $P$. For example, for 2 cotangent basis vector elements
\beq
dx\wedge dy=dx\otimes dy-dy\otimes dx
\label{2wedge}\eeq
The space of all linear
combinations of the basis elements (\ref{wedgeprod}),
\beq
\alpha=\frac{1}{k!}\alpha_{\mu_1\cdots\mu_k}(x)dx^{\mu_1}\wedge\cdots\wedge dx
^{\mu_k}
\label{formdef}\eeq
is the antisymmetrization ${\cal A}(T_x^*\man)^{\otimes k}$ of the $k$-th
tensor power of the cotangent bundle. The disjoint union, over all $x\in\man$,
of these vector spaces is called the $k$-th exterior power $\Lambda^k\man$ of
$\man$. Its elements (\ref{formdef}) are called differential $k$-forms
whose components are $C^\infty$ functions on $\man$ which are completely
antisymmetric in their indices $\mu_1,\dots,\mu_k$. Notice
that by the antisymmetry of the exterior product, if $\man$ is $n$-dimensional,
then $\Lambda^k\man=0$ for all $k>n$. Furthermore, $\Lambda^0\man=C^\infty
(\man)$, the space of smooth functions on
$\man\to\IR$, and $\Lambda^1\man=T^*\man$ is the cotangent bundle of $\man$.

The exterior product of a $p$-form $\alpha$ and a $q$-form $\beta$ is the
$(p+q)$-form $\alpha\wedge\beta=\frac{1}{(p+q)!}(\alpha\wedge\beta)_{\mu_1
\cdots\mu_{p+q}}(x)dx^{\mu_1}\wedge\cdots\wedge dx^{\mu_{p+q}}$ with local
components
\beq
(\alpha\wedge\beta)_{\mu_1\cdots\mu_{p+q}}(x)=\sum_{P\in S_{p+q}}{\rm sgn}(P)
\alpha_{\mu_{P(1)}\cdots\mu_{P(p)}}(x)\beta_{\mu_{P(p+1)}\cdots\mu_{P(p+q)}}
(x)
\eeq
The exterior product of differential forms makes the direct sum of the exterior
powers
\beq
\Lambda\man=\bigoplus_{k=0}^n\Lambda^k\man
\label{extalg}\eeq
into a graded-commutative algebra called the exterior algebra of $\man$. In
$\Lambda\man$, the exterior product of a $p$-form $\alpha$ and
a $q$-form $\beta$ obeys the graded-commutativity property
\beq
\alpha\wedge\beta=(-1)^{pq}\beta\wedge\alpha~~~~~,~~~~~\alpha\in\Lambda^p\man,
\beta\in\Lambda^q\man
\label{noncomm}\eeq

On the exterior algebra (\ref{extalg}), we define a linear operator
\beq
d:\Lambda^k\man\to\Lambda^{k+1}\man
\label{d}\eeq
on $k$-forms (\ref{formdef}) by
\beq
(d\alpha)_{\mu_1\cdots\mu_{k+1}}(x)=\sum_{P\in S_{k+1}}{\rm sgn}(P)\partial
_{\mu_{P(1)}}\alpha_{\mu_{P(2)}\cdots\mu_{P(k+1)}}(x)
\label{ddef}\eeq
and $d\alpha=\frac{1}{(k+1)!}(d\alpha)_{\mu_1\cdots\mu_{k+1}}(x)dx^{\mu_1}
\wedge\cdots\wedge dx^{\mu_{k+1}}$. The operator $d$ is called the exterior
derivative and it generalizes the notion of the differential of a function
\beq
df=\frac{\partial f(x)}{\partial x^\mu}dx^\mu~~~~~,~~~~~f\in\Lambda^0\man=
C^\infty(\man)
\label{df}\eeq
to generic differential forms. It is a graded derivation, i.e. it satisfies the
graded Leibniz property
\beq
d(\alpha\wedge\beta)=d\alpha\wedge\beta+(-1)^{p}\alpha\wedge d\beta~~~,~~~
\alpha\in\Lambda^p\man,\beta\in\Lambda^q\man
\label{dprod}\eeq
and it is nilpotent,
\beq
d^2=0
\label{d20}\eeq
which follows from the commutativity of multiple partial derivatives of
$C^\infty$ functions. Thus the exterior derivative allows one to generalize
the common notion of vector calculus to more general spaces other than
$\IR^n$. The collection of vector spaces $\{\Lambda^k\man\}_{k=0}
^n$ and nilpotent derivations $d$ form what is called the DeRham
complex $\Lambda^*(\man)$ of the manifold $\man$.

There are 2 important subspaces of the exterior algebra (\ref{extalg}) as far
as the map $d$ is concerned. One is the kernel of $d$,
\beq
{\rm ker}~d=\{\alpha\in\Lambda\man:d\alpha=0\}
\label{kerd}\eeq
whose elements are called closed forms, and the other is the image of $d$,
\beq
{\rm im}~d=\{\beta\in\Lambda\man:\beta=d\alpha~~{\rm for}~~{\rm some}~~\alpha
\in\Lambda\man\}
\label{imd}\eeq
whose elements are called exact forms. Since $d$ is nilpotent, we have
${\rm im}~d\subset{\rm ker}~d$. Thus we can consider the quotient of the kernel
of $d$ by its image. The vector space of closed $k$-forms modulo exact
$k$-forms is called the $k$-th DeRham cohomology group (or vector space) of
$\man$,
\beq
H^k(\man;\IR)={\rm ker}~d|_{\Lambda^k\man}/{\rm im}~d|_{\Lambda^{k-1}\man}
\label{derhamcoh}\eeq
The elements of the vector space (\ref{derhamcoh}) are the equivalence classes
of differential $k$-forms where 2 differential forms are equivalent if and only
if they differ only by an exact form, i.e. if the closed form $\alpha\in\Lambda
^k\man$ is a representative of the cohomology class $[\alpha]\in H^k(\man;
\IR)$, then so is the closed form $\alpha+d\beta$ for any differential form
$\beta\in\Lambda^{k-1}\man$.

One important theorem in the context of DeRham
cohomology is Poincar\'e's lemma. This states that if $d\omega=0$ in a
star-shaped region $S$ of the manifold $\man$ (i.e. one in which the affine
line segment joining any 2 points in $S$ lies in $S$), then one can write
$\omega=d\theta$ in that region for some other differential form $\theta$. Thus
each representative of a DeRham cohomology class can be {\it locally} written
as an exact form, but globally there may be an obstruction to extending the
form $\theta$ over the entire manifold in a smooth way depending on whether or
not $[\omega]\neq0$ in the DeRham cohomology group.

The DeRham cohomology groups are related to the topology of the manifold $\man$
as follows. Consider the following $q$-dimensional subspace of $\IR^{q+1}$,
\beq
\Delta^q=\left\{(x_0,x_1,\dots,x_q)\in\IR^{q+1}:x_i\geq0,\sum_{i=0}^qx_i=1
\right\}
\label{qsimplex}\eeq
which is called the standard $q$-simplex. Geometrically, $\Delta^q$ is the
convex hull generated by the vertices placed at unit distance along the axes
of $\IR^{q+1}$. We define the geometric boundary of the standard $q$-simplex as
\beq
\partial\Delta^q=\sum_{i=0}^q(-1)^i\hat\Delta_{(i)}^q
\label{boundarysimpl}\eeq
where $\hat\Delta_{(i)}^q$ is the $(q-1)$-simplex generated by all the
vertices of $\Delta^q$ except the $i$-th one, and the sum on the right-hand
side is the formal algebraic sum of simplices (where a minus sign signifies a
change of orientation). A singular $q$-simplex of the manifold $\man$ is
defined to be a continuous map $\sigma:\Delta^q\to\man$. A formal algebraic
sum of $q$-simplices with integer coefficients is called a $q$-chain, and the
collection of all $q$-chains in a manifold $\man$ is called the $q$-th chain
group $C_q(\man)$ of $\man$. It defines an abelian group under the formal
addition. The boundary of a $q$-chain is the $(q-1)$-chain
\beq
\partial\sigma=\sum_{i=0}^q(-1)^i\sigma\Bigm|_{\hat\Delta_{(i)}^q}
\label{boundarymap}\eeq
which is easily verified to give a nilpotent homomorphism
\beq
\partial:C_q(\man)\to C_{q-1}(\man)
\label{boundaryhom}\eeq
of abelian groups. The collection of abelian groups $\{C_q(\man)\}_{q\in\IZ^+}$
and nilpotent homomorphisms $\partial$ form the singular chain complex
$C_*(\man)$ of the manifold $\man$.

Nilpotency of the boundary map (\ref{boundaryhom}) means that every $q$-chain
in the image of $\partial|_{C_{q+1}}$, the elements of which are called the
$q$-boundaries of $\man$, lies as well in the kernel of $\partial|_{C_q}$,
whose elements are called the $q$-cycles of $\man$. The abelian group defined
as the quotient of the group of all $q$-cycles modulo the group of all
$q$-boundaries is called the $q$-th (singular) homology group of $\man$,
\beq
H_q(\man;\IZ)={\rm ker}~\partial|_{C_q}/{\rm im}~\partial|_{C_{q+1}}
\label{homgroup}\eeq
These groups are homotopy invariants of the manifold $\man$ (i.e. invariant
under continuous deformations of the space), and in particular they are
topological invariants and diffeomorphism invariants (i.e. invariant under
$C^\infty$ invertible bi-continuous mappings of $\man$). As such, they are
invariant under local deformations of the space and depend only on the global
characteristics of $\man$. Intuitively,
they measure whether or not a manifold has `holes' in it or not. If
$H_q(\man;\IZ)=0$, then every $q$-cycle (intuitively a closed $q$-dimensional
curve or surface) encloses a $q+1$-dimensional chain and $\man$ has no
`$q$-holes'. For instance, if $\man$ is simply-connected (i.e. every loop in
$\man$ can be contracted to a point) then $H_1(\man;\IZ)=0$. A star-shaped
region, such as a simplex, is simply-connected.

Given the abelian groups (\ref{homgroup}), we can form their duals using
the universal coefficient theorem
\beq
H^q(\man;\IZ)\simeq~{\rm Hom}_\IZ\left(H_q(\man;\IZ),\IZ\right)\oplus~{\rm
Ext}_\IZ(H_{q-1}(\man;\IZ),\IZ)
\label{cohgroup}\eeq
which is called the $q$-th singular cohomology group of $\man$ with integer
coefficients. Here ${\rm Hom}_\IZ(H_q(\man;\IZ),\IZ)=H_q(\man;\IZ)^*$ is the
free part of the cohomology group, and ${\rm Ext}_\IZ$ is the torsion
subgroup of $H^q(\man;\IZ)$. The DeRham theorem then states that the DeRham
cohomology groups are naturally isomorphic to the singular cohomology groups
with real coefficients,
\beq
H^q(\man;\IR)=H^q(\man;\IZ)\otimes\IR=H_q(\man;\IR)^*
\label{derhamsing}\eeq
where the tensor product with the reals means that $H^q$ is
considered as an abelian group with real instead of integer coefficients, i.e.
a vector space over $\IR$ (this eliminates the torsion subgroup in
(\ref{cohgroup})).

The crux of the proof of DeRham's theorem is Stokes' theorem,
\beq
\int_cd\omega=\oint_{\partial c}\omega~~~~~,~~~~~\omega\in\Lambda^q\man~~~,~~~
c\in C_{q+1}(\man)
\label{stokecycle}\eeq
which relates the integral of an exact $(q+1)$-form over a smooth $(q+1)$-chain
$c$ in $\man$ to an integral over the closed $q$-dimensional boundary $\partial
c$ of $c$. The map $(\omega,c)\to\int_c\omega$ on $H^q(\man;\IR)\otimes
H_q(\man;\IR)\to\IR$ defines a natural duality pairing between $H^q(\man;\IR)$
and $H_q(\man;\IR)$ and is the basis of the DeRham isomorphism. In particular,
(\ref{stokecycle}) generalizes to the global version
of Stokes' theorem,
\beq
\int_\man d\omega=\oint_{\partial\man}\omega~~~~~,~~~~~\omega\in\Lambda^{n-1}
\man
\label{stokes}\eeq
which relates the integral of an exact form over $\man$ to an integral over
the closed $(n-1)$-dimensional boundary $\partial\man$ of $\man$. Here
integration over a manifold is defined by partitioning the manifold up into
open sets homeomorphic to $\IR^n$, integrating a top form (i.e. a differential
form of highest degree $n$ on $\man$) locally over $\IR^n$ as
usual\footnote{\baselineskip=12pt The integral over $\man$ of a $p$-form with
$p<\dim\man$ is always understood here to be zero.}, and then summing up all of
these contributions. In this way, we see how the DeRham cohomology of a
manifold measures its topological (or global) features in an analytic way
suited for the differential calculus of $C^\infty$ manifolds. We refer to
\cite{massey} and \cite{bottu} for a more complete and leisurely introduction
to this subject.

\subsection{The Cartan Model of Equivariant Cohomology}

We shall now generalize the constructions of the last Subsection to the case
where there is a Lie group (i.e. a continuous group with a smooth structure
whose group multiplication is also smooth) acting on the space. Then the
construction of topological invariants for these spaces (i.e. structures that
are the same for homeomorphic spaces) will be the foundation for the derivation
of general integration formulas in the subsequent Sections.

Many situations in theoretical physics involve not only a differentiable
manifold $\man$, but also the action of some Lie group $G$ acting on $\man$,
which we denote symbolically by
\beq\new{\begin{array}{c}
G\times\man\to\man\\(g,x)\to g\cdot x\end{array}}
\label{gaction}\eeq
By a group action we mean that $g\cdot x=x,\forall x\in\man$ if $g$ is the
identity element of $G$, and the group action represents the multiplication
law of the group, i.e. $g_1\cdot(g_2\cdot x)=(g_1g_2)\cdot x,\forall g_1,g_2
\in G$. We shall throughout assume that $G$ is connected and that its
action on $\man$ is smooth, i.e. for fixed $g\in G$, the function $x\to g\cdot
x$ is a diffeomorphism of $\man$. Usually $G$ is taken to be the symmetry group
of the given physical problem. The common (infinite-dimensional) example in
topological field theory is where $\man$ is the space of gauge connections of a
gauge theory and $G$ is the group of gauge transformations. The space $\man$
modulo this group action is then the moduli space of gauge orbits. Another
example is in string theory where $\man$ is the space of metrics on a Riemann
surface (a connected orientable 2-manifold) and $G$ is the semi-direct product
of the Weyl and diffeomorphism groups of that 2-surface. Then $\man$ modulo
this group action is the moduli space of the Riemann surface. In such instances
we are interested in knowing the cohomology of the manifold $\man$ given this
action of the group $G$. This cohomology is known as the $G$-equivariant
cohomology of $\man$. Given the $G$-action on $\man$, the space of orbits
$\man/G$ is the set of equivalence classes where $x$ and $x'$ are equivalent if
and only if $x'=g\cdot x$ for some $g\in G$ (the topology of $\man/G$ is the
induced topology from $\man$). If the $G$-action on $\man$ is free, i.e.
$g\cdot x=x$ if and only if $g$ is the identity element of $G$, $\forall
x\in\man$, then the space of orbits $\man/G$ is also a differentiable manifold
of dimension $\dim\man-\dim G$ and the $G$-equivariant
cohomology is defined simply as the cohomology of the coset space $\man/G$,
\beq
H^k_G(\man)=H^k(\man/G)
\label{freecoh}\eeq

However, if the group action is not free and has fixed points on $\man$, the
space $\man/G$ can become singular. The dimension of the orbit $G\cdot
x=\{g\cdot x:g\in G\}$ of a point $x\in\man$ is $\dim G-\dim G_x$, where
$G_x=\{g\in G:g\cdot x=x\}$ is the isotropy subgroup of $x$. Consequently, in a
neighbourhood of a fixed point $x$, the dimension $\dim\man-\dim G\cdot x$ of
$\man/G$ can be larger than the dimension $\dim\man-\dim G$ of other
fixed-point free coordinate neighbourhoods (because then the isotropy subgroup
$G_x$ of that fixed point $x$ is non-trivial), and there is no smooth notion of
dimensionality for the coset
$\man/G$. A singular quotient space $\man/G$ is called an orbifold. In such
instances, one cannot define the equivariant cohomology of $\man$ in a smooth
way using (\ref{freecoh}) and more elaborate methods are needed to
define this cohomology. This is the ``right" cohomology theory that properly
accounts for the group action and it is always defined in a manner such that if
the group action is trivial, the cohomology reduces to the usual cohomological
ideas of the classical DeRham theory.

There are many approaches to defining the equivariant
cohomology of $\man$, but there is only one that will be used extensively in
this Review. This is the Cartan model of equivariant cohomology and it
is defined in a manner similar to the analytic DeRham cohomology which was
reviewed in the last Subsection. However, the other models of equivariant
cohomology are equally as important -- the Weil algebra formulation relates
the algebraic models to the topological definition of equivariant
cohomology using universal bundles of Lie groups
\cite{atiyahbott,cartan,kalkman1,mathai}, while the BRST model relates
the Cartan and Weil models and moreover is the basis for the superspace
formulation of topological Yang-Mills theory in 4-dimensions and other
cohomological field theories \cite{cordes,kalkman1,niemitirk3}. These other
models are outlined in Appendix B.

We begin by generalizing the notion of a differential form to the case where
there is a group action on $\man$ as above. We say that a map $f:\man_1\to
\man_2$ between 2 manifolds with $G$-actions on them is equivariant with
respect to the group action if
\beq
f(g\cdot x)=g\cdot f(x)~~~~~\forall x\in\man_1~~,~~\forall g\in G
\label{eqf}\eeq
We want to extend this notion of equivariance to differential forms. Consider
the symmetric polynomial functions from the Lie algebra $\bf g$ of
$G\equiv\exp({\bf g})$ into the exterior algebra $\Lambda\man$ of the manifold
$\man$. These maps
form the algebra $S({\bf g}^*)\otimes\Lambda\man$, where $S({\bf g}^*)$ is
called the symmetric algebra over the dual vector space ${\bf g}^*$ of $\bf g$
and it corresponds to the algebra of polynomial functions on $\bf g$. The
action of $g\in G$ on an element $\alpha\in S({\bf g}^*)\otimes\Lambda\man$ is
given by
\beq
(g\cdot\alpha)(X)=g\cdot\left(\alpha(g^{-1}Xg)\right)
\label{gacforms}\eeq
where $X\in{\bf g}$. Here we have used the natural coadjoint action of $G$ on
${\bf g}^*$ and the induced $G$-action on $\Lambda\man$ from that on $\man$
as dictated by the tensor transformation law (\ref{tenstransf}) with
$x'(x)=g\cdot x$. From this it follows immediately that the equivariance
condition (\ref{eqf}) is satisfied for the polynomial maps $\alpha:{\bf g}\to
\Lambda\man$ in the $G$-invariant subalgebra
\beq
\Lambda_G\man=\left(S({\bf g}^*)\otimes\Lambda\man\right)^G
\label{eqdiff}\eeq
where the superscript $G$ denotes the (infinitesimal) $G$-invariant part. The
elements of (\ref{eqdiff}) are called equivariant differential forms
\cite{berline1,berlinegetz}.

Elements of $G$ are represented in terms of elements of the Lie algebra $\bf g$
through the exponential map,
\beq
g=\e^{c^aX^a}
\label{glie}\eeq
where $c^a$ are constants and $X^a$ are the generators of $\bf g$
obeying the Lie bracket algebra
\beq
\left[X^a,X^b\right]=f^{abc}X^c
\label{liealg}\eeq
with $f^{abc}$ the antisymmetric structure constants of $\bf g$. Here and in
the following we shall assume an implicit sum over the Lie algebraic indices
$a,b,c,\dots$. The space where the $c^a$'s in (\ref{glie}) lie defines the
group manifold of $G$. The generators $X^a$ can be written as
$X^a=\frac{\partial}{\partial c^a}g|_{c=0}$ and so the Lie algebra $\bf g$ can
be regarded as the tangent space to the identity on the group
manifold of the Lie group $G$. The strucutre constants in (\ref{liealg}) define
a natural representation of $G$ of dimension $\dim G$, called the adjoint
representation, whose (Hermitian) generators have matrix elements $({\rm
ad}~X^a)_{bc}\equiv if^{abc}$.

The smooth $G$-action on $\man$ can be represented locally as the continuous
flow
\beq
g_t\cdot x=x(t)~~~~~,~~~~~t\in\IR^+
\label{flow}\eeq
where $g_t$ is a path in $G$ starting at the identity $g_{t=0}$. The induced
action on differential forms is defined by pullback, i.e. as
\beq
(g_t\cdot\alpha)(x)=\alpha(x(t))
\label{formflow}\eeq
For example, we can represent the group action on $C^\infty$ functions by
diffeomorphisms on $\man$ which are connected to the identity, i.e.
\beq
(g_t\cdot f)(x)=f(x(t))=\e^{tV(x(t))}f(x)~~~,~~~f\in\Lambda^0\man
\label{infaction}\eeq
The action (\ref{infaction}) represents the flow of the group
on $C^\infty$ functions on $\man$, where $V(x)=V^\mu(x)\frac{\partial}{\partial
x^\mu}$ is a vector field on $\man$ representing a Lie algebra element. It is
related to the flows (\ref{flow}) on the manifold by
\beq
\dot x^\mu(t)=V^\mu(x(t))
\label{intcurves}\eeq
which defines a set of curves in $\man$ which we will refer to as the integral
curves of the group action. If $V^a$ is the vector field representing the
generator $X^a$ of $\bf g$, then the Lie algebra (\ref{liealg}) is represented
on $C^\infty$ functions by
\beq
\left[V^a,V^b\right](h)=f^{abc}V^c(h)~~~~~,~~~~~\forall h\in\Lambda^0\man
\label{vliealg}\eeq
with Lie bracket represented by the ordinary commutator bracket. This defines a
representation of $G$ by vector fields in the tangent bundle $T\man$. In this
setting, the group $G$ is represented as a subgroup of the
(infinite-dimensional) connected diffeomorphism group of $\man$ whose Lie
algebra is generated by all vector fields of $\man$ with the commutator
bracket.

The infinitesimal ($t\to0$) action of the group on
$\Lambda^0\man$ can be expressed as
\beq
V(f)=i_Vdf
\label{vaction}\eeq
where
\beq
i_V:\Lambda^k\man\to\Lambda^{k-1}\man
\label{iv}\eeq
is the nilpotent contraction operator, or interior multiplication, with respect
to $V$ and it is defined locally on $k$-forms (\ref{formdef}) by
\beq
i_V\alpha=\frac{1}{(k-1)!}V^{\mu_1}(x)\alpha_{\mu_1\mu_2\cdots\mu_k}(x)
dx^{\mu_2}\wedge\cdots\wedge dx^{\mu_k}
\label{ivdef}\eeq
The operator $i_V$ is a graded derivation (c.f. (\ref{dprod})) and the quantity
$i_VT$ represents the component of a tensor $T$ along the vector field $V$. The
infinitesimal $G$-action on the higher-degree differential forms is
generated by the Lie derivative along $V$
\beq
\lie_V:\Lambda^k\man\to\Lambda^k\man
\label{lie}\eeq
where
\beq
\lie_V=di_V+i_Vd
\label{liederiv}\eeq
generates the induced action of $G$ on $\Lambda\man$, i.e.
\beq
\lie_V\alpha(x(0))=\frac{d}{dt}\alpha(x(t))\biggm|_{t=0}
\eeq
This can be verifed by direct computation from expanding (\ref{formflow}) about
$t=0$ using (\ref{tenstransf}) and (\ref{intcurves}), and by noting that
\beq
\left[\lie_{V^a},\lie_{V^b}\right](\alpha)=f^{abc}\lie_{V^c}(\alpha)~~~,~~~
\forall\alpha\in\Lambda\man
\label{liederivalg}\eeq
Thus the Lie derivative in general defines a representation of $G$ on $\Lambda
\man$. The local components of $\lie_VT$ for a general $(k,\ell)$ tensor field
$T$ are found by substituting into the tensor transformation law
(\ref{tenstransf}) the infinitesimal coordinate change $x'^\mu(x)=x^\mu(t)=
x^\mu+tV^\mu(x)$. For example, on a vector field
$W=W^\mu(x)\frac{\partial}{\partial x^\mu}$ we have
\beq
(\lie_VW)^\mu=W^\nu\partial_\nu V^\mu-V^\nu\partial_\nu W^\mu\equiv[W,V]^\mu
\label{lieVW}\eeq
Furthermore, the Lie derivative $\lie_V$ is an ungraded derivation and its
action on contractions is
\beq
\left[i_{V^a},\lie_{V^b}\right](\alpha)=f^{abc}i_{V^c}(\alpha)
\label{liecontr}\eeq

We are now ready to define the Cartan model for the $G$-equivariant cohomology
of $\man$ \cite{berlinegetz,cartan,mathai}. We assign a
$\IZ$-grading\footnote{\baselineskip=12pt A $\IZ$-grading is usually refered to
as a `ghost number' in the physics literature. The equivalence between the 2
notions will become clearer when we deal with path integrals in Section 4 --
see Appendices A and B for this algebraic correspondence.} to the
elements of (\ref{eqdiff}) by defining the degree of an equivariant
differential form to be the sum of its ordinary form degree and twice the
polynomial degree from the $S({\bf g}^*)$ part. Let $\{\phi^a\}_{a=1}^{\dim G}$
be a basis of ${\bf g}^*$ dual to the basis $\{X^a\}_{a=1}^{\dim G}$ of $\bf
g$, so that
\beq
\phi^a(X^b)=\delta^{ab}
\eeq
With the above grading, the basis elements $\phi^a$ have degree 2. We
define a linear map
\beq
D_{\bf g}:\Lambda_G^k\man\to\Lambda_G^{k+1}\man
\label{eqderiv}\eeq
on the algebra (\ref{eqdiff}) by
\beq
D_{\bf g}\phi^a=0~~~~~,~~~~~D_{\bf g}\alpha=({\bf 1}\otimes d-\phi^a\otimes
i_{V^a})\alpha~~~;~~~\alpha\in\Lambda\man
\label{donforms}\eeq
The operator $D_{\bf g}$ is called the equivariant exterior derivative and it
is a graded derivation. Its definition (\ref{donforms}) means that its action
on
forms $\alpha\in S({\bf g}^*)\otimes\Lambda\man$ is
\beq
(D_{\bf g}\alpha)(X)=(d-i_V)(\alpha(X))
\label{doneq}\eeq
where $V=c^aV^a$ is the vector field on $\man$ representing the Lie algebra
element $X=c^aX^a\in{\bf g}$. However, unlike the operators $d$ and $i_V$,
$D_{\bf g}$ is
not nilpotent in general, but its square is given by the Cartan-Weil identity
\beq
D_{\bf g}^2=-\phi^a\otimes(di_{V^a}+i_{V^a}d)=-\phi^a\otimes\lie_{V^a}
\label{weilid}\eeq
Thus the operator $D_{\bf g}$ is nilpotent on the algebra $\Lambda_G
\man$ of equivariant differential forms. The set of $G$-invariant algebras
$\{\Lambda_G^k\man\}_{k\in\IZ^+}$ and nilpotent derivations $D_{\bf g}$ thereon
defines the $G$-equivariant complex $\Lambda_G^*(\man)$ of the manifold $\man$.

Thus, just as in the last Subsection, we can proceed to define the cohomology
of
the operator $D_{\bf g}$. The space of equivariantly closed forms, i.e.
$D_{\bf g}\alpha=0$, modulo the space of equivariantly exact forms, i.e.
$\alpha
=D_{\bf g}\beta$, is called the $G$-equivariant cohomology group of $\man$,
\beq
H_G^k(\man)={\rm ker}~D_{\bf g}|_{\Lambda^k_G\man}/{\rm im}~D_{\bf g}|_{
\Lambda^{k-1}_G\man}
\label{eqcoh}\eeq
With this definition, the cohomology of the operator $D_{\bf g}$ for a
fixed-point free $G$-action on $\man$ reduces
to the DeRham cohomology of the quotient space $\man/G$, as in (\ref{freecoh}).
The definition (\ref{eqcoh}) of equivariant cohomology is known as the Cartan
model \cite{berlinegetz,cartan,mathai}. Note that the definition of $D_{\bf g}$
in (\ref{doneq}) resembles a gauge-covariant derivative.

We close this Subsection with a few remarks concerning the above
construction. First of all, it follows from these definitions that $H_G^k
(\man)$ coincides with the ordinary DeRham cohomology of $\man$ if $G$ is the
trivial group consisting of only the identity element (i.e. $V\equiv0$ in the
above), and that the $G$-equivariant cohomology of a point is the algebra of
$G$-invariant polynomials on $\bf g$, $H_G({\rm pt})=S({\bf g}^*)^G$, of the
given degree. Secondly, if a form $\alpha\in\Lambda_G\man$ is equivariantly
exact, $\alpha=D_{\bf g}\beta$, then its top-form component $\alpha^{(n)}\in
\Lambda^n\man$ is exact in the ordinary DeRham sense. This follows because the
$i_V$ part of $D_{\bf g}$ lowers the form-degree by 1 so there is no way to
produce a top-form by acting with $i_V$. Finally, in what follows we shall have
occasion to also consider the $C^\infty$ extension $\Lambda_G^\infty\man$ of
$\Lambda_G\man$ to include arbitrary $G$-invariant smooth functions from $\bf
g$ to $\Lambda\man$. In this extension we lose the $\IZ$-grading described
above, but we are left with a $\IZ_2$-grading corresponding to the differential
form being of even or odd degree \cite{berlinegetz} ($\IZ_2=\IZ/2\IZ$ is the
cyclic group of order 2).

\subsection{Fiber Bundles and Equivariant Characteristic Classes}

The `bundles' we introduced in Subsection 2.2 (tangent, cotangent, etc.) are
all examples of a more general geometric entity known as a fiber bundle. The
geometry and topology of fiber bundles will play an important role in the
development of equivariant localization theory, and in this Subsection we
briefly review the essential features that we shall need (see
\cite{birm,eguchi,cordes} for more detailed discussions). A fiber bundle
consists of a quadruple $(E,\man,F,\pi)$, where $E$ is a
topological space called the total space of the fiber bundle, $\man$ is a
topological space called the base space of the fiber bundle (usually we take
$\man$ to be a manifold), $F$ is a topological space called the fiber, and
$\pi:E\to\man$ is a surjective continuous map with $\pi^{-1}(x)=F$, $\forall x
\in\man$, which is called the projection of the fiber bundle. A fiber bundle is
also defined so that locally it is trivial, i.e. locally the bundle is a
product $U\times F$ of an open neighbourhood $U\subset\man$ of the base and the
fibers, and $\pi:U\times F\to U$ is the projection onto the first
coordinate\footnote{\baselineskip=12pt The topology
on the total spaces $E=\bigsqcup_{x\in\man}\pi^{-1}(x)$ of fiber bundles is
usually taken as the induced topology from the erection of points from
$\man$.}.
In the case of the tangent bundle, for instance, the fibers are
$F=T_x\IR^n\simeq\IR^n$ and the projection map is defined on $T\man\to\man$ by
$\pi:T_x\man\to x$. In fact, in this case the fibration spaces are vector
spaces, so that the tangent bundle is an example of a vector bundle. If the
fiber of a bundle is a Lie group $G$,
then the fiber bundle is called a principal fiber bundle with structure group
$G$. It has a right, smooth and free action of $G$ on the total space $E$ and
the base $\man$ which gives a local representation of the group in the fibers.
This action also embeds a copy of the group $G$ inside $E$.

The vector and tensor fields introduced in Subsection 2.2 should more precisely
be defined as `sections' of the associated bundles, i.e. smooth maps $s:\man\to
E$ which take a point $x\in\man$ into the fiber $\pi^{-1}(x)$ over $x$.
Although
we shall be a bit abusive in our discussion by considering these as genuine
functions on $\man$, for simplicity and ease of notation, it should be kept
in mind that it is only locally where these objects admit such a functional
interpretation. Thus, for instance, the tangent bundle is
$T\man=\{(x,V):x\in\man,V\in T_x\man\}$ and locally a section of $T\man$ can be
written as $(x,V^\mu(x)\frac{\partial}{\partial x^\mu})$.

The set of frames (i.e. bases) on the tangent bundle $T\man$ form a principal
$GL(n,\IR)$-bundle over $\man$, called the frame bundle, whose points are
$(x;(e_1,\dots,e_n))$ where $x\in\man$ and $(e_1,\dots,e_n)$ is a linear basis
for $T_x\man$. If $\man$ has a metric (i.e. a globally defined inner product on
each tangent space $T_x\man$), then we can restrict the basis to an orthonormal
basis and obtain a principal $O(n-N,N)$-bundle, where $(n-N,N)$ is the
signature of the metric. If $\man$ is furthermore orientable, then we can
further restrict to an oriented orthonormal basis, defined by the equivalence
classes with respect to the equivalence relation $e\equiv M\cdot f$ where $\det
M>0$, and get a principal $SO(n-N,N)$-bundle. When $\man$ is a space-time
manifold, the Lie group $SO(n-N,N)$ is then referred to as the local Lorentz
group of $\man$ (or of the tangent bundle $T\man$). The associated spin group
${\rm spin}(n-N,N)$ is defined as a double cover of the local Lorentz group,
i.e. $SO(n-N,N)\simeq~{\rm spin}(n-N,N)/\IZ_2$ (for instance, ${\rm
spin}(2)=U(1)$ and ${\rm spin}(3)=SU(2)$). A principal ${\rm
spin}(n-N,N)$-bundle over $\man$ whose fibers form a double cover of those of
the oriented orthonormal frame bundle is called a spin bundle and is said to
define a spin structure on the manifold $\man$.

Conversely, to any principal $G$-bundle $P\to\man$ there is an associated
vector bundle. Let $W$ be the representation space for a representation $\rho$
of $G$. Since $G$ acts smoothly and freely on the right of $P$, locally on
$P\times W$ there is the $G$-action $(p,v)\to(p\cdot g^{-1},\rho(g)\cdot v)$
where $g\in G$. This defines the associated vector bundle $(P\times W)/G\to
P/G$ for the representation $\rho$, which has fiber the vector space $W$. For
instance, for the trivial representation $\rho$, $(P\times W)/G=P/G\times W$.
In this way, we can naturally identify sections (e.g. differential forms) on
$(P\times W)/G$ with equivariant functions $f:P\to W$, i.e. $f(p\cdot
g)=\rho(g^{-1})\cdot f(p)$. Notice that for a vector bundle $E\to\man$, the
bundle of differential forms on $\man$ with values in $E$ is defined as
\beq
\Lambda^k(\man,E)\equiv\Lambda^k\man\otimes E
\eeq
where tensor products and direct (Whitney) sums of bundles are defined locally
by the corresponding algebraic combination of their fiber spaces.

Intuitively, a fiber bundle `pins' some geometrical or topological object over
each point of a manifold $\man$ (e.g. a vector space in the case of a vector
bundle). For instance, if $\man=\IR^n$, then the tangent bundle $T\IR^n$
associates the vector space $W=\IR^n$ to each point of $\IR^n$. In fact in this
case, the tangent bundle is globally given by $T\IR^n=\IR^n\times W$, the
product of its base and fibers. We then say that the bundle is trivial, in that
the erecting of points into vector spaces is done without any `twistings'
of the fibers. However, a general vector bundle is only {\it locally} trivial
and globally the fibers can twist in a very complicated fashion. One way
to characterize the non-triviality of fiber bundles is through
special cohomology classes of the base manifold $\man$ called characteristic
classes \cite{milnor}. A non-trivial characteristic class in this sense
signifies the non-triviality of the vector bundle.

As we shall see, all of these notions can be generalized to the case of the
equivariant cohomology of a manifold which signifies the non-triviality of an
equivariant bundle. First, we define what we mean by an equivariant bundle
\cite{berline1,berline2}. We say that a fiber bundle $E~{\buildrel\pi\over
\longrightarrow}~\man$ is a $G$-equivariant bundle if there are $G$-actions on
both $E$ and $\man$ which are compatible with each other in the sense that
\beq
g\cdot\pi(x)=\pi(g\cdot x)~~~~~\forall x\in E~~,~~\forall g\in G
\label{compatible}\eeq
This means that the bundle projection $\pi$ is a $G$-equivariant map.
The action of the group $G$ on differential forms with values in the bundle is
generated by the Lie derivatives $\lie_{V^a}$.

In the ordinary DeRham case, when there are `twists' in the given bundle one
needs to specify how to `connect' different fibers. This is done using a
connection $\Gamma$ which is a geometrical object (such as a 1-form) defined
over $\man$ with values in $E$ whose action on sections of the bundle specifies
their parallel transport along fibers, as required. The parallel
transport is generated by the covariant derivative associated with $\Gamma$,
\beq
\nabla=d+\Gamma
\label{covderiv}\eeq
The derivative operator (\ref{covderiv}) is a linear derivation which
associates to each section of the given vector bundle a 1-form in
$\Lambda^1(\man,E)$. If $V$ is a tangent vector on $\man$, its action on a
section $s$ is defined in local coordinates by
\beq
(\nabla s)^\alpha(V)=V^\mu(\partial_\mu s^\alpha+(\Gamma_\mu)^\alpha_\beta
s^\beta)
\eeq
where $(\Gamma_\mu)^\alpha_\beta dx^\mu$ is a 1-form on $\man$ with values in
${\rm End}(E)$ ($\alpha,\beta$ are the vector space indices in $E$). If $x(t)$
is a path in $\man$, then $\dot x(t)$ is a tangent vector along the path and
the equation
\beq
(\nabla s)(\dot x(t))=0
\label{partranspeq}\eeq
determines parallel transport along the path allowing us to connect different
fibers of the bundle. The first order differential equation (\ref{partranspeq})
can admit topologically non-trivial solutions if either the space $\man$ is
multiply-connected ($H_1(\man;\IZ)\neq0$), or if the connection $\Gamma$ has
non-trivial curvature $F\neq0$ (see below). The latter condition characterizes
the non-triviality of the bundle, so that $F=0$ on a trivial bundle and the
solutions to (\ref{partranspeq}) are straight lines. At each point $p$ of the
total space $E$, there is a natural vertical tangent space $V_p$ in the tangent
space $T_pE$ along the fiber of $E$. A choice of connection above is just
equivalent to a choice of horizontal component $H_p$ in the tangent space so
that $T_pE=V_p\oplus H_p$.

When the bundle $P\to\man$ is a principal $G$-bundle, we require further that
this splitting into horizontal and vertical components be $G$-equivariant (i.e.
$H_p\to H_{p\cdot g}$ under the action of $g\in G$). In this case,
$V_p\simeq{\bf g}$ so that the connection $\Gamma$ is a globally-defined 1-form
with values in the Lie algebra $\bf g$, i.e. $\Gamma\in\Lambda^1(P,{\bf g})$.
The horizontal subspace can then be taken to be $H_p={\rm ker}~\Gamma$.
Horizontality and $G$-equivariance mean, respectively, that $\Gamma$ satisfies
\beq
i_V\Gamma=X~~~~~,~~~~~\lie_V\Gamma=-~{\rm ad}(X)\Gamma\equiv-[X,\Gamma]
\label{conncondns}\eeq
where $V$ is the vector field on $P$ representing $X\in{\bf g}$ and ${\rm
ad}(X)$ denotes the infinitesimal adjoint action of $X$. This infinitesimal
action can be exponentiated to give the finite, total adjoint action of the
group on $\Gamma$ as
\beq
\Gamma_{p\cdot g}=~{\rm Ad}(g^{-1})\Gamma=g^{-1}\Gamma g
\eeq

Let us briefly look at some examples. If $G$ is a matrix group (e.g. $SU(N)$,
$SO(N)$, etc.), then we can regard matrix elements of $g\in G$ as functions on
$G$ and $G\to~{\rm pt}$ as a principal $G$-bundle. Then the unique solution to
the connection conditions (\ref{conncondns}) for a Lie group is called the
Cartan-Maurer (matrix) 1-form
\beq
\Xi=g^{-1}dg
\eeq
For a general $G$-bundle $P~{\buildrel\pi\over\longrightarrow}~\man$, in a
local trivialization $U\times G\to\pi^{-1}(U)$, the connection must look like
\beq
\Gamma_{(x,g)}=g^{-1}dg+g^{-1}A_\mu gdx^\mu
\label{gaugefielddef}\eeq
where $A=A_\mu dx^\mu=A_\mu^aX^a\otimes dx^\mu$ is a Lie algebra valued 1-form
on $\man$, which is usually refered to in this context as a gauge connection or
gauge field. The transformation laws across local patch boundaries on $\man$,
labelled by a $G$-valued transition matrix $g$ (the automorphisms of the
bundle), act on gauge connections via pull-back,
\beq
A\to A^g\equiv g^{-1}Ag+g^{-1}dg
\label{gaugetransfdef}\eeq
which is the familiar form of the gauge transformation law in a gauge field
theory \cite{birm}. Another example is where the bundle is the tangent bundle
$T\man$ equipped with a Riemannian metric $g$. Then $\Gamma$ is the (affine)
Levi-Civita-Christoffel connection $\Gamma^\lambda_{\mu\nu}(g)$ associated with
$g$. Its transformation law under local changes of coordinates is determined by
(\ref{gaugetransfdef}) when the diffeomorphisms of the tangent bundle are
regarded as the automorphisms of the associated frame bundle. In this case, the
parallel transport equation (\ref{partranspeq}) determines the geodesics of the
Riemannian manifold $(\man,g)$ (i.e. the ``straight lines", or paths of minimal
distance, with respect to the curved geometry $g$).

The failure of the covariant derivative $\nabla$ on a principal $G$-bundle to
define a complex is measured by its curvature
\beq
F\equiv\nabla^2=dA+[A~{\buildrel\wedge\over,}~A]/2
\label{curvprincgendef}\eeq
The curvature 2-form (\ref{curvprincgendef}) is horizontal,
\beq
i_VF=0
\label{Fvertical0}\eeq
and in a local trivialization of the bundle, $F$ transforms in the adjoint
representation of $G$,
\beq
F\to g^{-1}F_{\mu\nu}(x)gdx^\mu\wedge dx^\nu=(g^{-1}X^ag)F_{\mu\nu}^a\otimes
dx^\mu\wedge dx^\nu
\eeq
and so it can be regarded as an element of $\Lambda^2(\man,{\rm Ad}~P)$, where
${\rm Ad}~P$ is the vector bundle associated to $P$ by the adjoint
representation of $G$. Furthermore, from its definition
(\ref{curvprincgendef}), the curvature $F$ obeys the Bianchi identity
\beq
[\nabla,F]=dF+[A~{\buildrel\wedge\over,}~F]=0
\label{ordbianchi}\eeq

When the bundle being considered is a $G$-equivariant bundle, we assume that
the covariant derivative (\ref{covderiv}) is $G$-invariant,
\beq
\left[\nabla,\lie_{V^a}\right]=0
\label{ginvconn}\eeq
Mimicking the equivariant exterior derivative (\ref{eqderiv}), we define the
equivariant covariant derivative
\beq
\nabla_{\bf g}={\bf1}\otimes\nabla-\phi^a\otimes i_{V^a}
\label{eqcovderiv}\eeq
which is considered as an operator on the algebra $\Lambda_G(\man,E)$ of
equivariant differential forms on $\man$ with values in $E$. In a local
trivialization $E=U\times W$, $U\subset\man$, this algebra looks like
\beq
\Lambda_G(U,E)=(S({\bf g}^*)\otimes\Lambda U\otimes W)^G
\label{eqdiffE}\eeq
Recalling the Cartan-Weil identity (\ref{weilid}), we define the equivariant
curvature of the connection (\ref{eqcovderiv})
\beq
F_{\bf g}=(\nabla_{\bf g})^2+\phi^a\otimes\lie_{V^a}
\label{eqcurv}\eeq
which, using (\ref{ginvconn}), then satisfies the equivariant Bianchi identity
\beq
\left[\nabla_{\bf g},F_{\bf g}\right]=0
\label{eqbianchi}\eeq
Notice that if $G$ is the trivial group, these identities reduce to the usual
notions of curvature, etc. discussed above. Expanding out (\ref{eqcurv})
explicitly using (\ref{ginvconn}) gives
\beq
F_{\bf g}={\bf1}\otimes F+\mu
\label{fgexpl}\eeq
where
\beq
\mu=\phi^a\otimes\lie_{V^a}-\left[\phi^a\otimes i_{V^a},{\bf1}\otimes\nabla
\right]
\label{mommapdef}\eeq
is called the moment map of the $G$-action with respect to the connection
$\nabla$. The moment map $\mu$ is a $G$-equivariant extension of the ordinary
curvature 2-form (\ref{curvprincgendef}) from a covariantly-closed 2-form, in
the sense of (\ref{ordbianchi}), to an equivariant one in the sense
of (\ref{eqbianchi}).

When evaluated on an element $X\in{\bf g}$, represented by a
vector field $V\in T\man\otimes W$, we write
\beq
F_{\bf g}(X)=F+\mu(X)\equiv F+\mu_V\equiv F_V
\label{fgont}\eeq
where
\beq
\mu_V=\lie_V-[i_V,\nabla]
\eeq
generates the induced $G$-action on the fibers of the bundle.
The moment map in this way can be regarded locally as a function $\mu:\Lambda
U\otimes W\to{\bf g}^*$. Furthermore, using the equivariant Bianchi identity
(\ref{eqbianchi}) we see that it obeys the important property
\beq
\nabla\mu_V=i_VF
\label{momderiv}\eeq
so that a non-trivial moment map produces a non-zero vertical component of the
curvature of the connection $\nabla$ (c.f. (\ref{Fvertical0})). Later on, we
shall encounter 2 important instances of equivariant bundles on
$\man$, one associated with a Riemannian structure, and the other with a
symplectic structure. In the latter case the moment map is associated with the
Hamiltonian of a dynamical system.

Now we are ready to define the notion of an equivariant characteristic class.
First, we recall how to construct conventional characteristic classes
\cite{milnor}. Given a Lie group $H$ with Lie algebra $\bf h$, we say that a
real- or complex-valued function $P$ is an invariant polynomial on $\bf h$ if
it is invariant under the natural adjoint action of $H$ on $\bf h$,
\beq
P(h^{-1}Yh)=P(Y)~~~~~\forall h\in H,\forall Y\in{\bf h}
\label{polyinv}\eeq
An invariant polynomial $P$ can be used to define characteristic classes on
principal fiber bundles with structure group $H$. If we consider the polynomial
$P$ in such a setting as a function on $\bf h$-valued 2-forms on $\man$, then
the $H$-invariance (\ref{polyinv}) of $P$ implies that
\beq
dP(\alpha)=rP(\nabla\alpha)~~~~~,~~~~~\alpha\in\Lambda^2\man\otimes{\bf h}
\label{pinv2form}\eeq
where $r$ is the degree of $P$. In particular, taking the argument $\alpha$ to
be the curvature 2-form $\alpha=F=\nabla^2$ on the principal $H$-bundle $E~{
\buildrel\pi\over\longrightarrow}~\man$ (which is locally an $\bf h$-valued
2-form), we have
\beq
dP(F)=0
\label{dpf0}\eeq
as a consequence of the Bianchi identity (\ref{ordbianchi}) for $F$. This means
that $P(F)$ defines a (DeRham) cohomology class of $\man$.

What is particularly remarkable about this cohomology class is that it is
independent of the particular connection $\nabla$ used to define the curvature
$F$. To see this, consider the simplest case where the invariant polynomial is
just $P(\alpha)=\tr~\alpha^n$ \footnote{\baselineskip=12pt Occasionally, for
ease of notation, we
shall denote exterior products of differential forms as ordinary
multiplication. For instance, we define $\alpha^{\wedge
n}\equiv\alpha^n$.}, with $\tr$ the invariant Cartan-Killing linear form of the
Lie algebra $\bf h$ (usually the ordinary operator trace). Consider a
continuous one-parameter family of connections $\nabla_t$, $t\in\IR$, with
curvatures $F_t=\nabla_t^2$. Then
\beq
\frac{d}{dt}F_t=\left[\nabla_t,\frac{d}{dt}\nabla_t\right]
\label{dtft}\eeq
and applying this to the invariant polynomial $\tr~F_t^n$ gives
\beq
\frac{d}{dt}~\tr~F_t^n=n~\tr\left(\frac{d}{dt}F_t\right)F_t^{n-1}=n~\tr\left[
\nabla_t,\left(\frac{d}{dt}\nabla_t\right)F_t^{n-1}\right]=d~\tr\left(\frac{d}
{dt}\nabla_t\right)F_t^{n-1}
\label{varft}\eeq
where $d$ is the exterior derivative and in the last equality we have applied
(\ref{pinv2form}). This means that any continuous deformation of the $2n$-form
$\tr~F^n$ changes it by an exact form, so that the cohomology class determined
by it is independent of the choice of connection. In general, the invariant
polynomial $P(F)\in\Lambda\man$ is called a characteristic class of the given
$H$-bundle.

This notion and construction of characteristic classes can be generalized
almost verbatum to the equivariant case \cite{berlinegetz}. Taking instead the
$G$-equivariant curvature (\ref{eqcurv}) as the argument of the $G$-invariant
polynomial $P$, (\ref{dpf0}) generalizes to
\beq
D_{\bf g}P(F_{\bf g})=rP(\nabla_{\bf g}F_{\bf g})=0
\label{dgpfg0}\eeq
and now the resulting equivariant characteristic classes $P(F_{\bf g})$ of
the given $G$-equivariant bundle are elements of the algebra $\Lambda_G\man$.
These are denoted by $P_{\bf g}(F)$, or when evaluated on an element $X\in
{\bf g}$ with associated vector field $V\in TU\otimes W$, we write
\beq
P_{\bf g}(F)(X)=P(F_V)\equiv P_V(F)
\label{pvf}\eeq
The equivariant cohomology class of $P_{\bf g}(F)$ is independent of the chosen
connection on the bundle. Consequently, on a trivial vector bundle $\man\times
W$ we can choose a flat connection\footnote{\baselineskip=12pt A non-trivial
vector bundle can always be considered as a trivial one endowed with a
non-trivial curvature $F\neq0$. This point of view is quite useful in certain
applications to topological gauge theories.}, $F=0$, and then
\beq
P^{\man\times W}(F_{\bf g})(X)=P^{\man\times W}(\mu_V)=P(\rho(X))
\label{ptrivial}\eeq
where $\rho$ is the representation of $G$ defined by the $G$-action on the
fibers $W$.

There are 4 equivariant characteristic classes that commonly appear in the
localization formalism for topological field theories, all of which are to be
understood as elements of the completion $\Lambda_G^\infty\man$. These
can all be found and are extensively discussed in \cite{berlinegetz}. The
first one is related to the invariant polynomial $\tr\e^\alpha$ and is used for
$G$-equivariant complex vector bundles (i.e. one in which the fibers are vector
spaces over the complex numbers $\IC$). It is called the $G$-equivariant Chern
character
\beq
\ch_{\bf g}(F)=\tr\e^{F_{\bf g}}
\label{eqchern}\eeq
The other 3 are given by determinants of specific polynomials. On a
$G$-equivariant real vector bundle we define the equivariant Dirac $\hat
A$-genus
\beq
\hat A_{\bf g}(F)=\sqrt{\det\left[\frac{\frac{1}{2}F_{\bf g}}{\sinh(\frac{1}{2}
F_{\bf g})}\right]}
\label{eqahat}\eeq
where the inverse of an inhomogeneous polynomial of differential forms is
always to be understood in terms of the power series
\beq
(1+x)^{-1}=\sum_{k=0}^\infty (-1)^kx^k
\label{power1}\eeq
On a complex fiber bundle, the complex version of the equivariant $\hat
A$-genus is the equivariant Todd class
\beq
\td_{\bf g}(F)=\det\left[\frac{F_{\bf g}}{\e^{F_{\bf g}}-1}\right]
\label{eqtodd}\eeq
When $G$ is the trivial group, these all reduce to the conventional
characteristic classes \cite{milnor} defined by replacing $F_{\bf g}\to F$ in
the above. Just as for the ordinary $\hat A$-genus and Todd classes, their
equivariant generalizations inherit the multiplicativity property under Whitney
sums of bundles,
\beq
\hat A^{E\oplus F}_{\bf g}=\hat A^E_{\bf g}\hat A^F_{\bf g}~~~,~~~\td^{E
\oplus F}_{\bf g}=\td^E_{\bf g}~\td^F_{\bf g}
\label{multprop}\eeq

Finally, on an orientable real bundle we can define the equivariant
generalization of the Euler class,
\beq
E_{\bf g}(F)=\pfaff(F_{\bf g})
\label{eqeuler}\eeq
where the Pfaffian (or Salam-Mathiews determinant) of a $2N\times2N$
antisymmetric matrix $M=[M_{ij}]$ is defined as
\beq
\pfaff~M=\epsilon^{i_1\cdots i_{2N}}M_{i_1i_2}\cdots M_{i_{2N-1}i_{2N}}=
\frac{1}{2^NN!}\sum_{P\in S_{2N}}\sgn(P)\prod_{k=1}^NM_{P(2k-1),P(2k)}
\label{pfafdef}\eeq
with the property that
\beq
\det M=(\pfaff~M)^2
\label{detpfaf}\eeq
The sign of the Pfaffian when written as the square root of the determinant
as in (\ref{detpfaf}) is chosen so that it is the product of the upper
skew-diagonal eigenvalues in a skew-diagonalization of the antisymmetric
matrix $M$. In (\ref{pfafdef}), $\epsilon^{i_1\cdots i_N}$ is the antisymmetric
tensor with the convention $\epsilon^{123\cdots N}=+1$. Pfaffians arise
naturally, as we will see, as fermionic determinants from the integration of
fermion bilinears in supersymmetric and topological field theories.
Transformations which change the orientation of the bundle change the
sign of the Pfaffian.

When $F=R$ is the Riemann curvature 2-form associated
with the tangent bundle $T\man$ (which can be regarded as a principal
$SO(2n)$-bundle) of a closed manifold $\man$ of even dimension $2n$, the
integral over $\man$ of the ordinary Euler class is the integer
\beq
\chi(\man)=\frac{(-1)^n}{(4\pi)^nn!}\int_\man E(R)
\label{gaussbonchern}\eeq
where
\beq
\chi(\man)=\sum_{k=0}^{2n}(-1)^k{\dim}_{\IR}H^k(\man;\IR)
\label{eulerchar}\eeq
is the famous topological invariant called the Euler characteristic of the
manifold $\man$. That (\ref{eulerchar}) can be written as an integral
of a density in (\ref{gaussbonchern}) is a celebrated result of differential
topology known as the Gauss-Bonnet theorem. The generalization of
(\ref{gaussbonchern}) to the case of an arbitrary vector bundle with Euler
class $E(F)$ of a curvature 2-form $F\in\Lambda^2(\man,E)$ is called the
Gauss-Bonnet-Chern theorem. In that case, the cohomology groups appearing in
the alternating sum (\ref{eulerchar}) get replaced by the cohomology groups
$H^k(\man;E)$ of the twisted derivative operator
$\nabla:\Lambda^k(\man,E)\to\Lambda^{k+1}(\man,E)$.

Similarly, with $F=F_A$ the curvature of a gauge connection $A$ on a principle
$H$-bundle over a $2k$-dimensional manifold $\man$, the integral over $\man$ of
the $k$-th term in the expansion of the conventional version of the Chern class
(\ref{eqchern}) (which defines the $k$-th Chern class) is the number
\beq
c_k(\man)=\left(-\frac{1}{2\pi i}\right)^k\int_\man\tr~F_A^k
\label{chernnum}\eeq
which is a topological invariant of $\man$ called the $k$-th Chern number of
$\man$ (or, more precisely, of the complex vector bundle $(E,\man,W,\pi)$).
The Chern number is always an integer for closed orientable manifolds.
Thus the equivariant characteristic classes defined above lead to interesting
equivariant generalizations of some classical topological invariants. In the
next Section we will see that their topological invariance in both the ordinary
DeRham and the equivariant cases are a consequence of the topological
invariance of the integrations there. We shall see later on that they appear in
most interesting ways within the formalism of localization formulas and
topological field theory functional integration.

\subsection{The Equivariant Localization Principle}

We now discuss a very interesting property of equivariant cohomology which is
the fundamental feature of all localization theorems. It also introduces the
fundamental geometric constraint that will be one of the issues of focus in
what follows. In most of our applications we will be concerned with the
following situation. Let $\man$ be a compact orientable manifold without
boundary and let $V$ be a vector field over $\man$  corresponding to some
action of the circle group $G=U(1)\sim S^1$ on $\man$. In this case the role
of the multiplier $\phi\in S({\bf u(1)}^*)$, which is a linear functional on
the 1-dimensional Lie algebra of $U(1)$, will not be important for the
discussion that follows. Indeed, we can regard $\phi$ as just
some external parameter in this case and `localize' algebraically by setting
$\phi=-1$. As shown in \cite{atiyahbott} (see also Appendix B), the operations
of evaluating $\phi$ on Lie algebra elements and the formation of equivariant
cohomology commute for abelian group actions, so that all results below will
coincide independently of the interpretation of $\phi$. In particular, for a
free $U(1)$-action on $\man$ we have
\beq
(S({\bf u(1)}^*)\otimes\Lambda\man)^{U(1)}=S({\bf
u(1)}^*)\otimes\Lambda(\man/U(1))
\eeq
so that in this case the multipliers $\phi$ play no cohomological role and the
equivariant cohomology just restricts to the cohomology of the quotient space
$\man/U(1)$. The corresponding equivariant
exterior derivative is now denoted as
\beq
D_{\bf u(1)}\equiv D_V=d+i_V
\label{eqderivu1}\eeq
and it is now considered as an operator on the algebra
\beq
\Lambda_V\man=\{\alpha\in\Lambda\man:\lie_V\alpha=0\}
\label{u1invs}\eeq

It was Atiyah and Bott \cite{atiyahbott} and Berline and Vergne
\cite{berline1,berline2} who first noticed that equivariant cohomology is
determined by the fixed point locus of the $G$-action. In our simplified case
here, this is the set
\beq
\man_V=\{x\in\man:V(x)=0\}
\label{fixedptset}\eeq
This fact is at the very heart of the localization theorems in both the
finite dimensional case and in topological field theory, and it is known as the
equivariant localization principle. In this Subsection we shall establish this
property in 2 analytic ways. For a more algebraic description of this
principle using the Weil algebra and the topological definition of equivariant
cohomology, see \cite{atiyahbott}.

Our first argument for localization involves an explicit proof at the level
of differential forms. Given an integral $\int_\man\alpha$ over $\man$ of
an equivariantly closed differential form $\alpha\in\Lambda_V\man$,
$D_V\alpha=0$, we wish to show that this integral depends only on the
fixed-point set (\ref{fixedptset}) of the $U(1)$-action on $\man$. To show
this, we shall explicitly construct a differential form $\lambda$ on
$\man-\man_V$ satisfying $D_V\lambda=\alpha$. This is just the equivariant
version of the Poincar\'e lemma. Thus the form $\alpha$ is equivariantly exact
away from the zero locus $\man_V$, and we recall that this implies that the
top-form component of $\alpha$ is exact. Since integration over $\man$ picks up
the top-form
component of any differential form, and since $\partial\man=\emptyset$ by
hypothesis here, it follows from Stokes' theorem (\ref{stokes}) that the
integral $\int_\man\alpha$ only receives contributions from an arbitrarily
small neighbourhood of $\man_V$ in $\man$, i.e. the integral `localizes' onto
the smaller subspace $\man_V$ of $\man$.

To construct $\lambda$, we need to impose the following geometric restriction
on the manifold $\man$. We assume that $\man$ has a globally-defined
$U(1)$-invariant Riemannian structure on it, which means that it admits a
globally-defined metric tensor
\beq
g=\frac{1}{2}g_{\mu\nu}(x)dx^\mu\otimes dx^\nu
\label{metric}\eeq
which is invariant under the $U(1)$-action generated by $V$, i.e. for which
\beq
\lie_Vg=0
\label{lieg}\eeq
or in local coordinates on $\man$,
\beq
g_{\mu\lambda}\partial_\nu V^\lambda+g_{\nu\lambda}\partial_\mu V^\lambda
+V^\lambda\partial_\lambda g_{\mu\nu}=0
\label{liegcoord}\eeq
Alternatively, this Lie derivative constraint can be written as
\beq
g_{\nu\lambda}\nabla_\mu V^\lambda+g_{\mu\lambda}\nabla_\nu V^\lambda=0
\label{killingeq}\eeq
where $\nabla$ is the covariant derivative (\ref{covderiv}) constructed from
the Levi-Civita-Christoffel connection
\beq
\Gamma^\lambda_{\mu\nu}=\frac{1}{2}g^{\lambda\rho}\left(\partial_\mu g_{\nu
\rho}+\partial_\nu g_{\mu\rho}-\partial_\rho g_{\mu\nu}\right)
\label{levi}\eeq
associated with $g$ on the tangent bundle $T\man$. Here $g^{\mu\nu}$ is the
matrix inverse of $g_{\mu\nu}$ and the covariant derivative acts on the vector
field $V$ in the usual way as
\beq
\nabla_\mu V^\nu=\partial_\mu V^\nu+\Gamma^\nu_{\lambda\mu}V^\lambda
\label{covonv}\eeq
with a plus sign for $(0,k)$-tensors and a minus sign for $(k,0)$-tensors in
front of $\Gamma$, as in (\ref{covonv}). Notice that by construction the
Levi-Civita-Christoffel connection is torsion-free, $\Gamma^\lambda_{\mu\nu}=
\Gamma^\lambda_{\nu\mu}$, and it is compatible with the the metric $g$,
$\nabla_\lambda g_{\mu\nu}=0$, which together mean that $\nabla$ preserves the
inner product in the fibers of the tangent bundle.

The equivalent equations (\ref{lieg})--(\ref{killingeq}) are called the
Killing equations and in this case we say that $V$ is a Killing vector field
of the metric $g$. Since the map $V\to\lie_V$ is linear, the space of Killing
vectors of a Riemannian manifold $(\man,g)$ generate the Lie algebra of a Lie
group acting on $\man$ by diffeomorphisms which is called the isometry group of
$(\man,g)$. We shall describe this group in detail in Sections 5 and 6.
The Killing equations here are assumed to hold globally over the entire
manifold $\man$. If both $\man$ and $G$ are compact, then such a metric can
always be obtained from an arbitrary Riemannian metric $h$ on $\man$ by
averaging $h$ over the group manifold of $G$ in its ($G$-invariant) Haar
measure, i.e. $g=\int_GD\tilde g~(\tilde g\cdot h)$. However, we
shall have occasion to also consider more general vector field flows which
aren't necessarily closed or when the manifold $\man$ isn't compact, as are the
cases in many physical applications. In such cases the Lie derivative
constraint (\ref{lieg}) is a very stringent one on the manifold. This feature
of the localization formalism, that the manifold admit a globally defined
metric with the property (\ref{lieg}) whose components $g_{\mu\nu}(x)=g_
{\nu\mu}(x)$ are globally-defined $C^\infty$ functions on $\man$, is the crux
of all finite- and infinite-dimensional localization formulas and will be
analysed in detail later on in this Review. For now, we content ourselves with
assuming that such a metric tensor has been constructed.

Any metric tensor defines a duality between vector fields and differential
1-forms, i.e. we can consider the metric tensor (\ref{metric}) as a map
\beq
g:T\man\to T^*\man
\label{metricdual}\eeq
which takes a vector field $V$ into its metric dual 1-form
\beq
\beta\equiv g(V,\cdot)=g_{\mu\nu}(x)V^\nu(x)dx^\mu
\label{beta}\eeq
Non-degeneracy, $\det g(x)\neq0$, $\forall x\in
\man$, of the metric tensor implies that this defines an isomorphism between
the tangent and cotangent bundles of $\man$. The 1-form $\beta$ satisfies
\beq
D_V^2\beta=\lie_V\beta=0
\label{liebeta0}\eeq
since $\lie_VV=0$ and $V$ is a Killing vector of $g$. This means that $\beta$
is an equivariant differential 1-form. Furthermore, we have
\beq
D_V\beta=K_V+\Omega_V
\label{dvbeta}\eeq
where $K_V$ is the globally-defined $C^\infty$-function
\beq
K_V=g(V,V)=g_{\mu\nu}(x)V^\mu(x)V^\nu(x)
\label{kv}\eeq
and
\beq
\Omega_V=d\beta=dg(V,\cdot)
\label{omegav}\eeq
is the 2-form with local components
\beq
(\Omega_V)_{\mu\nu}=g_{\mu\lambda}\nabla_\nu V^\lambda-g_{\nu\lambda}\nabla_\mu
V^\lambda
\label{omegavcomp}\eeq
Consequently, away from zero locus $\man_V$ of the vector field $V$, the
0-form part $K_V$ of $D_V\beta$ is non-zero and hence $D_V\beta$ is invertible
on $\man-\man_V$. Again we understand here the inverse of an inhomogeneous
differential form with non-zero scalar term in analogy with the formula
(\ref{power1}).

We can now define an inhomogenous differential form by
\beq
\xi=\beta(D_V\beta)^{-1}
\label{xi}\eeq
on $\man-\man_V$, which satisfies $D_V\xi=1$ and $\lie_V\xi=0$
owing to the equivariance (\ref{liebeta0}) of $\beta$. Thus we can define
an equivariant differential form $\lambda=\xi\alpha$,
and since $\alpha$ is equivariantly closed it follows that
\beq
\alpha=1\cdot\alpha=(D_V\xi)\alpha=D_V(\xi\alpha)
\label{alphaexact}\eeq
Thus, as claimed above, any equivariantly closed form is equivariantly exact
away from $\man_V$, and in particular the top-form component of an
equivariantly closed form is exact away from $\man_V$. This establishes the
equivariant localization property mentioned above.

The other argument we wish to present here for equivariant localization is less
explicit and involves cohomological arguments. First, consider an ordinary
closed form $\omega$, $d\omega=0$. For any other differential form
$\lambda$, we have
\beq
\int_\man(\omega+d\lambda)=\int_\man\omega
\label{intcoh}\eeq
by Stokes' theorem (\ref{stokes}) since $\partial\man=\emptyset$. This means
that the integral $\int_\man\omega$ of a closed form $\omega$ depends only
on the cohomology class defined by $\omega$, not on the particular
representative. Since the map $\omega\to\int_\man\omega$ in general defines a
linear map on $\Lambda^k\man\to\Lambda^{n-k}({\rm pt})=\delta^{nk}\IR$, it
follows that this
map descends to a map on $H^n(\man;\IR)\to H^0({\rm pt};\IR)=\IR$. The same is
true for equivariant integration. Since, for a general $G$-action on $\man$,
integration of a differential form picks up the top-form component which for an
equivariantly exact form is exact, for any equivariantly-closed differential
form $\alpha$ we can again invoke Stokes' theorem to deduce
\beq
\int_\man\left(\alpha+D_{\bf g}\lambda\right)=\int_\man\alpha
\label{eqintcoh}\eeq
so that the integral of an equivariantly closed form depends only on the
equivariant cohomology class defined by it, and not on the particular
representative. Note, however, that equivariant integration for general Lie
groups $G$ takes a far richer form. In analogy with the DeRham case above,
the integration of equivariant differential forms defines a map on $H_G(\man)
\to H_G({\rm pt})=S({\bf g}^*)^G$. This we define by
\beq
\left(\int_\man\alpha\right)(X)=\int_\man\alpha(X)~~~~,~~~~X\in{\bf g}
\label{intmap}\eeq
with integration over the $\Lambda\man$ part of $\alpha$ in the ordinary
DeRham sense. Later on, we shall also consider the dual Lie
algebra elements $\phi^a$ in a more `dynamical' situation where they are a
more integral part of the cohomological description above. We shall see then
how
this definition of integration should be accordingly modified. In any case, the
arguments below which lead to the equivariant localization principle generalize
immediately to the non-abelian case as well.

Given that the integral $\int_\man\alpha$ depends only on the equivariant
cohomology class defined by $\alpha$, we can choose a particular representative
of the cohomology class making the localization manifest. Taking the
equivariant differential form $\beta$ defined in (\ref{beta}), we consider
the integral
\beq
{\cal Z}(s)=\int_\man\alpha\e^{-sD_V\beta}
\label{zs}\eeq
viewed as a function of $s\in\IR^+$. We assume that (\ref{zs}) is a regular
function of $s\in\IR^+$ and that its $s\to0$ and $s\to\infty$ limits exist.
Its $s\to0$ limit is the integral of interest, $\int_\man\alpha$, while from
the identities (\ref{dvbeta}) and (\ref{kv}) we see that the integrand of
(\ref{zs}) is an increasingly sharply Gaussian peaked form around $\man_V
\subset\man$ as $s\to\infty$. The crucial point here is that the equivariant
differential form which is the integrand of (\ref{zs}) is equivariantly
cohomologous to $\alpha$ for all $s\in\IR^+$. This can be seen by applying
Stokes' theorem to get
\beq\new{\begin{array}{ll}
\frac{d}{ds}{\cal Z}(s)&=-\int_\man\alpha(D_V\beta)\e^{-sD_V\beta}\\&=-\int_
\man\left\{D_V(\alpha\beta\e^{-sD_V\beta})+\beta D_V(\alpha\e^{-sD_V\beta}
)\right\}\\&=s\int_\man\alpha\beta(\lie_V\beta)\e^{-sD_V\beta}=0\end{array}}
\label{zsindeps}\eeq
where we have used the fact that $\alpha$ is equivariantly closed and the
equivariance property (\ref{liebeta0}) of $\beta$. Therefore the integral
(\ref{zs}) is independent of the parameter $s\in\IR^+$, and so its $s\to0$ and
$s\to\infty$ limits coincide. Hence, we may evaluate the integral of interest
as
\beq
\int_\man\alpha=\lim_{s\to\infty}\int_\man\alpha\e^{-sD_V\beta}
\label{locofint}\eeq
which establishes the localization of $\int_\man\alpha$ to $\man_V$.

It should be pointed out though that there is nothing particularly unique
about the choice of $\beta$ in (\ref{locofint}). Indeed, the same steps
leading to (\ref{locofint}) can be carried out for an {\it arbitrary}
equivariant differential form $\beta$, i.e. any one with the property
(\ref{liebeta0}). In this general case, the localization of $\int_\man\alpha$
is onto the subspace of $\man$ which is the support for the non-trivial
equivariant cohomology of $\alpha$, i.e. $\int_\man\alpha$ localizes to the
points where $D_V\beta=0$. Different choices of representatives $\beta$ for the
equivariant cohomology classes then lead to potentially different localizations
other than the one onto $\man_V$. This would lead to seemingly different
expressions for the integral in (\ref{locofint}), but of course these must
all coincide in some way. In principle this argument for localization could
also therefore work without the assumption that $V$ is a Killing vector for
some metric on $\man$, but it appears difficult to make general statements in
that case. Nonetheless, as everything at the end will be equivariantly closed
by our general arguments above, it is possible to reduce the
resulting expressions further to $\man_V$ by applying the above localization
arguments once more, now to the localized expression. We shall examine
situations in which $V$ isn't necessarily a Killing vector field in Section 7.

\subsection{The Berline-Vergne Theorem}

The first general localization formula using only the general equivariant
cohomological arguments presented in the last Subsection was derived by
Berline and Vergne \cite{berline1,berline2}. This formula, as well as some
of the arguments leading to the equivariant localization principle, have since
been established in many different contexts suitable to other finite
dimensional applications and also to path integrals
\cite{atiyahbott,audin,berlinegetz,bismut1,bismut2}. The proof presented here
introduces a method that will generalize to functional integrals.
For now, we assume the fixed-point set $\man_V$ of the $U(1)$-action on $\man$
consists of discrete isolated points, i.e. $\man_V$ is a submanifold of $\man$
of codimension $n=\dim\man$ \footnote{\baselineskip=12pt We shall assume here
that $n$ is even. This restriction is by no means necessary but it will allow
us to shorten some of the arguments in this Subsection.}. We shall discuss the
generalization to the case where $\man_V$ has non-zero dimension later on. If
we assume that $\man$ is compact, then $\man_V$ is a finite set of points.

We wish to evaluate explicitly the right-hand side of the localization formula
(\ref{locofint}). To do this, we introduce an alternative way of evaluating
integrals over differential forms which is based on a more algebraic
description of the exterior bundle of $\man$. We introduce a set of nilpotent
anticommuting
(fermionic) variables $\eta^\mu$, $\mu=1,\ldots,n$,
\beq
\eta^\mu\eta^\nu=-\eta^\nu\eta^\mu
\label{grassalg}\eeq
which generate the exterior algebra $\Lambda\man$. The variables $\eta^\mu$
are to be identified with the local basis vectors $dx^\mu$ of
$\Lambda^1\man=T^*\man$
with the exterior product of differential forms replaced by the ordinary
product of the $\eta^\mu$ variables with the algebra (\ref{grassalg}). The
$k$-th exterior power $\Lambda^k\man$ is then generated by the products $\eta^
{\mu_1}\cdots\eta^{\mu_k}$ and this definition turns $\Lambda\man$ into a
graded Grassmann algebra with the generators $\eta^\mu$ having grading 1. For
instance, suppose the differential form $\alpha$ is the sum
\beq
\alpha=\alpha^{(0)}+\alpha^{(1)}+\dots+\alpha^{(n)}~~~~,~~~~\alpha^{(k)}\in
\Lambda^k\man
\label{alphasum}\eeq
with $\alpha^{(k)}$ the $k$-form component of $\alpha$ and $\alpha^{(0)}(x)$
its
0-form component which is a $C^\infty$-function on $\man$. The $k$-form
component of $\alpha$ for $k>0$ then has the form
\beq
\alpha^{(k)}(x,\eta)=\alpha^{(k)}_{\mu_1\cdots\mu_k}(x)\eta^{\mu_1}\cdots
\eta^{\mu_k}~~~,~~~k>0
\label{kformeta}\eeq
and from this point of view differential forms are functions $\alpha(x,\eta)$
on the exterior bundle which is now the $2n$-dimensional supermanifold
$\man\otimes\Lambda\man$ with local coordinates $(x,\eta)$.

The integration of a differential form is now defined by introducing the
Berezin rules for integrating Grassmann variables \cite{berezin},
\beq
\int d\eta^\mu~\eta^\mu=1~~~~~,~~~~~\int d\eta^\mu~1=0
\label{grassint}\eeq
Since the $\eta^\mu$'s are nilpotent, any function of them is a
polynomial in $\eta^\mu$ and consequently the rules (\ref{grassint})
unambiguously define the integral of any function of the anticommuting
variables $\eta^\mu$. For instance, it is easily verified that with this
definition of integration we have\footnote{\baselineskip=12pt If we introduce a
second independent set $\{\bar\eta^\mu\}$ of Grassmann variables, then the
formula (\ref{pfafint}) generlizes to arbitrary (not necessarily even)
dimensions $n$ as
\bd
\int\prod_{\mu=1}^nd\bar\eta^\mu~d\eta^\mu~\e^{-\bar\eta^\mu
M_{\mu\nu}\eta^\nu}=\det M
\ed
}
\beq
\int d^n\eta~\e^{\frac{1}{2}\eta^\mu M_{\mu\nu}\eta^\nu}=\pfaff~M
\label{pfafint}\eeq
where $d^n\eta\equiv d\eta^n~d\eta^{n-1}\cdots d\eta^1$. Note that under a
local change of basis $\eta^\mu\to A^\mu_\nu\eta^\nu$ of the Grassmann algebra
the antisymmetry property (\ref{grassalg}) and the Berezin rules
(\ref{grassint}) imply that $\int d^n\eta\to\det A\int d^n\eta$. It follows
from this that the Berezin integral
in (\ref{pfafint}) is invariant under similarity transformations.
(\ref{pfafint}) is the fermionic analog of the Gaussian integration formula
(\ref{gaussintclass}). The differentiation of Grassmann variables, for which
the integration in (\ref{grassint}) is the antiderivative thereof, is defined
by the anticommutator
\beq
\left[\frac{\partial}{\partial\eta^\mu},\eta^\nu\right]_+=\delta_\mu^\nu
\label{grassdiff}\eeq
With these definitions, the integration by parts formula $\int
d\eta^\mu~\frac{d}{d\eta^\mu}f(\eta^\mu)=0$ always holds, since $\int
d\eta^\mu~f(\eta^\mu)=\frac{d}{d\eta^\mu}f(\eta^\mu)$.

Given these definitions, we can now alternatively write the integral of any
differential form over $\man$ as an integral over the cotangent bundle
$\man\otimes\Lambda^1\man$. Thus given the localization formula
(\ref{locofint})
with the 1-form $\beta$ in (\ref{beta}) and the identities
(\ref{dvbeta})--(\ref{omegavcomp}), we have
\beq
\int_\man\alpha=\lim_{s\to\infty}\int_{\man\otimes\Lambda^1\man}d^nx~d^n\eta~
\alpha(x,\eta)\exp\left(-sg_{\mu\nu}(x)V^\mu(x)V^\nu(x)-\frac{s}{2}(\Omega_V)
_{\mu\nu}(x)\eta^\mu\eta^\nu\right)
\label{intineta}\eeq
where the measure $d^nx~d^n\eta$ on $\man\otimes\Lambda^1\man$ is
coordinate-independent because the measures $d^nx\equiv dx^1\wedge\cdots\wedge
dx^n$ and $d^n\eta$ transform inversely to each other. To evaluate the
large-$s$ limit of (\ref{intineta}), we use the delta-function representations
\beq
\delta(V)=\lim_{s\to\infty}\left(\frac{s}{\pi}\right)^{n/2}\sqrt{\det g}\e^{-s
g_{\mu\nu}V^\mu V^\nu}
\label{deltav}\eeq
\beq
\delta(\eta)=\lim_{s\to\infty}(-s)^{-n/2}\frac{1}{\pfaff~\Omega_V}\e^{-\frac{s}
{2}(\Omega_V)_{\mu\nu}\eta^\mu\eta^\nu}
\label{deltaeta}\eeq
as can be seen directly from the respective integrations in local coordinates
on $\man$ and $\Lambda^1\man$. Notice that from the Killing equations
(\ref{killingeq}), the matrix $(\Omega_V)_{\mu\nu}$ is given by
\beq
(\Omega_V)_{\mu\nu}=2g_{\mu\lambda}\nabla_\nu V^\lambda
\label{omegakill}\eeq

Thus using (\ref{deltav}) and (\ref{deltaeta}) we can write (\ref{intineta}) as
\beq
\int_\man\alpha=(-\pi)^{n/2}\int_{\man\otimes\Lambda^1\man}d^nx~d^n\eta~\alpha
(x,\eta)\frac{\pfaff~\Omega_V(x)}{\sqrt{\det g(x)}}\delta(V(x))\delta(\eta)
\label{intloc1}\eeq
where we note the cancellation of the factors of $s^{n/2}$ between
(\ref{deltav}) and (\ref{deltaeta}). The integration over $\Lambda^1\man$ in
(\ref{intloc1}) kills off all $k$-form
components of the form $\alpha$ except its $C^\infty$-function part $\alpha^
{(0)}(x)\equiv\alpha(x,0)$, while the integration over $\man$ localizes it
onto a sum over the points in $\man_V$. This yields
\beq
\int_\man\alpha=(-\pi)^{n/2}\sum_{p\in\man_V}\frac{\alpha^{(0)}(p)}{|\det
dV(p)|}\frac{\pfaff~\Omega_V(p)}{\sqrt{\det g(p)}}
\label{intloc2}\eeq
where the factor $|\det dV(p)|$ comes from the Jacobian of the coordinate
transformation $x\to V(x)$ used to transform $\delta(V(x))$ to a sum of
delta-functions $\sum_{p\in\man_V}\delta(x-p)$ localizing onto the zero locus
$\man_V$. Substituting in the identity (\ref{omegakill}) and noting that
at a point $p\in\man_V$ we have $\nabla V(p)=dV(p)$, the expression
(\ref{intloc2}) reduces to
\beq
\int_\man\alpha=(-2\pi)^{n/2}\sum_{p\in\man_V}\frac{\alpha^{(0)}(p)}{\pfaff~
dV(p)}
\label{berlineloc}\eeq
where we emphasize the manner in which the dependence of orientation in the
Pfaffian has been transfered from the numerator to the denominator in going
from (\ref{intloc2}) to (\ref{berlineloc}). This is the
(non-degenerate form of) the Berline-Vergne integration formula, and it is our
first example of what we shall call a localization formula. It reduces the
original integral over the $n$-dimensional space $\man$ to a sum over a
discrete set of points in $\man$ and it is valid for any equivariantly-closed
differential form $\alpha$ on a manifold with a globally-defined circle
action (and Riemannian metric for which the associated diffeomorphism
generator is a Killing vector). In general, the localization formulas we shall
encounter will always at least reduce the dimensionality of the integration of
interest. This will be particularly important for path integrals, where we
shall see that localization theory can be used to reduce complicated
infinite-dimensional integrals to finite sums or finite-dimensional integrals.

We close this Section by noting the appearence of the operator in the
denominator of the expression (\ref{berlineloc}). For each $p\in\man_V$, it is
readily seen that the operator $dV(p)$ appearing in the argument of the
Pfaffian in (\ref{berlineloc}) is just the invertible linear transformation
$L_V(p)$ induced by the Lie derivative acting on the tangent spaces $T_p\man$,
i.e. by the induced infinitesimal group action on the tangent bundle (see
(\ref{lieVW})). Explicitly, this operator is defined on vector fields
$W=W^\mu(x)\frac{\partial}{\partial x^\mu}|_{x=p}\in T_p\man$ by
\beq
L_V(p)W=\partial_\nu V^\mu(p)W^\nu(p)\frac{\partial}{\partial x^\mu}\biggm|_
{x=p}
\label{lvpw}\eeq
Note however that $dV(p)$ is not covariant in general and so this is only true
right on the tangent space $T_p\man$ and not in general on the entire tangent
bundle $T\man$. A linear transformation on the whole of $T\man$ can only be
induced from the Lie derivative by introducing a (metric or non-metric)
connection $\Gamma^\lambda_{\mu\nu}$ of $T\man$ and inducing an operator from
$\nabla V$, as in the matrix (\ref{omegakill}). We shall return to this point
later on in a more specific setting.

\section{Finite-dimensional Localization Theory for Dynamical Systems}

We shall now proceed to study a certain class of integrals that can be
considered to be toy models for the functional integrals that we are
ultimately interested in. The advantage of these models is that they are
finite-dimensional and therefore rigorous mathematical theorems concerning
their
behaviour can be formulated. In the infinite-dimensional cases, although the
techniques used will be standard methods of supersymmetry and
topological field theory, a lot of rigor is lost due the ill-definedness of
infinite-dimensional manifolds and functional integrals. A lot can therefore
be learned by looking closely at some finite-dimensional cases.

We shall be interested in certain oscillatory integrals $\int_\man
d\mu\e^{iTH}$ representing the Fourier-Laplace transform of some
smooth measure $d\mu$ on a manifold $\man$ in terms of a smooth function $H$.
The common method of evaluating such integrals is the stationary phase
approximation which expresses the fact that for large-$T$ the main
contributions to the integral come from the critical points of $H$. The main
result of this Section is the Duistermaat-Heckman theorem \cite{dh} which
provides a criterion for the stationary phase approximation
to an oscillatory integral to be exact. Although this theorem was originally
discovered within the context of symplectic geometry, it turns out to have
its most natural explanation in the setting of equivariant cohomology and
equivariant characteristic classes
\cite{atiyahbott},\cite{berline1}--\cite{berlinegetz}. The Duistermaat-Heckman
theorem, and its various extensions that we shall discuss towards the end of
this Section, are precisely those which
originally motivated the localization theory of path integrals.

For physical applications, we shall be primarily interested in a special class
of differentiable manifolds known as `symplectic' manifolds. As we shall see in
this Section, the application of the equivariant cohomological ideas to these
manifolds leads quite nicely to the notion of a Hamiltonian from a mathematical
perspective, as well as some standard ideas in the
geometrical theory of classical integrability. Furthermore, the configuration
space of a topological field theory is typically an (infinite-dimensional)
symplectic manifold (or phase space) \cite{birm} and we shall therefore
restrict our attention for the remainder of this Review to the localization
theory for oscillatory integrals over symplectic manifolds.

\subsection{Symplectic Geometry}

Symplectic geometry is the natural mathematical setting for the geometrical
formulation of classical mechanics and the study of classical integrability
\cite{abmarsden,arnold}. It also has applications in other branches of physics,
such as geometrical optics \cite{guill}. In elementary classical mechanics
\cite{goldstein}, one is introduced to the Hamiltonian formalism of classical
dynamics as follows. For a dynamical system defined on some manifold $\man$
(usually $\IR^n$) with coordinates $(q^1,\dots,q^n)$, we introduce the
canonical momenta $p_\mu$ conjugate to each variable $q^\mu$ from the
Lagrangian of the system and then the Hamiltonian $H(p,q)$ is obtained by a
Legendre transformation of the Lagrangian. In this way one has a description
of the dynamics on the $2n$-dimensional space of the $(p,q)$ variables which
is called the phase space of the dynamical system. With this construction the
phase space is the cotangent bundle $\man\otimes\Lambda^1\man$ of the
configuration manifold $\man$. The equations of motion can be represented
through the time evolution of the phase space coordinates by Hamilton's
equations. For most elementary dynamical systems, this description is
sufficient. However, there are relatively few examples of mechanical systems
whose equations of motion can be solved by quadratures and it is desirable to
seek other more general formulations of this elementary situation in the hopes
of being able to formulate rigorous theorems about when a classical
mechanical system has solvable equations of motion, or is `integrable'.
Furthermore, the above notion of a `phase space' is very local and is strictly
speaking only globally valid when the phase space is $\IR^{2n}$, a rather
restrictive class of systems. Motivated by the search for more non-trivial
integrable models in both classical and quantum physics, theoretical physicists
have turned to the general theory of symplectic geometry which encompasses the
above local description in a coordinate-free way suitable to the methods of
modern differential geometry. In this Subsection we shall review the basic
ideas of symplectic geometry and how these descriptions tie in with the more
familiar ones of elementary classical mechanics.

A symplectic manifold is a differentiable manifold $\man$ of even dimension
$2n$
together with a globally-defined non-degenerate closed 2-form
\beq
\omega=\frac{1}{2}\omega_{\mu\nu}(x)dx^\mu\wedge dx^\nu
\label{symplform}\eeq
called the symplectic form of $\man$. By closed we mean as usual that
\beq
d\omega=0
\label{omclosed}\eeq
or in local coordinates
\beq
\partial_\mu\omega_{\nu\lambda}+\partial_\nu\omega_{\lambda\mu}+\partial_
\lambda\omega_{\mu\nu}=0
\label{omclosedloc}\eeq
Thus $\omega$ defines a DeRham cohomology class in $H^2(\man;\IR)$. By
non-degenerate we mean that the components $\omega_{\mu\nu}(x)$ of
$\omega$ define an invertible $2n\times2n$ antisymmetric matrix globally on
the manifold $\man$, i.e.
\beq
\det\omega(x)\neq0~~~~~\forall x\in\man
\label{nondeg}\eeq
The manifold $\man$ together with its symplectic form $\omega$ defines the
phase space of a dynamical system, as we shall see below.

Since $\omega$ is closed, it follows from the Poincar\'e lemma that locally
there exists a 1-form
\beq
\theta=\theta_\mu(x)dx^\mu
\label{theta}\eeq
such that
\beq
\omega=d\theta
\label{omdtheta}\eeq
or in local coordinates
\beq
\omega_{\mu\nu}=\partial_\mu\theta_\nu-\partial_\nu\theta_\mu
\label{omdthetacomp}\eeq
The locally-defined 1-form $\theta$ is called the symplectic potential or
canonical 1-form of $\man$. When $\omega$ is generated globally as above by a
symplectic potential $\theta$ it is said to be integrable. Diffeomorphisms of
$\man$ that leave the symplectic
2-form invariant are called canonical or symplectic transformations.
These are determined by $C^\infty$-maps that act on the symplectic potential as
\beq
\theta~{\buildrel F\over\longrightarrow}~\theta_F=\theta+dF
\label{cantransftheta}\eeq
or in local coordinates
\beq
\theta_\mu(x)~{\buildrel F\over\longrightarrow}~\theta_{F,\mu}(x)=
\theta_\mu(x)+\partial_\mu F(x)
\label{canloc}\eeq
so that by nilpotency of the exterior derivative it follows that $\omega$ is
invariant under such transformations,
\beq
\omega=d\theta~{\buildrel F\over\longrightarrow}~\omega_F=d\theta_F
\equiv\omega
\label{ominvcan}\eeq
The function $F(x)$ is called the generating function of the canonical
transformation.

The symplectic 2-form determines a bilinear function $\{\cdot,\cdot\}_\omega
:\Lambda^0\man\otimes\Lambda^0\man\to\Lambda^0\man$ called the Poisson
bracket. It is defined by
\beq
\{f,g\}_\omega=\omega^{-1}(df,dg)~~~~~,~~~~~f,g\in\Lambda^0\man
\label{poisson}\eeq
or in local coordinates
\beq
\{f,g\}_\omega=\omega^{\mu\nu}(x)\partial_\mu f(x)\partial_\nu g(x)
\label{poissonloc}\eeq
where $\omega^{\mu\nu}$ is the matrix inverse of $\omega_{\mu\nu}$.
Note that the local coordinate functions themselves have Poisson bracket
\beq
\left\{x^\mu,x^\nu\right\}_\omega=\omega^{\mu\nu}(x)
\label{poissoncoord}\eeq
The Poisson bracket is anti-symmetric,
\beq
\{f,g\}_\omega=-\{g,f\}_\omega~~~~~,
\label{antisympoisson}\eeq
it obeys the Leibniz property
\beq
\{f,gh\}_\omega=g\{f,h\}_\omega+h\{f,g\}_\omega
\label{leibpoisson}\eeq
and it satisfies the Jacobi identity
\beq
\left\{f,\{g,h\}_\omega\right\}_\omega+\left\{g,\{h,f\}_\omega\right\}_\omega
+\left\{h,\{f,g\}_\omega\right\}_\omega=0
\label{jacobipoisson}\eeq
This latter property follows from the fact (\ref{omclosedloc}) that $\omega$ is
closed. These 3 properties of the Poisson bracket mean that it defines a Lie
bracket. Thus the Poisson bracket makes the space of $C^\infty$-functions on
$\man$ into a Lie algebra which we call the Poisson algebra of $(\man,\omega)$.

The connection with the elementary formulation of classical mechanics
discussed above is given by a result known as Darboux's theorem \cite{guill},
which states that this connection is always possible locally. More precisely,
Darboux's theorem states that locally there exists a system of coordinates
$(p_\mu,q^\mu)_{\mu=1}^n$ on $\man$ in which the symplectic 2-form looks like
\beq
\omega=dp_\mu\wedge dq^\mu
\label{omdarboux}\eeq
so that they have Poisson brackets
\beq
\left\{p_\mu,p_\nu\right\}_\omega=\left\{q^\mu,q^\nu\right\}_\omega=0~~~,~~~
\left\{p_\mu,q^\nu\right\}_\omega=\delta^\nu_\mu
\label{poissoncan}\eeq
These coordinates are called canonical or Darboux coordinates on $\man$ and
from (\ref{poissoncan}) we see that they can be identified with the usual
canonical momentum and position variables on the phase space $\man$
\cite{goldstein}. In these coordinates the symplectic potential is
\beq
\theta=p_\mu dq^\mu
\label{thetadarboux}\eeq
and the transformation (\ref{cantransftheta}) becomes
\beq
\theta=p_\mu dq^\mu~{\buildrel F\over\longrightarrow}~\theta+dF=\theta_F=
P_\mu dQ^\mu
\label{candarboux}\eeq
where $(P_\mu,Q^\mu)_{\mu=1}^n$ are also canonical coordinates according to
(\ref{ominvcan}). It follows that
\beq
p_\mu dq^\mu-P_\mu dQ^\mu=dF
\label{usualcan}\eeq
where both $(p_\mu,q^\mu)$ and $(P_\mu,Q^\mu)$ are canonical momentum and
position variables on $\man$. (\ref{usualcan}) is the usual form of a
canonical transformation determined by the generating function $F$
\cite{goldstein}.

Smooth real-valued functions $H$ on $\man$ (i.e. elements of $\Lambda^0\man$)
will be called classical observables. Exterior products of $\omega$ with itself
determine non-trivial closed $2k$-forms on $\man$ (i.e. non-zero cohomology
classes $[\omega^k]\in H^{2k}(\man;\IR)$). In particular, the $2n$-form
\beq
d\mu_L=\omega^n/n!=\sqrt{\det\omega(x)}~d^{2n}x
\label{liouvillemeas}\eeq
defines a natural volume element on $\man$ which is invariant under canonical
transformations. It is called the Liouville measure, and in the local
Darboux coordinates (\ref{omdarboux}) it becomes the familiar phase space
measure \cite{goldstein}
\beq
(-1)^{n(n-1)/2}\omega^n/n!=dp_1\wedge\cdots\wedge dp_n\wedge dq^1\wedge
\cdots\wedge dq^n
\label{phasespmeas}\eeq

\subsection{Equivariant Cohomology on Symplectic Manifolds}

In this Subsection we shall specialize the discussion of Section 2 to the case
where the differentiable manifold $\man$ is a symplectic manifold of dimension
$2n$. Consider the action of some connected Lie group $G$ on $\man$ generated
by the vector fields $V^a$ with the commutator algebra (\ref{vliealg}). We
assume that the action of $G$ on $\man$ is symplectic so that it preserves
the symplectic structure,
\beq
\lie_{V^a}\omega=0
\label{lieom0}\eeq
or in other words $G$ acts on $\man$ by symplectic transformations. Since
$\omega$ is closed this means that
\beq
di_{V^a}\omega=0
\label{divom0}\eeq
Let $L\to\man$ be a complex line bundle with connection 1-form the
symplectic potential $\theta$. If $\theta$ also satisfies
\beq
\lie_{V^a}\theta=0
\label{lietheta0}\eeq
then the associated covariant derivative $\nabla=d+\theta$ is $G$-invariant,
and according to the general discussion of Subsection 2.4 this defines a
$G$-equivariant bundle. By definition (see Subsection 2.4) the structure group
of this symplectic line bundle acts by canonical transformations. As such,
$\omega$ represents the first Chern class of this $U(1)$-bundle, and, if $\man$
is closed, it defines an integer cohomology class in $H^2(\man;\IZ)$ (as the
Chern numbers generated by $\omega$ are then integers).

The associated moment map $H:\man\to{\bf g}^*$ evaluated on a Lie algebra
element $X\in{\bf g}$ with associated vector field $V$ is called
the Hamiltonian corresponding to $V$,
\beq
H_V=\lie_V-\left[i_V,\nabla\right]=i_V\theta=V^\mu\theta_\mu
\label{hamdef}\eeq
{}From (\ref{omdtheta}) and (\ref{lietheta0}) it then follows that
\beq
dH_V=-i_V\omega
\label{hameqv}\eeq
or equivalently this follows from the general property (\ref{momderiv}) of the
moment map since $\omega$ is the curvature of the connection $\theta$. In local
coordinates, this last equation reads
\beq
\partial_\mu H_V(x)=V^\nu(x)\omega_{\mu\nu}(x)
\label{hameqvloc}\eeq
In particular, the components $H^a$ of the moment map
\beq
H=\phi^a\otimes H^a
\label{momcomps}\eeq
satisfy
\beq
dH^a=-i_{V^a}\omega
\label{hameqa}\eeq
Comparing with the symplecticity condition (\ref{divom0}) on the group action,
we see that this is equivalent to the statement that the closed 1-forms
$i_{V^a}
\omega$ are exact. If $H^1(\man;\IR)=0$ this is certainly true, but in the
following we will want to consider multiply connected phase
spaces as well. We therefore impose this exactness requirement from the onset
on the action of $G$ on $\man$, i.e. the equivariance
requirement (\ref{lietheta0}) on the symplectic potential $\theta$. When such
a Hamiltonian function exists as a globally-defined $C^\infty$-map on $\man$,
we shall say that the group action is Hamiltonian. A vector field $V$ which
satisfies (\ref{hameqv}) is said to be the Hamiltonian vector field associated
with $H_V$, and we shall call the triple $(\man,\omega,H_V)$, i.e. a symplectic
manifold with a Hamiltonian $G$-action on it, a Hamiltonian system or a
dynamical system.

The integral curves (\ref{intcurves}) defined by the flows
(or time-evolution) of a Hamiltonian vector field $V$ as in (\ref{hameqvloc})
define the Hamilton equations of motion
\beq
\dot x^\mu(t)=\omega^{\mu\nu}(x(t))\partial_\nu H_V(x(t))=\left\{x^\mu,H_V
\right\}_\omega
\label{hameqmotion}\eeq
The Poisson bracket of the Hamiltonian with any other function $f$ determines
the (infinitesimal) variation (or time-evolution) of $f$ along the classical
trajectories of the dynamical system (compare with (\ref{lieVW})),
\beq
\{f,H_V\}_\omega=\lie_Vf=\frac{d}{dt}f(x(t))\biggm|_{t=0}
\eeq
In the canonical coordinates defined by (\ref{omdarboux}) the
equations (\ref{hameqmotion}) read
\beq
\dot q^\mu=\frac{\partial H}{\partial p_\mu}~~~~~,~~~~~\dot p_\mu=-\frac{
\partial H}{\partial q^\mu}
\label{hameqcan}\eeq
which are the usual form of the Hamilton equations of motion encountered in
elementary classical mechanics \cite{goldstein}. Thus we see that the above
formalisms for symplectic geometry encompass all of the usual ideas of
classical Hamiltonian mechanics in a general, coordinate-independent setting.

The equivariant curvature of the above defined equivariant bundle is given
by the equivariant extension of the symplectic 2-form,
\beq
\omega_{\bf g}={\bf1}\otimes\omega+\phi^a\otimes H^a
\label{eqextom}\eeq
and evaluated on $X\in{\bf g}$ we have
\beq
(D_{\bf g}\omega_{\bf g})(X)=(d-i_V)(\omega+H_V)=0
\label{omgclosed}\eeq
which is equivalent to the definition (\ref{hameqv}) of the Hamiltonian vector
field $V$. In fact, the extension (\ref{eqextom}) is the {\it unique}
equivariant extension of the symplectic 2-form $\omega$ \cite{niemitirk3}, i.e.
the unique  extension of $\omega$ from a closed 2-form to an
equivariantly-closed one. Thus, we see that finding an equivariantly-closed
extension of $\omega$ is equivalent to finding a moment map for the $G$-action.
If $\omega$ defines an integer cohomology class $[\omega]
\in H^2(\man;\IZ)$, then the line bundle $L\to\man$ introduced above can be
thought of as the prequantum line bundle of geometric quantization
\cite{woodhouse}, the natural geometric framework (in terms of symplectic
geometry) for the coordinate independent formulation of quantum mechanics.
Within this framework, the equivariant curvature 2-form $\omega_V=\omega_{\bf
g}
(X)$ above is refered to as the prequantum operator. We shall say more about
some of the general ideas of geometric quantization later on. Notice that if
(\ref{lietheta0}) (or (\ref{hamdef})) does hold, then $\theta$ is also the
equivariant symplectic potential for the equivariant extension (\ref{eqextom}),
i.e. $\omega_{\bf g}=D_{\bf g}\theta$.

{}From (\ref{hameqa}) it follows that the Poisson algebra of the Hamiltonians
$H^a$ is given by
\beq
\left\{H^a,H^b\right\}_\omega=\omega(V^a,V^b)=\omega_{\mu\nu}V^{a,\mu}V^{b,
\nu}=V^{a,\mu}\partial_\mu H^b=\lie_{V^a}H^b=-\lie_{V^b}H^a
\label{poissonham}\eeq
 From the Jacobi identity (\ref{jacobipoisson}) it follows that the map $H^a\to
V^a$ is a homomorphism of the Lie algebras $(\Lambda^0\man,\{\cdot,
\cdot\}_\omega)\to(T\man,[\cdot,\cdot])$ since
\beq
V^{\{H^a,H^b\}_\omega}=\left[V^a,V^b\right]
\label{homlie}\eeq
However, the inverse of this map does not necessarily define a homomorphism.
The Hamiltonian function which corresponds to the commutator of 2 group
generators may differ from the Poisson bracket of the pertinent Hamiltonian
functions as
\beq
\left\{H^a,H^b\right\}_\omega=f^{abc}H^c+c^{ab}
\label{hamcomm}\eeq
where $c^{ab}\equiv c(X^a,X^b)=-c^{ba}$ is a 2-cocycle in the Lie algebra
cohomology of $G$ \cite{jackiw} (see Appendix A), i.e.
\beq
c([X_1,X_2],X_3)+c([X_2,X_3],X_1)+c([X_3,X_1],X_2)=0~~~~~\forall X_1,X_2,X_3
\in{\bf g}
\label{2cocycle}\eeq
If $H^2(G)=0$ then we can set $c^{ab}=0$ and the map $X^a\to H^a$ determines
a homomorphism between the Lie algebra $\bf g$ and the Poisson algebra of
$C^\infty$-functions on $\man$.

The appearence of the 2-cocycle $c^{ab}$ in (\ref{hamcomm}) is in fact related
to the possible non-invariance of the symplectic potential under $G$ (c.f. eq.
(\ref{lietheta0})). From the symplecticity (\ref{lieom0}) of the group action
and (\ref{hameqa}) it follows that
\beq
\lie_{V^a}\theta=(i_{V^a}d+di_{V^a})\theta=dg^a
\label{liethetaha}\eeq
locally in a neighbourhood $\cal N$ in $\man$ wherein $\omega=d\theta$ and
$V^a\neq0$. Here the locally-defined linear functions $g^a\equiv g(X^a)=-H^a+
i_{V^a}\theta$ obey the consistency condition
\beq
\left\{H(X_1),g(X_2)\right\}_\omega-\left\{H(X_2),g(X_1)\right\}_\omega=
g([X_1,X_2])~~~~~\forall X_1,X_2\in{\bf g}
\label{consistencond}\eeq
which follows from (\ref{hamcomm}). However, if there exists a locally-defined
function $f$ such that
\beq
g^a=\left\{H^a,f\right\}_\omega~~~~~,~~~~~a=1,\dots,\dim G
\label{fdef}\eeq
then we can remove the functions $g^a$ by the canonical transformation $\theta
\to\theta_f=\theta+df$ so that the symplectic potential $\theta_f$ is
$G$-invariant. Indeed, the 1-form $\theta_f$ obeys
\beq
\lie_{V^a}\theta_f=0
\label{liethetaf0}\eeq
which implies that in the neighbourhood $\cal N$,
\beq
i_{V^a}\theta_f=H^a+C
\label{ivathetaf}\eeq
where $C$ is a constant. This constant is irrelevant here because we can
introduce a function $K$ in $\cal N$ such that
\beq
\left\{H^a,K\right\}_\omega=V^{a,\mu}\partial_\mu K=1
\label{hkpoisson}\eeq
and defining $F=f+CK$ we find
\beq
i_{V^a}\theta_F=H^a
\label{ivathetaF}\eeq

However, notice that the $G$-invariance (\ref{ivathetaF}) of the symplectic
potential in general holds only locally in $\man$, and furthermore the
canonical transformation $\theta\to\theta_f$ above does not remove the
functions $g^a$ for the entire Lie algebra $\bf g$, but only for a closed
subalgebra of $\bf g$ which depends on the function $f$ and on the phase space
$\man$ where $G$ acts \cite{guill,niemipas}. In this subspace, the symplectic
potential is $G$-invariant and the identity (\ref{hamdef}) relating the
Hamiltonians to the symplectic potential by $H^a=i_{V^a}\theta$ holds (so that
$\theta$ is a local solution to the equivariant Poincar\'e lemma). In
general though, on the entire Lie algebra $\bf g$, defining $h^a=-i_{V^a}dF$
in the above we have
\beq
i_{V^a}\theta=H^a+h^a
\label{ivathetaHh}\eeq
and then the Poisson bracket (\ref{poissonham}) implies that the 2-cocycle
appearing in (\ref{hamcomm}) is given by
\beq
c^{ab}=f^{abc}h^c-\lie_{V^a}h^b+\lie_{V^b}h^a
\label{cab}\eeq
Thus it is only when $c^{ab}=0$ for all $a,b$ that the $G$-action of the vector
fields $V^a$ lifts isomorphically to the Poisson action of the corresponding
Hamiltonians $H^a$ on $\man$. Notice that this is certainly true on the
Cartan subalgebra of the Lie algebra $\bf g$ (i.e. its maximal commuting
subalgebra), since $H^2(U(1))=H^2(S^1)=0$. We shall see in
Section 4 that the dynamical systems for which the equivariance condition
(\ref{hamdef}) holds determine a very special class of quantum theories.

\subsection{Stationary-phase Approximation and the Duistermaat-Heckman Theorem}

We now start examining localization formulas for a specific class of phase
space integrals. We shall concentrate for the time being on the case of an
abelian circle action on the manifold
$\man$, as we did in Subsection 2.6. We shall also assume that the Hamiltonian
$H$ defined as in the last Subsection is a Morse function. This means that the
critical points $p$ of the Hamiltonian, defined by $dH(p)=0$, are isolated
and the Hessian matrix of $H$,
\beq
\hess(x)=\left[\frac{\partial^2H(x)}{\partial x^\mu\partial x^\nu}\right]
\label{hessian}\eeq
at each critical point $p$ is a non-degenerate matrix, i.e.
\beq
\det\hess(p)\neq0
\label{nondegcond}\eeq
The Hamiltonian vector field $V$ is defined by (\ref{hameqvloc}) and it
represents the action of some 1-parameter group on the phase space $\man$. We
shall assume here that the orbits (\ref{intcurves}) of $V$ generate the circle
group $U(1)\sim S^1$. Later on we shall consider more general cases. Notice
that the critical points
of $H$ coincide with zero locus $\man_V$ of the vector field $V$.

There is an important quantity of physical interest for the statistical
mechanics of a classical dynamical system called the partition function. It is
constructed as follows. Each point $x$ of the phase space $\man$ represents a
classical state of the dynamical system which in canonical coordinates is
specified by its configuration $q$ and its momentum $p$. The energy of this
state is determined by the Hamiltonian $H$ of the dynamical system which as
usual is its energy function. According to the general principles of
classical statistical mechanics \cite{rief} the partition function is built by
attaching to each point $x\in\man$ the Boltzmann weight $\e^{iTH(x)}$ and
`summing' them over all states of the system. Here the parameter $iT$ is
`physically' to be identified with $-\beta/k_B$ where $k_B$ is Boltzmann's
constant and $\beta$ is the inverse temperature. However, for mathematical
ease in the following, we shall assume that the parameter $T$ is real. In the
canonical position and momentum coordinates we would just simply integrate up
the Boltzmann weights. However, we would like to obtain a quantity which is
invariant under transformations which preserve the (symplectic) volume of the
phase space $\man$ (i.e. those which preserve the classical equations of
motion (\ref{hameqmotion}) and hence the density of classical states),
and so we integrate using the Liouville measure (\ref{liouvillemeas}) to obtain
a canonically invariant quantity. This defines the classical partition
function of the dynamical system,
\beq
Z(T)=\int_\man\frac{\omega^n}{n!}\e^{iTH}=\int_\man
d^{2n}x~\sqrt{\det\omega(x)}
\e^{iTH(x)}
\label{classpart}\eeq
The partition function determines all the usual thermodynamic quantities of
the dynamical system \cite{rief}, such as its free energies and specific
heats, as well as all statistical averages in the canonical ensemble of the
classical system.

However, it is very seldom that one can actually obtain an exact closed form
for the partition function (\ref{classpart}) as the integrals involved are
usually rather complicated. But there is a method of approximating the integral
(\ref{classpart}), which is very familiar to both physicists and
mathematicians,
called the stationary-phase approximation \cite{guill,hormander,woodhouse}.
This method is often employed when one encounters oscillatory integrals such as
(\ref{classpart}) to obtain an idea of its behaviour, at least for large $T$.
It works as follows. Notice that for $T\to\infty$ the integrand of $Z(T)$
oscillates very rapidly and begins to damp to 0. The integral therefore has an
asymptotic expansion in powers of $1/T$. The larger $T$ gets the more the
integrand tends to localize around its stationary values wherever the function
$H(x)$ has extrema (equivalently where $dH(p)=0$)\footnote{\baselineskip=12pt
Usually one argues that the phase will concentrate around the points where $H$
is minimized (the ground state) since this should be the dominant contribution
for $T\to\infty$. However, the localization is properly determined by {\it all}
points where $dH(p)=0$ since the
contribution from other extrema turn out to be of the same order of magnitude
as those from the minima \cite{hormander}.}. To evaluate these contributions,
we expand both $H$ and the Liouville density in (\ref{classpart}) in a
neighbourhood $U_p$ about each critical point $p\in\man_V$ in a Taylor series,
where as usual integration in $U_p$ can be thought of as integration in the
more familiar $\IR^{2n}$. We expand the exponential of all derivative terms in
$H$ of order higher than 2 in the exponential power series, and in this way we
are left with an infinite series of Gaussian moment integrals with Gaussian
weight determined by the bilinear form defined by the Hessian matrix
(\ref{hessian}) of $H$ at $p$. The lowest order contribution is just the
normalization of the Gaussian (see (\ref{gaussintclass})), while the $k$-th
order moments are down by powers of $1/T^k$ compared to the leading term.
Carrying out these Gaussian integrations, taking into careful account the
signature of the Hessian at each point, and summing over all points
$p\in\man_V$, in this way we obtain the standard lowest-order stationary-phase
approximation to the integral (\ref{classpart}),
\beq
Z(T)=\left(\frac{2\pi i}{T}\right)^n\sum_{p\in\man_V}(-i)^{\lambda(p)}
\e^{iTH(p)}\sqrt{\frac{\det\omega(p)}{\det\hess(p)}}+{\cal O}(1/T^{n+1})
\label{lowstatphase}\eeq
where $\lambda(p)$ is the Morse index of the critical point $p$, defined as
the number of negative eigenvalues in a diagonalization of the symmetric
Hessian matrix of $H$ at $p$. We shall always ignore a possible regular
function of $T$ in the large-$T$ expansion (\ref{lowstatphase}). The
higher-order terms in (\ref{lowstatphase}) are
found from the higher-moment Gaussian integrals \cite{me} and they will be
analysed in Section 7. For now, we concern ourselves only with the
lowest-order term in the stationary-phase series of (\ref{classpart}).

The field of equivariant localization theory was essentially born in 1982
when Duistermaat and Heckman \cite{dh} found a general class of Hamiltonian
systems for which the leading-order of the stationary-phase approximation gives
the {\it exact} result for the partition function (\ref{classpart}) (i.e. for
which the ${\cal O}(1/T^{n+1})$ correction terms in (\ref{lowstatphase}) all
vanish). Roughly speaking, the Duistermaat-Heckman theorem goes as follows.
Let $\man$ be a compact symplectic manifold. Suppose that the vector field $V$
defined by (\ref{hameqvloc}) generates the global Hamiltonian action of a torus
group ${\cal T}=(S^1)^m$ on $\man$ (where we shall usually assume that $m=1$
for simplicity). Since the critical point set
of the Hamiltonian $H$ coincides with the fixed-point set $\man_V$ of the
$\cal T$-action on $\man$ we can apply the equivariant Darboux theorem to
the Hamiltonian system at hand \cite{guill}. This generalization of Darboux's
theorem tells us that not only can we find a local canonical system of
coordinates in a neighbourhood of each critical point in which the symplectic
2-form looks like (\ref{omdarboux}), but these coordinates can further be
chosen so that the origin $p_\mu=q^\mu=0$ of the coordinate neighbourhood
represents the fixed point $p$ of the given compact group action on $\man$.
This means that in these canonical coordinates the torus action is (locally)
linear and has the form of $n$ canonical rotation generators (rotations in each
$(p_\mu,q^\mu)$ plane) \cite{dh}
\beq
V=\sum_{\mu=1}^n\frac{\lambda_\mu(p)}{i}\left(p_\mu\frac{\partial}{\partial
q^\mu}-q^\mu\frac{\partial}{\partial p_\mu}\right)~~~~~,~~~~~p\in\man_V
\label{eqdarboux}\eeq
where $\lambda_\mu(p)$ are weights that will be specified shortly. From the
Hamilton equations (\ref{hameqvloc}) it follows that the Hamiltonian near each
critical point $p$ can be written in the quadratic form
\beq
H(x)=H(p)+\sum_{\mu=1}^n\frac{i\lambda_\mu(p)}{2}\left(p_\mu^2+q_\mu^2\right)
\label{hamquadr}\eeq
In these coordinates the flows determined by the Hamilton equations of motion
(\ref{hameqcan}) are the circles $p_\mu(t),q^\mu(t)\sim\e^{i\lambda_\mu t}$
about the critical points, which gives an explicit representation of the
Hamiltonian $\cal T$-action locally on $\man$ and the group action preserves
the Darboux coordinate neighbourhood. Thus each neighbourhood integration above
is purely Gaussian and so all higher-order terms in the stationary-phase
evaluation of (\ref{classpart}) vanish and the partition function is given
exactly by the leading term in (\ref{lowstatphase}) of its stationary-phase
series\footnote{\baselineskip=12pt Of course the proof is completed by showing
that there is no regular function of $T$ contributing in this case to
(\ref{lowstatphase}) -- for details we refer to \cite{dh}.}. This theorem
therefore has the potential of supplying a large class of
dynamical systems whose partition function (and hence all thermodynamic and
statistical observables) can be evaluated exactly.

Atiyah and Bott \cite{atiyahbott} pointed out that the basic principle
underlying the Duistermaat-Heckman theorem is not that of stationary-phase, but
rather of the more general localization properties of equivariant cohomology
that we discussed in the last Section. Suppose that the Hamiltonian vector
field $V$ generates a global, symplectic circle action on the phase space
$\man$. Suppose further that $\man$ admits a globally defined
Riemannian structure for which $V$ is Killing vector, as in Subsection 2.5.
Recall from the last Subsection that the symplecticity of the circle action
implies that
$\omega+H$ is the equivariant extension of the symplectic 2-form $\omega$, i.e.
$D_V(\omega+H)=0$. Since integration over the $2n$-dimensional manifold $\man$
picks up the $2n$-degree component of any differential form, it follows that
the
partition function (\ref{classpart}) can be written as
\beq
Z(T)=\int_\man\alpha
\label{ztalpha}\eeq
where $\alpha$ is the inhomogeneous differential form
\beq
\alpha=\frac{1}{(iT)^n}\e^{iT(H+\omega)}=\frac{1}{(iT)^n}\e^{iTH}\sum_{k=0}^n
\frac{(iT)^k}{k!}\omega^k
\label{alphadh}\eeq
whose $2k$-form component is $\alpha^{(2k)}=\e^{iTH}\omega^k/(iT)^{n-k}k!$.
Since $H+\omega$ is equivariantly closed, it follows that $D_V\alpha=0$. Thus
we can apply the Berline-Vergne localization formula (\ref{berlineloc}) to
the integral (\ref{ztalpha}) to get
\beq
Z(T)=\left(\frac{2\pi i}{T}\right)^n\sum_{p\in\man_V}\frac{\e^{iTH(p)}}{
\pfaff~dV(p)}
\label{ztberline}\eeq

In the case at hand the denominator of (\ref{ztberline}) at a critical point
$p$ is found from the Hamilton equations (\ref{hameqvloc}) which give
\beq
dV(p)=\omega^{-1}(p)\hess(p)
\label{dvpomhess}\eeq
and so we see how the determinant factors appear in the formula
(\ref{lowstatphase}). However, we have to remember that the Pfaffian also
encodes a specific choice of sign when taking the square root determinant.
The sign of the Pfaffian $\pfaff~dV(p)$ can be determined by examining it in
the equivariant Darboux coordinates above in which the matrix $\omega(p)$ is
skew-diagonal with skew-eigenvalues 1 and the Hessian $\hess(p)$ which comes
from (\ref{hamquadr}) is diagonal with eigenvalues $i\lambda_\mu(p)$ each of
multiplicity 2. It follows that in these coordinates the matrix $dV(p)$ is
skew-diagonal with skew-eigenvalues $i\lambda_\mu(p)$. Introducing the
eta-invariant $\eta(\hess(p))$ of $\hess(p)$, defined as the difference
between the number of positive and negative eigenvalues of the Hessian of $H$
at $p$, i.e. its spectral asymmetry, we find
\beq
\eta(\hess(p))=2\sum_{\mu=1}^n\sgn~i\lambda_\mu(p)
\label{etahess}\eeq
which is related to the Morse index of $H$ at $p$ by
\beq
\eta(\hess(p))=2n-2\lambda(p)
\label{etamorse}\eeq
Using the identity $\pm1=\e^{i\frac{\pi}{2}(\pm1-1)}$ it follows that
\beq
\sgn~\pfaff~dV(p)=\prod_{\mu=1}^n\sgn~i\lambda_\mu(p)=\e^{i\frac{\pi}{2}(
\frac{1}{2}\eta(\hess(p))-n)}=\e^{-i\frac{\pi}{2}\lambda(p)}=(-i)^{\lambda(p)}
\label{sgnpfaff}\eeq
and so substituting (\ref{dvpomhess}) and (\ref{sgnpfaff}) into
(\ref{ztberline}) we arrive finally at the Duistermaat-Heckman integration
formula
\beq
Z(T)=\left(\frac{2\pi i}{T}\right)^n\sum_{p\in\man_V}(-i)^{\lambda(p)}\e^{iT
H(p)}\sqrt{\frac{\det\omega(p)}{\det\hess(p)}}
\label{dhformula}\eeq

Recall from Subsection 2.6 that $dV(p)$ is associated with the
anti-self-adjoint
linear operator $L_V(p)$ which generates the infinitesimal circle (or torus)
action on the tangent space $T_p\man$. From the above it then follows that the
complex numbers $\lambda_\mu(p)$ introduced in (\ref{eqdarboux}) are just the
weights (i.e. eigenvalues of the Cartan generators) of the complex linear
representation of the circle (or torus) action in the tangent space at $p$ and
the determinant
factors from (\ref{ztberline}) appear in terms of them as the products
\beq
e(p)=(-1)^{\lambda(p)/2}\prod_{\mu=1}^n\lambda_\mu(p)
\label{weightprod}\eeq
as if each unstable mode contributes a factor of $i$ to the integral for
$Z(T)$ above. In fact, the Pfaffian $\pfaff~dV(p)$ which appears in
(\ref{ztberline}) is none other than the equivariant Euler characteristic
class $E_V({\cal N}_p)=~\pfaff~dV(p)$ (see (\ref{eqeuler})) of the normal
bundle ${\cal N}_p$ in $\man$ of each critical point $p\in\man_V$. The normal
bundle is defined as the bundle of points normal to the directions of the
critical point set $\man_V$, so that in a neighbourhood near $\man_V$ we can
write the local coordinates as $x=p+p_\perp$ with $p\in\man_V$ and
$p_\perp\in{\cal N}_p$. By its construction, $\pfaff~dV(p)$ is taken over
${\cal N}_p$ (see Subsection 2.6). Thus the terms in (\ref{ztberline}) define
an
equivariant cohomology class in $H_{U(1)}^{2n}(\man)$. From (\ref{eqdarboux})
it follows that the induced circle action on ${\cal N}_p$ is through
non-trivial irreducible representations and we can therefore decompose the
normal bundle at $p\in\man_V$ into a direct (Whitney) sum of 2-plane bundles
with respect to this group action,
\beq
{\cal N}_p=\bigoplus_{\mu=1}^nN_p^{(\mu)}
\label{npwhitney}\eeq
(\ref{eqdarboux}) then implies that the equivariant Euler class of
$N_p^{(\mu)}$ is simply $E_V(N_p^{(\mu)})=i\lambda_\mu(p)/2$. Taking into
account the proper orientation of ${\cal N}_p$ induced by the Hamiltonian
vector field near $x=p$ and the Liouville measure, and using the
multiplicativity of the Euler class under Whitney sums of bundles
\cite{berlinegetz}, we find that the equivariant Euler class of the normal
bundle at $p$ is
\beq
E_V({\cal N}_p)=\prod_{\mu=1}^nE_V(N_p^{(\mu)})\equiv e(p)
\label{eqeulerp}\eeq
which is just the weight product (\ref{weightprod}). Thus, for Hamiltonians
that generate circle actions, the 1-loop contribution to the classical
partition function (i.e. the Duistermaat-Heckman formula in the form
(\ref{ztberline})) describes the equivariant cohomology of the phase space
with respect to the Hamiltonian circle action on $\man$. The particular value
of the Duistermaat-Heckman formula depends on the equivariant cohomology
group $H_{U(1)}^{2n}(\man)$ of the manifold $\man$. All the localization
formulas we shall derive in this Review will be represented by equivariant
characteristic classes, so that the partition functions of the physical systems
we consider provide representations for the equivariant cohomology of the phase
space $\man$. This is a consequence of the cohomological localization principle
of Subsection 2.5.

Given the remarkable cohomological derivation
of the Duistermaat-Heckman formula above which followed from the quite
general principles of equivariant cohomology of the last Section, one could
try to develop more general types of localization formulas from these general
principles in the hopes of being able to generate more general types of
integration formulas for the classical partition function. Moreover, given
the localization criteria of the last Section this has the possibility of
expanding the set of dynamical systems whose partition functions are exactly
solvable. We stress again that the crucial step in this cohomological
derivation is the assumption that the Hamiltonian flows of the dynamical
system globally generate isometries of a metric $g$ on $\man$, i.e. the
Hamiltonian vector field $V$ is a global Killing vector of $g$ (equivalently,
as we will see, for $\man$ compact, the classical flows $x(t)$ trace out a
torus $\cal T$ in $\man$). This geometric condition and a classification of the
dynamical systems for which these localization constraints do hold true will be
one of our main topics in what follows. The extensions and applications of the
Duistermaat-Heckman localization formula and the general formalism of
equivariant cohomology for dynamical systems will be the focus of the remainder
of this Section.

\subsection{Morse Theory and Kirwan's Theorem}

There is a very interesting and useful connection between the
Duistermaat-Heckman theorem and the Morse theory determined by the
non-degenerate Hamiltonian $H$. Morse theory relates the structure of the
critical points of a Morse function $H$ to the topology of the manifold $\man$
on which it is defined. We very briefly now review some of the basic ideas in
Morse theory (see \cite{nash} for a comprehensive introduction).
Given a Morse function $H$ as above, we define its Morse series
\beq
M_H(t)=\sum_{p\in\man_V}t^{\lambda(p)}
\label{morseseries}\eeq
which is a finite sum because the non-degeneracy of $H$ implies that its
critical points are all discrete and the compactness of $\man$ implies that the
critical point set $\man_V$ is finite. The topology of the manifold $\man$
now enters the problem through the Poincar\'e series of $\man$, which is
defined by
\beq
P_\man(t;\IF)=\sum_{k=0}^{2n}\dim_\IF H^k(\man;\IF)t^k
\label{poinseries}\eeq
where $\IF$ is some algebraic field (usually $\IR$ or $\IC$). The fundamental
result of Morse theory is the inequality
\beq
M_H(t)\geq P_\man(t;\IF)
\label{morseineq}\eeq
for all fields $\IF$. If equality holds in (\ref{morseineq}) for all fields
$\IF$, then we say that $H$ is a perfect Morse function. The inequality
(\ref{morseineq}) leads to various relations between the critical
points of $H$ and the topology of $\man$. These are called the Morse
inequalities, and the only feature of them that we shall really need in the
following is the fact that the number of critical points of $H$ of a given
Morse index $k\geq0$ is always at least the number $\dim_\IR H^k(\man;\IR)$.
This puts a severe restriction on the types of non-degenerate functions that
can exist as $C^\infty$-maps on a manifold of a given topology.

Another interesting relation is obtained when we set $t=-1$ in the Morse and
Poincar\'e series. In the former series we get
\beq
M_H(-1)=\sum_{p\in\man_V}\sgn\det\hess(p)
\label{mh-1}\eeq
while (\ref{eulerchar}) shows that in the latter series the result is the Euler
characteristic $\chi(\man)$ of $\man$. That these 2 quantities are equal is
known as the Poincar\'e-Hopf theorem, and employing further the
Gauss-Bonnet-Chern theorem (\ref{gaussbonchern}) we find
\beq
\sum_{p\in\man_V}\sgn\det\hess(p)=\frac{(-1)^n}{(4\pi)^nn!}\int_\man E(R)
\label{phgbc}\eeq
with $E(R)$ the Euler class constructed from a Riemann curvature 2-form $R$ on
$\man$. This relation gives a very interesting connection between the structure
of the critical point set of a non-degenerate function and the topology and
geometry of the phase space $\man$. We remark that one can also define
equivariant versions of the Morse and Poincar\'e series using the topological
definition of equivariant cohomology \cite{nash} which is suitable to the
equivariant cohomological ideas that we formulated earlier on. These
equivariant generalizations which localize topological integrals such as
(\ref{phgbc}) onto the zero locus of a vector field is the basis of
the Mathai-Quillen formalism and its application to the construction of
topological field theories
\cite{blau2,blauthom,cordes,kalkman1,mathai,niemipalo2}. We shall discuss some
of these ideas in Section 8.

In regards to the Duistermaat-Heckman theorem, there is a very interesting
Morse theoretical result due to Kirwan \cite{kirwan}. Kirwan showed that
the only Morse functions for which the stationary phase approximation can
be exact are those which have only even Morse indices $\lambda(p)$. This
theorem includes the cases where the Duistermaat-Heckman integration formula
is exact, and under the assumptions of the Duistermaat-Heckman theorem it is a
consequence of the circle action (see the previous Subsection). However, this
result is even stronger -- it means that when one constructs the
full stationary-phase series as described in the last Subsection \cite{paniak},
if that series converges uniformly in $1/T$ to the exact partition function
$Z(T)$, then the Morse index of every critical point of $H$ must be even.
 From the Morse inequalities mentioned above this furthermore gives a relation
between equivariant localization and the topology of the phase space of
interest -- if the manifold $\man$ has non-trivial cohomology groups of odd
dimension, then the stationary phase series diverges for any Morse function
defined on $\man$ and in particular the Duistermaat-Heckman localization
formula for such phase spaces can never give the exact result for $Z(T)$.
In this way, Kirwan's theorem rules out a large number of dynamical systems
for which the stationary phase approximation could be exact in terms of the
topology of the underlying phase space where the dynamical system lives.
Moreover, an application of the Morse lacunary principle \cite{nash} shows
that, when the stationary-phase approximation is exact so that $H$ has only
even Morse indices, $H$ is in fact a perfect Morse function and
its Morse inequalities become equalities. We shall not go into the rather
straightforward proof of Kirwan's theorem here, but refer to \cite{kirwan}
for the details. In the following we can therefore use Kirwan's theorem as an
initial test using the topology of the phase space to determine which
dynamical systems will localize in the sense of the Duistermaat-Heckman
theorem. In Section 7 we shall see the direct connection between the higher
order terms in the saddle-point series for the partition function and Kirwan's
theorem, and more generally the geometry and topology of the manifold $\man$.

\subsection{Examples: The Height Function of a Riemann Surface}

We now present some concrete examples of
the equivariant localization formalism presented above. One of the most
common examples in both Morse theory and localization theory is the
dynamical system whose phase space is a compact Riemann surface $\Sigma^g$ of
genus $g$ (i.e. a closed surface with $g$ `handles') and whose Hamiltonian
$h_{\Sigma^g}$ is the height function on $\Sigma^g$
\cite{blauthom,keski,nash,semsz,me}. For instance, we have already encountered
the case of the Riemann sphere $\Sigma^0=S^2$ in Subsection 2.1 with the height
function $h_{\Sigma^0}$ given by (\ref{heights2}). The symplectic 2-form is the
usual volume form
\beq
\omega_{\Sigma^0}=d\cos\theta\wedge d\phi
\label{oms2}\eeq
induced by the Euclidean metric $g_{\mu\nu}=\delta_{\mu\nu}$ of $\IR^3$ from
the embedding of $S^2$ in 3-dimensional space. The partition function
\beq
Z_{\Sigma^0}(T)=\int_{\Sigma^0}\omega_{\Sigma^0}\e^{iTh_{\Sigma^0}}
\label{ZSig0}\eeq
is given by the expression (\ref{zs2eval}) which is precisely the value
anticipated from the Duistermaat-Heckman theorem. The relative
minus sign in the last line of (\ref{zs2eval}) comes from the fact that the
Morse index of the maximum $\theta=\pi$ is 2 while that of $\theta=0$ is 0. The
vector field generating the compact group action on $S^2$ associated with rigid
rotations of the sphere is $V=\frac{\partial}
{\partial\phi}$, and the corresponding moment map is just $h_{\Sigma^0}$.

The Poincar\'e series of the 2-sphere is\footnote{\baselineskip=12pt In
general,
if $\man$ is path-connected, as we always assume here, then $H_0(\man;\IZ)=\IZ$
and if $\man$ is closed, then $H_{2n}(\man;\IZ)=\IZ$. The intermediate
homology groups depend on whether or not $\man$ has `holes' in it or not.}
\beq
P_{S^2}(t;\IF)=\sum_{k=0}^2\dim_\IF H^k(S^2;\IF)t^k=1+t^2
\label{poins2}\eeq
which coincides with the Morse series (\ref{morseseries}) for the height
function $h_{\Sigma^0}$. Thus, consistent with Kirwan's theorem, we see that
$h_{\Sigma^0}$ is a perfect Morse function with even Morse indices.
Notice that the Hamiltonian vector field $V=\frac{\partial}{\partial\phi}$
here generates an isometry of the standard round metric $d\theta\otimes d
\theta+\sin^2\theta~d\phi\otimes d\phi$ induced by the flat Euclidean metric
of $\IR^3$. The differential form (\ref{xi}) with this metric is
$\xi=d\phi$, which as expected is ill-defined at the 2 poles of $S^2$. Now the
partition function can be written as
\beq
Z_{\Sigma^0}(T)=-\frac{1}{iT}\int_{\Sigma^0}d\left(\e^{iTh_{\Sigma^0}}d\phi
\right)
\label{zs2exact}\eeq
thus receiving contributions from only the critical points $\theta=0,\pi$, the
endpoints of the integration range for $\cos\theta$, in agreement with the
explicit evaluation in Subsection 2.1. The partition function (\ref{zs2eval})
represents the equivariant cohomology classes in\footnote{\baselineskip=12pt
Equivariant cohomology groups are usually
computed using so-called classifying bundles of Lie groups (the topological
definition of equivariant cohomology) -- see \cite{nash}, for example.}
\beq
H_{U(1)}^2(S^2)=\IZ\oplus\IZ
\label{equivs2}\eeq
Intuitively, (\ref{equivs2}) follows from the fact that the single Lie algebra
generator $\Phi\in\IR$ and the invariant volume form (\ref{oms2}) of $S^2$ are
linearly independent, i.e. for any 2 functions $f_1,f_2\in S({\bf u(1)}^*)$,
the equivariant cohomology classes are spanned by the linearly independent
generators $f_1(\Phi)(\omega_{\Sigma^0}+\Phi h_{\Sigma^0})$ and $f_2(\Phi)$.

As we shall see later on, the above example for the Riemann sphere is
essentially the only Hamiltonian system to which the geometric equivariant
localization constraints apply on a simply-connected phase space (i.e.
$H_1(\man;\IZ)=0$). The situation is much different on a multiply-connected
phase space, which as we shall see is due to the fact that the non-trivial
first homology group of the phase space severely restricts the allowed $U(1)$
group actions on it and hence the Morse functions thereon. For example,
consider the case of a genus 1 Riemann surface \cite{keski,semsz,me}, i.e.
$\Sigma^1$ is the 2-torus $T^2=S^1\times S^1$. The torus can be viewed as a
parallelogram in the complex plane with its opposite edges identified. We take
as horizontal edge the line segment from 0 to 1 along the real axis and the
other slanted edge the line segment from 0 to some complex number $\tau$ in
the complex plane. The number $\tau$ is called the modular parameter of the
torus and we can take it to lie in the upper complex half-plane
\beq
\IC^+=\{z\in\IC:{\rm Im}~z>0\}
\label{upperhalf}\eeq
Geometrically, $\tau$ determines the inner and outer radii of the 2 circles of
the torus, and it labels the inequivalent complex structures of $\Sigma^1$
\footnote{\baselineskip=12pt In algebraic geometry one would therefore say that
$\IC^+$ is the
Teichm\"uller space of the torus. The Teichm\"uller space of a simply-connected
Riemann surface is a point, so that there is a unique complex structure
(i.e. a unique way of defining complex coordinates) in genus 0. This is
a consequence of the celebrated Riemann uniformization theorem. We refer to
\cite{nash} and \cite{schlich} for an elementary introduction to Teichm\"uller
spaces in algebraic geometry, while a more extensive treatment can be found
in \cite{imayoshi}.}.

We view the torus embedded in 3-space as a doughnut standing on end on the
$xy$-plane and centered symmetrically about the $z$-axis. If $(\phi_1,\phi_2)$
are the angle coordinates on $S^1\times S^1$, then the height function on
$\Sigma^1$ can be written as
\beq
h_{\Sigma^1}(\phi_1,\phi_2)=r_2-(r_1+{\rm Im}~\tau\cos\phi_1)\cos\phi_2
\label{heightt2}\eeq
where $r_1=|{\rm Re}~\tau|+{\rm Im}~\tau$ and $r_2=|{\rm Re}~\tau|+2~{\rm
Im}~\tau$ label the inner and outer radii of the torus. The symplectic volume
form on $T^2$ is just that induced by the identification of $\Sigma^1$ as a
parallelogram in the plane with its opposite edges identified, i.e. the
Darboux 2-form
\beq
\omega_D=d\phi_1\wedge d\phi_2
\label{omdarbouxt2}\eeq
The associated Hamiltonian vector field for this dynamical system has
components
\beq
V_{\Sigma^1}^1=-(r_1+{\rm Im}~\tau\cos\phi_1)\sin\phi_2~~~,~~~V_{\Sigma^1}^2=
{\rm Im}~\tau\sin\phi_1\cos\phi_2
\label{hamvect2}\eeq

The Hamiltonian (\ref{heightt2}) has 4 isolated non-degenerate critical points
on $S^1\times S^1$ -- a maximum at $(\phi_1,\phi_2)=(0,\pi)$ (top of the
outer circle), a minimum at $(0,0)$ (bottom of the outer circle), and 2
saddle points at $(\pi,0)$ and $(\pi,\pi)$ (corresponding to the bottom and
top of the inner circle, respectively). The Morse index of the maximum is 2,
that of the minimum is 0, and those of the 2 saddle points are both 1.
According to Kirwan's theorem, the appearence of odd Morse indices, or
equivalently the fact that
\beq
H_1(\Sigma^1;\IZ)=\IZ\oplus\IZ
\label{homologyt2}\eeq
with each $\IZ$ labelling the windings around the 2 independent
non-contractable loops associated with each $S^1$-factor, implies that the
Duistermaat-Heckman integration formula should
fail in this case. Indeed, evaluating the right-hand side of the
Duistermaat-Heckman formula (\ref{dhformula}) gives
\beq\new{\begin{array}{l}
\frac{2\pi i}{T}\sum_{p\in\man_{V_{\Sigma^1}}}\frac{\e^{iT
h_{\Sigma^1}(p)}}{e(p)}\\=\frac{2\pi i}{T\sqrt{{\rm Im}~\tau}}\left[r_2^{-1/2}
\left(1+\e^{2iTr_2}\right)+|{\rm Re}~\tau|^{-1/2}\e^{2iT{\rm Im}\tau}
\left(1-\e^{2iT|{\rm Re}\tau|}\right)\right]\end{array}}
\label{dht2}\eeq
which for the parameter values $iT=1$ and $\tau=1+i$ gives the numerical value
\beq
2\pi\e^3\left(\frac{2}{\sqrt{3}}\sinh3+2\cosh1\right)\sim1849.33
\label{dht2num}\eeq
On the other hand, an explicit evaluation of the partition function gives
\beq
Z_{\Sigma^1}(T)=\int_0^{2\pi}\!\!\int_0^{2\pi}d\phi_1~d\phi_2~\e^{iTh_{\Sigma
^1}(\phi_1,\phi_2)}=2\pi\e^{r_2}\int_0^{2\pi}d\phi_1~J_0\left(iT(r_1+{\rm Im}~
\tau\cos\phi_1)\right)
\label{ztt2expl}\eeq
with $J_0$ the regular Bessel function of order 0 \cite{gradryz}. For the
parameter values above, a numerical integration in (\ref{ztt2expl}) gives $Z_
{\Sigma^1}\sim2117.13$ \footnote{\baselineskip=12pt All numerical integrations
in this Review
were performed using the mathematical software package {\it MATHEMATICA}.},
contradicting the result (\ref{dht2num}). Thus even
though in this case the Hamiltonian $h_{\Sigma^1}$ is a perfect Morse function,
it doesn't generate any torus action on the phase space here.

This argument can be extended to the case where the phase space is a
hyperbolic Riemann surface $\Sigma^g$, $g>1$ \cite{semsz}. For $g>1$,
$\Sigma^g=\Sigma^1\#\cdots\#\Sigma^1$ is the $g$-fold connected sum of 2-tori
and therefore its first homology group is
\beq
H_1(\Sigma^g;\IZ)=\bigoplus_{i=1}^{2g}\IZ
\label{homologysig}\eeq
It can be viewed in $\IR^3$ as $g$ doughnuts stuck together on end and
standing on the $xy$-plane. The height function on $\Sigma^g$ now has $2g+2$
critical points consisting of 1 maximum, 1 minimum and $2g$ saddle points.
Again the maximum and minimum have Morse indices 2 and 0, respectively, while
those of the $2g$ saddle points are all 1. As a consequence the perfect Morse
function $h_{\Sigma^g}$ generates no torus action on $\Sigma^g$.

The above non-exactness of the stationary-phase approximation (and even worse
the divergence of the stationary-phase series for (\ref{heightt2})) is a
consequence of the fact that the orbits of the vector field (\ref{hamvect2})
do not generate a global, compact group action on $\Sigma^1$. Here the orbits
of the Hamiltonian vector field bifurcate at the saddle points (like the
classical trajectories of the simple pendulum which cross each other in figure
eights), and we shall see explicitly in Section 7 why its flows cannot generate
isometries of any metric on $\Sigma^1$ as well as how this makes the stationary
phase series diverge. The extensions of the equivariant localization principle
to non-compact group actions and to non-compact phase spaces are not always
immediate \cite{blauthom}. A version of the Duistermaat-Heckman theorem
appropriate to both abelian and non-abelian group actions on non-compact
manifolds has been presented by Prato and Wu in \cite{prato}. This non-compact
version of the Duistermaat-Heckman theorem assumes that there is a component of
the moment map which is regular and bounded from below (so that the
Fourier-Laplace transform $Z(T)$ exists). The above examples illustrate the
strong topological dependence of the dynamical systems to which equivariant
localization is applicable. The height function restricted to a compact
Riemann surface can only be used for Duistermaat-Heckman localization in genus
0, and the introduction of more complicated topologies restricts even further
the class of Hamiltonian systems to which the localization constraints apply.
We shall investigate this phenomenon in a more detailed geometric setting
later on when we consider quantum localization techniques.

\subsection{Equivariant Localization and Classical Integrability}

In this Subsection we discuss an interesting connection between the
equivariant localization formalism and integrable Hamiltonian systems
\cite{karki,keski}. By an integrable dynamical system we mean this in the
sense of the Liouville-Arnold theorem which is a generalized, coordinate
independent version of the classical Liouville theorem that dictates when
a given Hamiltonian system will have equations of motion whose solutions can
be explicitly found by integrating by quadratures \cite{das,goldstein}. The
Liouville-Arnold theorem is essentially a global version of Darboux's theorem
and it states that a Hamiltonian is integrable if one can find canonically
conjugated action-angle variables $(I_\mu,\phi^\mu)_{\mu=1}^n$,
\beq
\left\{I_\mu,\phi^\nu\right\}_\omega=\delta_\mu^\nu
\label{actionanglebra}\eeq
defined almost everywhere on the phase space $\man$, such that the Hamiltonian
$H=H(I)$ is a functional of only the action variables \cite{arnold}. The action
variables themselves are supposed to be functionally-independent and in
involution,
\beq
\left\{I_\mu,I_\nu\right\}_\omega=0
\label{actioninv}\eeq
and from the Hamilton equations of motion (\ref{hameqmotion}) it follows that
\beq
\dot I_\mu(t)=\left\{I_\mu,H(I)\right\}_\omega=0
\label{actionconst}\eeq
so that the time-evolution of the action variables is constant. Consequently,
(\ref{actioninv}) implies that the action variables generate a Cartan
subalgebra $(S^1)^n$ of the Poisson algebra of the phase space, and the $I_\mu$
therefore label a set of canonically invariant tori on the phase space which
are called Liouville tori. The motion of $H(I)$ is constrained to the Liouville
tori, and the system is therefore integrable in the sense that we have found
$n$
independent degrees of freedom for the classical motion. The $I_\mu$'s in
simple problems are conserved quantities such as the total energy or angular
momentum which generate a particular symmetry of the dynamics, such as
time-independence or radial symmetry. The symplectic 2-form in the action-angle
variables is
\beq
\omega=dI_\mu\wedge d\phi^\mu
\label{omactionang}\eeq
and the corresponding symplectic potential which generates the Hamiltonian as
the moment map of a global $U(1)$ group action on $\man$ as in
(\ref{ivathetaF}) is
\beq
\theta_F\equiv\theta+dF=I_\mu d\phi^\mu
\label{thetafactionang}\eeq

The connection between integrability and equivariant localization now becomes
rather transparent. The above integrability requirement that $H$ be a
functional of some torus action generators is precisely the requirement of the
Duistermaat-Heckman theorem. The global solutions to the Hamilton equations of
motion in this case are
\beq
I_\mu(t)=I_\mu(0)~~~~~,~~~~~\phi^\mu(t)=\phi^\mu(0)+\omega^\mu(I)t
\eeq
where $\omega^\mu(I)=\partial_{I_\mu}H(I)$. The classical trajectories of the
dynamical system therefore move along the Liouville tori with constant angular
velocity $\dot\phi^\mu=\omega^\mu$ (equivalently they share a common period)
which represents a large symmetry of the classical mechanics. The Hamiltonian
vector field $\frac{\partial}{\partial\phi^\mu}$ associated with the action
variable $I_\mu$ generates the $\mu$-th circle action component of the full
torus action on $\man$, and consequently any Hamiltonian which is a {\it
linear} combination of the action variables will generate a torus action on
$\man$ and meet the criteria of the Duistermaat-Heckman theorem. For quadratic
and higher-order functionals of the action variables the associated Hamiltonian
vector field in general generates a circle action which does not have a
constant angular velocity on the phase space and the Duistermaat-Heckman
formula will not hold. We shall see, however, that modified versions of the
Duistermaat-Heckman localization formula can still be derived, so that {\it
any} integrable model will provide an example of a partition function that
localizes. For the height function of $S^2$ above, the action-angle variables
are $I_1=h_{\Sigma^0}=a-\cos\theta$, $\phi^1=\phi$. We shall see some more
general (higher-dimensional) examples in Section 5.

Recall that one of the primary assumptions in the
localization framework above was that the phase space admit a Riemannian
metric $g$ which is {\it globally} invariant under the $U(1)$ action on it. The
existence of a Riemannian geometry which is globally invariant under the
classical dynamics of a given Hamiltonian system is a very strong requirement.
A $U(1)$-invariant metric tensor always exists {\it locally} in the regions
where $H$ has no critical points. To see this, introduce local equivariant
Darboux coordinates $(p_1,\dots,p_n,q^1,\dots,q^n)$ in that region in which the
Hamiltonian vector field generates translations in $q^1$. This means that
$H=p_1$ is taken as the radius of this equivariant Darboux coordinate system.
The $U(1)$-invariant metric tensor can then be taken to be any metric tensor
whose components are independent of the coordinate $q^1$ (e.g. $g_{\mu\nu}=
\delta_{\mu\nu}$), which follows from the Killing equations (\ref{liegcoord}).
However, there may be global obstructions to extending
these local metrics to metrics defined globally on the entire phase space in
a smooth way. This feature is just equivalent to the well-known fact that any
Hamiltonian system is locally integrable. This is easily seen from the local
representation (\ref{eqdarboux}),(\ref{hamquadr}) where we can define $p_\mu
=I_\mu\cos\phi^\mu,q^\mu=I_\mu\sin\phi^\mu$. Then $H\sim\sum_{\mu=1}^nI_\mu^2$
and $V\sim\sum_{\mu=1}^n\frac{\partial}{\partial\phi^\mu}$ generates
translations in the local angle variables $\phi^\mu$ (rigid rotations of the
local coordinate neighbourhood). Then locally the metric tensor components
$g_{\mu\nu}$ should be taken to depend only on the action variables $I_\mu$
(i.e. $g$ is radially symmetric in the coordinate neighbourhood).

However, local integrability does
not necessarily ensure global integrability. For the latter to follow, it is
necessary that the neighbourhoods containing the conserved charges $I_\mu$ be
patched together in such a way as to yield a complete set of conserved charges
defined almost everywhere on the phase space $\man$. Furthermore, global
integrability also implements strong requirements on the behaviour of $H$ in
the vicinity of its critical points. As we shall see later on, the isometry
group of a compact Riemannian manifold is also compact, so that the {\it
global} existence of an invariant
metric tensor in the above for a compact phase space is equivalent to the
requirement that $H$ generates the global action of a circle (or more generally
a torus). This means that the Hamiltonian vector field $V$ is a Cartan element
of the algebra of isometries of the metric $g$ (or equivalently $H$ is a Cartan
element of the corresponding Poisson algebra). In other words, $H$ is a
globally-defined action variable (or a functional thereof), so that the
applicable Hamiltonians within the framework of equivariant localization
determine integrable dynamical systems. Thus it is the isometry condition that
puts a rather severe restriction on the Hamiltonian functions which generate
the circle action through the relation (\ref{hameqv}). These features also
appear in the infinite-dimensional generalizations of the localization
formalism
above and they will be discussed at greater length in Sections 5 and 6.

We note also that for an integrable Hamiltonian $H$ we can construct an
explicit representation of the function $F$ which appears in (\ref{ivathetaF})
and (\ref{thetafactionang}) above. Indeed, the function $K$ in
(\ref{hkpoisson})
can be constructed locally outside of the critical point set of $H$ by
assuming that a given action variable $I_\mu$ is such that
\beq
\frac{\partial H(I)}{\partial I_\mu}\neq0
\label{critnot0}\eeq
In this case, the function $K$ can be realized explicitly by
\beq
K(I,\phi)=\phi^\mu\cdot\left(\frac{\partial H}{\partial I_\mu}\right)^{-1}
\label{kexpl}\eeq
and the condition (\ref{ivathetaF}) becomes
\beq
i_V\theta_F=I_\mu\frac{\partial H}{\partial I_\mu}+\{H,F\}_\omega=H
\label{ivthetafexpl}\eeq
which is satisfied by
\beq
F=K\cdot\left(H-I_\mu\frac{\partial H}{\partial I_\mu}\right)+G(I)
\label{Fexpl}\eeq
where $G(I)$ is an arbitrary function of the action variables. Consequently,
in a neighbourhood where action-angle variables can be introduced and where
$H$ does not admit critial points, we get an explicit realization of the
function $F$ in (\ref{ivathetaF}) and thus a locally invariant symplectic
potential $\theta_F$.

In fact, given the equivariantly closed 2-form $K_V+\Omega_V$ introduced in
(\ref{dvbeta}), we note that $\Omega_V$ is a closed 2-form (but not
necessarily non-degenerate) and that the function $K_V$ satisfies
\beq
dK_V=-i_V\Omega_V
\label{dkv}\eeq
as a consequence of (\ref{omegav}) and (\ref{liebeta0}), respectively. It
follows that
\beq
V^\mu=\Omega_V^{\mu\nu}\partial_\nu K_V=\omega^{\mu\nu}\partial_\nu H
\label{bihampair}\eeq
and so the classical equations of motion for the 2 Hamiltonian systems
$(\man,\omega,H)$ and $(\man,\Omega_V,K_V)$
coincide\footnote{\baselineskip=12pt Here we assume
that $\Omega_V$ is non-degenerate on $\man$ except possibly on submanifolds
of $\man$ of codimension at least 2, since when it is degenerate some of the
equations in (\ref{dkv}) should be considered as constraints. On these
submanifolds, the Hamiltonian $K_V$ must then vanish in order to keep the
equations of motion non-singular \cite{arnold}.},
\beq
\dot x^\mu(t)=\left\{x^\mu,H\right\}_\omega=\left\{x^\mu,K_V\right\}_
{\Omega_V}
\label{sameeqmotion}\eeq
This means that these 2 dynamical systems determine a bi-Hamiltonian structure.
There are 2 interesting consequences of this structure. The
first follows from the fact that if $H=H(I)$ as above is integrable, then
these action-angle variables can be chosen so that in addition $K_V=K_V(I)$ is
an integrable Hamiltonian. We can therefore replace $H$ everywhere in
(\ref{critnot0})--(\ref{Fexpl}) by the function $K_V$ and $\omega$ by
$\Omega_V$, and after a bit of algebra we find that the 1-form $\theta_F^{(V)}$
above which generates $\Omega_V$ satisfies
\beq
K_V+\Omega_V=D_V\theta_F^{(V)}
\label{eqpoinlemma}\eeq
and likewise
\beq
H+\omega=D_V\theta_F
\eeq
Since both $H+\omega$ and $K_V+\Omega_V$ are equivariantly closed, we see that
for an integrable bi-Hamiltonian system we can solve explicitly the equivariant
version of the Poincar\'e lemma. The global existence of the 1-forms $\theta_F$
and $\theta_F^{(V)}$ is therefore connected not only to the non-triviality of
the DeRham cohomology of $\man$, but also to the non-triviality of the
equivariant cohomology associated with the equivariant exterior derivative
$D_V$. Note that this derivation could also have been carried out for an
arbitrary equivariant differential 1-form $\beta$ with the definition
(\ref{dvbeta}) (c.f. eq. (\ref{alphaexact})). This suggests an intimate
relationship between the localization
formalism, and more generally equivariant cohomology, and the existence of
bi-Hamiltonian structures for a given phase space.

Furthermore, it is well-known that the existence alone of a bi-Hamiltonian
system is directly connected to integrability \cite{arnold,das}. If the
symplectic 2-forms $\omega$ and $\Omega_V$ are such that the rank (1,1) tensor
\beq
L=\Omega_V\cdot\omega^{-1}
\label{nontrivtensor}\eeq
is non-trivial, then one can straightforwardly show \cite{karki} that
\beq
\dot L=V^\mu\partial_\mu L=[L,dV]
\label{laxeq}\eeq
which is just the Lax equation, so that $(L,dV)$ determines a Lax pair
\cite{das}. Under a certain additional assumption on the tensor $L$ it can
then be shown \cite{karki} that the quantities
\beq
I_\mu=\frac{1}{\mu}~\tr~L^\mu
\label{conscharges}\eeq
give variables which are in involution and which are conserved, i.e. which
commute with the Hamiltonian $H$. If these quantities are in addition complete,
i.e. the number of functionally independent variables (\ref{conscharges}) is
half the phase space dimension, then the Hamiltonian system $(\man,\omega,H)$
is
integrable in the sense of the Liouville-Arnold theorem. We refer to
\cite{karki} for more details of how this construction works. Therefore the
equivariant localization formalism for classical dynamical systems presents
an alternative, geometric approach to the problem of integrability.

\subsection{Degenerate Version of the Duistermaat-Heckman Theorem}

In these last 3 Subsections of this Section we shall quickly run through some
of
the generalizations of the Duistermaat-Heckman theorem which can be applied to
more general dynamical systems. The first generalization we consider is to the
case where $H$ isn't necessarily non-degenerate and its critical point set
$\man_V$ is now a submanifold of $\man$ of co-dimension $r=\dim\man-\dim
\man_V$ \cite{atiyahbott,berline1,berlinegetz,bismut1,bismut2,dh,niemipalo2}.
In this case some modifications are required in the evaluation of
the canonical localization integral (\ref{intineta}) which was used in the
derivation of the Berline-Vergne theorem with the differential form $\alpha$
given in (\ref{alphadh}). The Hessian of $H$ now vanishes everywhere on $\man_
V$ (because $dH=0$ everywhere on $\man_V$), but we assume that it is
non-vanishing in the directions normal to the critical submanifold $\man_V$
\cite{nash}. This defines the normal bundle ${\cal N}_V$ of $\man_V$ in $\man$,
and the phase space is now locally the disjoint union
\beq
\man=\man_V\sqcup{\cal N}_V
\label{phasenorm}\eeq
so that in a neighbourhood near $\man_V$ we can decompose the local coordinates
on $\man$ as
\beq
x^\mu=x^\mu_0+x^\mu_\perp
\label{xdecomp}\eeq
where $x_0$ are local coordinates on $\man_V$, i.e. $V(x_0)=0$, and $x_\perp$
are local coordinates on ${\cal N}_V$. Similarly, the tangent space at any
point $x$ near $\man_V$ can be decomposed as
\beq
T_x\man=T_x\man_V\oplus T_x{\cal N}_V
\label{tangentdecomp}\eeq
where $T_x{\cal N}_V$ is the space of vectors orthogonal to those in $T_x
\man_V$. We can therefore decompose the Grassmann variables $\eta^\mu$ which
generate the exterior algebra of $\man$ as
\beq
\eta^\mu=\eta_0^\mu+\eta_\perp^\mu
\label{etadecomp}\eeq
where $\eta_0^\mu$ generate the exterior algebra $\Lambda\man_V$ and $\eta
_\perp^\mu$ generate $\Lambda{\cal N}_V$.

Under the usual assumptions used in deriving the equivariant localization
principle, it follows that the tangent bundle, equipped with a
Levi-Civita-Christoffel connection $\Gamma$ associated with a $U(1)$-invariant
metric tensor $g$ as in (\ref{levi}), is an equivariant vector bundle. Recall
that the Lie derivative $\lie_V$ induces a non-trivial action of the group
on the fibers of the tangent bundle which is mediated by the matrix $dV$.
More precisely, this action is given by
\beq
\lie_V=V^\mu\partial_\mu+dV^\mu i_{V^\mu}-dV
\label{lieontangent}\eeq
and so the moment map associated with this equivariant bundle is the Riemann
moment map \cite{berlinegetz}
\beq
\mu_V=\nabla V
\label{riemmommap}\eeq
which as always is regarded as a matrix acting on the fiber spaces. Given the
Killing equations for $V$, this moment map is related to the 2-form
$\Omega_V$ by
\beq
(\Omega_V)_{\mu\nu}=2g_{\mu\lambda}(\mu_V)^\lambda_\nu
\label{omegavmom}\eeq
and the equivariant curvature of the bundle is
\beq
R_V=R+\mu_V
\label{taneqcurv}\eeq
where the Riemann curvature 2-form of the tangent bundle is
\beq
R_\nu^\mu=\frac{1}{2}R^\mu_{\nu\lambda\rho}(x)\eta^\lambda\eta^\rho
\label{curv2form}\eeq
and
\beq
R^\rho_{\sigma\mu\nu}=\partial_\mu\Gamma^\rho_{\nu\sigma}-\partial_\nu\Gamma
^\rho_{\mu\sigma}+\Gamma^\rho_{\mu\lambda}\Gamma^\lambda_{\nu\sigma}-\Gamma
^\rho_{\nu\lambda}\Gamma^\lambda_{\mu\sigma}
\label{riemanncurv}\eeq
are the components of the associated Riemann curvature tensor $R=d\Gamma+
\Gamma\wedge\Gamma$. Note that, from the decomposition (\ref{tangentdecomp}),
the normal bundle inherits a $U(1)$-invariant connection from $T\man$, and the
curvature and moment map on $T{\cal N}_V$ are just the restrictions of the
corresponding objects defined on $T\man$.

Given these features of the 2-form $\Omega_V$, it follows that
the generators $\eta_0^\mu$ of $\Lambda\man_V$ satisfy
\beq
(\Omega_V)_{\mu\nu}(x_0)\eta_0^\nu=2(g_{\mu\lambda}\partial_\nu V^\lambda)(x_0)
\eta_0^\nu=0
\label{eta0rel}\eeq
since $\eta_0^\mu\sim dx_0^\mu$ lie in a direction cotangent to $\man_V$.
For large $s\in\IR^+$ in (\ref{intineta}) the integral will localize
exponentially to a neighbourhood of $\man_V$, and so, in the linearization
(\ref{xdecomp}) of the coordinates perpendicular to $\man_V$ wherein we
approximate this neighbourhood with a neighbourhood of the normal bundle
${\cal N}_V$, we can extend the integration over all values of $x_\perp$ there.
We now introduce the scaled change of integration variables
\beq
x^\mu=x_0^\mu+x_\perp^\mu\to x_0^\mu+x_\perp^\mu/\sqrt{s}~~~,~~~\eta^\mu=
\eta_0^\mu+\eta_\perp^\mu\to\eta_0^\mu+\eta^\mu_\perp/\sqrt{s}
\label{coordchange}\eeq
and expand the argument of the large-$s$ exponential in (\ref{intineta}) using
the decompositions (\ref{coordchange}). The Jacobian determinants from the
anticommuting $\eta_\perp^\mu$ variables and the commuting $x_\perp^\mu$
variables cancel each other, and so the integral (\ref{intineta}) remains
unchanged under this coordinate rescaling. A tedious but straightforward
calculation using observations such as (\ref{eta0rel}) shows that the
large-$s$ expansion of the argument of the exponential in
(\ref{intineta}) is given by \cite{niemipalo2}
\beq\new{\begin{array}{l}
\frac{s}{2}\Omega_V~{\buildrel{s\to\infty}\over\longrightarrow}~\frac{1}{2}
(\Omega_V)_{\mu\nu}(x_0)\eta_\perp^\mu\eta^\nu_\perp+\frac{1}{2}(\Omega_V)_{\mu
\sigma}(x_0)R^\sigma_{\nu\lambda\rho}(x_0)x_\perp^\mu x_\perp^\nu\eta^\lambda
_0\eta_0^\rho+{\cal O}(1/\sqrt{s})\\sK_V~{\buildrel{s\to\infty}\over
\longrightarrow}~\frac{1}{2}(\mu_V)_\mu^\rho(x_0)(\Omega_V)_
{\rho\nu}(x_0)x_\perp^\mu x_\perp^\nu+{\cal O}(1/\sqrt{s})\end{array}}
\label{largesomvkv}\eeq
where we have expanded the $C^\infty$-functions in (\ref{largesomvkv}) in
their respective Taylor series about the critical points.

Thus with the coordinate change (\ref{coordchange}), the integration over
the normal part of the full integration domain
\beq
\man\otimes\Lambda^1\man=\left(\man_V\otimes\Lambda^1\man_V\right)\sqcup\left(
{\cal N}_V\otimes\Lambda^1{\cal N}_V\right)~~~~~,
\label{domaindecomp}\eeq
i.e. over $(x_\perp,\eta_\perp)$ in (\ref{intineta}), is Gaussian and can be
carried out explicitly. The result is an integral over the critical
submanifold
\beq\new{\begin{array}{l}
Z(T)\\=\left(\frac{2\pi i}{T}\right)^{\frac{r}{2}}\int_{\man_V\otimes\Lambda^1
\man_V}d^rx_0~d^r\eta_0\e^{iT(H(x_0)+\omega(x_0,\eta_0))}\frac{\pfaff~\Omega_V
(x_0)}{\sqrt{\det\Omega_V(x_0)(\mu_V(x_0)+R(x_0,\eta_0))}}\\=\left(\frac{2\pi
i}{T}\right)^{\frac{r}{2}}\int_{\man_V}\frac{\ch_V(iT\omega)|_{\man_V}}
{E_V(R)|_{{\cal N}_V}}\end{array}}
\label{degdh}\eeq
where we have identified the equivariant Chern and Euler characters
(\ref{eqchern}) and (\ref{eqeuler}) of the respective fiber bundles. In
(\ref{degdh}) the equivariant Chern and Euler characters are restricted to the
critical submanifold $\man_V$, and the determinant and Pfaffian there are taken
over the normal bundle ${\cal N}_V$. Note that the above derivation has assumed
that the critical submanifold $\man_V$ is connected. If $\man_V$ consists
of several connected components, then the formula (\ref{degdh}) means a sum
over the contributions from each of these components. Notice also the role that
the large equivariant cohomological symmetry of the dynamical system has played
here -- it renders the Jacobian for the rescaling transformation
(\ref{coordchange}) trivial and reduces the required integrations to Gaussian
ones. This symmetry appears as a sort of supersymmetry here (i.e. a symmetry
between the scalar $x^\mu$ and Grassmann $\eta^\mu$ coordinates).

There are several comments in order here. First of all, if
$\man_V$ consists of discrete isolated points, so that $r=2n$, then, since the
curvature of the normal bundle of a point vanishes and so the Riemann moment
map
$\mu_V$ coincides with the usual moment map $dV$ on $T\man$ calculated at that
point, the formula (\ref{degdh}) reduces to the non-degenerate localization
formula (\ref{ztberline}) and hence to the Duistermaat-Heckman theorem.
Secondly, we recall that the equivariant characteristic classes in
(\ref{degdh})
provide representatives of the equivariant cohomology of $\man$ and the
integration formula (\ref{degdh}) is formally independent of the chosen metric
on $\man$. Thus the localization formulas are topological invariants of
$\man$, as they should be, and they represent types of `index theorems'. This
fact will have important implications later on in the formal applications to
topological field theory functional integration. Finally, we point out that
Kirwan's theorem generalizes to the degenerate case above \cite{kirwan}. In
this
case, since the Hessian is a non-singular symmetric matrix along the directions
normal to $\man_V$, we can orthogonally decompose the normal bundle, with the
aid of some locally-defined Riemannian metric on $\man_V$, into a direct
sum of the positive- and negative-eigenvalue eigenspaces of $\hess$. The
dimension of the latter subspace is now defined as the index of $\man_V$ and
Kirwan's theorem now states that the index of every connected component of
$\man_V$ must be even when the localization formula (\ref{degdh}) holds.
The Morse inequalities for this degenerate case \cite{nash} then relate
the exactness or failure of (\ref{degdh}) as before to the homology of the
underlying phase space $\man$. One dynamical system to which the formula
(\ref{degdh}) could be applied is the height function of the torus when the
torus is now viewed in 3-space as a doughnut sitting on a dinner plate (the
$xy$-plane). This function has 2 extrema but they are now circles, instead of
points, which are parallel to each other and one is a minimum and the other is
a maximum. The critical submanifold of $T^2$ in this case consists of 2
connected components, $T^2_V=S^1\sqcup S^1$.

\subsection{The Witten Localization Formula}

We have thus far only applied the localization formalism to abelian group
actions on $\man$. The first generalization of the Duistermaat-Heckman theorem
to non-abelian group actions was presented by Guilleman and Prato
\cite{guillprato} in the case where the induced action of the Cartan subgroup
(or maximal torus) of $G$ has only a finite number of isolated fixed points
$p_i$ and the stabalizer $\{g\in G:g\cdot p_i=p_i\}$ of all these fixed points
coincides with the Cartan subgroup. The Guilleman-Prato localization formula
reduces the integrals over the dual Lie algebra ${\bf g}^*$ to integrals over
the dual of the Cartan subalgebra using the so-called Weyl integral formula
\cite{blauthom}. With this reduction one can apply the standard abelian
localization formalism above. This procedure of abelianization thus reduces
the problem to the consideration of localization theory for functions of Cartan
elements of the Lie group $G$, i.e. integrable Hamiltonian systems. Witten
\cite{witten3} proposed a more general non-abelian localization
formalism and used it to study 2-dimensional Yang-Mills theory. In this
Subsection we shall outline the basic features of Witten's localization theory.

Given a Lie group $G$ acting on the phase space $\man$, we wish to evaluate
the partition function with the general equivariant extension
(\ref{eqextom}),
\beq
Z_G=\int_\man\frac{\omega^n}{n!}\e^{-\phi^a\otimes H^a}
\label{zg}\eeq
where as usual the Boltzmann weights are given by the symplectic moment map of
the $G$-action on $\man$. There are 2 ways to regard the dual algebra functions
$\phi^a$ in (\ref{zg}). We can give the $\phi^a$ fixed values, regarding them
as the values of elements of $S({\bf g}^*)$ acting on algebra elements, i.e.
the $\phi^a$ are complex-valued parameters, as is unambiguously the case if
$G$ is abelian \cite{atiyahbott} (in which case we set $\phi=-iT$ in
(\ref{zg})). In this case we are integrating with a fixed element of the Lie
group $G$, i.e. we are essentially in the abelian case. We shall see that
various localization schemes reproduce features of character formulas for the
action of the Lie group $G$ on $\man$ at the quantum level. The other
possibility is to regard the $\phi^a$ as dynamical variables and integrate over
them. This case allows a richer intepretation and is the basis of non-abelian
localization formulas and the localization formalism in topological field
theory.

To employ this latter interpretation for the symmetric algebra elements, we
need a definition for equivariant integration. The definition (\ref{intmap})
gives a map on $\Lambda_G\man\to S({\bf g}^*)^G$, but in analogy with ordinary
DeRham integration we wish to obtain a map on $\Lambda_G\man\to\IC$. The group
$G$ has a natural $G$-invariant measure on it, namely its Haar measure. Since
$\bf g$ is naturally isomorphic to the tangent space of $G$ at the identity,
it inherits from the Haar measure a natural translation-invariant measure.
Given this measure, the definition we take for equivariant integration is
\cite{witten3}
\beq
\oint_{\man\otimes{\bf g}^*}\alpha\equiv\lim_{s\to\infty}\frac{1}{{\rm vol}(G)}
\int_{{\bf g}^*}\prod_{a=1}^{\dim G}\frac{d\phi^a}{2\pi}~\e^{-\frac{1}{2s}
(\phi^a)^2}\int_\man\alpha
\label{eqintdef}\eeq
for $\alpha\in\Lambda_G\man$, where ${\rm vol}(G)=\int_GDg=\int_{{\bf
g}^*}\prod_ad\phi^a/2\pi$ is the volume of the group
$G$ in its Haar measure. The parameter $s\in\IR^+$ in (\ref{eqintdef}) is used
to regulate the possible divergence on the completion $\Lambda_G^\infty\man$.
The definition (\ref{eqintdef}) indeed gives a map on $\Lambda_G\man\to\IC$,
and
the $\phi^a$'s in it can be regarded as local Euclidean coordinates on
${\bf g}^*$ such that the measure there coincides with the chosen Haar measure
at the identity of $G$. Setting $\alpha=\e^{\omega_{\bf g}}$ in
(\ref{eqintdef}), with $\omega_{\bf g}$ the equivariant extension
(\ref{eqextom}) of the symplectic 2-form of $\man$, and performing the
Gaussian $\phi^a$-integrals, we arrive at Witten's localization formula for
the partition function (\ref{zg}),
\beq
Z_G=\lim_{s\to\infty}\left(\frac{s}{2\pi}\right)^{\dim G/2}\frac{1}{{\rm vol}
(G)}\int_\man\frac{\omega^n}{n!}\e^{-\frac{s}{2}\sum_a(H^a)^2}
\label{loczg}\eeq

The right-hand side of (\ref{loczg}) localizes onto the extrema of the square
of the moment map $\sum_a(H^a)^2$. The absolute minima of this function are
the solutions to $H=\phi^a\otimes H^a=0$. The contribution of the absolute
minimum to $Z_G$ (the dominant contribution for $s\to\infty$) is given by
a simple cohomological formula \cite{witten3}
\beq
Z_G^{\rm min}=\lim_{s\to\infty}\frac{(2\pi)^{\dim G/2}}{{\rm vol}(G)}\int
_{\man_0}\e^{\omega+\frac{1}{2s}\Theta}\biggm|_{\man_0}
\label{zgmin}\eeq
where $\man_0=H^{-1}(0)/G$ is the Marsden-Weinstein reduced phase space
\cite{marsden} (or symplectic quotient) and the localization of the global
minima onto $\man_0$ is a consequence of the $G$-equivariance of the
integration in (\ref{eqintdef}) (as one can then integrate over the fibers of
the bundle $H^{-1}(0)\to H^{-1}(0)/G$, with bundle projection $\pi$ that takes
$x\in H^{-1}(0)$ into its equivalence class $[x]\in\man_0$). Here we have
assumed that $G$ acts freely on $H^{-1}(0)$, and $\Theta$ is a certain element
of the cohomology group $H^4(\man_0;\IR)$ that is defined as follows. In
integrating over the $\phi^a$'s in (\ref{eqintdef}) we note that
$\sum_a(\phi^a)^2/2\in H_G^4(\man)$, so that when restricted to $H^{-1}(0)$ it
is the pullback of some $\Theta\in H^4(\man_0;\IR)$. Therefore the equivariant
cohomology class of $\sum_a(\phi^a)^2/2\in H_G^4(\man)$ is determined by this
form $\Theta$ which then serves as a characteristic class of the principal
$G$-bundle $H^{-1}(0)\to\man_0$. The Witten localization formula can in this
way be used to describe the cohomology of the reduced phase space $\man_0$ of
the given symplectic $G$-action on $\man$. We refer to \cite{witten3} for
further details of this construction.

However, the contributions from the other local extrema of $\sum_a(H^a)^2$,
which correspond to the critical points of $H$ as in the Duistermaat-Heckman
integration formula, are in general very complicated functions of the limiting
parameter $s\in\IR^+$. For instance, in the simple abelian example of
Subsections 2.1 and 3.5 above where $G=U(1)$, $\man=S^2$ and $H=h_{\Sigma^0}$
is the height function (\ref{heights2}) of the sphere, the Witten localization
formula (\ref{loczg}) above becomes
\beq
Z_{\Sigma^0}=\lim_{s\to\infty}\left(\frac{s}{2\pi}\right)^{1/2}\int_{-1}^{+1}
d\cos\theta~\e^{-s(a-\cos\theta)^2/2}=\lim_{s\to\infty}\left(1-I_+(s)-I_-(s)
\right)
\label{zu1loc}\eeq
where we have assumed that $|a|<1$, and $I_\pm(s)$ are the transcendental
error functions \cite{gradryz}
\beq
I_\pm(s)=\pm\left(\frac{s}{2\pi}\right)^{1/2}\int_{\pm1}^{\pm\infty}dx~\e
^{-s(a-x)^2/2}
\label{errorfns}\eeq
The 3 final terms in (\ref{zu1loc}) are the anticipated contributions from the
3 critical points of $h_{\Sigma^0}^2=(\cos\theta-a)^2$ -- the absolute
minimum at $\cos\theta=a$ contributes $+1$, while the local maxima at $\cos
\theta=\pm1$ contribute negative terms $-I_\pm$ to the localization formula.
The complicated error functions arise because here the critical point at
$\cos\theta=a$ is a degenerate critical point of the canonical localization
integral in (\ref{intineta}). The appearence of these error functions is
in marked contrast with the elementary functions that appear as the
contributions from the critical points in the usual Duistermaat-Heckman
formula.

Another interesting application of the Witten localization formalism is that
it can be used to derive integration formulas when the argument of the
Boltzmann weight in the partition function is instead the square of the
moment map. This can be done by reversing the arguments which led to the
localization formula (\ref{loczg}), and further localizing the
Duistermaat-Heckman type integral (\ref{zg}) using the localization principle
of Subsection 2.5. The result (for finite $s$) is then a sum of local
contributions $\sum_mZ_m(s)$, where the functions $Z_m(s)$ can only be
determined explicitly in appropriate instances \cite{jeffrey,witten3,wu} (see
the simple abelian example above).
Combining these ideas together, we arrive at the localization formula
\beq\new{\begin{array}{ll}
\frac{1}{{\rm vol}(G)}&\left(-\frac{iT}{\pi}\right)^{\dim G/2}\int_\man\frac{
\omega^n}{n!}\e^{iT\sum_a(H^a)^2}\\&=\frac{1}{{\rm vol}(G)}\int_{{\bf g}^*}
\prod_{a=1}^{\dim G}\frac{d\phi^a}{2\pi}\e^{-\frac{1}{4iT}(\phi^a)^2}\int
_\man\frac{\omega^n}{n!}\e^{-\phi^a\otimes H^a}
\\&=\frac{1}{{\rm vol}(G)}\lim_{s\to\infty}\int_{{\bf g}^*}\prod_{a=1}^{\dim G}
\frac{d\phi^a}{2\pi}~\e^{-\frac{1}{4iT}(\phi^a)^2}\int_\man\frac{\omega^n}{n!}
\e^{-\phi^a\otimes H^a-sD_{\bf g}\lambda}\end{array}}
\label{sqloc}\eeq
where $\lambda\in\Lambda^1_G\man$ and we have applied the localization
principle
to the Duistermaat-Heckman type integral over $\man$ on the right-hand side
of the first equality in (\ref{sqloc}). The localization 1-form $\lambda$ is
chosen just as before using a $G$-invariant metric on $\man$ and the
Hamiltonian vector field associated with the square of the moment map. In
Section 8 we shall outline how the formal infinite-dimensional generalization
of this last localization formula can be used to evaluate the partition
function of 2-dimensional Yang-Mills theory \cite{witten3}.

Finally, we point out the work of Jeffrey and Kirwan \cite{jeffrey} who
rigorously derived, in certain special cases, the contribution to $Z_G$ from
the reduced phase space $\man_0=H^{-1}(0)/G$ in (\ref{zgmin}). Let $H_C
\subset G$ be the Cartan subgroup of $G$, and assume that the fixed points
$p$ of the induced $H_C$-action on $\man$ are isolated and non-degenerate.
Then for any equivariantly-closed differential form $\alpha$ of degree $\dim
\man_0$ in $\Lambda_G\man$, we have the so-called residue formula
\cite{jeffrey}
\beq
\int_{\man_0}\alpha|_{\man_0}=\sum_{p\in\man_{H_C}}{\rm Res}\left[\e^{-\phi^a
\otimes H^a(p)}\frac{\left(\prod_\beta\beta\right)\alpha^{(0)}(p)}{e(p)}\right]
\label{resform}\eeq
where $\beta$ are the roots associated to $H_C\subset G$ (the eigenvalues of
the generators of $H_C$ in the adjoint representation of $G$), and Res is
Jeffrey-Kirwan-Kalkman residue, defined as the coefficient of $\frac{1}{\phi}$
where $\phi$ is the element of the symmetric algebra $S({\bf g}^*)$
representing the induced $H_C$-action on $\man$ (see \cite{jeffrey} and
\cite{kalkman2} for its precise definition). This residue, whose explicit form
was computed by Kalkman \cite{kalkman2}, depends on the fixed-point set
$\man_{H_C}$ of the $H_C$-action on $\man$ and it can be expressed in terms of
the weight determinants $e(p)$ in (\ref{weightprod}) of the $H_C$-action and
the values $H(p)=\phi^a\otimes H^a(p)$. It is in forms similar to
(\ref{resform}) that the first non-abelian generalizations of the
Duistermaat-Heckman theorem due to Guilleman and Prato appeared
\cite{guillprato}. The residue formula can explicitly be used to obtain
information about the cohomology ring of the reduced phase space $\man_0$ above
\cite{jeffrey,kalkman2}. This is particularly useful in applications to
topological gauge theories (see Section 8).

\subsection{The Wu Localization Formula}

The final generalization of the Duistermaat-Heckman theorem that we shall
present here is an interesting application, due to Wu \cite{wu}, of Witten's
localization formula in the form (\ref{sqloc}) when applied to a global
$U(1)$-action on $\man$. This yields a localization formula for Hamiltonians
which are not themselves the associated symplectic moment map, but are
functionals of such an observable $H$. This is accomplished via the
localization formula
\beq
Z_{U(1)}(T)=\int_\man\frac{\omega^n}{n!}\e^{iTH^2}=\left(-\frac{1}{4\pi iT}
\right)^{1/2}\lim_{s\to\infty}\int_0^{2\pi}d\phi~\e^{-\frac{1}{4iT}\phi^2}
\int_\man\frac{\omega^n}{n!}\e^{-\phi\otimes H-sD_{\bf u(1)}\lambda}
\label{squ1loc}\eeq
The final integral on the right-hand side of (\ref{squ1loc})
is just that which appears in the canonical localization integral (\ref{zs})
used in the derivation of the Duistermaat-Heckman formula. Working this out
just as before and performing the resulting Gaussian $\phi$-integral yields
Wu's localization formula for circle actions \cite{wu}
\beq
Z_{U(1)}(T)=\frac{(2\pi)^n}{(n-1)!}\sum_{p\in\man_V}\frac{1}{e(p)}\int_0^\infty
ds~s^{n-1}\e^{iT(s+|H(p)|)^2}+\int_{\man_0}\e^{\omega+iF/4T}\Bigm|_{\man_0}
\label{wuloc}\eeq
where $F=dA$ is the curvature of an abelian gauge connection on the
(non-trivial) principal $U(1)$-bundle $H^{-1}(0)\to\man_0$. The formula
(\ref{wuloc}) can be used to determine the symplectic volume of the
Marsden-Weinstein reduced phase space $\man_0$ \cite{wu}. This gives an
alternative localization for Hamiltonians which themselves do not generate an
isometry of some metric $g$ on $\man$, but are quadratic in such isometry
generators. As we shall see, the path integral generalizations of Wu's formula
are rather important for certain physical problems.

\section{Quantum Localization Theory for Phase Space Path Integrals}

In quantum mechanics there are not too many path integrals that can be
evaluated explicitly and exactly, while the analog of the stationary phase
approximation, i.e. the semi-classical approximation, can usually be
obtained quite readily. In this Section we shall investigate the possibility
of obtaining some path integral analogs of the Duistermaat-Heckman formula
and its generalizations. A large class of examples where one has an underlying
equivariant cohomology which could serve as a structure responsible for
localization is provided by phase space path integrals, i.e. the direct
loop space analogs of (\ref{classpart}). Of course, as path integrals in
general are mathematically awkward objects, the localization formulas that we
will obtain in this way are not really definite predictions but rather
suggestions for what kind of results to expect. Because of the lack of rigor
that goes into deriving these localization formulas it is perhaps surprising
then that some of these results are not only conceptually interesting but
also physically reasonable.

Besides these there are many other field-theoretic analogies with the
functional integral generalization of the Duistermaat-Heckman theorem, the
common theme being always some underlying geometrical or topological structure
which is ultimately responsible for localization. We have already mentioned
one of these in the last Section, namely the Witten localization formula which
is in principle the right framework to apply equivariant localization to
a cohomological formulation of 2-dimensional quantum Yang-Mills theory (see
Section 8). Another
large class of quantum models for which the Duistermaat-Heckman theorem
seems to make sense is $N=\frac{1}{2}$ supersymmetric quantum mechanics
\cite{atiyah}. This formal application, due to Atiyah and Witten, was indeed
the first encouraging evidence that such a path integral generalization of
the rigorous localization formulas of the last Section exists. Strictly
speaking though, this example really falls into the category of the
Berline-Vergne localization of Subsection 2.6 as the free loop space of a
configuration manifold is not quite a symplectic manifold in general
\cite{berlinegetz}. More generally, the Duistermaat-Heckman localization can
be directly generalized to the infinite-dimensional case within the
Lagrangian formalism, if the loop space defined over the {\it configuration}
space has on it a natural symplectic structure. This is the case, for example,
for geodesic motion on a Lie group manifold, where the space of based loops
is a K\"ahler manifold \cite{pres} and the stationary phase approximation is
well-known to be exact \cite{dowker,schul1}. This formal localization has
been carried out by Picken \cite{picken,picken1}.

We will discuss these specific applications in more detail, but we are
really interested in obtaining some version of the equivariant localization
formulas available which can be applied to non-supersymmetric models and when
the partition functions cannot be calculated directly by some other means.
The Duistermaat-Heckman theorem in this context would now express something
like the exactness of the one-loop approximation to the path integral.
These functional integral formulas, and their connections to the
finite-dimensional formulas of Section 3, will be discussed at length in this
Section. The formal techniques we shall employ throughout use ideas from
supersymmetric and topological field theories, and indeed we shall see how to
interpret an arbitrary phase space path integral quite naturally both as a
supersymmetric and as a topological field theory partition function. In the
Hamiltonian approach to localization, therefore, topological field theories fit
quite  naturally into the loop space equivariant localization framework. As we
shall see, this has deep connections with the integrability properties of these
models. In all of this, the common mechanism will be a fundamental
cohomological nature of the model which can be understood in terms of a
supersymmetry allowing one to deform the integrand without changing the
integral.

\subsection{Phase Space Path Integrals}

We begin this Section by deriving the quantum mechanical path integral for
a bosonic quantum system with no internal degrees of freedom. For simplicity,
we shall present the calculation for $n=1$ degree of freedom in Darboux
coordinates on $\man$, i.e. we essentially carry out the calculation on the
plane $\IR^2$. The extension to $n>1$ will then be immediate, and then we
simply add the appropriate symplectic quantities to obtain a
canonically-invariant object on a general symplectic manifold $\man$ to
ensure invariance under transformations which preserve the density of states.

To transform the classical theory of the last Section into a quantum mechanical
one (i.e. to `quantize' it), we replace the phase space coordinates $(p,q)$
with operators $(\hat p,\hat q)$ which obey an operator algebra that is
obtained by replacing the Poisson algebra of the Darboux coordinates
(\ref{poissoncan}) by allowing the commutator bracket of the basis operators
$(\hat p,\hat q)$ to be simply equal to the Poisson brackets of the same
objects as elements of the Poisson algebra of $C^\infty$-functions on the phase
space, times an additional factor of $i\hbar$ where $\hbar$ is Planck's
constant,
\beq
[\hat p,\hat q]=i\hbar
\label{commalg}\eeq
The operators $(\hat p,\hat q)$ with the canonical commutation relation
(\ref{commalg}) make the space of $C^\infty$-functions on $\man$ into an
infinite-dimensional associative operator algebra called the Heisenberg
algebra\footnote{\baselineskip=12pt More precisely, the operators $(\hat p,\hat
q)$ generate the
universal enveloping algebra of an extended affine Lie algebra which is
usually identified as the Heisenberg algebra.}. This algebra  can be
represented on the space $L^2(q)$ of square integrable functions of the
configuration space coordinate $q$ by letting the operator $\hat q$ act as
multiplication by $q$ and $\hat p$ as the derivative
\beq
\hat p=i\hbar\frac{\partial}{\partial q}
\label{momenderiv}\eeq
This representation of the Heisenberg algebra is called the Schr\"odinger
picture and the elements of the Hilbert space $L^2(q)$ are called the
wavefunctions or physical quantum states of the dynamical
system\footnote{\baselineskip=12pt Strictly
speaking, these function spaces should be properly defined as distribution
spaces in light of the discussion which follows.}.

The eigenstates of the (Hermitian) position and momentum operators are denoted
by the usual Dirac bra-ket notation
\beq
\hat q|q\rangle=q|q\rangle~~~~~,~~~~~\hat p|p\rangle=p|p\rangle
\label{posmomeigen}\eeq
These states are orthonormal,
\beq
\langle q|q'\rangle=\delta(q-q')~~~~~,~~~~~\langle p|p'\rangle=\delta(p-p')
\label{orthoqp}\eeq
and they obey the momentum and position space completeness relations
\beq
\int_{-\infty}^\infty dp~|p\rangle\langle p|=\int_{-\infty}^\infty dq~|q
\rangle\langle q|={\bf1}
\label{completeness}\eeq
with $\bf1$ the identity operator on the respective space.
In the representation (\ref{momenderiv}) on $L^2$-functions the momentum and
configuration space representations are related by the usual Fourier
transformation
\beq
|q\rangle=\int_{-\infty}^\infty\frac{dp}{\sqrt{2\pi\hbar}}~\e^{-ipq/\hbar}|p
\rangle
\label{momposfourier}\eeq
which identifies the matrix element
\beq
\langle p|q\rangle=\langle q|p\rangle^*=\frac{1}{\sqrt{2\pi\hbar}}
\e^{-ipq/\hbar}
\label{pqmatrix}\eeq
The basis operators have the matrix elements
\beq
\langle p|\hat q|q\rangle=q\langle p|q\rangle~~~~~,~~~~~\langle p|\hat p|q
\rangle=p\langle p|q\rangle=i\hbar\frac{\partial}{\partial q}\langle p|q\rangle
\label{pqmatrix}\eeq
All observables (i.e. real-valued $C^\infty$-functions of $(p,q)$) now become
Hermitian operators acting on the Hilbert space. In particular, the Hamiltonian
of the dynamical system now becomes a Hermitian operator $\hat H\equiv H(\hat
p,\hat q)$ with the matrix elements
\beq
\langle p|\hat H|q\rangle=H(p,q)\langle p|q\rangle=H(p,q)\frac{\e^{-ipq/\hbar}}
{\sqrt{2\pi\hbar}}
\label{hammatrix}\eeq
and the eigenvalues of this operator determine the energy levels of the
physical system.

The time evolution of any quantum operator is
determined by the quantum mapping above of the Hamilton equations of motion
(\ref{hameqmotion}). In particular, the time evolution of the position operator
is determined by
\beq
\dot{\hat q}(t)=\frac{1}{i\hbar}\left[\hat q,\hat H\right]
\label{opeqmotion}\eeq
which may be solved formally by
\beq
\hat q(t)=\e^{i\hat Ht/\hbar}\hat q(0)\e^{-i\hat Ht/\hbar}
\label{qtevol}\eeq
so that the time evolution is determined by a unitary transformation of the
position operator $\hat q(0)$. In the Schr\"odinger representation, we treat
the operators as time-independent quantities using the unitary transformation
(\ref{qtevol}) and consider the time-evolution of the quantum states. The
configuration of the system at a time $t$ is defined using the unitary
time-evolution operator in (\ref{qtevol}) acting on an initial configuration
$|q\rangle$ at time $t=0$,
\beq
|q,t\rangle=\e^{i\hat Ht/\hbar}|q\rangle
\label{qtstate}\eeq
which is an eigenstate of (\ref{qtevol}) for all $t$.

An important physical quantity is the quantum propagator
\beq
{\cal K}(q',q;T)=\langle q',T|q,0\rangle=\langle q'|\e^{-i\hat HT/
\hbar}|q\rangle
\label{prop}\eeq
which, according to the fundamental principles of quantum mechanics
\cite{messiah}, represents the probability of the system evolving from a state
with configuration $q$ to one with configuration $q'$ in a time interval $T$.
The propagator (\ref{prop}) satisfies the Schr\"odinger wave equation
\beq
i\hbar\frac{\partial}{\partial T}{\cal K}(q',q;T)=\hat H{\cal K}(q',q;T)
\label{schroeq}\eeq
where the momentum operators involved in the Hamiltonian $\hat H$ on the
right-hand side of (\ref{schroeq}) are represented in the Schr\"odinger
polarization (\ref{momenderiv}). The Schr\"odinger equation is to be solved
with
the Dirac delta-function initial condition
\beq
{\cal K}(q',q;T=0)=\delta(q'-q)
\label{incond}\eeq
The function ${\cal K}(q',q;T)$ acts as an integration kernel which determines
the time-evolution of the wavefunctions as
\beq
\Psi(q';T)=\int_{-\infty}^\infty dq~{\cal K}(q',q;T)\Psi(q;0)
\eeq
where $\Psi(q;t)\equiv\langle q,t|\Psi\rangle$ are the time-dependent
configuration space representations of the physical states $|\Psi\rangle$ of
the system. Thus the propagator represents the fundamental quantum dynamics of
the system and the stationary state solutions to the Schr\"odinger equation
(\ref{schroeq}) determine the energy eigenvalues of the dynamical system.

The phase space path integral provides a functional representation of the
quantum propagator in terms of a `sum' over continuous trajectories on the
phase space. It is constructed as follows \cite{schulman}. Between the
initial and final configurations $q$ and $q'$ we introduce $N-1$ intermediate
configurations $q_0,\ldots,q_N$ with $q_0\equiv q$ and $q_N\equiv q'$, and
each separated by the time interval
\beq
\Delta t=T/N
\label{timeint}\eeq
Introducing intermediate momenta $p_1,\ldots,p_N$ and inserting the
completeness
relations
\beq
\int_{-\infty}^\infty dq_{j-1}~dq_j~dp_j~|q_j\rangle\langle q_j|p_j\rangle
\langle p_j|q_{j-1}\rangle\langle q_{j-1}|={\bf1}~~~~~,~~~~~j=1,\ldots,N
\label{completej}\eeq
into the matrix element (\ref{prop}) we obtain
\beq\new{\begin{array}{ll}
{\cal K}(q',q;T)&=\langle q'|(\e^{-i\hat H\Delta t/\hbar})^N|q\rangle\\&=
\int_{-\infty}^\infty\prod_{j=1}^Ndq_{j-1}~dq_j~dp_j~\langle q'|q_j\rangle
\langle q_j|\e^{-i\hat H\Delta t/\hbar}|p_j\rangle\langle p_j|q_{j-1}\rangle
\langle q_{j-1}|q\rangle\\&=\int_{-\infty}^{\infty}\prod_{j=1}^{N-1}dq_j~
\prod_{j=1}^N\frac{dp_j}{2\pi\hbar}~\exp\left\{\frac{i}{\hbar}\sum_{i=1}^N
\left(p_i\frac{q_i-q_{i-1}}{\Delta t}-H(p_i,q_i)\right)\Delta t\right\}
\end{array}}
\label{pathintderiv}\eeq
\bd
\times\delta(q_0-q)\delta(q_N-q')
\ed
where we have used the various identities quoted above. In the limit $N\to
\infty$, or equivalently $\Delta t\to0$, the discrete points $(p_j,q_j)$
describe paths $(p(t),q(t))$ in the phase space between the configurations $q$
and $q'$, and the sum in (\ref{pathintderiv}) becomes the continuous limit
of a Riemann sum representing a discretized time integration. Then
(\ref{pathintderiv}) becomes
\beq\new{\begin{array}{ll}
{\cal K}(q',q;T)=\lim_{N\to\infty}\int_{-\infty}^\infty&\prod_{j=1}^{N-1}dq_j~
\prod_{j=1}^N\frac{dp_j}{2\pi\hbar}~\exp\left\{\frac{i}{\hbar}\int_0^Tdt~
\left(p(t)\dot q(t)-H(p,q)\right)\right\}\\&
\times\delta(q(0)-q)\delta(q(T)-q')\end{array}}
\label{propathint}\eeq
Note that the argument of the exponential in (\ref{propathint}) is
just the classical action of the dynamical system, because its integrand
is the usual Legendre transformation between the Lagrangian and Hamiltonian
descriptions of the classical dynamics \cite{goldstein}. Notice also that,
in light of the Heisenberg uncertainty principle $\Delta q\Delta p\sim2\pi
\hbar$, the normalization factors $2\pi\hbar$ there can be physically
interpreted as the volume of an elementary quantum state in the phase space.

The integration measure in (\ref{propathint}) formally gives an
integral over all phase space paths defined in the time interval $[0,T]$.
This measure is denoted by
\beq
[dp~dq]\equiv\lim_{N\to\infty}\prod_{j=1}^N\frac{dp_j}{2\pi\hbar}~\prod_{j=1}
^{N-1}dq_j\equiv\prod_{t\in[0,T]}\frac{dp(t)}{2\pi\hbar}~dq(t)
\label{feynmeas}\eeq
and it is called the Feynman measure. The last equality means that it is to
be understood as a `measure' on the infinite-dimensional functional space of
phase space trajectories $(p(t),q(t))$, where for each fixed time slice
$t\in[0,T]$, $dp(t)~dq(t)$ is ordinary Riemann-Lebesgue measure. Being an
infinite-dimensional quantity, it is not rigorously defined, and some special
care must be taken to determine the precise meaning of the limit $N\to\infty$
above. This has been a topic of much dispute over the years and we shall make
no attempt in this Review to discuss the ill-defined ambiguities associated
with the Feynman measure. Many rigorous attempts at formulating the path
integral have been proposed in constructive quantum field theory. For instance,
it is possible to give the limit (\ref{feynmeas}) a somewhat precise meaning
using the so-called Lipschitz functions of order $\frac{1}{2}$ which assumes
that the paths which contribute in (\ref{propathint}) grow no faster than
${\cal O}(\sqrt{t})$ (these functional integrals are called Wiener
integrals)\footnote{\baselineskip=12pt Note that the transition from the
multiple integral representation in (\ref{pathintderiv}) to the representation
(\ref{propathint}),(\ref{feynmeas}) in terms of phase space paths requires
that these trajectories can at least be approximated by piecewise-linear
functions.}. We shall at least assume that the integration measure
(\ref{feynmeas}) is supported on $C^\infty$ phase space paths and that the
quantum mechanical propagator given by (\ref{propathint}) is a tempered
distribution, i.e. it can diverge with at most a polynomial growth. This latter
restriction on the path integral is part of the celebrated Wightman axioms for
quantum field theory which allows one to at least carry out certain formal
rigorous manipulations from the theory of distributions.

However, a physicist will typically proceed without worry and succeed in
extracting a surprising amount of information from formulas such as
(\ref{propathint}) without the need to investigate in more detail the
implications of the limit $N\to\infty$ above. To actually carry out functional
integrations such as (\ref{propathint}) one uses formal functional analogs of
the usual rules of Riemann-Lebesgue integration in the straightforward sense,
where all time integrals are treated as continuous sums on the functional space
(i.e. the time parameter $t$ is regarded as a continuous index).

If we set $q=q'$ and integrate over all $q$, then the left-hand
side of (\ref{propathint}) yields
\beq
\int_{-\infty}^\infty dq~\langle q|\e^{-i\hat HT/\hbar}|q\rangle\equiv
{}~\tr~\|\e^{-i\hat HT/\hbar}\|=\int\!\!\!\!\!\!\!\sum dE~\e^{-iET/\hbar}
\label{spectrum}\eeq
where $E$ are the energy eigenvalues of $\hat H$ and the symbol $\|\cdot\|$
will be used to emphasize that the matrix of interest is considered as an
infinite dimensional one over either the Hilbert space of physical states or
the functional trajectory space. On the other hand, the right-hand side of
(\ref{propathint}) becomes
\beq
Z(T)=\int[dp~dq]~\exp\left\{\frac{i}{\hbar}\int_0^Tdt~\left(p(t)\dot q(t)-
H(p,q)\right)\right\}\delta(q(0)-q(T))
\label{partdarboux}\eeq
which is called the quantum partition function. From (\ref{spectrum}) we see
that the quantum partition function describes the spectrum of the quantum
Hamiltonian of the dynamical system and that the poles of its
Fourier transform
\beq
G(E)=\int_0^\infty dT~\e^{iET/\hbar}Z(T)
\label{energygf}\eeq
give the bound state spectrum of the system \cite{messiah}. The quantity
(\ref{energygf}) is none other than the energy Green's function
$G(E)=~\tr\|(E-\hat H)^{-1}\|$ which is associated with the Schr\"odinger
equation (\ref{schroeq}). Thus the quantum partition function is in some sense
the fundamental quantity which describes the quantum dynamics (i.e. the energy
spectrum) of a Hamiltonian system.

Finally, the generalization to an arbitrary symplectic manifold $(\man,\omega)$
of dimension $2n$ is immediate. The factor $p\dot q$ becomes simply $p_\mu\dot
q^\mu$ in higher dimensions, and, in view of (\ref{thetadarboux}), the
canonical form of this is $\theta_\mu(x)\dot x^\mu$ in an arbitrary coordinate
system on $\man$. Likewise, the phase space measure $dp\wedge dq$ according to
(\ref{phasespmeas}) should be replaced by the canonically-invariant Liouville
measure (\ref{liouvillemeas}). Thus the quantum partition function for a
generic dynamical system $(\man,\omega,H)$ is defined as
\beq
Z(T)=\int_{L\man}[d\mu_L(x)]~\e^{iS[x]}=\int_{L\man}[d^{2n}x]~\prod_{t\in[0,T]}
\sqrt{\det\|\omega(x(t))\|}\e^{iS[x]}
\label{quantpart}\eeq
where
\beq
S[x]=\int_0^Tdt~\left(\theta_\mu(x)\dot x^\mu-H(x)\right)
\label{classaction}\eeq
is the classical action of the Hamiltonian system. Here and in the following
we shall set $\hbar\equiv1$ for simplicity, and the functional integration in
(\ref{quantpart}) is taken over the loop space $L\man$ of $\man$, i.e. the
infinite-dimensional space of paths $x(t):[0,T]\to\man$ obeying periodic
boundary conditions $x^\mu(0)=x^\mu(T)$. Although much of the formalism which
follows can be applied to path integrals over the larger trajectory space of
all paths, we shall find it convenient to deal mostly with the loop space
over the phase space. The partition function (\ref{quantpart}) can be regarded
as the formal infinite-dimensional analog of the classical integral
(\ref{classpart}), or, as mentioned before, the prototype of a topological
field theory functional integral regarded as a (0 + 1)-dimensional quantum
field theory. In the latter application the discrete index sums over $\mu$
contain as well integrals over the manifold on which the fields are defined.

Notice that the symplectic potential $\theta$ appearing in
(\ref{quantpart}),(\ref{classaction}) is only locally defined, and so some
care must be taken in defining (\ref{quantpart}) when $\omega$ is not globally
exact. We shall discuss this procedure later on. Note also that the Liouville
measure in (\ref{quantpart}), which is defined by the last equality in
(\ref{feynmeas}), differs from that of (\ref{partdarboux}) in that in the
latter case there is one extra momentum integration in the phase space Feynman
measure (\ref{feynmeas}), so that the endpoints are fixed and we integrate
over all intermediate momenta. Thus one must carefully define appropriate
boundary conditions for the integrations in (\ref{quantpart}) for the
Schr\"odinger path integral measure in order to maintain a formal analogy
between the finite and infinite dimensional cases. We shall elaborate on this
point in the next Section. Further discussion of this and the proper
discretizations and ordering prescriptions that are needed to define the
functional integrations
that appear above can be found in \cite{schulman} and \cite{langouche}.

\subsection{Example: Path Integral Derivation of the Atiyah-Singer Index
Theorem}

As we did at the start of Section 2 above, we shall motivate the formal
manipulations that will be carried out on the phase space path integral
(\ref{quantpart}) with an explicit example which captures the essential ideas
we shall need. At the same time, this particular example sets the stage for the
analogies with topological field theory functional integrals which will follow
and will serve as a starting point for some of the applications which will be
discussed in Section 8. We will consider the derivation, via the evaluation of
a path integral for supersymmetric quantum mechanics \cite{alvgaume,friedan},
of the Atiyah-Singer index theorem which expresses the fact that the analytical
index of a Dirac operator is a topological invariant of the background fields
in the quantum field theory in which it is defined. This theorem and its
extensions have many uses in quantum field theory, particularly for the study
of anomalies and the fractional fermion number of solitons
\cite{eguchi,niemisem,szsem1}.

Consider a Dirac operator $i\nabla\slash$ on an even-dimensional compact
orientable Riemannian manifold $\man$ with metric $g$ of Minkowski signature,
\beq
i{\nabla\slash}\equiv i\gamma^\mu\nabla_\mu\equiv i\gamma^\mu
\left(\partial_\mu+\frac{1}{8}
\omega_{\mu jk}\left[\gamma^j,\gamma^k\right]+iA_\mu\right)
\label{diracop}\eeq
Here $\gamma^\mu(x)$ are the Dirac matrices which generate the Clifford algebra
of $\man$,
\beq
\gamma^\mu\gamma^\nu+\gamma^\nu\gamma^\mu=2g^{\mu\nu}(x)
\label{cliffalg}\eeq
and $A_\mu$ is a connection on a principal fiber bundle $E\to\man$ (i.e. a
gauge field). We shall assume for simplicity that the structure group of the
principal bundle is $G=U(1)$, so that $A$ is a connection on a line bundle
$L\to\man$. The spin-connection $\omega^j_{\mu i}$ is defined as follows. At
each point $x\in\man$ we introduce a local basis of orthonormal tangent vectors
$e_\mu^i(x)$, called a vielbein, where $\mu$ labels the basis components in
$T\man$ and $i=1,\dots,\dim\man$ parametrizes the fibers of $T\man$ (i.e. the
local rotation index in the tangent space). Orthonormality means that
$g^{\mu\nu}(x)e_\mu^i(x)e^j_\nu(x)=\eta^{ij}$ is the flat Minkowski metric in
$T\man$, or equivalently
\beq
\eta_{ij}e_\mu^i(x)e_\nu^j(x)=g_{\mu\nu}(x)
\label{metricviel}\eeq
In this vielbein formalism, $\gamma^i\equiv e_\mu^i(x)\gamma^\mu(x)$ in
(\ref{diracop}) and the spin-connection (i.e. connection on the spin bundle
$S\man$ of $\man$, defined by the $\dim\man$-dimensional spinor representations
of the local Lorentz group of the tangent bundle) is
\beq
\omega^i_{\mu j}=e_\nu^i(\partial_\mu
E_j^\nu+\Gamma^\nu_{\mu\lambda}E_j^\lambda)
\label{spinconn}\eeq
and $E_i^\mu(x)$ are the inverse vielbein fields, i.e. $E_i^\mu
e_\mu^j=\delta_i^j$. The spin-connection (\ref{spinconn}) is a gauge field of
the local Lorentz group of the tangent bundle, i.e. under a local Lorentz
transformation $e_\mu^i(x)\to\Lambda^i_j(x)e_\mu^j(x)$, $\Lambda(x)\in
SO(2n-1,1)$, on the frame bundle of $\man$, the gauge field $\omega_\mu$
transforms in the usual way as
$\omega_\mu\to\Lambda\omega_\mu\Lambda^{-1}-\partial_\mu\Lambda\cdot
\Lambda^{-1}$. It is defined so that the covariant derivative in
(\ref{diracop}) coincides with the Levi-Civita-Christoffel connection, i.e.
$\nabla_\mu e_\nu^i\equiv\partial_\mu
e_\nu^i-\Gamma^\lambda_{\mu\nu}e_\lambda^i+\omega_{\mu j}^ie_\nu^j=0$. The
covariant derivative in (\ref{diracop}) is in general regarded as a connection
on the bundle $T\man\otimes S\man\otimes L$ which together define the twisted
spin complex of $\man$ (the `twisting' being associated with the presence of
the gauge field $A$).

The chiral representation of the Dirac matrices is that in which the chirality
matrix $\gamma^c\equiv i\gamma^1\gamma^2\cdots\gamma^{2n}$, with the properties
$(\gamma^c)^2={\bf1}$ and $(\gamma^c)^\dagger=\gamma^c$, is diagonal. Since the
Dirac operator (\ref{diracop}) commutes with $\gamma^c$, in this representation
of the Clifford algebra (\ref{cliffalg}) these 2 operators can be written in
the block forms
\beq
\gamma^c=\pmatrix{{\bf1}&0\cr0&-{\bf1}\cr}~~~~~,~~~~~i{\nabla\slash}=
\pmatrix{0&{\cal D}\cr{\cal D}^\dagger&0\cr}
\label{chiralrep}\eeq
The analytical index of $i\nabla\slash$ is then defined as the difference
between the dimensions of the kernel and co-kernel of the elliptic operator
$\cal D$,
\beq
{\rm index}(i{\nabla\slash})\equiv\dim~{\rm ker}~{\cal D}-\dim~{\rm
coker}~{\cal D}=\dim~{\rm ker}~{\cal D}-\dim~{\rm ker}~{\cal D}^\dagger
\label{anindex}\eeq
In the chiral representation (\ref{chiralrep}), the Dirac spinors, i.e. the
solutions $\Psi$ of the Dirac equation
\beq
i{\nabla\slash}\Psi=\pmatrix{0&{\cal D}\cr{\cal
D}^\dagger&0\cr}\pmatrix{\Psi_+\cr\Psi_-\cr}=E\Psi
\label{diraceq}\eeq
are determined by their positive and negative chirality spin components
$\Psi_\pm$. Since the zero mode solutions, $E=0$, satisfy ${\cal
D}^\dagger\Psi_+={\cal D}\Psi_-=0$, the index (\ref{anindex}) is just the
difference between the number of positive and negative chirality zero-mode
solutions of the Dirac equation (\ref{diraceq}), i.e.
\beq
{\rm index}(i{\nabla\slash})=~{\rm Tr}_{E=0}\gamma^c
\label{indexchiral}\eeq
Moreover, since  $[i\nabla\slash,\gamma^c]=0$, the chirality operator provides
a one-to-one mapping between positive and negative non-zero energy states. Thus
the index (\ref{indexchiral}) can be written as a trace over the full Hilbert
space $\cal H$ spanned by the Dirac spinors as
\beq
{\rm index}(i{\nabla\slash})=~{\rm tr}_{\cal H}\|\gamma^c\e^{-T({\cal D}{\cal
D}^\dagger+{\cal D}^\dagger{\cal D})}\|
\label{indextrfull}\eeq
where we have used the fact that the spinors satisfying the eigenvalue equation
(\ref{diraceq}) also obey the Schr\"odinger equations
\beq
{\cal D}{\cal D}^\dagger\Psi_+=E^2\Psi_+~~~~~,~~~~~{\cal D}^\dagger{\cal
D}\Psi_-=E^2\Psi_-
\label{spinschreq}\eeq
The parameter $T>0$ in (\ref{indextrfull}) regulates the operator trace.

The representation (\ref{indexchiral}) of the index of a Dirac operator is
known as the Witten index \cite{birm,witten1,witten2}. We can identify the
positive and negative chirality spinors as bosons and fermions, respectively,
and then the chirality operator can be written as $\gamma^c=(-1)^F$ where $F$
is the fermion (or ghost) number operator. The operator ${\cal D}^\dagger$ can
then be identified with a supersymmetry generator $Q$, which provides a mapping
between fermions and bosons, associated with a supersymmetric theory with
Hamiltonian given by the graded BRST commutator (see (\ref{indextrfull}) above)
\beq
H=\{Q,Q^\dagger\}\geq0
\label{susyham}\eeq
as is standard in a supersymmetric model. This is equivalent to the statement
above that bosonic and fermionic states of non-zero energy in the
supersymmetric theory are always paired. Since $QQ^\dagger$ and $Q^\dagger Q$
are positive-definite Hermitian operators, the zero modes $|0\rangle$ of $H$
are supersymmetric, $Q|0\rangle=Q^\dagger|0\rangle=0$, and thus they provide a
(trivial) 1-dimensional representation of the supersymmetry. Small
perturbations of the background gravitational and gauge fields $g$ and $A$ may
excite the $E=0$ states, but bosonic and fermionic states must always be lifted
in pairs. Consequently, the Witten index
\beq
{\rm index}(i{\nabla\slash})=~{\rm tr}_{\cal H}\|(-1)^F\e^{-TH}\|\equiv~{\rm
str}\|\e^{-TH}\|
\label{indexstr}\eeq
is a topological invariant that is independent of the choice of spin and gauge
connections. Here str denotes the supertrace and, since $\man$ is compact by
assumption, there are only finitely many modes which contribute in
(\ref{indexstr}). Thus all these quantities are independent of the parameter
$T$. In the low temperature limit ($T\to\infty$) only zero modes contribute to
(\ref{indexstr}) according to their chirality.

The physical relevence of the Witten index is immediate. As the zero energy
states in general need not be paired, the non-vanishing of the Witten index
(\ref{indexstr}) implies that there is at least one zero energy state which is
then an appropriate supersymmetric ground state of the underlying
supersymmetric theory. Thus the non-vanishing of (\ref{indexstr}) is a
sufficient condition for the presence of supersymmetric ground states.
Conversely, a necessary criterion for dynamical supersymmetry breaking is that
${\rm Tr}_{E=0}(-1)^F$ should vanish.

Using the standard path integral techniques of the last Subsection, it is
straightforward to write down a path integral representation of
(\ref{indexstr}) \cite{alvgaume}. The collection of all fields $\Phi$ will
clearly involve both bosonic and fermionic degrees of freedom which will be
connected by a supersymmetry, i.e. the appropriate path integral representation
will be that of a supersymmetric field theory. Furthermore, the integral over
the function space of fields on $\man$ will be restricted to fields which
satisfy periodic boundary conditions for both the space and time coordinates,
$\Phi(t+T)=\Phi(t)$. This restriction is necessary for the pairing of states
discussed above, and the reason for this condition in the time direction for
the fermionic fields is because of the presence of the Klein operator $(-1)^F$
in the supertrace. The path integral representation of the index
(\ref{indexstr}) is then \cite{alvgaume,friedan}
\beq
{\rm index}(i{\nabla\slash})=\int_{L\man\otimes
L\Lambda^1\man}[d^{2n}x]~[d^{2n}\psi]~\e^{iTI_{1/2}[x,\psi]}
\label{indexpathint}\eeq
where $\psi^\mu(t)$ are anticommuting periodic paths on $\man$ (which,
according to Subsection 2.6, can be taken to lie in $L\Lambda^1\man$) with path
integration defined using functional analogs of the Berezin integration rules
discussed in Subsection 2.6. The action in (\ref{indexpathint}) is that of
$N=\frac{1}{2}$ (Dirac) supersymmetric quantum mechanics, i.e. the invariant
action for a spinning particle in background gravitational and gauge
fields\footnote{\baselineskip=12pt In general an $N$-component supersymmetric
model contains $N$ fermion chiral conjugate pairs $(\bar\psi_i,\psi_i)$ with
$2N$ associated superpartner bosonic fields $F_i$, and $N$ corresponding
supersymmetry charges $(Q_i^\dagger,Q_i)$ which mix the fields with their
superpartners.}
\beq
I_{1/2}[x,\psi]=\oint_{S^1}d\tau~\left(\frac{1}{2}g_{\mu\nu}\dot x^\mu\dot
x^\nu+\dot x^\mu
A_\mu+\frac{1}{2}g_{\mu\nu}\psi^\mu\nabla_\tau\psi^\nu-\frac{1}{2}\psi^\mu
F_{\mu\nu}\psi^\nu\right)
\label{susy1/2action}\eeq
where $\psi^\mu$ are the Grassmann (superpartner) coordinates for the particle
configurations $x^\mu\in\man$, and we have rescaled the time by $T$ so that
time integrations lie on the unit circle $S^1$. Here
\beq
\nabla_\tau W^\mu(x(\tau))=\partial_\tau
W^\mu(x(\tau))+\Gamma^\mu_{\nu\lambda}(x(\tau))\dot
x^\nu(\tau)W^\lambda(x(\tau))
\label{covderxt}\eeq
is the covariant derivative along the loop $x(\tau)$ induced by the Riemannian
connection $\nabla$ on $\man$, and $F_{\mu\nu}=\partial_\mu A_\nu-\partial_\nu
A_\mu$ is the gauge field strength tensor. In (\ref{susy1/2action}), the
particle current $\dot x^\mu$ is minimally coupled to the gauge field $A_\mu$,
and its spinor degrees of freedom couple to the electromagnetic field of
$A_\mu$ by the usual Pauli magnetic moment interaction. The action
(\ref{susy1/2action}) has the (infinitesimal) supersymmetry
\beq
{\cal S}x^\mu(\tau)=\psi^\mu(\tau)~~~~~,~~~~~{\cal S}\psi^\mu(\tau)=\dot
x^\mu(\tau)
\label{susy1/2alg}\eeq
The action (\ref{susy1/2action}) arises from the standard supersymmetric
non-linear (Wess-Zumino) sigma-model, and we shall see in Section 8 how to
write this supersymmetric model in a more conventional fashion using superspace
coordinates and superfields.

Let us briefly describe how one arrives at the path integral representation
(\ref{indexpathint}) (see \cite{hietmor} for details). The local Dirac algebra
of the spin bundle is represented by the anticommutator
$[\psi^i,\psi^j]_+=\eta^{ij}$,
so that the Clifford algebra of $\man$ is represented is represented as
\beq
\left[\psi^\mu,\psi^\nu\right]_+=g^{\mu\nu}(x)
\label{cliffalgpsi}\eeq
where as always $\psi^\mu=E_i^\mu\psi^i$. The zero-mode equation for the Dirac
operator $i\nabla\slash$ can therefore be realized as the graded constraint
equation
\beq
{\cal S}\equiv\psi^\mu\left(\partial_\mu+\frac{1}{4}\omega_{\mu
ij}\psi^i\psi^j+iA_\mu\right)=0
\label{diraceqconstr}\eeq
where the supersymmetry generator $\cal S$ associated with (\ref{diracop})
generates the (graded) $N=\frac{1}{2}$ supersymmetry
algebra\footnote{\baselineskip=12pt See Appendix A for the convention for the
graded commutator $\{\cdot,\cdot\}$.}
\beq\new{\begin{array}{c}
\{{\cal S},{\cal S}\}=H=g^{\mu\nu}\left(\partial_\mu+\frac{1}{4}\omega_{\mu
ij}\psi^i\psi^j+iA_\mu\right)\left(\partial_\nu+\frac{1}{4}\omega_{\nu
k\ell}\psi^k\psi^\ell+iA_\nu\right)+\frac{1}{2}\psi^\mu F_{\mu\nu}\psi^\nu
\\ \{{\cal S},H\}=\{H,H\}=0\end{array}}
\label{calSalg}\eeq
In arriving at (\ref{calSalg}) we have used the various symmetry properties of
the Riemann curvature tensor. Notice that the Hamiltonian $H$ vanishes on
physical (supersymmetric) ground states, so that there are no local propagating
degrees of freedom and the model can only describe the global topological
characteristics of the manifold $\man$, i.e. this supersymmetric model defines
a topological field theory.

The constraint algebra (\ref{calSalg}) contains first class constraints, i.e.
it defines a closed algebra between $H$ and $\cal S$ (see Appendix A) such that
$H$ is supersymmetric under the infinitesimal supersymmetry transformations
generated by $\cal S$. The constraints $H={\cal S}=0$ ensure the
reparametrization invariance of the trajectories $x^\mu(\tau)$. It is
straightforward to now construct the BRST gauge fixed path integral associated
with this constraint algebra (in the proper time gauge). Since the various
ghost degrees of freedom only couple to world line quantities and not to the
metric structure of $\man$, (\ref{indexpathint}) coincides with the canonical
BRST gauge-fixed path integral describing the propagation of a Dirac particle
on the configuration space $\man$ (with the identification
$p_\mu\sim\partial_\mu$ as the canonical momentum conjugate to $x^\mu$). The
gauge-fixed quantum action (\ref{susy1/2action}) is written only modulo the
ghost field and other contributions that decouple from the background metric of
$\man$, as these fields only contribute to the overall normalization in
(\ref{indexpathint}). The necessity to use periodic boundary conditions in the
path integral follows from the identification $\psi^i\sim\gamma^c\gamma^i$ of
the Dirac matrices.

There are several ways to evaluate explicitly the supersymmetric path integral
(\ref{indexpathint}). The traditional method is to exploit the $T$-independence
and use, in the high-temperature limit ($T\to0$), either a heat kernel
expansion of the trace in (\ref{indextrfull}) \cite{eguchi} or a normal
coordinate expansion to evaluate the partition function (\ref{indexpathint})
\cite{alvgaume,friedan}. Here, however, we wish to emphasize the observation of
Atiyah and Witten \cite{atiyah} (and the later generalizations to twisted
Dirac operators by Bismut \cite{bismut1,bismut2} and Jones and Petrack
\cite{jones}) that the path integral for $N=\frac{1}{2}$ supersymmetric quantum
mechanics admits a formal equivariant cohomological structure on the superloop
space $L\man\otimes L\Lambda^1\man$. To see how this structure arises, we
introduce a geometric framework for manipulating the path integral
(\ref{indexpathint}). These geometric manipulations will be the starting point
for the general analyses of generic phase space path integrals which will
follow. Given any functional $F[x]$ of closed paths in the loop space $L\man$,
we define functional differentiation, for which functional integration is the
anti-derivative thereof, by the rule
\beq
\frac{\delta}{\delta x^\mu(\tau)}F[x(\tau')]=\delta(\tau-\tau')F'[x(\tau')]
\label{funderivdef}\eeq
and the rules for functional differentiation of the periodic Grassmann-valued
paths by the anti-commutator
\beq
\left[\frac{\delta}{\delta\psi^\mu(\tau)},\psi^\nu(\tau')\right]_+=
\delta_\mu^\nu\delta(\tau-\tau')
\label{grassfunderivdef}\eeq

The crucial point is that the fermionic part of the supersymmetric action
(\ref{susy1/2action}) is bilinear in the fermion fields so that the functional
Berezin integration induces a determinant factor ${\det}^{1/2}\|\hat\Omega\|$
which makes the remaining integration over $L\man$ in (\ref{indexpathint})
resemble the phase space path integral (\ref{quantpart}). More precisely, the
loop space fermionic bilinear form appearing in (\ref{susy1/2action}) is
\beq
\hat\Omega[x,\psi]=\oint_{S^1}d\tau~\frac{1}{2}\psi^\mu(\tau)\left(g_{\mu\nu}
\nabla_\tau-F_{\mu\nu}(x(\tau))\right)\psi^\nu(\tau)
\label{susyomega}\eeq
which, after Berezin integration, induces a loop space Liouville measure
$[d^{2n}x]~\sqrt{\det\|\hat\Omega\|}$. Introducing the nilpotent graded
derivative operator
\beq
D=\oint_{S^1}d\tau~\psi^\mu(\tau)\frac{\delta}{\delta x^\mu(\tau)}
\label{susyextderiv}\eeq
we see that (\ref{susyomega}) can be expressed as a $D$-exact quantity
\beq
\hat\Omega[x,\psi]=D\hat\Sigma[x,\psi]
\label{susyomexact}\eeq
where
\beq
\hat\Sigma[x,\psi]=\oint_{S^1}d\tau~\left\{g_{\mu\nu}(x(\tau))\dot
x^\nu(\tau)+A_\mu(x(\tau))\right\}\psi^\mu(\tau)\equiv\oint_{S^1}d\tau~\hat
\Sigma_\mu(x(\tau))\psi^\mu(\tau)
\label{susysigma}\eeq
The functional (\ref{susyomega}) can be interpreted as a loop space symplectic
structure. Strictly speaking though, it is properly termed a `pre-symplectic'
structure because although it is $D$-closed, $D\hat\Omega[x,\psi]=0$, it is not
necessarily non-degenerate on the loop space. It is this interpretation of
supersymmetric theories in general that makes infinite-dimensional
generalizations of the equivariant localization formalisms of Section 3 very
powerful tools.

In particular, the $N=\frac{1}{2}$ supersymmetry (\ref{susy1/2alg}) can be
represented by a loop space equivariant derivative operator. To see this,
introduce the nilpotent graded contraction operator
\beq
I_{\dot x}=\oint_{S^1}d\tau~\dot x^\mu(\tau)\frac{\delta}{\delta\psi^\mu(\tau)}
\label{IW}\eeq
and define the corresponding graded equivariant exterior derivative operator
\beq
D_{\dot x}=D+I_{\dot
x}=\oint_{S^1}d\tau~\left(\psi^\mu(\tau)\frac{\delta}{\delta x^\mu(\tau)}+\dot
x^\mu(\tau)\frac{\delta}{\delta\psi^\mu(\tau)}\right)
\label{DW}\eeq
Then the supersymmetry (\ref{susy1/2alg}) is immediately recognized as the
action of the derivative $D_{\dot x}\sim{\cal S}$ on $L\man\otimes
L\Lambda^1\man$. The square of $D_{\dot x}$ is the generator of time
translations on the superloop space $L\man\otimes L\Lambda^1\man$,
\beq
D_{\dot x}^2=\oint_{S^1}d\tau~\left(\dot x^\mu(\tau)\frac{\delta}{\delta
x^\mu(\tau)}+\dot\psi^\mu(\tau)\frac{\delta}{\delta\psi^\mu(\tau)}\right)
=\oint_{S^1}d\tau~\frac{d}{d\tau}
\label{susyDsquare}\eeq
so that its action on a loop space functional $W[x,\psi]$ is
\beq
D_{\dot x}^2W[x,\psi]=\oint_{S^1}d\tau~\frac{d}{d\tau}W[x,\psi]=W[x(1),\psi(1)]
-W[x(0),\psi(0)]
\label{singlevalued}\eeq
Consequently, (\ref{DW}) is a nilpotent operator provided we restrict to
single-valued loop space functionals $W[x,\psi]$. Hence the action of
$N=\frac{1}{2}$ supersymmetric quantum mechanics defines an equivariant
structure on $L\man\otimes L\Lambda^1\man$ and on the basis of the general
arguments of Section 2 we expect its path integral to localize to an integral
over $\man$, the zero locus of the vector field $\dot x^\mu(\tau)$ (i.e. the
constant paths $x(t)=x(0)\in\man~~\forall t$). This is well-known to be the
case \cite{alvgaume,friedan}.

In fact, the full action (\ref{susy1/2action}) is $D_{\dot x}$-exact,
\beq
I_{1/2}[x,\psi]=\oint_{S^1}d\tau~\hat\Sigma_\mu(x(\tau))\dot
x^\mu(\tau)+\hat\Omega[x,\psi]=I_{\dot
x}\hat\Sigma[x,\psi]+\hat\Omega[x,\psi]\equiv D_{\dot x}\hat\Sigma[x,\psi]
\label{I1/2exact}\eeq
and its bosonic part resembles the general phase space action functional
(\ref{classaction}) with $H\equiv0$ there. As mentioned before, the vanishing
of the Hamiltonian is the topological feature of such supersymmetric field
theories. Now the equivariant localization principle applied to the case at
hand would imply on its own that the path integral
\beq
{\rm index}(i\nabla\slash)=\int_{L\man\otimes
L\Lambda^1\man}[d^{2n}x]~[d^{2n}\psi]~\e^{iTD_{\dot x}\hat\Sigma[x,\psi]}
\label{susylocprinc}\eeq
is formally independent of the parameter $T$, and thus it manifestly localizes
onto $\dot x^\mu(\tau)=0$ (for $T\to\infty$). Of course, we cannot simply set
$T=0$ in (\ref{susylocprinc}) because the bosonic integration would yield
$\infty$ while the fermionic one would give 0, leading to an ill-defined
quantity. In any case, if we think of the coefficient $T$ in front of the
action as Planck's constant $\hbar$, then this is just another way of seeing
that the semi-classical approximation is exact. The $T$-independence can be
understood from the point of view that if we differentiate the right-hand side
of (\ref{susylocprinc}) with respect to $T$, then we obtain the vacuum
expectation value $\langle0|D_{\dot x}\hat\Sigma|0\rangle$ in the
supersymmetric quantum field theory above. If the vacuum itself is invariant
under the $N=\frac{1}{2}$ supersymmetry of the model, then $D_{\dot
x}|0\rangle=0$ and the vacuum expectation value of this operator vanishes. It
is these same sorts of arguments which establish the topological invariance of
BRST-exact path integrals (also known as cohomological or Witten-type
topological field theories) in general \cite{birm}. The above connection
between the formalisms of the previous Sections and the Atiyah-Singer index
theorem is the usual intimate connection between standard supersymmetric models
(for instance those which arise in the Duistermaat-Heckman interpretation of
the quantum mechanics of spin \cite{stone,alvarez}) and equivariant cohomology
\cite{blau1}.

We now use the fact that (\ref{susylocprinc}) can be evaluated for $T\to\infty$
and use a trick similar to that in Subsection 3.7. We introduce based loops on
$L\man\otimes L\Lambda^1\man$,
\beq
x^\mu(\tau)=x_0^\mu+\hat
x^\mu(\tau)~~~~~,~~~~~\psi^\mu(\tau)=\psi_0^\mu+\hat\psi^\mu(\tau)
\label{susydecomp}\eeq
with $(x_0,\psi_0)\in\man\otimes\Lambda^1\man$ the constant modes of the fields
and $(\hat x,\hat\psi)$ the non-constant fluctuations about these zero-modes,
and define the path integral measure by\footnote{\baselineskip=12pt In
Subsection 4.6 we shall be a bit more precise about this decomposition over a
general superloop space $L\man\otimes L\Lambda^1\man$. For now, we are just
concerned with using this to evaluate the index.}
\beq
[d^{2n}x]~[d^{2n}\psi]\equiv d^{2n}x_0~d^{2n}\psi_0~\prod_{\tau\in
S^1}d^{2n}\hat x(\tau)~d^{2n}\hat\psi(\tau)
\label{susypathintdecomp}\eeq
We then rescale the non-constant modes as
\beq
\hat x^\mu(\tau)\to\hat
x^\mu(\tau)/\sqrt{T}~~~~~,~~~~~\hat\psi^\mu(\tau)\to\hat\psi^\mu(\tau)/\sqrt{T}
\label{susyrescale}\eeq
 With this rescaling, we find after some algebra that in the limit $T\to\infty$
the action (\ref{susy1/2action}) becomes
\beq\new{\begin{array}{c}
T\cdot
I_{1/2}[x,\psi]~{\buildrel{T\to\infty}\over\longrightarrow}~\oint_{S^1}d\tau~
\left(\frac{1}{2}g_{\mu\nu}(x_0)\dot{\hat{x}}^\mu(\tau)\dot{\hat{x}}^\nu(\tau)
+\frac{1}{2}\hat\psi^i(\tau)\eta_{ij}\partial_\tau\hat\psi^j(\tau)-\frac{1}{2}
\psi_0^\mu
F_{\mu\nu}(x_0)\psi_0^\nu\right.\\\left.+\frac{1}{2}R_{ij\mu\nu}(x_0)
\psi_0^i\psi_0^j\dot{\hat {x}}^\mu(\tau)\dot{\hat{x}}^\nu(\tau)\right)
+{\cal O}(1/\sqrt{T})\end{array}}
\label{I1/2rescale}\eeq
where we have Taylor expanded the quantities in (\ref{susy1/2action}) about
$(x_0,\psi_0)$ using (\ref{susydecomp}) and (\ref{susyrescale}). The Jacobians
for the scaling by $\sqrt{T}$ cancel out from the bosonic and fermionic
integration measures in (\ref{susypathintdecomp}), and the remaining functional
integrations over non-constant modes are Gaussian. This illustrates the strong
role that supersymmetry plays here in reducing the complicated integrations in
(\ref{indexpathint}) to Gaussian ones.

Evaluating these Gaussian integrations in (\ref{indexpathint}) leads to
\beq
{\rm index}(i\nabla\slash)=\int_{\man\otimes\Lambda^1\man}d^{2n}x_0~d^{2n}
\psi_0~\e^{\frac{i}{4\pi}F_{\mu\nu}(x_0)\psi_0^\mu\psi_0^\nu}\left(
{\det}'\|\delta^\mu_\nu\partial_\tau-R^\mu_\nu(x_0,\psi_0)\|\right)^{-1/2}
\label{indexanalogy}\eeq
where we have ignored (infinite) constant factors arising from the Gaussian
functional integrations and normalized the $U(1)$ connection. Here the prime on
the determinant means that it is taken over the fluctuation modes with periodic
boundary conditions (i.e. the determinant with zero modes excluded). The
exponential factor in (\ref{indexanalogy}) is immediately seen to be the
(ordinary) Chern character $\ch(F)$ of the given complex line bundle
$L\to\man$, while the functional determinant coincides (modulo overall signs to
be discussed below) with the Euler form of the normal bundle to $\man$ in
$L\man$ (this is the bundle spanned by the non-constant modes of $x(\tau)$).
Thus (\ref{indexanalogy}) coincides with a formal application of the degenerate
Duistermaat-Heckman integration formula (\ref{degdh}) (more precisely the
degenerate version of the Berline-Vergne theorem) to the infinite-dimensional
integral (\ref{indexpathint}).

Finally, we discuss how to calculate the Euler form in (\ref{indexanalogy}). A
regularization scheme in general must always be chosen to evaluate
infinite-dimensional determinants \cite{miet}. Notice first that here the
infinite-dimensional Pfaffian arising from the fermionic integration cancels
from the result of the infinite-dimensional Gaussian integral over the bosonic
fluctuation modes. Thus, just as in the
finite-dimensional case, the sign dependence of the Pfaffian gets transfered
to the inverse square root of the determinant. The spectral asymmetry
associated with the sign of the infinite-dimensional Pfaffian (see
(\ref{sgnpfaff})) has to be regulated and is given by the Atiyah-Patodi-Singer
eta-invariant \cite{eguchi,szsem1} of the Dirac operator $\partial_\tau-R$,
\beq\new{\begin{array}{ll}
\eta(\partial_\tau-R)&=\lim_{s\to0}\int\!\!\!\!\!\!\!\sum
d\lambda~\sgn(\lambda)
|\lambda|^{-s}+\dim~{\rm
ker}(\partial_\tau-R)\\&=\lim_{s\to0}\frac{1}{\Gamma\left(\frac{s+1}{2}\right)
}\int_0^\infty
dt~t^{(s-1)/2}~\tr\left\|(\partial_\tau-R)\e^{-t(\partial_\tau-R)^2}\right\|
\end{array}}
\label{atiyahetainv}\eeq
where the integration (and/or sum) is over all non-zero eigenvalues $\lambda$
of
$\partial_\tau-R$ and
\beq
\Gamma(x)=\int_0^\infty dt~t^{x-1}\e^{-t}~~~~~,~~~~~x>0
\label{Eulergammafn}\eeq
is the Euler gamma-function.

Next, we evaluate the determinant using standard supersymmetry regularizations
\cite{alvgaume,friedan} for first-order differential operators defined on a
circle. The most convenient such choice is Riemann
zeta-function regularization. The non-constant single-valued eigenfunctions of
the operator $\partial_\tau$ on $S^1$ are
$\e^{2\pi ik\tau}$, where $k$ are non-zero integers. Since the matrix $R$ is
antisymmetric, it can be skew-diagonalized into $n$ $2\times2$ skew-diagonal
blocks $R^{(j)}$ with skew eigenvalues $\lambda_j$, where $j=1,\dots,n$.
For each such block $R^{(j)}$, we get the formal contribution to the
determinant in (\ref{fluctint}),
\beq
{\det}'\|\partial_\tau-R^{(j)}\|=\prod_{k\neq0}\left(2\pi ik+\lambda
_j\right)\left(2\pi ik-\lambda_j\right)=g(\lambda_j/2\pi i)g(-
\lambda_j/2\pi i)\prod_{k\neq0}\left(2\pi i\right)^2
\label{detblock}\eeq
where we have defined the function $g(z)$ as the formal product
\beq
g(z)=\prod_{k\neq0}(k+z)
\label{gdef}\eeq
We can determine the regulated form of the function $g(z)$ by examining its
logarithmic derivative $g'(z)/g(z)$ \cite{friedan}. This is, as a function of
$z\in\IC$, a function with simple poles of residue 1 at $z=k$ a non-zero
integer. Thus we take $g'(z)/g(z)=\pi\cot\pi z-1/z+b$ and integrating this we
get
\beq
g(z)=\sin\pi z\e^{bz}/\pi z
\label{greg}\eeq
where we have normalized $g(z)$ so that $g(0)=1$. The arbitrary phase in
(\ref{greg}) appears because the zeroes of the function (\ref{gdef}) occur at
$z=k\in\IZ$ which determine it up to a function without zeroes, i.e. an
exponential function. When substituted into (\ref{detblock}), it is related to
the sign of the determinant, and hence to the eta-invariant
(\ref{atiyahetainv}).  In certain instances (see Subsection 5.4) it is
necessary to make a specific choice for the regularization of this phase
\cite{miet} (i.e. a choice for $b$). In our case here, however, the phase $b$
will cancel out explicitly in (\ref{detblock}) and so we can neglect its
effect.

The infinite prefactor in (\ref{detblock}) is regularized using the Riemann
zeta-function
\beq
\zeta(s)=\sum_{k=1}^\infty\frac{1}{k^s}
\label{zetafn}\eeq
which is finite for $s\geq0$ with $\zeta(0)=-1/2$ \cite{gradryz}. We find that
\beq
\prod_{k\neq0}\left(2\pi i\right)^2=\prod_{k>0}\left(2\pi i
\right)^4=\left(2\pi i\right)^{4(\sum_{k=1}^\infty\frac{1}{k^s})
|_{s=0}}=\left(2\pi i\right)^{4\zeta(0)}=\left(2\pi i\right)^{-2}
\label{infprefact}\eeq
and thus the block contribution (\ref{detblock}) to the functional determinant
in (\ref{indexanalogy}) is
\beq
{\det}'\|\partial_\tau-R^{(j)}\|=\frac{1}{\pi^2}\left(\frac{
\sin\frac{i\lambda_j
}{2}}{\lambda_j}\right)^2=\left(\frac{1}{2\pi i}\right)^2\det\left[\frac{
\sinh\frac{1}{2}R^{(j)}}{\frac{1}{2}R^{(j)}}\right]
\label{ahatblock}\eeq
Multiplying the blocks together we see that the fluctuation determinant
appearing in (\ref{indexanalogy}) is just given by the ordinary $V=0$ $\hat
A$-genus (\ref{eqahat}) with respect to the curvature $R$, and thus the index
is
\beq
{\rm index}(i\nabla\slash)=\int_\man\ch(F)\wedge\hat A(R)
\label{atiyahsingthm}\eeq

The result (\ref{atiyahsingthm}) is the celebrated Atiyah-Singer index theorem
for a twisted spin complex \cite{eguchi}. We see then that a formal application
of the Berline-Vergne theorem yields the well-known Atiyah-Singer geometrical
representation of the index of $i\nabla\slash$. This result can be generalized
to include the coupling of fermions to a non-abelian gauge field $A$ on a
vector bundle $E\to\man$. Now the functional $\hat\Omega[x,\psi]$ above is no
longer closed ($F=dA+[A~{\buildrel\wedge\over,}~A]/2$ obeys the Bianchi
identity), but the construction above can still be carried through using the
coadjoint orbit representation of the structure group of the principal fiber
bundle \cite{alek2,hietmor} (see Subsection 5.1). The above representation of
the Witten index in terms of a supersymmetric path integral can also be
generalized to other differential operators, not just the Dirac operator
(\ref{diracop}). For instance, the Witten index for the DeRham exterior
derivative operator $d$ describes the DeRham complex of the manifold $\man$
\cite{witten1}. The index is now the Euler characteristic of $\man$, the
supersymmetric path integral is that of $N=1$ (DeRham) supersymmetric quantum
mechanics, and the localization formula reproduces the Gauss-Bonnet-Chern
theorem. An equivariant generalization then yields the Poincar\'e-Hopf theorem
of classical Morse theory \cite{witten1}. We shall elaborate on some of these
ideas, as well as how they extend to infinite-dimensional cases relevant to
topological field theories, in Section 8.

Finally, we point out that the equivariant cohomological interpretation above
is particularly well-suited to reproduce the Callias-Bott index theorems
\cite{hietniemi}, i.e. the analog of the Atiyah-Singer index theorem for a
Dirac operator on an odd-dimensional non-compact manifold \cite{eguchi}. The
supersymmetric model is now that of $N=1$ supersymmetric quantum mechanics with
background monopole and soliton configurations. The trace over zero modes
representing the Witten index can in this case be infinite-dimensional, and it
is simply not true that the partition function is independent of the parameter
$T$ (the index being obtained for $T\to\infty$) so that one cannot simplify
matters by taking the $T\to0$ limit. The canonical realization of the $N=1$
supersymmetry by considering an equivariant structure over an extended
superloop space, defined by a larger mixing of bosonic and fermionic
coordinates, preserves the contributions of the zero modes which would
otherwise be lost \cite{morniemi1,morniemi2} and the localization tricks used
above become directly applicable. We shall discuss this a bit more in Section
8. Furthermore, the index in these cases can be computed from a
higher-dimensional Atyah-Singer index theorem by introducing a simple first
class constraint (i.e. one that is a symmetry of the Hamiltonian, or
equivalently a constant of the motion) that eliminates the extra dimensions. We
refer to \cite{hietniemi} for more details about this approach to index
theorems in general.

\subsection{Loop Space Symplectic Geometry and Equivariant Cohomology}

The example of the last Subsection has shown that a formal generalization of
symplectic geometry and equivariant cohomology to the loop space of a physical
problem can result in a (correct) localization formula in the same spirit as
those of Sections 2 and 3. The localization principle in this context was just
a manifestation of the supersymmetry of that model. It has also provided us
with some important functional space tools that will be used throughout this
Review, as well as hints on how to proceed to loop space generalizations of the
results of the earlier Sections. Following these lessons we have learned, we
shall now focus on developing some geometric methods of determining quantum
partition functions of generic (not necessarily supersymmetric) dynamical
systems. Given the formulation of the path integral in Subsection 4.1 above on
a general symplectic manifold, we wish to treat the problem of its exact
evaluation within the geometric context of Section 3. As exemplified by the
example of the previous Subsection, for this we need a formulation of exterior
and symplectic differential geometry on the loop space $L\man$ over the phase
space $\man$. This will ultimately lead to a formal, infinite-dimensional
generalization of the equivariant localization priniciple for path integrals,
and thus formal conditions and methods for evaluating exactly these functional
integrations which in general are far more difficult to deal with than their
classical counterparts. As with the precise definition of the functional
integrals above, we shall be rather cavalier here about the technicalities of
infinite-dimensional manifolds. The loop space $L\man\to\man$ is an
infinite-dimensional vector bundle -- the fiber over a point $x\in\man$ is the
space of all loops $x(t)$ based at $x$, $x(T)=x(0)=x$, which is an
infinite-dimensional non-abelian group with group multiplication of loops
$(x_1x_2)(t)$ defined by first traversing the loop $x_1(t)$, and then the loop
$x_2(-t)$ in the opposite direction. These quantities should therefore be
properly defined using Sobolev completions of the infinite-dimensional groups
and spaces involved. This can always be done in an essentially straightforward
and routine manner \cite{birm}.

We define the exterior algebra $L\Lambda\man$ of the loop space by lifting the
Grassmann generators $\eta^\mu$ of $\Lambda\man$ to anti-commuting periodic
paths $\eta^\mu(t)$ which generate $L\Lambda\man$ and which are to be
identified as the basis $dx^\mu(t)$ of loop space 1-forms. With this, we can
define loop space differential $k$-forms
\beq
\alpha=\int_0^Tdt_1\cdots dt_k~\frac{1}{k!}\alpha_{\mu_1\cdots\mu_k}[x;t_1,
\dots,t_k]\eta^{\mu_1}(t_1)\cdots\eta^{\mu_k}(t_k)
\label{loopkform}\eeq
and the loop space exterior derivative is defined by lifting the exterior
derivative of the phase space $\man$,
\beq
d_L=\int_0^Tdt~\eta^\mu(t)\frac{\delta}{\delta x^\mu(t)}
\label{loopspderiv}\eeq

The loop space symplectic geometry is determined by a loop space symplectic
2-form
\beq
\Omega=\int_0^Tdt~dt'~\frac{1}{2}\Omega_{\mu\nu}[x;t,t']\eta^\mu(t)\eta^\nu(t')
\label{loopsympl}\eeq
which is closed
\beq
d_L\Omega=0
\label{omloopclosed}\eeq
or in local coordinates $x^\mu(t)$ on $L\man$,
\beq
\frac{\delta}{\delta
x^\mu(t)}\Omega_{\nu\lambda}[x;t',t'']+\frac{\delta}{\delta
x^\nu(t)}\Omega_{\lambda\mu}[x;t',t'']+\frac{\delta}{\delta x^\lambda(t)}
\Omega_{\mu\nu}[x;t',t'']=0
\label{locloopclosed}\eeq
Thus we can apply the infinite-dimensional version of Poincar\'e's lemma to
represent $\Omega$ locally in terms of the exterior derivative of a loop space
1-form
\beq
\vartheta=\int_0^Tdt~\vartheta_\mu[x;t]\eta^\mu(t)
\label{vartheta}\eeq
as
\beq
\Omega=d_L\vartheta
\label{looppoin}\eeq
We further assume that (\ref{loopsympl}) is non-degenerate, i.e. the matrix
$\Omega_{\mu\nu}[x;t,t']$ is invertible on the loop space.

The canonical choice of symplectic structure on $L\man$ which coincides with
the loop space Liouville measure introduced in (\ref{quantpart}) is that which
is induced from the symplectic structure of the phase space,
\beq
\Omega_{\mu\nu}[x;t,t']=\omega_{\mu\nu}(x(t))\delta(t-t')
\label{loopsympllift}\eeq
which is diagonal in its loop space indices $t,t'$. We shall use similar
liftings of other quantities from the phase space to the loop space. In this
way, elements $\alpha(x)$ of $L\Lambda_x\man$ (or $LT_x\man$) at a loop
$x\in L\man$ are regarded as deformations of the loop, i.e. as elements of
$\Lambda\man$ (or $T\man$) restricted to the loop $x^\mu(t)$ such that
$\alpha[x;t]\in \Lambda_{x(t)}\man$ (or $T_{x(t)}\man$). This means that these
vector bundles over $L\man$ are infinite-dimensional spaces of

            sections of
the pull-back of the phase space bundles to $[0,T]$ by the map $x(t):[0,T]\to
\man$. In particular, we define loop space canonical transformations as
loop space changes of variable $F[x(t)]$ that leave $\Omega$ invariant. These
are the transformations of the form
\beq
\vartheta~{\buildrel F\over\longrightarrow}~\vartheta_F=\vartheta+d_LF
\label{loopcantransf}\eeq
Thus in the context of the loop space symplectic geometry determined by
(\ref{loopsympllift}), the quantum partition function is an integral over the
infinite-dimensional symplectic manifold $(L\man,\Omega)$ with the loop space
Liouville measure there determined by the canonically-invariant closed form on
$L\man$ given by exterior products of $\Omega$ with itself,
\beq
[d\mu_L(x)]=[d^{2n}x]~\sqrt{\det\|\Omega\|}
\label{loopliouville}\eeq

The loop space Hamiltonian vector field associated with the action
(\ref{classaction}) has components
\beq
V_S^\mu[x;t]=\int_0^Tdt'~\Omega^{\mu\nu}[x;t,t']\frac{\delta S[x]}{\delta x^
\nu(t')}=\dot x^\mu(t)-V^\mu(x(t))
\label{loophamvec}\eeq
with $V^\mu=\omega^{\mu\nu}\partial_\nu H$ as usual the Hamiltonian vector
field on $\man$. The zeroes of $V_S$
\beq
L\man_S=\{x(t)\in L\man:V_S[x(t)]=0\}
\label{lszeroes}\eeq
are the extrema of the action (\ref{classaction}) and coincide with the
classical trajectories of the dynamical system, i.e. the solutions of the
classical Hamilton equations of motion. The loop space contraction operator
with respect to a loop space vector field $W^\mu[x;t]$ is given by
\beq
i_W=\int_0^Tdt~W^\mu[x;t]\frac{\delta}{\delta\eta^\mu(t)}
\label{loopcontr}\eeq
Thus we can define a loop space equivariant exterior derivative
\beq
Q_W=d_L+i_W
\label{loopeqderiv}\eeq
whose square is the Lie derivative along the loop space vector field $W$,
\beq
Q_W^2=d_Li_W+i_Wd_L=\lie_W=\int_0^Tdt\left(W^\mu\frac{\delta}{\delta x^\mu}
+\partial_\nu W^\mu\eta^\nu\frac{\delta}{\delta\eta^\mu}\right)
\label{lieloop}\eeq
When $W=V_S$ is the loop space Hamiltonian vector field, we shall for ease
denote the corresponding operators above as $i_{V_S}\equiv i_S$, etc.

The partition function can be written as in the finite-dimensional case using
the functional Berezin integration rules to absorb the determinant factor
into the exponential in terms of the anti-commuting periodic fields
$\eta^\mu(t)$,
\beq\new{\begin{array}{ll}
Z(T)&=\int_{L\man\otimes L\Lambda^1\man}[d^{2n}x]~[d^{2n}\eta]~\exp\left\{iS[x]
+\frac{i}{2}\int_0^Tdt~\omega_{\mu\nu}(x(t))\eta^\mu(t)\eta^\nu(t)\right\}\\&=
\int_{L\man\otimes L\Lambda^1\man}[d^{2n}x]~[d^{2n}\eta]~\e^{i(S[x]+\Omega
[x,\eta])}\end{array}}
\label{partexp}\eeq
so that in this way $Z(T)$ is written in terms of an augmented action $S+
\Omega$ on the super-loop space $L\man\otimes L\Lambda^1\man$. From this we can
now formally describe the $S^1$-equivariant cohomology of the loop space.

The operator $Q_S$ is nilpotent on the subspace
\beq
L\Lambda_S\man=\{\alpha\in L\Lambda\man:\lie_S\alpha=0\}
\label{qsnilpot}\eeq
of equivariant loop space functionals. The loop space observable $S[x]$
defines the loop space Hamiltonian vector field through
\beq
d_LS=-i_S\Omega
\label{hameqloopsp}\eeq
from which it follows that the integrand of the quantum partition function
(\ref{partexp}) is equivariantly closed,
\beq
Q_S(S+\Omega)=(d_L+i_S)(S+\Omega)=0
\label{actioneqclosed}\eeq
and so the augmented action $S+\Omega$ can be locally represented as the
equivariant exterior derivative of a 1-form $\hat\vartheta$,
\beq
S+\Omega=Q_S\hat\vartheta=\int_0^Tdt~\left(V_S^\mu\hat\vartheta_\mu+\frac{1}{2}
\Omega_{\mu\nu}\eta^\mu\eta^\nu\right)
\label{someqexact}\eeq
 From (\ref{actioneqclosed}) we find that
\beq
Q_S^2\hat\vartheta=\lie_S\hat\vartheta=0
\label{liehattheta}\eeq
and so $\hat\vartheta$ lies in the subspace (\ref{qsnilpot}). If $\Phi_S$ is
some globally defined loop space 0-form with
\beq
\lie_S(d_L\Phi_S)=0
\label{phislie}\eeq
then we see that $\hat\vartheta$ is not unique but the augmented action
(\ref{someqexact}) is invariant under the loop space canonical transformation
\beq
\hat\vartheta\to\hat\vartheta+d_L\Phi_S
\label{hatthetacan}\eeq
Thus the partition function (\ref{partexp}) has a very definite interpretation
in terms of the loop space equivariant cohomology $H_S(L\man)$ determined by
the operator $Q_S$ on $L\Lambda_S\man$.

\subsection{Hidden Supersymmetry and the Loop Space Localization Principle}

The fact that the integrand of the partition function above can be interpreted
in terms of a loop space equivariant cohomology suggests that we can
localize it by choosing an appropriate representative of the loop space
equivariant cohomology class determined by the augmented action $S+\Omega$.
However, the arguments which showed in the finite-dimensional cases that
the partition function integral is invariant under such topological
deformations cannot be straightforwardly applied here since there is no direct
analog of Stokes' theorem for infinite-dimensional manifolds. Nonetheless, the
localization priniciple can be established by interpreting the equivariant
cohomological structure on $L\man$ as a ``hidden" supersymmetry of the quantum
theory. In this way one has a sort of Stokes' theorem in the form of a Ward
identity associated with this supersymmetry (as was the case in Subsection 4.2
above), where we interpret the fundamental localization property
(\ref{eqintcoh}) as an infinitesimal change of variables in the integral. The
partition function (\ref{partexp}) can be interpreted as a BRST gauge-fixed
path integral \cite{birm} with the $\eta^\mu(t)$ viewed as fermionic ghost
fields and $x^\mu(t)$ as the fundamental bosonic fields of the model. The
supersymmetry is suggested by the ungraded structure of $Q_S$ on
$L\Lambda_S\man$ which maps even-degree, commuting loop space forms (bosons)
into odd-degree, anti-commuting forms (fermions)\footnote{\baselineskip=12pt In
this interpretation the form degree can be thought of as a ghost number, so
that the physical observables of the system (i.e. those with ghost number 0)
are the smooth functions on $\man$. Furthermore, at this point it is useful to
recall the analogy between $Q_S=d_L+i_S$ and the gauge-covariant derivative in
a gauge theory for the following analogies with BRST quantization of gauge
theories. See Appendix A for a brief review of some of the ideas of BRST
quantization.}. Since the fermion fields $\eta^\mu(t)$ appear by themselves
without a conjugate partner, this determines an $N=\frac{1}{2}$ supersymmetry.
The
$N=\frac{1}{2}$ supersymmetry algebra $Q_S^2=\lie_S$ implies that $Q_S$ is
a supersymmetry charge on the subspace $L\Lambda_S\man$, and the augmented
action is supersymmetric, $Q_S(S+\Omega)=0$. Thus here $L\Lambda_S\man$
coincides with the BRST complex of physical (supersymmetric) states, and the
BRST transformations of the fundamental bosonic fields $x^\mu(t)$ and their
superpartners $\eta^\mu(t)$ are given by the action of the infinitesimal
supersymmetry generator $Q_S$ \footnote{\baselineskip=12pt In supersymmetric
quantum field theories the BRST transformations of operators and fields are
represented by a graded BRST commutator $\{Q_S,\cdot\}$. This commutator in the
case at hand can be represented by the
Poisson structure of the phase space as follows. We introduce periodic
trajectories $\lambda_\mu(t)$ in $L\man$ conjugate to $x^\mu(t)$ and
anticommuting periodic paths $\bar\eta_\mu(t)$ conjugate to $\eta^\mu(t)$, i.e.
\bd
\{\lambda_\mu(t),x^\nu(t')\}_\omega=\{\bar\eta_\mu(t),\eta^\nu(t')\}_\omega=
\delta^\nu_\mu\delta(t-t')
\ed
which are to be identified as the Poisson algebra realization of the operators
$\lambda_\mu(t)\sim\frac{\delta}{\delta x^\mu(t)}$ and $\bar\eta_\mu(t)\sim
\frac{\delta}{\delta\eta^\mu(t)}$ acting in the usual way. This gives a
Poisson bracket realization of the actions of the operators $d_L$ and $i_S$,
and then the action of $Q_S$ is represented by the BRST commutator $\{Q_S,
\cdot\}_\omega$. In the following, one can keep in mind this representation
which maintains a complete formal analogy with supersymmetric theories.}
\beq
Q_Sx^\mu(t)=\eta^\mu(t)~~~~~,~~~~~Q_S\eta^\mu(t)=V_S^\mu[x;t]
\label{brst}\eeq

This formal identification of the equivariant cohomological structure as a
hidden supersymmetry allows one to interpret the (non-supersymmetric) quantum
theory as a supersymmetric or topological field theory. It was Blau,
Keski-Vakkuri and Niemi \cite{blaukesk} who pointed out that a quite general
localization principle could be formulated for path integrals using rather
formal functional techniques introduced in the BRST quantization of first class
constrained systems \cite{niemi}. In these theories a BRST transformation
produces a super-Jacobian on the super-loop space $L\man\otimes L\Lambda^1\man$
whose corrections are related to anomalies and BRST supersymmetry breaking. The
arguments below are therefore valid {\it provided} that the $Q_S$-supersymmetry
above is not broken in the quantum theory.

The argument for infinite-dimensional localization proceeds as follows.
Consider the 1-parameter family of phase space path integrals
\beq
{\cal Z}(\lambda)=\int_{L\man\otimes L\Lambda^1\man}[d^{2n}x]~[d^{2n}\eta]~
\e^{i(S[x]+\Omega[x,\eta]+\lambda Q_S\psi[x,\eta])}
\label{zspath}\eeq
where $\lambda\in\IR$ and $\psi\in L\Lambda^1_S\man$ is a gauged fermion
field which is homotopic to 0 under the supersymmetry transformation
generated by $Q_S$ (i.e. $\psi\equiv\psi_{s=1}$, where $\psi_s$, $s\in[0,1]$,
is a 1-parameter family of gauge fermions with $Q_S^2\psi_s=0$ and
$\psi_{s=0}=0$). As in the finite-dimensional case, we wish to establish the
$\lambda$-independence of this path integral, i.e. that (\ref{zspath}) depends
only on the BRST cohomology class determined by the augmented action, so that a
choice of $\lambda\neq0$ amounts to a choice of
representative of $S+\Omega$ in its loop space equivariant cohomology class
and different choices of non-trivial representatives then lead to the desired
localization schemes. Consider an infinitesimal variation $\lambda\to\lambda
+\delta\lambda$ of the argument of (\ref{zspath}), i.e. let
$\psi\to\psi+\delta\psi$
with
\beq
\delta\psi=\delta\lambda\cdot\psi
\label{deltapsi}\eeq
and consider the infinitesimal supersymmetry transformation on the super-loop
space parametrized by the gauge fermion $\delta\psi\in L\Lambda_S^1\man$,
\beq\new{\begin{array}{l}
x^\mu\to\bar x^\mu=x^\mu+\delta x^\mu=x^\mu+\delta\psi\cdot Q_Sx^\mu=x^\mu+
\delta\psi\cdot\eta^\mu\\\eta^\mu\to\bar\eta^\mu=\eta^\mu+\delta\eta^\mu=\eta^
\mu+\delta\psi\cdot Q_S\eta^\mu=\eta^\mu+\delta\psi\cdot V_S^\mu\end{array}}
\label{supertransf}\eeq
Since $Q_S(S+\Omega)=\lie_S\psi=0$, the argument of the path integral
(\ref{zspath}) is BRST-invariant.

However, the corresponding super-Jacobian arising in the Feynman measure in
(\ref{zspath}) on $L\man\otimes L\Lambda^1\man$ is non-trivial and it has
precisely the same functional form as that in a standard BRST transformation
\cite{niemi}. The pertinent super-Jacobian here is given by the
super-determinant
\beq
[d^{2n}\bar x]~[d^{2n}\bar\eta]=~{\rm sdet}\left\|\new{\begin{array}{cc}
\frac{\delta\bar x}{\delta x}&\frac{\delta\bar x}{\delta\eta}\\\frac{\delta
\bar\eta}{\delta x}&\frac{\delta\bar\eta}{\delta\eta}\end{array}}\right\|
[d^{2n}x]~[d^{2n}\eta]
\label{superjacdet}\eeq
and the path integral (\ref{zspath}) is invariant under arbitrary smooth
changes of variables. For infinitesimal $\delta\lambda$, the identity
\beq
\tr\log\|A\|=\log\det\|A\|
\label{trlogdet}\eeq
implies that the super-determinant in (\ref{superjacdet}) can be computed in
terms of the super-trace, the super-loop space sum of the diagonal entries in
(\ref{superjacdet}), as ${\rm sdet}\|A\|=1+~{\rm str}\|A\|$. This gives
\beq\new{\begin{array}{ll}
[d^{2n}\bar x]~[d^{2n}\bar\eta]&=\left\{1+\int_0^Tdt~\left(\frac{\delta}{
\delta x^\mu}(\delta\psi)\eta^\mu-\frac{\delta}{\delta\eta^\mu}(\delta\psi)
V_S^\mu\right)\right\}[d^{2n}x]~[d^{2n}\eta]\\&=\left\{1-\int_0^Tdt~\left(
\eta^\mu\frac{\delta}{\delta x^\mu}+V_S^\mu\frac{\delta}{\delta\eta^\mu}\right)
\delta\psi\right\}[d^{2n}x]~[d^{2n}\eta]\\&=(1-Q_S\delta\psi)[d^{2n}x]~[d^{2n}
\eta]\sim\e^{-\delta\lambda\cdot Q_S\psi}[d^{2n}x]~[d^{2n}\eta]\end{array}}
\label{supjactr}\eeq

Thus substituting the change of variables (\ref{supertransf}) with
super-Jacobian (\ref{supjactr}) into the path integral (\ref{zspath}) we
immediately see that
\beq
{\cal Z}(\lambda)={\cal Z}(\lambda-\delta\lambda)
\label{zspathindep}\eeq
which establishes the independence of the path integral (\ref{zspath}) under
homotopically-trivial deformations which live in the subspace (\ref{qsnilpot}).
This proof of the $\lambda$-independence (or the $\psi$-independence more
generally) of (\ref{zspath}) is a specialization of the Fradkin-Vilkovisky
theorem \cite{batfrad,fradvil,niemi} to the supersymmetric theory above, which
states that local supersymmetric variations of gauge fermions in a
supersymmetric BRST gauge-fixed path integral leave it invariant. Indeed, the
addition of the BRST-exact term
$Q_S\psi$ can be regarded as a gauge-fixing term (the reason why $\psi$ is
termed here a `gauge fermion') which renormalizes the theory but leaves
it invariant under these perturbative deformations. The addition of this
term to the action of the quantum theory above is therefore regarded as a
topological deformation, in that it does not change the value of the original
partition function which is the $\lambda\to0$ limit of (\ref{zspath}) above.
This is consistent with the general ideas of topological field theory, in
which a supersymmetric BRST-exact action is known to have no local propagating
degrees of freedom and so can only describe topological invariants of the
underlying space. We shall discuss these more topological aspects of
BRST-exact path integrals, also known as Witten-type topological field
theories \cite{birm}, in due course. In any case, we can now write down
the loop space localization principle
\beq
Z(T)=\lim_{\lambda\to\infty}\int_{L\man\otimes L\Lambda^1\man}[d^{2n}x]~[d^{2n}
\eta]~\e^{i(S[x]+\Omega[x,\eta]+\lambda Q_S\psi[x,\eta])}
\label{loopsploc}\eeq
so that the quantum partition function localizes onto the zeroes of the
gauge fermion field $\psi$.

Given the localization property (\ref{loopsploc}) of the quantum theory, we
would now like to pick a suitable representative $\psi$ making the localization
manifest. As in the finite dimensional cases, the localizations of interest
both physically and mathematically are usually the fixed point locuses of
loop space vector fields $W$ on $L\man$. To translate this into a loop space
differential form, we introduce a metric tensor $G$ on the loop space and
take $\psi$ to be the associated metric-dual form
\beq
\psi=\int_0^Tdt~dt'~G_{\mu\nu}[x;t,t']W^\mu[x;t]\eta^\nu(t')
\label{psiw}\eeq
of the loop space vector field $W$. The supersymmetry condition $\lie_S\psi=0$
is then equivalent to the Killing equation $\lie_SG=0$ and the additional
requirement $\lie_SW=0$ on $W$ \footnote{\baselineskip=12pt We also require
that the combination
(\ref{psiw}) be such that it determines a homotopically trivial element as
above, so that it introduces no extra topological effects into the path
integral (\ref{loopsploc}) when evaluated on contractable loops. For the
most part, we shall be rather cavalier about this requirement and discuss it
only towards the end of this Review.}, where
\beq
\lie_SW=\int_0^Tdt~\left(\frac{d}{dt}-\lie_{V(x(t))}\right)W[x;t]
\eeq
In principle there are many useful choices for $W$ obeying such a restriction,
but we shall
be concerned mostly with those which can be summarized in
\beq
W^\mu[x;t]=r\dot x^\mu(t)-sV^\mu(x(t))
\label{wchoice}\eeq
where the parameters $r,s$ are chosen appropriate to the desired localization
scheme.

As for the metric in (\ref{psiw}), there are also in principle many
possibilities. However,
there only seems to be 1 general class of loop space metric tensors
to which general arguments and analyses can be applied. To motivate these,
we note first that the equivariant exterior derivative $Q_S$ can be written as
\beq
Q_S=Q_{\dot x}-i_V=d_L+i_{\dot x}-i_V
\label{qsxdot}\eeq
and the square of the operator $Q_{\dot x}$ is just the generator of time
translations
\beq
Q_{\dot x}^2=\lie_{\dot x}=\int_0^Tdt~\left(\dot x^\mu(t)\frac{\delta}{\delta
x^\mu(t)}+\dot\eta^\mu(t)\frac{\delta}{\delta\eta^\mu(t)}\right)
=\int_0^Tdt~\frac{d}{dt}
\label{liexdot}\eeq
This operator arises when we assume that the loop space Hamiltonian vector
field
generates an $S^1$-flow on the loop space, parametrized by a parameter
$\tau\in[0,1]$ so that the flow is $x^\mu(t)\to x^\mu(t;\tau)$ with $x^\mu(t;0)
=x^\mu(t;1)$, such that in the selected loop space coordinates $x^\mu(t)$ the
flow parameter $\tau$ also shifts the loop (time) parameter $t\to t+\tau$. In
this case we have
\beq
V_S^\mu[x;t]=\frac{\partial x^\mu(t;\tau)}{\partial\tau}\biggm|_{\tau=0}=\dot
x^\mu(t)
\label{vsxdot}\eeq
and the supersymmetry transformation (\ref{brst}) becomes
\beq
Q_{\dot x}x^\mu(t)=\eta^\mu(t)~~~~~,~~~~~Q_{\dot x}\eta^\mu(t)=\dot x^\mu(t)
\label{brstxdot}\eeq
which we recall is the infinitesimal supersymmetry discussed in Subsection 4.2
above.
In particular, the effective action is now (locally) of the functional form
\beq
S+\Omega=\int_0^Tdt~\left(\theta_\mu(x)\dot x^\mu+\frac{1}{2}\omega_{\mu\nu}
(x(t))\eta^\mu\eta^\nu\right)=(d_L+i_{\dot x})\vartheta=Q_{\dot x}\hat\vartheta
\label{newsom}\eeq
and the topological invariance of the quantum theory, i.e. the invariance of
(\ref{newsom}) under BRST-deformations by elements $\psi$ of the subspace
$L\Lambda^1_S\man$, is according to (\ref{liexdot}) determined by
{\it arbitrary} globally defined single-valued functionals on $L\man$, i.e.
$\psi(0)=\psi(T)$. This form of the $U(1)$-equivariant cohomology on the loop
space is called the model-independent circle action.

We shall therefore demand that the localization functionals in (\ref{psiw})
be invariant under the model-independent $S^1$-action on $L\man$ (i.e.
rigid rotations $x(t)\to x(t+\tau)$ of the loops). This requires that the loop
space metric tensor above obey $\lie_{\dot x}G=0$, or equivalently that
$G_{\mu\nu}[x;t,t']=G_{\mu\nu}[x;t-t']$ is diagonal in its loop space indices.
Since the quantum theory is to describe the dynamics of a given Hamiltonian
system for which we know the underlying manifold $\man$, the best way to
pick the Riemannian structure on $L\man$ is to lift a metric tensor $g$ from
$\man$ so that $G$ takes the ultra-local form
\beq
G_{\mu\nu}[x;t,t']=g_{\mu\nu}(x(t))\delta(t-t')
\label{gultraloc}\eeq
and its action on loop space vector fields is given by
\beq
G(V_1,V_2)=\int_0^Tdt~g_{\mu\nu}(x(t))V_1^\mu[x;t]V_2^\nu[x;t]
\label{glooponvec}\eeq
Because of the reparametrization invariance of the integral (\ref{glooponvec}),
the metric tensor $G$ is invariant under the canonical flow on $L\man$
generated by $\dot x$. The Lie derivative condition on $G$ is then equivalent
to the Lie derivative condition (\ref{lieg}) with respect to the Hamiltonian
vector field $V$ on $\man$. Thus infinite-dimensional localization requires as
well that the phase space $\man$ admit a globally-defined $U(1)$-invariant
Riemannian structure on $\man$ with respect to the {\it classical} dynamics of
the given Hamiltonian system. As discussed before, the condition that the
Hamiltonian $H$ generates an isometry of a metric $g$ on $\man$ (through the
induced Poisson structure on $(\man,\omega)$) is a very restrictive condition
on the Hamiltonian dynamics. Essentially it means that $H$ must be related to
the global action
(\ref{gaction}) of a group $G$ on $\man$, so that the classical mechanics
generates a very large degree of symmetry. As mentioned before, the
infinite-dimensional results above, in particular
the evaluation of the super-Jacobian in (\ref{supjactr}), are as reliable
as the corresponding calculations in standard BRST quantization, provided that
the boundary conditions in (\ref{zspath}) are also supersymmetric. Provided
that the assumptions on the classical properties of the Hamiltonian are
satisfied (as for the finite-dimensional cases), the above derivation will
stand correct unless the supersymmetry $Q_S^2=\lie_S$ is broken in the
quantum theory, for instance by a scale anomaly in the rescaling of the
metric $G_{\mu\nu}\to\lambda\cdot G_{\mu\nu}$ above. See Appendices A and B for
the precise correspondence between BRST quantization, equivariant cohomology
and localization.

\subsection{The WKB Localization Formula}

We shall now begin examining the various types of localization formulas
that can be derived from the general principles of the last Subsection. The
first infinite-dimensional localization formula that we shall present is the
formal generalization of the Duistermaat-Heckman integration formula, whose
derivation follows the loop space versions of the steps used in Subsections 2.6
and 3.3. We assume that the action $S$ has (finitely-many) isolated and
non-degenerate critical trajectories, so that the zero locus (\ref{lszeroes})
consists of isolated classical loops in $L\man$, i.e. we assume that the
determinant of the associated Jacobi fields arising from a second-order
variation of $S$ is non-vanishing on these classical trajectories. Under these
assumptions, we set $r=s=1$ in (\ref{wchoice}), so that
\beq
\psi=\int_0^Tdt~g_{\mu\nu}V_S^\mu\eta^\nu~~~,~~~Q_S\psi=\int_0^Tdt~\left[
g_{\mu\nu}V_S^\mu V_S^\nu+\eta^\mu\left(g_{\mu\nu}\partial_t-g_{\nu\lambda}
\partial_\mu V^\lambda+V_S^\lambda\partial_\mu g_{\nu\lambda}\right)\eta^\nu
\right]
\label{wkbpsi}\eeq

Proceeding just as in the finite-dimensional case, the
evaluation of the localization integral (\ref{loopsploc}) gives
\beq\new{\begin{array}{ll}
Z(T)&\sim\int_{L\man}[d^{2n}x]~\sqrt{\det\|\Omega\|}\sqrt{\det\|\delta V_S\|}
{}~\delta(V_S)\e^{iS[x]}\\&\sim\int_{L\man}[d^{2n}x]~\sqrt{\det\|\Omega\|}
\sqrt{\det\|\delta^\mu_\nu\partial_t-\partial_\nu(\omega^{\mu\lambda}\partial
_\lambda H)\|}~\delta(\dot x^\mu-\omega^{\mu\nu}\partial_\nu
H)\e^{iS[x]}\\&\sim
\sum_{x(t)\in L\man_S}\frac{\sqrt{\det\|\omega(x(t))\|}\e^{iS[x]}}{\sqrt{
\det\|\delta^\mu_\nu\partial_t-\partial_\nu(\omega^{\mu\lambda}\partial_
\lambda H)\|}}\end{array}}
\label{wkb}\eeq
where here and in the following the symbol $\sim$ will be used to signify
the absorption of infinite prefactors into the determinants which arise
from the functional Gaussian integrations. The functional determinant in the
denominator of (\ref{wkb}) can be evaluated in the same manner as in Subsection
4.2 above with $R\to dV$ there and the eigenfunctions of $\partial_t$ are now
the periodic functions on $[0,T]$ instead of $S^1$ so that the eigenvalues are
replaced as $2\pi ik\to2\pi ik/T$ in (\ref{detblock}). The result can be
written in terms of the Dirac $\hat A$-genus of the tangent bundle of $\man$
\beq
\det\|\delta_\nu^\mu\partial_t-\partial_\nu V^\mu\|=\left(\frac{T}{2\pi
i}\right)^{2n}\det\left[\frac{\sinh\frac{T}{2}dV}{\frac{T}{2}dV}\right]=\hat
A(T\cdot dV)^{-2}
\label{fluctjacobi}\eeq
using the ordinary moment map for the $U(1)$-action here.

This result is the famous WKB approximation to the partition function
\cite{schulman}, except that it is summed over all classical paths and not
just those which minimize the action $S$. If we reinstate the factors of
$\hbar$, then it is formally the leading term of the stationary phase expansion
of the partition function in powers of $\hbar$ as $\hbar\to0$. The limit
$\hbar\to0$ is called the classical limit of the quantum mechanics problem
above, since then according to (\ref{commalg}) the operators $\hat p$ and $\hat
q$ behave as ordinary commuting c-numbers as in the classical theory. For
$\hbar\to0$ we can naturally evaluate the path integral by the stationary-phase
method discussed in Subsection 3.3, i.e. we expand the trajectories
$x(t)=x_0(t)
+\delta x(t)$ in the action, with $x_0(t)\in L\man_S$ and $\delta x(t)$ the
fluctuations about the classical paths $x_0(t)$ with $\delta x(0)=\delta x(T)
=0$, and then carry out the leading Gaussian functional integration over these
fluctuations. Indeed, this was the way Feynman originally introduced the path
integral to describe quantum mechanics as a sum over trajectories which
fluctuate around the classical paths of the system. This presentation of
quantum mechanics thus leads to the dynamical Hamilton action principle of
classical mechanics \cite{goldstein}, i.e. the classical paths of motion of a
dynamical system are those which minimize the action, as a limiting case. If
the classical trajectories were unique, then we would only obtain the factors
$\e^{iS[x]/\hbar}$ above as $\hbar\to0$. Quantum mechanics can then be
interpreted as implying fluctuations (the one-loop determinant factors in
(\ref{wkb})) around these classical trajectories.

We should point out here that the standard WKB formulas are usually given for
configuration space path integrals where the fluctuation determinant
$(\det\|L_S(x(t))\|)
^{-1/2}$ appearing in (\ref{wkb}) is the so-called Van Vleck determinant which
is essentially the Hessian of $S$ in configuration space coordinates $q$.
Here the determinant is the functional determinant of the Jacobi operator
which arises from the usual Legendre transformation to phase space coordinates
$(p,q)$. This operator is important in the Hamilton-Jacobi theory of classical
mechanics \cite{arnold,goldstein}, and this determinant can be interpreted
as the density of classical trajectories. The result (\ref{wkb}) and the
assumptions that went into deriving it, such as the non-vanishing of the
determinant of the Jacobi fields and the existence of an invariant phase
space metric, are certainly true for the classic examples in quantum
mechanics and field theory where the semi-classical approximation is known
to be exact, such as for the propagator of a particle moving on a group
manifold \cite{dowker,picken,schul1}. The above localization principle yields
sufficient, geometric conditions for when a given path integral is given
exactly by its WKB approximation, and it therefore has the possibility of
expanding the set of quantum systems for which the Feynman path integral is WKB
exact and localizes onto the classical trajectories of the system.

\subsection{Degenerate Path Integrals and the Niemi-Tirkkonen Localization
Formula}

There are many instances in which the WKB approximation is unsuitable for
a quantum mechanical path integral, such as a dynamical system whose
classical phase space trajectories coalesce at some point. It is therefore
desirable to seek alternative, more general localization formulas which can
be applied to larger classes of quantum systems. Niemi and Palo
\cite{niemipalo2} have investigated the types of degeneracies that can
occur for phase space path integrals and have argued that for Hamiltonians
which generate circle actions the classical trajectories can be
characterized as follows. In general, the critical point set of the action
$S$ with non-constant periodic solutions $x^\mu(T)=x^\mu(0)=x_0^\mu$ lie on
a compact submanifold $L\man_S$ of the phase space $\man$. In this context,
$L\man_S$ is refered to as the moduli space of $T$-periodic classical
solutions and it is in general a non-isolated set for only some discrete
values of the propagation time $T$. For generic values of $T$ the periodic
solutions with $x^\mu(T)=x^\mu(0)=x_0^\mu$ exist only if $x_0^\mu$ lies on
the critical submanifold $\man_V$ of the Hamiltonian $H$. Then the classical
equations of motion reduce to $\dot x^\mu=V^\mu=\omega^{\mu\nu}\partial_\nu
H=0$ and so the moduli space $L\man_S$ coincides with the critical point set
$\man_V\subset\man$. Notice that in this case the functional determinants
involving the symplectic 2-form in (\ref{wkb}) cancel out and one is left with
only the regularized determinant in (\ref{fluctjacobi}). We shall see some
specific examples of this later on.

With this in mind we can derive a loop space analog of the degenerate
Duistermaat-Heckman formula of Subsection 3.7. We decompose $L\man$ and
$L\Lambda^1\man$ into classical modes and fluctuations about the classical
solutions and scale the latter by $1/\sqrt{\lambda}$,
\beq
x^\mu(t)=\bar x^\mu(t)+x_f^\mu(t)/\sqrt{\lambda}~~~~~,~~~~~\eta^\mu(t)=\bar
\eta^\mu(t)+\eta_f^\mu(t)/\sqrt{\lambda}
\label{classdecomp}\eeq
where $\bar x(t)\in L\man_S$ are the solutions of the classical equations of
motion, i.e. $V_S^\mu(\bar x(t))={\dot{\bar x}}^\mu-\omega^{\mu\nu}(\bar
x)\partial_\nu H(\bar x)=0$, and $\bar\eta^\mu(t)\sim d\bar
x^\mu(t)\in\Lambda^1 L\man_S$ span the kernel of the loop space Riemann moment
map,
\beq
(\Omega_S)_{\mu\nu}(\bar x)\bar\eta^\nu=0
\label{omseta0}\eeq
where
\beq
\Omega_S=d_L\psi=\int_0^Tdt~\frac{\delta}{\delta x^\mu}\left(g_{\nu
\lambda}V_S^\lambda\right)\eta^\mu\eta^\nu
\label{loopmom}\eeq
with $\psi$ given in (\ref{wkbpsi}). In particular, this implies that $\bar
\eta^\mu(t)$ are Jacobi fields, i.e. they obey the fluctuation equation
\beq
\left(\delta^\mu_\nu\partial_t-\partial_\nu V^\mu(\bar x)\right)\bar\eta^\nu=0
\label{jacobieq}\eeq
The fluctuation modes in (\ref{classdecomp}) obey the boundary conditions
$x_f^\mu(0)=x_f^\mu(T)=0$ and $\eta_f^\mu(0)=\eta_f^\mu(T)=0$.

The super-loop space measure with this decomposition is then
\beq
[d^{2n}x]~[d^{2n}\eta]=d^{2n}\bar x(t)~d^{2n}\bar\eta(t)~\prod_{t\in[0,T]}
d^{2n}x_f(t)~d^{2n}\eta_f(t)
\label{classloopmeas}\eeq
where as usual the change of variables (\ref{classdecomp}) has unit Jacobian
because the determinants from the bosonic and fermionic fluctuations cancel
(this is the powerful manifestation of the ``hidden" supersymmetry in these
theories). The calculation now proceeds analogously to that in Subsection 3.7,
so that evaluating the Gaussian integrals over the fluctuation modes localizes
the path integral to a finite-dimensional integral over the moduli space
$L\man_S$ of classical solutions,
\beq
Z(T)\sim\int_{L\man_S}d^{2n}\bar x(t)~\frac{\sqrt{\det\omega(\bar
x)}\e^{iS[\bar
x]}}{\pfaff\|\delta^\mu_\nu\partial_t-(\mu_S)^\mu_\nu(\bar x)-R^\mu_\nu(\bar x)
\|\Bigm|_{{\cal N}L\man_S}}
\label{wkbdeg}\eeq
where $\mu_S=g^{-1}\cdot\Omega_S$ and $R$ is as usual the Riemann curvature
2-form of the metric $g$ evaluated on $L\man_S$. In (\ref{wkbdeg}) the
Pfaffian is taken over the fluctuation modes $x_f^\mu(t)$ about the classical
trajectories $\bar x^\mu(t)\in L\man_S$ (i.e. along the normal bundle
${\cal N}L\man_S$ in $L\man$), and the measure there is an invariant measure
over the moduli space of classical solutions which is itself a symplectic
manifold. The localization formula (\ref{wkbdeg}) is the loop space version of
the degenerate localization formula (\ref{degdh}) in which the various factors
can be interpreted as loop space extensions of the equivariant characteristic
classes. In particular, in the limit where the solutions to the
classical equations of motion $V_S^\mu(x(t))=0$ become isolated and
non-degenerate paths the integration formula (\ref{wkbdeg}) reduces to the
standard WKB localization formula (\ref{wkb}).

However, the degenerate localization formula (\ref{wkbdeg}) is hard to use
in practise because in general the moduli space of classical solutions has
a complicated, $T$-dependent structure, i.e. it is usually a highly non-trivial
problem to solve the classical equations of motion for the $T$-periodic
classical trajectories of a dynamical system\footnote{\baselineskip=12pt Some
features of the space of $T$-periodic classical trajectories for both energy
conserving and non-conserving Hamiltonian systems have been discussed recently
by Niemi and Palo in \cite{niemi1,niemipalo3}.}. We would therefore like to
obtain alternative degenerate localization formulas which are applicable
independently of the structure of the moduli space $L\man_S$ above. Given
the form of (\ref{wkbdeg}), we could then hope to obtain a localization onto
some sort of equivariant characteristic classes of the manifold $\man$. The
first step in this direction was carried out by Niemi and
Tirkkonen in \cite{niemitirk1}. Their localization formula can be derived by
setting $s=0$, $r=1$ in (\ref{wchoice}) so that
\beq
\psi=\int_0^Tdt~g_{\mu\nu}\dot x^\mu\eta^\nu~~~,~~~Q_S\psi=\int_0^Tdt~\left[
g_{\mu\nu}\dot x^\mu(\dot x^\nu-V^\nu)+\eta^\mu\left(g_{\mu\nu}
\partial_t+\dot x^\lambda g_{\mu\rho}\Gamma^\rho_{\lambda\nu}\right)\eta^\nu
\right]
\label{ntirkpsi}\eeq
Here the zero locus of the vector field (\ref{wchoice}) consists of the
constant
loops $\dot x^\mu=0$, i.e. points on $\man$, so that the canonical
localization integral will reduce to an integral over the finite-dimensional
manifold $\man$ (as in Subsection 4.2), rather than a sum or integral over the
moduli space of classical solutions as above.

To evaluate the right-hand side of (\ref{loopsploc}) with (\ref{ntirkpsi}), we
use the standard trick. We decompose $L\man$ and $L\Lambda^1\man$ into constant
modes and fluctuation modes and scale the latter by $1/\sqrt{\lambda}$,
\beq
x^\mu(t)=x_0^\mu+\hat x^\mu(t)/\sqrt{\lambda}~~~~~,~~~~~\eta^\mu(t)=\eta_0^\mu
+\hat\eta^\mu(t)/\sqrt{\lambda}
\label{xetadecomp}\eeq
where
\beq\new{\begin{array}{c}
x_0^\mu=\frac{1}{T}\int_0^Tdt~x^\mu(t)~~~~~,~~~~~\eta_0^\mu=\frac{1}{T}\int_0^T
dt~\eta^\mu(t)\\\partial_tx_0^\mu=\partial_t\eta_0^\mu=0~~~~~,~~~~~\int_0^Tdt~
\hat x^\mu(t)=\int_0^Tdt~\hat\eta^\mu(t)=0\end{array}}
\label{decompdef}\eeq
The decomposition (\ref{xetadecomp}) is essentially a Fourier decomposition in
terms of some complete sets of states $\{x_k^\mu(t)\}_{k\in\IZ}$ and $\{\eta^
\mu_k(t)\}_{k\in\IZ}$, so that
\beq
\hat x^\mu(t)=\sum_{k\neq0}s_k^\mu x_k^\mu(t)~~~~~,~~~~~\hat\eta^\mu(t)=
\sum_{k\neq0}\sigma_k^\mu\eta_k^\mu(t)
\label{decompfour}\eeq
and the Feynman measure in the path integral is then defined just as before as
\beq
[d^{2n}x]~[d^{2n}\eta]=d^{2n}x_0~d^{2n}\eta_0\prod_{t\in[0,T]}d^{2n}\hat x(t)
{}~d^{2n}\hat\eta(t)=d^{2n}x_0~d^{2n}\eta_0\prod_{k\neq0}d^{2n}s_k~d^{2n}
\sigma_k
\label{measdecomp}\eeq
With the rescaling in (\ref{xetadecomp}) of the fluctuation modes, the gauge
fixing term $Q_S\psi$ is
\beq
Q_S\psi=\int_0^Tdt~\left[x^\mu\left((\Omega_V)_{\mu\nu}\partial_t-g_{\mu\nu}
\partial_t^2\right)x^\nu+\frac{1}{2}R_{\mu\nu}x^\mu\dot x^\nu+
\hat\eta^\mu g_{\mu\nu}\partial_t\hat\eta^\nu\right]+{\cal O}(1/\sqrt{\lambda})
\label{rescqspsi}\eeq
where we have integrated by parts over $t$ and used the periodic boundary
conditions.

In (\ref{rescqspsi}) we see the appearence of the equivariant curvature of
the Riemannian manifold $(\man,g)$. Since $\Omega_V$ and $R$ there act on
the fluctuation modes, as usual they can be interpreted as forming the
equivariant curvature of the normal bundle of $\man$ in $L\man$. With the
above rescaling the fluctuation and zero modes decouple in the localization
limit $\lambda\to\infty$, just as before. The integrations over the
fluctuations are as usual Gaussian, and the result of these integrations is
\beq
Z(T)\sim\int_\man\ch_V(-iT\omega)\wedge\left({\det}'\|\delta^\mu_\nu\partial_t
-(R_V)^\mu_\nu\|\right)^{-1/2}
\label{fluctint}\eeq
This form of the partition function is completely analogous
to the degenerate localization formula of Subsection 3.7, and it is also
similar
to the formula (\ref{wkbdeg}), except that now the domain of
integration has changed from the moduli space $L\man_S$ of classical solutions
to the entire phase space $\man$. This makes the formula (\ref{fluctint}) much
more appealing, in that there is no further reference to the $T$-dependent
submanifold  $L\man_S$ of $\man$. Note that (\ref{fluctint}) differs from the
classical partition function for the dynamical system $(\man,\omega,H)$ by a
one-loop determinant factor which can be thought of as encoding the
information due to quantum fluctuations. The classical Boltzmann weight
$\e^{-iTH}$ comes from evaluating the action $S[x]$ on the constant loops
$x\in\man\subset L\man$, so that (\ref{fluctint}) is another sort of
semi-classical localization of the Feynman path integral.

The fluctuation determinant in (\ref{fluctint}) can be again evaluated just as
in Subsection 4.2 and it now yields the equivariant $\hat A$-genus
(\ref{eqahat}) with respect to the equivariant curvature $R_V$. The
localization formula (\ref{fluctint}) is therefore
\beq
Z(T)\sim\int_\man\ch_V(-iT\omega)\wedge\hat A_V(TR)
\label{niemitirkloc}\eeq
This is the Niemi-Tirkkonen localization formula \cite{niemitirk1} and it
expresses the quantum partition function as a (finite-dimensional)
integral over the phase space $\man$ of equivariant characteristic classes
in the $U(1)$-equivariant cohomology generated by the Hamiltonian vector
field $V$ on $\man$. The huge advantage of this formula over the localization
formula of the last Subsection is that no assumptions appear to have gone into
its derivation (other than the standard localization constraints). It thus
applies not only to the cases covered by the WKB localization theorem, but also
to those where the WKB approximation breaks down (e.g. when classical paths
coalesce in $L\man$). Indeed, being a localization onto time-independent loops
it does not detect degenerate types of phase space trajectories that a
dynamical system may possess.

In fact, the localization formula (\ref{niemitirkloc}) can be viewed as an
integral over the equivariant generalization of the Atiyah-Singer index density
of a Dirac operator with background gravitational and gauge fields, and it
therefore represents a sort of equivariant generalization of the Atiyah-Singer
index theorem for a twisted spin complex. Indeed, when $H,V\to0$ the effective
action in the canonical localization integral is
\beq
(S+\Omega+\lambda Q_S\psi)|_{H=V=0}=\int_0^Tdt~\left[\lambda g_{\mu
\nu}\dot x^\mu\dot x^\nu +\theta_\mu\dot x^\mu+\lambda g_{\mu\nu}
\eta^\mu\nabla_t\eta^\nu+\frac{1}{2}\eta^\mu\omega_{\mu\nu}\eta^\nu\right]
\label{supersymaction}\eeq
On the other hand, the left-hand side of the localization formula
(\ref{loopsploc}) becomes
\beq
Z(T)|_{H=0}=~\tr\|\e^{-i\hat HT}\|\Bigm|_{T=0}=\dim{\cal H}_\man
\label{pathh0}\eeq
which is an integer representing the dimension of the free Hilbert space
associated with $S(H=0)$ and which can therefore only describe the topological
characteristics of the manifold $\man$. Recalling the discussion of Subsection
4.2, we see that the action (\ref{supersymaction}) is the
supersymmetric action for a bosonic field $x^\mu(t)$ and its Dirac fermion
superpartner field $\eta^\mu(t)$ in the background of a gauge field
$\theta_\mu$ and a gravitational field $g_{\mu\nu}$, i.e. the action of
$N=\frac{1}{2}$ Dirac supersymmetric quantum mechanics. Moreover, the integer
(\ref{pathh0}) coincides with the $V=0$ limit of (\ref{niemitirkloc}) which
is the ordinary Atiyah-Singer index for a twisted spin complex (the `twisting'
here associated with the usual symplectic line bundle $L\to\man$). Thus the
localization formalism here is just a more general case of the localization
example of Subsection 4.2 above which reproduced quite beautifully the
celebrated Atiyah-Singer index theorem\footnote{\baselineskip=12pt This
analogy, as well as the localization of the quantum partition function in
general, requires that boundary conditions for the path integral be selected
which respect the pertinent supersymmetry. We shall say more about this
requirement later on.}. We shall describe some more of these cohomological
field theoretical aspects of equivariant localization in Subsection 4.10, and
the connections between the localization formalism and other supersymmetric
quantum field theories in Section 8. The equivariant cohomological structure of
these theories is consistent with the topological nature of supersymmetric
models (the basic topological field theories -- see Subsection 4.10 below) and
they always yield certain topological invariants of the underlying manifolds
such as the Atiyah-Singer index.

\subsection{Connections with the Duistermaat-Heckman Integration Formula}

In this Subsection we shall point out some relations between the path integral
localization formulas derived thus far and their relations to
the finite-dimensional Duistermaat-Heckman formula. Since the
localization formulas are all derived from the
same fundamental geometric constraints, one would expect that, in some limits
at least, they are all related to each other. In particular, when 2
localization formulas hold for a certain quantum mechanical path integral,
they must both coincide somehow. We can relate the various localization
formulas by noting that the integrand of (\ref{niemitirkloc}) is an
equivariantly closed differential form on $\man$ (being an equivariant
characteristic class) with respect to the finite-dimensional equivariant
cohomology defined by the ordinary Cartan derivative $D_V=d+i_V$. Thus we can
apply the Berline-Vergne theorem (in degenerate form -- compare with Subsection
3.7) of Subsection 2.6 to localize the equivariant Atiyah-Singer index onto the
critical points of the Hamiltonian $H$ to obtain
\beq
Z(T)\sim\int_{\man_V}\frac{\ch_V(-iT\omega)}{E_V(R)|_{{\cal N}_V}}\wedge\hat
A_V(TR)\biggm|_{\man_V}
\label{pathlochamdeg}\eeq
so that a (degenerate) Hamiltonian gives a localization of the path integral
onto $\man_V$ in terms of the equivariant Chern class restricted to $\man_V$,
and the equivariant Euler class and $\hat A$-genus of the normal bundle ${\cal
N}_V$. Note that this differs from the finite-dimensional localization formula
(\ref{degdh}) only in the appearence of the equivariant $\hat A$-genus which
arises from the evaluation of the temporal determinants which occur. This
factor therefore encodes the quantum fluctuations about the classical values,
and its appearence is quite natural according to the general supersymmetry
arguments above (as the Dirac $\hat A$-genus quite frequently arises from
supersymmetric field theory path integrals). Furthermore, the localization
formula (\ref{pathlochamdeg}) follows from the moduli space formula
(\ref{wkbdeg}) for certain values of the propagation time $T$ (see the
discussion at the beginning of the last Subsection).

The connection between the WKB and Niemi-Tirkkonen localization formulas is
now immediate if we assume that the critical point set $\man_V$ of the
Hamiltonian consists of only isolated and non-degenerate points (i.e. the
Hamiltonian $H$ is a Morse function). Then in the canonical
localization vector field (\ref{wchoice}) we can set $r=0$ and $s=-1$ so that
\beq
\psi=i_Vg=\int_0^Tdt~g_{\mu\nu}V^\mu\eta^\nu~~~,~~~Q_S\psi=\int_0^Tdt~\left[
\frac{1}{2}(\Omega_V)_{\mu\nu}\eta^\mu\eta^\nu+V^\mu g_{\mu\nu}\left(\dot
x^\nu-V^\nu\right)\right]
\label{hlocpsi}\eeq
We use the rescaled decomposition (\ref{xetadecomp}) again which decouples
the zero modes from the fluctuation modes because of the ``hidden"
supersymmetry. The Gaussian integrations over the
fluctuation modes then yields
\beq
Z(T)\sim\int_\man d^{2n}x_0~\e^{-iTH}\sqrt{\frac{\det\Omega_V}{{\det}'\|
\partial_t-\Omega_V\|}}~\delta(V)\sim\sum_{p\in\man_V}\frac{\e^{-iTH(p)}}{
\sqrt{\det\Omega_V}}\hat A(T\Omega_V)
\label{pathloch}\eeq
where the (ordinary) Dirac $\hat A$-genus arises from evaluating the temporal
determinant in (\ref{pathloch}) as described before and we recall that
$\Omega_V(p)=2dV(p)=2\omega^{-1}(p)\hess(p)$ at a critical point $p\in\man_V$.
Thus under these circumstances we can localize the partition function path
integral onto the time-independent classical trajectories of the dynamical
system, yielding a localization formula that differs from the standard
Duistermaat-Heckman formula (\ref{dhformula}) only by the usual quantum
fluctuation term.

The localization formula (\ref{pathloch}) of course also follows directly from
the degenerate formula (\ref{pathlochamdeg}) in the usual way, and it can be
shown \cite{keski} to also follow from the WKB formula (\ref{wkb}) using the
Weinstein action invariant \cite{blau1,weinstein} which probes the first
cohomology group of the symplectomorphism group of the symplectic manifold
(i.e. the diffeomorphism subgroup of canonical transformations). This latter
argument requires that $\man$ is compact, the classical trajectories are
non-intersecting and each classical trajectory can be contracted to a critical
point of $H$ through a family of classical trajectories (for instance when
$H^1(\man;\IR)=0$), and that the period $T$ is such that the boundary
condition $x^\mu(0)=x^\mu(T)$ admits only constant loops as solutions to the
classical equations of motion. The localization onto the critical points of the
Hamiltonian is not entirely surprising, since as discussed at the beginning
of the last Subsection for Hamiltonian circle actions on $\man$ the zero
locuses
$L\man_S$ and $\man_V$ in general coincide. Drawing from the analogy of
(\ref{pathloch}) with the Duistermaat-Heckman theorem (i.e. that the
equivariant Atiyah-Singer index (\ref{niemitirkloc}) is given exactly by its
stationary phase approximation), one can, in particular, in this case conclude
from Kirwan's theorem that the Hamiltonian $H$ is a perfect Morse
function that admits only even Morse indices \cite{keski}. We see therefore
that localization formulas and various Morse theoretic
arguments (such as Kirwan's theorem) follow (formally) for path
integrals in exactly the same way that they followed for ordinary
finite-dimensional phase space integrals.

\subsection{Equivariant Localization and Quantum Integrability}

For the remainder of this Section we shall discuss
some more formal features of the localization formalism for path integrals,
as well as some extensions of it. We have shown in Section 3 that there is an
intimate connection between classical integrability and the localization
formalism for dynamical systems. With this in mind, we can use the localization
formalism to construct an alternative, geometric formulation of the problem of
{\it quantum} integrability \cite{francoise,niemipalo1} (in the sense that the
quantum partition function can be evaluated exactly) which differs from the
usual approaches to this problem \cite{das}. As in Subsection 3.6 we consider a
generic integrable Hamiltonian which is a functional $H=H(I)$ of action
variables $I^a$ which are in involution as in (\ref{actioninv}). From the point
of view of the localization constraints above, the condition that $H$ generates
a circle action which is an isometry of some Riemannian geometry on $\man$
means that the action variables $I^a$ generate the Cartan subalgebra of the
associated
isometry group of $(\man,g)$ in its Poisson bracket realization on
$(\man,\omega)$.

For such a dynamical system, we use a set of generating functionals $J^a(t)$
to write the quantum partition function as
\beq\new{\begin{array}{l}
Z(T)\\=\exp\left(-i\int_0^Tdt~H\left[\frac{\delta}{i\delta J(t)}\right]\right)
\int_{L\man}[d^{2n}x]~\sqrt{\det\|\Omega\|}\exp\left\{i\int_0^Tdt~\left(
\theta_\mu\dot x^\mu-J^aI^a\right)\right\}\biggm|_{J=0}\end{array}}
\label{partintqu}\eeq
To evaluate the path integral in (\ref{partintqu}), we consider an
infinitesimal variation of its action
\beq
\delta(\theta_\mu\dot x^\mu-J^aI^a)=\delta x^\mu(\omega_{\mu\nu}\dot x^\nu-
J^a\partial_\mu I^a)
\label{infvar}\eeq
with the infinitesimal Poisson bracket variation
\beq
\delta x^\mu=\epsilon^a\left\{I^a,x^\mu\right\}_\omega=-\epsilon^a\omega^{\mu
\nu}\partial_\nu I^a
\label{poissonvar}\eeq
where $\epsilon^a$ are infinitesimal coordinate-independent parameters.
The transformation (\ref{infvar}),(\ref{poissonvar}) corresponds to the
leading order infinitesimal limit of the canonical transformation
\beq
x^\mu\to
x^\mu+\left\{\e^{-\epsilon^aI^a},x^\mu\e^{\epsilon^aI^a}\right\}_\omega=
x^\mu+\epsilon^a\left
\{x^\mu,I^a\right\}_\omega+\frac{1}{2}\epsilon^a\epsilon^b\left\{\left\{x^\mu,
I^a\right\}_\omega,I^b\right\}_\omega+\dots
\label{cantransfcoord}\eeq
and it gives
\beq
\delta(\theta_\mu\dot x^\mu-J^aI^a)=-\dot\epsilon^aI^a
\label{explshift}\eeq
after an integration by parts over time.
Since the Liouville measure in (\ref{partintqu}) is invariant under canonical
transformations, it follows that the only effect of the variation
(\ref{explshift}) on the loop space coordinates in (\ref{partintqu}) is to
shift the external sources as $J^a\to J^a+\dot\epsilon^a$. Note that if we
identify $J^a(t)$ as the temporal component $A_0^a$ of a gauge field
then this shift has the same functional form as a time-dependent abelian gauge
transformation \cite{niemipalo1}. Thus if for some reason the quantum theory
breaks the invariance of the Liouville measure under these coordinate
transformations, we would expect to be able to relate the non-trivial Jacobian
that arises to conventional gauge anomalies \cite{sem}.

Thus if we Fourier decompose the fields $J^a(t)$ into their zero modes $J^a_0$
and fluctuation modes $\hat J^a(t)$ as in (\ref{xetadecomp}), we can use this
canonical transformation to `gauge' away the time-dependent parts of $J^a$
in (\ref{partintqu}) so that the path integral there depends only on the
constant modes $J_0^a$ of the generating functionals and the partition
function is given by
\beq
Z(T)=\exp\left(-iTH\left[\frac{1}{i}\frac{\partial}{\partial J_0}\right]
\right)\int_{L\man}[d^{2n}x]~\sqrt{\det\|\Omega\|}\exp\left\{i\int_0^Tdt~
\left(\theta_\mu\dot x^\mu-J^a_0I^a\right)\right\}\biggm|_{J_0=0}
\label{part0modes}\eeq
Since the Hamiltonian $J_0^aI^a$ in the action in (\ref{part0modes}) generates
an abelian group action on $\man$, we can localize it using the
Niemi-Tirkkonen formula (\ref{niemitirkloc}) to arrive at
\beq
Z(T)\sim\exp\left(-iTH\left[\frac{1}{i}\frac{\partial}{\partial J_0}\right]
\right)\int_\man\ch_{J_0^aI^a}(-iT\omega)\wedge\hat A_{J_0^aI^a}(TR)
\biggm|_{J_0=0}
\label{locgenint}\eeq
and so the path integral now localizes to a derivative expansion of
equivariant characteristic classes. The localization formula (\ref{locgenint})
is valid for {\it any} integrable
Hamiltonian system whose conserved charges $J_0^aI^a$ generate a global
isometry on $\man$, and consequently the localization formalism can be used
to establish the exact quantum solvability of generic integrable models.

Indeed, there are several non-trivial examples of integrable models where
the WKB localization formula (\ref{wkb}) is known to be valid, and this has
led to the conjecture that for a large class of integrable field theories
a ``proper" version of the semi-classical approximation should yield a
reliable reproduction of the features of the exact quantum theory \cite{zinn}.
The formula (\ref{locgenint}) is one such candidate, and thus it yields an
explicit realization of this conjecture. However, one may also hope that the
localization principle of Subsection 4.4
could be used to derive weaker versions of the localization formulas
above for some dynamical systems which are not necessarily completely
integrable \cite{keski} (in the sense that the localization formalism above
does not carry through). For this, we consider a Hamiltonian with $r<n$
conserved charges $I^a$ which are in involution as in
(\ref{actioninv}),(\ref{actionconst}), and which have the classical equations
of motion $\dot I^a=0$. We then set
\beq
\psi=\int_0^Tdt~I^a\partial_\mu I^a\eta^\mu~~~,~~~Q_S\psi=\int_0^Tdt~\left(\dot
I^a\right)^2
\label{nonintpsi}\eeq
in the canonical localization integral (\ref{loopsploc}). The cohomological
relation $Q_S^2\psi=\lie_S\psi=0$ follows from the involutary property
of the charges $I^a$. Then the right-hand side of (\ref{loopsploc}) yields
a localization of the path integral onto the constant values of the
conserved charges $I^a$,
\beq
Z(T)\sim\int_{L\man}[d^{2n}x]~\sqrt{\det\|\Omega\|}~\prod_{a=1}^r\delta(\dot
I^a)\e^{iS[x]}
\label{locnonint}\eeq

The formula (\ref{locnonint}) is a weaker version of the above localization
formulas which is valid for any non-integrable system that admits conserved
charges. It can be viewed as a quantum generalization of the classical
reduction theorem \cite{arnoldnov} which states that conserved charges in
involution reduce the dynamics onto the symplectic subspace of the original
phase space determined by the constant (classical) values of the integrals of
motion $I^a$. When $H$ is completely integrable this subspace coincides with
the
invariant Liouville tori discussed in Subsection 3.6. Thus even when there are
corrections to the various localization formulas above (e.g. the WKB
approximation), the supersymmetry arguments of Subsection 4.4 can be used to
derive weaker versions of the localization formulas. Notice that, as
anticipated, the localization formula (\ref{locnonint}) does not presume any
isometric structure on the phase space (see the discussion of Subsection 3.6).
Equivariant cohomology might therefore provide a natural geometric framework
for understanding quantum integrability, and the localization formulas
associated with general integrable models represent equivariant characteristic
classes of the phase space. For more details about this and other connections
between equivariant localization and integrability, see
\cite{francoise,karki,keski}.

\subsection{Localization for Functionals of Isometry Generators}

In the last Subsection we considered a particular class of Hamiltonians which
were functionals of action variables and we were able to derive a quite
general localization formula for these dynamical systems. It is natural
to explore now whether or not localization formulas could be derived
for Hamiltonians which are more general types of functionals. We begin
with the case where the Hamiltonian of a dynamical system is an {\it a priori}
arbitrary functional ${\cal F}(H)$ of an observable $H$ which generates an
abelian isometry through the Hamiltonian equations for $H$ in the usual sense.
Thus we want to evaluate the path integral \cite{niemitirk2}
\beq
Z(T|{\cal F}(H))=\int_{L\man}[d^{2n}x]~\sqrt{\det\|\Omega\|}\exp\left\{i\int
_0^Tdt~\left(\theta_\mu\dot x^\mu-{\cal F}(H)\right)\right\}
\label{partfh}\eeq
We shall see that such path integrals are important for certain
physical applications. Note, however, that although such
functionals may seem arbitrary, we must at least require that ${\cal F}(H)$
be a semi-bounded functional of the observable $H$ \cite{szsem2}. Otherwise,
a Wick rotation off of the real time axis to imaginary time may produce a
propagator $\tr\|\e^{-iT{\cal F}(H)}\|$ which is not a tempered distribution
and thus eliminating any rigorous attempts to make the path integral a
well-defined mathematical entity.

The formalism used to treat path integrals such as (\ref{partfh}) is the
auxilliary field formalism for supersymmetric theories
\cite{hietmor,morniemi1,morniemi2} which enables one to relate the loop space
equivariant cohomology determined by the derivative $Q_S$ to the more general
model-independent $S^1$ loop space formalism, i.e. that determined by the
equivariant exterior derivative $Q_{\dot x}$. We recall from Subsection 4.4
that
in this formulation the path integral action is BRST-exact, as required for
supersymmetric field theories. Here the auxilliary fields that are introduced
turn out to coincide with those used to formulate generic Poincar\'e
supersymmetric theories in terms of the model-independent $S^1$ loop space
equivariant cohomology which renders their actions BRST-exact. These
supersymmetric models will be discussed in Section 8.

To start, we {\it assume} that there is a function $\phi(\xi)$ such that ${\cal
F}(H)$ is a Gaussian functional integral transformation of it,
\beq
\exp\left(-i\int_0^Tdt~{\cal F}(H)\right)=\int_{L\IR}[d\xi]~\exp\left\{i
\int_0^Tdt~\left(\frac{1}{2}\xi^2-\phi(\xi)H\right)\right\}
\label{fngausstransf}\eeq
Because of the local integrability of ${\cal F}(H)$, locally such a function
$\phi(\xi)$ can always be constructed, but there may
be obstructions to constructing $\phi(\xi)$ globally on the loop space
$L\man$, for the reasons discussed before. The transformation $\xi\to\phi$
which maps the Gaussian in $\xi$ to a non-linear functional of $\phi$ is
just the Nicolai transformation in supersymmetry theory \cite{birm,nicolai},
i.e. the change of variables that maps the bosonic part of the supersymmetric
action into a Gaussian such that the Jacobian for this change of variables
coincides with the determinant obtained by integrating over the bilinear
fermionic part of the supersymmetric action. This observation enables one to
explicitly construct a localization for the path integral (\ref{partfh}).

Notice that when ${\cal F}(H)$ is either linear or quadratic in the observable
$H$, the Nicolai transform $\xi$ is directly related to the functional Fourier
transformation of ${\cal F}(H)$,
\beq
\exp\left(-i\int_0^Tdt~{\cal F}(H)\right)=\int_{L\IR}[d\phi]~\exp\left(-i
\int_0^Tdt~\hat F(\phi)\right)\exp\left(-i\int_0^Tdt~\phi H\right)
\label{fnfourier}\eeq
However, for more complicated functionals ${\cal F}(H)$ this connection is
less straightforward. In particular, if we change variables $\xi\to\phi$ in
the Gaussian transformation (\ref{fngausstransf}), we find
\beq
\exp\left(-i\int_0^Tdt~{\cal F}(H)\right)=\int_{L\IR}[d\phi]~\prod_{t\in[0,T]}
\xi'(\phi)~\exp\left\{i\int_0^Tdt~\left(\frac{1}{2}\xi^2(\phi)-\phi H\right)
\right\}
\label{xiphichange}\eeq
so that the effect of this transformation is to isolate the isometry
generator $H$ and make it contribute linearly to the effective action in
(\ref{partfh}) (as we did in the last Subsection). This allows one to localize
(\ref{partfh}) using the general prescriptions of Subsection 4.4 above.

Substituting (\ref{xiphichange}) into (\ref{partfh}), we then carry out the
same steps which led to the Niemi-Tirkkonen localization formula
(\ref{niemitirkloc}). However, now there is an auxilliary, time-dependent field
$\phi$ which appears in the path integral action which must be incorporated
into the localization procedure. These fields appear in the terms $\phi H$
above and
are therefore interpreted as the {\it dynamical} generators of $S({\bf u(1)}
^*)$. We introduce a superpartner $\eta$ for the auxilliary field $\phi$
whose Berezin integration absorbs the Jacobian factor in (\ref{xiphichange}).
The path integral (\ref{partfh}) thus becomes a functional integral over an
extended superloop space. As discussed in Appendix B, one can now introduce an
extended BRST-operator incorporating the super-multiplet $(\phi,\eta)$ such
that the partition function is evaluated with a BRST-exact action whose
argument lies in the BRST-complex of physical states and, as was the case in
Subsection 4.2, the Niemi-Tirkkonen localization onto constant modes becomes
manifest. This extended BRST-operator
is the so-called Weil differential whose cohomology defines the BRST model
for the $U(1)$-equivariant cohomology \cite{mathai,niemitirk3}. This more
sophisticated technique is required whenever the basis elements $\phi^a$ of
the symmetric algebra $S({\bf g}^*)$ are made dynamical and are integrated out,
as is the case here.

We shall not enter into the cumbersome details of this extended superspace
evaluation of (\ref{partfh}), but merely refer to \cite{niemitirk2} for the
details (see also Appendix B for a sketch of the idea). The final result is the
integration formula
\beq
Z(T|{\cal F}(H))\sim\int_{-\infty}^\infty d\phi_0~\xi_0'(\phi_0)\e^{iT\xi_0^2
(\phi_0)/2}\int_\man\ch_{\phi_0V}(-iT\omega)\wedge\hat A_{\phi_0V}(TR)
\label{partfhloc}\eeq
where $\phi_0$ are the zero modes of the auxilliary field $\phi$.
(\ref{partfhloc}) is valid (formally) for any semi-bounded functional ${\cal
F}(H)$ of an
isometry generator $H$ on $\man$. Thus even for functionals of Hamiltonian
isometry generators the localization formula is a relatively simple expression
in terms of equivariant characteristic classes. The only computational
complication in these formulas is the identification of the function
$\xi(\phi)$ (or the functional Fourier transform $\hat F(\phi)$). We note that
when ${\cal F}(H)=H$, we have $\phi(\xi)=1$ and (\ref{partfhloc}) reduces
consistently to the Niemi-Tirkkonen localization formula (\ref{niemitirkloc}).
In the important special case ${\cal F}(H)=H^2$, we find $\phi(\xi)=\xi$
(i.e. $\hat F(\phi)=\phi^2$) and the localization formula (\ref{partfhloc})
becomes
\beq
Z(T|H^2)\sim\int_{-\infty}^\infty d\phi_0~\e^{iT\phi_0^2/2}\int_\man\ch_
{\phi_0V}(-iT\omega)\wedge\hat A_{\phi_0V}(TR)
\label{quadrloc}\eeq
which is the formal path integral generalization of the Wu localization
formula (\ref{wuloc}).

In fact, the above dynamical treatment of the multipliers $\phi$ suggests
a possible non-abelian generalization of the localization formulas and hence
a path integral generalization of the Witten localization formula of Subsection
3.8 \cite{tirk2}. At the same time we generalize the localization formalism
of Subsection 4.8 above to the case where the Hamiltonian is a functional of
the
generators of the full isometry group of $(\man,g)$, and not just simply the
Cartan subgroup thereof. We consider a general non-abelian Hamiltonian moment
map (\ref{momcomps}) where the component functions $H^a$ are assumed to
generate a Poisson algebra realization of the isometry group $G$ of some
Riemannian metric $g$ on $\man$. As mentioned in Subsection 3.8, when the
$\phi^a$ are fixed we are essentially in the abelian situation above and this
case will be discussed in more detail in what follows. Here we assume that the
multipliers $\phi^a$ are time-dependent and we integrate over them in the path
integral following the same prescription for equivariant integration introduced
in Subsection 3.8. This corresponds to modelling the $G$-equivariant cohomology
of $\man$ in the Weil algebra using the BRST formalism \cite{niemitirk3,tirk2}
(see Appendix B). When the $\phi^a$ are fixed parameters, the action functional
(\ref{classaction}) generates the action of $S^1$ on $L\man$ in the model
independent circle action described in Subsection 4.4 above. However, when the
$\phi^a$ are dynamical quantities, $S$ generates the action of the semi-direct
product $LG\semi S^1$, where the action of $S^1$ corresponds to translations
of the loop parameter $t$ and $LG=C^\infty(S^1,G)$ is the loop group of the
isometry group $G$. These actions are generated, respectively, by the loop
space vector fields
\beq\new{\begin{array}{c}
V_{S^1}=\int_0^Tdt~\dot x^\mu(t)\frac{\delta}{\delta x^\mu(t)}\\V_{LG}=\int
_0^Tdt~\phi^a(t)\omega^{\mu\nu}(x(t))\left(\frac{\delta}{\delta x^\nu(t)}H^a
\right)\frac{\delta}{\delta
x^\mu(t)}\equiv\int_0^Tdt~\phi^a(t)V^a(t)\end{array}}
\label{semiaction}\eeq
The commutator algebra of the vector fields (\ref{semiaction}) is that of
$LG\semi S^1$ on $L\man$,
\beq
\left[V_{S^1},V_{LG}\right]=\int_0^Tdt~\dot\phi^aH^a~~~~~,~~~~~\left[V^a(t),
V^b(t')\right]=f^{abc}V^c(t)\delta(t-t')
\label{semialg}\eeq
The equivariant extension of the symplectic 2-form $\Omega$ on $L\man$ is
therefore $S+\Omega$.

If the multipliers $\phi^a$ (now regarded as local coordinates on $L{\bf g}^*$)
are integrated over directly, then the isometry functions $H^a$ become
constraints because the $\phi^a$ appear linearly in the action and so act as
Lagrange multipliers. In this case we are left with a topological quantum
theory (i.e. there are no classical degrees of freedom) with vanishing
classical action, in parallel to the finite-dimensional case of Subsection 3.8.
Alternatively, we can add a functional $F=F(\phi^a)$ to the argument of the
exponential term in the partition function such that the quantity $S+\Omega+F$
is equivariantly closed. We then introduce a non-abelian generalization of the
procedure outlined above \cite{tirk2} (see Appendix B for details). Introducing
 an extended equivariant BRST operator $Q_T$ for the semi-direct product action
of $LG\semi S^1$ on $L\man$ (the non-abelian version of that above), it turns
out that $S+\Omega+F$ is equivariantly closed with respect to $Q_T$ only for
either $F=0$ or $F=\frac{1}{2}(\phi^a)^2$, where the latter is the invariant
polynomial corresponding to the quadratic Casimir element of $G$. Note that
this is precisely the choice that was made in our definition of equivariant
integration in Subsection 3.8. As shown in Appendix B, within this framework we
can reproduce loop space generalizations of the cohomological formulation of
Subsection 3.2 for the Hamiltonian dynamics. The rest of the localization
procedure now carries through parallel to that above and in the Niemi-Tirkkonen
localization, and it yields the localization formula \cite{tirk2}
\beq
Z(T)\sim\int_{{\bf g}^*}\prod_{a=1}^{\dim G}d\phi_0^a~\e^{iT(\phi_0^a)^2/2}
\int_\man\ch_{\phi_0^aV^a}(-iT\omega)\wedge\hat A_{\phi_0^aV^a}(TR)
\label{nonabpartloc}\eeq
which is a non-abelian version of the quadratic localization formula
(\ref{quadrloc}) and is the path integral generalization of the Witten
localization formula presented in Subsection 3.8 \footnote{\baselineskip=12pt
The procedure outlined above could also be employed in the discussion of the
Witten localization formalism in Subsection 3.8. This has been implicitly done
in carrying out the equivariant integrations there (see Appendix B).}. Notice
that the primary difference between this non-abelian localization and its
abelian counterpart is that in the latter the functional $F(\phi)$ is {\it a
priori} arbitrary.

\subsection{Topological Quantum Field Theories}

In this last Subsection of this Section we return to the case where the
dual basis elements of $S({\bf g}^*)$ are fixed numbers. We wish to study the
properties of the quantum theory when the effective action is BRST-exact as
in (\ref{someqexact}) locally on the loop space \cite{keski,niemipas}. In this
case the quantum theory is said to be topological, in that there are no local
physical degrees of freedom and the remaining partition function can only
describe topological invariants of the space on which it is defined
\cite{birm}. We shall see this explicitly below, and indeed we have already
seen hints of this in the expressions for the path integral in terms of
equivariant characteristic classes above. To get a flavour for this, we first
consider a quantum theory that admits a model independent circle action
globally on the loop space, i.e. whose loop space Hamiltonian vector field
generates a global constant velocity $U(1)$ action on $L\man$, so that its
action functional is given locally by (\ref{newsom}). In this case, the
determinant that
appears in the denominator of the WKB localization formula (\ref{wkb}) is
\beq
\det\|\delta^2S\|\Bigm|_{\dot x=0}=\det\|\delta(\Omega\cdot\dot x)\|\Bigm|_
{x=x_0}=\det\|\Omega\partial_t\|\Bigm|_{x=x_0}
\label{topdet}\eeq
where the localization is now onto the constant loops $x_0\in\man$. Since the
determinants on the right-hand side of (\ref{wkb}) now cancel modulo the
factor $\det\|\partial_t\|$, only the zero modes of $\partial_t$ can
contribute. Thus the (degenerate) WKB localization formula in this case becomes
\beq
Z(T)\sim\int_\man d^{2n}x_0~\sqrt{\det\|\Omega|_
{\partial_t=0}\|}\Bigm|_{x=x_0}
\label{partzeromode}\eeq
and only the zero modes of the symplectic 2-form contribute. Since this path
integral yields the topological Witten index of the corresponding
supersymmetric theory \cite{witten2}, the localization formula identifies the
loop space characteristic class which corresponds to the Witten index of which
the ensuing Atiyah-Singer index counts the zero modes of the associated Dirac
operator. This is one of the new insights gained into supersymmetric theories
from the equivariant localization formalism. (\ref{partzeromode}) is a purely
cohomological representative of the manifold $\man$ which contains no physical
information.

Next, consider the more general case of an equivariantly-exact action
(\ref{someqexact}). Note that this is precisely the solution to the problem
of solving the loop space equivariant Poincar\'e lemma for $S+\Omega$. If
we assume that the symplectic potential is invariant under the global
$U(1)$-action on $\man$, as in (\ref{lietheta0}), then the Hamiltonian is
given by $H=i_V\theta$ and the loop space 1-form $\hat\vartheta$ in
(\ref{someqexact}) is given by
\beq
\hat\vartheta=\int_0^Tdt~\theta_\mu(x(t))\eta^\mu(t)
\label{hatthetasympl}\eeq
The loop space localization principle naively implies that the
resulting path integral should be trivial. Indeed, since the 1-form
(\ref{hatthetasympl}) lies in the subspace (\ref{qsnilpot}), the partition
function can be written as
\beq
Z(T)=\int_{L\man\otimes L\Lambda^1\man}[d^{2n}x]~[d^{2n}\eta]~\e^{i\lambda
Q_S\hat\vartheta}
\label{parttopglobal}\eeq
and it is independent of the parameter $\lambda\in\IR$. In particular, it
should be independent of the action $S$.

However, the above argument for the triviality of the path integral assumes
that $\theta$ is homotopic to 0 in the subspace (\ref{qsnilpot}) under the
supersymmetry generated by $Q_S$, i.e. that (\ref{someqexact}) holds globally
for all loops. For the remainder of this Section we will assume that the
manifold $\man$ is simply connected, so that $H^1(\man;\IR)=0$. Then the above
argument presumes that the second DeRham cohomology group $H^2(\man;\IR)=0$ is
trivial. If this is not the case, then one must be careful about arguing the
$\lambda$-independence of the path integral (\ref{parttopglobal}). Consider
the family of symplectic 2-forms
\beq
\omega^{(\lambda)}=\lambda d\theta=\lambda\omega
\label{omlambda}\eeq
associated with the action in (\ref{parttopglobal}). We consider a closed loop
$\gamma(x)$ in the phase space $\man$ parametrized by the periodic trajectory
$x(t):[0,T]\to\man$. Since by assumption $\gamma(x)$ is the
boundary of a 2-surface $\Sigma_1$ in $\man$, Stokes' theorem implies that the
kinetic term $\theta^{(\lambda)}=\lambda\theta$ in (\ref{parttopglobal}) can
be written as
\beq
\int_0^Tdt~\theta^{(\lambda)}_\mu(x(t))\dot x^\mu(t)=
\oint_{\gamma(x)}\theta^{(\lambda)}=\int_{\Sigma_1}\omega^{(\lambda)}
\label{stokestheta}\eeq
For consistency of the path integral (\ref{parttopglobal}), which is expressed
as a sum over closed loops in $\man$, the phase (\ref{stokestheta}) must be
independent of the representative surface $\Sigma_1$ spanning $\gamma(x)$,
owing
to the topological invariance of the partition function $Z(T)$ over $L\man$.
Thus if we introduce another surface $\Sigma_2$ (of opposite orientation to
$\Sigma_1$) with boundary $\gamma(x)$ and let
$\Sigma$ be the closed surface (sphere) which is divided into 2 halves
$\Sigma_1$ and $\Sigma_2$ by $\gamma(x)$, then we have
\beq
\e^{i\oint_\Sigma\omega^{(\lambda)}}=\e^{i\int_{\Sigma_1}\omega^{(\lambda)}}
\e^{-i\int_{\Sigma_2}\omega^{(\lambda)}}
\label{conscond}\eeq
and consequently the integral of $\omega^{(\lambda)}$ over any closed
orientable surface $\Sigma$ in $\man$ must satisfy a version of the
Dirac or Wess-Zumino-Witten (flux) quantization condition \cite{witten4}
\beq
\frac{1}{2\pi}\oint_\Sigma\omega^{(\lambda)}=\frac{\lambda}{2\pi}\oint_\Sigma
\omega\in\IZ
\label{wesszum}\eeq
This means that $\omega^{(\lambda)}$ is an integral element of $H^2(\man;\IR)$,
i.e. it defines an integer cohomology class in $H^2(\man;\IZ)$, which is
possible only for certain discrete values of $\lambda\in\IR$. It follows
that a continuous variation $\delta\lambda$ of $\lambda$ cannot leave the path
integral (\ref{parttopglobal}) invariant and it depends non-trivially on the
localization 1-form $\psi\equiv\vartheta$ and thus also on the action $S$.

Thus the path integral (\ref{parttopglobal}) defines a consistent quantum
theory only when the symplectic 2-form (\ref{omlambda}) defines an integral
curvature on $\man$. However, if we introduce a variation $\theta\to\theta+
\delta\theta$ of the symplectic potential in (\ref{parttopglobal})
corresponding to a variation $\omega\to\omega+\delta\omega$ with $\delta\omega
=d\alpha$ a trivial element of $H^2(\man;\IR)$ in the subspace
(\ref{qsnilpot}), then the localization principle implies that the path
integral remains unchanged (using Stokes' theorem for $\delta\omega$ in
(\ref{wesszum})). Thus the path integral depends only on the
cohomology class of $\omega$ in $H^2(\man;\IR)$, not on the particular
representative $\omega=d\theta$, which means that the partition function
(\ref{parttopglobal}) determines a cohomological topological quantum field
theory on the phase space $\man$.

Furthermore, we note that within the framework of the Niemi-Tirkkonen
localization formula, the BRST-exact term $Q_S(\lambda\psi+\hat\vartheta)$,
with $\hat\vartheta$ given by (\ref{hatthetasympl}) and $\psi$ given in
(\ref{ntirkpsi}), gives the effective action in the canonical localization
integral (\ref{zspath}). We saw earlier that the $Q_{\dot x}$-exact
piece of this action corresponds to the Atiyah-Singer index of a Dirac operator
$i\nabla\slash$ in the background of a $U(1)$ gauge field $\theta_\mu$ and a
gravitational field $g_{\mu\nu}$. The remaining terms there, given by the
$i_V$-exact pieces, then coincide with the terms that one expects in a
supersymmetric path integral representation of the infinitesimal Lefschetz
number (also known as a character index or equivariant $G$-index)
\beq
{\rm
index}_H(i\nabla\slash;T)=\lim_{\lambda\to\infty}\tr\|\e^{iTH}(\e^{-\lambda
{\cal D}^\dagger{\cal D}}-\e^{-\lambda{\cal D}{\cal D}^\dagger})\|
\label{lefnum}\eeq
generated by the Hamiltonian $H$
\cite{berlinegetz,bismut1,bismut2,niemipalo1,niemitirk1}. This follows from
arguments similar to those in Subsection 4.2 which arrived at the
supersymmetric path integral representation of the (ordinary) Atiyah-Singer
index. In general, when the Dirac operator is invariant under the action of the
isometry group $G$ on $\man$, $[V^a,i\nabla\slash]=0$, then the eigenstates of
$i\nabla\slash$ which correspond to a fixed eigenvalue $E$ define a
representation of the Lie algebra of $G$. It is possible to show just as before
that the right-hand side of the Lefschetz number (\ref{lefnum}) is independent
of $\lambda\in\IR^+$, and, therefore, when either ${\cal D}^\dagger{\cal D}$ or
${\cal D}{\cal D}^\dagger$ has no zero modes, we can take the limit
$\lambda\to0$ there and only the zero modes of $i\nabla\slash$ contribute to
(\ref{lefnum}). Consequently, the equivariant index coincides with the
character
\beq
{\rm index}_H(i\nabla\slash;T)=~{\rm str}_R\e^{iTH}
\label{charindex}\eeq
of the (reducible or irreducible) representation $R$ of the Cartan element $H$
of $G$ determined by the zero modes of $i\nabla\slash$.

Consequently, in the case of Hamiltonian systems for which
$\lie_Vg=\lie_V\theta=0$, the Niemi-Tirkkonen localization formula
(\ref{niemitirkloc}) reproduces the Lefschetz fixed point
formulas of Bismut \cite{bismut1,bismut2} and Atiyah, Bott and Singer
\cite{eguchi}, provided that boundary conditions for the path integral have
been properly selected. Thus a purely bosonic theory can be related to the
properties of a (functional) Dirac operator defined in the canonical phase
space of the bosonic theory, and this analogy leads one to the hope that the
above localization prescriptions can be made quite rigorous in a number of
interesting infinite-dimensional cases. Note also that the path integral
(\ref{parttopglobal}) has the precise form of a Witten-type or cohomological
quantum field theory, which is characterized by a classical action which is
BRST-exact with the BRST charge $Q_S$ representing gauge and other symmetries
of the classical theory. These types of topological field theories are known to
have partition functions which are given exactly by their semi-classical
approximation -- more precisely, they admit Nicolai maps which trivialize the
action and restrict to the moduli space of classical solutions \cite{birm}.
Thus the topological and localization properties of
supersymmetric and topological field theories find their natural explanation
within the framework of loop space equivariant localization.

Of course, the above results rely heavily on the $G$-invariance condition
(\ref{lietheta0}) for the symplectic potential $\theta$. In the general case,
we recall from Subsection 3.2 that we have the relation (\ref{ivathetaF}) which
holds locally in a neighbourhood $\cal N$ in $\man$ away from the critical
points of $H$ and in which $\omega=d\theta$. In this case, (\ref{ivathetaF})
gives a solution to the equivariant Poincar\'e lemma and although the action
is locally BRST-exact, globally the quantum theory is non-trivial and may not
be given exactly by a semi-classical approximation. Then the path integral
(\ref{zspath}) has the form of a gauge-fixed topological field theory,
otherwise known as a Schwarz-type or quantum topological field theory
\cite{birm}, with $Q_S$ the BRST charge representing the gauge degrees of
freedom. With $\hat\vartheta$ as in (\ref{hatthetasympl}), the loop space
equivariant
symplectic 2-form can be written in the neighbourhood $L{\cal N}$ as
\beq
S+\Omega=Q_S(\hat\vartheta+d_LF)-\int_0^TdF(x(t))
\label{schwarzaction}\eeq
and the path integral can be represented locally as
\beq
Z(T)=\int_{L\man\otimes L\Lambda^1\man}[d^{2n}x]~[d^{2n}\eta]~\e^{iQ_S(\hat
\vartheta+d_LF)-i\oint_{\gamma(x)}dF}
\label{schwarzpart}\eeq
If we assume that $\man$ is simply connected, so that $H^1(\man;\IR)=0$, then,
by Stokes' theorem, the $dF$ term in (\ref{schwarzpart}) can be ignored for
closed trajectories on the phase space\footnote{\baselineskip=12pt This term is
analogous to the instanton term $F\wedge F$ in 4-dimensional Yang-Mills theory
which can be represented in terms of a locally exact form and is therefore
non-trivial only for space-times which have non-contractable loops
\cite{sem}.}. Since from (\ref{schwarzaction}) we have
\beq
\lie_S(\hat\vartheta+d_LF)=Q_S(S+\Omega)=0
\label{schwarzlie0}\eeq
it follows that $\hat\vartheta+d_LF\in L\Lambda_S^1\man$ and the effective
classical action $S+\Omega$ is equivariantly-exact in the neighbourhood $L
{\cal N}$.

The non-triviality of the path integral now depends on the non-triviality
that occurs when the local neighbourhoods $\cal N$ above are patched together.
In particular, we can invoke the above argument to conclude that the
partition function (\ref{schwarzpart}) depends only on the cohomology class of
$\omega$ in $H^2(\man;\IR)$, in addition to the critical point set of the
action $S$. Thus the partition function in the general case locally determines
a cohomological topological quantum field theory. From the discussion of
Subsection 3.6 we see that this is consistent with the fact that the theory is
locally integrable outside of the critical point set of $H$. We recall also
from that discussion that in a neighbourhood $\cal N$ where action-angle
variables can be introduced and where $H$ does not have any critical points,
we can construct an explicit realization of the function $F$ above and hence
an explicit realization of the topological quantum theory (\ref{schwarzpart}).
For integrable models where action-angle variables can be defined almost
everywhere on the phase space $\man$, the ensuing theory is topological, i.e.
it can be represented by a topological action of the form (\ref{schwarzaction})
almost everywhere on the loop space $L\man$. Notice that all of the above
arguments stem from the assumption that $H^1(\man;\IR)=0$. In Section
6 we shall encounter a cohomological topological quantum field theory defined
on a multiply-connected phase space which obeys all of the equivariant
localization criteria. We also remark that in the general case, when
$\omega$ is not globally exact, the Wess-Zumino-Witten prescription above for
considering the action (\ref{classaction}) in terms of surface integrals
as in (\ref{stokestheta}) makes rigorous the definition of the partition
function on a general symplectic manifold, a point which up until now we have
ignored for simplicity. In this case the required consistency condition
(\ref{wesszum}) means that $\omega$ itself defines an integral curvature,
which is consistent with the usual ideas of geometric quantization
\cite{woodhouse}. We shall see how this prescription works on a
multiply-connected phase space in Section 6.

\section{Equivariant Localization on Simply Connected Phase Spaces:
Applications to Quantum Mechanics, Group Theory and Spin Systems}

When the phase space $\man$ of a dynamical system is compact, the condition
that the Hamiltonian vector field $V$ generate a global isometry of some
Riemannian geometry on $\man$ automatically implies that its orbits must be
closed circles (see ahead Subsection 5.2). This feature is usually
essential for the finite-dimensional localization theorems, but within the
loop space localization framework, where the arguments for localization are
based on formal supersymmetry arguments on the infinite-dimensional manifold
$L\man$, the flows generated by $V$ need not be closed and indeed many of
the formal arguments of the last Section will still apply to non-compact
group actions. For instance, if we wanted to apply the localization formalism
to an $n$-dimensional potential problem, i.e. on the non-compact phase space
$\man=\IR^{2n}$, then we would expect to be allowed to use a Hamiltonian
vector field which generates non-compact global isometries. As we have
already emphasized, the underlying feature of quantum equivariant localization
is the interpretation of an equivariant cohomological structure of the model
as a supersymmetry among the physical, auxilliary or ghost variables. But as
shown in Subsection 4.3, this structure is exhibited quite naturally by {\it
arbitrary} phase space path integrals, so that, under the seemingly weak
conditions outlined there, this formally results in the equivariant
localization of these path integrals. This would in turn naively imply the
exact
computability of any phase space path integral.

Of course, we do not really expect this to be the case, and there is therefore
the need to explore the loop space equivariant localization formalism
in more detail to see precisely what sort of dynamical systems will localize.
In this Section we shall explore the range of applicability of the
equivariant localization formulas \cite{dlr,szsem2} by presenting a more
detailed analysis of the meaning and implications of the required localization
symmetries, and we shall work out numerous explicit mathematical and physical
applications of the formalisms of the previous Sections. As we shall see, the
global isometry condition on the Hamiltonian dynamics is a very restrictive
one,
essentially meaning that $H$ is related to a global group action
(\ref{gaction}). The natural examples of such situations are the harmonic
oscillator and free particle Hamiltonians on $\IR^{2n}$ (the trivial Gaussian,
free field theories), and the quantization
of spin \cite{nielsen} (i.e. the height function on the sphere), or more
generally the quantization of the coadjoint
orbits of Lie groups \cite{alvarez,blau1,keski,niemitirk2,stone,szsem2} and the
equivalent Kirillov-Kostant geometric quantization of homogeneous phase space
manifolds \cite{alek1,alek2}. Indeed, the exactness of the semi-classical
approximation (or the Duistermaat-Heckman formula) for these classes of phase
space path integrals was one of the most important inspirations for the
development of quantum localization theory and these systems will be
extensively studied in this Section, along with some generalizations of them.
We shall see that the Hamiltonian systems whose phase space path integrals can
be equivariantly localized essentially all fall into this general framework,
and that the localization formulas in these cases always represent important
group-theoretical invariants called characters, i.e. the traces $\tr_Rg
=\tr_R\e^{c^aX^a}$ evaluated in an irreducible representation $R$ of a group
$G$
which are invariant under similarity transformations representing equivalent
group representations, and they reproduce, in certain instances, some
classical formulas for these characters \cite{kirillov}. In our case the group
$G$ will be the group of isometries of a Riemannian structure on $\man$.

As it is essentially the isometry group $G$ that determines the integrable
structure of the Hamiltonian system in the equivariant localization
framework, we shall study the localization formalism from the point of view
of what the possible isometries can be for a given phase space manifold. A
detailed analysis of this sort will lead to a geometrical characterization
of the integrable dynamical systems from the viewpoint of localization and
will lead to topological field theoretical interpretations of integrability,
as outlined in Subsection 4.10. It also promises deeper insights into what one
may consider to be the geometrical structure of the quantum theory. This latter
result is a particularly interesting characterization of the quantum theory
because the partition functions considered are all {\it ab initio} independent
of any Riemannian geometry on the underlying phase space (as are usually the
classical and quantum mechanics). Nonetheless, we shall see that for a given
Riemannian geometry, the localizable dynamical systems depend on this
geometry in such a way so that they determine Hamiltonian isometry actions.

Strictly speaking, most of this general geometric analysis in this Section and
the next will only carry through for a 2-dimensional phase space. The reason
for this is that the topological and geometrical
classifications of Riemann surfaces is a completely solved problem from a
mathematical point of view. We may therefore invoke this classification scheme
to in turn classify the Hamiltonian systems which fit the localization
framework. Such a neat mathematical characterization of higher dimensional
manifolds is for the most part an unsolved problem (although much progress
has been made over the last 7 years or so in the classification of 3- and
4-manifolds), so that a classification scheme such as the one that follows
does not generalize to higher-dimensional models. We shall, however, illustrate
how these situations generalize to higher dimensions via some explicit examples
which will show that the 2-dimensional classifications do indeed tell us about
the properties of general localizable dynamical systems. In particular,
we shall see that from certain points of view all the localizable
Hamiltonians represent ``generalized" harmonic oscillators, a sort of
feature that is anticipated from the previous integrability arguments and
the local forms of Hamiltonians which generate circle actions. These seemingly
trivial behaviours are a reflection of the large degree of symmetry that is the
basis for the large reduction of the complicated functional integrals to
Gaussian ones. We will also analyse in full detail the localization formulas of
the last Section, which will therefore give explicit examples of the
cohomological and integrable models that appear quite naturally in (loop space)
equivariant localization theory. This analysis will also provide {\it new}
integrable quantum systems, as we shall see, which fall into the class of the
generalized localization formulas (e.g. the Niemi-Tirkkonen formula
(\ref{niemitirkloc})), but not the more traditional WKB approximation. Such
examples represent a major, non-trivial advance of localization theory and
illustrate the potential usefulness of the localization formulas as reliable
calculational tools.

At the same time we can address some of the issues that
arise when dealing with phase space path integrals, which are generally
regarded as rather disreputable because of the unusual discretization of
momentum and configuration paths that occurs (in contrast to the more
conventional configuration space (Lagrangian) path integral \cite{schulman}).
For instance, we recall from Subsection 4.1 that the general identification
between the Schr\"odinger picture path integral and loop space Liouville
measures was done rather artificially, basically by drawing an analogy between
them. For a generic phase space path integral to represent the actual energy
spectrum of the quantum Hamiltonian, one would have to carry out the usual
quantization of generic Poisson brackets $\{x^\mu,x^\nu\}_\omega=\omega^{\mu
\nu}(x)$. However, unlike the Heisenberg canonical commutation relations
(\ref{commalg}), the Lie algebra generated by this procedure is not necessarily
finite-dimensional (for $\man$ compact) and so the representation problem has
no straightforward solution when the phase space is not a cotangent bundle
$\man\otimes\Lambda^1\man$ \cite{marinov}, as is the case for a Euclidean phase
space. This approach is therefore hopelessly complicated
and in general hardly consistent. One way around this, as we shall see, is
to use instead {\it coherent state} path integrals. This enables one
to obtain the desired identification above while maintaining the original phase
space path integral, and therefore at the same time keeping a formal analogy
between the finite-dimensional and loop space localization formulas.
Furthermore, because of their classical properties, coherent states are
particularly well-suited for semi-classical studies of quantum dynamics.
We shall see that all the localizable dynamical systems in 2-dimensions
have phase space path integrals that can be represented in terms of coherent
states, thus giving an explicit evaluation of the quantum propagator and
the connection with some of the conventional coadjoint orbit models.

In this Section we shall in addition confine our attention to the case of a
simply-connected phase space, leaving the case where $\man$ can have
non-contractible loops for the next Section. In both cases, however, we
shall focus on the construction of localizable Hamiltonian systems starting
from a {\it generic} phase space metric, which will illustrate explicitly the
geometrical dependence of these dynamical systems and will therefore give
a further probe into the geometrical nature of (quantum) integrability. In
this way, we will get a good general idea of what sort of phase space path
integrals will localize and a detailed description of the symmetries
responsible for localization, as well as what sort of topological field
theories the
localization formulas will represent.

\subsection{Coadjoint Orbit Quantization and Character Formulas}

There is a very interesting class of cohomological quantum theories which
arise quite naturally within the framework of equivariant localization. These
will set the stage for the discussion of this Section wherein we shall focus
on the generic equivariant Hamiltonian systems with simply connected phase
spaces and thus present numerous explicit examples of the localization
formalism. For a (compact or non-compact) semi-simple Lie group $G$ (i.e. one
whose Lie algebra $\bf g$ has no abelian invariant subalgebras), we are
interested in the coadjoint action of $G$ on the coset space
$\man_G=G/H_C=\{gh_C:g\in G\}$, where $H_C\sim(S^1)^r$ is the Cartan subgroup
of $G$. The coset obtained by quotienting a Lie group by a maximal torus is
often called a `flag manifold'. The coadjoint orbit
\beq
O_{\Lambda'}=\{{\rm Ad}^*(g)\Lambda':g\in G\}\simeq\man_G~~~~~,~~~~~\Lambda'\in
{\bf h}^*
\label{coadorb}\eeq
is the orbit of maximal dimensionality of $G$. Here ${\rm Ad}^*(g)\Lambda'$
denotes the coadjoint action of $G$ on $\Lambda'$, i.e.
\beq
\left({\rm Ad}^*(g)\Lambda'\right)(\gamma)=\Lambda'(g^{-1}\gamma g)~~~~~,~~~~~
\forall\gamma\in{\bf g}
\label{coadaction}\eeq
and $\bf h$ is the Cartan subalgebra of $\bf g$.
The natural isomorphism in (\ref{coadorb}) between the flag manifold
$\man_G=G/H_C$ and the coadjoint orbit $O_{\Lambda'}$ is $gH_C\to~{\rm
Ad}^*(g)\Lambda'$ with the maximal torus $H_C$ identified as the stabalizer
group of the point $\Lambda'\in{\bf h}^*$. We assume henceforth that
$H^1(G)=H^2(G)=0$. There is a natural $G$-invariant symplectic structure on the
coadjoint orbit (\ref{coadorb}) which is defined by the Kirillov-Kostant 2-form
\cite{alek1,alek2}. This 2-form at the point $\Lambda\in{\bf g}^*$ is given by
\beq
\omega_{\Lambda}=\frac{1}{2}\Lambda([{\cal T}~{\buildrel\wedge\over,}~
{\cal T}])
\label{kirillov2form}\eeq
where $\cal T$ is a 1-form with values in the Lie algebra $\bf g$ which
satisfies the equation
\beq
d\Lambda(\gamma)={\rm ad}^*({\cal T})\Lambda(\gamma)\equiv\Lambda([\gamma,{\cal
T}])~~~~~,~~~~~\forall\gamma\in{\bf g}
\label{infcoadaction}\eeq
and ${\rm ad}^*({\cal T})$ denotes the infinitesimal coadjoint action of the
element ${\cal T}\in{\bf g}$.

The 2-form (\ref{kirillov2form}) is closed and non-degenerate on the orbit
(\ref{coadorb}), and by construction the group $G$ acts on $O_{\Lambda'}$
by symplectic (canonical) transformations with respect to the Kirillov-Kostant
2-form. Its main characteristic is that the Poisson algebra with
respect to (\ref{kirillov2form}) isomorphically represents the group $G$,
\beq
\left\{X_1(\Lambda),X_2(\Lambda)\right\}_{\omega_{\Lambda}}=[X_1,X_2](\Lambda)
\label{isogroupaction}\eeq
where $X_i\in{\bf g}$ are regarded as linear functionals on the orbit
$O_{\Lambda'}$ with $X_i(\Lambda)\equiv\Lambda(X_i)$. Alekseev, Faddeev and
Shatashvili \cite{alek1,alek2} have studied the phase space path integrals for
such dynamical systems with Hamiltonians defined on the coadjoint orbit
(\ref{coadorb}) (e.g. Cartan generators of $\bf g$) and have shown that, quite
generally, the associated quantum mechanical matrix elements correspond to
matrix elements of the Hamiltonian generator of $\bf g$ in some irreducible
representation of the group $G$. We shall see this feature explicitly later on.

However, for our purposes here, there is a much nicer description of the orbit
space (\ref{coadorb})
using its representation as the quotient space $\man_G=G/H_C$ \cite{helgason}.
As a smooth space, $\man_G$ is an example of a complex manifold of complex
dimension $n$ (real dimension $2n$), i.e. a manifold which is covered by open
sets each homeomorphic to $\IC^n$ and for which the coordinate transformations
on the overlap of 2 open sets are given by holomorphic functions. Here the
complexification of the group $G$ is defined by exponentiating the
complexification ${\bf g}\otimes\IC$ of the finite-dimensional vector space
$\bf g$. Let us quickly review some facts about the differential geometry of
complex manifolds. In local coordinates $x=(z^1,\dots,z^n)\in\IC^n$, we can
define the tangent space $T_x^{(0,1)}\man_G$ at $x\in\man_G$ as the complex
vector space spanned by the $\bar z$ derivatives
$\{\frac{\partial}{\partial\bar z^{\bar\mu}}\}_{\bar\mu=1}^n$, and analogously
$T_x^{(1,0)}\man_G$ is the complex vector space spanned by the $z$ derivatives
$\{\frac{\partial}{\partial z^{\mu}}\}_{\mu=1}^n$. The key feature is that
barred and unbarred vectors do not mix under a holomorphic change of
coordinates, and therefore it makes sense (globally) to consider tensors with
definite numbers $k$ and $\ell$ of holomorphic and anti-holomorphic indices of
either covariant or contravariant type. We refer to these as tensors of type
$(k,\ell)$. The vector space of $(p+q)$-forms of type $(p,q)$ is denoted
$\Lambda^{(p,q)}\man_G$ and the exterior algebra of $\man_G$ now refines to
\beq
\Lambda\man_G=\bigoplus_{p,q=0}^n\Lambda^{(p,q)}\man_G
\label{extalghol}\eeq

The DeRham exterior derivative operator $d\sim\eta^\mu\frac{\partial}{\partial
x^\mu}:\Lambda^k\man_G\to\Lambda^{k+1}\man_G$ now decomposes into holomorphic
and anti-holomorphic exterior derivative operators as
\beq
d=\partial+\bar\partial
\label{dolbeaultdef}\eeq
where $\partial\sim\eta^\mu\frac{\partial}{\partial
z^\mu}:\Lambda^{(p,q)}\man_G\to\Lambda^{(p+1,q)}\man_G$ ($\eta^\mu\sim dz^\mu$)
and $\bar\partial\sim\bar\eta^{\bar\mu}\frac{\partial}{\partial\bar
z^{\bar\mu}}:\Lambda^{(p,q)}\man_G\to\Lambda^{(p,q+1)}\man_G$
($\bar\eta^{\bar\mu}\sim d\bar z^{\bar\mu}$). The anti-holomorphic exterior
derivative $\bar\partial$ is called the Dolbeault operator, and the nilpotency
of $d$ now translates into the set of conditions
\beq
0=\partial^2=\bar\partial^2=\partial\bar\partial+\bar\partial\partial
\label{dolnilpotent}\eeq
Finally, let us note that a complex manifold always possesses a globally
defined rank (1,1) tensor field $J$ (i.e. an endomorphism of the space
$T^{(1,0)}\man_G\oplus T^{(0,1)}\man_G$) with $J^2=-{\bf1}$. It can be defined
locally by
\beq
J^\mu_\nu=i\delta^\mu_\nu~~~~~,~~~~~J^{\bar\mu}_{\bar\nu}=-i\delta^{\bar\mu}_
{\bar\nu}
\label{complexstrucdef}\eeq
with all other components vanishing, and it is known as a complex structure.

Given this important property of the coadjoint orbit, we now introduce local
complex coordinates $(z^\mu,\bar z^{\bar\mu})$ on $\man_G$ which are generated
by a complex structure $J$. The orbit (\ref{coadorb}) has the topological
features $H^1(\man_G;\IZ)=0$ and $H^2(\man_G;\IZ)=H^1(H_C;\IZ)=\IZ^r$, where
$r=\dim H_C$ is the rank of $G$ and $\IZ^r$ corresponds to the lattice of roots
of $H_C$ \cite{varadarajan}. The cohomology classes in $H^2(\man_G;\IZ)$ are
then represented by $r$ closed non-degenerate 2-forms of type (1,1)
\cite{helgason}
\beq
\omega^{(i)}=\frac{i}{2}g_{\mu\bar\nu}^{(i)}(z,\bar z)dz^\mu\wedge d\bar
z^{\bar\nu}
\label{kahlerclasses}\eeq
The components $g_{\mu\bar\nu}$ of (\ref{kahlerclasses}) define Hermitian
matrices, $g_{\bar\nu\mu}^*=g_{\mu\bar\nu}$, and the non-degeneracy condition
implies that they define metrics on $T^{(1,0)}\man_G\oplus T^{(0,1)}\man_G$ by
\beq
g^{(i)}=g_{\mu\bar\nu}^{(i)}(z,\bar z)dz^\mu\otimes d\bar z^{\bar\nu}
\label{kahlermetric}\eeq
The closure condition on the 2-forms (\ref{kahlerclasses}) can be written
in terms of the holomorphic and anti-holomorphic components of the exterior
derivative (\ref{dolbeaultdef}) as
\beq
\partial\omega^{(i)}=\bar\partial\omega^{(i)}=0
\label{dolclosed}\eeq
The analogue of the Poincar\'e lemma for the Dolbeault operator $\bar\partial$
is the Dolbeault-Grothendieck lemma. Since the 2-forms $\omega^{(i)}$ in the
case at hand are closed under both $\partial$ and $\bar\partial$, the
Dolbeault-Grothendieck lemma implies that locally they can be expressed in
terms of $C^\infty$-functions $F^{(i)}$ on $\man_G$ as
\beq
\omega^{(i)}=-i\partial\bar\partial F^{(i)}
\label{lockahlerrep}\eeq
or in local coordinates
\beq
g_{\mu\bar\nu}^{(i)}(z,\bar z)=\frac{\partial^2 F^{(i)}(z,\bar z)}{\partial
z^\mu\partial\bar z^{\bar\nu}}
\label{kahlerpotdef}\eeq

In general, a complex manifold with a symplectic structure such as
(\ref{kahlerclasses}) is called a K\"ahler manifold. The closed 2-forms
(\ref{kahlerclasses}) are then refered to as K\"ahler classes or
K\"ahler 2-forms, the associated metrics (\ref{kahlermetric}) are called
K\"ahler metrics, and the locally-defined functions $F^{(i)}$ in
(\ref{kahlerpotdef}) are called K\"ahler potentials. For an elementary,
comprehensive introduction to complex manifolds and K\"ahler structures, we
refer to \cite{eguchi} and \cite{greensw}. In the case at hand here, the above
construction yields a $G$-action on $\man_G$ by symplectic (canonical)
transformations \cite{helgason}, i.e. holomorphic functions $f(z)$ on $\man_G$
which act on the K\"ahler potentials by
\beq
F^{(i)}(z,\bar z)~{\buildrel f\over\longrightarrow}~\tilde F^{(i)}(z,\bar z)
=F^{(i)}(z,\bar z)+f(z)+\bar f(\bar z)
\label{kahpotcan}\eeq
This follows from the fact that the cotangent bundle of $G$ is
\beq
T^*G=G\times{\bf g}^*
\label{cotbundleLie}\eeq
so that the natural symplectic action of $G$ on $T^*G$ is $g\cdot(\tilde
g,\Lambda)=(g\cdot\tilde g,\Lambda)$. Consequently, the closed 2-forms
$\omega^{(i)}$ define $G$-invariant integral symplectic structures on $\man_G$.
Since $H^2(G)=0$, the 2-cocycles in (\ref{hamcomm}) vanish and this $G$-action
determines group isomorphisms into the Poisson algebras of $\man$. This also
follows directly from the property (\ref{isogroupaction}) of the
Kirillov-Kostant 2-form above. Notice that the only non-vanishing components
(up to permutation of indices and complex conjugation) of the Riemannian
connection and curvature associated with the K\"ahler metric
(\ref{kahlermetric}) are
\beq
\Gamma^\mu_{\nu\lambda}=g^{\mu\bar\rho}\partial_\nu
g_{\lambda\bar\rho}~~~~~,~~~~~R^{\lambda}_{\mu\nu\bar\rho}=-\partial_{\bar
\rho}\Gamma^\lambda_{\mu\nu}
\label{kahlergeomids}\eeq

The Cartan basis of $\bf g$ is defined by the root space decomposition
\beq
{\bf g}={\bf h}\oplus\left(\bigoplus_\alpha {\bf g}_\alpha\right)
\label{rootspacedecomp}\eeq
of $\bf g$, where $\alpha=(\alpha_1,\dots,\alpha_r)$ are the roots of $\bf g$
(i.e. the eigenvalues of the Cartan generators in the adjoint representation of
$G$) and ${\bf g}_\alpha$ are one-dimensional subspaces of $\bf g$
\cite{georgi,varadarajan}. In this basis, the generators have the non-vanishing
Lie brackets
\beq
\left[H_i,E_\alpha\right]=\alpha_iE_\alpha~~~,~~~\left[E_\alpha,E_\beta
\right]=\left\{\new{\begin{array}{ll}N_{\alpha\beta}E_{\alpha+\beta}
~~&,~~\alpha
+\beta\neq0\\\sum_{i=1}^r\alpha_iH_i~~&,~~\beta=-\alpha\end{array}}\right.
\label{hiealpha}\eeq
where $\alpha,\beta$ are the roots of $\bf g$, $H_i=H_i^\dagger$, $i=1,\dots,
r$, are the generators of the Cartan subalgebra ${\bf h}\otimes\IC$ of ${\bf g}
\otimes\IC$, and $E_\alpha=E_{-\alpha}^\dagger$ are the step operators of $\bf
g\otimes\IC$ which, for each $\alpha$, span ${\bf g}_\alpha$ in
(\ref{rootspacedecomp}) and which act as raising operators by $\alpha>0$
(relative to some Weyl chamber hyperplane in root space) on the representation
states $|\lambda\rangle$ which diagonalize the Cartan generators (the weight
states), i.e. $H_i|\lambda\rangle\propto|\lambda\rangle$ and
$E_\alpha|\lambda\rangle\propto|\lambda+\alpha\rangle$ for $\alpha>0$.
The unitary irreducible representations of $G$ are characterized by highest
weights $\lambda_i$, $i=1,\dots,r$. For each $i$, $\lambda_i$ is an eigenvalue
of $H_i$ whose eigenvector is annihilated by all the $E_\alpha$ for $\alpha>0$.
Corresponding to each highest weight vector $\lambda=(\lambda_1,\dots,\lambda
_r)$ we introduce the $G$-invariant symplectic 2-form
\beq
\omega^{(\lambda)}=\sum_{i=1}^r\lambda_i\omega^{(i)}
\label{symplweight}\eeq
The symplectic potentials associated with (\ref{symplweight}) are
\beq
\theta^{(\lambda)}=\sum_{i=1}^r\lambda_i\left(\frac{\partial F^{(i)}}{\partial
z^\mu}dz^\mu-\frac{\partial F^{(i)}}{\partial\bar z^{\bar\mu}}d\bar z^{\bar\mu}
\right)+dF
\label{potweight}\eeq

To construct a topological path integral from this symplectic structure, we
need to construct a Hamiltonian satisfying (\ref{hamdef}), i.e. a
Hamiltonian which is given by generators of the subalgebra of ${\bf
g}\otimes\IC$ which leave the symplectic potential (\ref{potweight}) invariant.
These are the canonical choices that give well-defined functions on the
coadjoint orbit (\ref{coadorb}). As remarked at the end of Subsection 3.2,
there usually exists a choice of function $F(z,\bar z)$ in (\ref{potweight})
for which this subalgebra contains the Cartan subalgebra ${\bf h}\otimes\IC$ of
${\bf g}\otimes\IC$. Let $H_i^{(\lambda)}$ be the generators of ${\bf
h}\otimes\IC$ in the representation with highest weight vector $\lambda$. Then
the Hamiltonian
\beq
H^{(\lambda)}=\sum_{i=1}^rh_iH_i^{(\lambda)}
\label{cartanham}\eeq
satisfies the required conditions and the corresponding path integral will
admit the topological form (\ref{parttopglobal}). Note that this is also
consistent with the integrability arguments of the previous Sections, which
showed that the localizable Hamiltonians were those given by the Cartan
generators of an isometry group $G$. Thus the path integral for the above
dynamical system determines a cohomological topological quantum field theory
which depends only on the second cohomology class of the symplectic 2-form
(\ref{symplweight}), i.e. on the representation with highest weight vector
$\lambda=(\lambda_1,\dots,\lambda_r)$.

To apply the equivariant localization formalism to these dynamical systems, we
note that since the K\"ahler metrics $g^{(i)}$ above are $G$-invariant, the
metric
\beq
g^{(\lambda)}=\sum_{i=1}^r\lambda_ig^{(i)}
\label{kahlermetweight}\eeq
obeys the usual localization criteria. We shall soon see that these group
theoretic structures are in fact {\it implied} by the localization constraints,
in that they are the only equivariant Hamiltonian systems associated with
homogeneous symplectic manifolds as above. Through numerous examples of such
systems and others we shall verify the localization formulas of the last
Section and discuss the common features that these quantum theories all
represent. For now, however, we will just explore what the localization
formulas will represent for the propagators $\tr\e^{iTH^{(\lambda)}}$ assuming
that they admit the phase space path integral form of the last Section with the
above symplectic structure. This will introduce some important group theoretic
notions, and later we shall show more precisely how to arrive at these path
integral representations and discuss some of the intricacies involved in
evaluating the localization formulas.

To apply the Niemi-Tirkkonen localization formula (\ref{niemitirkloc}) to the
dynamical system above, we first observe that the tangent and normal bundles of
$O_{\Lambda'}$ in ${\bf g}^*$ are related by \cite{berlinegetz}
\beq
T_{\Lambda'}{\bf g}^*=TO_{\Lambda'}\oplus{\cal N}O_{\Lambda'}=O_{\Lambda'}
\times{\bf g}^*
\label{tannorcoad}\eeq
{}From the construction of the coadjoint orbit it follows that the normal
bundle
${\cal N}O_{\Lambda'}$ in ${\bf g}^*$ is a trivial bundle with trivial
$G$-action on the fibers, and the product $O_{\Lambda'}\times{\bf g}^*$ is a
trivial bundle over ${\bf g}^*$ with the coadjoint action of $G$ in the fibers.
Then using (\ref{ptrivial}) and the multiplicativity property (\ref{multprop}),
we can write the $G$-equivariant $\hat A$-genus of the orbit $O_{\Lambda'}$ as
\beq
\hat A_V=\sqrt{\det\left[\frac{{\rm ad}~X}{\sinh({\rm ad}~X)}\right]}\equiv
\frac{1}{\sqrt{j({\rm ad}~X)}}
\label{ahatcoad}\eeq
where ${\rm ad}~X$ is the Cartan element $X\in{\bf h}$ in the adjoint
representation of $\bf g$. We now choose the radius of the orbit to be the
Weyl shift of the weight vector $\lambda$, i.e. $\Lambda'=\lambda+\rho$ where
\beq
\rho=\frac{1}{2}\sum_{\alpha>0}\alpha
\label{halfsumroots}\eeq
is the Weyl vector (the half-sum of positive roots of $G$), where $\lambda$ and
$\alpha$ are regarded as linear functions on $\bf g$ by returning the total
value of the weight or root associated to $X\in{\bf h}$. Then the localization
formula (\ref{niemitirkloc}) is none other than the celebrated Kirillov
character formula \cite{kirillov,picken1}
\beq
\tr_\lambda\e^{iTX}=\frac{1}{\sqrt{j({\rm ad}~TX)}}
\int_{O_{\lambda+\rho}}\frac{(\omega^{(\lambda)})^n}{n!}\e^{iTH^{(\lambda)}}
\label{kirchar}\eeq
where $\tr_\lambda$ denotes the trace in the representation with highest
weight vector $\lambda$ and $H^{(\lambda)}$ is the Hamiltonian
(\ref{cartanham})
associated with the Cartan element $X\in{\bf h}$. If we further apply the
finite-dimensional Duistermaat-Heckman theorem to the Fourier transform of the
orbit on the right-hand side of (\ref{kirchar}) (i.e. the localization formula
(\ref{pathloch})) we arrive at the famous Harish-Chandra formula
\cite{berlinegetz,harish,picken1}.

The resulting character formula associated with the Harish-Chandra formula
for the Fourier transform of the orbit is the classical Weyl character formula
of $G$ \cite{alek1}--\cite{alvarez},
\cite{funahashi,kirillov,nielsen,rajeev,stone,varadarajan}.
Let $W(H_C)=N(H_C)/H_C$ be the Weyl group of $H_C$, where $N(H_C)$ is the
normalizer subgroup of $H_C$, i.e. the subgroup of $g\in G$ with $hgH_C=gH_C$,
$\forall h\in H_C$, so that $N(H_C)$ is the subgroup of fixed points of the
left action of $H_C$ on the orbit $\man_G=G/H_C$. Given $w=nH_C\in W(H_C)$,
with $n=\e^{iN}\in N(H_C)$, let $X^{(w)}=n^{-1}Nn$ be the respective
adjoint representation $\e^{iX^{(w)}}=n^{-1}\e^{iX}n$.
The Weyl character formula can then be written as
\beq
\tr_\lambda\e^{iTX}=\sum_{w\in W(H_C)}\e^{iT(\lambda+\rho)(X^{(w)})}\prod_
{\alpha>0}\frac{1}{2i\sin\frac{T}{2}\alpha(X^{(w)})}
\label{weylchar}\eeq
where $\alpha(X^{(w)})$ are the roots associated to the Cartan elements
$X^{(w)}$. We shall see explicitly later on how these character formulas arise
from the equivariant localization formulas of the last Section, but for now
we simply note here the deep group theoretical significance that the
localization formulas will represent for the path integral representations of
the characters $\tr_\lambda\e^{iTX}$ in that the equivariant localization
formalism reproduces some classical results of group theory. Note that the Weyl
character formula writes the character of a Cartan group element as a sum of
terms, one for each element of the Weyl group, the group of symmetries of the
roots of the Lie algebra $\bf g$. In the context of the formalism of Section 4,
the Weyl character formula will follow from the coadjoint orbit path integral
over $LO_{\lambda+\rho}$. It was Stone \cite{stone} who first related this
derivation of the Weyl character formula to the index of a Dirac operator from
a supersymmetric path integral and hence to the semi-classical WKB evaluation
of the spin partition function, as we did quite generally in Subsection 4.2
above. The path integral quantization of the coadjoint orbits of
semi-simple Lie groups is essential to the quantization of spin systems. One
important feature of the above topological field theories is that there is a
one-to-one correspondence between the points on the orbits $G/H_C$ and the
so-called coherent states associated with the Lie group $G$ in the
representation with highest weight vector $\lambda$ \cite{perel}. The
above character formulas can therefore be represented in complex polarizations
using coherent state path integrals. We shall discuss these and other
aspects of the path integral representations of character formulas as we go
along in this Section.

{}From the point of view of path integral quantization, the necessity of
performing a Weyl shift $\lambda\to\lambda+\rho$ in the above is rather
unsatisfactory. This Weyl shift problem has been a point of some controversy in
the literature \cite{fujii}. As we shall see, the Weyl character formula
follows {\it directly} from the WKB formula for the spin partition function
\cite{stone}, and a proper discretization of the trace in (\ref{kirchar})
really does give the path integral over the orbit $O_\lambda$
\cite{alek2,nielsen}. The Weyl shift is in fact an artifact of the
regularization procedure
\cite{alek2,miet,nielsen,rajeev,tirk1} discussed in Subsection 4.2 in
evaluating the fluctuation determinant there which led to the Niemi-Tirkkonen
localization
formula (\ref{niemitirkloc}) and which leads directly to the Kirillov character
formula (\ref{kirchar}). As the $\hat A$-genus is inherently related to
tangent bundles of real manifolds, the problem here essentially is that the
regularization discussed in Subsection 4.2 does not respect the complex
structure defined on the orbit. We shall see later on how a coherent state
formulation avoids this problem and leads to a correct localization formula
without the need to introduce an explicit Weyl shift.

\subsection{Isometry Groups of Simply Connected Riemannian Spaces}

Given the large class of localizable dynamical systems of the last Subsection
and their novel topological and group theoretical properties, we now turn
to an opposite point of view and begin examining what Hamiltonian systems
in general fit within the framework of equivariant localization. For
this we shall analyse the fundamental isometry condition on the physical
theory in a quite general setting, and show that the localizable systems
``essentially" all fall into the general framework of the coadjoint orbit
quantization of the last Subsection. Indeed, this will be consistent with the
integrability features implied by the equivariant localization criteria.

We consider a simply-connected, connected and orientable Riemannian manifold
$(\man,g)$ of dimension $d$ (not necessarily symplectic for now) and with
metric $g$ of Euclidean signature, for definiteness. The isometry group
${\cal I}(\man,g)$ is the diffeomorphism subgroup of $C^\infty$ coordinate
transformations $x\to x'(x)$ which preserve the metric distance on $\man$, i.e.
for which $g_{\mu\nu}'(x')=g_{\mu\nu}(x')$. The generators $V^a$ of the
connected component of ${\cal I}(\man,g)$ form the vector field Lie algebra
\beq
{\cal K}(\man,g)=\{V\in T\man:\lie_Vg=0\}
\label{isomliealg}\eeq
and obey the commutation relations (\ref{vliealg}). For a
generic simply-connected space, the Lie group ${\cal I}(\man,g)$ is locally
compact in the compact-open topology induced by $\man$ \cite{helgason}. In
particular, if $\man$ is compact then so is ${\cal I}(\man,g)$. When $\dim{\cal
K}(\man,g)\neq0$, we shall say that the Riemannian manifold $(\man,g)$ is a
symmetric space.

We shall now quickly run through some of the basic facts concerning isometries
of simply-connected Riemannian manifolds, all of whose proofs can be found in
\cite{eisen1,eisen2,helgason,szsem2,weinberg}. First of all, by analysing the
possible solutions of the first order linear partial differential equations
$\lie_Vg=0$, it is possible to show that the number of linearly independent
Killing vectors (i.e. generators of (\ref{isomliealg})) is bounded as
\beq
\dim{\cal K}(\man,g)\leq d(d+1)/2
\label{dimbound}\eeq
when $\man$ has dimension $d$, so that the infinitesimal isometries of
$(\man,g)$ are therefore characterized by finitely-many linearly independent
Killing vectors in ${\cal K}(\man,g)$. There are 2 important classes of metric
spaces $(\man,g)$ characterized by their possible isometries. We say that a
metric space $(\man,g)$ is homogeneous if there exists infinitesimal isometries
$V$ that carry any given point $x\in\man$ to any other point in its immediate
neighbourhood. $(\man,g)$ is said to be isotropic about a point $x\in\man$ if
there exists infinitesimal isometries $V$ that leave the point $x$ fixed, and,
in particular, if $(\man,g)$ is isotropic about all of its points then we say
that it is isotropic. The homogeneity condition means that the metric $g$ must
admit Killing vectors that at any given point of $\man$ take on all possible
values (i.e. any point on $\man$ is geometrically like any other point). The
isotropy condition means that an isotropic point $x_0$ of $\man$ is always
a fixed point of an ${\cal I}(\man,g)$-action on $\man$, $V(x_0)=0$ for some
$V\in{\cal K}(\man,g)$, but whose first derivatives take on all possible
values,
subject only to the Killing equation $\lie_Vg=0$.

It follows that a homogeneous metric space always admits $d=\dim\man$ linearly
independent Killing vectors (intuitively generating translations in the $d$
directions), and a space that is isotropic about some point admits $d(d-1)/2$
Killing vector fields (intuitively generating rigid rotations about that
point). The connection between isotropy and
homogeneity of a metric space lies in the fact that any metric space which is
isotropic is also homogeneous. The spaces which have the maximal number
$d(d+1)/2$ of linearly independent Killing vectors enjoy some very special
properties, as we shall soon see. We shall refer to such spaces as maximally
symmetric spaces. The above dimension counting shows that a homogeneous
metric space which is isotropic about some point is maximally symmetric, and,
in particular, any isotropic space is maximally symmetric. The converse is
also true, i.e. a maximally symmetric space is homogeneous and isotropic.
In these cases, there is only one orbit under the ${\cal I}
(\man,g)$-action on $\man$, i.e. $\man$ can be represented as the orbit
$\man={\cal I}(\man,g)\cdot x$ of any element $x\in\man$, and the space of
orbits $\man/{\cal I}(\man,g)$ consists of only a single point. In this case
we say that the group ${\cal I}(\man,g)$ acts transitively on $\man$.

Conversely, if a Lie group $G$ acts transitively on a $C^\infty$-manifold
$\man$, then $\man$ is a homogeneous space and the stabalizer $G_x=\{g\in
G:g\cdot x=x\}$ of any point $x\in\man$ is a closed subgroup of $G$. The map
$hG_x\to h\cdot x$ defines a homeomorphism $G/G_x\simeq\man$ with the quotient
topology on $G/G_x$ induced by the natural (continuous and surjective)
projection map $\pi:G\to G/G_x$. On the other hand, if $G$ is locally compact
and $H$ is any closed subgroup of $G$, then there is a natural action of $G$ on
$G/H$ defined by $g\cdot\pi(h)=\pi(gh)$, $g,h\in G$, which is transitive and
for which $H$ is the stabalizer of the point $\pi({\bf1})$. In other words,
homogeneous spaces are essentially coadjoint orbits $G/H$ of Lie groups
\cite{helgason}, with $H=G_\Lambda$ the stabalizer group of a point
$\Lambda\in{\bf g}^*$ under the coadjoint action $O_\Lambda\equiv~{\rm
Ad}^*(G)\Lambda\subset{\bf g}^*$ of $G$ on ${\bf g}^*$. A sufficient condition
for the coset space $G/H=\{gH:g\in G\}$ to be a symmetric space is that $\bf g$
admit a reductive decomposition, i.e. an orthogonal decomposition ${\bf g}={\bf
h}\oplus{\bf h}^\perp$ such that $[{\bf h}^\perp,{\bf h}^\perp]\subset{\bf h}$
\cite{helgason}. Furthermore, it is possible to introduce K\"ahler structures
(the Kirillov-Kostant 2-form introduced in the previous Subsection) on the
group orbits for which $G$ is the associated isometry group. These spaces
therefore generalize the maximal coadjoint orbit models of the last Subsection
where $H$ was taken as the Cartan subgroup $H_C$ and which can in fact be shown
to be maximally symmetric \cite{helgason}. We shall see explicit examples later
on.

We shall now describe the rich features of maximally symmetric spaces. It
turns out that these spaces are uniquely characterized by a special curvature
constant $K$. Specifically, $(\man,g)$ is a maximally symmetric Riemannian
manifold if and only if there exists a constant $K\in\IR$ such that the Riemann
curvature tensor of $g$ can be written locally almost everywhere as
\beq
R_{\lambda\rho\sigma\nu}\equiv g_{\lambda\mu}R^\mu_{\rho\sigma\nu}=K(g_{\sigma
\rho}g_{\lambda\nu}-g_{\nu\rho}g_{\lambda\sigma})
\label{maxcurvtensor}\eeq
In dimension $d\geq3$, Schur's lemma \cite{helgason} states that the existence
of such a form for the curvature tensor
automatically implies the constancy of $K$. For $d=2$, however, this is not
the case, and indeed dimension counting shows that the curvature of a
Riemann surface always takes the form (\ref{maxcurvtensor}). In this case
$K$ is called the Gaussian curvature of $(\man,g)$ and it is in general not
constant. The above result implies that the Gaussian curvature $K$ of a
maximally symmetric simply connected Riemann surface is constant.

The amazing result here is the isometric correspondence between maximally
symmetric spaces. Any 2 maximally symmetric spaces $(\man_1,g_1)$ and
$(\man_2,g_2)$ of the same dimension and with the same curvature constant $K$
are isometric, i.e. there exists a diffeomorphism $f:\man_1\to\man_2$ between
the 2 manifolds relating their metrics by $g_1(x)=g_2(f(x))$.
Thus given any maximally symmetric space we can map it isometrically onto any
other one with the same curvature tensor (\ref{maxcurvtensor}). We can
therefore model maximally symmetric spaces by some ``standard" spaces, which
we now proceed to describe. Consider a flat $(d+1)$-dimensional
space with coordinates $(x^\mu,z)$ and metric
\beq
\eta_{d+1}=\frac{1}{|K|}dx_\mu\otimes dx^\mu+\frac{1}{K}dz\otimes dz
\label{euclmetricd1}\eeq
where $K$ is a real-valued constant. A $d$-dimensional space
can be embedded into this larger space by restricting the variables $x^\mu$
and $z$ to the surface of a (pseudo-)sphere,
\beq
\sgn(K)x^2+z^2=1
\label{psphereeq}\eeq
Using (\ref{psphereeq}) to solve for $z(x)$ and substituting this into
(\ref{euclmetricd1}), the metric induced on the surface by this embedding is
then
\beq
g_K=\left\{\new{\begin{array}{l}\frac{1}{K}\left(dx_\mu\otimes dx^\mu+\frac{
x_\mu x_\nu}{1-x^2}dx^\mu\otimes dx^\nu\right)~~~{\rm for}~~~K>0\\\frac{1}{|K|}
\left(dx_\mu\otimes dx^\mu-\frac{x_\mu x_\nu}{1-x^2}dx^\mu\otimes dx^\nu
\right)~~~{\rm for}~~~K<0\\dx_\mu\otimes dx^\mu~~~{\rm for}~~~K=0\end{array}
}\right.
\label{gk}\eeq
These 3 cases represent, respectively, the standard metrics on the $d$-sphere
$S^d$ of radius $K^{-1/2}$, the hyperbolic Lobaschevsky space ${\cal H}^d$ of
constant negative curvature $K$, and Euclidean $d$-space $\IR^d$ with its
usual flat metric $\eta_{E^d}$.

{}From the embedding condition (\ref{psphereeq}) and the manifest invariances
of
the embedding space geometry (\ref{euclmetricd1}) it is straightforward to
show that the above spaces all admit a $d(d+1)/2$-parameter group of
isometries. These consist of $d(d-1)/2$ rigid rotations about the origin and
$d$ (quasi-)translations. The first set of isometries always leave some points
on the manifold fixed, while the second set translate any point on $\man$ to
any other point in its vicinity. The 3 spaces above are therefore the 3 unique
(up to isometric equivalence) maximally symmetric spaces in $d$-dimensions,
and {\it any} other maximally symmetric space will be isometric to one of
these spaces, depending on whether $K=0$, $K>0$ or $K<0$. It is this feature
of maximally symmetric spaces that allows the rather complete isometric
correspondence which will follow. The Killing vector fields that generate the
above stated isometries are, respectively,
\beq
V_K=\left\{\new{\begin{array}{l}\left(\Omega_\nu^\mu x^\nu+\alpha^\mu\left[
1-~\sgn(K)x^2\right]^{1/2}\right)\frac{\partial}{\partial x^\mu}~~~{\rm for}
{}~~~K\neq0\\\left(\Omega_\nu^\mu
x^\nu+\alpha^\mu\right)\frac{\partial}{\partial
x^\mu}~~~{\rm for}~~~K=0\end{array}}\right.
\label{maxkilling}\eeq
where $\Omega_\nu^\mu=-\Omega_\mu^\nu$ and $\alpha^\mu$ are real-valued
parameters. These Killing vectors generate the respective isometry groups
\beq
{\cal I}(S^d)=SO(d+1)~~~,~~~{\cal I}({\cal H}^d)=SO(d,1)~~~,~~~{\cal I}(\IR^d)
=E^d
\label{maxisomgroups}\eeq
where $E^d$ denotes the Euclidean group in $d$-dimensions, i.e. the semi-direct
product of the rotation and translation groups in $\IR^d$, $SO(d+1)$ is the
rotation group of $\IR^{d+1}$, and $SO(d,1)$ is the Lorentz group in
$(d+1)$-dimensional Minkowski space. From this we see
therefore what sort of group actions should be considered within the
localization framework for maximally symmetric spaces. Note that the
maximal symmetry of the spaces $S^d$ and ${\cal H}^d$ are actually implied by
that of $\IR^d$, because $S^d$ can be regarded as the one-point
compactification of $\IR^d$, i.e. $S^d=\IR^d\cup\{\infty\}$ (also known as
stereographic projection), and ${\cal H}^d$ can be obtained from $S^d$ by Wick
rotating one of its coordinates to purely imaginary values.

The next situation of interest is the case where $(\man,g)$ is not itself
maximally symmetric, but contains a smaller, $d_H$-dimensional maximally
symmetric subspace $\man_0$ (e.g. a homogeneous but non-isotropic space). The
general theorem that governs the structure of such spaces  is as follows. We
can distinguish $\man_0$ from $\man-\man_0$ by $d-d_H$ coordinates $v^\alpha$,
and locate points within the subspace $\man_0$ with $d_H$ coordinates $u^i$. It
can then be shown \cite{helgason,weinberg} that it is possible to choose the
local $u$-coordinates so that the metric of the entire space $\man$ has the
form
\beq
g=\frac{1}{2}g_{\mu\nu}(x)dx^\mu\otimes dx^\nu=\frac{1}{2}g_{\alpha\beta}(v)
dv^\alpha\otimes dv^\beta+\frac{1}{2}f(v)\tilde g_{ij}(u)du^i\otimes du^j
\label{hommetric}\eeq
where $g_{\alpha\beta}(v)$ and $f(v)$ are functions of the $v$-coordinates
alone, and $\tilde g_{ij}(u)$ is a function of only the $u$-coordinates that is
itself the metric of $\man_0$. As $(\man_0,\tilde g)$ is a $d_H$-dimensional
maximally symmetric space, it is isometric to one of the 3 standard spaces in
$d_H$-dimensions above and $\tilde g$ can be represented in one of the forms
given in (\ref{gk}) depending on the curvature of the maximally symmetric
subspace $\man_0$.

Our final general result concerning Killing vectors on generic $d$-dimensional
simply connected manifolds is for the cases where the isometry group of
$(\man,g)$ has the opposite feature of maximal symmetry, i.e. when ${\cal I}(
\man,g)$ is 1-dimensional. Consider a 1-parameter group of isometries acting on
the metric space $(\man,g)$. Let $V=V^\mu(x)\frac{\partial}{\partial x^\mu}\in
T\man$ be a generator of ${\cal I}(\man,g)$, and let $\chi^\mu(x)$ be
differentiable functions on $\man$ such that the change of variables $x'^\mu=
\chi^\mu(x)$ has non-trivial Jacobian
\beq
\det\left[\frac{\partial\chi^\mu}{\partial x^\nu}\right]\neq0
\label{nontrivjac}\eeq
For $\mu=2,\dots,d$ we can choose the diffeomorphisms $\chi^\mu(x)$ to in
addition be the $d-1$ linearly independent solutions of the first order linear
homogeneous partial differential equation
\beq
V(\chi^\mu)=\lie_V\chi^\mu=V^\nu\partial_\nu\chi^\mu=0~~~~~,~~~~~\mu=2,\dots,d
\label{partialdiffchi}\eeq
given by the constant coordinate lines $\chi^\mu(x)={\rm constant}$ embedded
into $\man$ from $\IR^{d-1}$. The functions $\chi^\mu(x)$ for $\mu=2,\dots,d$
also have an invertible Jacobian matrix since then
\beq
{\rm rank}_{2\leq\mu,\nu\leq d}\left[\frac{\partial\chi^\mu}
{\partial x^\nu}\right]=d-1
\label{rankchi}\eeq
which owes to the existence of paths under the flow of the isometry group
such that
\beq
\frac{dx^1}{V^1}=\frac{dx^2}{V^2}=\cdots=\frac{dx^d}{V^d}
\label{constpaths}\eeq
as implied by (\ref{partialdiffchi}) and the flow equation (\ref{intcurves}).

If we now choose the function $\chi^1(x)$ so that $\frac{\partial\chi^1}
{\partial x^\mu}\neq0$ for $\mu=1,\dots,d$, then the coordinate transformation
$x^\mu\to x'^\mu(x)=\chi^\mu(x)$ changes the components of the vector field $V$
to
\beq
V'^\mu=V^\nu\frac{\partial}{\partial x^\nu}\chi^\mu~~~~~,~~~~~\mu=1,\dots,d
\label{vectortransf}\eeq
It follows from (\ref{partialdiffchi}) that in these new $x'$-coordinates $V$
therefore has components $V'^1\neq0$ and $V'^\mu=0$ for $\mu=2,\dots,d$. Now
further change coordinates $x'\to x''$ defined by
\beq
x''^1=\int_{x_0'}^{x'}\frac{dx'^1}{V'^1(x')}~~~~~,~~~~~x''^\mu=x'^\mu=\chi^\mu
{}~~~{\rm for}~~~\mu=2,\dots,d
\label{xpprimecoords}\eeq
where $x_0$ is a fixed basepoint in $\man$. In this way we have shown that, in
the case of a 1-parameter isometry group action on $(\man,g)$, there exists
a local system of $x''$-coordinates defined almost everywhere on $\man-\man_V$
in which the Killing vector of the isometry group has components
\beq
V''^1=1~~~~~,~~~~~V''^\mu=0~~~{\rm for}~~~\mu=2,\dots,d
\label{vectransl}\eeq
Furthermore, an application of the Killing equation (\ref{liegcoord}) shows
that $(\man,g)$ admits a Killing vector if and only if there are local
coordinates $x''$ on $\man$ in which the metric tensor components $g''_{\mu\nu}
(x'')$ are independent of the coordinate $x''^1$,
\beq
\frac{\partial g''_{\mu\nu}(x'')}{\partial x''^1}=0
\label{metricindepx1}\eeq
and then the integral curves of $x''^1$ parametrize the paths of the
infinitesimal isometry and of the finite total isometry according to
(\ref{intcurves}). Moreover, the above derivation also shows that 2 distinct
isometries $V_1$ and $V_2$ of $(\man,g)$ cannot have the same path, since they
can be independently chosen to have the single non-vanishing components
$V_1''^1=V_2''^2=0$. These results mean that locally any isometry of $g$ looks
like translations in a single coordinate, and this therefore gives the
representation of a 1-parameter isometry as an explicit $\IR^1$-action on
$(\man,g)$ (which is either bounded or is a $U(1)$-action when $\man$ is
compact). We shall refer to this system of coordinates as a prefered set of
coordinates with respect to a Killing vector field $V$.

For simplicity, we shall now concentrate on the cases where $\man$ is a
simply connected 2-dimensional symplectic manifold with metric $g$. Notice that
the standard Cauchy-Riemann equations of complex analysis imply that a Riemann
surface is always a complex manifold. The
advantage of this insofar as the localization formalism is concerned is that
the Riemann uniformization theorem \cite{imayoshi,nash,schlich} tells us
that $g_{\mu\nu}(x)$ has globally only 1 independent component. This situation
is therefore amenable to a detailed analysis of the equivariant localization
constraints in terms of the single degree of freedom of the metric $g$.
Defining complex coordinates $z,\bar z=x^1\pm ix^2$, we can represent the
metric as
\beq
g=\lambda(dz+\mu d\bar z)\otimes(d\bar z+\bar\mu dz)
\label{gz}\eeq
where
\beq
\lambda=(~\tr~g+2\sqrt{\det g})/4~~~~~,~~~~~\mu=(g_{11}-g_{22}+2ig_{12})/4
\lambda
\label{lambdamu}\eeq
Orientation-preserving diffeomorphisms of $\man$ which only change the function
$\lambda>0$ above are called conformal transformations. The function $\mu$
determines the complex structures of $\man$, and therefore the set of
inequivalent complex structures of $\man$ is in a one-to-one correspondence
with the space of conformal equivalence classes of metrics on $\man$. A
complex coordinate $w$ is said to be an isothermal coordinate for $g$ if
$g=\rho~dw\otimes d\bar w$ for some function $\rho>0$. Using the tensor
transformation law for $g$, it follows from (\ref{gz}) that an isothermal
coordinate $w$ for $g$ exists if and only if the Beltrami partial differential
equation
\beq
\frac{\partial w}{\partial\bar z}=\mu\frac{\partial w}{\partial z}
\label{beltrami}\eeq
has a $C^\infty$-solution $w(z,\bar z)$. Such a solution always exists provided
that the function $\mu(z,\bar z)$ is uniformly bounded as $\|\mu\|_\infty<1$.
A complex structure on $\man$ can therefore be identified with the conformal
structure represented by the Riemannian metric $g$.

The simple-connectivity of a 2-manifold $\man$ implies that via a
diffeomorphism and Weyl rescaling $g\to\e^\varphi g$ of the coordinates the
metric can be put globally\footnote{\baselineskip=12pt The fact that this holds
{\it globally} follows from an application of the classical Riemann-Roch
theorem, or the more modern Atiyah-Singer index theorem
\cite{eguchi,greensw,nash}.} into the isothermal form
\beq
g_{\mu\nu}(x)=\e^{\varphi(x)}\delta_{\mu\nu}~~~~~{\rm or}~~~~~g=\e^{\varphi
(z,\bar z)}dz\otimes d\bar z
\label{isotherm}\eeq
where $\varphi(x)$ is a globally-defined real-valued function on $\man$
which we shall refer to as the conformal factor of the metric. This means that
there is a unique complex structure on the Riemann surface $\man$
which we can define by the standard local complex coordinates $z,\bar z=x^1\pm
ix^2$. Notice that these remarks are {\it not} true if $H^1(\man;\IZ)\neq0$,
because then the metric has additional degrees of freedom from moduli
parameters (see the torus example in Subsection 3.5), i.e. (\ref{isotherm})
should be replaced by
\beq
g=\e^{\varphi}\hat g(\tau)
\label{ghattau}\eeq
where $\tau$ labels the additional modular degress of freedom of the metric.
We shall discuss the case
of multiply-connected phase spaces in the next Section.

With this complex structure we define $V^{z,\bar z}=V^1\pm iV^2$ for any
vector field $V$, and we set $\partial,\bar\partial=\frac{1}{2}(\frac{\partial}
{\partial x^1}\mp i\frac{\partial}{\partial x^2})$. The Killing equations
(\ref{liegcoord}) in these complex coordinates can be written as
\beq
\bar\partial V^z=\partial V^{\bar z}=0~~~~~,~~~~~\partial V^z+\bar\partial
V^{\bar z}+V^z\partial\varphi+V^{\bar z}\bar\partial\varphi=0
\label{killingeqcomplex}\eeq
The first set of equations in (\ref{killingeqcomplex}) are the Cauchy-Riemann
equations and they imply that in these local coordinates the Killing vector
field $V^z$ is a holomorphic function on $\man$. The other equation is a
source equation for $V^z$ and $V^{\bar z}$ that explicitly determines the
Killing fields in terms of the single degree of freedom of the metric $g$
(i.e. the conformal factor $\varphi$).

The Gaussian curvature scalar $K(x)$ of $(\man,g)$, which is always defined
by (\ref{maxcurvtensor}) in 2 dimensions, can be written in these isothermal
coordinates as
\beq
K(x)=-\frac{1}{2}\e^{-\varphi(x)}\nabla^2\varphi(x)
\label{gausscurv}\eeq
where $\nabla^2=g^{\mu\nu}\nabla_\mu\nabla_\nu=\partial\bar\partial$ is the
2-dimensional scalar Laplacian on $\man$ associated with the metric
(\ref{isotherm}). This follows from noting that the only non-vanishing
connection coefficients of the metric (\ref{isotherm}) are
\beq
\Gamma_{zz}^z=\partial\varphi~~~~~,~~~~~\Gamma^{\bar z}_{\bar z\bar z}=\bar
\partial\varphi
\label{connkahler}\eeq
The Gaussian curvature of $(\man,g)$ then uniquely characterizes the
isometry group acting on the phase space. If $K$ is constant, then $(\man,g)$
is maximally symmetric with 3 linearly independent Killing vectors. Moreover,
in this case $(\man,g)$ is isometric to either the 2-sphere $S^2$, the
Lobaschevsky plane ${\cal H}^2$ or the Euclidean plane $\IR^2$. We shall soon
examine these 3 distinct Riemannian spaces in detail. Notice, however, that
if $\man=\Sigma^h$ is a compact Riemann surface of genus $h$, then the
Gauss-Bonnet-Chern theorem (\ref{gaussbonchern}) in the case at hand reads
\beq
\int_{\Sigma^h}d{\rm vol}(g(x))~K(x)=4\pi(1-h)
\label{gaussbon2d}\eeq
where $d{\rm vol}(g(x))=d^2x~\sqrt{\det{g(x)}}$ is the metric volume form of
$(\man,g)$. Thus a maximally symmetric compact Riemann surface of constant
negative curvature must have genus $h\geq2$. It follows, under the
simple-connectivity assumption of this Section, that when $K=0$ or $K>0$ the
phase space $\man$ can be either compact or non-compact, but when $K<0$ it
is necessarily non-compact.

The other extremal case is where $(\man,g)$ admits only a 1-parameter group
of isometries. From the above general discussion it follows that in this
case there exist 2 differentiable functions $\chi^1$ and $\chi^2$ on $\man$
and local coordinates $x'$ on $\man$ such that
\beq
V^\mu\frac{\partial}{\partial x^\mu}\chi^2(x^1,x^2)=0~~~~~,~~~~~x'^2=
\chi^2(x^1,x^2)
\label{xprime2d}\eeq
and in these coordinates the Killing vector field has components $V'^1=1,V'^2
=0$. Moreover, the characteristic curves of the coordinate $x'^2=\chi^2$,
defined by the initial data surfaces of the partial differential equation in
(\ref{xprime2d}), can be chosen to be orthogonal to the paths defined by the
isometry generator $V$, i.e. we can choose the initial conditions for the
solutions of (\ref{xprime2d}) to lie on a non-characteristic surface. This
means that in these new coordinates $g_{12}'(x')=0$. Thus in this case the
metric can
be written locally as
\beq
g=g_{11}'dx'^1\otimes dx'^1+g_{22}'dx'^2\otimes dx'^2
\label{1parmetric}\eeq
and from (\ref{metricindepx1}) it follows that $g_{11}'$ and $g_{22}'$ are
functions only of $x'^2$. The phase space therefore describes a surface of
revolution, for example a cylinder or the `cigar-shaped' geometries that
are described in typical black hole theories \cite{weinberg}.

The only other case left to consider here is when $(\man,g)$ has a
2-dimensional isometry group. In this case we have 2 independent vector fields
$V_1=V_1^\mu\frac{\partial}{\partial x^\mu}$ and $V_2=V_2^\mu\frac{\partial}
{\partial x^\mu}$ which obey the Lie algebra (\ref{vliealg}) with $a,b,c=1,2$.
There are 2 possibilities for this Lie algebra -- either the isometry group
is abelian, $f^{abc}=0$, or it is non-abelian, $f^{abc}\neq0$ for some $a,b,c$.
Since $V_1$ and $V_2$ cannot have the same path in $\man$, we can choose paths
for the constant coordinate lines so that $V_1^2=V_2^1=0$. In the abelian
case, the commutativity of $V_1$ and $V_2$
\beq
\left[V_1,V_2\right]=0
\label{abelianv12}\eeq
implies that $V_1^1$ is a function of $x^1$ alone and $V_2^2$ is a function
only of $x^2$. As above, we can choose local coordinates almost everywhere
on $\man$ in which $V_1^1=V_2^2=1$. In these coordinates, the Killing equations
imply that the metric components $g_{\mu\nu}(x)$ are all constant. Thus in
this case $(\man,g)$ is isometric to flat Euclidean space, which contradicts
the standard maximal symmetry arguments above.

In the non-abelian case, we can choose linear combinations of the isometry
generators $V_1$ and $V_2$ so that their Lie algebra is
\beq
\left[V_1,V_2\right]=V_1
\label{nonabv12}\eeq
which implies that
\beq
\partial_1V_2^2=0~~~~~,~~~~~\partial_2\log V_1^1=-1/V_2^2
\label{nonabrel}\eeq
and so we can choose local coordinates almost everywhere on $\man$ in which
$V_2^2=1$ and $V_1^1=\e^{-x^2}$. The Killing equations then become
\beq
\partial_2g_{\mu\nu}=\partial_1g_{11}=0~~~,~~~\partial_1g_{12}=g_{11}~~~,~~~
\partial_1g_{22}=2g_{12}
\label{killeqv12}\eeq
which have solutions
\beq
g_{11}=\alpha~~~,~~~g_{12}=\alpha x^1+\beta~~~,~~~g_{22}=\alpha(x^1)^2+2\beta
x^1+\gamma
\label{killeqsolv12}\eeq
where $\alpha$, $\beta$ and $\gamma$ are real-valued constants. It is then
straightforward to compute the Gaussian curvature of $g$ from the identity
\beq
K(x)=-R_{1212}(x)/\det g(x)
\label{gausscurvr12}\eeq
which gives $K(x)$ as the constant $K=\alpha/(\beta^2-\alpha\gamma)$, again
contradicting the maximal symmetry theorems quoted above.

Thus a 2-dimensional phase space is {\it either} maximally symmetric with a
3-dimensional isometry group, or it admits a 1-parameter group of
isometries (or, equivalently, has a single 1-dimensional maximally symmetric
subspace), because the above arguments show that it clearly cannot have a
2-dimensional isometry group. The fact that there are only 2 distinct classes
of isometries in 2 dimensions is another very appealing feature of these
cases for the analysis which follows. For the remainder of this Section we
shall analyse the equivariant Hamiltonian systems which can be studied on
the various isometric types of spaces discussed in this Subsection and discuss
the features of the integrable quantum models that arise from the localization
formalism. We shall primarily develop these systems in 2 dimensions, and
present higher-dimensional examples in Subsections 5.7 and 5.8. This will
provide a large set of explicit examples of the
formalism developed thus far and at the same time clarify some other issues
that arise within the formalism of path integral quantization.

\subsection{Euclidean Phase Spaces and Holomorphic Quantization}

We begin our study of general localizable Hamiltonian systems with the case
where the phase space $\man$ is locally flat, i.e. $K=0$. The conformal factor
$\varphi$ in (\ref{isotherm}) and (\ref{gausscurv}) then satisfies the
2-dimensional Laplace equation
\beq
\nabla^2\varphi(z,\bar z)=\partial\bar\partial\varphi(z,\bar z)=0
\label{laplaceeq}\eeq
whose general solutions are
\beq
\varphi(z,\bar z)=f(z)+\bar f(\bar z)
\label{laplacesol}\eeq
where $f(z)$ is {\it any} holomorphic function on $\man$. The Riemannian
manifold $(\man,g)$ is isometric to the flat Euclidean space $(\IR^2,\eta_{E^2}
)$ and from the metric tensor transformation law it
follows that this coordinate change $z\to w$ taking the metric (\ref{isotherm})
to $dw\otimes d\bar w$ satisfies
\beq
\frac{\partial w}{\partial z}\frac{\partial\bar w}{\partial\bar z}+\frac{
\partial\bar w}{\partial z}\frac{\partial w}{\partial\bar z}=\e^{\varphi
(z,\bar z)}=\e^{f(z)}\e^{\bar f(\bar z)}~~~,~~~\frac{\partial w}{\partial z}
\frac{\partial\bar w}{\partial z}=\frac{\partial w}{\partial\bar z}\frac{
\partial\bar w}{\partial\bar z}=0
\label{metricchange}\eeq
It follows from (\ref{metricchange}) that this isometric transformation is the
2-dimensional conformal transformation $z\to w_f(z)$ (i.e. an analytic
rescaling of the standard flat Euclidean metric of the plane) where
\beq
w_f(z)=\int_{C_z}d\xi~\e^{f(\xi)}
\label{conftransf}\eeq
and $C_z\subset\man$ is a simple curve from some fixed basepoint in $\man$ to
$z$. From the last Subsection (eq. (\ref{maxkilling})) we know that the Killing
vectors of $(\IR^2,\eta_{E^2})$ in the complex coordinates $(w,\bar w)$ take
on the general form
\beq
V_{\IR^2}^w=-i\Omega w+\alpha~~~~~,~~~~~V_{\IR^2}^{\bar w}=i\Omega\bar w+
\bar\alpha
\label{killingr2}\eeq
where $\Omega\in\IR$ and $\alpha\in\IC$ are constants. The Killing vectors
(\ref{killingr2}) follow directly from (\ref{killingeqcomplex}) with
$\varphi=0$ there, and they generate the groups of 2-dimensional rotations $w
\to\e^{i\Omega}w$ and translations $w\to w+\alpha$ whose semi-direct product
forms the Euclidean group $E^2$ of the plane.

In these local complex coordinates on $\IR^2$ the Hamiltonian equations
$dH=-i_V\omega$ take the form
\beq
\partial H=\frac{i}{2}\omega(w,\bar w)V^{\bar w}~~~~~,~~~~~\bar\partial H=-
\frac{i}{2}\omega(w,\bar w)V^w
\label{hameqcomplex}\eeq
where
\beq
\omega=\frac{i}{2}\omega(w,\bar w)dw\wedge d\bar w
\label{symplcomplex}\eeq
The symplectic 2-form (\ref{symplcomplex}) can be explicitly determined here
by recalling that the Hamiltonian group action on the phase space is
symplectic so that $\lie_V\omega=0$. In local coordinates this means that
\beq
\partial_\mu(V^\lambda\omega_{\nu\lambda})-\partial_\nu(V^\lambda\omega_{\mu
\lambda})=0
\label{symplcondloc}\eeq
for each $\mu$ and $\nu$. Requiring this symplecticity condition for the full
isometry group action of $E^2$ on $\IR^2$, we substitute into
(\ref{symplcondloc}) each of the 3 linearly independent Killing vectors
represented by (\ref{killingr2}) (corresponding to $\Omega=0$, $\alpha^1=0$
and $\alpha^2=0$ there). The differential equations (\ref{symplcondloc}) for
the function $\omega(w,\bar w)$ now easily imply that it is constant on $\IR^2$
with these substitutions. Thus $\omega(w,\bar w)$ is the Riemannian volume
(and in this case the Darboux) 2-form globally on $\IR^2$. Substituting
the Darboux value $\omega(w,\bar w)=1$ and the Killing vectors
(\ref{killingr2}) into the Hamiltonian equations above and integrating them up
to get $H(w,\bar w)$, we see that the most general equivariant Hamiltonian
on a planar phase space $\man$ is
\beq
H_0(z,\bar z)=\Omega w_f(z)\bar w_f(\bar z)+\bar\alpha w_f(z)+\alpha\bar w_f(
\bar z)+C_0
\label{h0}\eeq
where $C_0\in\IR$ is a constant of integration and $w_f(z)$ is the conformal
transformation (\ref{conftransf}) from the flat Euclidean space back onto
the original phase space.

The fact that the symplectic 2-form here is uniquely determined to be the
volume form associated with the phase space geometry is a general feature of
any homogeneous symplectic manifold. Indeed, when a Lie group $G$ acts
transitively on a symplectic manifold there is a unique $G$-invariant measure
\cite{helgason}, i.e. a unique solution for the $d(d-1)/2$ functions $\omega_{
\mu\nu}$ from the $d(d-1)\cdot d(d+1)/4$ differential equations
(\ref{symplcondloc}). Thus $\omega^n/n!$ is necessarily the maximally symmetric
volume form of $(\man,g)$ and the phase space is naturally a K\"ahler manifold,
as in Subsection 5.1. We shall soon see the precise connection between
maximally
symmetric phase spaces and the coadjoint orbit models of Subsection 5.1. In the
present context, this is one of the underlying distinguishing features between
the maximally symmetric and inhomogeneous cases. In the latter case $\omega$ is
not uniquely determined from the requirement of symplecticity of the isometry
group action on $\man$, leading to numerous possibilities for the equivariant
Hamiltonian systems. In the case at hand here, the Darboux 2-form on $\IR^2$ is
the unique 2-form which is invariant under the full Euclidean group, i.e.
invariant under rotations and translations in the plane, and on $\man$ it is
the K\"ahler form associated with the K\"ahler metric (\ref{isotherm}) and
(\ref{laplacesol}).

The form (\ref{h0}) for the planar equivariant Hamiltonian systems illustrates
how the integrable dynamical systems which obey the localization criteria
depend on the phase space geometry which needs to be introduced in this
formalism. These systems are all, however, holomorphic copies of the same
initial dynamical system on $\IR^2$ defined by the Darboux Hamiltonian
\beq
H^D_0(z,\bar z)=\Omega z\bar z+\bar\alpha z+\alpha\bar z+C_0~~~;~~~z\in\IC
\label{h0darboux}\eeq
or identifying $z,\bar z=p\pm iq$ with $(p,q)$ canonical momentum and position
variables, these dynamical Hamiltonians are of the form
\beq
H_0^D(p,q)=\Omega(p^2+q^2)+\alpha_1p+\alpha_2q+C_0
\label{h0cancoord}\eeq
Thus the dependence on the phase space Riemannian geometry is trivial in the
sense that these systems all lift to families of holomorphic copies of the
planar dynamical systems (\ref{h0darboux}). This sort of trivial dependence
is to be expected since the (classical or quantum) dynamical problem is
initially independent of any Riemannian geometry of the phase space. It is
also anticipated from the general topological field theory arguments that we
presented earlier. Nonetheless, the general functions $H_0(z,\bar z)$ in
(\ref{h0}) illustrate how the geometry required for equivariant localization
is determined by the different dynamical systems, and vice versa, i.e. the
geometries that make these dynamical systems integrable. This probes
into what one may consider to be the geometry of the classical or quantum
dynamical system, and it illustrates the strong interplay between the
Hamiltonian and Riemannian symmetries that are responsible for localization.

Thus essentially the only equivariant Hamiltonian system on a planar symplectic
manifold is the displaced harmonic oscillator Hamiltonian
\beq
H_0^D=\Omega(z+a)(\bar z+\bar a)=\Omega\left\{(p+a_1)^2+(q+a_2)^2\right\}+C_0
\label{displharm}\eeq
and in this case we can replace the requirement that $H$ generate a circle
action with the requirement that it generate a semi-bounded group action. To
compare the localization formulas with some well-known results from elementary
quantum mechanics, we note that the Hamiltonian (\ref{h0cancoord}) can only
describe 2 distinct 1-dimensional quantum mechanical models. These are the
harmonic oscillator $\frac{1}{2}z\bar z=\frac{1}{2}(p^2+q^2)$ wherein we
take $\Omega=\frac{1}{2}$ and $\alpha=0$ in (\ref{h0darboux}) and apply either
the WKB or the Niemi-Tirkkonen localization formulas of the last Section,
and the free particle $\frac{1}{2}p^2$ where we take $\Omega=0$ and
$\alpha=1/2\sqrt{2}$ in (\ref{h0darboux}) and apply the quadratic localization
formula (\ref{quadrloc}) (or equivalently (\ref{locgenint})). In fact, these
are the original classic examples, which were for a long time the {\it only}
known examples, where the Feynman path integral can be evaluated exactly
because then their functional (and classical statistical mechanical) integrals
are Gaussian. For the same reasons, these are also the basic examples where the
WKB approximation is known to be exact \cite{schulman}.

It is straightforward to verify the Niemi-Tirkkonen localization formula
(\ref{niemitirkloc}) for the harmonic oscillator. In polar coordinates
$z=r\e^{i\theta}$ with $r\in\IR^+$ and $\theta\in[0,2\pi]$, we have
$\omega_{r\theta}=r$, $(\Omega_V)_{\theta r}=-2r$ and $R=0$ on flat $\IR^2$,
so that the integral in (\ref{niemitirkloc}) gives
\beq
Z_{\rm harm}(T)\sim\int_0^\infty dr~\frac{T}{2\sin\frac{T}{2}}\e^{-iTr^2/2}=
\frac{1}{2i\sin\frac{T}{2}}
\label{harmpart}\eeq
That this is the correct result can be seen by noting that the energy spectrum
determined by the Schr\"odinger equation for the harmonic oscillator is
$E_k=k+\frac{1}{2}$, $k\in\IZ^+$ \cite{messiah}, so that
\beq
\tr\|\e^{-iT(\hat p^2+\hat q^2)/2}\|=\sum_k\e^{-iTE_k}=\sum_{k=0}^\infty
\e^{-iT(k+\frac{1}{2})}=\frac{1}{2i\sin\frac{T}{2}}
\label{harmexact}\eeq
This result also follows from the WKB formula (\ref{wkb}) after working out
the regularized fluctuation determinant as described there. Here the classical
trajectories determined by the flows of the vector field $V^z=iz/2$ are the
circular orbits $z(t)=z(0)\e^{it/2}$. Note that the only way these orbits can
be defined on the loop space $L\IC$ is to regard $z(t)=z(0)\e^{it/2}$ and $\bar
z(t)=\bar z(T)\e^{i(T-t)/2}$ as independent complex variables. This means that
the functional integral should be evaluated in a holomorphic polarization. We
shall return to this point shortly.

Alternatively, we note that for $T\neq2\pi n$ the only $T$-periodic
critical trajectories of this dynamical system are the critical points $z,\bar
z=0$ of the harmonic oscillator Hamiltonian $z\bar z$ and
(\ref{harmexact}) also follows from (\ref{pathloch}) which gives
\beq
Z_{\rm
harm}(T)\sim\frac{1}{\pfaff\pmatrix{0&1\cr-1&0\cr}}\frac{\frac{1}{2i}}{\sin
\frac{T}{2}}=\frac{1}{2i\sin\frac{T}{2}}
\label{harmlocham}\eeq
For the discretized values $T=2\pi n$
any initial condition $z(0)\in\IC$ leads to $T$-periodic orbits, and the
moduli space of critical trajectories is non-isolated and coincides with the
entire phase space $\man=\IR^2$. In that case the degenerate path integral
formula (\ref{wkbdeg}) yields the correct result. These
results therefore all agree with the general assertions made at the beginning
of Subsection 4.6 concerning the structure of the moduli space of $T$-periodic
classical trajectories for a Hamiltonian circle action on the phase space.

For the free particle partition function, we have $R=\Omega_V
=0$, and so the $\hat A$-genus term in the localization formula
(\ref{quadrloc}) contributes 1. The $\phi_0$-integral in (\ref{quadrloc}) is
thus a trivial Gaussian one and we find
\beq
Z_{\rm free}(T)\sim\int_{-\infty}^\infty dp~dq~\int_{-\infty}^\infty d\phi_0~
\e^{iT\phi_0^2-iT\phi_0p/2\sqrt{2}}\sim\int_{-\infty}^\infty
dp~dq~\e^{-iTp^2/2}
\label{freepart}\eeq
which also coincides with the exact propagator $\tr\|\e^{-iT\hat p^2/2}\|$ in
the phase space representation. In this case the Hamiltonian $\frac{1}{2}p^2$
is degenerate on $\IR^2$, so that the WKB localization formula is unsuitable
for this dynamical system and the result (\ref{freepart}) follows from the
degenerate formula (\ref{wkbdeg}) by noting that $L\man_S=\IR^2$ in this case.
Notice also that (\ref{freepart}) coincides exactly with the classical
partition function of this dynamical system as there are no quantum
fluctuations.

There is another way to look at the path integral quantization of the Darboux
Hamiltonian system (\ref{h0darboux}) which ties in with some of the general
ideas of Subsection 5.1 above. The Heisenberg-Weyl algebra ${\bf g}_{HW}$
\cite{messiah} is the algebra generated
by the usual harmonic oscillator creation and annihilation operators
\beq
\hat a,\hat a^\dagger=\frac{1}{\sqrt{2}}\left(\hat p\pm i\hat q\right)
\label{agens}\eeq
in the canonical quantum theory associated with the phase space $\IR^2$ and
the operator algebra (\ref{commalg}). The Lie algebra ${\bf g}_{HW}$ is
generated by the operators $\hat a^\dagger$, $\hat a$ and $\hat N\equiv\hat a^
\dagger\hat a=\frac{1}{2}(\hat p^2+\hat q^2-1)$ with the commutation relations
\beq
\left[\hat a^\dagger,\hat a\right]=1
\label{harmcomm}\eeq
The (infinite-dimensional) Hilbert space which defines a representation of
these operators is spanned by the bosonic number basis $|n\rangle$, $n\in\IZ
^+$, which form the complete orthonormal system of eigenstates of the number
operator $\hat N$ with eigenvalue $n$,
\beq
\hat N|n\rangle=\hat a^\dagger\hat a|n\rangle=n|n\rangle
\label{numberbasis}\eeq
and on which $\hat a^\dagger$ and $\hat a$ act as raising and lowering
operators, respectively,
\beq
\hat a^\dagger|n\rangle=\sqrt{n+1}|n+1\rangle~~~~~,~~~~~\hat a|n\rangle=
\sqrt{n}|n-1\rangle
\label{raislow}\eeq

We now define the canonical coherent states \cite{faddeev,messiah,perel}
associated with this representation of the Heisenberg-Weyl group $G_{HW}$ as
\beq
|z)\equiv\e^{z\hat a^\dagger}|0\rangle=\sum_{n=0}^\infty\frac{z^n}{\sqrt{n!}}
|n\rangle~~~;~~~z\in\IC
\label{cancoherent}\eeq
These states are normalized as
\beq
(z|z)=\e^{z\bar z}
\label{cancohnorm}\eeq
with $(z|\equiv|z)^\dagger$, and they obey the completeness relation
\beq\new{\begin{array}{ll}
\int\frac{d^2z}{2\pi}~|z)\!)(\!(z|&=\int\frac{d^2z}{2\pi}~\e^{-z\bar z}
\sum_{n,m}\frac{z^n\bar z^m}{\sqrt{n!m!}}|n\rangle\langle m|\\&=\frac{1}{2\pi}
\int_0^\infty dr~r\e^{-r^2}\sum_{n,m}\frac{r^{n+m}}{\sqrt{n!m!}}\int_0^{2\pi}d
\theta~\e^{i(n-m)\theta}|n\rangle\langle m|\\&=\frac{1}{2\pi}\int_0^\infty dr~r
\e^{-r^2}\sum_{n,m}\frac{r^{n+m}}{\sqrt{n!m!}}\cdot2\pi\delta_{nm}|n\rangle
\langle m|=\sum_{n=0}^\infty|n\rangle\langle n|={\bf1}\end{array}}
\label{cohcompl}\eeq
where we have as usual written $z=r\e^{i\theta}$ and
\beq
|z)\!)~\equiv|z)/\sqrt{(z|z)}=\e^{-z\bar z/2}|z)
\label{cancohnormalized}\eeq
are the normalized coherent states.

The normalized matrix
elements of the algebra generators in these states are
\beq
(\!(z|\hat a^\dagger\hat a|z)\!)=z\bar z~~~,~~~(\!(z|\hat a|z)\!)
=\bar z~~~,~~~(\!(z|\hat a^\dagger|z)\!)=z
\label{genmatrixcoh}\eeq
Thus the 3 independent terms in the Darboux Hamiltonian (\ref{h0darboux}) are
none other than the normalized canonical coherent state matrix elements of
the Heisenberg-Weyl group generators. These 3 observables represent the
Poisson Lie group action of the Euclidean group $E^2$ on the coadjoint orbit
$G_{HW}/H_C=G_{HW}/U(1)=\IC^1$ with the Darboux Poisson bracket
\beq
\{z,\bar z\}_{\omega_D}=1
\label{hwpoissonaction}\eeq
which is the Poisson algebra representation of the Heisenberg-Weyl algebra
(\ref{harmcomm}). This correspondence with the coset space $G_{HW}/H_C$ and the
general framework of Subsection 5.1 is not entirely surprising, since
homogeneous symplectic manifolds are in {\it general}
essentially coadjoint orbits of Lie groups \cite{helgason}, i.e. they can be
represented as the quotient of their isometry groups by a maximal torus
according to the general discussion of the last Subsection. The
integrable Hamiltonian systems in this case are functionals of Cartan elements
of ${\bf g}_{HW}$ (e.g. the harmonic oscillator $\hat a^\dagger\hat a$ or the
free particle $(\hat a+\hat a^\dagger)^2$).

The canonical coherent states (\ref{cancoherent}) are those quantum states
which minimize the Heisenberg uncertainty principle $\Delta q\cdot\Delta p\geq
\frac{1}{2}$ \cite{messiah}, because they diagonalize the annihilation
operator $\hat a$, $\hat a|z)=z|z)$, and they can be generalized to arbitrary
Lie groups \cite{perel}, as we shall soon see. The Darboux 2-form
\beq
\omega_D=\frac{i}{2}dz\wedge d\bar z
\label{darboux2formz}\eeq
is defined globally on $\IC$ and, since $\IR^2$ is contractable and hence
$H^2(\IR^2;\IZ)=0$, it can be be generated {\it globally} by the symplectic
potential
\beq
\theta_D=-\frac{i}{2}\left(\bar zdz-zd\bar z\right)
\label{thetadarbouxz}\eeq
The canonical 1-form (\ref{thetadarbouxz}) and the flat K\"ahler
metric associated with (\ref{darboux2formz}) on $\IR^2$ can be written in terms
of the coherent states (\ref{cancoherent}) as
\beq
\theta_D=\frac{i}{2}(\!(z|d|z)\!)
\label{thetadcoh}\eeq
\beq
g_D=dz\otimes d\bar z=\|~d|z)\!)~\|\otimes\|~d|z)\!)~\|
-(\!(z|d|z)\!)\otimes(\!(z|d|z)\!)^*
\label{kahlermetcoh}\eeq
and the globally-defined K\"ahler potential associated with
(\ref{thetadarbouxz}) is
\beq
F_{\IR^2}(z,\bar z)=z\bar z
\label{r2kahlerpot}\eeq

The path integral here then coincides with the standard coherent state path
integral
\beq\new{\begin{array}{ll}
\tr\e^{-iT\hat{\cal H}}&=\int\frac{d^2z}{2\pi}~(\!(z|\e^{-iT\hat{\cal H}}|z)\!)
\\&=\int_{L\IR^2}\prod_{t\in[0,T]}\frac{dz(t)~d\bar z(t)}{2\pi}~\exp\left\{
i\int_0^Tdt~\left(z\dot{\bar z}-\bar z\dot z-H(z,\bar z)\right)\right\}
\end{array}}
\label{cohpathintr2}\eeq
where
\beq
H(z,\bar z)=(\!(z|\hat{\cal H}|z)\!)
\label{cohham}\eeq
is the coherent state matrix element of some operator $\hat{\cal H}=\hat{\cal
H}(\hat a,\hat a^\dagger)$ on the underlying representation space of the
Heisenberg-Weyl algebra. The derivation of (\ref{cohpathintr2}) is identical to
that in Subsection 4.1 except that now we use the modified completeness
relation
(\ref{cohcompl}) for the coherent state representation. This manner of
describing the quantum dynamics goes under the names of holomorphic,
coherent state or K\"ahler polarization. One of its nice features in general is
that it provides a natural identification of the path integral and loop space
Liouville measures. We recall from (\ref{feynmeas}) that in the former measure
there is one unpaired momentum in general and, besides the periodic boundary
conditions, there is a formal analog between the measures in (\ref{feynmeas})
and (\ref{quantpart}) only if the initial configuration of the propagator
depends on the position variables $q$ and the final configuration on the
momentum variables $p$, or vice versa. In the holomorphic polarization above,
however, the initial configuration depends on the $z$ variables, the final one
on the $\bar z$ variables, and the path integral measure is the formal $N\to
\infty$ limit of $\prod_{i=1}^Ndz_i~d\bar z_i/2\pi$. Since the number of $z$
and $\bar z$ integrations are the same, we obtain the desired formal
identifications. Besides providing one with a formal analog between the path
integral localization formulas and the Duistermaat-Heckman theorem and its
generalizations, this enables one to also ensure that the loop space
supersymmetry encountered in Subsection 4.4, which is intimately connected with
the definition of the path integral measure (as are the boundary conditions for
the propagator), is consistent with the imposed boundary conditions.

Thus on a planar phase space essentially the only equivariant Hamiltonian
systems are harmonic oscillators, generalized as in (\ref{h0}) to the inclusion
of a generic flat geometry so that the remaining Hamiltonian systems
are merely holomorphic copies of these displaced oscillators defined by the
analytic coordinate transformation (\ref{conftransf}). These systems generate
a topological quantum theory of the sort discussed in Subsections 4.10 and
5.1, with the Darboux Hamiltonian (\ref{h0darboux}) related to the
symplectic potential (\ref{thetadarbouxz}) by the usual topological condition
$H_0^D=i_{V_{\IR^2}}\theta_D$ reflecting the fact that (\ref{thetadarbouxz})
is invariant under the action of the rotation group of the plane. It is not,
however, invariant under the translation group action, so that the translation
generators do not determine a Witten-type topological field theory like the
harmonic oscillator Hamiltonians do. This means that there are no
$E^2$-invariant symplectic potentials on the plane, i.e. it is impossible to
find a function $F$ in (\ref{potweight}) that gives an invariant potential
simultaneously for all 3 of the independent generators in (\ref{h0darboux}).
The harmonic oscillator nature of these systems is consistent with
their global integrability properties. The holomorphic polarization of the
quantum theory associates the canonical quantum theory above with the
topological coadjoint orbit quantum theory of Subsection 5.1 and the coherent
state path integral (\ref{cohpathintr2}) yields character formulas for the
isometry group of the phase space. This will be the general characteristic
feature of {\it all} localizable systems we shall find. In the case at hand,
the character formulas are associated with the Cartan elements of the
Heisenberg-Weyl group.

\subsection{Coherent States on Homogeneous K\"ahler Manifolds and Holomorphic
Localization Formulas}

Before carrying on with our geometric determination of the localizable
dynamical systems and their path integral representations, we pause to
briefly discuss how the holomorphic quantization introduced above on the
coadjoint orbit $\IR^2$ can be generalized to the action of an arbitrary
semi-simple Lie group $G$ \cite{berezin,klaudskag,perel}. This representation
of the quantum dynamics proves to be the most fruitful on homogeneous spaces
$G/H_C$, and later on we shall generalize this construction to apply to
non-homogeneous phase spaces and even non-symmetric multiply connected phase
spaces. As the coherent states are those which are closest to ``classical"
states, in that they are the most tightly peaked ones about their locations,
they are the best quantum states in which to study the semi-classical
localizations for quantum systems. We shall also see that they are related to
the geometric quantization of dynamical systems \cite{woodhouse}.

Given any irreducible unitary representation $D(G)$ of the group $G$ and some
normalized state $|0\rangle$ in the representation space, we define the
(normalized) state $|g\rangle$ by
\beq
|g\rangle=D(g)|0\rangle
\label{gstatedef}\eeq
If $Dg$ denotes Haar measure of $G$, then Schur's lemma \cite{varadarajan}
and the completeness of the representation $D(G)$ implies the completeness
relation
\beq
\frac{\dim D(G)}{{\rm vol}(G)}\int_GDg~|g\rangle\langle g|={\bf 1}
\label{gcompl}\eeq
Following the derivation of Subsection 4.1, it follows that the partition
function associated with an operator $\hat{\cal H}$ acting on the
representation
space of $D(G)$ can be represented by the path integral
\beq\new{\begin{array}{ll}
\tr_{D(G)}\e^{-iT\hat{\cal H}}&=\int_GDg~\langle g|\e^{-iT\hat{\cal
H}}|g\rangle
\\&=\int_{LG}\prod_{t\in[0,T]}\frac{\dim D(G)}{{\rm vol}(G)}~Dg(t)~\exp\left\{
i\oint_{\gamma(g)}\langle g|d|g\rangle-i\int_0^Tdt~\langle g|\hat{\cal H}|g
\rangle\right\}\end{array}}
\label{charg}\eeq
However, if we take $|0\rangle$ to be a simultaneous eigenstate of the
generators of $H_C\subset G$ (i.e. a weight state), then the `coherent' states
$|g\rangle$ associated with any one coset of $G/H_C$ are all phase multiples of
one another. Thus the set of coherent states form a principal $H_C$-bundle
$L\to\man_G$ over $G/H_C$ and the coherent state path integral (\ref{charg}) is
in fact taken over paths in the homogeneous space $G/H_C$. This geometrical
method for constructing irreducible representations of semi-simple Lie groups
as sections of a line bundle $L\to\man_G$ associated to the principal fiber
bundle $G\to G/H_C$ is known as the Borel-Weil-Bott method \cite{pres}. The
holomorphic sections of this complex line bundle (the coherent states) form a
basis for the irreducible representation.

What is most interesting about the character representation (\ref{charg}) is
that it is closely related to the K\"ahler geometry of the homogeneous space
$G/H_C$. To see this, we first define the Borel subgroups $B_\pm$ of $G$ which
are the exponentiations of the subalgebras ${\cal B}_\pm$ spanned by $H_i\in{
\bf h} \otimes\IC$ and $E_\alpha$ for $\alpha>0$ and $\alpha<0$, respectively
(see (\ref{rootspacedecomp})). The complexification of the coadjoint orbit
$\man_G$ is then provided by the isomorphism $G/H_C\simeq G^c/B_\pm$, where
$G^c$ is the complexification of $G$ \cite{varadarajan}. Almost any $g\in G$
can be factored as a Gauss decomposition
\beq
g=\zeta_+h\zeta_-
\label{ggaussdecomp}\eeq
where $h\in H_C^c$ and
\beq
\zeta_+=\e^{\sum_{\alpha>0}z^\alpha E_\alpha}~~~~~,~~~~~\zeta_-=\e^{\sum_
{\alpha<0}z^\alpha E_\alpha}
\label{zetas}\eeq
Here $z^\alpha\in\IC$, and if we now apply the representation operator $D(g)$
to a lowest weight state, then $\zeta_-$ acts as the identity and the set of
physically distinct states are in a one-to-one correspondence with the coset
space $G^c/B_+$. Since ${\cal B}_+$ is a closed subalgebra of ${\bf g}\otimes
\IC$, the parameters $z^\alpha\in\IC$ define a complex structure on $\man_G$.
In this way, we can now write down coherent state path integral representations
of the character formulas of Subsection 5.1 above. The choice of
$|0\rangle\equiv|\lambda\rangle$ above as a lowest weight state ensures that
the coherent states
\beq
|z)\equiv\zeta_+|\lambda\rangle
\label{cohgendef}\eeq
are holomorphic. Note that their coherency follows from the fact
that $E_{-\alpha}|z)=z^\alpha|z)$ for $\alpha>0$.

The coherent state representation can be used to provided the one-to-one
identification of the given representation space with the coadjoint orbit of
Subsection 5.1. This is provided by the injective mapping on ${\bf g}\to\IR$
defined by $X\to\langle g|X|g\rangle$. The Kirillov-Kostant 2-form on the
coadjoint orbit is then mapped onto the flag manifold $\man_G$ as
\beq
\omega_\lambda=\langle\lambda|D(g^{-1}dg)\wedge
D(g^{-1}dg)^\dagger|\lambda\rangle
\label{kirkostflag}\eeq
and likewise for the canonical $G$-invariant orbit metric
\beq
g_\lambda=\langle\lambda|D(g^{-1}dg)\otimes D(g^{-1}dg)^\dagger|\lambda\rangle
\label{invmetricflag}\eeq
in terms of the Cartan-Maurer 1-form $g^{-1}dg$ in the representation $D(G)$.
(\ref{kirkostflag}) coincides with the appropriate symplectic potential
$\theta_\lambda$ associated with the kinetic term $\langle g|\dot
g\rangle=\langle g|\frac{d}{dt}|g\rangle$ in (\ref{charg}). Furthermore, the
matrix element $\langle g|\hat{\cal H}|g\rangle$ in (\ref{charg}) for
$\hat{\cal H}\in D(H_C)$ generates the Hamiltonian flow $(g,t)\to\e^{t\hat{\cal
H}}g$ for the action of the direct product $LS^1\otimes H_C$ on $LG$
\cite{perret}. This follows from the identity $i_V(g^{-1}dg)=g^{-1}\hat{\cal
H}g$, so that the action in (\ref{charg}) admits the usual topological form
$i_{\dot g+V}\hat\theta_\lambda$.
The K\"ahler potentials are thus given by the normalizations
\beq
\e^{F^{(h)}(z,\bar
z)}=(z|z)\equiv\langle\lambda|g|\lambda\rangle=\langle\lambda|h|\lambda\rangle
\label{kahlerpotcoh}\eeq
with the potentials $F^{(i)}(z,\bar z)$, $i=1,\dots,r$, each associated with
the Cartan generator $H_i$ in (\ref{kahlerpotcoh}) (compare eqs.
(\ref{cancohnorm}) and (\ref{r2kahlerpot})). From this it follows that the
associated K\"ahler metrics $g^{(i)}$ and symplectic potentials $\theta^{(i)}$
can be represented as coherent state matrix elements as in (\ref{thetadcoh})
and (\ref{kahlermetcoh}). In this way the kinetic term in the coherent state
path integral (\ref{charg}) coincides with the usual one induced by the
symplectic K\"ahler structures of the homogeneous space $\man_G$ and the path
integral measure becomes the loop space Liouville measure.

A particularly important aspect of the holomorphic quantization scheme above is
that it resolves some ambiguities in the localization formulas when compared
with the classical group character formulas \cite{miet}. For this, we consider
the usual symplectic line bundle $L^{(\lambda)}\to\man_G$ associated with the
principal $G$-bundle $G\to\man_G$ with connection 1-form $\theta^{(\lambda)}$
given in (\ref{potweight}). On this line bundle we then define the twisted
(covariant) Dolbeault derivative
\beq
\bar\partial_\lambda=\bar\partial+\theta^{(\lambda)}_{\bar z}
\label{twistdol}\eeq
The Riemann-Roch-Hirzebruch index theorem (the analog of the Atiyah-Singer
index theorem for the twisted elliptic Dolbeault exterior derivative)
\cite{berlinegetz,miet,stone} relates the analytical index of (\ref{twistdol})
to the topological invariant
\beq
{\rm
index}(\bar\partial_\lambda)=\int_{O_\lambda}\ch(\omega^{(\lambda)})\wedge\td
(R^{(\lambda)})
\label{RRHindex}\eeq
Furthermore, the index of the twisted Dolbeault complex is
\beq
{\rm index}(\bar\partial_\lambda)=\sum_{k=0}^n(-1)^k\dim
H^{(0,k)}(\man_G;L^{(\lambda)})
\label{arithgenus}\eeq
where we recall that
$\Lambda^{(p,q)}(\man_G;L^{(\lambda)})\equiv\Lambda^{(p,q)}\man_G\otimes
L^{(\lambda)}$. When the line bundle $L^{(\lambda)}\to\man_G$ is trivial, the
index (\ref{arithgenus}) is called the arithmetic genus of the complex manifold
$\man_G$ and the expression (\ref{RRHindex}) for it in terms of the Todd class
is called the Riemann-Roch theorem \cite{eguchi,nash}. This is the complex
analog of the Gauss-Bonnet theorem for the Euler characteristic (the index of
the DeRham complex) and it can also be generalized to higher-dimensional
complex vector bundles.

The crucial point here is that the twisted Dolbeault operator (\ref{twistdol})
annihilates the normalized coherent states $|z)\!)=|z)/(z|z)$, so that
$|z)\!)\in H^{(0,0)}(\man_G;L^{(\lambda)})$. An application of the Lichnerowicz
vanishing theorem \cite{alvarez,berlinegetz} shows that all other cohomology
groups $H^{(0,q)}(\man_G;L^{(\lambda)})$ for $q>0$ are trivial, so that the
dimension of the representation ${\cal R}_\lambda$ with highest weight vector
$\lambda$ coincides with $\dim H^{(0,0)}(\man_G;L^{(\lambda)})$ which is just
the index of the twisted Dolbeault complex,
\beq
\dim{\cal R}_\lambda=~\tr_\lambda{\bf1}=~{\rm index}(\bar\partial_\lambda)
\label{dimreplambda}\eeq
Since \cite{varadarajan}
\beq
\dim{\cal R}_\lambda=\prod_{\alpha>0}\frac{\alpha(\lambda+\rho)}{\alpha(\rho)}
\label{dimRexpl}\eeq
where $\alpha(\tilde\lambda)\equiv\sum_{i=1}^r\alpha_i\tilde\lambda_i$, we see
that the holomorphic polarization above naturally incorporates the Weyl shift
$\lambda\to\lambda+\rho$ of the highest weight vector $\lambda$.
Furthermore, in this case the Dirac operator $\bar\partial_\lambda^\dagger$ has
no zero modes and according to the arguments used in Subsection 4.10 above the
Lefschetz number
\beq
{\rm
index}_{H^{(\lambda)}}(\bar\partial_\lambda;T)=\lim_{\beta\to\infty}\tr\|\e
^{iTH^{(\lambda)}}(\e^{-\beta\bar\partial_\lambda^\dagger\bar\partial_
\lambda}-\e^{-\beta\bar\partial_\lambda\bar\partial_\lambda^\dagger})\|
\label{eqindexdol}\eeq
coincides with the character of the zero mode representation defined by
$\bar\partial_\lambda$, i.e.
\beq
{\rm index}_{H^{(\lambda)}}(\bar\partial_\lambda;T)=~{\rm
str}_\lambda\e^{iTH^{(\lambda)}}
\label{charindexrep}\eeq
The character of the representation ${\cal R}_\lambda$ is therefore the
equivariant index of the twisted Dolbeault complex.

This identification with the holomorphic properties of the complex manifold
$\man_G$ motivates another way to regularize the fluctuation determinants
discussed in Subsection 4.2. The proper way in this case to carry out the
regularization procedure is to attach different signs to the factors of
$b$ in Subsection 4.2 (see (\ref{greg})) corresponding to the holomorphic and
antiholomorphic sectors in the regularization (\ref{detblock}),(\ref{greg})
\cite{miet}. This restores the holomorphic properties of the path integral
wherein the skew-eigenvalues $(\lambda_j,-\lambda_j)$ of the block $R_V^{(j)}$
correspond, respectively, to the holomorphic and anti-holomorphic components of
the equivariant curvature 2-form. The above argument implies that the correct
way to treat the complex tangent bundle here is to then restrict to the
holomorphic component of this curvature. In doing this, the fluctuation
determinant in (\ref{fluctint}) is not under a square root in this complex
case, because now it arises from Berezin integration over {\it complex}
Grassmann variables, i.e. the localization symmetries in a K\"ahler
polarization are determined now by a larger $N=1$ ``hidden" supersymmetry. Now
the determinant (\ref{detblock}) for the case at hand is replaced with
\beq
{\det}_{\rm
hol}'\|\partial_t-R_{V^{(\lambda)}}^{(j)}\|=\prod_{k\neq0}\left(\frac{2\pi
ik}{T}-\lambda_j\right)
\label{kahlerpoldet}\eeq
where the determinant in (\ref{kahlerpoldet}) is taken only over those
eigenvalues corresponding to the holomorphic indices. To regularize
(\ref{kahlerpoldet}) properly we now have to take into account that the Dirac
operator $\partial_t-R_{V^{(\lambda)}}^{(j)}$ has an infinite number of
negative eigenvalues, i.e. its spectral asymmetry must be regularized using the
eta-invariant (\ref{atiyahetainv}). Thus we take the fluctuation regularization
factor of Subsection 4.2 to be $b=\frac{\pi}{2}$, which explicitly takes into
account the spectral asymmetry determined by the Atiyah-Patodi-Singer
eta-invariant and maintains the original symmetry of the path integral under
the ``large gauge transformations" $R_{V^{(\lambda)}}\to R_{V^{(\lambda)}}+2\pi
n/T$, $n\in\IZ$. The evaluation of the fluctuation determinant in
(\ref{fluctint}) now leads instead to the equivariant Todd class (\ref{eqtodd})
of the complex tangent bundle,
\beq
\frac{1}{{\det}'_{\rm hol}\|\partial_t-R_{V^{(\lambda)}}^{(j)}\|}=\prod_{j=1}^n
\frac{\lambda_j/2}{\sinh(T\lambda_j/2)}\e^{iT\lambda_j/2}=\frac{1}{T^n}~\td_{V^
{(\lambda)}}(TR^{(\lambda)})
\label{holdettodd}\eeq

Then, using the usual Niemi-Tirkkonen localization prescription, we arrive at
the Kirillov character formula {\it without} an explicit Weyl shift,
\beq
\tr_\lambda\e^{iTX}\sim\int_{O_\lambda}\ch_{V^{(\lambda)}}
(-iT\omega^{(\lambda)})\wedge\td_{V^{(\lambda)}}(TR^{(\lambda)})
\label{kircomplex}\eeq
which can be derived as well from the coadjoint orbit path integral over
$LO_\lambda$ (as opposed to $LO_{\lambda+\rho}$ as in
(\ref{kirchar}))\footnote{\baselineskip=12pt The analog of the Weyl shift
ambiguity for the coadjoint orbit models of the last Subsection is associated
with the appearence of the correct zero-point energy $E_0=\frac{1}{2}$ for the
harmonic oscillator Hamiltonian $\hat a^\dagger\hat a$ in certain ordering
prescriptions. See \cite{miet} for details.}. The corresponding localization
onto the critical points of the Hamiltonian $H^{(\lambda)}$ (c.f. Subsection
4.7), or equivalently the Weyl group, is
\beq
\tr_\lambda\e^{iTX}\sim\sum_{z\in\man_{V^{(\lambda)}}}
\frac{\e^{-iTH^{(\lambda)}(z,\bar z)}}{{\det}^+_{\rm hol}dV^{(\lambda)}(z,\bar
z)}~\td(dV^{(\lambda)}(z,\bar z))
\label{weylcomplex}\eeq
where we have identified the Pfaffian in the real polarization with the
determinant over the holomorphic eigenvalues of the matrix $dV^{(\lambda)}$. We
recall from Subsection 4.10 that the Atiyah-Singer index contribution to
(\ref{kirchar}) evaluates the spectral asymmetry of the zero mode
representation of $\bf g$ determined by the pertinent Dirac operator, while
the Lefschetz number coincides with the character of that representation of
the spin complex. From the formulas (\ref{kircomplex}),(\ref{eqindexdol}),
however, we see that the character of a Lie group $G$ is a Lefschetz number
related to the $G$-index theorem of the holomorphic Dolbeault complex, rather
than to the Atiyah-Singer index theorem of the spin complex. The localizations
onto the (equivariant) Todd classes hold quite generally for any phase space
path integral in a holomorphic representation. One just replaces the $\hat
A$-genus factors everywhere in Section 4 with the corresponding Todd classes.

\subsection{Spherical Phase Spaces and Quantization of Spin Systems}

We are now ready to continue with our general isometric classification and
hence work out some more explicit examples. The next case we consider is when
the phase space $\man$ has a positive constant Gaussian curvature $K>0$. In
this case the conformal factor solves the Liouville field equation
\beq
\nabla^2\varphi(z,\bar z)=-2K\e^{\varphi(z,\bar z)}
\label{liouvilleeq}\eeq
which is a completely integrable system \cite{das} whose general solutions are
\beq
\varphi(z,\bar z)=\log\left[\frac{\partial f(z)\bar\partial\bar f(\bar z)}{
\left(\frac{K}{4}+f(z)\bar f(\bar z)\right)^2}\right]
\label{s2phi}\eeq
By the essential uniqueness of maximally symmetric spaces, we know that
in this case $(\man,g)$ is isometric to the sphere $S^2$ of radius $K^{-1/2}$
with its standard round metric given in (\ref{gk}). From the transformation law
of the metric tensor $g$ and (\ref{s2phi}) it is
straightforward to work out the explicit diffeomorphism $(z,\bar z)\to(w(z,
\bar z),\bar w(z,\bar z))$ which accomplishes this isometric correspondence.

First of all, we rewrite the spherical metric in (\ref{gk}) in complex
coordinates $w,\bar w=x^1\pm ix^2$, with $x^\mu$ the spherical coordinates
defined in (\ref{gk}), to get
\beq
g_{S^2}=\frac{1}{4K}\left[\frac{\bar w^2}{1-w\bar w}dw\otimes dw+\frac{w^2}
{1-w\bar w}d\bar w\otimes d\bar w+2\left(2+\frac{w\bar w}{1-w\bar w}\right)
dw\otimes d\bar w\right]
\label{gs2w}\eeq
where $w\bar w\leq1$. If we view the unit sphere as centered in the
$x'y'$-plane in $\IR^3$ and symmetrically about the $z'$-axis, then
we can map $S^2$ onto the complex plane via the standard stereographic
projection from the south pole $z'=-1$,
\beq
w=\frac{2w'}{1+w'\bar w'}~~~~~,~~~~~z'=\sqrt{1-w\bar w}=\frac{1-w'\bar w'}
{1+w'\bar w'}
\label{stereo}\eeq
This gives a diffeomorphism of $S^2$ with the compactified plane $\IC\cup\{
\infty\}$. From (\ref{gs2w}), the metric tensor transformation law and
(\ref{s2phi}) we find after some algebra that the coordinate transformation
above must satisfy
\beq
\frac{1}{(1+w'\bar w')^2}\left(\frac{\partial w'}{\partial z}\frac{\partial\bar
w'}{\partial\bar z}+\frac{\partial w'}{\partial\bar z}\frac{\partial\bar w'}
{\partial z}\right)=K\e^{\varphi(z,\bar z)}=K\frac{\partial f(z)\bar\partial
\bar f(\bar z)}{\left(\frac{K}{4}+f(z)\bar f(\bar z)\right)^2}
\label{s2coordchange}\eeq
{}From (\ref{s2coordchange}) and (\ref{stereo}) it then follows that the
desired
coordinate transformation from $(\man,g)$ to $S^2$ with the standard round
metric (\ref{gs2w}) is given by
\beq
w(z,\bar z)=\frac{4K^{-1/2}f(z)}{1+4K^{-1}f(z)\bar f(\bar z)}
\label{ws2z}\eeq
The mapping (\ref{ws2z}) is just a generalized stereographic projection from
the south pole of $S^2$ where $f(z)$ maps $(\man,g)$ onto the entire complex
plane with the usual K\"ahler geometry of $S^2$ defined by the coordinates in
(\ref{stereo}),
\beq\new{\begin{array}{l}
g_{S^2}=4\partial\bar\partial F_{S^2}(z,\bar z)dz\otimes d\bar z=\frac{4}{
(1+z\bar z)^2}dz\otimes d\bar z\\\omega_{S^2}=2i\partial\bar\partial F_{S^2}
(z,\bar z)dz\wedge d\bar z=\frac{2i}{(1+z\bar z)^2}dz\wedge d\bar z\end{array}}
\label{s2kahler}\eeq
where the associated K\"ahler potential is
\beq
F_{S^2}(z,\bar z)=\log(1+z\bar z)
\label{s2kahlerpot}\eeq
Notice that the diffeomorphism (\ref{ws2z}) obeys $w\bar w\leq1$, as
required for $(w,\bar w)\in S^2$, and that the K\"ahler metric $g_{S^2}$ in
(\ref{s2kahler}) coincides with the original phase space geometry
(\ref{isotherm}) when $f(z)=\frac{1}{2}K^{1/2}z$ in (\ref{s2phi}) above.

{}From the general considerations of Subsection 5.2 above we know that the
Killing
vectors of the metric (\ref{gs2w}) are
\beq
V_{S^2}^w=-i\Omega w+\alpha(1-w\bar w)^{1/2}~~~~~,~~~~~V_{S^2}^{\bar w}=i
\Omega\bar w+\bar\alpha(1-w\bar w)^{1/2}
\label{s2killingw}\eeq
The Killing vectors (\ref{s2killingw}) generate the rigid rotations $w\to\e^
{i\Omega}w$ of the sphere and the quasi-translations $w\to w+\alpha(1-w\bar w)
^{1/2}$ (i.e. translations along the geodesical great circles of $S^2$), and
they together generate the Lie group $SO(3)$. Requiring the symplecticity
condition (\ref{symplcondloc}) again under the full $SO(3)$ group action
generated by (\ref{s2killingw}) on the symplectic 2-form (\ref{symplcomplex}),
we find after some algebra that
the equations (\ref{symplcondloc}) are uniquely solved by
\beq
\omega_{S^2}(w,\bar w)=1/K(1-w\bar w)^{1/2}
\label{omegas2w}\eeq
This symplectic 2-form is again the volume form associated with (\ref{gs2w}).
It is a non-trivial element of $H^2(S^2;\IZ)=\IZ$ and it coincides with the
K\"ahler classes in (\ref{s2kahler}) in the stereographic coordinates
(\ref{stereo}). We now substitute (\ref{s2killingw}) and (\ref{omegas2w}) into
the Hamiltonian equations (\ref{hameqcomplex}), which are easily solved on
$S^2$ in the $w$-coordinates above, and then apply the generalized
stereographic projection (\ref{ws2z}) to get the most general equivariant
Hamiltonian on a spherical phase space as
\beq
H_+(z,\bar z)=\frac{\Omega\left(\frac{K}{4}-f(z)\bar f(\bar z)\right)}{
\frac{K}{4}+f(z)\bar f(\bar z)}+\frac{\alpha\bar f(\bar z)+\bar\alpha f(z)}
{\frac{K}{4}+f(z)\bar f(\bar z)}+C_0
\label{hplus}\eeq

Thus, again the Riemannian geometry of the phase space $\man$ is realized (or
even determined) by the equivariant Hamiltonian systems which can be defined on
$\man$. The transformation to Darboux coordinates on $\man$,
defined as usual as those coordinates $(v,\bar v)$ in which the symplectic
2-form is locally $\omega_{S^2}=\frac{i}{2}dv\wedge d\bar v$, can be found
from the fact that $\omega_{S^2}$ is the (K\"ahler) volume form associated with
(\ref{s2phi}) and applying the tensor transformation law (\ref{tenstransf}) for
$\omega$. After some algebra we find that the local Darboux coordinates
on $\man$ are defined by the diffeomorphism $(z,\bar z)\to(v(z,\bar z),\bar
v(z,\bar z))$, where the function
\beq
v(z,\bar z)=\frac{f(z)}{\left(\frac{K}{4}+f(z)\bar f(\bar z)\right)^{1/2}}
\label{darbouxdiffs2}\eeq
maps $\man$ onto the unit disc
\beq
D^2=\{z\in\IC:z\bar z\leq1\}
\label{disc}\eeq
which is the Darboux phase space associated with a general spherical phase
space geometry. Thus, applying the transformation (\ref{darbouxdiffs2}) to
(\ref{hplus}), we see that the general Darboux Hamiltonians in the present
case are
\beq
H_+^D(z,\bar z)=\Omega z\bar z+(\bar\alpha z+\alpha\bar z)(1-z\bar z)^{1/2}
+C_0~~~;~~~z\in D^2
\label{hplusdarboux}\eeq
which correspond to the quasi-displaced harmonic oscillators
\beq
H_+^D(z,\bar z)=\Omega\left[z+a(1-z\bar z)^{1/2}\right]\left[\bar z+\bar a
(1-z\bar z)^{1/2}\right]
\label{hplusdispl}\eeq
with compactified position and momentum ranges. Thus here the criterion of a
(compact) circle action {\it cannot} be removed, in contrast to the case of
the planar geometries of Subsection 5.3 where the Darboux phase space was
the entire complex plane $\IC$. Notice that all translations in the
planar case become quasi-translations in the spherical case, which is a measure
of the presence of a curved Riemannian geometry on $\man$.

The mapping onto Darboux coordinates above shows that once again all the
general spherical Hamiltonians are holomorphic copies of each other, as they
all define the same Darboux dynamics. We shall therefore focus our attention
to the quantum dynamics defined on the phase space $S^2$ (i.e. $f(z)=K^{1/2}
z/2$ above), and for simplicity we normalize the coordinates so that now
$K=1$, i.e. $S^2$ has unit radius. First of all, we write the 3 independent
observables appearing in (\ref{hplus}) above as
\beq
J_3^{(j)}(z,\bar z)=-j\frac{1-z\bar z}{1+z\bar z}~~~,~~~J_+^{(j)}(z,\bar z)=
2j\frac{\bar z}{1+z\bar z}~~~,~~~J_-^{(j)}(z,\bar z)=2j\frac{z}{1+z\bar z}
\label{su2poissongens}\eeq
where the parameter $j$ will be specified below. Using (\ref{s2kahler}) we
define the K\"ahler 2-form
\beq
\omega^{(j)}=j\omega_{S^2}
\label{omegaj}\eeq
and working out the associated Poisson algebra of the functions
(\ref{su2poissongens})
\beq
\left\{J_3^{(j)},J_\pm^{(j)}\right\}_{\omega^{(j)}}=\pm J_\pm^{(j)}~~~~~,~~~~~
\left\{J_+^{(j)},J_-^{(j)}\right\}_{\omega^{(j)}}=2J_3^{(j)}
\label{su2poisson}\eeq
shows that they realize the $SU(2)$ (angular momentum) Lie algebra
\cite{varadarajan}. The functions (\ref{su2poissongens}) therefore generate
the Poisson-Lie group action of the $S^2$ isometry group $SO(3)$ on the
coadjoint orbit
\beq
G/H_C=SU(2)/U(1)\simeq S^3/S^1=S^2
\label{su2coadorb}\eeq
and we obtain the usual coadjoint orbit topological quantum theory by choosing
the Hamiltonian to be an element of the Cartan subalgebra ${\bf u(1)}$ of
${\bf su(2)}$. The homogeneous space $SU(2)\to SU(2)/U(1)=S^2$ is often called
the magnetic monopole bundle. For the basic representation space $W=\IC$ of
$U(1)$, the associated vector bundle $(SU(2)\times\IC)/U(1)$ is the usual
symplectic line bundle over $S^2$. The Borel-Weil-Bott wavefunctions in the
presence of a magnetic monopole take values in this bundle.

Notice that, comparing (\ref{su2poissongens}) with the stereographic
coordinates (\ref{stereo}), we see that these observables just describe the
Larmor precession of a classical spin vector of unit length $J=\pm1$. The
coadjoint orbit path integral associated with the observables
(\ref{su2poissongens}) will therefore describe the quantum dynamics of a
classical spin system, e.g. the system with Hamiltonian $H=J_3$ describes the
Pauli magnetic moment interaction between a spin $\vec J$ and a uniform
magnetic field directed along the $z$-axis. Thus in this case $S^2$ is actually
naturally the {\it configuration} space for a spin system \cite{stone}, which
has on it a natural symplectic structure and so the corresponding path integral
can be regarded as one for the Lagrangian formulation of the theory, rather
than the Hamiltonian one \cite{picken,stone}. This is also immediate from
noting that the stereographic complex coordinates above can be written as
\beq
z=\e^{-i\phi}\tan(\theta/2)
\label{stereoangle}\eeq
in terms of the usual spherical polar coordinates $(\theta,\phi)$, so that the
observable $J_3$ in (\ref{su2poissongens}) coincides with the height
function (\ref{heights2}) of $S^2$ with $a=1$ (up to an additive constant), the
K\"ahler geometry above becomes the standard round geometry of $S^2$, and the
kinetic term in the action is
\beq
\theta_\mu(x)\dot x^\mu=\cos\theta\dot\phi
\eeq
after an integration by parts over $t$. The classical partition function,
evaluated in Subsection 2.1, yields the usual Langevin formula for the
classical statistical mechanics of a spin system. Alternatively, the motion of
the precessing spin can be reduced to that of a charged particle around a
monopole which is isomorphic to the problem for the motion of an excited
diatomic molecule where the electrons are in a state with angular momentum $j$
about the axis joining the nuclei (i.e. a rigid rotator with fixed angular
momentum $j$ about its axis) \cite{stone}. It is the balance between the the
Lorentz force on the particle, due to the fictitious magnetic field
$\omega=d\theta$ of the monopole located at the center of the sphere, and the
potential force on the moment, due to the real magnetic field, that leads to
the characteristic Larmor precession about the direction of the field.

To construct a topological Hamiltonian along the lines of the theory of
Subsection 5.1, we consider an irreducible spin-$j$ representation of $SU(2)$,
where $j=\frac{1}{2},1,\frac{3}{2},2,\dots$ \cite{varadarajan}. The state
space for this representation with heighest weight $j$ is spanned by the
complete set of orthonormal basis states $|j,m\rangle$, where $m$ are the
magnetic quantum numbers with the range $m=-j,-j+1,\dots,j-1,j$. The
$SU(2)$ generators act on these states as
\beq
\hat J_3|j,m\rangle=m|j,m\rangle~~~~~,~~~~~\hat J_\pm|j,m\rangle=\sqrt{(j\mp m)
(j\pm m+1)}~|j,m\pm1\rangle
\label{jrepgendef}\eeq
Following the last Subsection, we define the $SU(2)$ coherent states by
successive applications of the raising operator $\hat J_+$ to the lowest
weight (vacuum) state $|j,-j\rangle$ \cite{faddeev,perel},
\beq
|z)=\e^{-ij\rho}\e^{z\hat J_+}|j,-j\rangle=\e^{-ij\rho}\sum_{m=-j}^j\pmatrix{
2j\cr j+m\cr}^{1/2}z^{j+m}|j,m\rangle~~~;~~~z\in\IC
\label{su2coh}\eeq
where for $n,m\in\IZ^+$ with $n\geq m$ the binomial coefficient is defined by
\beq
\pmatrix{n\cr m\cr}=\frac{n!}{m!(n-m)!}
\label{binomialcoeff}\eeq
and where the function $\rho(z,\bar z)$ is an arbitrary phase which as we shall
see is related to the function $F(z,\bar z)$ in (\ref{potweight}).
It is easily verified that then the $SU(2)$ generators (\ref{su2poissongens})
are the normalized matrix elements of the operators $\hat J_3,\hat J_\pm$ in
the
coherent states (\ref{su2coh}), respectively\footnote{\baselineskip=12pt The
dimension $\dim{\cal R}_j=2j+1$ of the spin-$j$ representation of $SU(2)$ can
also be derived from the index theorems of the last Subsection (see
(\ref{dimreplambda})).}.

The coherent states (\ref{su2coh}) are normalized as
\beq
(z_2|z_1)=(1+z_1\bar z_2)^{2j}\e^{ij[\rho(z_2,\bar z_2)-\rho(z_1,\bar z_1)]}
\label{su2cohnorm}\eeq
where we have used the binomial theorem
\beq
(x+y)^n=\sum_{k=0}^n\pmatrix{n\cr k\cr}x^ky^{n-k}
\label{binthm}\eeq
They obey the completeness relation
\beq
\int d\mu^{(j)}(z,\bar z)~|z)\!)(\!(z|={\bf1}^{(j)}
\label{su2compl}\eeq
where ${\bf1}^{(j)}$ is the identity operator in the spin-$j$ representation
of $SU(2)$ and the coherent state measure is
\beq
d\mu^{(j)}(z,\bar z)=\frac{i}{2\pi}\frac{2j+1}{(1+z\bar z)^2}dz\wedge d\bar z
\label{su2meas}\eeq
which coincides with the symplectic 2-form of the spin system above. The
identity (\ref{su2compl}) follows from a calculation analogous to that in
(\ref{cohcompl}). Note that, as explained in the last Subsection, the K\"ahler
structure is generated through the identity $(z|z)=\e^{2jF_{S^2}(z,\bar z)}$.

We want to evaluate the propagator
\beq
{\cal K}(z_2,z_1;T)=(\!(z_2|\e^{-iT\hat{\cal H}}|z_1)\!)
\label{su2prop}\eeq
for some $SU(2)$ operator $\hat{\cal H}$ given the one-to-one correspondence
between the points on the coadjoint orbit $SU(2)/U(1)=S^2\simeq\IC\cup\{
\infty\}$ and the $SU(2)$ coherent states (\ref{su2coh}). Dividing the time
interval in (\ref{su2prop}) up into $N$ segments and letting $N\to\infty$,
following the analogous steps as in Subsection 4.1 using the completeness
relation (\ref{su2compl}) we arrive at the coherent state path integral
\bd
{\cal K}(z_2,z_1;T)={\cal N}\int_{L\IR^2}\prod_{t\in[0,T]}dz(t)~d\bar z(t)~
\sqrt{\det\|\Omega^{(j)}\|}\exp\left\{j\log(1+z_2\bar z_2)+j\log(1+z_1\bar z_1)
\right.
\ed
\beq
\left.+i\int_0^Tdt~\left[\frac{ij}{1+z\bar z}\left(\bar z\dot z-z\dot{\bar z}
\right)-ij\left(\frac{\partial\rho}{\partial z}\dot z+\frac{\partial\rho}
{\partial\bar z}\dot{\bar z}\right)-H(z,\bar z)\right]\right\}
\label{su2cohpath}\eeq
where
\beq
{\cal N}=\lim_{N\to\infty}\prod_{k=1}^{N-1}\frac{2j+1}{2j\pi}
\label{su2norm}\eeq
is a normalization constant and $H(z,\bar z)$ denotes the matrix elements
(\ref{cohham}) in the coherent states (\ref{su2coh}). Here we see once again
the formal equivalence of the path integral and Liouville measures defined by
the K\"ahler polarization above. In particular, the local symplectic potential
generating the K\"ahler structures (\ref{omegaj}) are
\beq
\theta^{(j)}=\frac{ij}{1+z\bar z}\left(\bar zdz-zd\bar z\right)-ijd\rho
\label{s2symplpot}\eeq
and they coincide with the standard coherent state canonical 1-forms
(\ref{thetadcoh}). Similarly, the K\"ahler structure (\ref{s2kahler}) can be
represented in the standard coherent state form (\ref{kahlermetcoh}).

The Wess-Zumino-Witten quantization condition (\ref{wesszum}) applied to
$\omega^{(j)}$ implies that $j$ must be a half-integer, since $\int_{S^2}
\omega_{S^2}=4\pi$, corresponding to the unitary irreducible representations of
$G=SU(2)$ \cite{varadarajan}. This is the topological (or Dirac) quantization
of spin. The quantization of the magnetic quantum numbers $m$ above then
follows from an application of the semi-classical Bohr-Sommerfeld quantization
condition \cite{messiah} for the spin system. To construct a topological
quantum theory (or equivalently an integrable quantum system) as described in
Subsections 4.10 and 5.1, we need to choose the phase function $\rho(z,\bar z)$
in the above so that $i_V\theta^{(j)}=H$. This
problem was analysed in detail by Niemi and Pasanen \cite{niemipas} who showed
that it is impossible to satisfy this integrability requirement simultaneously
for all 3 of the generators in (\ref{su2poissongens}). Again, this means that
there are no $SU(2)$-invariant symplectic potentials on the sphere $S^2$.
However, such 1-forms do exist on the cylindrical representation of $SU(2)$
\cite{niemipas}, i.e. the complex plane with the origin removed, which is
conformally equivalent to the K\"ahler representation of $S^2$ above under the
transformation $z=\e^{s^1+is^2}$ which maps $(s^1,s^2)\in\IR\times S^1$ to
$z\in\IC-\{0\}$. In this latter representation, the Hamiltonian in
(\ref{su2cohpath}) can be taken to be an arbitrary linear combination of the
$SU(2)$ generators, and the coherent state path integral (\ref{su2cohpath})
determines a topological quantum field theory with $\rho=0$ in
(\ref{s2symplpot}). This is not true, however, in the K\"ahler representation
above, but we do find, for example, that the symplectic invariance condition
can be fulfilled by choosing the basis $H(z,\bar z)=J_3^{(j)}(z,\bar z)$ of
the Cartan subalgebra ${\bf u(1)}$ and $\rho(z,\bar z)=\frac{1}{2}\log(z/\bar
z)$. The ensuing topological path integral (\ref{su2cohpath}) then describes
the
quantization of spin.

To evaluate this spin partition function, we set $\rho=0$ above. Although
the ensuing quantum theory now does not have the topological form in
terms of a BRST-exact action, it still maintains the Schwarz-type topological
form described in Subsection 4.10, since the Hamiltonian then satisfies
(\ref{ivathetaf}) with $C=j$ and the function $K$ in (\ref{hkpoisson}) is
\beq
K(z,\bar z)=\frac{i}{2}\log\left(\frac{z}{\bar z}\right)
\label{kspin}\eeq
so that (\ref{su2cohpath}) is a topological path integral of the form
(\ref{schwarzpart}), i.e. the quantum theory determines a Schwarz-type
topological field theory, as opposed to a Witten-type one as above. We first
analyse the WKB localization formula (\ref{wkb}) for the coadjoint orbit path
integral (\ref{su2cohpath}). We note first of all that the boundary conditions
in (\ref{su2cohpath}) are $z(0)=z_1$ and $\bar z(T)=\bar z_2$. In particular,
the final value $z(T)$ and the initial value $\bar z(0)$ are not specified,
and the boundary terms in (\ref{su2cohpath}) ensure that with these boundary
conditions there is no boundary contribution to the pertinent classical
equations of motion
\beq
\dot z+iz=0~~~~~,~~~~~\dot{\bar z}-i\bar z=0
\label{su2eqmotion}\eeq
In general, if $z(t)$ and $\bar z(t)$ are complex conjugates of each other,
then there are no classical trajectories that connect $z(0)=z_1$ with
$\bar z(T)=\bar z_2$ on the sphere $S^2$. But if we view the path integral
(\ref{su2cohpath}) instead as a matrix element between 2 configurations in
{\it different} polarizations, then there is always the following solution to
the equations of motion (\ref{su2eqmotion}) with the required boundary
conditions for arbitrary $z_1$ and $\bar z_2$,
\beq
z(t)=z_1\e^{-it}~~~~~,~~~~~\bar z(t)=\bar z_2\e^{-i(T-t)}
\label{su2classsols}\eeq
The solution (\ref{su2classsols}) is complex, and hence $z(t)$ and $\bar z(t)$
must be regarded as independent variables. This is one of the characteristic
features behind the holomorphic quantization formalism that makes it suitable
to describe topological field theories. The trajectories (\ref{su2classsols})
are therefore regarded as describing a {\it complex} saddle-point of the path
integral \cite{ercolessi,keski,rajeev}. We shall see other forms of this
feature
later on.

Substituting the solutions (\ref{su2classsols}) into the WKB formula
(\ref{wkb}) we find the propagator\footnote{\baselineskip=12pt In this case the
fluctuation determinant in (\ref{wkb}) is regulated using the generic
non-periodic boundary conditions discussed above.}
\beq
{\cal K}(z_2,z_1;T)=\frac{(1+z_1\bar z_2\e^{-iT})^{2j}\e^{-ijT}}{(1+z_1\bar
z_1)^j(1+z_2\bar z_2)^j}
\label{su2wkbprop}\eeq
The exact propagator from a direct calculation is
\beq
(\!(z_2|\e^{-iT\hat J_3}|z_1)\!)=\frac{1}{(1+z_1\bar z_1)^j(1+z_2\bar
z_2)^j}\sum_{m=-j}^j\pmatrix{2j\cr j+m\cr}(z_1\bar z_2\e^{-iT})^{j+m}\e^{ijT}
\label{exactsu2prop}\eeq
which coincides with (\ref{su2wkbprop}) upon application of the binomial
theorem (\ref{binthm}).
In particular, setting $z_1=z_2=z$ and integrating over $z\in\IC$ using the
coherent state measure (\ref{su2meas}), we find the partition function
\beq\new{\begin{array}{ll}
Z_{SU(2)}(T)&=\int d\mu^{(j)}(z,\bar z)~(\!(z|\e^{-iT
\hat J_3}|z)\!)\\&=\int_0^\infty dr~r~\frac{(2j+1)(1+r^2\e^{-iT})^{2j}
\e^{-ijT}}{(1+r^2)^{2j}}=\frac{\sin\left(\frac{T}{2}(2j+1)\right)}{\sin
\frac{T}{2}}\end{array}}
\label{su2wkbpart}\eeq
which also coincides with the exact result
\beq
\tr_j\e^{-iT\hat J_3}=\sum_{m=-j}^j\e^{-iTm}=\frac{\e^{iTj}}{1-\e^{-iTj}}+
\frac{\e^{-iTj}}{1-\e^{iTj}}
\label{su2weylchar}\eeq
The right-hand side of (\ref{su2weylchar}) is precisely what one
anticipates from the Weyl character formula (\ref{weylchar}). The roots of
$SU(2)$ are $\alpha=\pm1$ \cite{varadarajan}, and the Cartan subalgebra is
${\bf u(1)}$ consisting of the single element $\hat J_3$. The Weyl group is
$W=\IZ_2$ and it has 2 elements, the identity map and the reflection map $
\hat J_3\to-\hat J_3$. Thus the formula (\ref{su2weylchar}) is simply the
Weyl character formula (\ref{weylchar}) for the spin-$j$ representation of
$SU(2)$.

Within the framework of the Duistermaat-Heckman theorem, the terms
summed in (\ref{su2weylchar}) are each associated with one of the poles of
the sphere $S^2$, i.e. with the critical points of the height function on
$S^2$. Indeed, since this Hamiltonian is a perfect Morse function with even
Morse indices, we expect that the Weyl character formula above coincides with
the pertinent stronger version (\ref{pathloch}) of the localization formulas.
Because of the K\"ahler structure (\ref{s2kahler}) on $S^2$ (see
(\ref{connkahler})), the Riemann moment map has the non-vanishing components
\beq
(\mu_{V^{(j)}})^z_z=-(\mu_{V^{(j)}})^{\bar z}_{\bar z}=iJ_3^{(j)}(z,\bar z)/j
\label{s2mommap}\eeq
and consequently the Dirac $\hat A$-genus is
\beq
\hat A(T\Omega_{V^{(j)}})=\frac{T}{2j}\frac{J_3^{(j)}}{\sin\left(\frac{T}
{2j}J_3^{(j)}\right)}
\label{s2ahat}\eeq
Substituting these into the localization formula (\ref{pathloch}) yields
precisely the Weyl character formula (\ref{su2weylchar}). This
localization onto the critical points of the Hamiltonian, as for the
harmonic oscillator example of Subsection 5.3, agrees with
the general arguments at the beginning of Subsection 4.6. Substituting the
stereographic projection map (\ref{stereoangle}) into the classical equations
of motion (\ref{su2eqmotion}) gives
\beq
\dot\theta\sin\theta=0~~~~~,~~~~~\dot\phi+1=0
\label{eqmotionangle}\eeq
For $T\neq2\pi n$, $n\in\IZ$, the only $T$-periodic critical trajectories
coincide with the critical points of the Hamiltonian $j(1-\cos
\theta)$, i.e. $\theta=0,\pi$, and in this case the critical point set of
the action is isolated and non-degenerate. However, for $T=2\pi n$,
$n\in\IZ$, we find $T$-periodic classical solutions for any initial value of
$\theta$ and $\phi$ in (\ref{eqmotionangle}) and the critical point set of
the classical action coincides with the original phase space $S^2$. Thus the
moduli space of classical solutions in this case is $L\man_S=S^2$, and the
localization onto this moduli space is now easily verified from (\ref{wkbdeg})
to give the correct anticipated result above. From the discussion of
Subsection 4.10, it also follows that the sum of the terms in
(\ref{su2weylchar}) describes exactly the properly normalized period group of
the symplectic 2-form $\omega^{(j)}$ on the sphere \cite{keski}, i.e. the
integer-valued surface integrals of $\omega^{(j)}$ as in (\ref{wesszum}). We
shall see in the next Section that quantizations of the propagation time $T$ as
above lead to interesting quantum theories in certain other instances of the
localization framework.

It is an instructive exercise to work out the Niemi-Tirkkonen localization
formula (\ref{niemitirkloc}) for the above dynamical system. For this we
note that, again because of the K\"ahler geometry of $S^2$, the Riemann
curvature 2-form has the non-vanishing components
\beq
R_z^z=-R_{\bar z}^{\bar z}=-i\omega^{(j)}/j
\label{curvsu2}\eeq
and so combined with (\ref{s2mommap}) we see that the equivariant
$\hat A$-genus here is
\beq
\hat A_{V^{(j)}}(TR)=\frac{T}{2j}\frac{J_3^{(j)}-\omega^{(j)}}{\sin\left(
\frac{T}{2j}(J_3^{(j)}-\omega^{(j)})\right)}
\label{s2eqahat}\eeq
The equivariant extension of $\omega^{(j)}$ is
\beq
J_3^{(j)}-\omega^{(j)}=j\left(\frac{1-z\bar z}{1+z\bar z}-\frac{2i}{(1+z\bar
z)^2}\eta\bar\eta\right)=j\left(\frac{1-z\bar z-\eta\bar\eta}{1+z\bar z+\eta
\bar\eta}\right)
\label{s2eqext}\eeq
where we have redefined the Grassmann variables $\eta^\mu\to\sqrt{i}\cdot\eta^
\mu$. The Niemi-Tirkkonen localization formula (\ref{niemitirkloc}) can then be
written as
\beq
Z_{SU(2)}(T)\sim\frac{i}{\pi T}\int_{\IR^2\otimes\Lambda^1\IR^2}dz~d\bar z~d
\eta~d\bar\eta~L(z\bar z+\eta\bar\eta)
\label{ntirks2}\eeq
where
\beq
L(y)=\frac{\frac{T}{2}\frac{1-y}{1+y}}{\sin\left(\frac{T}{2}\left(\frac{1-y}
{1+y}\right)\right)}\exp\left[-ijT\left(\frac{1-y}{1+y}\right)\right]
\label{lydef}\eeq
Using the Parisi-Sourlas integration formula \cite{parisi}
\beq
\frac{1}{\pi}\int_{\IR^2\otimes\Lambda^1\IR^2}d^2x~d\eta~d\bar\eta~L(x^2+\eta
\bar\eta)=\int_0^\infty du~\frac{dL(u)}{du}=L(\infty)-L(0)
\label{parisi}\eeq
we obtain from (\ref{ntirks2}) the partition function
\beq
Z_{SU(2)}(T)\sim\sin(Tj)/\sin(T/2)
\label{s2partbfweyl}\eeq
Introducing the Weyl shift $j\to j+\frac{1}{2}$ in (\ref{s2partbfweyl}) then
yields the correct Weyl character formula (\ref{su2wkbpart}) for $SU(2)$
\footnote{\baselineskip=12pt Of course, we could alternatively obtain the Weyl
character formula using instead the $G$-index localization formula
(\ref{kircomplex}) without having to perform this Weyl shift \cite{miet}.}.
Note that (\ref{parisi}) shows explicitly how the localization in
(\ref{ntirks2}) comes directly from the extrema of the height function at
$z=\infty$ and $z=0$.

As a final application for the above dynamical system, we examine the
quadratic localization formula (\ref{quadrloc}). Now the (degenerate)
Hamiltonian is
\beq
{\cal F}(J_3^{(j)})=(J_3^{(j)})^2=j^2\left(\frac{1-z\bar z}{1+z\bar z}\right)^2
\label{spinjquadr}\eeq
Following the same steps as above, the localization formula (\ref{quadrloc})
can be written as
\beq
Z_{SU(2)}(T|(J_3^{(j)})^2)\sim\frac{i}{\sqrt{4\pi iT}}\int_{-\infty}^\infty
\frac{d\phi_0}{\phi_0}~\int_{\IR^2\otimes\Lambda^1\IR^2}dz~d\bar z~d\eta~d\bar
\eta~L(\phi_0,z\bar z+\eta\bar\eta)
\label{s2quadrloc}\eeq
where
\beq
L(\phi_0,y)=\frac{\frac{T\phi_0}{2}\frac{1-y}{1+y}}{\sin\left(\frac{T\phi_0}{2}
\left(\frac{1-y}{1+y}\right)\right)}\exp\left[\frac{iT}{4}\phi_0^2-ijT\phi_0
\left(\frac{1-y}{1+y}\right)\right]
\label{lphiydef}\eeq
and we have redefined $\eta^\mu\to\sqrt{i/\phi_0}\cdot\eta^\mu$. Using the
Parisi-Sourlas integration formula (\ref{parisi}) again and introducing the
Weyl shift $j\to j+\frac{1}{2}$, we find
\beq\new{\begin{array}{ll}
Z_{SU(2)}(T|(J_3^{(j)})^2)&\sim\sqrt{\frac{T}{4\pi i}}\int_{-\infty}^\infty
d\phi_0~\e^{iT\phi_0^2/4}~\frac{\sin[(j+\frac{1}{2})T\phi_0]}{\sin(T\phi_0/2)}
\\&=\sum_{m=-j}^j\sqrt{\frac{T}{4\pi i}}\int_{-\infty}^\infty d\phi_0~\e^{-iT
m\phi_0}\e^{iT\phi_0^2/4}=\sum_{m=-j}^j\e^{-iTm^2}\end{array}}
\label{s2quadrchar}\eeq
which is again the correct character $\tr_j\e^{-iT\hat J_3^2}$.

Thus on a spherical phase space geometry the equivariant Hamiltonian systems
provide a rich example of the topological quantum field theories discussed
in Subsection 4.10, and they are the natural framework for the study of the
quantum properties of classical spin systems. The character formula path
integrals above describe the quantization of the harmonic oscillator on the
sphere, and therefore the only integrable quantum system, up to holomorphic
equivalence (i.e. modification by the general geometry of the phase space),
that exists within the equivariant localization framework on a general
spherical geometry is the harmonic oscillator defined on the reduced compact
phase space $D^2$.

\subsection{Hyperbolic Phase Spaces}

The situation for the case where the phase space is endowed with a
Riemannian geometry of constant negative Gaussian curvature $K<0$ parallels
that of the last Subsection, and we only therefore briefly discuss the
essential differences \cite{szsem2}. The phase space $\man$ is now necessarily
a non-compact manifold, and we can map it onto the maximally symmetric space
${\cal H}^2$, the Lobaschevsky plane (or pseudo-sphere) of constant negative
curvature, with its standard curved hyperbolic metric $g_{{\cal H}^2}$
\cite{eisen1,eisen2,helgason}. The Killing vectors of this metric have the
general form
\beq
V_{{\cal H}^2}^w=-i\Omega w+\alpha(1+w\bar w)^{1/2}~~~~~,~~~~~V_{{\cal H}^2}
^{\bar w}=i\Omega\bar w+\bar\alpha(1+w\bar w)^{1/2}
\label{h2killingw}\eeq
and they generate the isometry group $SO(2,1)$. The rest of the analysis
at the beginning of the last Subsection now carries through analogously to the
case at hand here, where we replace the $K$ factors everywhere by $-|K|$ and
the $K^{1/2}$ factors by $|K|^{1/2}$.

In particular, with these changes the generalized stereographic coordinate
transformation (\ref{ws2z}) is the same except that now the holomorphic
function $f(z)$ there maps the phase space onto the Poincar\'e disk of radius
$\frac{1}{2}|K|^{1/2}$, i.e. the disk $D^2$ with the Poincar\'e metric
\beq
g_{{\cal H}^2}=\frac{4}{(1-z\bar z)^2}dz\otimes d\bar z
\label{poincmetric}\eeq
which defines a K\"ahler geometry on the disk for which the associated
symplectic 2-form is the unique invariant volume form under the transitive
$SO(2,1)$-action. The Poincar\'e disk is the stereographic projection image
for the Lobaschevsky plane when we regard it through its embedding in $\IR^3$
as
the pseudo-sphere, so that we can represent it by pseudo-spherical
coordinates $(\tau,\phi)\in\IR\times[0,2\pi]$ as $x^1=\sinh\tau\cos\phi$,
$x^2=\sinh\tau\sin\phi$ and $z=\cosh\tau$. The stereographic projection
is again taken from the projection center $z'=-1$, and the boundary of the
Poincar\'e disc corresponds to points at infinity of the hyperboloid ${\cal
H}^2$. The pseudo-sphere itself is represented by the interior of the disk.
The explicit transformation in terms of pseudo-spherical coordinates is
\beq
z=\frac{w'}{1+z'}=\e^{-i\phi}\tanh(\tau/2)
\label{stereopseudo}\eeq
We also note here that the Poincar\'e disc is conformally equivalent to
the upper half plane $\IC^+$ via the Cayley transform $\xi\to z=(\xi-i)/(\xi
+i)$ which takes $\xi\in\IC^+$ onto the Poincar\'e disk, and the Poincar\'e
metric (\ref{poincmetric}) on the (Poincar\'e) upper-half plane is
\beq
g_{{\cal H}^2}={\rm Im}(\xi)^{-2}d\xi\otimes d\bar\xi
\label{upmetric}\eeq
The path integral over such hyperbolic geometries arises in
string theory and studies of quantum chaos \cite{bohm}.

The most general localizable Hamiltonian in a hyperbolic phase space geometry
is therefore
\beq
H_-(z,\bar z)=\frac{\Omega\left(\frac{|K|}{4}+f(z)\bar f(\bar z)\right)}{
\frac{|K|}{4}-f(z)\bar f(\bar z)}+\frac{\alpha\bar f(\bar z)+\bar\alpha f(z)}
{\frac{|K|}{4}-f(z)\bar f(\bar z)}+C_0
\label{h2eqham}\eeq
The transformation to Darboux coordinates on $\man$ is now accomplished by
the diffeomorphism
\beq
v(z,\bar z)=\frac{f(z)}{\left(\frac{|K|}{4}-f(z)\bar f(\bar z)\right)^{1/2}}
\label{h2dardiff}\eeq
which maps $\man$ onto the complement of the unit disk $\IC-{\rm int}(D^2)$
in $\IR^2$. The general Darboux Hamiltonians are therefore
\beq
H_-^D(z,\bar z)=\Omega z\bar z+(\bar\alpha z+\alpha\bar z)(1+z\bar z)^{1/2}
{}~~~;~~~z\in\IC-{\rm int}(D^2)
\label{eqhamdarh2}\eeq
We note that here there are 2 inequivalent Hamiltonians, corresponding to a
choice of ``spacelike" and ``timelike" Killing vectors, but the generic
hyperbolic Hamiltonians are again all holomorphic copies of one another, again
reducing to a quasi-displaced harmonic oscillator. However, given that
the Darboux phase space is now non-compact, we can again weaken the
requirement of a global circle action on the phase space to a semi-bounded
group action.

Considering therefore the quantum problem defined on the Poincar\'e disk
of unit radius, we write the 3 independent observables in (\ref{h2eqham}) as
\beq
S_3^{(k)}(z,\bar z)=k\frac{1+z\bar z}{1-z\bar z}~~~,~~~S_+^{(k)}(z,\bar z)
=2k\frac{\bar z}{1-z\bar z}~~~,~~~S_-^{(k)}(z,\bar z)=2k\frac{z}{1-z\bar z}
\label{su11gens}\eeq
Defining the K\"ahler 2-form $\omega^{(k)}=k\omega_{{\cal H}^2}$, we see that
the associated Poisson algebra of these observables is just the $SU(1,1)$
Lie algebra
\beq
\left\{S_3^{(k)},S_\pm^{(k)}\right\}_{\omega^{(k)}}=\pm S_\pm^{(k)}~~~~~,~~~~~
\left\{S_+^{(k)},S_-^{(k)}\right\}_{\omega^{(k)}}=-2S_3^{(k)}
\label{su11liealg}\eeq
The Hamiltonians in (\ref{h2eqham}) are therefore functions on the
coadjoint orbit
\beq
SU(1,1)/U(1)\simeq{\cal H}^2
\label{h2coadorb}\eeq
of the non-compact Lie group $SU(1,1)$, and the generators (\ref{su11gens})
are the normalized matrix elements of the $SU(1,1)$ generators in the
$SU(1,1)$ coherent states
\beq
|z)=\e^{z\hat S_+}|k,0\rangle=\sum_{n=0}^\infty\pmatrix{2k+n+1\cr n\cr}^{1/2}
z^n|k,n\rangle~~~;~~~z\in~{\rm int}(D^2)
\label{su11coh}\eeq
for the discrete irreducible representation of $SU(1,1)$ characterized by
$k=1,\frac{3}{2},2,\frac{5}{2},\dots$ \cite{perel}. The representation spaces
are now infinite-dimensional because of the non-compactness of the group
manifold of $SU(1,1)$, and the representation states $|k,n\rangle$ defined here
are the eigenstates of the generator $\hat S_3$ with eigenvalues
\beq
\hat S_3|k,n\rangle=(k+n)|k,n\rangle
\label{su11eigen}\eeq
The coherent states (\ref{su11coh}) have the normalization
\beq
(z_2|z_1)=(1-z_1\bar z_2)^{-2k}
\label{su11cohnorm}\eeq
where we have used the binomial series expansion
\beq
\frac{1}{(1-x)^n}=\sum_{m=0}^\infty\pmatrix{m+n-1\cr m\cr}x^m
\label{binseries}\eeq
which is valid for $n\in\IZ^+$ and $|x|<1$.

Again, the integrable Hamiltonian systems are obtained by taking $H=S_3^{(k)}$,
which is the height function on ${\cal H}^2$, and the corresponding coherent
state path integral describes the quantization of the harmonic oscillator on
the open infinite space ${\cal H}^2$ (and up to holomorphic equivalence
these are the only integrable systems on a general hyperbolic phase space).
It is straightforward to analyse the localization formulas for the
coherent state path integral just as in the last Subsection. For instance,
the WKB localization formula for the coadjoint orbit path integral
\beq
Z_{SU(1,1)}(T)=\int_{L{\cal H}^2}[d\cosh\tau]~[d\phi]~\exp\left\{i\int_0^Tdt~
\left(k\cosh\tau\dot\phi-k(1+\cosh\tau)\right)\right\}
\label{su11part}\eeq
can be shown to coincide with the exact Weyl character formula for $SU(1,1)$
\cite{funahashi,rajeev}
\beq
Z_{SU(1,1)}(T)=~\tr_k\e^{-iT\hat S_3}=\sum_{n=0}^\infty\e^{-iT(k+n)}=
2i\frac{\e^{-iT(k-\frac{1}{2})}}{\sin\frac{T}{2}}
\label{su11weylchar}\eeq

\subsection{Localization of Generalized Spin Models and Hamiltonian Reduction}

The explicit examples we have given thus far of the localization formalism in
both the classical and quantum cases have, for simplicity, focused on dynamical
systems with 2-dimensional phase spaces. Our main examples have been the
harmonic oscillator, where the localization is trivial because the Hamiltonian
is a quadratic function, and the spin partition function, where the exactness
of the stationary-phase approximation is a consequence of the conspiracy
between the phase space volume and energy which makes this dynamical system
resemble a harmonic oscillator. In Section 8 we shall present some true field
theoretical applications of equivariant localization, but in this Subsection
and the next we wish to overview the results which concern the exactness of the
localization formulas for some higher-dimensional coadjoint orbit models which
can be considered as generalizations of the spin models of this Section (and
the previous ones) to larger Lie groups. We have already established quite
generally that these are always examples of localizable dynamical systems, and
here we shall explicitly examine their features in some special instances.

The generalization of the classical partition function for $SU(2)$ is what is
commonly refered to as the Itzykson-Zuber integral \cite{iz}
\beq
I[X,Y;T]=\int_{U(N)}DU~\e^{iT~\tr(UXU^\dagger Y)}
\label{izintegral}\eeq
where
\beq
DU\equiv\prod_{i,j=1}^NdU_{ij}~\delta\left(\sum_{k=1}^NU_{ik}U^*_{jk}-\delta
_{ij}\right)
\label{haarUN}\eeq
is Haar measure on the group $U(N)$ of $N\times N$ unitary matrices
$U^\dagger=U^{-1}$ (or on $SU(N)=U(N)/(U(1)\times\IZ_2)$). Here
$(X,Y)=(X^\dagger,Y^\dagger)$ are Hermitian $N\times N$ matrices (i.e. elements
of the $U(N)$ Lie algebra) and which can be therefore diagonalized with
eigenvalues $x_i,y_i\in\IR$ by unitary transformations $(X,Y)\to(V^\dagger
XV,W^\dagger YW)$. By the invariance of the Haar measure in (\ref{izintegral})
under the left-right action $U\to WUV^\dagger$ of $U(N)$, we can thus assume
without loss of generality that the matrices $X$ and $Y$ in (\ref{izintegral})
are diagonal so that
\beq
I[X,Y;T]=\int_{U(N)}DU~\exp\left(iT\sum_{i,j=1}^Nx_iy_j|U_{ij}|^2\right)
\label{izintdiag}\eeq
The Itzykson-Zuber integral is a fundamental object that appears in matrix
models of string theories, low dimensional quantum gravity and
higher-dimensional lattice gauge theories \cite{fgz,kmsw,mormatrix}.

The integration over unitary matrices in (\ref{izintegral}) can be carried out
using the Duistermaat-Heckman theorem via the following observation. If we
define the Hermitian matrix
\beq
\bar\Lambda\equiv UYU^\dagger
\label{orbitvar}\eeq
then we can explicitly compute the Jacobian for the change of integration
variables $U\to \bar\Lambda$ in (\ref{izintegral}) to get \cite{fgz,mormatrix}
\beq
D\bar\Lambda\equiv\prod_{i=1}^Nd\bar\Lambda_{ii}~\prod_{1\leq j<k\leq N}d~{\rm
Re}~\bar\Lambda_{jk}~d~{\rm Im}~\bar\Lambda_{jk}=\Delta[y]^2~DU
\label{Lambdameas}\eeq
where
\beq
\Delta[y]\equiv\det_{i,j}\left[y_i^{j-1}\right]=\prod_{1\leq i<j\leq
N}\left(y_i-y_j\right)
\label{vandermonde}\eeq
is the Vandermonde determinant. The Itzykson-Zuber integral can therefore be
written as
\beq
I[X,Y;T]=\frac{1}{\Delta[y]^2}\int_{O_Y}D\bar\Lambda~\e^{iT~\tr(X\bar\Lambda)}
\label{izorbitint}\eeq
Notice that the diagonal components of $U$ do not act on a diagonal matrix $Y$
under unitary transformation. The integration in (\ref{izorbitint}) is
therefore over the coadjoint orbit of $Y$ under the action of the unitary
group, and as such it is an integral over the symmetric space $G/H_C$ where
$H_C=U(1)^N$ is the Cartan subgroup of $G=U(N)$. This can also be noted
directly from the definition (\ref{izintegral}) in which the integrand is
unchanged if $U$ is multiplied on the right by a diagonal matrix so that the
integration is really over the coset space obtained by quotienting $U(N)$ by
the subgroup of diagonal unitary matrices (this extra integration then produces
a factor $[{\rm vol}(U(1))]^N=(2\pi)^N$ in front of the coadjoint orbit
integral). This coset space has (even) dimension
\beq
\dim U(N)-\dim U(1)^N=N^2-N
\label{dimuncoset}\eeq
and $D\bar\Lambda$ is the standard symplectic measure on the coadjoint orbit.
The integral (\ref{izorbitint}) was explicitly evaluated in \cite{alek2} using
the so-called Gelfand-Tseytlin parametrization of the orbit.

To apply the Duistermaat-Heckman integration formula to the integral
(\ref{izintdiag}), we note that the extrema of the Hamiltonian
\beq
H_I[U]=\sum_{i,j=1}^Nx_iy_j|U_{ij}|^2
\label{izham}\eeq
as a function of $U\in U(N)$, given by differentiating (\ref{izham}) with
respect to $U_{ij}$ and setting it equal to zero, satisfy the stationary
conditions
\beq
\left[X,UYU^\dagger\right]=0
\label{izstatcond}\eeq
It is easy to see that the solutions of (\ref{izstatcond}) for $X,Y$ diagonal
are of the form $U=U_d\cdot P$ where $U_d$ is a diagonal matrix and $P$ is a
permutation matrix that permutes the diagonal entries of a matrix when acting
by unitary conjugation. Diagonal matrices do not contribute to the Hamiltonian
(\ref{izham}) and thus the sum over extrema in the Duistermaat-Heckman formula
is over permutations $P\in S_N$. To evaluate the pertinent fluctuation
determinants, we write $U=P\cdot\e^{iL}$ in (\ref{izham}) where $L$ is an
infinitesimal Hermitian matrix. Then Taylor expanding (\ref{izham}) to
quadratic order in $L$ we find
\beq
H_I[U=P\cdot\e^{iL}]=\sum_{i=1}^Nx_iy_{P(i)}+\frac{1}{2}\sum_{i,j=1}^N|L_{ij}|
^2(x_i-x_j)(y_{P(i)}-y_{P(j)})+\dots
\label{izhamquorder}\eeq
and the Duistermaat-Heckman formula yields
\beq\new{\begin{array}{ll}
I[X,Y;T]&=\sum_{P\in
S_N}\frac{\e^{iT\sum_ix_iy_{P(i)}}}{N!}\int_{-\infty}^\infty
\prod_{i,j}dL_{ij}~\e^{(iT/2)\sum_{i,j}|L_{ij}|^2(x_i-x_j)(y_{P(i)}-y_{P(j)})}
\\&=\left(\frac{2\pi i}{T}\right)^{N(N-1)/2}\frac{1}{N!}\sum_{P\in S_N}{\rm
sgn}(P)\frac{\e^{iT\sum_ix_iy_{P(i)}}}{\Delta[x]\Delta[y]}\\&=\left(\frac{2\pi
i}{T}\right)^{N(N-1)/2}\frac{1}{N!}\frac{\det_{i,j}\left[\e^{iTx_i
y_j}\right]}{\Delta[x]\Delta[y]}\end{array}}
\label{izformula}\eeq
where we have used
\beq
\prod_{i<j}(y_{P(i)}-y_{P(j)})=~{\rm sgn}(P)\prod_{i<j}(y_i-y_j)=~{\rm
sgn}(P)\Delta[y]
\label{vanPid}\eeq
The sign of the permutation $P$ in (\ref{vanPid}) arises from the eta-invariant
of the Hessian of (\ref{izham}).

The localization formula (\ref{izformula}), which is a special case of the
Harish-Chandra formula, was discovered by Itzykson and Zuber \cite{iz} in the
context of 2-matrix models which describe conformal matter coupled to
2-dimensional quantum gravity \cite{fgz} (e.g. the Ising model on a random
surface). It was originally derived using heat kernel and $U(N)$ group
character expansion methods \cite{iz}, and orthogonal polynomial techniques for
the corresponding 2-matrix model \cite{fgz}. In matrix models, saddle-point
approximations are always employed in the large-$N$ limit where the models
describe the relevant continuum physical theories. Notice that the classical
spin partition function (\ref{zs2eval}) is a special case of the above result
where $N=2$ and $X$ and $Y$ are both proportional to the $SU(2)$ Pauli spin
matrix
\beq
\sigma^3=\pmatrix{1&0\cr0&-1\cr}
\label{paulispin3}\eeq
In the (defining or vector) spin-$\frac{1}{2}$ representation, we can represent
an arbitrary matrix $D\in SU(2)$ (obtained from a 3-dimensional rotation matrix
in $SO(3)$) as
\beq
D=\pmatrix{-i\e^{\frac{i}{2}(\psi-\phi)}\sin\frac{\theta}{2}&i\e^{\frac{i}{2}
(\psi+\phi)}\cos\frac{\theta}{2}\cr
i\e^{-\frac{i}{2}(\psi+\phi)}\cos\frac{\theta}{2}&i\e^{-\frac{i}{2}(\psi-\phi)}
\sin\frac{\theta}{2}\cr}
\label{eulerang}\eeq
where $\theta$, $\phi$ and $\psi\in[0,4\pi]$ are the usual Euler angles. The
projection map of the principal fiber bundle $SU(2)\to SU(2)/U(1)$ is then the
Hopf map $(\theta,\phi,\psi)\in S^3\to(\theta,\phi)\in S^2$ which effectively
sets $\psi=0$ in (\ref{eulerang}). Substituting these identifications into the
Itzykson-Zuber integral (\ref{izintegral}), we recover (\ref{zs2}) and
(\ref{zs2eval}) now follows from (\ref{izformula}) (for $a=0$ in Subsection
2.1).

There are various extensions of the above generalized classical spin model
which are relevant in matrix model theories \cite{kmsw}. First of all, we can
consider a unitary matrix integral of the form
\beq
I[A;T]=\int_{U(N)}DU~\exp\left(iT\sum_{i,j=1}^NA_{ij}|U_{ij}|^2\right)
\label{izgen}\eeq
The stationary conditions for the Hamiltonian in (\ref{izgen}) are
\beq
\sum_{j=1}^NU_{ij}(A_{ji}-A_{jk})U_{jk}^\dagger=0~~~~~,~~~~~i,k=1,\dots,N
\label{statcondgen}\eeq
whose solutions are again the permutation matrices $U=P$. The saddle-point
expansion of (\ref{izgen}) thus yields (c.f. eq. (\ref{lowstatphase}))
\beq
I[A;T]\sim\sum_{P\in
S_N}\frac{\e^{iT\sum_kA_{k,P(k)}}}{\prod_{i<j}(A_{i,P(i)}+A_{j,P(j)}-A_{i,P(j)}
-A_{j,P(i)})}\left(1+{\cal O}(1/T\cdot A)\right)
\label{izgenloworder}\eeq
In general, the higher-order corrections in (\ref{izgenloworder}) to the
lowest-order stationary-phase approximation do not vanish because the
Hamiltonian in (\ref{izgen}) is not defined on any coadjoint orbit in general
but on the entire group manifold which is not even a symplectic space. The
corrections do, however, vanish in some interesting exceptions where
(\ref{izgen}) is a coadjoint orbit integral. One case is that discussed above,
namely when $A_{ij}$ is of rank 1, so that $A_{ij}=x_iy_j$. Another interesting
case is when $N=2$, so that (\ref{izgen}) is slight modification of the spin
partition function. For the case of $SU(2)$ one can check explicitly that the
leading order term in (\ref{statcondgen}) is the exact result for the integral
so that
\beq
I_{SU(2)}[A;T]\sim\frac{\e^{iT(A_{11}+A_{22})}-\e^{iT(A_{12}+A_{21})}}
{A_{11}+A_{22}-A_{12}-A_{21}}
\label{izgensu2}\eeq

The unitary matrix integrals of the form (\ref{izgen}) are generating functions
for the correlation functions in the Itzykson-Zuber model (\ref{izintegral}),
\beq
I[X,Y;T]^{k_1l_1\cdots k_pl_p}_{i_1j_1\cdots
i_pj_p}=\int_{U(N)}DU~U_{i_1j_1}\cdots
U_{i_pj_p}(U^\dagger)^{k_1l_1}\cdots(U^\dagger)^{k_pl_p}\e^{iT~\tr(UXU^\dagger
Y)}
\label{izcorrs}\eeq
which are the only non-vanishing correlators because of the $U(1)$ phase
invariance $U_{ij}\to\e^{i\theta}U_{ij}$ of the integral (\ref{izintegral}).
The evaluation of the unitary matrix integrals (\ref{izcorrs}) is very
important for matrix models of induced gauge theories and string theories
\cite{kmsw} and has been a difficult problem that has received much attention
in the last few years. For instance, using Gelfand-Tseytlin coordinates on the
group manifold, Shatashvili \cite{shatash} has shown that some of the
correlators (\ref{izcorrs}) are explicitly given by very complicated formulas,
such as
\beq\new{\begin{array}{l}
I[X,Y;T]^{k_1,1,k_2,1\cdots k_p,1}_{1,j_1,1,j_2,\cdots
1,j_p}\\=\frac{\delta_{j_1k_1}\delta_{j_2k_2}\cdots\delta_{j_pk_p}}
{(iT)^{(N-2)(N-1)}}\frac{\prod
_{\ell=1}^{N-1}\ell!}{\Delta[x]\Delta(y_2,\dots,y_N)}\int_{-\infty}^\infty\prod
_{k=1}^{N-1}d\lambda_k~\prod_{l=1}^n\frac{\prod_{q=1}^{N-1}(\lambda_q-x_{j_l})}
{\prod_{q\neq j_l}(x_q-x_{j_l})}\\~~~~~~
\times\exp\left\{iTy_1\left(\sum_{k=1}^Nx_k-\sum_{k=1}^{N-1}\lambda_k
\right)\right\}\det_{1\leq i,j-1\leq
N-1}\left[\e^{iT\lambda_iy_j}\right]\end{array}}
\label{shatashcorr}\eeq
where the delta-functions in (\ref{shatashcorr}) arise from the $U(N)$ gauge
invariance $U\to VUV^\dagger$ of the Itzykson-Zuber integral
(\ref{izintegral}).
{}From the point of view of equivariant localization theory, these types of
integrals fall into the category of the problem of developing a description of
the corrections to the Duistermaat-Heckman formula in a universal way. This
problem will be discussed in Section 7.

There is also a more geometric generalization of the spin partition function in
terms of higher-dimensional generalizations of the K\"ahler structure of $S^2$.
These examples will also introduce another interpretation of the localization
symmetries which is directly tied to the integrability properties of these spin
systems. Consider the complex $N$-dimensional projective space $\IC P^N$,
defined as the space of all complex lines through the origin in $\IC^{N+1}$. A
homogeneous K\"ahler structure on the bundle $\IC P^N\simeq S^{2N+1}/S^1$ is
obtained by the symplectic reduction of that from $S^{2N+1}$, i.e. the
restriction of the unique K\"ahler structure of $S^{2N+1}$ to $\IC P^N$ (note
that $\IC P^1=S^2$). The embedding of $S^{2N+1}$ in $\IC^{N+1}$ is defined by
the constraint
\beq
P(z,\bar z)\equiv\sum_{\mu=1}^{N+1}z^\mu\bar z^{\bar\mu}-1=0
\label{sNconstr}\eeq
and the symplectic structure on the maximally symmetric space $\IC^{N+1}$ is as
usual defined by the Darboux 2-form
\beq
\omega_D^{(N+1)}=2iJ\sum_{\mu=1}^{N+1}d\bar z^{\bar\mu}\wedge dz^\mu
\label{darbouxN+1}\eeq
The standard complex structure on $\IC P^N$ is defined by the complex
coordinates $\xi^{\mu-1}=z^\mu/z^1$ for $\mu=2,\dots,N+1$, where $z^1=\bar
z^1\neq0$ solves the constraint (\ref{sNconstr}). Solving for $z^1$ and
substituting it along with $z^\mu=\xi^\mu z^1$ into (\ref{darbouxN+1}), the
descendent symplectic structure on $\IC P^N$ from $\IC^{N+1}$ is
\beq
\omega_{N+1}^{(J)}=2iJ\left[\sum_{\mu=1}^N\frac{d\bar\xi^{\bar\mu}\wedge
d\xi^\mu}{1+\sum_{\lambda=1}^N\xi^\lambda\bar\xi^{\bar\lambda}}-
\sum_{\mu,\nu=1}^N \frac{\xi^\mu\bar\xi^{\bar\nu}d\bar\xi^{\bar\mu}\wedge
d\xi^\nu}{(1+\sum_{\lambda=1}^N\xi^\lambda\bar\xi^{\bar\lambda})^2}\right]
\label{fubinistudy}\eeq

The K\"ahler metric associated with (\ref{fubinistudy}) is usually refered to
as the Fubini-Study metric \cite{eguchi} of $\IC P^N$, and the associated
K\"ahler potential is
\beq
F_{N+1}^{(J)}=J\log\left(1+\sum_{\mu=1}^N\xi^\mu\bar\xi^{\bar\mu}\right)
\label{CPNkahlerpot}\eeq
Since the symplectic 2-forms (\ref{fubinistudy}) define non-trivial elements of
$H^2(\IC P^N;\IZ)$, for $N>1$ the spaces $\IC P^N$ are not homeomorphic to any
of the maximally symmetric spaces discussed in Subsection 5.2 above and this
symplectic manifold leads to our first example of localization on a homogeneous
space which is not maximally symmetric. This space for $N>1$ has $N$
independent isometries which are the rotations in each of the $N$ 2-planes (see
below). In the next Subsection we shall see how to explicitly relate the space
$\IC P^N$ to a (non-maximal) coadjoint orbit of $SU(N+1)$ so that, physically,
this example describes the dynamics of a particle with internal $SU(N+1)$
isospin degrees of freedom in an external magnetic field. There we shall also
encounter other examples of this situation, i.e. where the phase space is not
maximally symmetric but has a maximally symmetric subspace. For now, we shall
concentrate on the properties of the $SU(N+1)$ classical spin partition
function associated with an integrable model.

An action of the torus $T^{N+1}=(S^1)^{N+1}$ on $\IC^{N+1}$ is given by
$z^\mu\to\e^{i\theta^\mu}z^\mu$, $\theta^\mu\in[0,2\pi)$, which is generated by
the Darboux Hamiltonian
\beq
H_D^{(N+1)}(z,\bar z;\theta)=2J\sum_{\mu=1}^{N+1}\theta^\mu z^\mu\bar
z^{\bar\mu}
\label{darbouxhamN+1}\eeq
that describes the dynamics of $N+1$ independent simple harmonic oscillators
which each correspond to a conserved charge of this integrable system. The
classical partition function is of course given by a trivial Gaussian
integration
\beq
Z_D^{(N+1)}(T)=\int_{\IR^{2N+2}}\prod_{\mu=1}^{N+1}\frac{2J~dz^\mu~d\bar
z^{\bar\mu}}{\pi}~\e^{iTH_D^{(N+1)}(z,\bar
z;\theta)}=\frac{1}{(-iT)^{N+1}\prod_{\mu=1}^{N+1}\theta^\mu}
\label{partharmN+1}\eeq
which coincides with the Duistermaat-Heckman formula as always because the only
fixed point of the torus action generated by (\ref{darbouxhamN+1}) is at
$z^\mu=0$. More interesting though is the Hamiltonian $T^N$-action on $\IC
P^N$, $\xi^\mu\to\e^{i\tilde\theta^\mu}\xi^\mu$. For each $\mu$, the
Hamiltonian function which generates the circle action on the $\xi^\mu$-plane
in $\IC P^N$ is the conserved charge
\beq
I_\mu(\xi,\bar\xi;\tilde\theta)=\frac{2J\tilde\theta^\mu\xi^\mu\bar\xi^
{\bar\mu}}{1+\sum_{\nu=1}^N\xi^\nu\bar\xi^{\bar\nu}}
\label{CPNconscharge}\eeq
and an integrable Hamiltonian can be constructed as the sum of the $N$
height-functions (\ref{CPNconscharge}),
\beq
\tilde H_{N}(\xi,\bar\xi;\tilde\theta)=\sum_{\mu=1}^NI_\mu(\xi,\bar\xi
;\tilde\theta)
\label{CPNintham}\eeq

The conserved charges (\ref{CPNconscharge}) are the action variables of the
above integrable model and the associated angle variables $\phi^\mu\in[0,2\pi]$
are the usual polar angles of the $\xi^\mu$-planes (as for the $\IC^{N+1}$
example above). The Liouville-Arnold integrability of this dynamical system can
be made more explicit by considering the generalized stereographic projection
onto $\IC P^N$ in terms of the spherical coordinates $(\theta^\mu,\phi^\mu)\in
S^{2N+1}$ (where $0\leq\theta^\mu\leq\pi$),
\beq\new{\begin{array}{c}
\xi^1=\tan(\theta^1/2)\cos(\theta^2/2)\e^{-i\phi^1}~~~,~~~
\xi^2=\tan(\theta^1/2)
\sin(\theta^2/2)\cos(\theta^3/2)\e^{-i\phi^2}\\\dots\dots\\\xi^{N-1}=\tan
(\theta^1/2)\sin(\theta^2/2)\cdots\sin(\theta^{N-1}/2)
\cos(\theta^N/2)\e^{-i\phi
^{N-1}}\\\xi^N=\tan(\theta^1/2)\sin(\theta^2/2)\cdots\sin(\theta^{N-1}/2)
\sin(\theta^N/2)\e^{-i\phi^N}\end{array}}
\label{CPNstereo}\eeq
The action variables (\ref{CPNconscharge}) in these spherical coordinates are
then
\beq\new{\begin{array}{c}
I_\mu=2J\tilde\theta^\mu\sin^2(\theta^1/2)\cdots\sin^2(\theta^\mu/2)\cos^2
(\theta^{\mu+1}/2)~~~~~,~~~~~\mu<N\\I_N=2J\tilde\theta^N\sin^2(\theta^1/2)
\cdots\sin^2(\theta^{N-1}/2)\sin^2(\theta^N/2)\end{array}}
\label{CPNactionstereo}\eeq
and it is straightforward to see that the symplectic 2-form (\ref{fubinistudy})
takes the usual integrable form $dI_\mu\wedge d\phi^\mu$ (for the oscillator
above $I_\mu\sim z^\mu\bar z^{\bar\mu}$ as usual). The explicit torus action on
$\IC P^N$ generated by the Hamiltonian vector field associated with
(\ref{CPNintham}) is thus
\beq
\tilde
V_{N}=\sum_{\mu=1}^Ni\tilde\theta^\mu\left(\xi^\mu\frac{\partial}{\partial
\xi^\mu}-\bar\xi^{\bar\mu}\frac{\partial}{\partial\bar\xi^{\bar\mu}}\right)
=\sum_{\mu=1}^N\tilde\theta^\mu\frac{\partial}{\partial\phi^\mu}
\label{CPNhamvec}\eeq
This integrable system on the phase space $\IC P^N$ is therefore once again
isomorphic to a linear combination of independent harmonic oscillators as
above, generalizing the dynamics of our standard spin example.

The relation between such generalized spin systems and systems of harmonic
oscillators, and hence the exactness of the localization formulas, can be
understood from a slightly different perspective than the usual localization
symmetries provided by the underlying equivariant cohomological or
supersymmetric structures of the dynamical system. To see this, we evaluate the
classical partition function explicitly for the above dynamical system by
extending the integration over $\IC P^N$ to the whole of $\IC^{N+1}$ via a
Lagrange multiplier which enforces the constraint (\ref{sNconstr}). This leads
to
\beq\new{\begin{array}{l}
Z_N(T)\\=\int_{S^{2N+1}/S^1}\frac{\prod_{\mu=1}^N2J~d\bar
\xi^{\bar\mu}~d\xi^\mu}
{\pi^N(1+\sum_{\nu=1}^N\xi^\nu\bar\xi^{\bar\nu})^{N+1}}~\e^{iT\tilde
H_N(\xi,\bar\xi;\tilde\theta)}\\
=\int_{-\infty}^\infty\frac{d\lambda}{2\pi}~\int_{\IR^{2N+2}}\prod_{\mu=1}^
{N+1}\frac{d\bar z^{\bar\mu}~dz^\mu}{\pi}~\exp\left\{-TH_D^{(N+1)}(z,\bar
z;\theta)+i\lambda\left(\sum_{\mu=1}^{N+1}z^\mu\bar z^{\bar\mu}-1
\right)\right\}\end{array}}
\label{intlinred}\eeq
where the angular parameters in (\ref{intlinred}) are related by
$\tilde\theta^\mu=\theta^\mu-\theta^{N+1}$. The Gaussian integral over
$\IC^{N+1}$ in (\ref{intlinred}) can be performed to yield
\beq
Z_N(T)=\int_{-\infty}^\infty
d\lambda~\e^{-i\lambda}\prod_{\mu=1}^{N+1}\frac{i}{T\theta^\mu+\lambda}
\label{residueint}\eeq
The remaining integration in (\ref{residueint}) can be carried out by
continuing the integration over a large contour in the complex plane and using
the residue theorem to pick up the $N+1$ simple poles of the integrand at
$\lambda=-T\theta^\mu$, each of which have residue 1. Thus the classical
partition function is
\beq
Z_N(T)=\sum_{\mu=1}^{N+1}\e^{iT\theta^\mu}\prod_{\nu\neq\mu}\frac{i}{T(
\theta^\nu-\theta^\mu)}
\label{CPNDH}\eeq
It is readily verified that (\ref{CPNDH}) coincides the Duistermaat-Heckman
integration formula for the Hamiltonian (\ref{CPNintham}) which has precisely
$N+1$ stationary points \cite{funahashi1}. (\ref{CPNDH}) therefore represents
the $T^N$-equivariant cohomology of the manifold $\IC P^N$.

The explicit evaluation above illustrates an interesting feature of the
localization in these cases, as was first pointed out in
\cite{ercolessi,funahashi1}. This is seen in (\ref{intlinred}) -- the dynamical
system describing the $SU(N+1)$ isospin is the Hamiltonian reduction of a
larger dynamical system which is described by a quadratic Hamiltonian on the
Darboux symplectic manifold $\IR^{2N+2}$ and for which the localization is
therefore trivial. The reduction constraint function (\ref{sNconstr}) commutes
with the Darboux oscillator Hamiltonian (\ref{darbouxhamN+1}), i.e.
\beq
\left\{P(z,\bar z),H_D^{(N+1)}\right\}_{\omega_D^{(N+1)}}=0
\label{1stclassconstr}\eeq
so that the constraint function $P(z,\bar z)$ determines a first class
constraint on the Darboux dynamical system and is therefore a symmetry (or
conserved charge) of the classical dynamics. This commutativity property is in
fact the crucial mechanism for the correspondence between the integrations in
(\ref{intlinred}), as then the dynamics on $\IC^{N+1}$ is restricted to the
symplectic subspace determined by the constant values of the conserved charges
$P(z,\bar z)$ (according to the reduction theorem \cite{arnoldnov} -- see
(\ref{locnonint})). Another important feature of the reduction mechanism here
is the action of $SU(N+1)$ on $\IC^{N+1}$ defined by the usual matrix-vector
multiplication by matrices in the defining (vector) representation of
$SU(N+1)$. This defines a symplectic Hamiltonian action on the Darboux system
above \cite{ercolessi} which leaves invariant the sphere $S^{2N+1}$ and hence
restricts the dynamics to this submanifold of $\IR^{2N+2}$. From the $U(1)$
invariance of the harmonic oscillator Hamiltonian on $\IC^{N+1}$ above, this
restricts further to the coset space $\IC P^N=S^{2N+1}/U(1)$ and the conserved
charges under the mapping by the symplectic constraint function $P(z,\bar z)$
on $\IC^{N+1}$ become the components of the projection map of the generalized
Hopf bundle $U(1)\to S^{2N+1}\to\IC P^N$ \cite{ercolessi}. It is easily seen
that the critical points of the quadratic Hamiltonian in (\ref{intlinred}) in
the space of the $\lambda$ and $z$ variables coincide exactly with the poles in
(\ref{residueint}) so that (\ref{CPNDH}) coincides precisely with the
Duistermaat-Heckman formula for this Darboux system. The localization of the
isospin partition function is thus the reduction of that from $\IR^{2N+2}$.

We recall that such a reduction method in equivariant localization is also that
which is used to derive the Callias-Bott index theorems from higher-dimensional
Atiyah-Singer index theorems \cite{hietniemi} (see comments at the end of
Subsection 4.2). The above reduction procedure is a standard method of
integrating dynamical systems by associating to a given Hamiltonian system  a
related one on a lower dimensional symplectic manifold \cite{abmarsden}. In
this scheme one reduces the rank of the differential equations of motion or the
number of degrees of freedom using invariant relations and constants of the
motion onto a coadjoint orbit as above. Furthermore, in the context of generic
integrable models, there is the conjecture that every integrable system is the
Hamiltonian reduction of a larger linear dynamical system by first class
constraints \cite{zinn}. If this conjecture were true, then the localization of
generic integrable systems could be cast in another formalism giving an even
stronger connection between the Duistermaat-Heckman theorem and the
integrability properties of a dynamical system. Conversely, as the integration
formulas we have encountered also always correspond to a sort of Hamiltonian
reduction of the original dynamical model, they yield realizations of this
reduction conjecture as well as the quantum mechanical conjecture mentioned
earlier about the exactness of the semi-classical approximation for the
description of the quantum dynamics of generic integrable systems.

\subsection{Quantization of Isospin Systems}

Let us now quickly describe the quantum generalizations of the results of the
last Subsection. First, we note that, for a general group $G$, a point particle
in an irreducible representation $\cal R$ of the internal symmetry group $G$
has its time-dependent isospin vectors ${\cal R}(t)$ living in a fixed orbit of
$G$ in the adjoint representation \cite{bal}. As described earlier, this fixed
coadjoint orbit determines a unitary irreducible representation of $G$ and if
we let $\Lambda'$ denote a (time-independent) fixed fiducial point on this
orbit, then the internal isospin vectors can be written as
\beq
{\cal R}(t)=~{\rm Ad}^*(\Lambda')g(t)=g(t)\Lambda'g(t)^{-1}
\label{isospinorbit}\eeq
with $g(t)\in G$ time-dependent group elements. In the absence of any
``external" motion of the charged particle, the isospin vector is conserved,
$\dot{\cal R}(t)=0$. Thus the point particle action describing its motion in a
generalized external magnetic field $B=B^a{\cal R}(X^a)$ is \cite{bal}
\beq
S[g,\Lambda']=\int_0^Tdt~\left(~\tr(\Lambda'g^{-1}\dot g)+~\tr({\cal R}\cdot
B)\right)
\label{genspinaction}\eeq
When $\Lambda'$ is taken to be a Cartan element, this action generalizes the
spin action before which described the dynamics of a particle moving in both
monopole and external fields (Hamiltonianly reduced from a free particle action
on $\IR^4$) \cite{bal,ercolessi}. Thus again we see that the kinetic term in
(\ref{genspinaction}) is of first order in time derivatives and consequently
the action (\ref{genspinaction}) is already cast in phase space. The phase
space variables are the isospin vectors ${\cal R}(X^a)$ and the Poisson algebra
of them is just the Lie algebra of the internal symmetry group $G$ \cite{bal}.
Thus these generalized dynamical systems all fall into the class of those
systems whose configuration and phase spaces coincide so that their path
integral interpretations on the respective spaces are the same.

To describe the quantum dynamics of these models, we start with the
quantization of the classical $SU(N+1)$ isospin model of the last Subsection
over the phase space $\IC P^N$. Since \cite{rajeev}
\beq
SU(N+1)\simeq S^{2N+1}\times SU(N)\simeq S^{2N+1}\times
S^{2N-1}\times\cdots\times S^3
\label{SUN+1sphereiso}\eeq
it follows that this phase space is the coadjoint orbit
$O_{\{N,1\}}=SU(N+1)/(SU(N)\times U(1))$ of $SU(N+1)$ \cite{helgason}, and an
integrable Hamiltonian on it will be a combination of the $N$ Cartan elements
of $SU(N+1)$ which leave the maximal torus $H_C\sim(S^1)^N\subset O_{\{N,1\}}$
invariant. Although the orbit space is not a flag manifold as before, it is
still possible to generalize the construction of the coherent states in
Subsection 5.4. They are constructed by taking the highest weight vector
$|0\rangle=\sum_\ell n_\ell\lambda^{(f)}_\ell$ where $n_\ell$ is a non-negative
integer, $\lambda^{(f)}_\ell$ is a highest weight vector of the fundamental
representation, and the sum over $\ell$ is taken in such a way that the maximum
stabalizer group of $|0\rangle$ is $SU(N)\times U(1)$ (so that the loop space
path integral will be canonically associated with
$O_{\{N,1\}}=O_{\Lambda'}\simeq~{\rm Ad}^*(SU(N+1))\Lambda'\simeq
SU(N+1)/SU(N+1)_{\Lambda'}$) and so that $|0\rangle$ is appropriate to the
geometry of the given coadjoint orbit (e.g. the normalization of the coherent
states generates the K\"ahler potentials of the orbits in the manner described
in Subsection 5.4). The existence of such a weight state is in general always
guaranteed by the Borel-Weil-Bott theorem \cite{helgason,pres}.

We now define the coherent states associated with the irreducible $[2J]$
representation of the $SU(N+1)$ group, where $2J\in\IZ^+$, $2J\leq N+1$
\cite{georgi,kirillov} (i.e. the representation with fundamental highest weight
$J$ corresponding to a Young tableau representation by a single column of $2J$
boxes). In this representation, the highest weight vector is denoted as
$|J;N+1\rangle$, there are precisely $N$ simple roots $\alpha=1,2,\dots,N$
(i.e. roots which cannot be decomposed into the sum of 2 other positive roots)
with $E^{(N+1)}_\alpha|J;N+1\rangle=0$, and the Cartan generators in the
Gell-Mann basis have the (diagonal) matrix elements
\beq
[H_m]_{ij}=\frac{1}{\sqrt{2m(m+1)}}\left(\sum_{k=1}^m\delta_{ik}\delta_{jk}
-m\delta_{i,m+1}\delta_{j,m+1}\right)~~~~~,~~~~~m=1,\dots,N
\label{suN+1cartangens}\eeq
where we have normalized the generators so that
$\tr(X^aX^b)=\frac{1}{2}\delta^{ab}$. The dimension of this representation is
\beq
\dim~[2J]=\pmatrix{N+1\cr2J\cr}=\frac{(N+1)!}{(2J)!(N+1-2J)!}
\label{dim2J}\eeq
The generalized coherent states associated with this representation of
$SU(N+1)$ are then \cite{perel}
\beq\new{\begin{array}{ll}
|\xi)&\equiv\e^{\sum_{\alpha=1}^N\xi^\alpha
E^{(N+1)}_{-\alpha}}|J;N+1\rangle~~~~~,~~~~~\xi=(\xi^1,\dots,\xi^N)\in\IC
P^N\\&=\sum_{n_1+\dots+n_{N+1}=2J}\sqrt{\frac{(2J)!}{n_1!\cdots
n_{N+1}!}}~(\xi^1)^{n_1}\cdots(\xi^N)^{n_N}|\{n_k\}_{k=1}^{N+1};N+1\rangle
\end{array}}
\label{SUN+1cohstates}\eeq
These coherent states have the normalizations
\beq
(\xi_2|\xi_1)=\left(1+\sum_{\alpha=1}^N\xi_1^\alpha\bar\xi_2^
{\bar\alpha}\right)^{2J}
\label{SUN+1cohnorm}\eeq
and they obey the completeness relation
\beq
\hat P_J\equiv\int d\mu^{(J)}(\xi,\bar\xi)~|\xi)\!)(\!(\xi|={\bf1}^{(J)}
\label{PJdef}\eeq
where
\beq
d\mu^{(J)}(\xi,\bar\xi)=\frac{(2J+N)!}{(2J)!}\frac{\prod_{\alpha=1}^Nd\bar\xi
^{\bar\alpha}\wedge
d\xi^\alpha}{\pi^N(1+\sum_{\mu=1}^N\xi^\mu\bar\xi^{\bar\mu})^{N+1}}
\label{SUN+1meas}\eeq
is the associated coherent state measure.

Consider the quantum propagator
\beq
{\cal K}^{(J)}(\xi_2,\xi_1;T)=(\!(\xi_2|\e^{-iT\hat{\cal H}}|\xi_1)\!)
\label{SUN+1prop}\eeq
defined in terms of the invariant $SU(N+1)$ operator that is a linear
combination of the Cartan generators (\ref{suN+1cartangens})
\beq
\hat{\cal H}=2J\sum_{m=1}^N\tilde\theta^mH_m+2JC_0
\label{SUN+1op}\eeq
which leads as usual to a topological and integrable quantum theory. Then the
standard calculation yields the coherent state path integral
\beq\new{\begin{array}{l}
{\cal K}^{(J)}(\xi_2,\xi_1;T)\\=\int_{L(S^{2N+1}/S^1)}[d\mu^{(J)}(\xi,\bar\xi)]
{}~\exp\left\{2J\log\left(1+\sum_{\alpha=1}^N\xi_2^\alpha\bar\xi_2^{
\bar\alpha}\right)+2J\log\left(1+\sum_{\alpha=1}^N\xi_1^\alpha
\bar\xi_1^{\bar\alpha}\right)\right.\\~~~~~~\left.
+i\int_0^Tdt~\left[\sum_{\alpha=1}^N\frac{2iJ(\bar\xi^{\bar\alpha}
\dot\xi^\alpha-\xi^\alpha\dot{\bar\xi}{}^{\bar\alpha})}{1+\sum_\mu\xi^\mu\bar
\xi^{\bar\mu}}-H(\xi,\bar\xi)\right]\right\}\end{array}}
\label{SUN+1cohpathint}\eeq
where the boundary conditions in the path integral are $\xi(0)=\xi_1$ and
$\bar\xi(T)=\bar\xi_2$, and
\beq
H(\xi,\bar\xi)\equiv(\!(\xi|\hat{\cal H}|\xi)\!)=2JC_0+2J\sum_{\alpha=1}^N
\frac{\tilde\theta^\alpha\xi^\alpha\bar\xi^{\bar\alpha}}{1+\sum_\mu\xi^\mu
\bar\xi^{\bar\mu}}
\label{genheightfn}\eeq
is the generalized height function (\ref{CPNintham}) on $\IC P^N$. The WKB
approximation for the coherent state path integral has been discussed
extensively in \cite{funahashi1,oh}. In fact, all of the localization formulas
of Section 4 can be explicitly verified as the above dynamical system is just a
multi-dimensional generalization of the K\"ahler polarization for the spin
propagator in Subsection 5.5 above. The classical equations of motion are
\beq
\dot\xi^\alpha+2iJ\tilde\theta^\alpha\xi^\alpha=0~~~~~,~~~~~\dot{\bar\xi}{}^{
\bar\alpha}-2iJ\tilde\theta^\alpha\bar\xi^{\bar\alpha}=0
\label{CPNeqmotion}\eeq
whose solutions in the stereographic coordinates (\ref{CPNstereo}) are the
conditionally periodic motions
\beq
\theta^\alpha(t)=\theta^\alpha(0)~~~~~,~~~~~\phi^\alpha(t)=\phi^\alpha(0)
+2J\tilde\theta^\alpha t
\label{CPNcondperiodic}\eeq
generalizing the results of Subsection 5.5.

In any case, we arrive at the propagator
\beq
{\cal K}^{(J)}(\xi_2,\xi_1;T)=\frac{\left(1+\sum_{\alpha=1}^N\xi_1^\alpha
\bar\xi_2^{\bar\alpha}\e^{-2iJ\tilde\theta^\alpha
T}\right)^{2J}}{(1+\sum_\mu\xi_1^\mu\bar\xi^{\bar\mu}_1)^J(1+\sum_\mu\xi_2
^\mu\bar\xi_2^{\bar\mu})^J}\e^{2iJC_0T}
\label{CPNprop}\eeq
and the associated quantum partition function
\beq
Z_{SU(N+1)}(T)=\int d\mu^{(J)}(\xi,\bar\xi)~(\!(\xi|\e^{-i\hat{\cal
H}T}|\xi)\!)=\sum_{\alpha=1}^{N+1}\e^{-2iJ\tilde\theta^\alpha
T}\prod_{\beta\neq\alpha}\frac{1}{1-\e^{-2iJ(\tilde\theta^\beta-\tilde\theta^
\alpha)T}}
\label{SUN+1qupart}\eeq
where $\tilde\theta^{N+1}\equiv C_0$. These generalize the previous results for
$SU(2)$ and, in particular, (\ref{SUN+1qupart}) coincides with the anticipated
Weyl character formula for the $[2J]$ representation of $G=SU(N+1)$. What is
also interesting here is that the Hamiltonian reduction mechanism of the last
Subsection has a quantum counterpart which can be used to interpret the quantum
localization of this dynamical system. To see this, we recall from the last
Subsection that the Hamiltonian (\ref{genheightfn}) is the reduction of an
$(N+1)$-dimensional simple harmonic oscillator Hamiltonian, which in the
quantum case is the Hermitian operator
\beq
\hat{\cal H}^{(D)}=2J\sum_{\alpha=1}^{N+1}\tilde\theta^\alpha\hat
a_\alpha^\dagger\hat a_\alpha
\label{harmN+1ham}\eeq
acting on a multi-dimensional Heisenberg-Weyl algebra
$\bigoplus_{\alpha=1}^{N+1}{\bf g}_{HW}$, where $\hat a_\alpha^\dagger,\hat
a_\alpha$ are $N+1$ mutually commuting copies of the raising and lowering
operators (\ref{agens}). Relating the coordinates on $\IC^{N+1}$ and $\IC P^N$
as usual via the constraint function (\ref{sNconstr}), it is straightforward to
show that the projection operator in (\ref{PJdef}) is \cite{funahashi1}
\beq
\hat P_J=\int_0^{2\pi}\frac{d\lambda}{2\pi}~\e^{i\lambda(\sum_\alpha\hat
a_\alpha^\dagger\hat a_\alpha-2J)}=\sum_{n_1+\dots
+n_{N+1}=2J}|n_1,\dots,n_{N+1}\rangle\langle n_1,\dots,n_{N+1}|
\label{projN+1osc}\eeq
where
\beq
|n_1,\dots,n_{N+1}\rangle\equiv\frac{1}{\sqrt{n_1!\cdots n_{N+1}!}}(\hat
a_1^\dagger)^{n_1}\cdots(\hat a_{N+1}^\dagger)^{n_{N+1}}|0\rangle
\label{N+1numberbasis}\eeq
are the states of the orthonormal number basis for
$\bigoplus_{\alpha=1}^{N+1}{\bf g}_{HW}$. This identifies the weight states of
the $SU(N+1)$ representation above as
$|\{n_k\}_{k=1}^{N+1};N+1\rangle=|n_1,\dots,n_{N+1}\rangle$ in terms of the
bosonic Fock space states (\ref{N+1numberbasis}). This method of constructing
coherent states is known as the Schwinger boson formalism
\cite{fujii}--\cite{funahashi1}.

The trace formula (\ref{SUN+1qupart}) can now be represented in terms of the
canonical coherent states
\beq
|z)\!)\equiv\e^{\sum_{\alpha=1}^{N+1}\hat a_\alpha^\dagger
z^\alpha}|0\rangle/\e^{\frac{1}{2}\sum_\mu z^\mu\bar
z^{\bar\mu}}~~~,~~~z=(z^1,\dots,z^{N+1})\in\IC^{N+1}
\label{N+1cancohstates}\eeq
as
\beq
Z_{SU(N+1)}(T)=\int\frac{\prod_{\alpha=1}^{N+1}d\bar
z^{\bar\alpha}~dz^\alpha}{\pi^{N+1}}~(\!(z|\hat P_J\e^{-i\hat{\cal
H}^{(D)}T}|z)\!)
\label{N+1quparta}\eeq
Then using the important property
\beq
\left[\hat{\cal H}^{(D)},\hat P_J\right]=0
\label{impcomm0}\eeq
it is straightforward to show that (\ref{N+1quparta}) becomes \cite{funahashi1}
\beq\new{\begin{array}{l}
Z_{SU(N+1)}(T)\\=\int_0^{2\pi}\frac{d\lambda}{2\pi}~\e^{-2iJ\lambda}
\int_{L\IR^{2N+2}}\prod_{t\in[0,T]}\frac{\prod_\mu d\bar
z^{\bar\mu}(t)~dz^\mu(t)}{\pi^{N+1}}\\~~~~~\times\exp\left\{-\int_0^Tdt~
\sum_{\alpha=1}^{N+1}\left[\e^{-2iJT\tilde\theta^\alpha+i\lambda}\left(\bar
z^{\bar\alpha}\dot z^\alpha-z^\alpha\dot{\bar z}{}^{\bar\alpha}\right)-z^\alpha
\bar z^{\bar\alpha}\right]\right\}\end{array}}
\label{SUN+1harm}\eeq
where we have also used the resolution of unity for the canonical coherent
states (\ref{N+1cancohstates}). The Gaussian functional integration in
(\ref{SUN+1harm}) is now trivial to carry out and we find
\beq
Z_{SU(N+1)}(T)=\int_0^{2\pi}\frac{d\lambda}{2\pi}~\e^{-2iJ\lambda}
\prod_{\alpha=1}^{N+1}\frac{1}{1-\e^{-2iJ\tilde\theta^\alpha
T+i\lambda}}=\oint_{S^1}\frac{dw}{2\pi}~w^{2J+N}\prod_{\alpha=1}^{N+1}\frac{1}
{w-\e^{-2iJ\tilde\theta^\alpha T}}
\label{ZSUN+1cont}\eeq
where we have transformed the angular integration in (\ref{ZSUN+1cont}) into a
contour integral over the unit circle $S^1$. Carrying out the contour
integration picks up $N+1$ simple poles each of residue 1 and leads to Weyl
character formula (\ref{SUN+1qupart}).

Thus the classical Hamiltonian reduction mechanism discussed in the last
Subsection also implies localization at the quantum level, and again the terms
summed over in the WKB formula represent the singular points of the symplectic
reduction of $\IC^{N+1}$ to $\IC P^N$. In the quantum case the first class
constraint algebra (\ref{impcomm0}) is determined by the projection operator
$\hat P_J$ which coincides with the identity operator on the group
representation space when restricted to the $SU(N+1)$ coherent states as above.
This Hamiltonian reduction is always just a manifestation of the fact that
these localizable dynamical systems are always in some way just a set of
harmonic oscillators, because of the large ``hidden" supersymmetry in these
problems. The above analysis can also be straightforwardly generalized to the
non-compact hyperbolic space $D^{N,1}$ (the complex open $(N+1)$-dimensional
Poincar\'e ball \cite{helgason}) with the associated $SU(N,1)$ coherent states,
generalizing the results of Subsection 5.6 for $SU(1,1)$ \cite{funahashi1}. In
particular, the localization formulas all lead to the Weyl character formula
for $SU(N,1)$
\beq
Z_{SU(N,1)}(T)=\e^{-iC_0KT}\prod_{\alpha=1}^N\frac{1}{1-\e^{-i\tilde\theta^
\alpha T}}
\label{SUN1weylchar}\eeq

The $\IC P^N$ model above is a special case of the more general K\"ahler space
called a Grassmann manifold. Geometrically, this is defined as the
$k\cdot(N-k)$ complex-dimensional space ${\rm Gr}(N,k)$ of $k$-planes through
the origin of $\IC^N$ (note that ${\rm Gr}(N,1)=\IC P^{N-1}$). Algebraically,
it is the space of $N\times N$ Hermitian matrices obeying a quadratic
constraint,
\beq
{\rm Gr}(N,k)=\{P:P^\dagger=P,P^2=P,~\tr~P=k\}
\label{grassmfldalg}\eeq
from which it can be shown to be isomorphic to the $U(N)$ coadjoint orbit
\cite{helgason,rajeev}
\beq
{\rm Gr}(N,k)\simeq U(N)/(U(k)\times U(N-k))
\label{grassmfldcoset}\eeq
with the transitive $U(N)$ (coadjoint) action $P\to UPU^\dagger$ on ${\rm
Gr}(N,k)$. Note that the quadratic constraint in (\ref{grassmfldalg}) implies
that the operators $P$ have eigenvalues 0 or 1. They can therefore be
interpreted as fermionic occupation number operators and the trace condition in
(\ref{grassmfldalg}) can be interpreted as the total number of fermions in a
given state. The Grassmann manifolds are therefore intimately related to an
underlying free fermion theory \cite{rajeev}. The symplectic structure on
(\ref{grassmfldalg}) is the associated Kirillov-Kostant 2-form
\beq
\omega^{(N,k)}=i~\tr(PdP\wedge dP)
\label{kirkostgrass}\eeq
whose explicit form can be written using the local coordinatization provided by
the diffeomorphism (\ref{SUN+1sphereiso}) and $U(N)\simeq SU(N)\times S^1$
\cite{fujii}. Similarly, the non-compact hyperbolic analog of the Grassmann
manifold is known as the Siegel disc $D^{N,k}$ \cite{helgason}. It is defined
algebraically as
\beq
D^{N,k}=\{\tilde P:\tilde P^\dagger=\tilde P,\tilde
P^\dagger\eta_{(N,k)}P=\eta_{(N,k)}\}
\label{siegeldiscalg}\eeq
where $\eta_{(N,k)}$ is the flat Minkowski metric with diagonal elements
consisting of $N-k$ entries of $-1$ and $k$ entries of $+1$. It is isomorphic
to the coadjoint orbit
\beq
D^{N,k}\simeq U(N,k)/(U(N)\times U(k))
\label{siegeldiskalg}\eeq
of the non-compact Lie group $U(N,k)$, and its K\"ahler structure is defined by
\beq
\tilde\omega^{(N,k)}=~\tr\left(\tilde P\eta_{(N,k)}d\tilde
P\wedge\eta_{(N,k)}d\tilde P\eta_{(N,k)}\right)
\label{kirkostsiegel}\eeq

The Grassmann and Siegel spaces above are the representative spaces for
homogeneous manifolds \cite{helgason}. The coherent state quantization and
semi-classical exactness of these dynamical systems has been discussed
extensively in \cite{fujii,rajeev}. The integrable Hamiltonians on
(\ref{grassmfldalg}) are parametrized by a constant $N\times N$ Hermitian
matrix $X$ and are defined by
\beq
H_X(P)=~\tr(XP)
\label{HXgrass}\eeq
so that the partition function generalizes the Itzykson-Zuber integral
(\ref{izintegral}). Their Poisson algebra is
\beq
\left\{H_X,H_Y\right\}_{\omega^{(N,k)}}=H_{[X,Y]}
\label{grasspoissonalg}\eeq
so that the associated partition functions define topological theories. The
Duistermaat-Heckman formula
\beq
Z_{\rm cl}^{{\rm Gr}(N,k)}(T)=\left(\frac{2\pi i}{T}\right)^{k(N-k)}\sum_{1\leq
i_1<\dots<i_k\leq N}\frac{\e^{iT\sum_{l=1}^kh_{i_l}}}
{\prod_{l=1}^k\prod_{j\neq i_l}(h_j-h_{i_l})}~~~~~,
\label{grasspartcl}\eeq
with $h_i\in\IR$ the eigenvalues of the Hermitian matrix $X$ in (\ref{HXgrass})
parametrizing the Hamiltonian, has been verified for this dynamical system and
shown to be associated with a Hamiltonian reduction mechanism as described
above for the case of $\IC P^N$. The construction of coherent states via the
algebraic or Schwinger boson formalisms parallels that above for the $\IC P^N$
coherent states, although in the present case it is somewhat more involved. We
shall not go into the technical details here, but refer to \cite{fujii} where
it was also shown that the WKB localization formula
\beq
Z_{U(N)}^{{\rm Gr}(N,k)}(T)\sim\sum_{1\leq i_1<\dots<i_k\leq
N}\frac{\e^{-ikT\sum_{l=1}^kh_{i_l}}}{\prod
_{l=1}^k\prod_{j\neq
i_l}(1-\e^{-iT(h_j-h_{i_l})})}=\frac{\det_{i,j}\left[\e^{-iTh_in_j}\right]}
{\Delta[\e^{-iTh_i}]}
\label{grasspartqu}\eeq
yields the anticipated Weyl character formula in the representation defined by
the coset space (\ref{grassmfldcoset}). Here $k\in\IZ^+$ characterizes the
highest weight of the pertinent $U(N)$ representation, the non-negative
integers $n_i>n_{i+1}$ are the components of the vector $\lambda+\rho$ (defined
as $n_i=N-i+b_i$ where $b_i$ is the number of boxes in the $i$-th row of the
associated Young tableau representation), and here we have taken the basis of
the Cartan subalgebra in which $(H_i)_{jk}=\delta_{ij}\delta_{jk}$. Similarly,
the integrable Hamiltonians on the hyperbolic space (\ref{siegeldiscalg}) are
$H_{\tilde X}(\tilde P)=-~\tr(\eta_{(N,k)}\tilde X\eta_{(N,k)}\tilde P)$ with
$\tilde X$ a constant real-valued diagonal matrix.

These constructions of coherent states can also be generalized to the coadjoint
orbit spaces \cite{hahnoh,kimoh,oh1}
\beq
O_{\{n_1,n_2,\dots,n_\ell\}}\equiv SU(N)/(SU(n_1)\times\cdots\times
SU(n_\ell)\times U(1)^{\ell-1})
\label{coadorbgen}\eeq
of dimension $N^2-\sum_{i=1}^\ell n_i^2$, where $\sum_{i=1}^\ell n_i=N$ and the
dimension of the $SU(N)$ subgroup $SU(n_1)\times\dots\times SU(n_\ell)\times
U(1)^{\ell-1}$ is $N-1$. The complex structure on the orbit (\ref{coadorbgen})
is defined by the isomorphism $O_{\{n_1,n_2,\dots,n_\ell\}}\simeq
SL(N,\IC)/P_{\{n_1,n_2,\dots,n_\ell\}}$, where $SL(N,\IC)$ is the
complexification of $SU(N)$ and $P_{\{n_1,n_2,\dots,n_\ell\}}$ is a parabolic
subgroup of $SL(N,\IC)$ which is the subgroup of block upper triangular
matrices in the $(n_1+\dots+n_\ell)\times(n_1+\dots+n_\ell)$ block
decomposition of elements of $SL(N,\IC)$. Note that for $n_1=\dots=n_N=1$ the
coadjoint orbit (\ref{coadorbgen}) is the $SU(N)$ flag manifold (the maximal
orbit) with $P_{\{1,1,\dots,1\}}$ the Borel subgroup $B_N$, while for $n_1=N-1$
and $n_2=1$ non-zero only, (\ref{coadorbgen}) coincides with the $\IC P^N$
manifold discussed above (the minimal orbit). The isospin degrees of freedom on
the coadjoint orbit (\ref{coadorbgen}) are defined as \cite{fujii,oh1} (c.f.
eq. (\ref{isospinorbit}))
\beq
Q=~{\rm Ad}^*(X)g=gXg^{-1}~~~~~,~~~~~g\in SU(N)
\label{isospinSUNcoad}\eeq
where $X$ is a diagonal matrix with entries $x_i\in\IR$ satisfying
$\sum_{i=1}^Nx_i=0$ and
$x_1=x_2=\dots=x_{n_1}>x_{n_1+1}=x_{n_1+2}=\dots=x_{n_2}>\dots=x_{n_\ell-1}
=x_{n_\ell}$.

The complexification of any element $g\in SU(N)$ can be parametrized by $N$
column vectors $Z_i\in\IC^N$, $i=1,\dots,N$, with
\beq
Z_i^\dagger Z_j=\delta_{ij}~~~~~,~~~~~\det[Z_1,\dots,Z_N]=1
\label{SUNcomplstruc}\eeq
The canonical 1-form on the orbit is then
\beq
\theta^{\{n_i\}}=~\tr\left(Xg^{-1}dg\right)=i\sum_{i=1}^{N-1}J_iZ_i^\dagger
dZ_i
\label{gencoadsymplpot}\eeq
where $J_i=x_1+\dots+2x_i+\dots+x_{N-1}\geq0$ and we have used the determinant
constraint in (\ref{SUNcomplstruc}). The associated K\"ahler 2-form is
\beq
\omega^{\{n_i\}}=d\theta^{\{n_i\}}=-i\sum_{i=1}^{N-1}J_idZ_i\wedge dZ_i^\dagger
\label{kahler2formgencoad}\eeq
and the isospin element (\ref{isospinSUNcoad}) can be expressed as
\beq
Q=i\sum_{i=1}^{N-1}\left(J_iZ_iZ^\dagger_i-\frac{J_i}{N}{\bf1}\right)
\label{isospincompl}\eeq
Defining the isospin generators $Q^a=~\tr(QX^a)$ with the normalization of the
$SU(N)$ generators defined as above for the $\IC P^N$ case, it is
straightforward to see that these functions generate the $SU(N)$ Lie algebra in
the Poisson bracket generated by the symplectic structure
(\ref{kahler2formgencoad}) once the first set of orthonormal constraints in
(\ref{SUNcomplstruc}) are substituted into the above relations. These functions
generalize the angular momentum generators in the $SU(2)$ case (where the only
coadjoint orbit is the maximal one). The constraint functions in
(\ref{SUNcomplstruc}) are easily seen to be first class constraint functions
\cite{fujii,oh1} so that the coadjoint orbit is determined again as the
symplectic reduction of a larger space. These constraint functions generate the
subgroup $SU(n_1)\times\dots\times SU(n_\ell)\times U(1)^{\ell-1}$ of $SU(N)$.
The localization properties of the topological Hamiltonians generated by the
orbit functions (\ref{isospincompl}) can therefore be interpreted again in
terms of generalized sorts of harmonic oscillator dynamical systems, the
characteristic feature of the equivariant localization mechanism. The coherent
states in these generic cases are defined analogously to those over $\IC P^N$
above, with the restrictions discussed at the beginning of this Subsection.

The structure of the integrable models defined on these coset spaces has been
discussed in \cite{hahnoh,kimoh}. The Poisson-Lie relations of the $SU(N)$
symmetry of each orbit (\ref{coadorbgen}) lead to a maximal number $N^2-N$ of
mutually commuting functions, which is equal to half the (real) dimension of
the maximal coadjoint orbit. In particular, the dynamical properties of the
flag manifold $SU(N)/U(1)^{N-1}$ have been related to ideas in the theory of
non-commutative integrability. The coherent state quantization of this flag
manifold in the case of $SU(3)$ has also been worked out in detail in
\cite{hahnoh,kimoh} (see also \cite{johnson,mcmullan}). In that case, the
K\"ahler potential is \cite{picken}
\beq
F_{SU(3)}(z,\bar z)=\log\left[(1+z_1\bar z_1+z_2\bar z_2)^p(1+z_3\bar
z_3+(z_2-z_1z_3)(\bar z_2-\bar z_1\bar z_3))^q\right]
\label{SU3kahlerpot}\eeq
where $p$ and $q$ are integers. Using the usual integrable Hamiltonians $Q^3$
and $Q^8$ defined as above, it is straightforward to construct a coherent state
path integral representation for the quantum dynamics of this localizable
system and verify the localization formulas of Section 4 above. For instance,
the semi-classical approximation has been verified for the dynamical system
with topological Hamiltonian function $H=\sum_{i=\pm}\tilde\theta_iQ_i$, where
$Q_\pm=Q^3\pm\sqrt{3}Q^8$ generate the symplectic circle actions
$(z_1,z_2,z_3)\to(\e^{i\theta_1}z_1,\e^{i\theta_1}z_2,z_3)$ and
$(z_1,z_2,z_3)\to(\e^{i\theta_2}z_1,z_2,\e^{-i\theta_2}z_3)$, respectively
\cite{hahnoh,kimoh} (so that $H$ generates the symplectic action of the 2-torus
group $T^2=S^1\times S^1$ on $SU(3)/U(1)^2$). The WKB localization formula for
the quantum propagator in the coherent state representation for this integrable
spin model can then be worked out to be
\beq\new{\begin{array}{l}
{\cal K}_{SU(3)}(\bar z',z;T)\\=(1+\bar
z_1'z_1\e^{i(\tilde\theta_++\tilde\theta_-)T}+\bar
z_2'z_2\e^{i\tilde\theta_+T})^p(1+\bar z_3'z_3\e^{-i\tilde\theta_-T}+(\bar
z_2'-\bar z_1'\bar z_3')(z_2-z_1z_3)\e^{i\tilde\theta_+T})^q\end{array}}
\label{SU3quprop}\eeq
which can be shown to coincide with the exact result from a direct calculation.

Finally, we point out that one can also apply the nonabelian localization
formalisms of Subsections 3.8 and 4.9 to these generalized spin models
regarding the associated partition functions as defined on the coordinate
manifolds. As such, they can be applied to problems such as geodesic motion on
group manifolds, and in particular we can reproduce the results of Picken
\cite{picken} in the Hamiltonian framework \cite{tirk2}. More precisely, we
note that in these cases there is the natural $G$-invariant metric
\beq
g=~\tr(\Xi\otimes\Xi^\dagger)
\label{geometric}\eeq
defined on the group manifold of $G$, where $\Xi=h^{-1}dh$ is the adjoint
representation of the Cartan-Maurer 1-form which takes values in the Lie
algebra $\bf g$ of $G$. For free geodesic motion on $G$, the Hamiltonian
operator in the Schr\"odinger polarization is given by the Laplace-Beltrami
operator
\beq
\hat H^{(0)}=-\frac{1}{2}\nabla_g^2
\label{lapbelop}\eeq
with respect to the metric (\ref{geometric}). The invariant measure on $G$ is
taken to be the Riemannian volume form associated with (\ref{geometric}), and
it is possible to show that the space of trajectories over $G$ can be made into
a K\"ahler manifold by absorbing the Riemannian volume form into an effective
action in the usual way \cite{picken}.

The generic classical trajectories generated by the Hamiltonian
(\ref{lapbelop}) are straight lines in the Weyl alcove ${\cal A}=H_C/W(H_C)$ of
the Lie group $G$ because of the decomposition \cite{varadarajan}
\beq
g_1g_2^{-1}=vhv^{-1}~~~~~\forall g_1,g_2\in G
\label{alcovedecomp}\eeq
where $h\in H_C$ and the elements $v$ can be parametrized by the coset space
$G/H_C$. The geodesic distance between $g_1$ and $g_2$ defined by
(\ref{geometric}) is independent of $v$. Then the sum over extrema in the
semi-classical approximation is given by a sum over the lattice $\IZ^r$ in
$r$-dimensional Euclidean root space generated by the simple roots of $\bf g$.
The time-evolution kernel is therefore a function of the $r$-dimensional vector
$\vec a=(a_1,\dots,a_r)\in{\cal A}$ and the localization formulas for it can be
shown to yield the anticipated result \cite{dowker,picken,schul1,tirk2}
\beq
{\cal K}(\vec a;T)\sim\left(\frac{1}{2\pi iT}\right)^{\dim G/2}\sum_{\vec
k=(k_1,\dots,k_r)\in\IZ^r}\e^{i\sum_{j=1}^r(a_j+2\pi k_j)^2/2T}\prod_{\alpha>0}
\frac{\alpha(\vec a+2\pi\vec k)}{2\sin\frac{1}{2}\alpha(\vec a+2\pi\vec k)}
\label{geoprop}\eeq
where $\alpha(\vec a+2\pi\vec k)=\sum_{j=1}^r\alpha_j(a_j+2\pi k_j)$.
(\ref{geoprop}) is the configuration space analog of the Weyl character formula
(\ref{weylchar}). Applying the Poisson resummation formula \cite{marinov} leads
to a spectral expansion of the quantum propagator which is a series over the
unitary irreducible representations of $G$ given by
\beq
{\cal K}(\vec a;T)=\sum_{\lambda\in\IZ^r}\dim{\cal
R}_\lambda\left(\tr_\lambda\vec a\right)\e^{-ic(\lambda)T}
\label{geopropcharexp}\eeq
where
\beq
c(\lambda)=\sum_{i=1}^r\left((\lambda_i+\rho_i)^2-\rho_i^2\right)
\label{quadrcaseigen}\eeq
are the eigenvalues of the quadratic Casimir operator $\sum_a(X^a)^2$ in the
representation with highest weight vector $\lambda$.

Similar considerations also apply to $n$-spheres $S^n\simeq SO(n+1)/SO(n)$
\cite{marinov,mcmullan} and their hyperbolic counterparts ${\cal H}^n\simeq
SO(n-1,1)/SO(n)$ obtained by the usual analytical continuation of $S^n$
\cite{bohm}. The case of $S^2$ we saw was associated with the Dirac monopole,
that of $S^3\simeq SU(2)$ describes the emergence of spin, and $S^4$
corresponds to the BPST instanton-antiinstanton pair with 2 chiral spins
\cite{mcmullan}. Notice that these localizations also apply to the basic
integrable models which are well-known to be equivalent to the group geodesic
motion problems above, such as 2-dimensional Yang-Mills theory, supersymmetric
quantum mechanics and Calegoro-Moser type theories. These describe the quantum
mechanics of integrable models related to Hamiltonian reduction of free field
theories \cite{ercolessi,gorsky} and will be discussed again in Section 8. Note
that these free theory reductions again illustrate the isomorphism between the
localizable models and (trivial) harmonic oscillator type theories.

\subsection{Quantization on Non-homogeneous Phase Spaces}

Thus far in this Section we have examined the localizable dynamical systems on
both those phase spaces which are maximally symmetric and those with
multi-dimensional maximally-symmetric subspaces. This exhausts all spaces with
constant curvature and which are symmetric, and we outlined both the physical
and group theoretical features of these dynamical systems. In this final
Subsection of this Section we consider the final remaining
possible class of Riemannian geometries on the phase space $\man$, i.e. those
with a Gaussian curvature $K(x)$ which is a non-constant function of the
coordinates on $\man$, so that $\dim{\cal K}(\man,g)=1$. For simplicity we
restrict attention again to 2-dimensional phase spaces. The geometries which
admit only a single Killing vector are far more numerous than the maximally
symmetric or homogeneous ones and it is here that one could hope to
obtain more non-trivial applications of the localization formulas. Another
nice feature of these spaces is that the corresponding Hamiltonian Poisson
algebra will be abelian, so that the Hamiltonians so obtained will
automatically be Cartan elements, in contrast to the previous cases where the
Lie algebra ${\cal K}(\man,g)$ was non-abelian. Thus the abelian localization
formulas of the last Section can be applied straightforwardly, and the
resulting propagators will yield character formulas for the isometry group
elements defined in terms of a topological field theory type path integral
describing the
properties of integrable quantum systems corresponding to Cartan element
Hamiltonians.

Given a 1-parameter isometry group $G^{(1)}$ acting on $(\man,g)$, we begin by
introducing a set of prefered coordinates $(x'^1,x'^2)$ defined in terms of
2 differentiable functions $\chi^1$ and $\chi^2$ as described in Subsection
5.2, so that in these coordinates the Killing vector $V$ has components
$V'^1=1,V'^2=0$. For now, the function $\chi^1$ is any non-constant function
on $\man$, but we shall soon see how, once a given isometry of the dynamical
system is identified, it can be fixed to suit the given problem. For
a Hamiltonian system $(\man,\omega,H)$ which generates the flows of the
given isometry in the usual way via Hamilton's equations, the defining
condition (\ref{xprime2d}) for the coordinate function $\chi^2$ now reads
\beq
\left\{H,\chi^2\right\}_{\omega}=\lie_V\chi^2=0
\label{chi2eqham}\eeq
which is assumed to hold away from the critical point set of $H$ (i.e. the
zeroes of $V$) almost everywhere on $\man$. This means that $\chi^2$ is
a conserved charge of the given dynamical system, i.e. a $G^{(1)}$-invariant
function of action variables. In higher dimensions there would be many such
possibilities for the conserved charges depending on the integrability
properties of the system. However, in 2-dimensions this requirement fixes the
action variable to be simply a functional of the Hamiltonian $H$,
\beq
\chi^2={\cal F}(H)
\label{chi2fnham}\eeq
and so even in the non-homogeneous cases we see the intimate
connection here between the equivariant localization formalism and the
integrability of a (classical or quantum) dynamical system. We note that
this only fixes the requirement (\ref{chi2eqham}) that the coordinate
transformation function be constant along the integral curves of the Killing
vector field $V$. The isometry condition (\ref{symplcondloc}) on the
symplectic 2-form now only implies that, in the new $x'$-coordinates,
$\omega_{\mu\nu}(x')$ is independent of $x'^1$ (just as for the metric). The
Hamiltonian equations with $V'^1=1,V'^2=0$ must be solved consistently now
using (\ref{chi2fnham}) and an associated symplectic structure. Notice that
this construction is explicitly independent of the other coordinate
transformation function $\chi^1$ used in the construction of the prefered
coordinates for $V$ (c.f. Subsection 5.2).

Thus for a general metric (\ref{isotherm}) that admits a sole isometry,
the general ``admissible" Hamiltonians within the framework of equivariant
localization are given by the functionals in (\ref{chi2fnham}) determined
by the transformation $x\to x'$ to coordinates in which the (circle or
translation) action of the corresponding Killing vector is explicit. The
rich structure now arises because the integrability condition $\lie_V\omega=0$
for the Hamiltonian equations does not uniquely determine the symplectic 2-form
$\omega$, as it did in the case of a homogeneous symmetric geometry. The above
construction could therefore be started with any {\it given} symplectic 2-form
obeying this requirement, with the hope of being able to analyse quite general
classes of Hamiltonian systems. This has the possibility of largely expanding
the known examples of quantum systems where the Feynman path integral could be
evaluated exactly, in contrast to the homogeneous cases where we saw
that there was only a small number of few-parameter Hamiltonians which fit the
localization framework. However, it has been argued that the set of Hamiltonian
systems in general for which the localization criteria apply is still rather
small \cite{dlr,szsem2}. For instance, we could from the onset take $\omega$ to
be the Darboux 2-form on $\man=\IR^2$ and hope to obtain localizable examples
of 1-dimensional quantum mechanical problems with static potentials. These
are defined by the Darboux Hamiltonians
\beq
H_{QM}(p,q)=\frac{1}{2}p^2+U(q)
\label{qmham}\eeq
where $U(q)$ is some potential which is a $C^\infty$-function of the
position $q\in\IR^1$. It was Dykstra, Lykken and Raiten \cite{dlr} who first
pointed out that the formalism in Section 4 above, which naively seems
like it would imply the exact solvability of any phase space path
integral, does not work for arbitrary potentials $U(q)$.

To see this, we consider a generic potential $U(q)$ which is bounded from
below. By adding an irrelevant constant to the Hamiltonian (\ref{qmham})
if necessary, we can assume that $U(q)\geq0$ without loss of generality.
We introduce a ``harmonic" coordinate $y\in\IR$ and polar coordinates $(r,
\theta)\in\IR^+\times S^1$ by
\beq
p=r\sin\theta~~~~~,~~~~~U(q)=\frac{1}{2}y^2=\frac{1}{2}r^2\cos^2\theta
\label{harmpolar}\eeq
where we further assume that $U(q)$ is a monotone function.
In these coordinates the Hamiltonian (\ref{qmham}) takes the usual integrable
harmonic oscillator form $H=\frac{1}{2}r^2$, so that the function $\chi^2$
above defines the radial coordinate $r$ in (\ref{harmpolar}) and
${\cal F}(H)=\sqrt{2H}$ in (\ref{chi2fnham}). The Hamiltonian vector field
in these polar coordinates has the single non-vanishing component
\beq
V^\theta=-\frac{dy}{dq}
\label{vtheta}\eeq
The metric tensor (\ref{isotherm}) will have in general have 3 components
$g_{rr}$, $g_{\theta\theta}$ and $g_{\theta r}$ under the coordinate
transformation (\ref{harmpolar}), and the Killing equations (\ref{liegcoord})
become
\beq
V^\theta\partial_\theta g_{\theta\theta}+2g_{\theta\theta}\partial_\theta
V^\theta=0~~~,~~~\partial_\theta(g_{r\theta}V^\theta)+g_{\theta\theta}\partial
_r V^\theta=0~~~,~~~V^\theta\partial_\theta g_{rr}+2g_{r\theta}\partial_r
V^\theta=0
\label{killingrtheta}\eeq
The 3 equations in (\ref{killingrtheta}) can be solved in succession by
integrating them and the general solution has the form
\beq
g_{\theta\theta}=\frac{f(r)}{(V^\theta)^2}~~~,~~~g_{r\theta}=\frac{f(r)}{V^
\theta}
\int_{\theta_0}^\theta d\theta'~\partial_r\left(\frac{1}{V^{\theta'}}\right)+
\frac{h(r)}{V^\theta}~~~,~~~g_{rr}=\frac{(V^\theta)^2}{f(r)}g_{r\theta}^2+k(r)
\label{metsolns}\eeq
where $f(r)$, $h(r)$ and $k(r)$ are arbitrary $C^\infty$-functions that are
independent of the angular coordinate $\theta$.

Note that, as expected, there is no unique solution for the conformal
factor $\varphi$ in (\ref{isotherm}), only the requirement that it be
radially symmetric (i.e. independent of $\theta$). However, the equations
(\ref{metsolns}) impose a much stronger requirement, this time on the actual
coordinate transformation (\ref{harmpolar}). If we impose the required
single-valuedness property on the metric components above, then the
requirement that $g_{r\theta}(r,\theta)=g_{r\theta}(r,\theta+2\pi)$ is
equivalent to the condition
\beq
\frac{\partial}{\partial r}\int_0^{2\pi}\frac{d\theta}{V^\theta}=0
\label{singlecond}\eeq
or equivalently that
\beq
\int_0^{2\pi}d\theta~\frac{dq}{dy}=~{\rm constant}
\label{constcond}\eeq
However, the only solution to (\ref{constcond}) is when the function $\frac{
dq}{dy}$ is independent of the radial coordinate $r$, which from
(\ref{harmpolar}) is possible only when $y=-q$, so that $U(q)=\frac{1}{2}q^2$
and $H_{QM}$ is the harmonic oscillator Hamiltonian. Thus, with the
exception of the harmonic oscillator, equivariant localization fails for
all 1-dimensional quantum mechanical Hamiltonians with static potentials which
are bounded below, due to the non-existence of a single-valued metric
satisfying the Lie derivative constraint in this case.

It is instructive to examine the localization formulas for the harmonic
oscillator, which is considered trivial from the
point of view of localization theory, to see what role is played by the degree
of freedom remaining in the metric tensor which is not determined by the
equivariant localization constraints. The Hamiltonian vector field
(\ref{vtheta}) in this case is $V^\theta=1$ which
generates a global $S^1$-action on $\man=\IR^2$ given by translations of the
angle coordinate $\theta$. Thus the localization formulas should
be exact for the harmonic oscillator using any radially symmetric geometry
(\ref{isotherm}) to make manifest the localization principle. This is
certainly true of the WKB formula (\ref{wkb}) which does not involve the
metric tensor at all, but the more general localization formulas, such as
the Niemi-Tirkkonen formula (\ref{niemitirkloc}), are explicitly metric
dependent through, e.g. the $\hat A$-genus terms, although not manifestly so.
Explicitly, the non-vanishing components of the metric tensor (\ref{isotherm})
under the coordinate transformation (\ref{harmpolar}) in the case at hand are
\beq
g_{rr}=\e^{\varphi(r)}~~~~~,~~~~~g_{\theta\theta}=r^2\e^{\varphi(r)}
\label{metricrtheta}\eeq
and it is straightforward to work out the Riemann moment map and curvature
tensor which with $V^\theta=1$ lead to the non-vanishing components
\beq
(\Omega_V)_{\theta r}=-(\Omega_V)_{r\theta}=\frac{r}{2}\e^{\varphi(r)}\left(2+r
\frac{d\varphi(r)}{dr}\right)~~~,~~~R_{\theta r\theta r}=-\frac{1}{2}
(\Omega_V)_{\theta r}\frac{d}{dr}\log\lambda(r)
\label{momcurvphi}\eeq
where we have introduced the function
\beq
\lambda(r)=\e^{-\varphi(r)}(\Omega_V)_{\theta r}/2r
\label{lambdar}\eeq

Substituting the above quantities into the Niemi-Tirkkonen formula
(\ref{niemitirkloc}) with $\omega_{r\theta}=r$ and working out the Grassmann
and $\theta$ integrals there, after some algebra we find the following
expression for the harmonic oscillator partition function,
\beq
Z_{\rm harm}(T)\sim\frac{1}{i}\int_0^\infty dr~\frac{d}{dr}\left(\frac{\lambda
(r)}{\sin T\lambda(r)}\e^{-iTr^2/2}\right)=\frac{1}{i}\lim_{r\to0}\frac{
\lambda(r)}{\sin T\lambda(r)}
\label{harmpartnonhom}\eeq
Comparing with (\ref{harmpart}), we see that this result coincides with the
exact result for the harmonic oscillator partition function only if the
function (\ref{lambdar}) behaves at the origin $r=0$ as
\beq
\lim_{r\to0}\lambda(r)=\frac{1}{2}
\label{lambdar0}\eeq
which using (\ref{momcurvphi}) and (\ref{lambdar}) means that the phase space
metric must satisfy, in addition to the radial symmetry constraint, the
additional constraint
\beq
\lim_{r\to0}r\frac{d}{dr}\varphi(r)=0
\label{radmetcond}\eeq
The requirement (\ref{radmetcond}) means that the conformal factor $\varphi(r)$
of the Riemannian geometry must be an analytic function of $r$ about $r=0$,
and this restriction on the general form of the metric (\ref{isotherm}) (i.e.
on the functional properties of the conformal factor $\varphi$) ensures that
the partition function is independent of this phase space metric, as it should
be.

This analyticity requirement, however, simply means that the metric should
be chosen so as to eliminate the singularity at the origin of the coordinate
transformation to polar coordinates $(r,\theta)$ on the plane. That this
transformation is singular at $p=q=0$ is easily seen by computing the Jacobian
for the change of variables (\ref{harmpolar}) with the harmonic oscillator
potential (or by noting that $\omega$ and $g$ are degenerate at $r=0$ in these
coordinates). Since the equivariant Atiyah-Singer index which appears as the
Niemi-Tirkkonen formula for the quantum mechanical path integral is an integral
over characteristic classes, it is manifestly invariant under $C^\infty$
deformations of the metric on $\man$. The transformation to polar coordinates
is a diffeomorphism only on the punctured plane $\IR^2-\{0\}$, which destroys
the manifest topological invariance of the partition function (at $r=0$
anyway).
For $\lambda(0)\neq0$, the metric tensor describes a conical geometry
\cite{miet} for which the parameter $\lambda(0)$ represents the tip angle of
the cone. This example shows that for the localization to work the choice of
metric tensor is not completely arbitrary, since it has to respect the topology
of the phase space on which the problem is defined. As discussed in \cite{dlr}
and \cite{szsem2}, this appears to be a general feature of the generalized
localization formalisms, in that they detect explicitly the topology of the
phase space and this can be used to eliminate some of the arbitrariness of the
metric (\ref{isotherm}). Indeed, in the set of prefered coordinates for $V$ it
has no zeroes and so the critical points are ``absorbed" into the symplectic
2-form $\omega$ and in general also the metric $g$. Thus the prefered
coordinate transformation for $V$ is a diffeomorphism only on $\man-\man_V$
in general. Nonetheless, this simple example illustrates that quite
general, non-homogeneous geometries can still be used to carry out the
equivariant localization framework for path integrals and describe the
equivariant Hamiltonian systems which lead to topological quantum theories
in terms of the generic phase space geometry.

Although the above arguments appear to have eliminated a large number of
interesting physical problems, owing to the fact that their Hamiltonian vector
fields do not generate well-defined orbits on the $\theta$-circle, it is still
possible that quantum mechanical Hamiltonians with {\it unbounded} static
potentials could fit the localization framework. Such dynamical systems indeed
do represent a rather large class of physically interesting quantum systems.
The first such attempt was carried out by Dykstra, Lykken and Raiten \cite{dlr}
who showed that the Niemi-Tirkkonen localization formula for such models can be
reduced to a relatively simple contour integral. For example, consider the
equivariant localization formalism applied to the
1-dimensional hydrogen atom Hamiltonian \cite{loudon}
\beq
H_h(p,q)=\frac{1}{2}p^2-\frac{1}{|q|}
\label{hydrham}\eeq
The eigenvalues of the associated quantum Hamiltonian form a discrete spectrum
with energies
\beq
E_n=-1/2n^2~~~~~,~~~~~n=1,2,\dots
\label{hydrspec}\eeq
which resembles the bound state spectrum of the more familiar 3-dimensional
hydrogen atom \cite{messiah}. What is even more interesting about this
dynamical system is that the classical bound state orbits all coalesce at
the phase space points $q=0$, $p=\pm\infty$ on $\IR^2$, so that a localization
onto classical trajectories (like the WKB formula) is highly unsuitable for
this quantum mechanical problem. This problem could therefore provide an
example wherein although the standard WKB approximation cannot be employed,
the more general localization formulas, like the Niemi-Tirkkonen formula,
which seem to have no constraints on them other than the usual isometry
restrictions on the phase space $\man$, could prove of use in describing the
exact quantum theory of the dynamical system.

The key to evaluating the localization formulas for the Darboux Hamiltonian
(\ref{hydrham}) is the transformation to the hyperbolic
coordinates $(r,\tau)$ with $-\infty\leq r,\tau\leq\infty$,
\beq
p=|r|\sinh\tau~~~~~,~~~~~q=2/r|r|\cosh^2\tau
\label{hypercoord}\eeq
so that the Hamiltonian is again $H_h=-\frac{1}{2}r^2$ and the Hamiltonian
vector field has the single non-vanishing component
\beq
V^\tau=-\frac{1}{4}r^3\cosh^3\tau
\label{vtau}\eeq
Now the Killing equations have precisely the same form as in
(\ref{killingrtheta}), with $(r,\theta)$ replaced by $(r,\tau)$ there, and thus
the general solutions for the metric tensor have precisely the same form as in
(\ref{metsolns}). However, because of the non-compact range of the hyperbolic
coordinate $\tau$ in the case at hand, we do not encounter a single-valuedness
problem in defining the components $g_{r\tau}$ as $C^{\infty}$ functions on
$\IR^2$ and from (\ref{metsolns}) and (\ref{vtau}) we find that it is given
explicitly by the perfectly well-defined function
\beq
g_{r\tau}=\frac{12f(r)}{r^4V^\tau}\left(\frac{\sinh\tau}{2\cosh^2\tau}+
\frac{1}{2}\arctan(\sinh\tau)\right)+\frac{h(r)}{V^\tau}
\label{grtau}\eeq

In the context of our isometry analysis above, we again choose the coordinate
transformation function $\chi^2$ so that ${\cal F}(H)=\sqrt{-2H}$ in
(\ref{chi2fnham}). The other coordinate function $\chi^1\equiv x''^1$ is
determined by noting that the above $(r,\tau)$ coordinates are the
$x'$-coordinates in (\ref{xpprimecoords}) from which we wish to define the
prefered set of $x''$-coordinates for the Hamiltonian vector field $V$. There
we identify $(x'^1,x'^2)=(\tau,r)$ according to that prescription.
Carrying out the explicit integration over $x'^1=\tau$ using (\ref{vtau}), and
then substituting in the transformation (\ref{hypercoord}) back to the original
Darboux coordinates, after some algebra we find
\beq
\chi^1(p,q)=-\left|\frac{2}{|q|}-p^2\right|^{-3/2}\left[p|q|\left|\frac{2}
{|q|}-p^2\right|^{1/2}+2\arctan\left(\frac{p}{\left|\frac{2}{|q|}-p^2
\right|^{1/2}}\right)\right]
\label{chi1hydr}\eeq
Thus the Hamiltonian (\ref{hydrham}) is associated with the phase space metric
tensor (\ref{isotherm}) which is invariant under the translations $\chi^1\to
\chi^1+a_0$ of the coordinate (\ref{chi1hydr}). The analysis above shows
explicitly that the phase space indeed does admit a globally well-defined
metric which is translation invariant in the variable (\ref{chi1hydr}). It is
also possible to evaluate the Niemi-Tirkkonen localization formula for this
quantum problem in a similar fashion as the harmonic oscillator example above.
We shall not go into this computation here, but refer to \cite{dlr} for the
technical details. The only other point we wish to make here is that one
needs to impose again certain regularity requirements on the
conformal factor of the metric (\ref{isotherm}). These conditions are far more
complicated than above because of the more complicated form of the translation
function (\ref{chi1hydr}), but they are again associated with the cancelling of
the coordinate singularities in (\ref{hypercoord}) which make the equivariant
Atiyah-Singer index in (\ref{niemitirkloc}) an explicitly metric dependent
quantity. With these appropriate geometric restrictions it is enough to
argue that the quantum partition function for the Darboux
Hamiltonian (\ref{hydrham}) has the form \cite{dlr}
\beq
Z_h(T)\sim\sum_{n=1}^\infty\e^{iT/2n^2}
\label{hydrpart}\eeq
which from (\ref{hydrspec}) we see is indeed the exact spectral propagator for
the 1-dimensional hydrogen atom \cite{loudon}.

This example shows that more complicated quantum systems can be studied
within the equivariant localization framework on a simply connected phase
space, but only for those phase spaces which admit Riemannian geometries
which have complicated and unusual symmetries, such as translations in the
coordinate (\ref{chi1hydr}) above. Thus besides having to find a metric tensor
appropriate to the geometry and topology of a phase space, there is the further
general problem as to whether or not a geometry can in fact possess the
required symmetry (e.g. for Hamiltonians associated with bounded potentials,
there is no such geometry). It is not expected, of course, that any Hamiltonian
will
have an exactly solvable path integral, and from the point of view of this
Section the cases where the Feynman path integral fails to be effectively
computable within the framework of equivariant localization will be those cases
where a required symmetry of the phase space geometry does not lead to
a globally well-defined metric tensor appropriate to the given topology.
Nonetheless, the analysis in \cite{dlr} for the 1-dimensional hydrogen atom is
a highly non-trivial success of the equivariant localization formulas for path
integrals which goes beyond the range of the standard WKB method.

We conclude this Section by showing that it is possible to relate the
path integrals for generic dynamical systems on non-homogenous phase spaces
which fall into the framework of loop space equivariant localization to
character formulas for the associated 1-parameter isometry groups $G^{(1)}$
\cite{szsem2}. For this, we need to introduce a formalism for constructing
coherent states associated with non-transitive group actions on manifolds
\cite{klauder,szsem2}. We consider the isothermal metric (\ref{isotherm})
in the prefered $x'$-coordinates for a Hamiltonian vector field $V$ on
$\man$. Using these coordinates, we define the complex coordinates
$z=x'^2\e^{ix'^1}$, in analogy with the case where $V$ defines a rotationally
symmetric geometry (as for the harmonic oscillator). Let $f(z\bar z)$ be a
$G^{(1)}$-invariant analytic solution of the ordinary differential equation
\beq
\frac{d}{d(z\bar z)}z\bar z\frac{d}{d(z\bar z)}\log f(z\bar z)=\frac{1}{2}
\e^{\varphi(z\bar z)}
\label{fdiff}\eeq
For the symplectic 2-form of the phase space, we take the $G^{(1)}$-invariant
volume form associated with $(\man,g)$,
\beq
\omega^{(\varphi)}=i\frac{d}{d(z\bar z)}z\bar z\frac{d}{d(z\bar z)}\log f(z
\bar z)~dz\wedge d\bar z
\label{omegaphi}\eeq
whose associated symplectic potential is
\beq
\theta^{(\varphi)}=\frac{i}{2}\frac{d}{d(z\bar z)}\log f(z\bar z)\left(\bar zdz
-zd\bar z\right)
\label{thetaphi}\eeq
This definition turns the phase space into a non-homogeneous K\"ahler
manifold with K\"ahler potential
\beq
F^{(\varphi)}(z,\bar z)=\log f(z\bar z)
\label{nonhompot}\eeq
such that $\omega^{(\varphi)}$ determines the first Chern class of the usual
symplectic line bundle $L^{(\varphi)}\to\man$.

Let $N_\varphi$, $0<N_\varphi\leq\infty$, be the integer such that the
function $f(z\bar z)$ admits the Taylor series expansion
\beq
f(z\bar z)=\sum_{n=0}^{N_\varphi}(z\bar z)^nf_n
\label{taylorf}\eeq
and let $\rho(z\bar z)$ be a $G^{(1)}$-invariant integrable function whose
moments are
\beq
\int_0^Pd(z\bar z)~(z\bar z)^n\rho(z\bar z)=\frac{1}{f_n}~~~~~,~~~~~0\leq n
\leq N_\varphi
\label{rhodef}\eeq
where $P$ is a real number with $0<P\leq\infty$. Let $\hat a^\dagger$ and
$\hat a$ be bosonic creation and annihilation operators on some representation
space of the isometry group (as in Subsection 5.3 above), and let $|n\rangle$,
$n\in\IZ^+$, be the complete system of orthonormal eigenstates of the
corresponding number operator, $\hat a^\dagger\hat a|n\rangle=n|n\rangle$. The
desired coherent states are then defined as
\beq
|z)=\sum_{n=0}^{N_\varphi}\sqrt{f_n}~z^n|n\rangle
\label{noncoh}\eeq
The states (\ref{noncoh}) have the normalization
\beq
(z|z)=f(z\bar z)=\e^{F^{(\varphi)}(z,\bar z)}
\label{nonnorm}\eeq
and they obey a completeness relation analogous to (\ref{su2compl}) in the
isometry invariant measure
\beq
d\mu^{(\varphi)}(z,\bar z)=\frac{i}{2\pi}f(z\bar z)\rho(z\bar z)\Theta(P-z\bar
z)~dz\wedge d\bar z
\label{noncohmeas}\eeq
where $\Theta(x)$ denotes the step function for $x\in\IR$. The completeness
of the coherent states (\ref{noncoh}) follows from a calculation analogous to
that in (\ref{cohcompl}) using the definitions (\ref{taylorf})--(\ref{noncoh})
above.

Notice that for the functional values $f(z\bar z)=\e^{z\bar z}$, $(1+z\bar z)
^{2j}$ and $(1-z\bar z)^{-2k}$, (\ref{noncoh}) reduces to, respectively, the
Heisenberg-Weyl group, spin-$j$ $SU(2)$ and level-$k$ $SU(1,1)$ coherent states
that we described earlier. Moreover, in that case we consistently find,
respectively, the weight functions $\rho(z\bar z)=\e^{-z\bar z}$ with $P=
\infty$, $\rho(z\bar z)=(2j+1)(1+z\bar z)^{-2(j+1)}$ with $P=\infty$, and
$\rho(z\bar z)=(2k-1)(1-z\bar z)^{2(k-1)}$ with $P=1$. This is anticipated
from (\ref{fdiff}), as then the isothermal metrics in (\ref{isotherm})
correspond to the standard maximally symmetric K\"ahler geometries. Here the
isometry group acts on the states (\ref{noncoh}) as
$h^{(\tau)}|z)=|\e^{i\tau}z)$, $h^{(\tau)}\in{\cal I}(\man,g)\equiv G^{(1)}$,
$\tau\in\IR^1$, which ensures that a Hamiltonian exists (as we shall see
explicitly below) such that a time-evolved coherent state remains coherent in
this sense, regardless of the choice of $\rho$ \cite{klauder}. The
(holomorphic)
dependence of the non-normalized coherent state vectors $|z)$ on only the
single complex variable $z$ is, as usual, what makes them amenable to the
study of the isometry situation at hand. Notice also that the metric tensor
(\ref{isotherm}) and canonical 1-form (\ref{thetaphi}) can as usual be
represented in the standard coherent state forms (\ref{kahlermetcoh}) and
(\ref{thetadcoh}), respectively.

Considering as usual the coherent state matrix elements (\ref{cohham}) with
respect to (\ref{noncoh}), using (\ref{thetaphi}) and (\ref{noncohmeas})
we can construct the usual coherent state path integral
\beq\new{\begin{array}{ll}
Z^{(\varphi)}(T|{\cal F}(H))=&\int_{L\man}\prod_{t\in[0,T]}d\mu^{(\varphi)}
(z(t),\bar z(t))\\&\times\exp\left\{i\int_0^Tdt~\left[\frac{1}{2}\frac{d}
{d(z\bar z)}\log f(z\bar z)\left(z\dot{\bar z}-\bar z\dot z\right)-{\cal F}(H)
\right]\right\}\end{array}}
\label{nonpart}\eeq
where we have again allowed for a possible functional ${\cal F}(H)$ of the
isometry generator $H$. The observable $H(z,\bar z)$ in (\ref{nonpart}) can
be found by substituting (\ref{omegaphi}), written back in the $x'$-coordinates
using the standard radial form for $z=x'^2\e^{ix'^1}$ given in
(\ref{metricrtheta}), and $V'^1=a_0$, $V'^2=0$ into the Hamiltonian equations.
Thus the equivariant localization constraints in these cases determine
$H$ in terms of the phase space metric as
\beq
H^{(\varphi)}(z,\bar z)=a_0\cdot z\bar z\frac{d}{d(z\bar z)}\log f(z\bar
z)+C_0=a_0\cdot
(\!(z|\hat a^\dagger\hat a|z)\!)+C_0=i_V\theta^{(\varphi)}
\label{noncohham}\eeq
where the function $f(z\bar z)$ is related to the metric (\ref{isotherm}) by
(\ref{fdiff}). Notice that (\ref{noncohham}) reduces to the usual harmonic
oscillator height functions in the maximally symmetric cases of Subsections
5.3, 5.5 and 5.6 above. Thus (\ref{noncohham}) can be considered as the general
localizable Hamiltonian valid for {\it any} phase space Riemannian geometry,
be it maximally symmetric or otherwise (the same is true, of course, for the
coherent state path integral (\ref{nonpart})). This is to be expected, because
the localizable Hamiltonian functions in the case of a homogeneous symmetry are
simply displaced harmonic oscillators, and these oscillator Hamiltonians
correspond to the rotation generators of the isometry groups, i.e. translations
in ${\rm arg}(z)=x'^1$ (this also agrees with the usual integrability
arguments). In fact, (\ref{noncohham}) shows explicitly that the function $H$
is essentially just a harmonic oscillator Hamiltonian written in terms of
some generalized phase space geometry.

The main difference in the present context between the homogeneous and
non-homogeneous cases lies in the path integral (\ref{nonpart}) itself. In
the former case the coherent state measure $d\mu^{(\varphi)}(z,\bar z)$ which
must be used in the Feynman measure in (\ref{nonpart}) coincides with the
volume form (\ref{omegaphi}), because as mentioned earlier if the
isometry group acts transitively on the Riemannian manifold $(\man,g)$ then
there is a unique left-invariant measure (i.e. a unique solution to
(\ref{symplcondloc})) and so $d\mu^{(\varphi)}=\omega^{(\varphi)}$ yields the
standard Liouville measure on the loop space $L\man$. In the latter case
$d\mu^{(\varphi)}\neq\omega^{(\varphi)}$, and (\ref{nonpart}) is not in
the canonical form (\ref{partfh}) for the quantum partition function associated
with the loop space symplectic geometry. Nonetheless, by a suitable
modification of the loop space supersymmetry associated with the dynamical
system by noting that the coherent state measure in (\ref{noncohmeas}) is
invariant under the action of the isometry group on $\man$, it is still
possible to derive appropriate versions of the standard localization formulas
with the obvious replacements corresponding to this change of integration
measure. Of course, we can alternatively follow the analysis of the former part
of this Subsection and use the standard Liouville path integral measure, but
then we lose the formal analogies with the Duistermaat-Heckman theorem and its
generalizations. It is essentially this non-uniqueness of an invariant
symplectic 2-form in the case of non-transitive isometry group actions which
leads to numerous possibilities for the localizable Hamiltonian systems defined
on such spaces, in marked contrast to the homogeneous cases where
everything was uniquely fixed. If one consistently makes the ``natural" choice
for $\omega$ as the K\"ahler 2-form (\ref{omegaphi}), then indeed the only
admissible Hamiltonian functions $H$ are generalized harmonic oscillators.

\section{Equivariant Localization on Multiply Connected Phase Spaces:
Applications to Homology and Modular Representations}

In the last Section we deduced the general features of the localization
formalism on a simply-connected symplectic manifold. We
found general forms for the Hamiltonian functions in terms of the underlying
phase space Riemannian geometry which is required for their Feynman path
integrals to manifestly localize. This feature is quite interesting from the
point of view that, as the quantum theory is always {\it ab initio}
metric-independent, this analysis probes the role that the geometry and
topology plays towards the understanding of quantum integrability. For
instance, we saw that the classical trajectories of a harmonic oscillator
must be embedded into a rotationally-invariant geometry and that
as such its orbits were always circular trajectories. For more
complicated systems these quantum geometries are less familiar and endow the
phase space with unusual Riemannian structures (i.e. complicated forms of the
localization supersymmetries). In any case, all the localizable Hamiltonians
were essentially harmonic oscillators (e.g. the height function for a spherical
phase space geometry) in some form or another, and their
quantum partition functions could be represented naturally using coherent state
formalisms associated with the Poisson-Lie group actions of the isometry
groups of the phase space. In the non-homogeneous cases we saw, in particular,
that to investigate equivariant localization in general one needs to
determine if a Riemannian geometry can possess certain symmetries imposed by
some rather ad-hoc restrictions from the dynamical system. In practice, the
introduction of such a definite geometry into the problem is highly
non-trivial, although we saw that it was possible in some non-trivial examples.
These results also impose restrictions on the classes of topological quantum
field theories and supersymmetric models which fall into the framework of
these geometric localization principles, as we shall discuss at greater length
in Section 8.

In this Section we shall extend the analysis of Section 5 to the case when
the phase space $\man$ is multiply-connected \cite{semsz}. We shall primarily
focus on the case where $\man$ is a compact Riemann surface of genus $h\geq1$,
again because of the wealth of mathematical characterizations that are
available for such spaces. We shall explore how the localization formalism
differs from that on a simply-connected manifold. Recall that much of the
formalism developed in Section 4, in particular that of Subsection 4.10, relied
quite heavily on this topological restriction. We shall see that now the
topological quantum field theories that appear also describe the non-trivial
first homology group of
the Riemann surface, and that it is completely independent of the geometrical
structures that are used to carry out the equivariant localization on $\man$,
such as the conformal factors and the modular parameters. This is typically
what a topological field theory should do (i.e. have only global features),
and therefore the equivariant Hamiltonian systems that one obtains in these
cases are nice examples of how the localization formalism is especially suited
to describe the characteristics of topological quantum field theories on
spaces with much larger topological degrees of freedom. Again the common
feature will be the description of the quantum dynamics using a coherent state
formalism, this time associated with a non-symmetric spin system and some
ideas from geometric quantization \cite{birm,woodhouse}. We shall
in addition see that the coherent states span a multi- but finite-dimensional
Hilbert space in which the wavefunctions carry a non-trivial representation of
the discrete first homology group of the phase space. We shall verify the
localization formulas of Section 4 in a slightly modified setting, pointing
out the important subtleties that arise in trying to apply them directly on
a multiply-connected phase space.

Although we shall attempt to give a quite general description of the
localizable dynamics on such spaces, most of our analysis will only be
carried out explicitly for genus 1, i.e. on the 2-torus $T^2=S^1\times S^1$. In
particular, we shall view the torus in a way best suited to describe its
complex algebraic geometry, i.e. in the parallelogram representation of
Subsection 3.5, so that we can examine the topological properties of the
quantum theory we find and get a good idea of the features of the localization
formalism on multiply-connected spaces in general. Another more explicit way to
view the torus is by embedding it in $\IR^3$ by revolving the circle
$(y-a)^2+x^2=b^2$ on the $xy$-plane around the $x$-axis, where $0<b<a$, i.e.
embedding $T^2$ in 3-space by $x=b\sin\phi_1$, $y=(a+b\cos\phi_1)\sin\phi_2$
and $z=(a+b\cos\phi_1)\cos\phi_2$. The induced metric on the surface from the
flat Euclidean metric of $\IR^3$ is then $b^2d\phi_1\otimes
d\phi_1+(a+b\cos\phi_1)^2d\phi_2\otimes d\phi_2$, and the modular parameter
$\tau\in\IC^+$ of the parallelogram representation of  $T^2$ is (c.f.
Subsection 3.5)
\beq
\tau=ib/\sqrt{a^2-b^2}
\label{taugeom}\eeq
If we now introduce the coordinate
\beq
\theta=\theta(\phi_1)=\int_0^{\phi_1}d\phi'_1~\frac{b}{a+b\cos\phi'_1}
\label{t2newcoord}\eeq
then it is straightforward to verify that $w=\phi_2+i\theta$ is an isothermal
coordinate for the induced metric on $T^2$ for which its isothermal form is
$\rho(\theta)(d\phi_2\otimes d\phi_2+d\theta\otimes d\theta)$. This defines
a complex structure on $T^2$. Since this metric is invariant under translations
in $\phi_2$, we could heuristically follow the analysis of Subsection 5.9 to
deduce that one class of localizable Hamiltonians are those which are functions
only of $\phi_1$. In order that these Hamiltonians be well-defined globally on
$T^2=S^1\times S^1$, we require in addition that these be periodic functions of
$\phi_1$. As we shall soon see, this is consistent with the general localizable
dynamical systems we shall find. Topological invariance of the associated
quantum theory in this context would be something like the invariance of it
under certain rescalings of the modular parameter (\ref{taugeom}), i.e. under
rescalings of the radius parameters $a$ or $b$ corresponding to a uniform
`shift' in the local geometry of $T^2$. A topological
quantum theory shouldn't detect such shifts which aren't considered as ones
modifying the topological properties of the torus. In other words, the
topological quantum theory should be independent of the phase space complex
structure. We shall see this in a more algebraic form later on in this Section.

\subsection{Isometry Groups of Multiply Connected Spaces}

To describe the isometries of a generic path connected, multiply-connected
Riemannian manifold $(\man,g)$, we lift these isometries up into what is known
as the universal covering space of the manifold. The multiple-connectivity of
$\man$ means that it has loops in it which cannot be contracted to a point
(i.e. $\man$ has `holes' in it). This is measured algebraically by
what is called the fundamental homotopy group $\pi_1(\man)$ of $\man$, a
similar
but rather different mathematical entity as the first homology group
$H_1(\man;\IZ)$. Roughly speaking, this group is defined as follows. We fix
a basepoint $x_0\in\man$ and consider the loop space of periodic maps
$\sigma:[0,1]\to\man$ with $\sigma(0)=\sigma(1)=x_0$. For any 2 loops
$\sigma$ and $\tau$ based at $x_0$ in this way, the product loop $\sigma\cdot
\tau$ is defined to be the loop obtained by first going around $\sigma$, and
then going around $\tau$. The set $\pi_1(\man)$ is the space of all
equivalence classes $[\sigma]$ of loops, where 2 loops are equivalent if
and only if they are homotopic to each other, i.e. there exists a continuous
deformation between the loops. It can be shown that the above multiplication
of loops then gives a well defined multiplication in $\pi_1(\man)$ and
turns it into a group with identity the homotopy class of the trivial loop
$[0,1]\to x_0$ and with inverse defined by reversing the orientation of
a loop. In general, this group is non-abelian and discrete, and it is related
to the first homology group $H_1(\man;\IZ)$ as follows. Let $[G,G]$ denote the
commutator subgroup of any group $G$, i.e. $[G,G]$ is the normal subgroup of
$G$ generated by the products $ghg^{-1}h^{-1}$, $g,h\in G$. The homology group
$H_1(\man;\IZ)$ is then the abelianization of the fundamental group,
\beq
H_1(\man;\IZ)=\pi_1(\man)_{\rm ab}\equiv\pi_1(\man)/\left[\pi_1(\man),
\pi_1(\man)\right]
\label{homolhomot}\eeq
If $\pi_1(\man)$ is itself abelian, then the
homology and homotopy of $\man$ coincide. We refer to \cite{massey} for a
more complete exposition of homotopy theory and how homology, in the sense
of (\ref{homolhomot}), is the natural approximation of homotopy.

The universal covering space of $\man$ is now defined as the smallest simply
connected manifold $\tilde\man$ covering $\man$. By a covering space
we mean that there is a surjective continuous projection map
$\pi:\tilde\man\to\man$ such that its restriction to any neighbourhood of
$\tilde\man$ defines a local diffeomorphism. This means that locally on $\man$
we can lift any quantity defined on it to its universal cover and study it on
the simply connected space $\tilde\man$. The manifold $\man$ and its universal
covering space $\tilde\man$ are related by the homeomorphism
\beq
\man\simeq\tilde\man/\pi_1(\man)
\label{covsphom}\eeq
where the fundamental group acts freely on $\tilde\man$ through what are known
as deck or covering transformations \cite{massey}, i.e. the diffeomorphisms
$\sigma:\tilde\man\to\tilde\man$ such that $\pi(\sigma(x))=\pi(x)$, $\forall
x\in\tilde\man$. Thus in this setting, the universal covering space is a
principal fiber bundle where the total space $\tilde\man$ is locally regarded
as the space of all pairs $(x,[C_x])$, where $C_x$ is a curve in $\man$ from
$x_0$ to $x$ and $[C_x]$ is its homotopy class\footnote{\baselineskip=12pt Here
a homotopy class of curves $[C_x]$ can be identified with an element of
$\pi_1(\man)$ by choosing another basepoint $x_0'$ and a grid of standard paths
from $x_0'$ to any other point in $\man$. Then the associated homotopy class is
represented by the loop $[x_0',x_0]\cup C_x\cup[x,x_0']$.}. The structure group
of the bundle is $\pi_1(\man)$ and the bundle projection
$\tilde\man~{\buildrel\pi\over\longrightarrow}~\man$ takes a homotopy class of
curves to their endpoint, $\pi:[C_x]\to x$. Clearly, $\man$ is its own
universal cover if it is simply connected, i.e. $\pi_1(\man)=0$. We shall see
some examples in due course.

Consider now a Riemannian metric $g$ defined on $\man$, and let $\pi^*g$
be its inverse image under the canonical bundle projection of $\tilde\man$
onto $\man$. Then $(\tilde\man,\pi^*g)$ is a simply-connected Riemannian
manifold, and from the analysis of the last Section we are well
acquainted with the structure of its isometry groups. It is possible to show
\cite{lich}, from the principal fiber bundle interpretation (\ref{covsphom})
above, that to every isometry $h\in{\cal I}(\man,g)$ one can associate
an isometry $\tilde h\in{\cal I}(\tilde\man,\pi^*g)$ which is compatible with
the universal covering projection in the sense that
\beq
\pi\circ\tilde h=h\circ\pi
\label{pih}\eeq
To prove this one needs to show that the lifting $\tilde h\equiv\pi^* h$
gives a diffeomorphism of $\tilde\man$ which is a well-defined function on
the homotopy classes of curves used for the definition of $\tilde\man$
\cite{lich}. Thus the isometries of the Riemannian manifold $(\man,g)$ lift
to isometries of the simply connected space $(\tilde\man,\pi^*g)$ of which we
have a complete description from the last Section. It should be kept in mind
though that there can be global obstructions from the homotopy of $\man$ to
extending an isometry of $\tilde\man$ projected locally down onto $\man$ by the
bundle projection $\pi$. We shall see how this works in the next Subsection.

\subsection{Equivariant Hamiltonian Systems in Genus One}

Our prototypical model for a multiply-connected symplectic manifold
will be the 2-torus $T^2=S^1\times S^1$ which we first studied in Subsection
3.5. Notice that the circle is multiply-connected with $\pi_1(S^1)=\IZ$ with
the integers labelling the number of times that a map $\sigma:S^1\to S^1$
`winds' around the circle, i.e. to each homotopy class $[\sigma]\in\pi_1(S^1)$
we can associated an integer which we call the winding number of the
loop $\sigma$ (where a change of sign signifies a change in the direction of
traversing the loop). We can describe the homotopy of the torus by introducing
2
loops $a$ and $b$, both fixed at the same basepoint on $S^1\times S^1$, with
$a$ encircling once the inner circle of the torus (i.e. $a:S^1\to(\phi_1,0)
\in S^1\times S^1$) and $b$ encircling once the outer circle (i.e.
$b:S^1\to(0,\phi_2)\in S^1\times S^1$). Since clearly any other loop in $T^2$
is homotopic to some combination of the loops $a$ and $b$, it follows that
they generate the fundamental group $\pi_1(T^2)$ of the torus, and furthermore
they obey the relation
\beq
aba^{-1}b^{-1}={\bf 1}
\label{abelian}\eeq
which is easily seen by simply tracing the loop product in (\ref{abelian})
around $S^1\times S^1$. (\ref{abelian}) means that $\pi_1(T^2)$ is abelian
and therefore coincides with the first homology group (\ref{homologyt2}).
Thus the loops $a$ and $b$ defined above are also generators of the first
homology group $H_1(\Sigma^1;\IZ)$, and they will henceforth be refered to as
the canonical homology cycles of the torus. Note that any homology cycle in
$\Sigma^1$ which defines the homology class of $a$ (respectively $b$) can be
labelled by the $\phi_1$ angle coordinates (respectively $\phi_2$). Thus any
homology class of a genus 1 compact Riemann surface is labelled by a pair of
integers $(n,m)$ which represents the winding numbers around the canonical
homology cycles $a$ and $b$.

Recall from Subsection 3.5 the description of the torus as a parallelogram with
its opposite edges identified in the plane, and with modular parameter
$\tau\in\IC^+$ which labels the inequivalent complex analytic structures on the
torus (or equivalently the conformal equivalence classes of metrics on $T^2$)
\cite{nash,schlich}. This means that it can be represented as the quotient
space
\beq
\Sigma^1=\IC/(\IZ\oplus\tau\IZ)
\label{t2quotient}\eeq
where the quotient is by the free bi-holomorphic action of the lattice group
$\IZ\oplus\tau\IZ$ on the simply-connected complex plane $\IC$. In other words,
the lattice group is the discrete automorphism group of the complex plane and
it
acts on $\IC$ by the translations\footnote{\baselineskip=12pt For an exposition
of the various
equivalent ways, such as above, of describing compact Riemann surfaces in
different geometric forms, see \cite{naras}.}
\beq
z\to z+2\pi(n+\tau m)~~~~~,~~~~~\bar z\to\bar z+2\pi(n+\bar\tau m)
{}~~~;~~~n,m\in\IZ
\label{t2transls}\eeq
under which the canonical bundle projection $\IC~{\buildrel\pi\over
\longrightarrow}~\Sigma^1$ is invariant.
That the plane is the universal cover of the torus is easily seen by observing
that the real line $\IR^1$ is the universal cover of the circle $S^1$ with the
bundle projection $\pi(x)=\e^{2\pi ix}$ for $x\in\IR^1$.

With the identification (\ref{t2quotient}), we can now consider the most
general Euclidean signature metric on $\Sigma^1$. From our discussion in
Subsection 5.2, we know that the most general metric on $\IC$ can be written in
the global isothermal form (\ref{isotherm}). The covering projection in
(\ref{t2quotient}) in this way induces the most general metric on the torus,
which can therefore be written in terms of a flat K\"ahler metric as
\beq
g_\tau=\frac{\e^{\varphi(z,\bar z)}}{{\rm Im}~\tau}dz\otimes d\bar z
\label{t2metric}\eeq
or in terms of the angle coordinates $(\phi_1,\phi_2)\in S^1\times S^1$
\beq
[g_{\phi_\mu\phi_\nu}]=\frac{\e^{\varphi(\phi_1,\phi_2)}}{{\rm Im}~\tau}
\pmatrix{1&{\rm Re}~\tau\cr{\rm Re}~\tau&|\tau|^2\cr}
\label{t2metricphi}\eeq
The complex structure on $\Sigma^1$ is now defined by the complex
coordinates $z=\phi_1+\tau\phi_2,\bar z=\phi_1+\bar\tau\phi_2$ which are
therefore considered invariant under the transformations
(\ref{t2transls}). The conformal factor $\varphi(z,\bar z)$ is now a globally
defined real-valued function on $\Sigma^1$ (i.e. invariant under the
translations (\ref{t2transls})), and the normalization in (\ref{t2metric}) is
chosen for simplicity so that the associated metric volume of the torus
\beq
{\rm vol}_{g_\tau}(\Sigma^1)=\int_{\Sigma^1}d^2\phi~\sqrt{\det g_\tau}=\int
_{\Sigma^1}d^2\phi~\e^{\varphi(\phi_1,\phi_2)}\equiv(2\pi)^2v
\label{t2vol}\eeq
is finite and independent of the complex structure of $\Sigma^1$ with $v\in
\IR$ a fixed volume parameter of the torus. The metric (\ref{t2metric}) is
further constrained by its Gaussian curvature scalar
\beq
K(g_\tau)=-\frac{1}{2}~{\rm Im}(\tau)\e^{-\varphi}\nabla^2_\tau\varphi
\label{t2curv}\eeq
which by the Gauss-Bonnet-Chern theorem (\ref{gaussbon2d}) for genus $h=1$
must obey
\beq
\int_{\Sigma^1}d^2\phi~\nabla^2_\tau\varphi(\phi_1,\phi_2)=0
\label{t2gaussbon}\eeq
where $\nabla_\tau^2=\partial\bar\partial$ is the scalar Laplacian
\beq
\nabla_\tau^2=\partial^2_{\phi_1}+|\tau|^{-2}\partial^2_{\phi_2}+2~{\rm Re}
(\tau)|\tau|^{-2}\partial_{\phi_1}\partial_{\phi_2}
\label{t2laplacian}\eeq
associated with the K\"ahler structure in (\ref{t2metric}).

Given this general geometric structure of the 2-torus, following the analysis
of the last Section we would like to find the most general
Hamiltonian system on it which obeys the localization criteria. First of all,
the condition that the Hamiltonian $H$ generates a globally integrable
isometry of the metric (\ref{t2metric}) implies that the associated
Hamiltonian vector fields $V^\mu(x)$ must be single-valued functions under
the windings (\ref{t2transls}) around the non-trivial homology cycles of
$\Sigma^1$. This means that these functions must admit convergent
2-dimensional harmonic mode expansions
\beq
V^\mu(\phi_1,\phi_2)=\sum_{n,m=-\infty}^\infty V_{n,m}^\mu\e^{i(n\phi_1+m
\phi_2)}
\label{t2harmexp}\eeq
In other words, the components of $V$ must be $C^\infty$-functions which
admit a 2-dimensional Fourier series plane wave expansion (\ref{t2harmexp})
appropriate to globally-defined periodic functions on $S^1\times S^1$. As
we shall now demonstrate, these topological restrictions from the underlying
phase space severely limit the possible Hamiltonian systems to which the
equivariant localization constraints apply.

{}From (\ref{liegcoord}) it follows that the Killing equations for the metric
(\ref{t2metricphi}) are
\beq\new{\begin{array}{c}
2\partial_{\phi_1}V^1+2~{\rm Re}(\tau)\partial_{\phi_1}V^2+V^\mu\partial_{\phi
_\mu}\varphi=0\\2~{\rm Re}(\tau)\partial_{\phi_2}V^1+2|\tau|^2\partial_{\phi_2}
V^2+|\tau|^2V^\mu\partial_{\phi_\mu}\varphi=0\\\partial_{\phi_2}V^1+~{\rm Re}
(\tau)(\partial_{\phi_2}V^2+\partial_{\phi_1}V^1)+|\tau|^2\partial_{\phi_1}V^2
+~{\rm Re}(\tau)V^\mu\partial_{\phi_\mu}\varphi=0\end{array}}
\label{t2killing}\eeq
Substituting in the harmonic expansions (\ref{t2harmexp}) and using the
completeness of the plane waves there to equate the various components of
the expansions in (\ref{t2killing}), we find after some algebra that
(\ref{t2killing}) generates 2 coupled equations for the Fourier components of
the Hamiltonian vector field,
\beq\new{\begin{array}{c}
(|\tau|^2n-~{\rm Re}(\tau)m)V_{n,m}^1=|\tau|^2(m-~{\rm Re}(\tau)n)V_{n,m}^2\\
(m-~{\rm Re}(\tau)n)V_{n,m}^1=\left[(~{\rm Re}(\tau)^2-~{\rm Im}(\tau)^2)n
-~{\rm Re}(\tau)m\right]V_{n,m}^2\end{array}}
\label{t2coupled}\eeq
which hold for all integers $n$ and $m$. It is straightforward to show from
the coupled equations (\ref{t2coupled}) that for $\tau\in\IC^+$, $V_{n,m}^1
=V_{n,m}^2=0$ unless $n=m=0$. Thus the only non-vanishing components of the
harmonic expansions (\ref{t2harmexp}) are the constant modes,
\beq
V_{\Sigma^1}^\mu(x)=V_0^\mu
\label{constmodes}\eeq
and the only Killing vectors of the metric (\ref{t2metric}) are the generators
of translations (by $V_0^\mu\in\IR$) along the 2 independent homology cycles of
$\Sigma^1$. Notice that this result is completely independent of the structure
of the conformal factor $\varphi$ in (\ref{t2metric}), and it simply means that
although the torus inherits {\it locally} 3 isometries from the maximally
symmetric plane, i.e. local rotations and translations, only the 2 associated
translations on $\Sigma^1$ are {\it global} isometries. The independence of
this result on the conformal factor is not too surprising, since this just
reflects the fact that given any metric on a compact phase space we can make
it invariant under a compact group action by averaging it over the group in
its Haar measure. The above derivation gives an explicit geometric view of how
the non-trivial topology of $\Sigma^1$ restricts the allowed global circle
actions on the phase space, and we see therefore that the isometry group of
any globally-defined Riemannian geometry on the torus is $U(1)\times U(1)$.

The invariance condition (\ref{symplcondloc}) for the symplectic structure can
be solved by imposing the requirement of invariance of $\omega_{\Sigma^1}$
independently under the 2 Killing vectors (\ref{constmodes}). This implies
that the components $\omega_{\phi_\mu\phi_\nu}$ must be constant functions,
i.e. that $\omega$ must be proportional to the Darboux 2-form $\omega_D$, and
thus we take
\beq
\omega_{\Sigma^1}=v~d\phi_1\wedge d\phi_2
\label{t2omega}\eeq
to be an associated metric-volume form on $\Sigma^1$ for the present Riemannian
geometry (c.f. (\ref{t2vol})). The symplectic structure here is thus the
symplectic reduction of that from the universal bundle projection
$\IC~{\buildrel\pi\over\longrightarrow}~\Sigma^1$. It is straightforward to
now integrate up the Hamiltonian equations with (\ref{constmodes}) and
(\ref{t2omega}), and we find that the Hamiltonian $H_{\Sigma^1}$ is given by
displacements along the homology cycles of $\Sigma^1$,
\beq
H_{\Sigma^1}(\phi_1,\phi_2)=h^1\phi_1+h^2\phi_2
\label{t2eqhams}\eeq
where
\beq
h^1=vV_0^2~~~~~,~~~~~h^2=-vV_0^1
\label{hmudef}\eeq
are real-valued constants. Note that, as anticipated from
(\ref{t2quotient}), the invariant symplectic structure here is uniquely
determined just as for 2-dimensional maximally symmetric phase spaces which
have 3 (as opposed to just 2 as above) linearly independent Killing vectors.
Thus we
see here that the localizable Hamiltonian systems in genus 1 are even more
severely restricted by the equivariant localization constraints as compared
to the simply-connected cases. Note
that the Hamiltonian (\ref{t2eqhams}) does not determine a globally-defined
single-valued function on $\Sigma^1$, a point which we shall return to shortly.

\subsection{Homology Representations and Topological Quantum Field Theory}

The Hamiltonian (\ref{t2eqhams}) defines a rather odd dynamical system on the
torus, but besides this feature we see that the allotted Hamiltonians as
determined from the geometric localization constraints are in effect
completely {\it independent} of the explicit form of the phase space
geometry and depend only on the topological properties of the manifold
$\Sigma^1$, i.e. (\ref{t2eqhams}) is explicitly independent of both the
complex structure $\tau$ and the conformal factor $\varphi$ appearing in
(\ref{t2metric}). From the analysis of the last Section, we see that this is in
marked contrast to what occurs in the case of a simply connected phase space,
where the conformal factor of the metric entered into the final expression for
the observable $H$ and the equivariant Hamiltonian systems so obtained depended
on the phase space geometry explicitly. In the present case the
partition function with Hamiltonian (\ref{t2eqhams}) and symplectic 2-form
(\ref{t2omega}) obtained as the unique solutions of the equivariant
localization constraints can be thought of in this way as defining a
topological quantum theory on the torus which is completely independent of any
Riemannian geometry on $\Sigma^1$. Furthermore, the symplectic potential
associated with (\ref{t2omega}) is
\beq
\theta_{\Sigma^1}=\frac{v}{2}\left(\phi_1d\phi_2-\phi_2d\phi_1\right)
\label{t2theta}\eeq
which we note is only locally defined because it involves multi-valued
functions in this local form, so that $\omega_{\Sigma^1}$ is a non-trivial
element of $H^2(\Sigma^1;\IZ)=\IZ$. The Hamiltonian (\ref{t2eqhams}) thus
admits the local topological form $H_{\Sigma^1}=i_{V_{\Sigma^1}}\theta_{\Sigma
^1}$, so that the corresponding partition function defines a cohomological
field theory and it will be a topological invariant of the manifold $\Sigma^1$.

To explore some of the features of this topological quantum field theory, we
note first that (\ref{t2eqhams}) is not defined as a global $C^\infty$-function
on $\Sigma^1$. However, this is not a problem from the point of view of
localization theory. Although for the classical dynamics the Hamiltonian can
be a multi-valued function on $\Sigma^1$, to obtain a well-defined
quantum theory we require single-valuedness, under the windings
(\ref{t2transls}) around the homology cycles of $\Sigma^1$, of the time
evolution operator $\e^{-iT\hat H_{\Sigma^1}}$ which defines the quantum
propagator (and also of the Boltzmann weight $\e^{iTH_{\Sigma^1}}$ if we wish
to have a well-defined classical statistical mechanics). This implies that the
constants $h^\mu$ in (\ref{t2eqhams}) must be quantized, i.e. $h^\mu\in h\IZ$
for some $h\in\IR$, and then time propagation in this quantum system can only
be defined in discretized intervals of the base time $h^{-1}$, i.e. $T=N_Th^
{-1}$ where $N_T\in\IZ^+$. Such quantizations of coupling parameters in
topological gauge theories is a rather common occurence to ensure the
invariance of a quantum theory under `large gauge transformations' when the
underlying space has non-trivial topology \cite{birm}.

In the quantum theory, the Hamiltonian (\ref{t2eqhams}) therefore represents
the winding numbers around the homology cycles of the torus, and therefore
to each homology class of $\Sigma^1$ we can associate a corresponding
Hamiltonian system obeying the equivariant localization constraints. The
partition function is now denoted as
\beq
Z_{\Sigma^1}^v(k,\ell;N_T)\sim\int_{L\Sigma^1}[d^2\phi]~\exp\left\{i\int_0^{N_T
h^{-1}}dt~\left(v\phi_2\dot\phi_1+h(k\phi_1+\ell\phi_2)\right)\right\}
\label{t2part}\eeq
where $k$ and $\ell$ are integers and we have integrated the kinetic term in
(\ref{t2part}) by parts. This path integral can be evaluated directly
by first integrating over the loops $\phi_2(t)$, which gives
\beq
Z_{\Sigma^1}^v(k,\ell;N_T)\sim\int_{LS^1}[d\phi_1]~\delta(v\dot\phi_1+h\ell)
\exp\left\{i\int_0^{N_Th^{-1}}dt~hk\phi_1(t)\right\}\sim\e^{-ik\ell N_T^2/2v}
\label{t2parttop}\eeq
Thus the partition function of this quantum system represents the
non-trivial homology classes of the torus, through the winding numbers $k$ and
$\ell$ and the time evolution integer $N_T$. In fact, (\ref{t2parttop})
defines a family of 1-dimensional unitary irreducible representations of
the first homology group of $\Sigma^1$ through the family of homomorphisms
\beq
Z_{\Sigma^1}^v(\cdot,\cdot;N_T):H_1(\Sigma^1;\IZ)\to U(1)\otimes U(1)
\label{t2homrep}\eeq
from the additive first homology group (\ref{homologyt2}) into a multiplicative
circle group. Notice that the associated homologically-invariant quantum
theory is trivial, in that the sum over all winding numbers of the partition
function (\ref{t2parttop}) vanishes,
\beq
\sum_{k,\ell=-\infty}^\infty Z_{\Sigma^1}^v(k,\ell;N_T)=\sum_{k=-\infty}^\infty
\left(\frac{1}{1-\e^{-ikN_T^2/2v}}+\frac{1}{1-\e^{ikN_T^2/2v}}-1\right)=0
\label{homtrivial}\eeq
This sum over all winding numbers is analogous to what one would do in
4-dimensional Yang-Mills theory to include all instanton sectors into the
quantum theory \cite{sem}.

However, it is possible to modify slightly the definition of the quantum
propagator on a multiply-connected phase space so that we obtain a partition
function which is independent of the homology class defined by the
Hamiltonian using a modification of the definition of the path integral over a
multiply connected space \cite{schulman}. In general, if the phase space $\man$
is multiply connected, i.e. $\pi_1(\man)\neq0$, then the Feynman path integral
representation of the quantum propagator can contain parameters
$\chi([\sigma])$
which are not present in the classical theory and which weight the homotopy
classes $[\sigma]$ of topologically inequivalent time evolutions of the
system\footnote{\baselineskip=12pt This
definition could also be applied to the full quantum propagator ${\cal K}(x',x
;T)$ between 2 phase space points. Then the sum in (\ref{chipart}) is over all
homotopy classes of curves $[C_{xx'}]$ from $x$ to $x'$ which are identified
with elements of $\pi_1(\man)$ using a standard mesh of paths.},
\beq
Z_{\rm hom}(T)=\sum_{[\sigma]\in\pi_1(\man)}\chi([\sigma])\int_{x(t)\in[\sigma]
}[d^{2n}x]~\sqrt{\det\|\Omega\|}\e^{iS[x]}
\label{chipart}\eeq
Unitarity and completeness of the quantum theory (i.e. of the propagator
(\ref{prop})) yield, respectively, the constraints that the parameters
$\chi([\sigma])$ are phases,
\beq
\chi([\sigma])^*\chi([\sigma])=1
\label{chiphases}\eeq
and that they form a 1-dimensional unitary representation of $\pi_1(\man)$,
\beq
\chi([\sigma])\chi([\sigma'])=\chi([\sigma\cdot\sigma'])
\label{chirep}\eeq
Note that the restriction of the path integration to homotopy classes as in
(\ref{chipart}) makes well-defined the representation of the partition
function action $S$ with a local symplectic potential following the
Wess-Zumino-Witten prescription of Subsection 4.10. In particular, we can
invoke
the argument there to conclude that over each homotopy class $[\sigma]\in\pi
_1(\man)$, the path integral depends only on the second cohomology class
defined by $\omega$.

In the case at hand, the partition function (\ref{t2parttop}) is regarded
as that obtained by restricting the path integration in (\ref{t2part}) to
loops in the homology class labelled by $(k,\ell)\in\IZ^2$. In particular,
we can add to the sum in
(\ref{homtrivial}) the phases $\chi(k,\ell)=\e^{i\alpha(k,\ell)}$ for
each $(k,\ell)\in\IZ^2$, which from (\ref{chirep}) would then have to satisfy
\beq
\alpha(k+k',\ell+\ell')=\alpha(k,\ell)+\alpha(k',\ell')
\label{alpharep}\eeq
The condition (\ref{alpharep}) means that the phase $\alpha(k,\ell)$
defines a ${\bf u(1)}$-valued 1-cocycle of the fundamental (or homology) group
$\IZ\oplus\IZ$ of $\Sigma^1$ (see Appendix A) as required for them to form a
representation of it in the circle group $S^1$. When they are combined with the
character representation (\ref{t2parttop}) and the resulting quantity is summed
as in (\ref{homtrivial}), we can obtain a propagator which is a non-trivial
homological invariant of $\Sigma^1$ and which yields a character formula for
the
non-trivial topological groups of the phase space. We shall see how to
interpret these character formulas in a group-theoretic setting, as we did in
the last Section, in the Subsection 6.5. Notice that, strictly speaking, the
volume parameter $v$ in (\ref{t2parttop}) should be quantized in terms of
$h$, $k$ and $\ell$ so that the partition function yields a non-zero result
when integrated over the moduli space of $T$-periodic trajectories. In this
way,
(\ref{t2parttop}) also represents the cohomology class defined by the
symplectic 2-form (\ref{t2omega}) through the parameter $v$. We recall from
Subsection 4.10 that for a simply-connected phase space, the localizable
partition functions depend only on the second cohomology class defined by
$\omega$. Here we find that the multiple-connectivity of the phase space makes
it depend in addition on the first homology group of the manifold. Thus the
partition function of the localizable quantum systems on the torus yield
topological invariants of the phase space representing its (co-)homology
groups.

\subsection{Integrability Properties and Localization Formulas}

We now turn to a discussion of the structure of the localization formulas for
these localizable Hamiltonian systems. Of course, since the canonical
$U(1)\times U(1)\sim T^2$ action on the torus generated by (\ref{t2eqhams}) has
no fixed points, this means that the classical partition function
\beq
\int_{T^2}v~d\phi_1~d\phi_2~\e^{iTH_{\Sigma^1}(\phi_1,\phi_2)}=\frac{1}
{T^2vV_0^1V_0^2}\left(\e^{-2\pi iTvV_0^2}-1\right)\left(1-\e^{2\pi
iTvV_0^1}\right)
\label{clt2dh}\eeq
isn't given by the Duistermaat-Heckman formula. The reason can be traced back
to the Poisson algebra $\{\phi_1,\phi_2\}_{\omega_{\Sigma^1}}=-1/v$ which shows
that the full Hamiltonian (\ref{t2eqhams}) is not a functional of action
variables in involution. Even if we choose a Hamiltonian which is a Morse
function on $T^2$ given by a functional of the components of the localizable
isometry generators on the torus (like the height function), Kirwan's theorem
forbids the exactness of the stationary phase approximation for the associated
classical partition functions.

We can also view this failure at the level of a Hamiltonian reduction as
discussed in Subsections 5.7 and 5.8 above. For instance, consider the
following partition function that does not fulfill the Duistermaat-Heckman
theorem,
\beq
Z_M^{T^2}(T)\equiv\int_{T^2}v~d\phi_1~d\phi_2~\e^{iT(a\cos\phi_1+b\cos\phi_2)}
=(2\pi)^2vJ_0(-iTa)J_0(-iTb)
\label{morset2dh}\eeq
We can write the left-hand side of (\ref{morset2dh}) as a reduction from a
larger integral on $\IC^2$ by introducing 2 complex coordinates
$z_1=|z_1|\e^{i\phi_1}$ and $z_2=|z_2|\e^{i\phi_2}$ with the constraints
\beq
P(z,\bar z)\equiv z\bar z-1=0
\label{t2constrfn}\eeq
for $z=z_1,z_2$. Introducing Lagrange multipliers $\lambda_1$ and $\lambda_2$
whose integration over $\IR$ produces delta-functions enforcing these
constraints, we can write the integration in (\ref{morset2dh}) as
\beq\new{\begin{array}{c}
Z_M^{T^2}(T)=v\int_{-\infty}^\infty
d\lambda_1~d\lambda_2~\int_{\IR^4}\frac{d^2z_1~d^2z_2}{\pi^2}~
\exp\left\{iT\left(\frac{a}{2}\left(z_1+\bar z_1\right)+\frac{b}{2}\left(z_2+\bar z_2\right)\right)\right.\\\left.+i\lambda_1(z_1\bar z_1-1)+i\lambda_2(z_2\bar z_2-1)\right\}\end{array}}
\label{morset2red}\eeq
This shows that, unlike the cases mentioned in Subsections 5.7 and 5.8, the
Hamiltonian in (\ref{morset2red}) from which the dynamical system in
(\ref{morset2dh}) is obtained by Hamiltonian reduction is not a bilinear
function and it does not commute with the constraint functions defined by
(\ref{t2constrfn}), i.e. the constraints of the reduction are not first class
constraints. These alternative integrability arguments therefore also serve as
an indication of the breakdown of the localization formalism when applied to
multiply connected phase spaces, in addition to the usual topological criteria
provided by Kirwan's theorem. We shall see again in Section 7 this sort of
interplay between integrability and Kirwan's theorem.

The situation is better for the quantum localizations, even as far as the
possibilities of using functionals ${\cal F}(H_{\Sigma^1})$ of the isometry
generator (\ref{t2eqhams}) for localization as in Subsection 4.9. Here the
arbitrariness of these functionals is not as great as it was in the simply
connected cases of Section 5. There we required generally only that $\cal F$ be
bounded from below, while in the case at hand the discussion of Subsection 6.3
above shows that we need in addition the requirement that $\cal F$ be formally
a periodic functional of the observable (\ref{t2eqhams}). In general, this will
not impose any quantization condition on the time translation $T$, as it did
before. For such functionals, however, it is in general rather difficult to
determine explicitly the Nicolai transform in (\ref{xiphichange}) required for
the localization (\ref{partfhloc}). Alternatively, one can try to localize
the system using (\ref{locgenint}) and the above description of the
quantum theory as a topological one, but then we lose the interpretation of the
independent Hamiltonians in (\ref{t2eqhams}) as conserved charges of some
integrable dynamical system with phase space the torus. These remarks imply,
for example, that one cannot equivariantly quantize a free particle or harmonic
oscillator (with compactified momentum and position ranges) on the torus, so
that the localizable dynamical systems do not represent generalized harmonic
oscillators as they did in the simply connected cases. The same is true of
the torus height function (\ref{heightt2}), as anticipated. However, in these
cases the periodicity of the Hamiltonian function leads to a much better
defined propagator in the sense of it being a tempered distribution represented
by a functional integral. Notice that this also shows explicitly, in a rather
transparent way, how the Hamiltonian functions on $T^2$ are restricted by
Kirwan's theorem, which essentially means in the above context that the
localization formalism loses its interpretation in terms of integrability
arguments on a multiply connected phase space. The topological field theory
interpretations do, however, carry through from the simply-connected cases but
with a much richer structure now.

The expression (\ref{t2parttop}) for the quantum partition function also
follows
directly from substituting into the Boltzmann weight $\e^{iS[x]}$ the value
of the action in (\ref{t2part}) evaluated on the classical trajectories
$\dot x^\mu(t)=V^\mu_{\Sigma^1}$ for the above quantum system, which here are
defined by
\beq
\dot\phi_1(t)=V_0^1~~~~~,~~~~~\dot\phi_2(t)=V_0^2
\label{t2class}\eeq
Thus the path integral (\ref{t2part}) (trivially) localizes onto the classical
loops as in the WKB localization formula (\ref{wkb}), except that now even
the 1-loop fluctuation term vanishes and the path integral is given exactly
by its tree-level value. This also independently establishes the quantizations
of the propagation time $T$ and the volume parameter $v$, in that
$T$-periodic solutions to the classical equations of motion with the
degenerate structure of the Hamiltonian (\ref{t2eqhams}) only exist with the
discretizations of the parameters $h^\mu$ and $T$ above. This is consistent
with the discussion at the beginning of Subsection 4.6 concerning the structure
of the moduli space of classical solutions, and again for these discretizations
the path integral can be evaluated using the degenerate localization formula
(\ref{wkbdeg}) while for the non-discretized values the critical trajectory set
(trivially) coincides with the critical point set $\man_V$ of the
Hamiltonian. Furthermore, the fact that the conformal factor $\varphi$ is not
involved at all in the solutions of the localization constraints just reflects
the fact that the torus is locally flat (as is immediate from its parallelogram
representation) and any global `curving' of its geometry represented by
$\varphi$ in (\ref{t2metric}) can only be done in a uniform periodic fashion
around the canonical homology cycles of $\Sigma^1$ (c.f. eq.
(\ref{t2gaussbon})). However, the Niemi-Tirkkonen formula (\ref{niemitirkloc})
does depend explicitly on $\varphi$. It is here that the geometry of the phase
space enters explicitly into the quantum theory, as it did in Section 5, if we
demand that the metric (\ref{t2metric}) obey the appropriate regularity
conditions and therefore make the equivariant localization
manifest. This ensures that the localization formula (\ref{niemitirkloc})
coincides with the exact result (\ref{t2parttop}), as it should.

In the case at hand (\ref{niemitirkloc}) becomes
\beq\new{\begin{array}{ll}
Z_{\Sigma^1}^v(k,\ell;N_T)&\sim\int_{\Sigma^1}\ch_{V_{\Sigma^1}}(-iN_T\omega_
{\Sigma^1}/h)\wedge\hat A_{V_{\Sigma^1}}(N_TR_\tau/h)\\&=\int_{\Sigma^1}
d^2\phi~\int d^2\eta~\exp\left[-\frac{iN_T}{h}\left(H_{\Sigma^1}(k,\ell)-\frac
{1}{2}(\omega_{\Sigma^1})_{\mu\nu}\eta^\mu\eta^\nu\right)\right]\\&~~~~~
\times\sqrt{\det\left[\frac{N_T(2(\nabla_\tau)_\mu
V_{\Sigma^1}^\nu+(R_\tau)^\nu
_{\mu\lambda\rho}\eta^\lambda\eta^\rho)/4h}{\sinh\left(N_T(2(\nabla_\tau)_\mu
V_{\Sigma^1}^\nu+(R_\tau)^\nu_{\mu\lambda\rho}\eta^\lambda\eta^\rho)/4h
\right)}\right]}\end{array}}
\label{t2ntirk}\eeq
Again, because of the K\"ahler structure of (\ref{t2metric}), the Riemann
moment map and curvature 2-form have the non-vanishing components
\beq
(\mu_{V_{\Sigma^1}})^z_{\bar z}=-(\mu_{V_{\Sigma^1}})^{\bar z}_z=V_{\Sigma^1}
^z\partial\varphi+V^{\bar z}_{\Sigma^1}\bar\partial\varphi~~~,~~~R_z^z=
-R_{\bar z}^{\bar z}=~{\rm Im}(\tau)\e^{-\varphi}\nabla_\tau^2\varphi
{}~\eta\bar\eta
\label{t2equivcurv}\eeq
We substitute (\ref{constmodes})--(\ref{hmudef}) and (\ref{t2equivcurv}) into
(\ref{t2ntirk}) and carry out the Berezin integrations there. Comparing the
resulting expression with the exact one (\ref{t2parttop}) for the partition
function, we arrive after some algebra at a condition
on the conformal factor of the metric (\ref{t2metric}),
\beq
\int_{\Sigma^1}d^2\phi~\e^{-iN_T(k\phi_1+\ell\phi_2)}\sqrt{1-\frac{N_T^2(\ell
\partial_{\phi_1}\varphi-k\partial_{\phi_2}\varphi)^2}{4v^2\sinh^2\left(\frac{
N_T}{2v}(\ell\partial_{\phi_1}\varphi-k\partial_{\phi_2}\varphi)\right)}}=
-\frac{2i}{N_Tv}\e^{-ik\ell N_T^2/2v}
\label{t2amb}\eeq
The Fourier series constraint (\ref{t2amb}) on the metric is rather
complicated and it represents a similar sort of metric regularity condition
that we encountered in Subsection 5.9 before. It fixes the harmonic modes of
the
square-root integrand in (\ref{t2amb}) which should have an expansion such as
(\ref{t2harmexp}). Notice, however, that (\ref{t2amb}) is
independent of the phase space complex structure $\tau$, and thus it only
depends on the representative of the conformal equivalence class of the metric
(\ref{t2metric}). This is typical of a topological field theory path integral
\cite{birm}.

The condition (\ref{t2amb}) can be used to check if a given phase space
metric really does result in the correct quantum theory (\ref{t2parttop}), and
this procedure then tells us what (representatives of the conformal equivalence
classes of) quantum geometries in this sense are applicable to the equivariant
localization of path integrals on the torus. For example, suppose we tried to
quantize a flat torus using equivariant localization. Then from (\ref{t2curv})
the conformal factor would have to solve the Laplace equation $\nabla_\tau^2
\varphi=0$ globally on $\Sigma^1$. Since $\varphi$ is assumed to be a
globally-defined function on $\Sigma^1$, it must admit a harmonic mode
expansion over $\Sigma^1$ as in (\ref{t2harmexp}). From (\ref{t2laplacian})
and this Fourier series for $\varphi$ we see that the Laplace equation implies
that all Fourier modes of $\varphi$ except the constant modes vanish, and so
the
left-hand side of (\ref{t2amb}) is zero. Thus a flat torus cannot be used to
localize the quantum mechanical path integral (\ref{t2part}) onto the
equivariant Atiyah-Singer index in (\ref{niemitirkloc}). This means that a
flat K\"ahler metric (\ref{t2metric}) on $\Sigma^1$ does not lead to a
homotopically trivial localization 1-form $\psi=i_{V_{\Sigma^1}}g_\tau$ on
the loop space $L\Sigma^1$ within any homotopy class (c.f. Subsection 4.4).
This simple example shows that the condition (\ref{t2amb}), along with the
Riemannian restrictions (\ref{t2vol}) and (\ref{t2gaussbon}), give a very
strong probe of the quantum geometry of the torus. Moreover, when (\ref{t2amb})
does hold, we can represent the equivariant characteristic classes in
(\ref{niemitirkloc}) in terms of the homomorphism (\ref{t2parttop}) of the
first homology group of $\Sigma^1$.

\subsection{Holomorphic Quantization and Non-symmetric Coadjoint Orbits}

In this Subsection we shall show that it is possible to interpret the
topological path integral (\ref{t2part}) as a character formula associated with
the quantization of a coadjoint orbit corresponding to some novel sort of spin
system described by $\Sigma^1$, as was the situation in all of the simply
connected cases of the last Section. For this, we examine the canonical quantum
theory defined by the symplectic structure (\ref{t2omega}) in the
Schr\"odinger picture representation. We first rewrite the symplectic
2-form (\ref{t2omega}) in complex coordinates to get the K\"ahler structure
\beq
\omega_{\Sigma^1}=\frac{v}{2i~{\rm Im}~\tau}dz\wedge d\bar z=-i\partial\bar
\partial F_{\Sigma^1}
\label{t2kahlerom}\eeq
with corresponding local K\"ahler potential
\beq
F_{\Sigma^1}(z,\bar z)=vz\bar z/2~{\rm Im}~\tau
\label{t2kahlerpot}\eeq
We then map the corresponding Poisson algebra onto the associated Heisenberg
algebra by the standard commutator prescription (c.f. beginning of Subsection
5.1). With this we obtain the quantum commutator
\beq
\left[\hat z,\hat{\bar z}\right]=2~{\rm Im}(\tau)/v
\label{t2comm}\eeq
We can represent the algebra (\ref{t2comm}) on the space ${\rm Hol}(\Sigma^1;
\tau)$ of holomorphic functions $\Psi(z)$ on $\Sigma^1$ by letting $\hat z$
act as multipication by the complex coordinate $z=\phi_1+\tau\phi_2$ and $\hat
{\bar z}$ as the derivative operator
\beq
\hat{\bar z}=-\frac{2~{\rm Im}~\tau}{v}\frac{\partial}{\partial z}
\label{zbarderiv}\eeq

With this holomorphic Schr\"odinger polarization, the operators
$\hat z$ and $\hat{\bar z}$ with the commutator algebra (\ref{t2comm})
resemble the creation and annihilation operators (\ref{agens}) of the
Heisenberg-Weyl algebra with the commutation relation (\ref{harmcomm}). In
analogy with that situation, we can construct the corresponding coherent states
\beq
|z)=\e^{(-v/2~{\rm Im}~\tau)z\hat{\bar z}}|0\rangle~~~;~~~z\in\Sigma^1
\label{t2coh}\eeq
which are normalized as
\beq
(z|z)=\e^{(v/2~{\rm Im}~\tau)z\bar z}=\e^{F_{\Sigma^1}(z,\bar z)}
\label{t2cohnorm}\eeq
and obey the completeness relation
\beq
\int_{\Sigma^1}\frac{d^2z}{(2\pi)^2}~|z)\!)(\!(z|={\bf1}
\label{t2cohcompl}\eeq
These coherent states are associated with the quantization of the
coadjoint orbit $U(1)\times U(1)=S^1\times S^1$. However, since $\Sigma^1$ is
a non-symmetric space, it cannot be considered as a K\"ahler manifold
associated with the coadjoint orbit of a semi-simple Lie group, as was the
case in the last Section. The orbits above are, however, associated with the
action of the isometry group $U(1)\times U(1)$ on $\Sigma^1$, which has
an interesting Lie algebraic structure that we shall discuss below.

In the Schr\"odinger representation (\ref{zbarderiv}), we consistently find
the action of the operator $\hat{\bar z}$ on the states (\ref{t2coh}) as
$\hat{\bar z}|z)$. The holomorphic representation space ${\rm Hol}(\Sigma^1;
\tau)$ in this context is then regarded as the space of entire functions
$\Psi(z)=(z|\Psi)$ for each state $|\Psi)$ in the span of the coherent states
(\ref{t2coh}). An inner product on ${\rm Hol}(\Sigma^1;\tau)$ is then
determined from the completeness relation (\ref{t2cohcompl}) and the
normalization (\ref{t2cohnorm}) as
\beq
(\Psi_1|\Psi_2)=\int_{\Sigma^1}\frac{d^2z}{(2\pi)^2}~\frac{(\Psi_1|z)
(z|\Psi_2)}{(z|z)}=\int_{\Sigma^1}\frac{d^2z}{(2\pi)^2}~\e^{-(v/2~{\rm
Im}~\tau)
z\bar z}\Psi_1^\dagger(\bar z)\Psi_2(z)
\label{t2cohinner}\eeq
With the inner product (\ref{t2cohinner}), we find that the operator
$\hat{\bar z}=\hat z^\dagger$ is the adjoint of $\hat z$, as it consistently
should be. An operator $\hat{\cal H}$ acting on the space of coherent states
(\ref{t2coh}) can now be represented on ${\rm Hol}(\Sigma^1;\tau)$ as usual by
an integral kernel as in (\ref{cohham}) with the identification of $\bar
z$ as the derivative operator (\ref{zbarderiv}). Furthermore, the quantum
propagator associated with such an operator
\beq
{\cal K}_{\Sigma^1}(\bar z',z;T)=(\!(z'|\e^{-iT\hat{\cal H}}|z)\!)
\label{sig1quprop}\eeq
determines the corresponding time-evolution of the coherent state wavefunctions
as
\beq
\Psi(z;T)=\int_{\Sigma^1}\frac{d^2z'}{(2\pi)^2}~\e^{-F_{\Sigma^1}(z',\bar
z')}{\cal K}_{\Sigma^1}(\bar z',z;T)\Psi(z')
\label{cohwaveev}\eeq
with $\Psi(z;t)\equiv(z,t|\Psi)$ and $\Psi(z)\equiv\Psi(z;0)$. In the following
we shall build up the initial states $\Psi(z)$, and then the associated time
evolution determined by the localizable Hamiltonians on $\Sigma^1$ (and hence
the solutions of the Schr\"odinger wave equation) are determined by the path
integral above for the propagator (\ref{sig1quprop}).

The advantage of working with the holomorphic representation space ${\rm Hol}
(\Sigma^1;\tau)$ is that we shall want to discuss the explicit structure of the
Hilbert space associated with the localizable quantum systems we found above.
With the K\"ahler structure defined by the symplectic 2-form $\omega_
{\Sigma^1}$ above, the Hilbert space of the quantum theory is then the space of
holomorphic sections of the usual symplectic complex line bundle
$L\rightarrow\Sigma^1$, which in this context is usually called
the prequantum line bundle over $\Sigma^1$. As such, $\omega_{\Sigma^1}$
represents the first Chern characteristic class of $L$, and so such a bundle
exists only if $\omega_{\Sigma^1}$ is an integral 2-form on $\Sigma^1$. This
method of quantizing the Hamiltonian dynamics in terms of the geometry of fiber
bundles is called geometric quantization \cite{woodhouse} and it is equivalent
to the Borel-Weil-Bott method of constructing coherent states that we
encountered in the last Section. In light of the requirement of
single-valuedness of the quantum propagator that we discussed in the Subsection
6.3, we require, from the point of view of equivariant localization, that the
wavefunctions $\Psi(z)$ change only by a unitary transformation under the
winding transformations (\ref{t2transls}) on $\Sigma^1$, so that all physical
quantities, such as the probability density $\Psi^\dagger\Psi$, are
well-defined $C^\infty$-functions on the phase space $\Sigma^1$ and respect the
symmetries of the quantum theory as defined by the quantum Hamiltonian, i.e. by
the supersymmetry making the dynamical system a localizable one. In this
setting, the multivalued wavefunctions, regarded as sections of the associated
line bundle $L\to\Sigma^1$ where the structure group
$\pi_1(\Sigma^1)=\IZ\oplus\IZ$ acts through a unitary representation, are
single-valued functions on the universal cover $\IC$ of the torus and so they
can be thought of as single-valued functions of homotopy classes $[\sigma]$ of
loops on $\Sigma^1$. This also ensures that the coherent states (\ref{t2coh})
remain coherent under the time evolution determined by the localizable
Hamiltonians of the last Subsection (i.e. under the action of ${\cal
I}(\Sigma^1;g_\tau)$) which will lead to a consistent coherent state path
integral representation of (\ref{t2part}).

To explore this in more detail, we need a representation for the discretized
equivariant Hamiltonian generators above of the isometry group ${\cal I}
(\Sigma^1;g_\tau)$ on the space ${\rm Hol}(\Sigma^1;\tau)$
\cite{bergeron,semsz}. This group action then coincides with the automorphisms
of the symplectic line bundle above in the usual way. Note that translations by
$a\in\IC$ on $z$ are generated on functions of $z$ by the action of the
operator $\e^{a\frac{\partial}{\partial z}}$, and
likewise on functions of $\bar z$ by $\e^{\bar a\frac{\partial}{\partial\bar
z}}$. On the holomorphic representation space ${\rm Hol}(\Sigma^1;\tau)$, we
represent the latter operator using the commutation relation (\ref{t2comm}) as
$\e^{(v/2~{\rm Im}~\tau)\bar az}$, in accordance with the coherent state
representation above. Thus the generators of large $U(1)$ transformations
around the homology cycles of $\Sigma^1$ in the holomorphic Schr\"odinger
polarization above are the unitary quantum operators
\beq
U(n,m)=\exp\left(2\pi(n+m\tau)\frac{\partial}{\partial z}+\frac{\pi v}
{{\rm Im}~\tau}(n+m\bar\tau)z\right)~~~;~~~n,m\in\IZ
\label{u1gens}\eeq
which generate simultaneously both of the winding transformations in
(\ref{t2transls}). By the above arguments, the quantum states should be
invariant (up to unitary equivalence) under their action on the Hilbert space.
Solving this invariance condition will then give a representation of the
equivariant localization symmetry constraints (i.e. of the pertinent
cohomological supersymmetry) and of the coadjoint orbit system directly in the
Hilbert space of the canonical quantum theory.

In contrast with their classical counterparts, the quantum operators
(\ref{u1gens}) do not commute among themselves in general and products of them
differ from their reverse-ordered products by a ${\bf u(1)}$-valued 2-cocycle
(see Appendix A). The Baker-Campbell-Hausdorff formula,
\beq
\e^{X+Y}=\e^{-[X,Y]/2}\e^X\e^Y~~~{\rm when}~~~[X,[X,Y]]=[Y,[X,Y]]=0
\label{bchform}\eeq
implies
\beq
\e^X\e^Y=\e^{[X,Y]}\e^Y\e^X
\label{bch2}\eeq
Applying (\ref{bch2}) to products of the operators (\ref{u1gens}) and using
the commutation relation (\ref{t2comm}) with (\ref{zbarderiv}), we find that
they obey what is called a clock algebra,
\beq
U(n_1,m_1)U(n_2,m_2)=\e^{2\pi iv(n_2m_1-n_1m_2)}U(n_2,m_2)U(n_1,m_1)
\label{clockalg}\eeq
To determine the action of the operators (\ref{u1gens}) explicitly on the
wavefunctions $\Psi(z)$, we apply the Baker-Campbell-Hausdorff formula
(\ref{bchform}) to get
\beq
U(n,m)=\exp\left[\frac{\pi v}{{\rm Im}~\tau}\left(\pi|n+m\tau|^2+(n+m\bar\tau)
z\right)\right]\e^{2\pi(n+m\tau)\frac{\partial}{\partial z}}
\label{unmderiv}\eeq
so that the action of (\ref{unmderiv}) on the quantum states of the theory is
\beq
U(n,m)\Psi(z)=\exp\left[\frac{\pi v}{{\rm Im}~\tau}\left(\pi|n+m\tau|^2
+(n+m\bar\tau)z\right)\right]\Psi(z+2\pi(n+m\tau))
\label{unmpsi}\eeq

If the volume parameter $v=~{\rm vol}_{g_\tau}(\Sigma^1)/(2\pi)^2$ is an
irrational number, then it follows from the clock algebra (\ref{clockalg})
that the $U(1)$ generators above act as infinite-dimensional raising operators
in (\ref{unmpsi}) and so the Hilbert space of quantum states in this case is
infinite-dimensional. However, we recall the necessary quantization
requirements for the parameters of the Hamiltonian system required for a
consistent quantum theory. With this in mind, we instead consider
the case where the volume of the torus is quantized so that
\beq
v=v_1/v_2~~~~~;~~~~~v_1,v_2\in\IZ^+
\label{volquant}\eeq
is rational-valued. Alternatively, such a discretization of $v$ is required
in order that the symplectic 2-form $\omega_{\Sigma^1}$ define an integer
cohomology class, as in (\ref{wesszum}). In this case, the cocycle relation
(\ref{clockalg}) shows that the operator $U(v_2n,v_2m)$ commutes with all of
the other $U(1)$ generators and the time evolution operator, and so they can be
simultaneously diagonalized over the same basis of states. This means that
their action (\ref{unmpsi}) on the wavefunctions must produce a state that lies
on the same ray in the Hilbert space as that defined by $\Psi(z)$, i.e.
\beq
U(v_2n,v_2m)\Psi(z)=\e^{i\eta(n,m)}\Psi(z)
\label{etadef}\eeq
for some phases $\eta(n,m)\in S^1$. The invariance condition (\ref{etadef}),
expressing the symmetry of the wavefunctions under the action of the
(non-simple) Lie group $U(1)\times U(1)$, is called a projective representation
of the symmetry group. It must obey a particular consistency
condition. The composition law for the group operations induces a composition
law for the phases in (\ref{etadef}),
\beq\new{\begin{array}{l}
U(v_2(n_1+n_2),v_2(m_1+m_2))\Psi(z)\\=U(v_2n_1,v_2m_1)U(v_2n_2,v_2m_2)\Psi(z)
\\=\e^{i\eta(n_1+n_2,m_1+m_2)}\Psi(z)\cdot\exp\left[i\{\eta(n_1,m_1)+
\eta(n_2,m_2)-\eta(n_1+n_2,m_1+m_2)\}\right]\end{array}}
\label{phasecons}\eeq
If the last phase in (\ref{phasecons}) vanishes, as in (\ref{alpharep}), then
the projective phase $\eta(n,m)$ is a 1-cocycle of the symmetry group
$U(1)\times U(1)$ and the wavefunctions carry a unitary representation of
the group, as required \cite{jackiw}. The determination of these 1-cocycles
explicitly below will then yield an explicit representation of the
homologically-invariant partition function (\ref{chipart}).

Comparing (\ref{etadef}) and (\ref{unmpsi}), we see that the invariance
of the quantum states under the $U(1)$ action on the phase space can be
expressed as
\beq
\Psi(z+2\pi v_2(n+m\tau))=\exp\left[i\eta(n,m)-\frac{\pi v_1}{{\rm Im}~\tau}
\left(\pi v_2|n+m\tau|^2+(n+m\bar\tau)z\right)\right]\Psi(z)
\label{quasiperiod}\eeq
The only functions which obey quasi-periodic conditions like
(\ref{quasiperiod}) are combinations of the Jacobi theta functions
\cite{greensw,mumford,schlich}
\beq
\Theta^{(D)}\pmatrix{c\cr d\cr}[z|\Pi]=\sum_{\{n^\ell\}\in\IZ^D}\exp\left[i\pi
(n^\ell+c^\ell)\Pi_{\ell p}(n^p+c^p)+2\pi i(n^\ell+c^\ell)(z_\ell+d_\ell)
\right]
\label{jacthetadef}\eeq
where $c^\ell,d_\ell\in[0,1]$. The functions (\ref{jacthetadef}) are
well-defined holomorphic functions of $\{z_\ell\}\in\IC^D$ for $D\times D$
complex-valued matrices $\Pi=[\Pi_{\ell p}]$ in the Siegal upper half-plane
(i.e. ${\rm Im}~\Pi>0$). They obey the doubly semi-periodic conditions
\beq
\Theta^{(D)}\pmatrix{c\cr d\cr}[z+s+\Pi\cdot t|\Pi]=\exp\left[2\pi ic^\ell
s_\ell-i\pi t^\ell\Pi_{\ell p}t^p-2\pi it^\ell(z_\ell+d_\ell)\right]\Theta
^{(D)}\pmatrix{c\cr d\cr}[z|\Pi]
\label{semiper1}\eeq
where $s=\{s_\ell\}$ and $t=\{t^\ell\}$ are integer-valued vectors, and
\beq
\Theta^{(D)}\pmatrix{c\cr d\cr}[z+\alpha\Pi\cdot
t|\Pi]=\exp\left[-i\pi\alpha^2t^\ell\Pi_{\ell p}
t^p-2\pi i\alpha t^\ell(z_\ell+d_\ell)\right]\Theta^{(D)}\pmatrix{c-\alpha t
\cr d\cr}[z|\Pi]
\label{semiper2}\eeq
for any non-integer constant $\alpha\in\IR$. We remark here that the
transformations in (\ref{semiper1}) can be applied in many different steps
with the same final result, but successive applications of (\ref{semiper1})
and (\ref{semiper2}) do not commute \cite{bergeron}. In the context of the
action of the unitary operators $U(n,m)$ above, when these transformations are
applied in different orders in (\ref{quasiperiod}), the final results differ by
a phase which forms a representation of the clock algebra (\ref{clockalg}). To
avoid this minor ambiguity, we simply {\it define} the operators $U(n,m)$ by
their action on the states $\Psi(z)$ with the convention that the
transformation (\ref{semiper1}) is applied before (\ref{semiper2}).

After some algebra, we find that the algebraic constraints (\ref{quasiperiod})
are uniquely solved by the $v_1v_2$ independent holomorphic wavefunctions
\beq
\Psi_{p,r}\pmatrix{c\cr d\cr}(z)=\e^{-(v/4~{\rm Im}~\tau)z^2}\Theta^{(1)}
\pmatrix{\frac{c+2\pi v_1p+v_2r}{2\pi v_1v_2}\cr d\cr}[v_1z|2\pi v_1v_2\tau]
\label{holwavefns}\eeq
where $p=1,2,\dots,v_2$ and $r=1,2,\dots,v_1$. The phases in (\ref{etadef}) are
then found to be the non-trivial 1-cocycles
\beq
\eta(n,m)/2\pi=cn-dm+\pi v_1v_2nm
\label{etaexpl}\eeq
of the global $U(1)\times U(1)$ group acting on $\Sigma^1$ here. Furthermore,
the winding transformations (\ref{unmpsi}) can be written as
\beq
U(n,m)\Psi_{p,r}\pmatrix{c\cr d\cr}(z)=\sum_{p'=1}^{v_2}[U(n,m)]_{pp'}
\Psi_{p',r}\pmatrix{c\cr d\cr}(z)
\label{windingrep}\eeq
where the finite-dimensional unitary matrices
\beq
[U(n,m)]_{pp'}=\exp\left[\frac{2\pi i}{v_2}\left(cn-dm+\pi v_1n(m+2p)\right)
\right]\delta_{p+m,p'}
\label{projrep}\eeq
form a $v_2$-dimensional projective representation, which is cyclic of
period $v_2$, of the clock algebra (\ref{clockalg}). The projective phase here
is the non-trivial $U(1)\times U(1)$ 1-cocycle
\beq
\eta^{(p)}(n,m)/2\pi=(cn-dm+\pi v_1n(m+2p))/v_2
\label{v2cocycle}\eeq
which could also therefore be used to construct an unambiguous partition
function as in the Subsection 6.3 by
\beq
Z_{\rm hom}(T)=\sum_{k,\ell=-\infty}^\infty\e^{i\eta(k,\ell)}Z_{\Sigma^1}^v(k,
\ell;N_T)
\label{parthometa}\eeq

Thus the Hilbert space is $v_1v_2$-dimensional
and the quantum states carry a $v_2$-dimensional projective representation
of the equivariant localization symmetries via the clock algebra
(\ref{clockalg}) which involves the ${\bf u(1)}$-valued 2-cocycle
\beq
\xi(n_1,m_1;n_2,m_2)/2\pi=v_1(n_2m_1-n_1m_2)/v_2
\label{clockalgcocycle}\eeq
of the $U(1)\times U(1)$ isometry group of $\Sigma^1$. This shows explicitly
how the $U(1)$ equivariant localization constraints and the topological
toroidal restrictions are realized in the canonical quantum theory, as then
these conditions imply that the only invariant operators on the Hilbert space
here are essentially combinations of the generators (\ref{u1gens}). In
particular, this implies, by construction, that the coherent state
wavefunctions (\ref{holwavefns}) are complete. This is much different than
the situation for the coherent states associated with simply-connected phase
spaces where there are no such topological symmetries to be respected for the
supersymmetric localization of the path integral and the Hilbert space is
1-dimensional. Intuitively, the finite-dimensionality of the Hilbert space
of physical states is expected from the compactness of the phase space
$\Sigma^1$.

Notice though that the wavefunctions (\ref{holwavefns}) contain the 2 free
parameters $c$ and $d$. We can eliminate one of them by requiring that
the Hamiltonian (\ref{t2eqhams}) in this basis of states does indeed lead
to the correct propagator (\ref{t2parttop}), i.e. that (\ref{t2parttop}) be
equal to the trace of the time evolution operator on the finite dimensional
vector space spanned by the coherent states (\ref{holwavefns}),
\beq
\tr\e^{-iN_T\hat H_{\Sigma^1}(k,\ell)/h}=\sum_{p=1}^{v_2}\sum_{r=1}^{v_1}(
\Psi_{p,r}|\e^{-iN_T\hat H_{\Sigma^1}(k,\ell)/h}|\Psi_{p,r})
\label{t2proptrace}\eeq
where the coherent state inner product is given by (\ref{t2cohinner}). With
this inner product, it is straightforward to show that the states
(\ref{holwavefns}) define an orthonormal basis of the Hilbert space,
\beq
(\Psi_{p_1,r_1}|\Psi_{p_2,r_2})=\delta_{p_1,p_2}\delta_{r_1,r_2}
\label{orthonormal}\eeq
Substituting the identity $\e^{-iN_T\hat H_{\Sigma^1}(k,\ell)/h}=[U(\ell,-k)]
^{N_Tv_2/2\pi v_1}$ into (\ref{t2proptrace}) and using (\ref{windingrep}),
(\ref{projrep}) and (\ref{orthonormal}) we find
\beq
\tr\e^{-iN_T\hat H_{\Sigma^1}(k,\ell)/h}=(-1)^{k\ell N_T}\e^{iN_T(ck+d\ell)
/v_1}
\label{propcohstates}\eeq
Comparing the result (\ref{propcohstates}) with the exact one
(\ref{t2parttop}), we find that the parameter $d$ appearing in the
wavefunctions (\ref{holwavefns}) can be determined as
\beq
d_{k\ell}=(k\ell N_T-2ck)/2\ell
\label{dexact}\eeq

Another way to eliminate the parameters $c$ and $d$ appearing in
(\ref{holwavefns}) is to regard the quantum theory as a topological field
theory. The above construction produces a Hilbert space ${\cal H}^\tau$ of
holomorphic sections of a complex line bundle $L^\tau\to\Sigma^1$ for each
modular parameter $\tau$. If we smoothly vary the complex structure $\tau$,
then this gives a family of finite-dimensional Hilbert spaces which can be
regarded as forming in this way a holomorphic vector bundle (i.e. one with a
holomorphic projection map) over the Teichm\"uller space $\IC^+$ of the torus
for which the projective representations above define a canonical
projectively-flat connection. This is a typical feature of the Hilbert space
for a Schwarz-type topological gauge theory \cite{birm}. Equivalent complex
structures (i.e. those which generate the same conformal equivalence classes as
(\ref{t2metric})) in the sense of the topological field theory of this Section
should be regarded as leading to the same quantum theory, and this should be
inherent in both the homological partition functions of the Subsection 6.3 and
in the canonical quantum theory above. It can be shown \cite{nash} that 2
toroidal complex structures $\tau,\tau'\in\IC^+$ define conformally equivalent
metrics (i.e. $g_\tau=\rho g_{\tau'}$ for some smooth function $\rho>0$) if and
only if they are related by the
projective transformation\footnote{\baselineskip=12pt That the 2 associated
tori are conformally isomorphic can be seen intuitively by representing each as
a
parallelogram in the complex plane and tracing out this transformation.}
\beq
\tau'=\frac{\alpha\tau+\beta}{\gamma\tau+\delta}~~~{\rm with}~~~\alpha,\beta,
\gamma,\delta\in\IZ~~,~~\alpha\delta-\beta\gamma=1
\label{discrmobius}\eeq
on $\IC^+\to\IC^+$. The transformations (\ref{discrmobius}) generate the
action of the group $SL(2,\IZ)/\IZ_2$ on $\IC^+$, which is a discrete
subgroup of the M\"obius group $SL(2,\IC)/\IZ_2$ of linear fractional
transformations of $\IC$ wherein we take $\alpha,\beta,\gamma,\delta\in\IC$ in
(\ref{discrmobius}). We call this discrete group the modular or mapping
class group $\Gamma_{\Sigma^1}$ of the Riemann surface $\Sigma^1$ and it
consists of the discrete automorphisms of $\Sigma^1$ (i.e. the conformal
diffeomorphisms of $\Sigma^1$ which aren't connected to the identity and
so cannot be represented as global flows of vector fields). The
Teichm\"uller space $\IC^+$ modulo this group action, i.e. the space of
inequivalent complex structures on $\Sigma^1$, is called the moduli space
$\man_{\Sigma^1}\equiv\IC^+/\Gamma_{\Sigma^1}$ of $\Sigma^1$. The topological
quantum theory above therefore should also reflect this sort of full
topological invariance on the torus, because it is independent of the conformal
factor $\varphi$ in (\ref{t2metric}).

Under the modular transformation (\ref{discrmobius}), it is possible to show
that, up to an overall phase, the 1-dimensional Jacobi theta functions in
(\ref{jacthetadef}) transform as \cite{mumford}
\beq
\Theta^{(1)}\pmatrix{c\cr d\cr}[z|\tau]\to\Theta^{(1)}\pmatrix{c'\cr d'\cr}
[z'|\tau']=\sqrt{\gamma\tau+\delta}\e^{i\pi z^2/(\gamma\tau+d)}\Theta^{(1)}
\pmatrix{c\cr d\cr}[z|\tau]
\label{thetamodular}\eeq
where
\beq
z'=\phi_1+\tau'\phi_2=z/(\gamma\tau+\delta)
\label{zprime}\eeq
is the new (but equivalent) complex structure defined by (\ref{discrmobius})
and the new parameters $c'$ and $d'$ are given by
\beq
c'=\delta c-\gamma d-\gamma\delta/2~~~~~,~~~~~d'=\alpha d-\beta c-\alpha\beta/2
\label{newcd}\eeq
Using (\ref{thetamodular}) we find after some algebra that the wavefunctions
(\ref{holwavefns}) transform under the modular transformation of isomorphic
complex structures as
\beq
\Psi_{p,r}\pmatrix{c\cr d\cr}(z)\to\frac{1}{\sqrt{\gamma\tau+\delta}}
\Psi_{p,r}\pmatrix{\tilde c'\cr\tilde d'\cr}(z')
\label{psimodular}\eeq
with
\beq
\tilde c'=\delta c-\gamma d-\pi v_1v_2\gamma\delta~~~~~,~~~~~\tilde d'=\alpha
d-\beta c-\pi v_1v_2\alpha\beta
\label{newtildecd}\eeq
It follows that a set of modular invariant wavefunctions can exist only when
the combination $v_1v_2$ is an even integer, in which case the
invariance condition requires $c=d=0$. For $v_1v_2$ an odd integer, we
can take $c,d\in\{0,\frac{1}{2}\}$, and then the holomorphic wavefunctions
carry a non-trivial spinor representation of the modular group as defined by
(\ref{psimodular}). These choices of $c$ and $d$ correspond to the 4
possible choices of spin structure on the torus \cite{greensw} (i.e.
representations of the 2-dimensional spinor group $U(1)$ in the tangent bundle
of $\Sigma^1$) which are determined by the mod 2 cohomology
$H^1(\Sigma^1;\IZ)\otimes\IZ_2=\IZ_2\oplus\IZ_2$ \cite{eguchi,greensw}. This
increases the number of basis wavefunctions (\ref{holwavefns}) by a factor of
4.

It is in this way that one may adjust the parameters $c$ and $d$ so that the
wavefunctions (\ref{holwavefns}) are modular invariants, as they should be
since the topological quantum theory defined by (\ref{t2part}) is independent
of the phase space complex structure. We note also that these specific
choices of the parameters in turn then fix the propagation time integers
$N_T$ by (\ref{dexact}), so that these topological requirements completely
determine the topological quantum field theory in this case. Thus one
can remove all apparent ambiguities here and obtain a situation
that parallels the topological quantum theories in the simply connected
cases, although now the emerging topological and group theoretical structures
are far more complicated. With these appropriate choices of
parameter values, the propagator (\ref{t2proptrace}) then coincides with
the coherent state path integral
\beq\new{\begin{array}{l}
Z_{\Sigma^1}^v(k,\ell;N_T)\sim\int_{L\Sigma^1}\prod_{t\in[0,T]}\frac{dz(t)~
d\bar z(t)}{(2\pi)^2}~\\\times\exp\left\{\frac{1}{2i~{\rm Im}~\tau}\int_0^{N_T
h^{-1}}dt~\left[\frac{v_1}{2v_2}\left(\bar z\dot z-z\dot{\bar z}\right)+ih
\left((\ell-\tau k)\bar z-(\ell-\bar\tau k)z\right)\right]\right\}\end{array}}
\label{t2cohpart}\eeq
The coherent state path integral (\ref{t2cohpart}) models the
quantization of some novel, unusual spin system defined by the Hamiltonians
(\ref{t2eqhams}) which are associated with the quantized, non-symmetric
coadjoint Lie group orbit $U(1)\times U(1)=S^1\times S^1$. This abelian
orbit is an unreduced one as it already is its own maximal torus, and we can
therefore think of this spin system as 2 independent planar spins each tracing
out a circle. The points on this orbit are in one-to-one
correspondence with the coherent state representations above of the
projective clock algebra (\ref{clockalg}) of the discrete first homology
group of the torus. The associated character formula represented by
(\ref{t2cohpart}) gives path integral representations of the homology classes
of $\Sigma^1$, in accordance with the fact that it defines a topological
quantum field theory, and these localizable quantum systems are exactly
solvable via both the functional integral and canonical quantization
formalisms, as above. In this latter formalism, the Hilbert space of physical
states is finite-dimensional and the basis states carry a non-trivial
projective representation of the first homology group of the phase space, in
addition to the usual representation of $H^2(\man;\IZ)$.

\subsection{Generalization to Hyperbolic Riemann Surfaces}

We conclude this Section by indicating how the above features of equivariant
localization could generalize to the case where the phase space is a hyperbolic
Riemann surface \cite{semsz}, although our conclusions are somewhat heuristic
and more care needs to be exercised in order to study these examples in
detail. Since for $h>1$, $\Sigma^h$ can be regarded as $h$ tori
stuck together, its homotopy can be described by the $2h$ loops $a_i,b_i$,
$i=1,\dots,h$, where each pair $a_i,b_i$ encircle the 2 holes
of the $i$-th torus in the connected sum representation of $\Sigma^h$.
The constraint (\ref{abelian}) on the fundamental homotopy generators now
generalizes to
\beq
\prod_{i=1}^ha_ib_ia_i^{-1}b_i^{-1}={\bf1}
\label{nonabelian}\eeq
and so the commutator subgroup of $\pi_1(\Sigma^h)$ for $h>1$ is non-trivial
and the fundamental group of a hyperbolic Riemann surface is non-abelian.
Its first homology group is given by (\ref{homologysig}), and, using an
abusive notation, we shall denote its generators as well by $a_i,b_i$, $i=1,
\dots,h$ and call them a canonical basis of homology cycles for $\Sigma^h$.

According to the Riemann uniformization theorem \cite{nash}, there are only
3 (compact or non-compact) simply-connected Riemann surfaces -- the 2-sphere
$S^2$, the plane $\IC$ and the Poincar\'e upper half-plane ${\cal H}^2$, each
equipped with their standard metrics as discussed in the last Section. The
sphere is its own universal cover (being simply-connected and having
a unique complex structure), while $\IC$ is the universal cover of the torus.
The hyperbolic plane ${\cal H}^2$ is always the universal cover of a Riemann
surface of genus $h\geq2$, which is represented as
\beq
\Sigma^h={\cal H}^2/F_h
\label{sigmahcover}\eeq
where $F_h=\pi_1(\Sigma^h)$ is in this context refered to as a discrete
Fuchsian group. The quotient in (\ref{sigmahcover}) is by the fixed-point
free bi-holomorphic action of $F_h$ on ${\cal H}^2$. The group of analytic
automorphisms of the upper half-plane ${\cal H}^2$ is $PSL(2,\IR)=SL(2,\IR)/
\IZ_2$, the group of projective linear fractional transformations as in
(\ref{discrmobius}) except that now the coefficients $\alpha,\beta,\gamma,
\delta$ are taken to be real-valued. Then $\pi_1(\Sigma^h)$ is taken as a
discrete subgroup of this $PSL(2,\IR)$-action on ${\cal H}^2$ and the different
isomorphism classes of complex analytic structures of $\Sigma^h$ are
essentially the different possible classes of discrete subgroups. Note that
this generalizes the genus 1 situation above, where the automorphism group of
$\IC$ was the group $PSL(2,\IC)$ of global conformal transformations in
2-dimensions and $\pi_1(\Sigma^1)$ was taken to be the lattice subgroup.
Indeed, it is possible to regard $\Sigma^h$ as a $4h$-gon in the plane with
edges identified appropriately to generate the $h$ `holes' in $\Sigma^h$.

It is difficult to generalize the explicit constructions of the last few
Subsections because of the complicated, abstract fashion in (\ref{sigmahcover})
that the complex coordinatization of $\Sigma^h$ occurs. For the various ways
of describing the Teichm\"uller space and Fuchsian groups of hyperbolic
Riemann surfaces without the explicit introduction of local coordinates, see
\cite{imayoshi}. The Teichm\"uller space of $\Sigma^h$ can be naturally
given the geometric structure of a non-compact complex manifold which is
homeomorphic to $\IC^{3h-3}$, so that the coordinatization of $\Sigma^h$ is
far more intricate for $h\geq2$ because it now involves $3h-3$ complex
parameters, as opposed to just 1 as before. Nonetheless, it is still possible
to deduce the unique localizable Hamiltonian system on a hyperbolic
Riemann surface and deduce some general features of the ensuing topological
quantum field theory just as we did above.

We choose a complex structure on $\Sigma^h$ for which the universal bundle
projection in (\ref{sigmahcover}) is holomorphic (as for the torus), and then
the metric $g_{\Sigma^h}$ induced on $\Sigma^h$ by this projection involves
a globally-defined conformal factor $\varphi$ as in (\ref{t2metric}) and a
constant negative curvature K\"ahler metric (the hyperbolic Poincar\'e metric
-- see Subsection 5.6). The condition now that the Killing vectors of this
metric be globally-defined on $\Sigma^h$ means that they must be single-valued
under windings around the canonical homology cycles $\{a_\ell,b_\ell\}_{\ell
=1}^h\in H_1(\Sigma^h;\IZ)$, or equivalently that
\beq
\oint_{a_\ell}dV^\mu=\oint_{b_\ell}dV^\mu=0~~~~~,~~~~~\ell=1,\dots,h
\label{vmusingle}\eeq
Using this single-valued condition and the Killing equations
\beq
g_{\Sigma^h}(dV,\cdot)=-i_Vdg_{\Sigma^h}
\label{killingsigh}\eeq
we can now deduce the general form of the Killing vectors of $\Sigma^h$. For
this, we apply the Hodge decomposition theorem \cite{birm,bottu} to the
metric-dual 1-form $g(V,\cdot)\in\Lambda^1\Sigma^h$,
\beq
g(V,\cdot)=d\chi+\star d\xi+\lambda_h
\label{hodgedecomp}\eeq
where $\chi$ and $\xi$ are $C^\infty$-functions on $\Sigma^h$ and $\lambda_h$
is a harmonic 1-form, i.e. a solution
of the zero-mode Laplace equation for 1-forms,
\beq
\Delta_1\lambda_h\equiv(\star d\star d+d\star d\star)\lambda_h=0
\label{laplaceeq1form}\eeq

In the above, $\star$ denotes the Hodge duality operator and, on a general
$d$-dimensional Riemannian manifold $(\man,g)$, it encodes the Riemannian
geometry directly into the DeRham cohomology. It is defined as the map
\beq
\star:\Lambda^k\man\to\Lambda^{d-k}\man
\label{hodgestar}\eeq
which is given locally by
\beq
\star\alpha=\frac{1}{(d-k)!}\sqrt{\det g(x)}~g_{j_1\lambda_1}(x)\cdots
g_{j_{d-k}\lambda_{d-k}}(x)\epsilon^{\lambda_1\dots\lambda_{d-k}i_1\dots
i_k}\alpha_{i_1\dots i_k}(x)~dx^{j_1}\wedge\cdots\wedge dx^{j_{d-k}}
\label{stardef}\eeq
and satisfies
\beq
(\star)^2=(-1)^{(d-1)k}
\label{starsq}\eeq
on $\Lambda^k\man$. It defines an inner product $\int_\man\alpha\wedge\star
\beta$ on each vector
space $\Lambda^k\man$. Using this inner product it is possible to show that
a differential form $\lambda_h$ as above is harmonic if and only if
\beq
d\lambda_h=d\star\lambda_h=0
\label{harmonicformdef}\eeq
and the Hodge decomposition theorem (\ref{hodgedecomp}) (which can be
generalized to arbitrary degree differential forms in the general case) implies
that the DeRham cohomology groups of $\man$ are spanned by a basis of harmonic
forms.

The Hodge decomposition (\ref{hodgedecomp}) is unique and the components
involved there are explicitly given by
\beq
\chi=\frac{1}{\nabla_{\Sigma^h}^2}\star d\star g(V,\cdot)~~~~~,~~~~~\xi=
\frac{1}{\nabla_{\Sigma^h}^2}\star d g(V,\cdot)
\label{chixiexpl}\eeq
where the scalar Laplacians $\nabla^2_{\Sigma^h}\equiv\star d\star d$ in
(\ref{chixiexpl}) are assumed to have their zero modes removed. The 1-form
$\lambda_h$ in (\ref{hodgedecomp}) can be written as a linear combination of
basis elements of the DeRham cohomology group $H^1(\Sigma^h;\IR)$. According to
the Poincar\'e-Hodge duality theorem \cite{bottu}, we can in particular choose
an orthonormal basis of harmonic 1-forms $\{\alpha_\ell,\beta_\ell\}_{\ell=1}
^h\in H^1(\Sigma^h;\IR)$ which are Poincar\'e-dual to the chosen canonical
homology basis $\{a_\ell,b_\ell\}_{\ell=1}^h\in H_1(\man;\IZ)$ above, i.e.
\beq
\oint_{a_\ell}\alpha_{\ell'}=\oint_{b_\ell}\beta_{\ell'}=\delta_{\ell\ell'}
{}~~~~~,~~~~~\oint_{b_\ell}\alpha_{\ell'}=\oint_{a_\ell}\beta_{\ell'}=0
\label{poincaredual}\eeq
We remark here that the local parts of the decomposition (\ref{hodgedecomp})
simply form the decomposition of the vector $g_{\Sigma^h}(V,\cdot)$ into its
curl-free, longitudinal and divergence-free, transverse pieces as $\nabla_{
\Sigma^h}\chi+\nabla_{\Sigma^h}\times\xi$. The harmonic part $\lambda_h$
accounts for the fact that this 1-form may sit in a non-trivial DeRham
cohomology class of $H^1(\Sigma^h;\IR)$.

We can now write the general form of the isometries of $\Sigma^h$. As before,
$\Sigma^h$ inherits 3 local isometries via the bundle projection in
(\ref{sigmahcover}) from the maximally symmetric Poincar\'e upper half-plane.
However, only the 2 quasi-translations on ${\cal H}^2$ become global isometries
of $\Sigma^h$, and they can be expressed in terms of the canonical homology
basis using the above relations. This global isometry condition along with
(\ref{vmusingle}) and (\ref{killingsigh}) imply that Hodge decomposition
(\ref{hodgedecomp}) of the metric-dual 1-form to the Hamiltonian vector
field $V_{\Sigma^h}$ is simply given by its harmonic part which can be
written as
\beq
g_{\Sigma^h}(V_{\Sigma^h},\cdot)=\sum_{\ell=1}^h\left(V_1^\ell\alpha_\ell+V_2^
\ell\beta_\ell\right)
\label{killingharm}\eeq
The harmonic decomposition (\ref{killingharm}) is the generalization of
(\ref{constmodes}). Indeed, on the torus we can identify the
canonical harmonic forms above as $\alpha=d\phi_1/2\pi$ and $\beta=d\phi_2
/2\pi$. The Killing vectors dual to (\ref{killingharm}) generate translations
along the homology cycles of $\Sigma^h$, and the isometry group of $\Sigma^h$
is $\prod_{i=1}^{2h}U(1)$. The usual equivariance condition
$\lie_{V_{\Sigma^h}}\omega=0$ on the symplectic 2-form of $\Sigma^h$ now
becomes
\beq
di_{V_{\Sigma^h}}\omega=\sum_{\ell=1}^hd\left(\bar\omega\star\left\{V_1^\ell
\alpha_\ell+V_2^\ell\beta_\ell\right\}\right)=0
\label{sighsympl}\eeq
where $\bar\omega(x)$ is the $C^\infty$-function on $\Sigma^h$ defined
by $\omega_{\mu\nu}(x)=\bar\omega(x)\epsilon_{\mu\nu}$, and (\ref{sighsympl})
implies that it is constant on $\Sigma^h$, just as in (\ref{t2omega}).

Integrating up the Hamiltonian equations we see therefore that the unique
equivariant (Darboux) Hamiltonians have the form
\beq
H_{\Sigma^h}(x)=\sum_{\ell=1}^h\int_{C_x}\left(h_1^\ell\alpha_\ell+h_2^\ell
\beta_\ell\right)
\label{sigheqhams}\eeq
where $h_\mu^\ell$ are real-valued constants and $C_x\subset\Sigma^h$ is a
simple curve from some fixed basepoint to $x$. The Hamiltonian
(\ref{sigheqhams}) is multi-valued because it depends explicitly on the
particular representatives $\alpha_\ell,\beta_\ell$ of the DeRham cohomology
classes in $H^1(\Sigma^h;\IR)=\IR^{2h}$. As before, single-valuedness of the
time-evolution operator requires that $h_\mu^\ell=n_\mu^\ell h$, for some
$n_\mu^\ell\in\IZ$ and $h\in\IR$, and the propagation times are again the
discrete intervals $T=N_Th^{-1}$. Thus the Hamiltonian (\ref{sigheqhams})
represents the windings around the non-trivial homology cycles of $\Sigma^h$
and the partition function defines a topological quantum field theory which
again represents the homology classes of $\Sigma^h$ through a family of
homomorphisms from $\bigoplus_{i=1}^{2h}\IZ$ into $U(1)^{\otimes2h}$. Again,
the
partition function path integral should be properly defined in the
homologically-invariant form (\ref{chipart}) to make the usual quantities
appearing in the associated action $S$ well-defined by restricting the
functional integrations to homotopically equivalent loops. We note that
again the general conformal factor involved in the metric $g_{\Sigma^h}$
obeys Riemannian restrictions from the Gauss-Bonnet-Chern theorem and
a volume constraint similar to those in Subsection 6.2 above. When the volume
parameter is quantized as in (\ref{volquant}), we expect that the Hilbert
space of physical states will be $(v_1v_2)^{3h-3}$ dimensional (one copy
of the genus 1 Hilbert spaces for each of the $3h-3$ modular degrees of
freedom in this case) and the coherent state wavefunctions, which can be
expressed in terms of $D=3h-3$ dimensional Jacobi theta functions
(\ref{jacthetadef}), will in addition carry a $(v_2)^{3h-3}$ dimensional
projective representation of the discrete first homology group of $\Sigma^h$
(i.e. of the equivariant localization constraint algebra). The explicit proofs
of all of the above facts appear to be difficult, because of the lack of
complex
coordinatization for these manifolds which is required for the definition
of coherent states associated with the isometry group action $\prod_{i=1}^{2h}
U(1)=\prod_{i=1}^{2h}S^1$ on the non-symmetric space $\Sigma^h=(S^1\times
S^1)^{\#h}$.

Thus the general feature of abelian equivariant localization of path integrals
on multiply connected compact Riemann surfaces is that it leads to a
topological quantum theory whose associated topologically invariant partition
function represents the non-trivial homology classes of the phase space. The
coherent states in the finite-dimensional Hilbert space also carry a
multi-dimensional representation of the discrete first homology group, and the
localizable Hamiltonians on these phases spaces are rather unusual and even
more restricted than in the simply-connected cases. The invariant
symplectic 2-forms in these cases are non-trivial elements of $H^2(\Sigma^h;
\IZ)=\IZ$, as in the maximally-symmetric cases, and it is essentially the
global topological features of these multiply-connected phase spaces which
leads to these rather severe restrictions. The coherent state quantization of
these systems shows that the path integral describes the coadjoint orbit
quantization of an unusual spin system described by the Riemann surface. These
spin systems are exactly solvable both from the point of view of path integral
quantization on the loop space and of canonical holomorphic quantization in
the Schr\"odinger polarization. The localizable systems that one obtains in
these cases are rather trivial in appearence and are associated with {\it
abelian} isometry groups acting on the phase spaces. However, these
quantum theories probe deep geometric and topological features of the phase
spaces, such as their complex algebraic geometry and their homology. This
is in contrast with the topological quantum field theories that we found in
the simply-connected cases, where at best the topological path integral
could only represent the possible non-trivial cohomology classes in
$H^2(\man;\IZ)$. It is not completely clear though how these path integral
representations correspond to analogs of the standard character formulas on
homogeneous symplectic manifolds which are associated with semi-simple Lie
groups, since, for instance, the usual K\"ahler structure between the
Riemannian and symplectic geometries is absent in these non-symmetric cases.

\section{Beyond the Semi-classical Approximation}

In this Section we shall examine a different approach to the problem of
localization \cite{me}. We return to the general finite-dimensional analysis of
Section 3 and consider a Hamiltonian system whose Hamiltonian function is a
Morse function\footnote{\baselineskip=12pt The extension to
degenerate Hamiltonians is fairly straightforward. In what follows all
statements made concerning the structure of the discrete critical point set
$\man_V$ of $H$ will then apply to the full critical submanifold.}. From this
we will construct the full $\frac{1}{T}$-expansion for the classical partition
function, as we described briefly in Subsection 3.3. A proper covariantization
of this expansion will then allow us to determine somewhat general geometrical
characteristics of dynamical systems whose partition functions localize, which
in this context will be the vanishing of all terms in the perturbative loop
expansion beyond 1-loop order. The possible advantages of this analysis are
numerous. For instance, we can analyse the fundamental isometry condition
required for equivariant localization and see more precisely what mechanism or
symmetry makes the higher-order terms disappear. This could then expand the set
of localizable systems beyond the ones we have already encountered that are
predicted from localization theory, and at the same time probe deeper into the
geometrical structures of the phase space or the whole dynamical system thus
providing richer examples of topological field theories. Indeed
we shall find some noteworthy geometrical significances of when a partition
function is given exactly by its semi-classical approximation as well as new
geometric criteria for localization which expand the previous isometry
conditions.

This approach to the Duistermaat-Heckman integration formula using the
perturbative loop-expansion has been discussed in a different context recently
in \cite{yasui} where the classical partition function was evaluated for the
dynamical system describing the kinematics of thin vortex tubes in a
3-dimensional fluid whose Hamiltonian is similar to that considered at the end
of Subsection 5.8 for geodesic motion on group manifolds. For such fluid
mechanics problems, the phase space $\man$ is neither finite-dimensional nor
compact and the Hamiltonian flows need not be periodic, but the dynamical
system admits an infinite sequence of constants of motion which are in
involution so that it should be localizable \cite{yasui}. The standard
localization analyses therefore do not apply and one needs to resort to an
analysis of the sort which will follow here. We shall indeed find extensions of
the localization formalisms which cover certain such cases.

Recalling that the isometry condition can always be satisfied at least
{\it locally} on $\man$, we then present some ideas towards developing a novel
geometric method for systematically constructing corrections to the
Duistermaat-Heckman formula. Given that a particular system does not localize,
the idea is that we can ``localize" in local neighbourhoods on $\man$ where the
Killing equation can be satisfied. The correction terms are then picked up when
these open sets are patched back together on the manifold, as then there are
non-trivial singular contributions to the usual 1-loop term owing to the fact
that the Lie derived metric tensor cannot be defined globally in a smooth way
over the entire manifold $\man$. Recalling from Subsection 3.6 that the
properties of such a metric tensor are intimately related to the integrability
properties of the dynamical system, we can explore the integrability problem
again in a (different) geometric setting now by closely examining these
correction terms. This will provide a highly non-trivial geometric
classification of the localizability of a dynamical system which is related to
the homology of $\man$, the integrability of the dynamical system, and is
moreover completely consistent with Kirwan's theorem. Although these ideas are
not yet fully developed, they do provide a first step to a full analysis of
corrections to path integral localization formulas (e.g. corrections to the WKB
approximation), and to uncovering systematically the reasons why these
approximations aren't exact for certain dynamical systems \cite{me}.
The generalizations of these ideas to path integrals are not yet known, but
we discuss the situation somewhat heuristically at the end of this Section.

\subsection{Geometrical Characterizations of the Loop Expansion}

Throughout this Section we return to the situation of Subsection 3.3 where the
Hamiltonian $H$ is a Morse function on a (usually compact) symplectic
manifold $\man$. For now we assume that $\partial\man=\emptyset$, but later
we shall also consider manifolds with boundary. We now explicitly work out
the full stationary phase series whose construction we briefly outlined in
Subsection 3.3 \cite{hormander}. We first expand the $C^\infty$-function $H$ in
a neighbourhood $U_p$ of a given critical point $p\in\man_V$ in a Taylor
series
\beq
H(x)=H(p)+\hess(p)_{\mu\nu}x_p^\mu x_p^\nu/2+g(x;p)~~~~~,~~~~~x\in U_p
\label{hxtaylor}\eeq
where $x_p=x-p\in U_p$ are the fluctuation modes about the extrema of $H$ and
$g(x;p)$ is the Gaussian deviation of $H(x)$ in the neighbourhood $U_p$ (i.e.
all terms in the Taylor series beyond quadratic order). The determinant of the
symplectic 2-form which appears in (\ref{classpart}) is similarly expanded in
$U_p$ as
\beq
\sqrt{\det\omega(x)}=\sqrt{\det\omega(p)}+\sum_{k=1}^\infty\frac{1}{k!}
x_p^{\mu_1}\cdots x_p^{\mu_k}\partial_{\mu_1}\cdots\partial_{\mu_k}\sqrt{\det
\omega(x)}\Bigm|_{x=p}~~~,~~~x\in U_p
\label{omegataylor}\eeq

We substitute (\ref{hxtaylor}) and (\ref{omegataylor}) into (\ref{classpart}),
expand the exponential function there in powers of the Gaussian deviation
function, and then integrate over each of the neighbourhoods
$U_p\simeq\IR^{2n}$. In this way we arrive at a series expansion of
(\ref{classpart}) for large-$T$ in terms of Gaussian moment integrals over the
fluctuations $x_p$, with Gaussian weight $\e^{iT\hess(p)_{\mu\nu}x_p^\mu
x_p^\nu/2}$, associated with each open neighbourhood $U_p$ for $p\in\man_V$.
The Gaussian moments $\langle x^{\mu_1}\cdots x^{\mu_k}\rangle$ can be found
from the Gaussian integration formula (\ref{gaussintclass}) in the usual way by
applying the operator
$\frac{\partial}{\partial\lambda_{\mu_1}}\cdots\frac{\partial}{
\partial\lambda_{\mu_k}}$ to both sides of (\ref{gaussintclass}) and then
setting all the $\lambda$'s equal to 0. The odd-order moments vanish, since
these integrands are odd functions, and the $2k$-th order moment contributes a
term of order ${\cal O}(1/T^{n+k})$. Rearranging terms carefully, taking
into account the signature of the Hessian at each critical point, and noting
that for large-$T$ the integral will localize around each of the disjoint
neighbourhoods $U_p$, we arrive at the standard stationary-phase
expansion\footnote{\baselineskip=12pt See \cite{hormander} for the
generalization of this formula to the case where $H$ is a degenerate function.}
\beq
Z(T)=\left(\frac{2\pi i}{T}\right)^n\sum_{p\in\man_V}(-i)^{\lambda(p)}
\e^{iTH(p)}\sum_{\ell=0}^\infty\frac{A_\ell(p)}{(-2T)^\ell}
\label{loopexp}\eeq
where
\beq
A_{\ell}(p)=\frac{1}{\sqrt{\det\hess(p)}}\sum_{j=0}^{2\ell}\frac{(-1)^j}{2^jj!
(\ell+j)!}\left(\hess(p)^{\mu\nu}\partial_\mu\partial_\nu\right)^{\ell+j}
\left(g(x;p)^j\sqrt{\det\omega(x)}\right)\biggm|_{x=p}
\label{aellp}\eeq
and $\hess(x)^{\mu\nu}$ is the matrix inverse of $\hess(x)_{\mu\nu}$.

If the stationary-phase series diverges (e.g. applying Kirwan's
theorem in appropriate instances), then (\ref{loopexp}) is to be understood
formally as an asymptotic expansion order by order in $\frac{1}{T}$.
Borrowing terminology from quantum field theory, we shall refer to the series
(\ref{loopexp}) as the loop-expansion of this zero-dimensional quantum field
theory, because each of the $2\ell+1$ terms in
(\ref{aellp}) can be understood from pairing fluctuation modes $x_p^{\mu}x_p^
\nu$ (i.e. a loop) associated with each derivative operator there. Indeed,
the expansion (\ref{loopexp}),(\ref{aellp}) is just the finite-dimensional
version of the perturbation expansion (for large-$T$) in quantum field theory.
We shall call the ${\cal O}(1/T^{n+\ell})$ contribution to the series
(\ref{loopexp}) the $(\ell+1)$-loop term.

In this Subsection we shall be interested in extracting information about the
dynamical system under consideration from the loop-expansion with the hope of
being able to understand the vanishing or non-vanishing of the $k$-loop
contributions for $k>1$ in terms of geometrical and topological features of the
phase space. Given our experience now with the Duistermaat-Heckman theorem, we
will remove any requirements on the flows of the Hamiltonian vector field and
leave these as quite arbitrary for now. When these orbits describe tori, we
already have a thorough understanding of the localization in terms of
equivariant cohomology, and we shall therefore look at dynamical systems which
do not necessarily obey this requirement. Thus any classification that we
obtain below that is described solely by the vanishing of higher-loop
contributions will for the most part be of a different geometrical nature than
the situation that prevails in Duistermaat-Heckman localization. This then has
the possibility of expanding the cohomological symmetries usually resposible
for localization.

The perturbative series (\ref{loopexp}), however, must be appropriately
modified before we can put it to use. The partition function (\ref{classpart})
is invariant under arbitrary smooth changes of local coordinates on $\man$
(i.e. $Z(T)$ is manifestly a topological invariant), and this should be
reflected order by order in the $1/T$-expansion (\ref{loopexp}). This is
explicitly observed in the lowest-order Duistermaat-Heckman term $A_0(p)$
above, but the higher-order terms (\ref{aellp}) in the loop expansion are not
manifestly scalar quantities under local diffeomorphisms of the coordinates.
This is a result of having to pick local coordinates on $\man$ to carry out
explicitly the Gaussian integrations in $\IR^{2n}$. At each
order of the $\frac{1}{T}$-expansion we should have a manifestly coordinate
independent quantity, i.e. a scalar. To write the contributions
(\ref{aellp}) in such a fashion so as to be manifestly invariant under local
diffeomorphisms of $\man$, we have to introduce a Christoffel connection
$\Gamma^\lambda_{\mu\nu}$ of the tangent bundle of $\man$ which makes the
derivative operators appearing in (\ref{aellp}) manifestly covariant
objects, i.e. we write them in terms of covariant derivatives
$\nabla=d+\Gamma$.
Because $dH(p)=0$ at a critical point $p\in\man_V$, the Hessian evaluated at
a critical point is automatically covariant, i.e. $\nabla\nabla H(p)=\hess(p)$.
This process, which we shall call `covariantization', will then ensure that
each term (\ref{aellp}) is manifestly a scalar. We note that the Morse index
of any critical point is a topological invariant in this sense.

First, we cycle out the symplectic factors in (\ref{aellp}) to get
\beq
A_\ell(p)=A_0(p)\sum_{j=0}^{2\ell}\frac{(-1)^j}{2^jj!(\ell+j)!}\left(\hess(p)
^{\mu\nu}{\cal D}_\mu{\cal D}_\nu\right)^{\ell+j}g(x;p)^j\biggm|_{x=p}
\label{aellcald}\eeq
where
\beq
A_0(p)=\sqrt{\frac{\det\omega(p)}{\det\hess(p)}}
\label{a0p}\eeq
is the Duistermaat-Heckman (1-loop) contribution to (\ref{loopexp}), and
\beq
{\cal D}=d+\gamma
\label{cald}\eeq
where we have introduced the one-component connection
\beq
\gamma=h_L^{-1}dh_L
\label{gammadef}\eeq
and
\beq
h_L(x)=\sqrt{\det\omega(x)}
\label{liouvilledens}\eeq
is the Liouville volume density. The derivative operator $\cal D$ transforms
like an abelian gauge connection under local diffeomorphisms $x\to x'(x)$
of $\man$,
\beq
{\cal D}_\mu(x)~{\buildrel\Lambda\over\longrightarrow}~{\cal D}_\mu'(x')
=\Lambda_\mu^\nu(x')\left[{\cal D}_\nu(x')+~\tr~\Lambda^{-1}(x')\partial_\nu'
\Lambda(x')\right]
\label{gaugetransf}\eeq
where
\beq
\Lambda^{-1}(x)=\left[\frac{\partial x'^\mu(x)}{\partial x^\nu}\right]\in
GL(2n,\IR)
\label{lambdadef}\eeq
is the induced change of basis transformation on the tangent bundle.

For example, consider the expression (\ref{aellcald}) for $\ell=1$. We expand
out the 3 terms there in higher-order derivatives of $H$ and the connection
$\gamma$, noting that only third- and higher-order derivatives of $g(x;p)$
when evaluated at $x=p$ are non-vanishing. After some algebra, we arrive at
\beq\new{\begin{array}{ll}
A_1(p)=&\frac{A_0(p)}{2}\hess(p)^{\mu\nu}\left\{{\cal D}_\mu(x)\gamma_\nu(x)
-\frac{\hess(p)^{\lambda\rho}}{4}\biggl(\partial_\mu\partial_
\lambda\partial_\nu
\partial_\rho H(x)+4\gamma_\mu(x)\partial_\lambda\partial_\nu\partial_\rho H(x)
\right.\biggr.\\&
\left.\left.+\frac{\hess(p)^{\alpha\beta}}{12}\Bigl[3\partial_\mu\partial_
\lambda\partial_\nu H(x)\partial_\rho\partial_\alpha\partial_\beta H(x)+2
\partial_\mu\partial_\lambda\partial_\alpha H(x)\partial_\nu\partial_\rho
\partial_\beta H(x)\Bigr]\right)\right\}\biggm|_{x=p}\end{array}}
\label{2loop}\eeq
It is readily checked, after some algebra, that this expression is indeed
invariant under local diffeomorphisms of $\man$. To manifestly covariantize it,
we introduce an arbitrary torsion-free connection $\Gamma_{\mu\nu}^\lambda$
of the tangent bundle $T\man$. For now, we need not assume that $\Gamma$ is
the Levi-Civita connection associated with a Riemannian metric on $\man$.
Indeed, as the original dynamical problem is defined only in terms of a
symplectic geometry, not a Riemannian geometry, the expression (\ref{2loop})
should be manifestly covariant in its own right without reference to any
geometry that is external to the problem. All that is required is some
connection that specifies parallel transport along the fibers of the tangent
bundle and allows us to extend derivatives of quantities to an entire
neighbourhood, rather than just at a point, in a covariant way.

The Hessian of $H$ can be written in terms of this connection and the
associated covariant derivative as
\beq
\hess(x)_{\mu\nu}=\nabla_\mu\nabla_\nu H(x)+\Gamma^\lambda_{\mu\nu}(x)\partial
_\lambda H(x)
\label{covhess}\eeq
and, using $d\equiv\nabla-\Gamma$, we can write the third and fourth order
derivatives appearing in (\ref{2loop}) in terms of $\nabla$ and $\Gamma$ by
taking derivatives of (\ref{covhess}). Substituting these complicated
derivative
expressions into (\ref{2loop}) and using the fact that $dH=0$ on $\man_V$,
after a long and quite tedious calculation we arrive at a manifestly covariant
and coordinate-independent expression for the 2-loop correction,
\beq\new{\begin{array}{ll}
A_1(p)=&\frac{A_0(p)}{8}\hess(p)^{\mu\nu}\left\{\frac{\hess(p)^{\lambda\rho}
\hess(p)^{\alpha\beta}}{3}\Bigl[3\nabla_\mu\nabla_\nu\nabla_\lambda H(x)\nabla
_\alpha\nabla_\beta\nabla_\rho H(x)\Bigr.\right.\\&\Bigl.+2\nabla_\mu\nabla_
\lambda\nabla_\alpha H(x)\nabla_\nu\nabla_\rho\nabla_\beta H(x)\Bigr]
-\hess(p)^{\lambda\rho}\nabla_\rho\nabla_\mu\nabla_\nu\nabla_\lambda H(x)\\&
\left.+4\left(\nabla_\mu+\Delta_\mu-\hess(p)^{\lambda\rho}\nabla_\rho\nabla
_\mu\nabla_\lambda H(x)\right)\Delta_\nu(x)+R_{\mu\nu}(\Gamma)\right\}\biggm
|_{x=p}\end{array}}
\label{2loopcov}\eeq
where
\beq
R_{\mu\nu}(\Gamma)\equiv R^\lambda_{\mu\lambda\nu}=\partial_\nu\Gamma^\lambda
_{\mu\lambda}-\partial_\lambda\Gamma^\lambda_{\mu\nu}+\Gamma^\alpha_{\mu
\lambda}\Gamma^\lambda_{\alpha\nu}-\Gamma^\alpha_{\mu\nu}\Gamma^\lambda_{
\alpha\lambda}
\label{ricci}\eeq
is the (symmetric) Ricci curvature tensor of $\Gamma$ and we have introduced
the
1-form $\Delta=\Delta_\mu(x)dx^\mu$ with the local components
\beq
\Delta_\mu(x)=\gamma_\mu(x)-\Gamma^\lambda_{\mu\lambda}(x)=\nabla_\mu\log
h_L(x)
\label{deltaform}\eeq
It is intriguing that the covariantization of the 2-loop expression simply
involves replacing ordinary derivatives $d$ with covariant ones $\nabla$,
non-covariant connection terms $\gamma$ with the 1-form $\Delta$, and then
the remainder terms from this process are simply determined by the curvature
of the Christoffel connection $\Gamma$ which realizes the covariantization.
Note that if $\Gamma$ is in addition chosen as the Levi-Civita connection
compatible with a metric $g$, i.e. $\nabla g=0$, then
$\Gamma^\lambda_{\mu\lambda}=\partial_\mu\log\sqrt{\det g}$ and the 1-form
components (\ref{deltaform})
become $\Delta_\mu=\partial_\mu\log\sqrt{\det(g^{-1}\cdot\omega)}$.

The covariantization of the higher-loop terms is hopelessly complicated. We
note that if, however, the 2-loop correction $A_1(x)$ vanishes in an entire
neighbourhood $U_p\subset\man$ of each critical point $p$, then this
is enough to imply the vanishing of the corrections to the Duistermaat-Heckman
formula to all orders in the loop-expansion. To see this, we exploit the
topological invariance of (\ref{2loopcov}) and apply the Morse lemma
\cite{nash} to the correction terms (\ref{aellcald}). This says that there
exists a sufficiently small neighbourhood $U_p$ about each critical point $p$
in which the Morse function $H$ looks like a ``harmonic oscillator",
\beq
H(x)=H(p)-(x^1)^2-(x^2)^2-\dots-(x^{\lambda(p)})^2+(x^{\lambda(p)+1})^2+\dots
+(x^{2n})^2~~~,~~~x\in U_p
\label{morsecoords}\eeq
so that the critical point $p$ is at $x=0$ in this open set in $\man$. We shall
call these `harmonic coordinates', and this result simply means that the
symmetric matrix $\hess(x)$ can be diagonalized constantly in an entire
neighbourhood of the critical point. Given that the quantity (\ref{aellcald})
must be independent of coordinates (although not manifestly), we can evaluate
it in a harmonic coordinate system. Then the Gaussian deviation function
$g(x;p)$ vanishes identically in the neighbourhood $U_p$ and only the $j=0$
term contributes to (\ref{aellcald}). Thus the series (\ref{loopexp}) is simply
\beq
Z(T)=\left(\frac{2\pi i}{T}\right)^n\sum_{p\in\man_V}(-i)^{\lambda(p)}
\e^{iTH(p)}A_0(p)\left(\e^{-\frac{\hess(p)^{\mu\nu}}{2T}{\cal D}_\mu(x){\cal D}
_\nu(x)}\right)\cdot1\Bigm|_{x=p}
\label{loopexpmorse}\eeq
It follows that if the 2-loop term vanishes in the entire {\it neighbourhood}
of the critical point $p$ (and not just {\it at} $x=p$), i.e.
\beq
\hess(p)^{\mu\nu}{\cal D}_\mu(x){\cal D}_\nu(x)\equiv0~~~{\rm for}~~~x\in U_p
\label{neighvan}\eeq
then, as all higher-loop terms in these coordinates can be written as
derivative operators acting on the 2-loop contribution $A_1(x)$ as prescribed
by (\ref{loopexpmorse}), all corrections to the semi-classical approximation
vanish. In the case of the Duistermaat-Heckman theorem, it is for this reason
that the Lie derivative condition $\lie_Vg=0$ is generally required to hold
globally on $\man$. In general, though, whether or not the vanishing of
$A_1(p)$ implies the vanishing of all loop orders is not that
clear, because there is a large ambiguity in the structure of a function
in a neighbourhood of $\man_V$ which vanishes at each critical point $p$ (any
functional of $\nabla H$ will do). The above vanishing property in an entire
neighbourhood therefore need not be true. It is hard to imagine though that the
vanishing of the 2-loop correction term would not imply the vanishing of all
higher-orders, because then there would be an infinite set of conditions that
a dynamical system would have to obey in order for its partition function to
be WKB-exact. This would then greatly limit the possibilities for localization.

In any case, from the point of view of localization, we can consider the
vanishing of the 2-loop contribution in (\ref{2loopcov}) as an infinitesimal
Duistermaat-Heckman localization of the partition function.
The expression (\ref{2loopcov}) in general is extremely complicated. However,
besides being manifestly independent of the choice of coordinates,
(\ref{2loopcov}) is independent of the chosen connection $\Gamma$, because by
construction it simply reduces to the original connection-independent term
(\ref{2loop}). We can exploit this degree of freedom by choosing a connection
that simplifies the correction (\ref{2loopcov}) to a form that is amenable to
explicit analysis. We shall now describe 2 geometrical characteristics of the
loop-expansion above which can be used to classify the localizable or
non-localizable dynamical systems \cite{me}.

The first such general geometric localization symmetry is a symmetry implied by
(\ref{2loopcov}) between Hessian and metric tensors, i.e. that the Hessian
essentially defines a metric on $\man$. This is evident in the correction term
(\ref{2loopcov}), where the inverse Hessians contract with the other tensorial
terms to form scalars, i.e. the Hessians in that expression act just like
metrics. This suggests that the non-degenerate Hessian of $H$ could be used to
define a metric which is compatible with the connection $\Gamma$ used in
(\ref{2loopcov}). This in general cannot be done globally on the
manifold $\man$, because the signature of $\hess(x)$ varies over $\man$ in
general, but for a $C^\infty$ Hamiltonian $H$ it can at least be done locally
in a sufficiently small neighbourhood surrounding each critical point. For
now, we concentrate on the case of a 2-dimensional phase space. We {\it define}
a Riemannian metric tensor $g$ that is proportional to the covariant Hessian in
the neighbourhood $U_p$ of each critical point $p$,
\beq
\nabla\nabla H={\cal G}g
\label{metrican}\eeq
where ${\cal G}(x)$ is some globally-defined $C^\infty$-function on $\man$, and
for which the connection $\Gamma$ used in the covariant derivatives $\nabla$
is the Levi-Civita connection for $g$. This means that, given a Hamiltonian
$H$ on $\man$, we try to locally solve the coupled system of non-linear partial
differential equations
\beq\new{\begin{array}{c}
{\cal G}(x)g_{\mu\nu}(x)=\partial_\mu\partial_\nu H(x)-\Gamma^\lambda_{\mu\nu}
(x)\partial_\lambda H(x)\\\Gamma^\lambda_{\mu\nu}=\frac{1}{2}g^{\lambda\rho}
\left(\partial_\mu g_{\rho\nu}+\partial_\nu g_{\rho\mu}-\partial_\rho
g_{\mu\nu}\right)\end{array}}
\label{ansatzeqs}\eeq
consistently for $g$ (or $\Gamma$).

This sort of ``metric ansatz" may seem somewhat peculiar, and indeed impossible
to solve in the general case. The covariant
constancy condition on $g$ in (\ref{ansatzeqs}) implies that
\beq
\partial_\mu{\cal G}=R^\lambda_\mu\partial_\lambda H=R\partial_\mu H
\label{gansatzeq}\eeq
where $R=g^{\mu\nu}R_{\mu\nu}$ is the scalar curvature of $g$.
(\ref{gansatzeq})
follows from the defining identity for the Riemann curvature tensor
$R=d\Gamma+\Gamma\wedge\Gamma$,
\beq
\nabla_\mu\nabla_\nu\nabla_\lambda H=\nabla_\nu\nabla_\mu\nabla_\lambda H
+R^\rho_{\lambda\mu\nu}\nabla_\rho H
\label{3covderivs}\eeq
Given $H$, (\ref{gansatzeq}) determines $\cal G$ locally in terms of $g$. This
means that the above ansatz can be written as an equation for the associated
connection coefficients $\Gamma^\lambda_{\mu\nu}$,
\beq
\nabla_\lambda\nabla_\mu\nabla_\nu H=R_{\mu\nu}\nabla_\lambda H
\label{conneqsan}\eeq
The existence of a local solution to (\ref{conneqsan}) `almost' all of the time
can now be argued by analysing this set of differential equations in local
isothermal coordinates (\ref{isotherm}) for the connection $\Gamma$ above
\cite{me}. Notice, in particular, that if the metric (\ref{metrican}) actually
has a constant curvature $R$, then the equation (\ref{gansatzeq}) can be
integrated in each neighbourhood $U_p$ to give
\beq
{\cal G}(x)=C_0+R\cdot H(x)
\label{Gconstcurv}\eeq
We shall see examples and other evidences for this sort of geometric structure
throughout this Section.

The main advantage of using the inductively-defined metric in (\ref{ansatzeqs})
is that it allows a relatively straightforward analysis of the 2-loop
correction terms to the Duistermaat-Heckman formula. With this definition,
(\ref{gansatzeq}) implies that the first order derivatives of $\cal G$ vanish
at each critical point $p\in\man_V$, and hence so do all third order covariant
derivatives of $H$ in (\ref{2loopcov}). The fourth order covariant derivatives
contribute curvature terms according to (\ref{conneqsan}), which are then
cancelled by the curvature tensor already present in (\ref{2loopcov}). The
final result is an expression involving only the Liouville and Levi-Civita
connections expressed in terms of the 1-form $\Delta$, which
after some algebra we find can be written in the simple form
\beq
A_1(p)=\frac{A_0(p)}{2{\cal
G}}g^{\mu\nu}\left[\nabla_\mu+\Delta_\mu(x)\right]\Delta_\nu(x)\biggm|_{x=p}
\label{A1Delta}\eeq
Requiring this correction term to vanish in an entire neighbourhood of each
critical point implies, from the definition (\ref{deltaform}), that the
connection $\gamma$ associated with the symplectic structure coincides with the
connection $\Gamma$ associated with the Riemannian structure which solves the
relation (\ref{ansatzeqs}). Thus the components of the symplectic 2-form
$\omega$ and the metric tensor $g$ are proportional to each other in local
complex isothermal coordinates for $g$. The proportionality factor must be a
constant so that this be consistent with the existence of local Darboux
coordinates for $\omega$ \cite{me}. In other words, $\omega$ and $g$ together
locally define a K\"ahler structure on the phase space $\man$.

Conversely, suppose that $\Gamma$ is the Levi-Civita connection
associated to some generic, globally-defined metric tensor $g$ on $\man$, and
consider the rank (1,1) tensor field
\beq
J_\mu^\nu=\sqrt{\frac{\det\omega}{\det g}}g_{\mu\lambda}\omega^{\lambda\nu}
\label{complexstruc}\eeq
In 2-dimensions, it is easily seen that (\ref{complexstruc}) defines a
linear isomorphism $J:T\man\to T\man$ satisfying $J^2=-{\bf1}$. In general, if
such a linear transformation $J$ exists then it is called an almost complex
structure of the manifold $\man$ \cite{eguchi,greensw}. This means that there
is a local basis of tangent vectors in which the only non-vanishing components
of $J$ are given by (\ref{complexstrucdef}), so that there is ``almost" a
separation of the tangent bundle into holomorphic and anti-holomorphic
components. However, an almost complex structure does not necessarily lead to a
complex structure -- there are certain sufficiency requirements to be met
before $J$ can be used to define local complex coordinates in which the overlap
transition functions can be taken to be holomorphic \cite{birm}. One such case
is when $J$ is covariantly constant, $\nabla J=0$ -- actually this condition
only ensures that a sub-collection of subsets of the differentiable structure
determine a local complex structure (but recall that any Riemann surface is a
complex manifold). Again in 2-dimensions this means that then $\nabla\omega=0$
and the pair $(g,\omega)$ define a K\"ahler structure on $\man$ (again note
that any 2-dimensional symplectic manifold is automatically a K\"ahler manifold
for {\it some} metric defined by $\omega$).

Given these facts, suppose now that $g$ and $\omega$ define a K\"ahler
structure on the $2n$-dimensional manifold $\man$ with respect to an
almost complex structure $J$, i.e. $\det\omega=\det g$, $g$ is Hermitian
with respect to $J$,
\beq
g_{\mu\nu}=J^\lambda_\mu g_{\lambda\rho}J^\rho_\nu~~~~~,
\label{gherm}\eeq
and $\omega$ is determined from $g$ by (\ref{complexstruc}). In the local
coordinates (\ref{complexstrucdef}), this means the usual K\"ahler conditions
that we encountered before, i.e. $g_{\mu\nu}=g_{\bar\mu\bar\nu}=0$,
$g_{\mu\bar\nu}=g^*_{\bar\nu\mu}$ and $\omega_{\mu\bar\nu}=-ig_{\mu\bar\nu}$.
In this case, the flows of $g$ under the action of the Hamiltonian vector field
$V$,
\beq
(\lie_V g)_{\mu\nu}=g_{\mu\lambda}\nabla_\nu\omega^{\lambda\rho}\partial_\rho
H+g_{\nu\lambda}\nabla_\mu\omega^{\lambda\rho}\partial_\rho H+\omega^{\lambda
\rho}(g_{\mu\lambda}\nabla_\nu\nabla_\rho H+g_{\nu\lambda}\nabla_\mu\nabla
_\rho H)
\label{lievgom}\eeq
can be written using the almost complex structure as the anti-commutator
\beq
\lie_Vg=\left[\nabla\nabla H,J\right]_+
\label{lieanticomm}\eeq
Thus if $V$ is a global Killing vector of a K\"ahler metric on $\man$, then
the covariant Hessian of $H$ is also Hermitian with respect to $J$, as in
(\ref{gherm}). Since the K\"ahler metric of a K\"ahler manifold is essentially
the unique Hermitian rank (2,0) tensor, it follows that the covariant Hessian
is related to the K\"ahler metric by a transformation of the form
\beq
\nabla_\mu\nabla_\nu H=K_\mu^\lambda g_{\lambda\rho}K^\rho_\nu
\label{hessgk}\eeq
where $K$ is some non-singular (1,1) tensor which commutes with $J$. In
2-dimensions, the Hermiticity conditions imply that both the Hessian and
$g$ have only 1 degree of freedom and (\ref{hessgk}) gets replaced by the
much simpler condition (\ref{ansatzeqs}). From the fundamental equivariant
localization principle we know that this implies the vanishing of the 2-loop
correction term, i.e. the Duistermaat-Heckman theorem. Indeed, from the
analysis of the last 2 Sections we have seen that most of the localizable
examples fall into these K\"ahler-type scenarios. Notice that the covariant
Hessian determined from the Hamiltonian equations is
\beq
\nabla_\mu\nabla_\nu H=V^\lambda\nabla_\mu \omega_{\nu \lambda} +
\omega_{\nu \lambda} \nabla_\mu V^\lambda
\label{HessFromHamil}\eeq
so that in the K\"ahler case, when $\nabla\omega=0$, the proportionality
function $\cal G$ is determined in terms of the Riemann moment map
$\mu_V=\nabla V$ as
\beq
{\cal G}(x)=\sqrt{\det\mu_V(x)}
\eeq
On a homogeneous K\"ahler manifold, when $\cal G$ is integrated to be
(\ref{Gconstcurv}), this relation generalizes that observed for the height
function of the sphere in Subsection 5.5.

What is particularly interesting here is that conversely the localization of
the partition function determines this sort of K\"ahler structure on $\man$.
The Lie derivative (\ref{lievgom}) of the metric (\ref{ansatzeqs}) is
easily seen to be zero in a neighbourhood of the critical point. Conversely,
if the Lie derivative of the metric in (\ref{ansatzeqs}) vanishes on
$\man$, then it induces a K\"ahler structure (i.e. $\nabla\omega=0$). Recalling
from Subsection 3.3 the proof of the Duistermaat-Heckman theorem
using solely symplectic geometry arguments, we see that the main feature of the
localization is the possibility of simultaneously choosing harmonic and Darboux
coordinates. This same feature occurs similarly above, when we map onto local
Darboux coordinates \cite{me}. The new insight gained here is the geometric
manner in which this occurs -- the vanishing of the loop expansion beyond
leading order gives the dynamical system a local K\"ahler structure (see
(\ref{neighvan})). Whether or not this extends to a {\it global} K\"ahler
geometry depends on many things. If the topology of $\man$ allows this to be
globally extended away from $\man_V$, then the Riemannian geometry so
introduced induces a global K\"ahler structure with respect to the canonical
symplectic
structure of $\man$ \footnote{\baselineskip=12pt The existence of an almost
complex structure $J$ for which the symplectic 2-form $\omega$ is Hermitian and
for which the associated K\"ahler metric $g=J\cdot\omega$ is positive-definite
is not really an issue for a symplectic manifold \cite{witten3}. Such a $J$
always exists (and is unique up to homotopy) because the Siegal upper-half
plane is contractible. Thus the existence of a K\"ahler structure for which
$\Delta
=\nabla\log h_L=0$ is not a problem. However, for the Killing equation for
$g=J\cdot\omega$ to hold, $J$ itself must be invariant under the flows of the
Hamiltonian vector field $V$, i.e. $\lie_VJ=0$.}. Furthermore, the
coefficient function ${\cal G}(x)$ in (\ref{metrican}) must be so that the
metric defined by that equation has a constant signature on the whole of
$\man$. If $H$ has odd Morse indices $\lambda(p)$, then it is impossible to
choose the function $\cal G$ in (\ref{metrican}) such that, say, $g$ has a
uniform Euclidean signature on the whole of $\man$. But if the correction terms
above vanish, then Kirwan's theorem implies that $H$ has only even Morse
indices and it may be possible to extend this geometry globally. In this way,
the examination of the vanishing of the loop expansion beyond leading order
gives insights into some novel geometrical structures on the phase space
representing symmetries of the localization. Moreover, if such a metric is
globally-defined on $\man$, then $\lie_Vg=0$, and these classes of localizable
systems fall into the same framework as those we studied before.

The second general geometric symmetry of the loop expansion that we wish to
point out here is based on the observation that the symplectic connection
(\ref{gammadef}) is reminescent of the connection that appears when one
constructs the Fubini-Study metric using the geometry of a holomorphic line
bundle $L\to\man$ \cite{eguchi}. If we choose such a line bundle as the
standard symplectic line bundle over $\man$ and view the Liouville density
(\ref{liouvilledens}) as a metric in the fibers of this bundle, then from it
one can construct a K\"ahler structure on $\man$ from the curvature 2-forms of
the associated connections (\ref{gammadef}), i.e.
\beq
\Omega=-i(\partial+\bar\partial)\gamma_{\bar z}=-i\partial\bar\partial\log h_L
\label{fubinistudygen}\eeq
where we have restricted to the holomorphic and anti-holomorphic components
of the connection (\ref{gammadef}). For instance, the symplectic structure
(\ref{fubinistudy}) is the Fubini-Study metric associated with the natural line
bundle $L\to\IC P^N$ which is a sub-bundle of the trivial bundle $\IC
P^N\times\IC^{N+1}$, i.e. the fiber of $L$ over a point $p\in\IC P^N$ is just
the set of points $(z^1,\dots,z^{N+1})\in\IC^{N+1}$ which belong to the line
$p$. In that case, the natural fiber metric induced by the canonical complex
Euclidean metric of $\IC^{N+1}$ is
$h(p;z^1,\dots,z^{N+1})=\sum_{\mu=1}^{N+1}|z^\mu|^2$ and (\ref{fubinistudy}) is
determined as the symplectic reduction of $h$ as in (\ref{fubinistudygen}).
This construction generalizes to any holomorphic line bundle once a fiber
metric has been chosen. In the general case, if $\man$ itself is already a
K\"ahler manifold, then whether or not (\ref{fubinistudygen}) agrees with the
original K\"ahler 2-form will depend on where these 2-forms sit in the DeRham
cohomology of $\man$. If we further adjust the Christoffel connection $\Gamma$
so that it is related to the Liouville connection by the boundary condition
$\Gamma^\lambda_{\mu\lambda}=\gamma_\mu$, i.e. $\Delta=0$, then the correction
term (\ref{2loopcov}) will involve only derivatives of the Hamiltonian, but now
the vanishing of the correction term can be related to the geometry of a line
bundle $L\to\man$.

This is as far as we shall go with a general geometric interpretation of
localization from the covariant loop expansion. The above discussion shows what
deep structures could be uncovered from further development of such analyses.
It is in this way that the localization of the partition function probes the
geometry and topology of the phase space $\man$ and thus can lead to
interesting topological quantum field theories.

\subsection{Conformal and Geodetic Localization Symmetries}

In this Subsection we shall examine some alternative geometric symmetries which
lead to a localization of the partition function of a dynamical system and
which extend the fundamental symmetry requirement of the localization
theorems we encountered earlier on \cite{karki,paniak}. Let $g$ be a
globally-defined metric tensor on $\man$, and consider
its flows under the Hamiltonian vector field $V$. Instead of the usual
assumption that $V$ be an infinitesimal isometry generator for $g$, we weaken
this requirement and assume that instead $V$ is globally an infinitesimal
generator of {\it conformal} transformations with respect to $g$, i.e.
\beq
\lie_Vg=\Upsilon g
\label{conflie}\eeq
where $\Upsilon(x)$ is some $C^\infty$-function on $\man$. Intuitively, this
means that the diffeomorphisms generated by $V$ preserve angles in the space,
but not lengths. The function $\Upsilon$ can be explicitly determined
by contracting both sides of (\ref{conflie}) with $g^{-1}$ to get
\beq
\Upsilon=\nabla_\mu V^\mu/n=\nabla_\mu\omega^{\mu\nu}\partial_\nu H/n
\label{upsilon}\eeq
which we note vanishes on the critical point set $\man_V$ of the Hamiltonian
$H$. This implies, in particular, that either $\Upsilon\equiv0$ almost
everywhere on $\man$ (in which case $V$ is an isometry of $g$) or $\Upsilon(x)$
is a non-constant function on $\man$ corresponding to non-homothetic
or `special conformal' transformations (i.e. constant rescalings, or
dilatations, of $g$ are not possible under the flow of a Hamiltonian vector
field). Ordinary Killing vector fields in this context arise as those which are
covariantly divergence-free, $\tr~\mu_V=\nabla_\mu V^\mu=0$.

We shall show that this conformal symmetry requirement on the dynamical system
also leads to the Duistermaat-Heckman integration formula. First, we establish
this at the level of the perturbative loop expansion of the last Subsection by
showing that the (infinitesimal) corrections to the 1-loop term in the
stationary phase series vanish. For simplicity we restrict our attention here
to the case of $n=1$ degree of freedom. The extension to arbitrary degrees of
freedom is immediate, and indeed we shall shortly see how the condition
(\ref{conflie}) explicitly implies the vanishing of the correction terms
to the Duistermaat-Heckman formula in arbitrary dimensions and in a much more
general framework. We insert everywhere into the covariant
connection-independent expression (\ref{2loopcov}) the covariant
derivative $\nabla$ associated with $g$ in (\ref{conflie}). Using the covariant
Hessian (\ref{HessFromHamil}) and the conformal Killing equation
(\ref{conflie}) we can solve for the covariant derivatives of the Hamiltonian
vector field. Notice that, in contrast to the ordinary Killing equations, one
of the 3 components of the conformal Killing equation (\ref{conflie}) will be
an identity since one of the Killing equations tells us that $V$ is covariantly
divergence-free with respect to $g$ (see (\ref{killingeqcomplex})). In
2-dimensions, after some algebra we find that the (symmetric) Hessian
(\ref{HessFromHamil}) can be written as
\beq
\nabla_\mu\nabla_\nu H=-\Omega~\Sigma~g_{\mu\nu}+\Bigl(\nabla_\mu H~\nabla_\nu
\log\Omega+\nabla_\nu H~\nabla_\mu\log\Omega\Bigr)/2
\label{ckvhess}\eeq
where we have introduced the $C^\infty$-functions
\beq
\Sigma(x)=\sqrt{\det\nabla V(x)-(\nabla_\lambda V^\lambda(x)/2)^2}~~~~~,~~~~~
\Omega(x)=\sqrt{\frac{\det\omega(x)}{\det g(x)}}
\label{Omega}\eeq

The covariant derivatives appearing in (\ref{2loopcov}) are now easily found
from (\ref{ckvhess}) to be
\beq
\nabla_\rho\nabla_\mu\nabla_\nu H=\frac{1}{\Omega\Sigma}\nabla_\rho(\Omega
\Sigma)\nabla_\mu\nabla_\nu H+\frac{1}{2}\nabla_\mu\log\Omega\nabla_\rho
\nabla_\nu H+\frac{1}{2}\nabla_\nu\log\Omega\nabla_\rho\nabla_\mu H+{\cal F}
(\nabla H)
\label{3covckv}\eeq
\beq\new{\begin{array}{l}
\nabla_\lambda\nabla_\rho\nabla_\mu\nabla_\nu H\\=\frac{1}{\Omega\Sigma}\left(
\nabla_\lambda\nabla_\rho(\Omega\Sigma)\nabla_\mu\nabla_\nu H+\frac{1}{2}
\nabla_\lambda(\Sigma\nabla_\mu\Omega)\nabla_\rho\nabla_\nu H-\frac{1}{2}
\nabla_\lambda(\Sigma\nabla_\nu\Omega)\nabla_\rho\nabla_\mu H\right)\\
+\frac{1}{2}\nabla_\lambda\nabla_\rho H\nabla_\mu\log\Omega\nabla_\nu\log\Omega
+\frac{1}{2}\nabla_\lambda\nabla_\nu H\left(\frac{1}{2}\nabla_\mu\log\Omega
\nabla_\rho\log\Omega+\nabla_\rho\nabla_\mu\log\Omega\right)\\+\frac{1}{2}
\nabla_\lambda\nabla_\mu H\left(\frac{1}{2}\nabla_\nu\log\Omega\nabla_\rho\log
\Omega+\nabla_\rho\nabla_\nu\log\Omega\right)+{\cal F}(\nabla H)\end{array}}
\label{4covckv}\eeq
where ${\cal F}(\nabla H)$ denotes terms involving single derivatives of $H$,
and which therefore vanish on $I(H)$. Substituting (\ref{3covckv}) into
(\ref{3covderivs}) and then contracting (\ref{3covderivs}) with
$\hess(x)^{\mu\lambda}$ leads to a relationship between the
symplectic structure and the flows generated by the Hamiltonian vector field,
\beq
\nabla_\mu\log\Omega+2\nabla_\mu\log\Sigma={\cal F}(\nabla H)
\label{reorderid}\eeq
We now use a covariant derivative of (\ref{3covderivs}) to rewrite the
curvature
term in (\ref{2loopcov}) in terms of covariant derivatives of the Hamiltonian,
substitute (\ref{3covckv}) and (\ref{4covckv}) into (\ref{2loopcov}), and
use everywhere the identity (\ref{reorderid}). Since $\Delta_\mu=\nabla_\mu
\log\Omega$ for the metric connection $\Gamma$, it follows after some algebra
that the generally-covariant expression (\ref{2loopcov}) for the first-order
correction to the Duistermaat-Heckman formula is ${\cal F}(\nabla H)$. This
establishes the claim above.

We shall now establish quite generally that this conformal localization
symmetry is in some sense the most general one that can be construed for a
classical partition function \cite{paniak}. For this, we return to the general
localization principle of Subsection 2.5 and re-evaluate the function ${\cal
Z}(s)$ in (\ref{zs}) for a generic (not necessarily equivariant) metric-dual
1-form $\beta=i_Vg$ of the Hamiltonian vector field $V$ (i.e. with no
assumptions for now about the symmetries of $g$) and the equivariantly-closed
form $\alpha=\e^{iT(H+\omega)}/(iT)^n$ with respect to $V$, i.e. $D_V\alpha=0$,
but $\lie_V\beta\neq0$ in general. We also assume, for full generality, that
the manifold $\man$ can have a boundary $\partial\man$. The derivative
(\ref{zsindeps}) need not vanish now and the second equality there can be
evaluated using Stokes' theorem and the Cartan-Weil identity. Then using the
identity
\beq
{\cal Z}(0)=\lim_{s\to\infty}{\cal Z}(s)-\int_0^\infty ds~\frac{d}{ds}{\cal
Z}(s)~~~~~
\label{derivid}\eeq
it follows that the partition function $Z(T)=\int_\man\alpha$ can be determined
in general by
\beq
Z(T)=\lim_{s\to\infty}\int_\man\alpha
\e^{-sD_V\beta}+\int_0^\infty ds~\left\{\oint_{\partial\man}\alpha\beta
\e^{-sD_V\beta}-s\int_\man\alpha\beta(\lie_V\beta)
\e^{-sD_V\beta}\right\}
\label{2ndhalf}\eeq
The large-$s$ limit integral in (\ref{2ndhalf}) can be worked out in the same
way as we did in Subsection 2.6 to arrive at the same expression
(\ref{intloc2}), except that now the 2-form $\Omega_V=d\beta$ there is given
quite arbitarily as
\beq
\Omega_V=2g\cdot\mu_V-\lie_Vg
\label{Omegalien0}\eeq
so that
\beq
\lim_{s\to\infty}\int_\man\alpha\e^{-sD_V\beta}=(-2\pi)^{n/2}
\sum_{p\in\man_V}\frac{\alpha^{(0)}(p)}{|\det
dV(p)|}~\pfaff\left(dV(p)-g(p)^{-1}\lie_Vg(p)/2\right)
\label{largesintgen}\eeq

Using (\ref{dvpomhess}) to rewrite derivatives of $V$ in (\ref{largesintgen})
in the usual way, substituting in the definitions in Subsection 2.5 for the
geometric quantities in (\ref{2ndhalf}), and then expanding exponentials in the
2-forms $\omega$ and $\Omega_V$ to the highest degrees that the manifold
integrations pick up, we arrive at an expression for the partition function in
terms of geometrical characteristics of the phase space \cite{paniak}
\beq\new{\begin{array}{ll}
Z(T)=&\left(\frac{2\pi i}{T}\right)^n\sum_{p\in\man_V}(-i)^{\lambda(p)}
\sqrt{\frac{\det\omega(p)}{\det\hess(p)}}\e^{iTH(p)}\sqrt{\det\left(
{\bf1}-\hess^{-1}\omega g^{-1}\lie_Vg/2\right)(p)}\\&+\frac{1}{(iT)^n}\int_0
^\infty ds~\oint_{\partial\man}\frac{\e^{iTH-sK_V}}{(n-1)!}g(V,\cdot)\wedge
\left(iT\omega-s\Omega_V\right)^{n-1}\\&-\frac{1}{(iT)^n}\int_0^\infty ds~s
\int_\man\frac{\e^{iTH-sK_V}}{(n-1)!}g(V,\cdot)\wedge(\lie_Vg)(V,\cdot)\wedge
\left(iT\omega-s\Omega_V\right)^{n-1}\end{array}}
\label{classpartgen}\eeq
which holds for an arbitrary Riemannian metric $g$ on $\man$. From the
expression (\ref{classpartgen}) and the fact that
$g(V,\cdot)^{\wedge2}\equiv0$, we can see explicitly now how the conformal Lie
derivative condition (\ref{conflie}) collapses this expression down to the
Duistermaat-Heckman formula (\ref{dhformula}) when $\partial\man=\emptyset$,
as we saw by more explicit means above. Note that we cannot naively carry out
the $s$-integrations quite yet, because the function $K_V=g(V,V)$ has zeroes on
$\man$. The expression (\ref{classpartgen}), although quite complicated, shows
explicitly how the Lie derivative conditions make the semi-classical
approximation to the partition function exact. This is in contrast to the
loop-expansion we studied earlier, where the corrections to the
Duistermaat-Heckman formula were not just some combinations of Lie derivatives.
(\ref{classpartgen}) therefore represents a sort of resummation of the
loop-expansion that explicitly takes into account the geometric symmetries that
make the 1-loop approximation exact. We shall see soon that it is quite
consistent with the results predicted from the loop-expansion, and moreover
that it gives many new insights.

There are several points to make at this stage. First of all we note the
appearence of the boundary contribution in (\ref{2ndhalf}). If we assume that
$\lie_V\beta=0$, that the group action represented by the flows of the
Hamiltonian vector field $V$ preserves the boundary of $\man$ (i.e.
$g\cdot\partial\man=\partial\man$) and that the action is free on
$\partial\man$, then the $s$-integral in the boundary term in (\ref{2ndhalf})
can be carried out explicitly and we find the extra contribution to the
Duistermaat-Heckman formula for manifolds with boundary,
\beq
Z_{\partial\man}(T)=-\oint_{\partial\man}\frac{\beta\wedge\alpha}{D_V\beta}
\label{bdrycontr}\eeq
In this context $\beta=i_Vg$ is the connection 1-form for the induced group
action on the boundary $\partial\man$, because as we have seen earlier $d\beta$
is the moment map for this action. This boundary term can be determined using
the Jeffrey-Kirwan-Kalkman residue that was introduced in Subsection 3.8, i.e.
the coefficient of $\frac{1}{\phi}$ in the quantity
$(\beta\wedge\alpha)/D_V\beta$, where $\phi$ is the element of the symmetric
algebra $S({\bf g}^*)$ representing the given circle action \cite{kalkman2}.

Secondly, notice that the conformal localization symmetry gives an explicit
realization in (\ref{ckvhess}) of the Hessian-metric ansatz which was discussed
in the last Subsection. In particular, (\ref{classpartgen}) establishes that
with this Hessian-metric substitution and the appropriate extension away from
the critical points of $H$ (the terms proportional to $\nabla H$ in
(\ref{ckvhess})) the corrections to the Duistermaat-Heckman integration formula
vanish to all orders of the loop-expansion (and not just to 2-loop order as was
established in the last Subsection). Notice, however, that because of
(\ref{upsilon}) any conformal Killing vector on a K\"ahler manifold is
automatically an isometry. In fact, the generic case of a non-vanishing scaling
function $\Upsilon(x)$ in (\ref{conflie}) is similar to the isometry case from
the point of view of the equivariant localization priniciple. Note that away
from the critical points of $H$ we can rescale $\beta\to\bar\beta\equiv
g(V,\cdot)/g(V,V)$. With this choice for the localization 1-form $\beta$ it is
easy to show from (\ref{conflie}) that away from the critical point set of the
Hamiltonian it satisfies $\lie_V\bar\beta = 0$, i.e. $\bar\beta$ is then an
equivariant differential form on $\man-\man_V$. Thus away from the subset
$\man_V\subset\man$ the conformal Killing condition can be cast into the same
equivariant cohomological framework as the isometry condition by a rescaling of
the metric tensor in (\ref{conflie}), $g_{\mu\nu}\to
G_{\mu\nu}=g_{\mu\nu}/g(V,V)$, for which $\lie_VG=0$. The rescaled metric
$G_{\mu\nu}(x)$ is only defined on $\man-\man_V$, but we recall from Subsection
2.5 that all that is needed to establish the localization of
(\ref{classpart}) onto the zeroes of the vector field $V$ (i.e. the
equivariant localization principle) is an invariant metric tensor (or
equivalently an equivariant differential form $\beta$) which is defined
everywhere on $\man$ except possibly in an arbitrarily small neighbourhood of
$\man_V$. The fact that the weaker conformal symmetry condition is equivalent
to the isometry condition in this respect is essentially a consequence of the
fact that the differential form $\beta=i_Vg$ above is still a connection 1-form
that specifies a splitting of the tangent bundle into a component over $\man_V$
(represented by the discrete sum over $\man_V$ in (\ref{classpartgen})) and a
component orthogonal to $\man_V$ (represented by the Lie derivative integral in
(\ref{classpartgen})). This is in fact implicit in the proof by Atiyah and Bott
in \cite{atiyahbott} using the Weil algebra.

One may ask as to the possibilities of using other localization forms to carry
out the localization onto $\man_V$, but it is readily seen that, up to
components orthogonal to $V$, $\beta=i_Vg$ is the most general localization
form up to multiplication by some strictly positive function. This follows from
the fact that in order to obtain finite results in the limit $s\to\infty$ in
(\ref{largesintgen}) we need to ensure
that the form $D_V\beta$ has a 0-form component to produce an exponential
damping factor, since higher-degree forms will contribute only polynomially in
the Taylor expansion of the exponential. This is guaranteed only if $\beta$ has
a 1-form component. Thus it is only the 1-form part of $\beta$ that is relevant
to the localization formula, and so without loss of generality the most general
localization principle follows from choosing $\beta$ to be a 1-form.
Furthermore, the 0-form part $V^\mu\beta_\mu$ of $D_V\beta$ must attain its
global minimum at zero so that the large-$s$ limit in
(\ref{largesintgen}) yields a non-zero result. This boundedness requirement is
equivalent to the condition that the component of $\beta$ along $V$ has the
same
orientation as $V$, i.e. that $\beta$ be proportional to the metric dual 1-form
of $V$ with respect to a globally-defined Euclidean signature Riemannian
structure on the phase space $\man$.

In addition, for a compact phase space $\man$ the conformal group is in general
non-compact, so that conformal Killing vectors need not automatically generate
circle actions in these cases as opposed to isometry generators where this
would be immediate. To explore whether this larger conformal group symmetry of
the Duistermaat-Heckman localization leads to {\it globally} different sorts of
dynamical systems, one would like to construct examples of systems with
non-trivial (i.e. $\Upsilon\neq0$) conformal symmetry. For this, one has to
look at spaces which have a Riemannian metric $g$ for which the Hamiltonian
vector field $V$ is a generator of both the conformal group ${\rm
Conf}(\man,g)$ {\it and} the symplectomorphism group ${\rm Sp}(\man,\omega)$ of
canonical transformations on the phase space. From the analysis thus far we
have a relatively good idea of what the latter group looks like. The conformal
group for certain Riemannian manifolds is also well-understood \cite{ginsparg}.
For instance, the conformal group of flat Euclidean space of dimension $d\geq3$
is locally isomorphic to $SO(d+1,1)$. The (global) conformal group of the
Riemann sphere $\IC\cup\{\infty\}$ was already encountered in Section 6 above
(in a different context), namely the group $SL(2,\IC)/\IZ_2\simeq SO(3,1)$ of
projective conformal transformations. In these cases, the conformal group
consists of the usual isometries of the space, along with dilatations or scale
transformations (e.g. translations of $r$ in $z=r\e^{i\theta}$) and the
$d$-dimensional subgroup of so-called special conformal transformations.

An interesting example is provided by the flat complex plane $\IC$. Here the
conformal algebra is infinite-dimensional and its Lie algebra is just the
classical Virasoro algebra \cite{ginsparg}. Indeed, the conformal Killing
equations in this case are just the first set of Cauchy-Riemann equations in
(\ref{killingeqcomplex}) (the other one represents the divergence-free
condition $\tr~\mu_V=0$). This means that the conformal Killing vectors in this
case are the holomorphic functions $V^z=f(z),V^{\bar z}=\bar f(\bar z)$. The
Hamiltonian flows of these vector fields are therefore the {\it arbitrary}
analytic coordinate transformations
\beq
\dot z(t)=f(z(t))~~~~~,~~~~~\dot{\bar z}(t)=\bar f(\bar z(t))
\label{confflows}\eeq
As an explicit example, consider the conformal Killing vector which describes
a Hamiltonian system with $n+1$ distinct stationary points,
\beq
V^z=i \beta z ( 1 - \alpha_1 z) \cdots ( 1- \alpha_n z)
\label{ckv}
\eeq
at $z=0$ and $z=1/\alpha_i$, where $\beta,\alpha_i\in\IC$. The associated
scaling function in (\ref{conflie}) is then
\beq
\Upsilon(z,\bar z)=\partial_z V^z+\partial_{\bar z}V^{\bar z}
\label{scalefn}\eeq
The symplecticity condition $\lie_V\omega=0$ leads to the first-order linear
partial differential equation
\beq
V^z\partial_z\omega_{z\bar z}+V^{\bar{z}}\partial_{\bar{z}}\omega_{z\bar z}
=-\Upsilon(z,\bar z)\omega_{z\bar z}
\label{heqns}\eeq
which is easily solved by separation of the variables $z,\bar z$.
The solution for the symplectic 2-form with arbitrary separation parameter
$\lambda\in\IR$ is
\beq
\omega_{z\bar z}^{(\lambda)}(z,\bar z)=w_\lambda(z)\bar w_\lambda(\bar
z)/V^zV^{\bar z}
\label{ckvomegax}\eeq
where
\beq
w_\lambda(z)=\e^{i\lambda\int dz/V^z}=\left(\frac{z}{ ( 1 -  \alpha_1 z )^{A_1}
\cdots ( 1 - \alpha_n z)^{A_n}}\right)^{\lambda/\beta}
\label{wlambda}\eeq
and the constants
\beq
A_i( \alpha_1,\dots,\alpha_n)=(\alpha_i)^{n-1}\prod_{j\neq
i}\frac{1}{\alpha_i-\alpha_j}
\eeq
are the coefficients of the partial fraction decomposition
\beq
(V^z)^{-1}=\frac{1}{i\beta}\left(\frac{1}{z}+\sum_{i=1}^n\frac{A_i}{1-
\alpha_iz}\right)
\eeq

To ensure that (\ref{ckvomegax}) is a single-valued function on $\IC$,
we restrict the $\alpha_k$'s to all have the same phase, so that
$A_i( \alpha_1,\dots,\alpha_n)\in\IR$, and the parameter $\beta$ to be
real-valued. The Hamiltonian equations (\ref{hameqcomplex}) can now be
integrated up with the vector field (\ref{ckv}) and the symplectic 2-form
(\ref{ckvomegax}), from which we find the family of Hamiltonians
\beq
H^{(\lambda)}_{\beta,\alpha_i}(z,\bar z)=\frac{1}{\lambda}\left(\frac{z}{ ( 1 -
 \alpha_1 z )^{A_1} \cdots ( 1 - \alpha_n z)^{A_n}}\right)^{\lambda/\beta}
\left(\frac{\bar z}
{ ( 1 -  \bar{ \alpha}_1 \bar{z} )^{A_1} \cdots ( 1 - \bar{ \alpha}_n
\bar{ z})^{A_n}} \right)^{\lambda/ \beta}
\label{ckvHx}\eeq
To ensure that this Hamiltonian has only non-degenerate critical points we set
$\lambda = \beta$. This also guarantees that the level (constant energy) curves
of this Hamiltonian coincide with the curves which are the solutions of the
equations of motion (\ref{confflows}) \cite{paniak}.

Since the Hamiltonian (\ref{ckvHx}) either vanishes or is infinite on its
critical point set, it is easy to show that the partition function
(\ref{classpart}) is independent of ${\alpha_k}$ and coincides with the
anticipated result from the Duistermaat-Heckman integration formula, namely
$Z(T)=2\pi i\beta/T$. This partition function coincides with that of the simple
harmonic oscillator $z\bar z/\beta$, as expected since for $\alpha_i=0$,
$\omega^{(\beta)}_{z \bar{z}}$ becomes the Darboux 2-form and
$H^{(\beta)}_{\beta,0}$ the harmonic oscillator Hamiltonian. In fact, we can
integrate up the flow equation (\ref{confflows}) in the general case and we
find that the classical trajectories $z(t)$ are determined by the equation
\beq
 \e ^{ i \beta (t - t_0) } = w_\beta(z(t))=\frac{ z(t)}{
( 1 -  \alpha_1 z(t) )^{A_1 } \cdots ( 1 - \alpha_n z(t))^{A_n} }
\label{conftrans}\eeq
The coordinate change $z\to w_\beta(z)$ is just the finite conformal
transformation generated by the vector field (\ref{ckv}) and it maps the
dynamical system $(\omega_{z\bar z}^{(\beta)},H_{\beta,\alpha_i}^{(\beta)})$
onto the harmonic oscillator $H\propto w\bar w$, $\omega\propto\omega_D$ with
the usual circular classical trajectories $w(t)=\e^{i\beta(t-t_0)}$ associated
with a $U(1)$ generator corresponding to an isometry. This transformation
is in general multi-valued and has singularities at the critical points
$z=1/\alpha_i$ of the Hamiltonian $H^{(\beta)}_{\beta,\alpha_i}$. It is
therefore not a diffeomorphism of the plane for $\alpha_i\neq0$ and the
Hamiltonian system $(\IR^2,\omega_{z\bar
z}^{(\beta)},H_{\beta,\alpha_i}^{(\beta)})$ is not globally isomorphic to the
simple harmonic oscillator. For more details about the global differences
between these systems and those associated with isometry generators, see
\cite{paniak}. The conformal group structures on phase spaces like $S^2$ yield
novel generalizations of the localizable systems which are associated with
coadjoint orbits of isometry groups as we discussed in Section 5. The
appearence of the larger (non-compact) conformal group may lead to interesting
new structures in other instances which usually employ the full isometry group
of the phase space, such as the Witten localization formalism of Subsection
3.8.

Other geometric alternatives to the Lie derivative condition $\lie_Vg=0$
have also been discussed by K\"arki and Niemi in \cite{karki}. For instance,
consider the alternative condition
\beq
V^\lambda\nabla_\lambda V^\mu=0
\label{geodetic}\eeq
to the Killing equation, which means that the Hamiltonian flows $\dot
x^\mu=V^\mu$ are geodetic to $g$ (see Subsection 2.4). From (\ref{geodetic}) it
follows that the Lie derivative of the localization 1-form $\beta=i_Vg$ can be
written as
\beq
\lie_V\beta=V^\rho(g_{\rho\lambda}\nabla_\mu
V^\lambda+g_{\mu\lambda}\nabla_\rho V^\lambda)dx^\mu=V^\rho
g_{\rho\lambda}\nabla_\mu V^\lambda dx^\mu=\frac{1}{2}di_V\beta
\eeq
Comparing this with the Cartan-Weil identity $\lie_V=di_V+i_Vd$ for the Lie
derivative acting on differential forms we find the relation
\beq
i_Vd\beta=-\frac{1}{2}di_V\beta
\eeq
which leads to the equivariance condition \cite{karki}
\beq
D_V(K_V/2+\Omega_V)=0
\label{eqclosed12}\eeq
so that the dynamical systems $(\frac{1}{2}K_V,\Omega_V)$ and $(H,\omega)$
determine a bi-Hamiltonian structure. Moreover, in this case it is also
possible to explicitly solve the equivariant Poincar\'e lemma \cite{karki},
just as we did in Subsection 3.6. Thus given that $\frac{1}{2}K_V+\Omega_V$ is
an
equivariantly-closed differential form, the condition (\ref{geodetic}) has
the potential of leading to possibly new localization formulas.

However, there are 2 things to note about the geometric condition
(\ref{geodetic}). The first is its connection with a non-trivial conformal
Killing equation $\lie_Vg=\Upsilon g$, which follows from the identity
\beq
V^\nu g^{\alpha\mu}(\lie_Vg)_{\mu\nu}=V^\lambda\nabla_\lambda V^\alpha+
g^{\alpha\mu}\nabla_\mu K_V/2
\label{geokilling}\eeq
Contracting both sides of (\ref{geokilling}) with $g_{\alpha\rho}V^\rho$ leads
to
\beq
\Upsilon K_V=2g_{\mu\nu}V^\nu V^\lambda\nabla_\lambda V^\mu
\eeq
when (\ref{conflie}) holds. This implies that if (\ref{geodetic}) is satisfied,
then $\Upsilon\equiv0$ away from the zeroes of $V$. Thus $\Upsilon\equiv0$
almost everywhere on $\man$ and so the geometric condition (\ref{geodetic}) can
only be compatible with the Killing equation, and not the inhomogeneous
conformal Killing equation.

Secondly, the exact 2-form $\Omega_V\equiv di_Vg$ is degenerate on $\man$,
because an application of the Leibniz rule and Stokes' theorem gives
\beq
n!\int_\man d^{2n}x~\sqrt{\det\Omega_V(x)}=\int_\man\Omega_V^n=\int_\man
d\left(i_Vg\wedge\Omega_V^{n-1}\right)=0
\label{stokeslieb}\eeq
when $\partial\man=\emptyset$. Thus $\det\Omega_V(x)=0$ on some submanifold
of $\man$, and so the Hamiltonian system determined by $(\frac{1}{2}K_V,
\Omega_V)$ is degenerate. As mentioned in Subsection 3.6, this isn't so crucial
so long as the support of $\det\Omega_V(x)$ is a submanifold of $\man$ of
dimension at least 2 (so that there exists at least 1 degree of freedom from
the classical equations of motion). It would be interesting to investigate
these
geometric structures in more detail and see what localization schemes they
lead to.

\subsection{Corrections to the Duistermaat-Heckman Formula: A Geometric
Approach}

The integration formula (\ref{classpartgen}) suggests a geometric approach
to the evaluation of corrections to the Duistermaat-Heckman formula in the
cases where it is known to fail. Recall that there is always {\it locally} a
metric tensor on $\man-\man_V$ for which $V$ is a Killing vector (see the
discussion at the beginning of Subsection 3.6). For the systems where the
semi-classical approximation is not exact, there are global obstructions to
extending these locally invariant metric tensors to globally-defined geometries
on the phase space which are invariant under the full group action generated by
the Hamiltonian vector field $V$ on $\man$, i.e. there are no {\it globally}
defined single-valued Riemannian geometries on $\man$ for which $V$ is globally
a Killing vector. This means that although the Killing equation $\lie_Vg=0$ can
be solved for $g$ locally on patches covering the manifold, there is no way to
glue the patches together to give a single-valued invariant geometry on the
whole of $\man$ (c.f. Subsection 5.9). In this Subsection we shall describe how
the expression (\ref{classpartgen}) could be used in this sense to evaluate the
corrections to the sum over critical points there \cite{me}, and we shall see
that not only does this method encompass much more of the loop-expansion than
the term-by-term analysis of Subsection 7.1 above, but it also characterizes
the non-exactness of the Duistermaat-Heckman formula in a much
more transparent and geometric way than Kirwan's theorem. In this way we can
obtain an explicit geometric picture of the failure of the Duistermaat-Heckman
theorem and in addition a systematic, geometric method for approximating the
integral (\ref{classpart}). Furthermore, this analysis will show
explicitly the reasons why for certain dynamical systems there are no globally
defined Riemannian metrics on the given symplectic manifold for which any
given vector field with isolated zeroes is a Killing vector, and as well this
will give another geometric description of the integrability properties of the
given dynamical system. The analysis presented here is by no means complete and
deserves a more careful, detailed investigation.

The idea is to define a set of patches covering $\man$ in each of which we
can solve the Killing equations for $g$, but for which the gluing of these
patches together to give a globally defined metric tensor is highly singular.
The non-triviality that occurs when these subsets are patched back together
will then represent the corrections to the Duistermaat-Heckman formula, and
from our earlier arguments we know that this will be connected with the
integrability of the Hamiltonian system. We introduce a set of preferred
coordinates $x''$ for the vector field $V$ following Subsection 5.2. In
general,
this diffeomorphism can only be defined locally on patches over $\man$ and the
failure of this coordinate transformation in producing globally-defined
$C^\infty$-coordinates on $\man$ gives an analytic picture of why the
Hamiltonian vector field fails to generate global isometries. Notice in
particular that these coordinates are only defined on $\man-\man_V$. In this
way we shall see geometrically how Kirwan's theorem restricts dynamical systems
whose phase spaces have non-trivial odd-degree homology and explicitly what
type of flow the Hamiltonian vector field generates.

Recall that the coordinate functions $x''$ map the constant coordinate lines
$(x_0^2,\dots,x_0^{2n})\in\IR^{2n-1}$ onto the integral curves of the isometry
defined by the classical Hamilton equations of motion $\dot x^\mu(t)=V^\mu
(x(t))$, i.e. in the coordinates $x''(x)$, the flows generated by the
Hamiltonian vector field look like
\beq
x^{\prime\prime1}(t)=x_0^1+t~~~~~;~~~~~x^{\prime\prime\mu}(t)=x_0^\mu~~~,~~~
\mu>1
\label{xprimeflows}\eeq
In general, this coordinate transformation function will have singularities
associated with the fact that there is no Riemannian metric tensor on $\man$
for which the Lie derivative condition $\lie_Vg=0$ holds. Otherwise, if these
transformation functions were globally defined on $\man-\man_V$, then we could
take the metric on $\man$ to be any one whose components in the
$x''$-coordinates are independent of $x^{\prime\prime1}$ thereby solving
the Killing equations directly, and hence from (\ref{classpartgen}) the
WKB approximation would be exact. For a non-integrable system, there must
therefore be some sort of obstructions to defining the
$x^{\prime\prime}$-coordinate system globally over $\man$. In light of the
above comments, these singularities will partition the manifold up into patches
$P$, each of which is a $2n$-dimensional contractable submanifold of $\man$
with boundaries $\partial P$ which are some other $(2n-1)$-dimensional
submanifolds of $\man$ induced by the constant coordinate line transformation
from $\IR^{2n-1}$ above. By dropping some of these coordinate surfaces if
necessary, we can assume that these patches induced from the singularities of
the above coordinate transformation form a disjoint cover of the manifold,
$\man=\bigsqcup_PP$ \footnote{\baselineskip=12pt Here we assume that $\man$ is
compact,  but we shall see that this formalism can also be extended to the
phase space $\IR^{2n}$.}. Then we can write the partition function as
\beq
Z(T)=\sum_P\int_P\alpha|_P
\label{partpatch}\eeq
where as usual $\alpha$ is the equivariantly-closed differential form
(\ref{alphadh}).

By the choice of the patches $P$, in their interior there is a well-defined
(bounded) translation action generated by $V^{\prime\prime\mu}$. Since the
patches $P$ are diffeomorphic to hypercubes in $\IR^{2n}$, we can place a
Euclidean metric on them,
\beq
g_P=\e^{\varphi_P(x'')}dx''_\mu\otimes dx^{\prime\prime\mu}
\label{gpatch}\eeq
where the conformal factor $\varphi_P(x'')$ is a globally-defined real-valued
$C^\infty$-function on $P$. If we choose it so that it is independent of the
coordinate $x^{\prime\prime1}$, then the metric (\ref{gpatch}) satisfies the
Killing equation on $P$. Thus on each patch $P$, by the given choice of
coordinates, we can solve the Lie derivative constraint, even though this
cannot be extended to the whole of $\man$. Then each integral over $P$ in
(\ref{partpatch}) can be written using the formula (\ref{classpartgen}),
restricted to the patch $P$, to get
\beq\new{\begin{array}{ll}
\int_P\alpha|_P=&\left(\frac{2\pi i}{T}\right)^n\sum_{p\in\man_V\cap P}(-i)
^{\lambda(p)}\sqrt{\frac{\det\omega(p)}{\det\hess(p)}}\e^{iTH(p)}\\&+
\frac{1}{(iT)^n}\int_0^\infty ds~\oint_{\partial P}\frac{\e^{iTH-sK_V}}{(n-1)!}
g(V,\cdot)\wedge\left(iT\omega-s\Omega_V\right)^{n-1}\Bigm|_{\partial P}
\end{array}}
\label{dhpatch}\eeq
The first term here, when (\ref{dhpatch}) is substituted back into
(\ref{partpatch}), represents the lowest-order term in the semi-classical
expansion of the partition function over $\man$, i.e. the Duistermaat-Heckman
term $Z_0(T)$ in (\ref{dhformula}), while the boundary terms give the
general corrections to this formula and represent the non-triviality that
occurs rendering inexact the stationary-phase approximation. The result is
\beq
Z(T)=Z_0(T)+\delta Z(T)
\label{partperturb}\eeq
where
\beq
\delta Z(T)=\frac{1}{(iT)^n}\int_0^\infty ds~\sum_P\oint_{\partial P}\frac{
\e^{iTH-sK_V}}{(n-1)!}g(V,\cdot)\wedge\left(iT\omega-s\Omega_V\right)^{n-1}
\Bigm|_{\partial P}
\label{patchcorrs}\eeq
The contributions from the patch terms in (\ref{patchcorrs}) therefore
represent an alternative geometric approach to the loop-expansion of Subsection
7.1 above.

To evaluate the correction term $\delta Z(T)$, we recall from Subsection 5.2
(eqs. (\ref{partialdiffchi}),(\ref{xpprimecoords})) that the coordinate
functions $\chi^\mu(x)$ for $\mu=2,\dots,2n$ are local conserved charges
of the Hamiltonian system, i.e.
\beq
\left\{\chi^\mu,H\right\}_\omega=V^\nu\partial_\nu\chi^\mu=0
\label{chiconserved}\eeq
Thus we can take one of them, say $\chi^2$, to be a functional of the
Hamiltonian, which we choose to be $x^{\prime\prime2}(x)=\chi^2(x)=\sqrt{H(x)}
$, where by adding an irrelevant constant to $H$ we may assume that it is
a positive function on the (compact) manifold $\man$. Then, using the metric
tensor transformation law, we find that the metric (\ref{gpatch}) when written
back into the original (unprimed) coordinates has the form
\beq
g_P=\e^{\varphi_P(x)}\left(\frac{1}{(V^\lambda\partial_\lambda\chi^1)^2}
\partial
_\mu\chi^1\partial_\nu\chi^1+\frac{1}{4H}\partial_\mu H\partial_\nu H
+\sum_{\alpha>2}\partial_\mu\chi^\alpha\partial_\nu\chi^\alpha\right)dx^\mu
\otimes dx^\nu
\label{gpatchx}\eeq
so that the metric-dependent quantities appearing in (\ref{patchcorrs}) can
be written as
\beq
g_P(V,\cdot)=\frac{\e^{\varphi_P(x)}}{V^\lambda(x)\partial_\lambda\chi^1(x)}
\partial_\mu\chi^1(x)dx^\mu~~~~~,~~~~~K_V(x)|_P=g_P(V,V)=\e^{\varphi_P(x)}
\label{gpatchvv}\eeq
\bd
\Omega_V|_P=\frac{\e^{\varphi_P(x)}}{2(V^\lambda\partial_\lambda\chi^1)^2}\left
\{\partial_\lambda\chi^1\left(\partial_\mu V^\lambda\partial_\nu\chi^1-
\partial_\nu V^\lambda\partial_\mu\chi^1\right)\right.
\ed
\beq
\left.+V^\lambda\partial_\lambda\chi^1\left(\partial_\mu\varphi_P\partial_\nu
\chi^1-\partial_\nu\varphi_P\partial_\mu\chi^1\right)\right\}dx^\mu\wedge
dx^\nu
\label{omegapatch}\eeq
When these expressions are substituted back into the correction term
(\ref{patchcorrs}), we find that the integrands of $\delta Z(T)$ depend only
on the coordinate function $\chi^1(x)$. This is not surprising, since the
only effect of the other coordinate functions, which define local action
variables of the dynamical system, is to make the effect of the partitioning
of $\man$ into patches above non-trivial, reflecting the fact that the system
is locally integrable, but not globally (otherwise, the partition function
localizes).

In general, the correction term (\ref{patchcorrs}) is extremely complicated,
but we recall that there is quite some freedom left in the choice of $\chi^1$.
All that is required of this function is that it have no critical points in
the given coordinate neighbourhood. We can therefore choose it appropriately
so as to simplify the correction $\delta Z(T)$ somewhat. Given this choice,
in general singularities will appear from the fact that it cannot be defined
globally on $\man$, and we can use these identifications to identify the
specific regions $P$ above. The form of the function $\chi^1$ is at the
very heart of this approach to evaluating corrections to the
Duistermaat-Heckman formula. We shall see how this works in some explicit
examples in the next Subsection. Notice that a similar phenomenon to what
occured in Subsection 5.9 has happened here -- the function $K_V$ in
(\ref{gpatchvv}) is non-zero, as the zeroes of the vector field $V$ have been
absorbed into the metric term $g_P(V,\cdot)$ thereby making it singular. We
can therefore now carry out the explicit $s$-integral in (\ref{patchcorrs}),
as the singularities on $\man_V$ are already present in the integrand there.
Although this may seem to make everything hopelessly singular, we shall
see that they can be regulated with special choices of the function $\chi^1$
thereby giving workable forms. We shall see in fact that when such divergences
do occur, they are related to those predicted by Kirwan's theorem which we
recall dictates also when the full stationary-phase series diverges for a given
function $H$.

There does not seem to be any immediate way of simplifying the patch
corrections $\delta Z(T)$ above due to the complicated nature of the integrand
forms. However, as usual in 2-dimensions things can be simplified rather nicely
and the analysis reveals some very interesting properties of this formalism
which could be generalized to higher-dimensional symplectic manifolds.
To start, we notice that in 2-dimensions, if $\man$ is a compact manifold,
then the union above over all of the patch boundaries $\partial P\subset\man$
will in general form a sum over 1-cycles $a_\ell\in H_1(\man;\IZ)$. Next, we
substitute (\ref{gpatchvv}) and (\ref{omegapatch}) into (\ref{patchcorrs})
with $n=1$, and after working out the easy $s$-integration we find that
the 2-dimensional correction terms can be written as
\beq
\delta Z(T)=\frac{1}{iT}\sum_\ell\oint_{a_\ell}\frac{\e^{iTH(x)}}{V^\lambda
(x)\partial_\lambda\chi^1(x)}\partial_\mu\chi^1(x)dx^\mu
\label{2dcorrs}\eeq
As for the function $\chi^1$, we need to choose one which is independent of the
other coordinate transformation function $\chi^2$ to ensure that these 2
functions truly do define a (local) diffeomorphism of $\man$. The simplest
choice, as far as the evaluation of (\ref{2dcorrs}) is concerned, is to choose
$\chi^1$ as the solution of the first-order linear partial differential
equation
\beq
V^1(x)\partial_1\chi^1(x)=V^2(x)\partial_2\chi^1(x)
\label{chi1eq}\eeq
With this choice of $\chi^1$, the functions $\chi^1$ and $\chi^2$ are
independent of each other wherever $\partial_\mu\chi^\nu(x)\neq0$, $\mu,\nu=
1,2$, which follows from working out the Jacobian for the coordinate
transformation defined by $\chi^\mu$ and using their defining partial
differential equations above.

With this and the Hamiltonian equations $dH=-i_V\omega$, the correction terms
(\ref{2dcorrs}) become
\beq
\delta Z(T)=-\frac{1}{2iT}\sum_\ell\oint_{a_\ell}F|_{a_\ell}
\label{corrfaell}\eeq
where we have introduced the 1-form
\beq
F=\omega_{12}(x)\e^{iTH(x)}\left(\frac{1}{\partial_2H(x)}dx^1-\frac{1}{\partial
_1H(x)}dx^2\right)
\label{Fform}\eeq
The expression (\ref{corrfaell}) leads to a nice geometric interpretation of
the corrections above to the Duistermaat-Heckman formula. To each of the
homology cycles $a_\ell\in H_1(\man;\IZ)$, there corresponds a cohomology
class $\eta_\ell\in H^1(\man;\IR)$, called their Poincar\'e dual \cite{bottu},
which has the property that it localizes integrals of 1-forms $\alpha\in\Lambda
^1\man$ to $a_\ell$, i.e.
\beq
\oint_{a_\ell}\alpha|_{a_\ell}=\int_\man\alpha\wedge\eta_\ell
\label{poindualdef}\eeq
Defining
\beq
\eta=\sum_\ell\eta_\ell\in H^1(\man;\IR)
\label{dualtot}\eeq
we see that the correction term (\ref{corrfaell}) can be written as
\beq
\delta Z(T)=-\frac{1}{2iT}\int_\man F\wedge\eta
\label{corrdual}\eeq
Noting also that the original partition function itself can be written as
\beq
Z(T)=\frac{1}{2}\int_\man F\wedge dH
\label{partfdh}\eeq
it then follows from $Z(T)=Z_0(T)+\delta Z(T)$ that
\beq
\int_\man F\wedge\left(iTdH+\eta\right)=-4\pi\sum_{p\in\man_V}(-i)^{\lambda(p)}
\sqrt{\frac{\det\omega(p)}{\det\hess(p)}}\e^{iTH(p)}
\label{neatformula}\eeq
Thus in this sense, the partition function represents intersection numbers
of $\man$ associated to the homology cycles $a_\ell$.

This last equation is particularly interesting. It shows that the corrections
to the Duistermaat-Heckman formula generate the Poincar\'e duals to the
homology cycles which signify that the Hamiltonian vector field does not
generate a globally well-defined group action on $\man$. When the correction
1-form $\eta/iT$ is added to the 1-form $dH=-\omega(V,\cdot)$ which defines the
flow of the Hamiltonian vector field on $\man$, the resulting 1-form is enough
to render the Duistermaat-Heckman formula exact for the new ``effective"
partition function. This means that although the initial Hamiltonian flow $dH$
doesn't `close enough' to satisfy the conditions required for the
Duistermaat-Heckman theorem, adding the cohomological Poincar\'e dual to the
singular homology cycles of the flow is enough to close the flows so that the
partition function is now given exactly by the lowest-order term $Z_0(T)$ of
its
semi-classical expansion. One now can solve for the vector field $W$
satisfying the ``renormalized" Hamiltonian equations
\beq
dH+\eta/iT=-\omega(W,\cdot)
\label{newvec}\eeq
We can consider $W$ as a ``renormalization" of the Hamiltonian vector field
$V$ which renders the stationary-phase series convergent and the
Duistermaat-Heckman formula exact. Note that since the symplectic form $\omega$
defines a cohomology class in $H^2(\man;\IR)$, this just corresponds to
choosing a different, possibly non-trivial representative in $H^1(\man;\IR)$
for $\omega(V,\cdot)$ (recall $\eta\in H^1(\man;\IR)$). Thus in our approach
here, the corrections to the Duistermaat-Heckman formula compute (possibly)
non-trivial cohomology classes of the manifold $\man$ and express geometrically
what symmetry is missing from the original dynamical system that prevents its
saddle-point approximation from being exact. The explicit constructions of the
Poincar\'e duals above are well-known \cite{bottu} -- one takes the embedding
$\sigma_\ell:S^1\to\man$ of $S^1$ in $\man$ which corresponds to the loop
$a_\ell$, and constructs its DeRham current which is the Dirac delta-function
1-form $\delta^{(1,1)}(x,\sigma_\ell(y))\in\Lambda^1\man(x)\otimes\Lambda^1
\man(y)$ with the property (\ref{poindualdef}) \cite{bergeron}.

There is one crucial point that needs to be addressed before we turn to some
explicit examples. In general we shall see that there are essentially 2 types
of homology cycles that appear in the above when examining the singularities of
the diffeomorphisms $\chi^\mu$ that prevent them from being global coordinate
transformations of $\man$. The first type we shall call `pure singular
cycles'. These arise solely as a manifestation of the choice of equation
satisfied by $\chi^1$. The second type shall be refered to as `critical
cycles'. These are the cycles on which at least one of the components of the
Hamiltonian vector field vanish, $V^\mu(x)=0$ for $\mu=1$ or 2. On these
latter cycles the above integrals in $\delta Z(T)$ become highly singular
and require regularization. Notice in particular that if, say, $V^1(x)=0$ but
$V^2(x)\neq0$ on some cycle $a_\ell$, then the equations (\ref{chiconserved})
and (\ref{chi1eq}) which determine the functions $\chi^\mu$ imply that
$\partial_2\chi^\mu(x)=0$ while leaving the derivatives $\partial_1\chi^
\mu(x)$ undetermined. Recall that it was precisely at these points where the
Jacobian of the coordinate transformation defined by $\chi^\mu$ vanished.

In this case one must regulate the 1-form $F$ defined above by letting
$\partial_1\chi^1$ and $\partial_2\chi^1$ both approach zero on this
cycle $a_\ell$ in a correlated manner so as to cancel the resulting divergence
in the integrand of (\ref{2dcorrs}). Note that this regularization procedure
now requires that $x^1$ and $x^2$ transform identically, particularly under
rescalings, so that the tensorial properties of the differential form $F$
are unaffected by this definition. In this case, the 1-form $F$ which appears
above gets replaced by the 1-form
\beq
F|_{a_\ell}=-\frac{1}{V^2(x)}\left(dx^1+dx^2\right)\e^{iTH(x)}=\frac{\omega_
{12}(x)}{\partial_1 H(x)}\left(dx^1+dx^2\right)\e^{iTH(x)}
\label{fredef}\eeq
which follows from the general expression (\ref{2dcorrs}). This procedure for
defining $F$ can be thought of as a quantum field theoretic ultraviolet
regularization for the higher-loop corrections to the partition function. In
general, we shall always obtain such singularities corresponding to the
critical points of the Hamiltonian because, as mentioned before, the
diffeomorphism equations above become singular at the points where $V^\mu(x)
=0$. Note that (\ref{fredef}) will also diverge when the cycle $a_\ell$ crosses
a critical point, i.e. on $a_\ell\cap\man_V$. Such singularities, as we shall
see, will be just a geometric manifestation of Kirwan's theorem and the fact
that in general the stationary-phase expansion does not converge for the given
Hamiltonian system. We shall also see that in general the pure singular
cycles do not contribute to the corrections, as anticipated, as they are only
a manifestation of the particular coordinate system used, of which the
covariant corrections should be independent. It is only the critical cycles
that contribute to the corrections and mimick in some sense the sum over
critical points series for the partition function.

\subsection{Examples}

In this Subsection we illustrate some of the formalism of this Section with 2
classes of explicit examples. The first class we shall consider is the
height function of a Riemann surface, a set of examples with which we have
become well-acquainted. In the case of the Riemann sphere we have little to add
at this point since the height function (\ref{heights2}) localizes. The only
point we wish to make here is that the covariant Hessian in this case with
respect to the standard K\"ahler geometry of $S^2$ (see Subsection 5.5) is
related to the K\"ahler metric $g_{S^2}$ by
\beq
\nabla\nabla h_{\Sigma^0}=2\frac{1-z\bar z}{(1+z\bar z)^3}dz\otimes d\bar z=
2(1-h_{\Sigma^0})g_{S^2}
\eeq
which is in agreement with the analysis of Subsection 7.1 above. This shows
the precise mechanism (i.e. the Hessian of $h_{\Sigma^0}$ generates covariantly
the K\"ahler structure of $S^2$) that makes the loop corrections vanish.

An interesting check of the above formalisms is provided by a modified
version of the height function $h_{\Sigma^0}$ which is the quadratic functional
\beq
h_{\Sigma^0}^{(2)}=h_{\Sigma^0}-h_{\Sigma^0}^2=(1-\cos\theta)-(1-\cos\theta)^2
=-\frac{2z\bar z}{1+z\bar z}-\left(\frac{2z\bar z}{1+z\bar z}\right)^2
\label{heights2sq}\eeq
which has the same critical behaviour as $h_{\Sigma^0}$. Now we find that the
metric equations (\ref{ansatzeqs}) are solved by taking the isothermal solution
\cite{me}
\beq
\Gamma^z_{zz}\equiv g^{z\bar z}\partial_zg_{z\bar
z}=\partial_z^2h_{\Sigma^0}^{(2)}/\partial_zh_{\Sigma^0}^{(2)}~~~,~~~g_{z\bar
z}(z,\bar z)=H'(z\bar z)
\eeq
for $H=h_{\Sigma^0}^{(2)}$, which follows from (\ref{conneqsan}) written in
local isothermal coordinates for the implicitly defined metric. Thus the
solution to (\ref{ansatzeqs}) is
\beq
\nabla\nabla h_{\Sigma^0}^{(2)}=\frac{2z\bar z}{(1+z\bar z)^3}\left(z\bar z-1
\right)dz\otimes d\bar z=g_{z\bar z}dz\otimes d\bar z
\label{meths2sq}\eeq
As (\ref{meths2sq}) does not coincide with the standard K\"ahler geometry of
$S^2$, the 1-loop approximation to the partition function in this case is not
exact, as expected. However, the partition function still localizes, in the
sense that it can be computed via the Gaussian integral transform
\beq
Z(T)=\int_\man d\mu_L~\e^{iT(H-H^2)}=\int_{-\infty}^\infty\frac{d\phi}{\sqrt{
2\pi i}}~\e^{-i\phi^2/2}\int_\man d\mu_L~\e^{i(T-2i\sqrt{T}\phi)H}
\label{gaussinttransf}\eeq
of the usual equivariant characteristic classes. Thus since (\ref{heights2sq})
is a functional of an isometry generator (i.e. a conserved charge), it is still
localizable, as anticipated from the discussions in Subsections 4.8 and 4.9.
This is also consistent with the formalism of the previous Subsection. In this
case, the preferred coordinates for the Hamiltonian vector field are $\theta$
and $x=\phi/(1-\cos\theta)$. Although these coordinates are singular at the
poles of $S^2$ (i.e. the critical points of (\ref{heights2sq})), the correction
terms $\delta Z(T)$ do not localize onto any cycles and just represent the
terms in the characteristic class expansion for $Z(T)$ here. This just reflects
the fact that $S^2$ is simply connected, and also that the geometric terms
$\delta Z(T)$ detect the integrability features of a dynamical system (as
(\ref{heights2sq}) is an integrable Hamiltonian).

Next, we consider the height function on the torus, with the K\"ahler geometry
in Subsection 6.2 adjusted so that $\varphi=0$ in (\ref{t2metric}) and $v=1$
in (\ref{t2kahlerom}). The covariant Hessian of the Hamiltonian
(\ref{heightt2}) in this case is
\beq\new{\begin{array}{ll}
{\cal{H}}(\phi_1,\phi_2)=&{\rm Im}~\tau\cos\phi_1\cos\phi_2d\phi_1\otimes d
\phi_1-2~{\rm Im}~\tau\sin\phi_1\sin\phi_2d\phi_1\otimes d\phi_2\\&+(r_1+
{\rm Im}~\tau\cos\phi_1)\cos\phi_2d\phi_2\otimes d\phi_2
\end{array}}
\eeq
In the complex coordinatization used to define the K\"ahler structure
this Hessian is not of the standard Hermitian form and the analysis used to
show the exactness of the stationary phase approximation in the case of the
height function on $S^2$ using the loop-expansion will not work here. Indeed,
we do not expect that any metric on $T^2$ will be defined from the covariant
Hessian here as we did in Subsection 7.1, and we already know that the
Duistermaat-Heckman formula is not exact for this example. This is because of
the saddle-points at $(\phi_1,\phi_2)=(0,\pi)$ and $(\pi,\pi)$. The Hessian at
these points will always determine an indefinite metric which is not admissible
as a globally-defined geometry on the torus.

This is also apparent from examination of the connection (\ref{gammadef}) and
its associated Fubini-Study geometry defined by (\ref{fubinistudygen}). In this
case $\gamma\equiv0$ and the curvature (\ref{fubinistudygen}) is trivial. The
2-form $\Omega$ does not determine the same cohomology class as the K\"ahler
2-form of $\Sigma^1$ does, so that there is not enough ``mixing" of the
Hessian and Liouville terms in the loop expansion to cancel out higher-order
corrections. For the sphere, the Fubini-Study metric coincides with the
standard K\"ahler metric and thus the appropriate mixing is there to make the
dynamics integrable (recall that $\IC P^1=S^2$). It is the
lack of formation of a non-trivial K\"ahler structure on the torus here that
makes almost all dynamical systems on it non-integrable.

Although the failure of the Duistermaat-Heckman theorem in this case can be
understood in terms of the non-trivial first homology of $T^2$ via Kirwan's
theorem, we can examine analytically the obstructions in extending the
Hamiltonian vector field (\ref{hamvect2}) to a global isometry of the standard
K\"ahler metric (\ref{t2metric}) of $T^2$ which defines the unique Riemannian
geometry for equivariant localization on the torus. We shall find that the
local translation action defined by the vector field (\ref{hamvect2}) cannot
be extended globally in a smooth way to the whole of $T^2$. The set of
coordinates $(x,y)$ on the torus in which the components of the Hamiltonian
vector field are $V^x=1$ and $V^y=0$ as prescribed before are first defined by
taking $\chi^2(\phi_1,\phi_2)$ to be the square root of the height function
(\ref{heightt2}) and $\chi^1(\phi_1,\phi_2)$ to be the $C^\infty$-function
with non-vanishing first order derivatives which is the solution of the
partial differential equation (\ref{chi1eq}). In the case at hand
(\ref{chi1eq}) can be written as
\beq
-(r_1+~{\rm Im}~\tau\cos\phi_1)\frac{\partial\chi^1}{\partial\phi_1}=
{}~{\rm Im}~\tau\sin\phi_1\cot\phi_2\frac{\partial\chi^1}{\partial\phi_2}
\label{t2partial}
\eeq
which is solved by
\beq
\chi^1(\phi_1,\phi_2)=\log(r_1+~{\rm Im}~\tau\cos\phi_1)-\log(\cos\phi_2)
\label{t2phi1}
\eeq
and integrating (\ref{t2phi1}) as in (\ref{xpprimecoords}) yields the
desired set of coordinates $(x,y)$. This gives
\bd
x(\phi_1,\phi_2)=\frac{1}{2~{\rm Im}~\tau}\left[\frac{2~{\rm Im}~\tau}{
\sqrt{r_2|{\rm Re}~\tau|}\sin\phi_2}\arctan\left(\sqrt{\frac{|{\rm
Re}~\tau|}{r_2}}\tan\frac{\phi_1}{2}\right)-\frac{\log\left(\tan\frac{
\phi_2}{2}\right)}{\cos\phi_1}\right]
\ed
\beq
y(\phi_1,\phi_2)=\sqrt{r_2-(r_1+~{\rm Im}~\tau\cos\phi_1)\cos\phi_2}
\label{toruscoords}
\eeq
which hold provided that ${\rm Re}~\tau\neq0$.

In the coordinates defined by the diffeomorphism (\ref{toruscoords}) the
Hamiltonian vector field generates the local action of the group $\IR^1$ of
translations in $x$. However, this diffeomorphism cannot be
extended globally to the whole of $T^2$ because it has singularities along
the coordinate circles
\beq
a_1=\left\{(\pi/2,\phi)\in T^2\right\}~~~,~~~a_2=\left\{(3\pi/2,
\phi)\in T^2\right\}
\label{singt2}\eeq
\beq
b_1=\left\{(\phi,0)\in T^2\right\}~~~,~~~b_2=\left\{(\phi,\pi)\in T^2\right\}
\label{critt2}\eeq
This means that $V_{\Sigma^1}$ cannot globally generate isometries of any
Riemannian geometry on $T^2$. Although translations in the coordinate $x$
represent some unusual local symmetry of the torus, it shows that the existence
of non-trivial homology cycles on $T^2$ lead to singularities in the circle
action of the Hamiltonian vector field on $T^2$. These singularities do
not appear on the Riemann sphere because any closed loop on $S^2$ is
contractable, so that the singular circles above collapse to points which
can be identified with the critical points of the Hamiltonian function.
In fact, as we saw in Section 6, the only equivariant Hamiltonians on the
torus are precisely those which generate translations along the homology
cycles of $T^2$, and so we see that the Hamiltonian (\ref{heightt2}) generates
a circle action that is singular along those cycles which are exactly the ones
required for a globally equivariantly-localizable system on the torus. This is
equivalent to the fact that the flow generated by $V_{\Sigma^1}$ bifurcates at
the saddle points of $h_{\Sigma^1}$ (like the equations of motion for a
pendulum), and the above shows analytically why there is
no single-valued, globally-defined Riemannian geometry on the torus for which
the height function $h_{\Sigma^1}$ generates isometries.

The local circle action defined by the diffeomorphism (\ref{toruscoords})
however partitions the torus into 4 open sets $P_i$ which are the disjoint
sets that remain when one removes the 2 canonical homology cycles
discussed above. Each of these sets $P_i$ is diffeomorphic to an open
rectangle in $\IR^2$ on which the Hamiltonian vector field $V_{\Sigma^1}$
generates a global $\IR^1$-action. Thus the above formalism implies that
the corrections to the Duistermaat-Heckman formula for the partition function
in this case is given by (\ref{2dcorrs}) evaluated on the pure singular
cycles $a_1$ and $a_2$ above, and on the critical cycles $b_1$ and $b_2$
(see the previous Subsection). Summing the 2 contributions from the 1-form
$F$ in (\ref{Fform}) along the pure homology cycles shows immediately that
$\oint_{a_1}F|_{a_1}+\oint_{a_2}F|_{a_2}=0$, as anticipated. As for the
integrals along the critical cycles, taking proper care of orientations
induced by the contractable patches, we find that the contributions from
$b_1$ and $b_2$ are the same and that the corrections can be written as
\beq\new{\begin{array}{l}
\delta Z_{T^2}(T)\\=-\frac{1}{iT~{\rm
Im}~\tau}\left(\e^{iT(r_2-r_1)}\int_0^{\pi}
d\phi~\frac{\e^{-iT~{\rm Im}~\tau\cos\phi}}{\sin\phi}-\e^{iT(r_2+r_1)}\int
_0^\pi d\phi~\frac{\e^{iT~{\rm Im}~\tau\cos\phi}}{\sin\phi}\right)\end{array}}
\label{t2corr}\eeq
After a change of variables we find that the integrals in (\ref{t2corr}) can be
expressed in terms of the exponential integral function \cite{gradryz}
\beq
{\rm Ei}(x)=-~{\int_{-x}^\infty}\!\!\!\!\!\!\!\!\!\!\!-~~dt~\frac{\e^{-t}}{t}
\label{expint}\eeq
which diverges for $x\leq0$. Here the integral denotes a Cauchy principal
value integration. After some algebra we find
\beq\new{\begin{array}{l}
\delta Z_{T^2}(T)\\=-\frac{1}{iT~{\rm Im}~\tau}\left[\e^{iT(r_2-r_1)}\left\{
\frac{\e^{iT~{\rm Im}~\tau}}{2}\left({\rm Ei}(-2iT~{\rm Im}~\tau)-~{\rm Ei}
\left(-2iT~{\rm Im}~\tau\cos^2\frac{y}{2}\right)\right)\right.\right.\\\left.
-\frac{\e^{-iT~{\rm Im}~\tau}}{2}\left({\rm Ei}\left(2iT~{\rm Im}~\tau\sin^2
\frac{\epsilon}{2}\right)-~{\rm Ei}(2iT~{\rm Im}~\tau)\right)\right\}\\-\e
^{iT(r_2+r_1)}\left\{\frac{\e^{-iT~{\rm Im}~\tau}}{2}\left({\rm Ei}(2iT~
{\rm Im}~\tau)-~{\rm Ei}\left(2iT~{\rm Im}~\tau\cos^2\frac{y}{2}\right)\right)
\right.\\\left.\left.-\frac{\e^{iT~{\rm Im}~\tau}}{2}\left({\rm Ei}\left(-2iT
{}~{\rm Im}~\tau\sin^2\frac{\epsilon}{2}\right)-~{\rm Ei}(-2iT~{\rm Im}~\tau)
\right)\right\}\right]\end{array}}
\label{ztsei}\eeq
where $y=\pi-\epsilon$ and $\epsilon\to0$ is used to regulate the divergence
of the integrals in (\ref{t2corr}) at $\phi=0$ and $\phi=\pi$.

The correction term (\ref{ztsei}) tells us quite a bit. First of all, note that
it is a sum of 4 terms which can be identified with the contributions from the
critical points of the Hamiltonian $h_{\Sigma^1}$. However, these terms are
resummed, since the above correction terms take into account the full loop
corrections to the Duistermaat-Heckman formula. Next, the terms involving
$\epsilon$ are divergent, and the overall divergence of $\delta Z_{T^2}(T)$ is
anticipated from Kirwan's theorem, which says that the full saddle-point series
for this Hamiltonian diverges. The exponential integral function can be
expanded as the series \cite{gradryz}
\beq
{\rm Ei}(x)=\gamma+\log x+\sum_{n=1}^\infty\frac{x^n}{nn!}
\label{eiexp}\eeq
for $x$ small, where $\gamma$ is the Euler-Mascheroni constant. Thus the
divergent pieces in (\ref{ztsei})
can be explicitly expanded in powers of $\frac{1}{T}$, giving a much simpler
way to read off the coefficients of the loop-expansion (note the enormous
complexity of the series coefficients in (\ref{loopexp}) for this Hamiltonian
-- a direct signal of the messiness of its stationary-phase series). Finally,
the finite terms (those independent of the regulator $\epsilon$), can be
evaluated for $T=-i$ and $\tau=1+i$, and we find $\delta Z_{T^2}=123.086$.
In Subsection 3.5 we saw that the exact value of the partition function for
this dynamical system was 2117.12, while the Duistermaat-Heckman formula gave
$Z_0=1849.327$. Thus $Z_0+\delta Z_{T^2}=1972.41$, which is a better
approximation to the partition function than the Duistermaat-Heckman formula.
Of
course, given the large divergence of the stationary phase series, we do not
expect that the finite contributions in (\ref{ztsei}) will give the exact
result for the partition function, but we certainly do get much closer. As the
function $\chi^1$ which generates the set of preferred coordinates is by no
means unique, perhaps a refined definition of it could lead to a better
approximation $Z_0+\delta Z$. Then, however, we lose a lot of the geometrical
interpretation of the corrections that we gave in the last Subsection.

The second set of examples we consider here are the potential problems
(\ref{qmham}) defined on the plane $\IR^2$, where $U(q)$ is a $C^\infty$
potential which is a non-degenerate function. In this
case the equation (\ref{chi1eq}) becomes
\beq
p\frac{\partial\chi^1}{\partial q}=-U'(q)\frac{\partial\chi^1}
{\partial p}
\label{r2eq}\eeq
which is solved by
\beq
\chi^1(q,p)=p^2/2-U(q)
\label{phir2}\eeq
Then proceeding as above the local coordinates $(\bar x,\bar y)$ in which the
Hamiltonian vector field generates translations are
\beq
\bar x(q,p)=\frac{1}{pU'(q)}\left(qU'(q)-p^2\right)~~~,~~~\bar y(q,p)=
\sqrt{\frac{p^2}{2}+U(q)}
\label{coordsr2}\eeq
Thus here there are only critical `cycles' given by the infinite lines
\beq
P=\left\{(0,q)\in\IR^2\right\}~~~,~~~{\cal U}_i=\left\{(p,q_i)\in\IR^2\right\}
\label{critr2}\eeq
where $q_i$ are the extrema of the potential $U(q)$.

Since for the Darboux Hamiltonian (\ref{qmham}), $V^p$ and $V^q$ vanish on
the `cycles' $P$ and ${\cal U}_i$ respectively, we must use the renormalized
version of (\ref{Fform}), namely (\ref{fredef}). Combining (\ref{fredef}) with
(\ref{corrfaell}) we find, for an even potential function, that the corrections
are
\beq
\delta Z_{\IR^2}(T)=-\frac{1}{iT}\left\{\int_0^\infty dq~\frac{\e^{iTU(q)}}
{U^\prime(q)}-\left(\sum_{q_i}\e^{iTU(q_i)}\right)\int_0^\infty dp~
\frac{\e^{iTp^2/2}}{p}\right\}
\label{poteval}\eeq
and we note the manner in which the divergences are cancelled here. From this
we immediately see that for the harmonic oscillator potential $U(q)=aq^2$,
the corrections (\ref{poteval}) vanish (note that the integration measures
in (\ref{poteval}) contain implicit factors of $\omega_{12}$ that maintain
covariance). Similarly, it is easily verified, by a simple change of variables,
that for a potential of the form $U(q)=aq+bq^2$ these correction terms
vanish, again as expected. Finally, for a quartic potential $U(q)=\frac{q^2}
{2}+\frac{q^4}{4}$ (the anharmonic oscillator), a numerical integration of
(\ref{poteval}) for $T=i$ gives $\delta Z_{\IR^2}=-0.538$ and the
Duistermaat-Heckman formula yields $Z_0=2\pi$. A numerical integration of the
original partition function gives $Z=4.851$, which differs from the value
$Z_0+\delta Z_{\IR^2}=5.745$. The corrections do not give the exact value here,
but again at least they are a better approximation than the Duistermaat-Heckman
formula. Again, a refinement of the
preferred coordinates could lead to a better approximation. The method of the
last Subsection has therefore ``stripped" off any potentially divergent
contributions to the loop-expansion and at the same time approximated the
partition function in a much better way. These last
few examples illustrate the applicability and the complete consistency of the
geometric approach of the last Subsection to the saddle-point expansion.
Indeed,
we see that it reproduces the precise analytic features of the loop-expansion
but avoids many of the cumbersome calculations in evaluating (\ref{loopexp}).
It would be interesting to develop some of these ideas further.

We would next like to check, following the analysis of Subsection 5.9, if
there are any conformally-invariant geometries for this dynamical system when
the potential $U(q)\geq0$ is bounded from below. In the harmonic-polar
coordinates (\ref{harmpolar}), the conformal Killing equations (\ref{conflie})
can be determined by setting the right-hand sides of the Killing equations
(\ref{killingrtheta}) equal to instead $(\nabla_\theta V^\theta)g_{\mu\nu}=
(\partial_\theta V^\theta+\Gamma^\theta_{\theta\theta}V^\theta)g_{\mu\nu}$.
After some algebra, we find that they generate the 2 equations
\beq
\partial_\theta\log\left(\frac{(V^\theta)^2g_{\theta\theta}}{g_{rr}}\right)
=2\frac{g_{r\theta}}{g_{rr}}\partial_r\log V^\theta
\label{confharmlie1}\eeq
\beq
\partial_\theta\log\left(V^\theta\frac{g_{\theta\theta}}{g_{r\theta}}\right)
=\frac{g_{\theta\theta}}{g_{r\theta}}\partial_r\log V^\theta
\label{confharmlie2}\eeq
(\ref{confharmlie2}) can be formally solved as
\beq
g_{r\theta}=-V^\theta g_{\theta\theta}\int_{\theta_0}^\theta d\theta'~\partial
_rV^{\theta'}+f(r)
\label{confformsol}\eeq
from which we see that again single-valuedness $g_{r\theta}(r,\theta+2\pi)
=g_{r\theta}(r,\theta)$ holds only when (\ref{constcond}) is true, i.e. when
$U(q)$ is the harmonic oscillator potential with $V^\theta=1$. Even for the
harmonic oscillator, the equations (\ref{confharmlie1}) and
(\ref{confharmlie2}) only seem to admit radially-symmetric solutions
$g_{\mu\nu}=g_{\mu\nu}(r)$ so that $V^\theta=1$ is a global isometry of $g$.
Thus, even though we lose the third equation in (\ref{killingrtheta}) which
established the results of Subsection 5.9 using the Killing equations, we still
arrive at the conclusion that there are no single-valued metric tensors
obeying the conformal Lie derivative requirement for essentially all potentials
which are bounded from below (and the harmonic oscillator only seems to
generate isometries). Thus the conformal symmetry requirement in the case at
hand does not lead to any new localizable systems.

Finally, we examine what can be learned in these cases from the vanishing of
the 2-loop correction (\ref{neighvan}) in harmonic coordinates. In these
coordinates, the connection 1-form (\ref{gammadef}) has components
\beq
\gamma_p=0~~~~~,~~~~~\gamma_y=\frac{dq}{dy}
\label{gammaharm}\eeq
and the condition (\ref{neighvan}) reads
\beq
\frac{d}{dy}\gamma_y=-\gamma_y^2
\label{gammavaneq}\eeq
There are 2 solutions to (\ref{gammavaneq}). Either $\gamma_y=0$, in which case
$U(q)$ is the harmonic oscillator potential, or $\gamma_y=(y+a)^{-1}$, where
$a$ is an integration constant. This latter solution, however, yields $q(y)=
C_1y^2+ay+C_0$, which gives a potential $U(q)$ which is not globally defined
as a $C^\infty$-function on $\IR^2$. Thus the {\it only} potential which is
bounded from below that leads to a localizable partition function is that of
the simple-harmonic oscillator. This example illustrates how the deep
geometric analyses of this Section serve of use in examining the localizability
properties of dynamical systems. As for these potential problems, it could
prove of use in examining the localization features of other more complicated
integrable systems \cite{gorsky}.

\subsection{Heuristic Generalizations to Path Integrals: Supersymmetry
Breaking}

The generalization of the loop expansion to functional integrals is not yet
known, although some formal suggestive techniques for
carrying out the full semi-classical expansion can be found in \cite{langouche}
and \cite{schulman}. It would be of utmost interest to carry out an analysis
along the lines of this Section for path integrals for several reasons.
There the appropriate loop space expansion should again be covariantized, but
this time the functional result need not be fully independent of the loop
space coordinates. This is because the quantum corrections could cause
anomalies for many of the symmetries of the classical theory (i.e. of the
classical partition function). In particular, the larger conformal dynamical
structures discussed in Subsection 7.2 above could play an important role in
path integral localizations which are expressed in terms of trajectories on the
phase space \cite{paniak}. It would be very interesting to see if these
general conformal symmetries of the classical theory remain unbroken by quantum
corrections in a path integral generalization. The absence of such a conformal
anomaly could then lead to a generalization of the above extended localizations
to path integral localization formulas. As this symmetry in the
finite-dimensional case is not represented by a nilpotent operator, such as
an exterior derivative, one would need some sort of generalized supersymmetry
arguments to establish the localization with these sorts of symmetries. When
these supersymmetries are globally present, the vanishing of higher-loop terms
in the path integral loop expansion is a result of the usual
non-renormalizations of 1-loop quantities in supersymmetric quantum field
theories that arise from the mutual cancellations between bosonic and fermionic
loops in perturbation theory (where the fermionic loops have an extra minus
sign compared to the bosonic ones).

Quite generally though, one also has to keep in mind that the loop space
localization formulas are rather formal. We have overlooked several formal
functional aspects, such as difficulties associated with the definition of
the path integral measure. There may be anomalies associated with the
argument in Subsection 4.4 that the path integral is independent of the
limiting
parameter $\lambda\in\IR$, for instance the supersymmetry may be broken in
the quantum theory (e.g. by a scale anomaly in the rescaling of the phase
space metric $g\to\lambda\cdot g$). The same sort of anomalies could also
break the larger conformal symmetry we have found for the classical theory
above. However, even if the localization formulas are not
correct as they stand, it would then be interesting to uncover the reasons
for that. This could then provide one with a systematic geometric method
for analysing corrections to the WKB approximation.

The ideas presented in this Section are a small step forward in this direction.
In particular, it would be interesting to generalize the construction of
Subsection 7.3, as this is the one that is intimately connected to the
integrability features of the dynamical system. The Poincar\'e duality
interpretation there is one possible way that the construction could generalize
to path integrals. For path integrals, we would expect the feature of an
invariant metric tensor that cannot be extended globally to manifest itself as
a local (i.e. classical) supersymmetry of the theory which is dynamically
broken globally on the loop space. This has been discussed by Niemi and Palo
\cite{niemipalo3} in the context of the supersymmetric non-linear sigma-model
(see Section 8). Another place where the metric could enter into a breakdown of
the localization formulas is when the localization 1-form $\psi\sim i_Wg$ does
not lead to a homotopically-trivial element under the (infinitesimal)
supersymmetry transformation described by $Q_S$. Then additional input into the
localization formalism should be required on a topologically non-trivial phase
space to ensure that $Q_S\psi$ indeed does reside in the trivial homotopy
class. These inputs could follow from an appropriate loop space extension of
the correction terms $\delta Z(T)$ discussed above, which will then always
reflect global properties of the quantum theory. Other directions could also
entail examining the connections between equivariant localization and other
ideas we have discussed in this Review. One is the Parisi-Sourlas supersymmetry
that we encountered in the evaluation of the Niemi-Tirkkonen localization
formula for the height function on the sphere (Subsection 5.5), although this
feature seems to be more intimately connected to the K\"ahler geometry of
$S^2$, as we showed above. The K\"ahler symmetries we found in Subsection 7.1
would be a good probe of the path integral correction formulas, and it would
interesting to see if they could also be generalized to some sort of
supersymmetric structure.

\section{Equivariant Localization in Cohomological Field Theory}

We have seen that the equivariant localization formalism is an excellent,
conceptual geometric arena for studies of supersymmetric and topological field
theories, and more generally of (quantum) integrability. Given that the
Hamiltonians in an integrable hierarchy are functionals of action variables
alone \cite{mormatrix}, the equivariant localization formalism might yield a
geometric characterization of quantum integrability, and perhaps some deeper
connection between quantum-integrable bosonic theories and supersymmetric
quantum field theories. This is particularly interesting from the point of view
of examining corrections to the localization formulas, which in the last
Section we have seen reflect global properties of the theory. This would be of
particular interest to analyse more closely, as it could then lead to a unified
description of localization in the symplectic loop space, the supersymmetric
loop space and in topological quantum field theory.

In this final Section we shall discuss some of the true field theoretical
models to which the equivariant localization formalism can be applied. We shall
see that the quantum field theories which fall into this framework always have,
as anticipated, some large symmetry group (such as a topological gauge symmetry
or a supersymmetry) that serves to provide an equivariant cohomological
structure on the space of fields that can be understood as a ``hidden"
supersymmetry of the theory. Furthermore, the configuration spaces of these
models must always admit some sort of (pre-)symplectic structure in order that
the localization properties of phase space path integrals can be applied.
Because of space considerations we have not attempted to give a detailed
presentation here and simply present an overview of the various constructions
and applications, mainly just presenting results that have been obtained. The
interested reader is refered to the extensive list of references that are cited
throughout this Section for more details. We shall emphasize here the
connections eluded to throughout this Review between the localization formalism
for dynamical systems and genuine topological quantum field theories. Exploring
the connections between the topological field theories and more conventional
physical quantum field theories will then demonstrate how the equivariant
localization formalism for phase space path integrals serves as the correct
arena for studying the (path integral) quantization of real physical systems.

\subsection{Two-dimensional Yang-Mills Theory: Equivalences between Physical
and Topological Gauge Theories}

In Subsection 3.8 we first pointed out that, instead of circular actions, one
can consider the Poisson action of some non-abelian Lie group acting on the
phase space. Then the non-abelian generalizations of the equivariant
localization formulas, discussed in Subsections 3.8, 4.9 and 5.8, lead to
richer structures in the quantum representations discussed earlier and one
obtains intriguing path integral representations of the groups involved. In
this Subsection we shall demonstrate how a formal application of the Witten
localization formalism can be used to study a cohomological formulation of
2-dimensional QCD (equivalently the weak-coupling limit of 2-dimensional pure
Yang-Mills theory). This leads to interesting physical and mathematical
insights into the structures of these theories. We shall also discuss how these
results can be generalized to topological field theory limits of other models.

First, we briefly review some of the standard lore of 2-dimensional QCD. The
action for pure Yang-Mills theory on a 2-dimensional surface $\Sigma^h$ of
genus $h$ is
\beq
S_{YM}[A]=-\frac{1}{2e^2}\int_{\Sigma^h}\tr~F_A\star F_A
\label{YMaction}\eeq
where $A$ is a gauge connection of a trivial prinicipal $G$-bundle over
$\Sigma^h$, $F_A$ is its curvature 2-form (c.f. Subsection 2.4), and $e^2$ is
the coupling constant of the gauge field theory. Since $\phi\equiv\star F_A$ is
a scalar field in 2-dimensions, the action (\ref{YMaction}) depends on the
metric of $\Sigma^h$ only through its dependence on the area
$A(\Sigma^h)=\int_{\Sigma^h}\star1$ of the surface. A deformation of the metric
can therefore be compensated by a change in the coupling constant $e^2$. The
action (\ref{YMaction}) is invariant under the gauge transformations
(\ref{gaugetransfdef}). The corresponding quantum field theory is described by
the path integral
\beq
Z_{\Sigma^h}(e^2)\equiv\int_{\cal A}[dA]~\e^{iS_{YM}[A]}
\label{YMpartA}\eeq
where $\cal A$ is the space of gauge connections over $\Sigma^h$.

We can write the partition function in a much simpler (first order) form by
treating the $\bf g$-valued scalar field $\phi=\star F_A\in
C^\infty(\Sigma^h,{\bf g})$ as a Lagrange multiplier to write
\beq
Z_{\Sigma^h}(e^2)=\int_{\cal A}[dA]~\int_{C^\infty(\Sigma^h,{\bf
g})}[d\phi]~\e^{-i\int_{\Sigma^h}\tr\left(i\phi
F_A+\frac{e^2}{2}\phi\star\phi\right)}
\label{YMpartAp}\eeq
In the weak coupling limit $e^2\to0$, the action in (\ref{YMpartcoh}) reduces
to the topological one $\int_{\Sigma^h}\tr~i\phi F_A$ (i.e. one that is
independent of the metric of $\Sigma^h$ and which consequently determines a
cohomological quantum field theory)\footnote{\baselineskip=12pt This sort of
topological field theory is called a `BF theory' and it is the prototype of a
Schwarz-type topological gauge theory \cite{birm}.}. The gauge invariance of
the action $S[\phi,A]$ appearing in (\ref{YMpartAp}) is expressed as
$S[g^{-1}\phi g,A^g]=S[\phi,A]$ where $g\in{\bf g}$ and $\phi$ transforms under
the adjoint representation of the gauge group. Because of this gauge
invariance, it is necessary to fix a gauge and restrict the integration in
(\ref{YMpartAp}) to the equivalence classes ${\cal A}/G$ of gauge connections
modulo gauge transformations. This can done by the standard BRST gauge fixing
procedure (see Appendix A for a brief account).

For this, we introduce an auxilliary, $\bf g$-valued fermion field $\psi^\mu$,
which is an anti-commuting 1-form in the adjoint representation of $\bf g$, and
write (\ref{YMpartAp}) as
\beq\new{\begin{array}{ll}
Z_{\Sigma^h}(e^2)=&\frac{1}{{\rm vol}~C^\infty(\Sigma^h,{\bf g})}\int_{{\cal
A}\otimes\Lambda^1{\cal A}}[dA]~[d\psi]~\int_{C^\infty(\Sigma^h,{\bf
g})}[d\phi]\\&\times\exp\left\{-\int_{\Sigma^h}\tr\left(\phi
F_A-\frac{1}{2}\psi\wedge\psi\right)-i\frac{e^2}{2}\int_{\Sigma^h}\tr~\phi
\star\phi-i\int_{\Sigma^h}\star\{Q,\Psi\}\right\}\end{array}}
\label{YMpartcoh}\eeq
In (\ref{YMpartcoh}) we have introduced the usual BRST and Faddeev-Popov gauge
fixing terms defined by the graded BRST commutator of a gauge fermion
$\Psi=\psi^\mu\Pi_\mu(x)$ with the usual BRST charge $Q$. The square of $Q$ is
$Q^2=-i\delta_\phi$ where $\delta_\phi$ is the generator of a gauge
transformation with infinitesimal parameter $\phi$. Thus $Q$ is nilpotent on
the space of physical (i.e. gauge-invariant) states of the quantum field
theory. The system of fields $(A,\psi,\phi)$ is the basic multiplet of
cohomological Yang-Mills theory. The (infinitesimal) gauge invariance of
(\ref{YMpartcoh}) is manifested in its invariance under the infinitesimal BRST
supersymmetry transformations
\beq
\delta A_\mu=i\epsilon\psi_\mu~~~,~~~\delta\psi_\mu=-\epsilon(\nabla_A)_\mu\phi
=-\epsilon(\partial_\mu+[A_\mu,\phi])~~~,~~~\delta\phi=0
\label{YMBRST}\eeq
where $\epsilon$ is an anticommuting parameter. The supersymmetry
transformations (\ref{YMBRST}) are generated by the graded BRST commutator
$\delta\Phi=-i\{Q,\Phi\}$ for each field $\Phi$ in the multiplet
$(A,\psi,\phi)$. The ghost quantum numbers ($\IZ$-gradings) of the fields
$(A,\psi,\phi)$ are $(0,1,2)$.

We shall not enter here into a discussion of the physical characteristics of
2-dimensional Yang-Mills theory. It is a super-renormalizable quantum field
theory which is exactly solvable and whose simplicity therefore allows one to
explore the possible structures of more complicated non-abelian gauge theories
such as higher-dimensional cohomological field theories and other physical
models such as 4-dimensional QCD. It can be solved using group character
expansion methods \cite{cordes} or by diagonalization of the functional
integration in (\ref{YMpartAp}) onto the Cartan subalgebra using the elegant
Weyl integral formula \cite{blauthom}. Here we wish to point out the
observation of Witten \cite{witten3} that the BRST gauge-fixed path integral
(\ref{YMpartcoh}) is an infinite-dimensional version of the partition function
in the last line of (\ref{sqloc}) used for non-abelian localization. Indeed,
the integration over the auxilliary fermion fields $\psi$ acts to produce a
field theoretical analog of the super-loop space Liouville measure introduced
in Section 4. The ``Hamiltonian" here is the field strength tensor $F_A$ while
the Lagrange multiplier fields $\phi$ serve as the dynamical generators of the
symmetric algebra $S({\bf g}^*)$ used to generate the $G$-equivariant
cohomology. The ``phase space" $\man$ is now the space $\cal A$ of gauge
connections, and the Cartan equivariant exterior derivative
\beq
D=\int_{\Sigma^h}\star\left(\psi^\mu\frac{\delta}{\delta
A_\mu}-i\phi^aV^{a,\mu}\frac{\delta}{\delta\psi^\mu}\right)
\label{YMeqderiv}\eeq
in this case coincides with the action of the BRST charge $Q$, i.e.
$D\Phi=-\{Q,\Phi\}$. The gauge fermion $\Psi$ thus acts as the localization
1-form $\lambda$ and, by the equivariant localization principle, the
integration will localize onto the field configurations where
$\lambda(V^a)=V^{a,\mu}\Pi_\mu=0$ where $V^a=V^{a,\mu}\frac{\partial}{\partial
x^\mu}$ are the vector fields generating $\bf g$.

The equivalence between the first and last lines of (\ref{sqloc}) is the basis
of the mapping between ``physical" Yang-Mills theory with action
(\ref{YMaction}) and the cohomological Yang-Mills theory with action
$\int_{\Sigma^h}\tr~i\phi F_A$ which is defined essentially by the steps which
lead to the non-abelian localization principle, but now in reverse. The extrema
of the action (\ref{YMaction}) are the classical Yang-Mills field equations
$F_A=0$. Thus the localization of the partition function will be onto the
symplectic quotient $\man_0$ which here is the moduli space of flat gauge
connections modulo gauge transformations associated with the gauge group $G$.
This mapping between the physical gauge theory and the cohomological quantum
field theory is the basis for the localization of the 2-dimensional Yang-Mills
partition function. Thus the large equivariant cohomological symmetry of this
theory explains its strong solvability properties that have been known for
quite some time now. More generally, as mentioned at the end of Subsection 5.8,
the equivariant localization here also applies to the basic integrable models
which are related to free field theory reductions of 2-dimensional Yang-Mills
theory, such as Calegoro-Moser integrable models \cite{gorsky}.

To carry out the localization onto $\man_0$ explicitly, we choose a
$G$-invariant metric $g$ on $\Sigma^h$ and take the localization 1-form in
(\ref{sqloc}) to be
\beq
\lambda=\int_{\Sigma^h}\tr~\psi\wedge\star Df=\int_{\Sigma^h}d{\rm
vol}(g(x))~\tr~\psi^\mu D_\mu f
\label{YMlocform}\eeq
where $f=\star F_A$. The localization onto $\lambda(V)=0$ is then identical to
localization onto the solutions of the classical Yang-Mills equations. We shall
not enter into the cumbersome details of the evaluation of the partition
function (\ref{YMpartcoh}) at weak coupling $e^2\to0$ (the localization limit)
using the Witten non-abelian localization formalism. For details, the reader is
refered to \cite{witten3}. As in Subsection 5.8, the final integration formula
can be written as a sum over the unitary irreducible representations of $G$,
\beq
Z_{\Sigma^h}(e^2)=\e^{-\frac{e^2}{2}\sum_i\rho_i^2}\prod_{\alpha>0}\alpha
(\rho)^{2-2h}\sum_{\lambda\in\IZ^r}(\dim{\cal
R}_\lambda)^{2-2h}\e^{-\frac{e^2}{2}\sum_i(\lambda+\rho)_i^2}
\label{YMpartreps}\eeq
This result follows from expanding the various ($G$-invariant) physical
quantities appearing in the localization formula in characters of the group $G$
(c.f. Subsection 5.8). From a physical standpoint the localization formula
(\ref{YMpartreps}) is interesting because although it expresses the exactness
of a loop approximation to the partition function, it is a non-polynomial
function of the coupling constant $e^2$. This non-polynomial dependence arises
from the contributions of the unstable classical solutions to the functional
integral as described in Section 3. Such behaviours are not readily determined
using the conventional perturbative techniques of quantum field theory. Thus
the mapping provided above between the physical and topological gauge theories
(equivalently the generalization of the Duistermaat-Heckman integration formula
to problems with non-abelian symmetries) provides an unexpected and new insight
into the structure of the partition function of 2-dimensional Yang-Mills
theory. This simple mapping provides a clearer picture of this quantum field
theoretical equivalence which is analogous to the more mysterious equivalence
of topological and physical gravity in 2 dimensions. From a mathematical
perspective, the quantity (\ref{YMpartreps}) is the correct one to use for
determining the intersection numbers of the moduli space of flat
$G$-connections on $\Sigma^h$ \cite{jeffrey,kalkman2,witten3}. This approach to
2-dimensional Yang-Mills theory has also been studied for genus $h=0$ in
\cite{minahan}.

The intriguing mapping between a physical gauge theory, with propagating
particle-like local degrees of freedom, and a topological field theory with
only global degrees of freedom has also been applied to more complicated
models. In \cite{blauthom1}, similar considerations were applied to the
non-linear cousin of 2-dimensional topological Yang-Mills theory, the gauged
$G/G$ Wess-Zumino-Witten model. This model at level $k\in\IZ$ is defined by the
action
\beq\new{\begin{array}{c}
S_{G/G}[g,A]=-\frac{k}{8\pi}\int_{\Sigma^h}\tr~g^{-1}\nabla_Ag\wedge\star
g^{-1}\nabla_Ag+\frac{k}{12\pi}\int_M\tr(g^{-1}dg)^{\wedge3}\\-\frac{k}{4\pi}
\int_{\Sigma^h}\tr\left(A\wedge dgg^{-1}+A\wedge A^g\right)\end{array}}
\label{G/GWZWaction}\eeq
where $g\in C^\infty(\Sigma^h,{\bf g})$ is a smooth group-valued field, $M$ is
a 3-manifold with boundary the surface $\Sigma^h$, and $A$ is a gauge field for
the diagonal $G$ subgroup of the $G_{\rm L}\times G_{\rm R}$ symmetry group of
the ordinary (ungauged) Wess-Zumino-Witten model defined by the action
$S_G[g]=S_{G/G}[g,A=0]$ \cite{ginsparg}. Since the Hodge duality operator
$\star$ is conformally invariant when acting on 1-forms, the action
(\ref{G/GWZWaction}) depends only on the chosen complex structure of
$\Sigma^h$. As for the Yang-Mills theory above, the geometric interpretation of
the theory comes from adding to the bosonic action (\ref{G/GWZWaction}) the
term $\Omega(\psi)=\frac{1}{2\pi}\int_{\Sigma^h}\psi_z\psi_{\bar z}$ quadratic
in Grassmann-odd variables $\psi$ which represents the symplectic form
$\int_{\Sigma^h}\delta A\delta A$ on the space $\cal A$ of gauge fields on
$\Sigma^h$. Again the resulting theory is supersymmetric and the infinitesimal
supersymmetry transformations are
\beq
\delta A_z=\psi_z~~~,~~~\delta\psi_z=A_z^g-A_z~~~~;~~~~\delta A_{\bar
z}=\psi_{\bar z}~~~,~~~\delta\psi_{\bar z}=A_{\bar z}-(A_{\bar z}^g)^{-1}
\label{G/Gsusy}\eeq
with the supplemental condition $\delta g=0$. Unlike its Yang-Mills theory
counterpart, the square of this supersymmetry $\Delta=\delta^2$ does not
generate infinitesimal gauge transformations but rather `global' gauge
transformations (generated by the cohomological elements of the gauge group
which are not connected to the identity).

Thus the action (\ref{G/GWZWaction}) here admits a supersymmetry which does not
manifest itself as the local gauge symmetry of the quantum field theory.
Nonetheless, this implies a supersymmetric structure for equivariant cohomology
which can be used to obtain a localization of the corresponding path integral
in the usual way. The localization formula of \cite{bismut1} for equivariant
K\"ahler geometry has a field theoretic realization in this model
\cite{blauthom1} and the fixed-point localization formula is the algebraic
Verlinde formula for the dimension of the space of conformal blocks of the
ordinary Wess-Zumino-Witten model. For example, in the case $G=SU(2)$ the
localization formula gives
\beq\new{\begin{array}{ll}
Z_{\Sigma^h}(SU(2),k)&=\int_{C^\infty(\Sigma^h,{\bf su(2)})}[dg]~\int_{\cal
A}[dA]~\e^{iS_{SU(2)/SU(2)}[g,A]}\\&=\left(\frac{2}{k+2}\right)^{2-2h}~
\sum_{\ell=1}^{k+1}\left(\sin\frac{\pi\ell}{k+2}\right)^{2-2h}\end{array}}
\label{su2ver}\eeq
Thus the equivariant localization formalism can also be used to shed light on
some of the more formal structures of 2-dimensional conformal field theories.

Perret \cite{perret} has used the path integral for a version of the ordinary
Wess-Zumino-Witten model to give a field theoretical generalization of Stone's
derivation for the Weyl-Kac character formula for Kac-Moody algebras (i.e. loop
groups). This is done by exploiting the $G_{\rm L}\times G_{\rm R}$ Kac-Moody
symmetry associated with the quantum field theory with action $S_G[g]$ as a
supersymmetry of the model along the same lines as in Section 5 before. Let us
now briefly describe Perret's derivation. The Kac-Moody group $\tilde G$ is a
central extension of the loop group $S^1\to\tilde G\to LG$ of a compact
semi-simple Lie group $G$, and it looks locally like the direct product
$LG\otimes S^1$, i.e. an element $\tilde g\in\tilde G$ looks locally like
$\tilde g=(g(x),c)$ where $g:S^1\to G$ and $c\in S^1$. The coherent state path
integral for the character is
\beq
\tr_\lambda\e^{iH}q^{L_0}=\int_{L\tilde G}[d\tilde g]~\e^{\oint_{\tilde
g}\langle\tilde g|d+i(H+\tau L_0)dt|\tilde g\rangle}
\label{KMcharpathint}\eeq
where $H=\sum_ih_iH_i\in H_C$, $q=\e^{i\tau}$ and $L_0$ is the generator of
rotations of the loops. The action in (\ref{KMcharpathint}) depends on 2
coordinates, the coordinate $x$ along the loop in $L{\bf g}$ and the time
coordinate $t$ of the path integral. The central $S^1$-part of the coherent
states in the path integral drops out due to gauge invariance, and thus the
character representation (\ref{KMcharpathint}) will define a 2-dimensional
quantum field theory on the torus $T^2$ (i.e. the quotient of loops in the loop
group $LG$).

It can be shown \cite{perret} that the coherent state path integral
(\ref{KMcharpathint}) is the quantum field theory with action
\beq
S_{KM}=\frac{1}{2\pi}\int_{T^2}dx~dt~\tr\left(\lambda
g^{-1}(\bar\partial+H)g+\frac{k}{2}g^{-1}\partial_x(\bar\partial+2H)g\right)
+\frac{k}{12\pi}\int_M\tr(g^{-1}dg)^{\wedge3}
\label{KMaction}\eeq
Here $k\in\IZ$ is the given central extension of the loop group, $\lambda$ is a
dominant weight of $G$, and $\bar\partial=\partial_t-\tau\partial_x$ so that
the Cartan angle $\tau$ becomes the modular parameter of the torus. When
$\lambda=H=0$ the action (\ref{KMaction}) becomes the chiral Wess-Zumino-Witten
model with the single Kac-Moody symmetry $g\to k(z)g$. This symmetry is still
present for generic $\lambda\neq0$, and we can gauge the action $S_{KM}$ with
respect to an arbitrary subgroup of $G$ by replacing $H$ with a vector field
(see (\ref{G/GWZWaction})). One can now evaluate the infinite-dimensional
Duistermaat-Heckman integration formula for this path integral. The critical
points of the action (\ref{KMaction}) are in one-to-one correspondence with the
affine Weyl group $W_{\rm aff}(H_C)=W(H_C)\semi\hat H_C$, where $\hat H_C$ is
the set of co-roots. Let $r(w)$ denote the rank of an element $w\in W(H_C)$,
and let $h$ be the dual Coxeter number of the Lie group $G$ \cite{varadarajan}.
Using modular invariance of the character to fix the conventional zero-point
energy (associated with the usual $SL(2,\IR)$-invariance of the conformal field
theory vacuum \cite{ginsparg}), it can be shown that the WKB localization
formula for the coherent state path integral (\ref{KMcharpathint}) coincides
with the Weyl-Kac character formula \cite{perret} (for notation see Subsection
5.1)
\beq\new{\begin{array}{ll}
\tr_\lambda\e^{iH}q^{L_0}=&\sum_{w\in W(H_C),\eta\in\hat
H_C}(-1)^{r(w)}~q^{\sum_i((\lambda+\rho)_i+(k+h)\eta_i^2-\rho_i^2)/2(k+h)}
\\&~~~~~\times\e^{i(\lambda+\rho+(k+h)\eta)(H^{(w)})}\e^{-i\rho(H^{(w)})}
\\&~~~~~\times\prod
_{n>0}\left(1-q^n\right)^{-r}\prod_{\alpha>0}\left(1-\e^{i\alpha(H^{(w)})}q^n
\right)^{-1}\left(1-\e^{-i\alpha(H^{(w)})}q^{n-1}\right)^{-1}\end{array}}
\label{weylkacform}\eeq
which arises from the expansions of various quantities defined on the torus in
terms of Jacobi theta-functions. Thus infinite dimensional analogs of the
localization formulas in quantum field theory can also lead to interesting
generalizations of the character formulas that the topological field theories
of earlier Sections represented.

Finally, it is possible to use some of these ideas in the context of abelian
localization as well. For instance, in \cite{niemisr} the abelian gauge theory
with action
\beq
S_{CSP}[A]=\frac{k}{8\pi}\int_M A\wedge dA-\frac{m}{2}\int_MA\wedge\star A
\label{CSPaction}\eeq
defined on a 3-manifold $M$ was studied within the equivariant localization
framework. The first term in (\ref{CSPaction}) is the Chern-Simons action which
defines a topological field theory\footnote{\baselineskip=12pt The
Cherns-Simons action is in fact another prototype of a Schwarz-type topological
field theory \cite{birm}.}, while the second term is the Proca mass term for
the gauge field which gives a propagating degree of freedom with mass $m$ and
thus breaks the topological invariance of the quantum field theory. In a
canonical formalism where $M=\Sigma^h\times\IR^1$, one can naturally write the
model (\ref{CSPaction}) as a quantum mechanics problem on the phase space
$\Sigma^h$ and apply the standard abelian equivariant localization techniques
to evaluate the path integral from the ensuing supersymmetry generated by the
gauge invariance of (\ref{CSPaction}) (in the Lorentz gauge $\partial^\mu
A_\mu=0$) \cite{blau1}. The path integral localization formula coincides with
that of a simple harmonic oscillator of frequency $\omega_h=8\pi m/k$, i.e.
$Z=1/2\sin(T\cdot4\pi m/k)$, indicating a mapping once again to a topological
field theory using the equivariant localization framework. The infrared limit
$m\to0$ of the model leads to the usual topological quantum mechanical models
associated with Chern-Simons theory \cite{blau1} and the supersymmetry, which
is determined by a loop space equivariant cohomology, emerges from the
symplectic structure of the theory on $\Sigma^h$ and could yield interesting
results in the full 3-dimensional quantum field theory defined by
(\ref{CSPaction}).

\subsection{Symplectic Geometry of Poincar\'e Supersymmetric Quantum Field
Theories}

In the last Subsection we showed how loop space equivariant localization
provides a correspondence between certain physical gauge theories and
topological ones which makes manifest the localization properties of the
physical models. In these cases the original theory contains a large gauge
symmetry which leads to a localization supersymmetry and a limit where the
model becomes topological that gives the usual localization limit. It is now
natural to ask what happens when the original quantum field theory is
explicitly defined with a supersymmetry (i.e. one that is not ``hidden"). In
Subsection 4.2 we saw that $N=\frac{1}{2}$ supersymmetric quantum mechanics
admits a loop space equivariant cohomological structure as a result of the
supersymmetry which provides an alternative explanation for the well-known
localization properties of this topological field theory (where the topological
nature now arises because the bosonic and fermionic degrees of freedom mutually
cancel each other out). From the point of view of path integration, this
approach in fact led to a nice, geometric interpretation on the functional loop
space of the features of this theory. It has been argued
\cite{morniemi1,morniemi2,palo} that this interpretation can be applied to {\it
generic} quantum field theories with Poincar\'e supersymmetry. In this
Subsection we shall briefly discuss how this works and how the techniques of
equivariant localization could lead to new geometrical interpretations of such
models.

To start, let us quickly review some of the standard ideas in Poincar\'e
supersymmetric quantum field theories. The idea of supersymmetry was first used
to relate particles of different spins to each other (e.g. the elementary
representation theory of $SU(3)$ which groups particles of the same spin into
multiplets) by joining the internal (isospin) symmetries and space-time
(Poincar\'e) symmetries into one large symmetry group (see \cite{sohnius} for a
comprehensive introduction). This is not possible using bosonic commutation
relations because then charges of internal symmetries have to commute with
space-time transformations so that dynamical breaking of these symmetries
(required since in nature such groupings of particles are not observed) is not
possible. However, it is possible to consider anti-commutation relations, i.e.
supersymmetries, and then the imposition of the Jacobi identity for the
symmetry group leads to a very restricted set of commutation relations. Here we
shall be concerned only with those anti-commutation relations satisfied by the
infinitesimal supersymmetry generators $Q_\alpha^i,\bar Q_{\dot\alpha}^i$,
\beq
\{Q_\alpha^i,\bar Q_{\dot\beta}^j\}=2\delta^{ij}\Sigma^\mu_{\alpha\dot\beta}
P_\mu+Z_{\alpha\dot\beta}^{ij}~~~;~~~i,j=1,\dots,N
\label{Nsusyalg}\eeq
where $\Sigma^\mu=\gamma^\mu{\cal C}$ with $\gamma^\mu$ the Dirac matrices and
$\cal C$ the charge conjugation matrix, $P_\mu=-i\partial_\mu$ is the generator
of space-time translations, and $Z^{ij}$ is an antisymmetric matrix of
operators proportional to the generators of the internal symmetry group. For
the rest of the relations of the super-Poincar\'e group, see \cite{sohnius}. We
assume here that the spacetime has Minkowski signature.

We shall be interested in using the relations of the super-Poincar\'e algebra
to obtain a symplectic structure on the space of fields. For this, it turns out
that only the $i=j$ terms in (\ref{Nsusyalg}) are relevant. It therefore
suffices to consider an $N=1$ supersymmetry with no internal $Z^{ij}$ symmetry
group terms. The most expedient way to construct supersymmetric field theories
(i.e. those with actions invariant under the full super-Poincar\'e group) is to
use a superspace formulation. We introduce 2 Weyl spinors $\theta^\alpha$ and
$\bar\theta_{\dot\alpha}$ which parametrize the infinitesimal supersymmetry
transformations. Then the $N=1$ supersymmetry generators in 4 dimensions can be
written as
\beq
Q_\alpha=\frac{\partial}{\partial\theta^\alpha}-i\Sigma^\mu_{\alpha\dot\alpha}
\bar\theta^{\dot\alpha}\partial_\mu~~~,~~~\bar
Q_{\dot\alpha}=-\frac{\partial}{\partial\bar\theta^{\dot\alpha}}+i
\Sigma^\mu_{\alpha\dot\alpha}\theta^\alpha\partial_\mu
\label{Qsupersp}\eeq
Kinetic terms in the supersymmetric action are constructed from the covariant
superderivatives
\beq
D_\alpha=\frac{\partial}{\partial\theta^\alpha}+i\Sigma^\mu_{\alpha\dot\alpha}
\bar\theta^{\dot\alpha}\partial_\mu~~~,~~~\bar
D_{\dot\alpha}=-\frac{\partial}{\partial\bar\theta^{\dot\alpha}}-i\Sigma^\mu_{
\alpha\dot\alpha}\theta^\alpha\partial_\mu
\label{covsuperderivs}\eeq
It can be readily verified that with the representation (\ref{Qsupersp}) the
relations of the $N=1$ super-Poincar\'e algebra are satisfied. In this
superspace notation, a general group element of the supersymmetry algebra is
$\e^{\theta^\alpha Q_\alpha+\bar\theta^{\dot\alpha}\bar Q_{\dot\alpha}}$ and,
using the supersymmetry algebra along with the Baker-Campbell-Hausdorff
formula, its action on a field $A(x)$ is $\e^{\theta^\alpha
Q_\alpha+\bar\theta^{\dot\alpha}\bar Q_{\dot\alpha}}A(x)=\e^{\theta^\alpha
Q_\alpha}\e^{\bar\theta^{\dot\alpha}\bar Q_{\dot\alpha}}A(y)$ where
$y^\mu=x^\mu+i\theta^\alpha\Sigma^\mu_{\alpha\dot\alpha}\bar
\theta^{\dot\alpha}$ are coordinates in superspace. Thus the supersymmetry transformation parameters live in superspace and the fields of the supersymmetric field theory are defined on superspace.

To incorporate the supersymmetry algebra as the symmetry algebra of a physical
system, we need some representation of it in terms of fields defined over the
space-time. The lengthy algorithm to construct supermultiplets associated with
a given irreducible or reducible spin representation of the super-Poincar\'e
algebra can be found in \cite{sohnius}. For instance, in 4 dimensions chiral
superfields (satisfying $\bar D_{\dot\alpha}\Phi=0$) are given by
\beq
\Phi^\mu(x,\theta,\bar\theta)=\phi^\mu(y)+\theta^\alpha\psi_\alpha^\mu(y)+
\theta^\alpha\theta_\alpha F^\mu(y)
\label{chiralsupfield}\eeq
where $(\phi,\psi)$ are spin $(0,\frac{1}{2})$ fields, and $F$ are auxilliary
fields use to close the supersymmetry representation defined by the chiral
superfields. In the following we shall consider multiplets of highest spin 1.
Other multiplets can be obtained by imposing some additional constraints. The
most general $N=1$ supermultiplet in 4 space-time dimensions consists of a
scalar field $M$, 3 pseudoscalar fields $C$, $N$ and $D$, a vector field
$A_\mu$ and 2 Dirac spinor fields $\chi$ and $\lambda$. The supersymmetry
charges $Q_\alpha^i$ and $\bar Q_{\dot\alpha}^i$ respectively raise and lower
the (spin) helicity components of the mulitplets by $\frac{1}{2}$. The
super-Poincar\'e algebra can be represented in the Majorana representation
where $\gamma^0=-\sigma^2\otimes{\bf1}$, $\gamma^1=-i\sigma^3\otimes\sigma^1$,
$\gamma^2=i\sigma^1\otimes{\bf1}$ and $\gamma^3=-i\sigma^3\otimes\sigma^3$,
with $\sigma^i$ the usual Pauli spin matrices. Then the $4\times4$
$\Sigma$-matrices on the right-hand side of (\ref{Nsusyalg}) are
\beq
\Sigma^0=\pmatrix{-{\bf1}&0\cr0&-{\bf1}\cr}~~~,~~~\Sigma^1=\pmatrix{0&{\bf1}
\cr{\bf1}&0\cr}~~~,~~~\Sigma^2=\pmatrix{0&-i\cdot{\bf1}\cr
i\cdot{\bf1}&0\cr}~~~,~~~\Sigma^3=\pmatrix{{\bf1}&0\cr0&-{\bf1}\cr}
\label{Sigmamajrep}\eeq
The Majorana representation selects the preferred light-cone coordinates
$x^\pm=x^2\pm x^0$ for the translation generators $P_\mu$ in (\ref{Nsusyalg})
above. Different representations of the Dirac gamma-matrices would then define
different preferred light-cone directions.

The general supersymmetry transformations of the complex $N=1$ supermultiplet
$V=(C;\chi;M,N,A_\mu;\lambda;D)$ and the action of the supersymmetry charges
$Q_\alpha^i,\bar Q_{\dot\alpha}^i$ on $V$ can be found in \cite{palo}. There it
was shown that the infinitesimal supersymmetry transformations can be written
in a much simpler form using the auxilliary fields
\beq\new{\begin{array}{c}
M'=M+A_3+\partial_1C~~~,~~~N'=N+A_1-\partial_3C~~~,~~~D'=D+\partial_1A_3-
\partial_3A_1\\\lambda_1'=\lambda_1-\partial_3\chi_1~~~,~~~\lambda_2'=
\lambda_2-\partial_1\chi_1~~~,~~~\lambda_3'=2\lambda_3-\partial_-\chi_1
\end{array}}
\label{susyauxfields}\eeq
These are precisely the (non-standard) auxilliary fields introduced in
\cite{morniemi1,morniemi2} which, as we discussed in Subsection 4.9, form the
basis for equivariant localization in supersymmetric quantum field theories and
phase space path integrals whose Hamiltonians are functionals of isometry
generators. Using these auxilliary fields we now define 2 functional derivative
operators on the space of fields,
\beq\new{\begin{array}{c}
{\cal D}=\int d^3x~\oint_0^Tdt~\left\{\chi_2\frac{\delta}{\delta
C}+iA_+\frac{\delta}{\delta\chi_1}+iM'\frac{\delta}{\delta\chi_3}+iN'
\frac{\delta}{\delta\chi_4}\right.\\\left.-\lambda_3'\frac{\delta}{\delta
A_-}-\lambda_2'\frac{\delta}{\delta A_1}-\lambda_1'\frac{\delta}{\delta
A_3}-iD'\frac{\delta}{\delta\lambda_4}\right\}\end{array}}
\label{dextsusy}\eeq
\beq\new{\begin{array}{c}
{\cal I}_{V^+}=\int
d^3x~\oint_0^Tdt~\left\{i\partial_+C\frac{\delta}{\delta\chi_2}+
\partial_+\chi_3\frac{\delta}{\delta M'}+\partial_+\chi_4\frac{\delta}{\delta
N'}+\partial_+\chi_1\frac{\delta}{\delta
A_+}\right.\\\left.-\partial_+A_3\frac{\delta}{\delta\lambda_1'}-\partial_+A_1
\frac{\delta}{\delta\lambda_2'}-\partial_+A_-\frac{\delta}{\delta\lambda_3'}
-\partial_+\lambda_4\frac{\delta}{\delta D'}\right\}\end{array}}
\label{contrsusy}\eeq
Here we have imposed the boundary conditions on the fields $\Phi(x,t)$ that
they vanish at spatial infinity and that they be periodic in time $t=x^0$,
\beq
\lim_{|x|\to\infty}\Phi(x,t)=0~~~~~,~~~~~\Phi(x,t+T)=\Phi(x,t)
\label{bdrycondsfields}\eeq
so that the space of fields can be thought of as a loop space.

The operators (\ref{dextsusy}) and (\ref{contrsusy}) are nilpotent. If we now
define
\beq
Q_+={\cal D}+{\cal I}_{V^+}
\label{Q+def}\eeq
then it can be checked that
\beq
Q_+^2={\cal D}{\cal I}_{V^+}+{\cal I}_{V^+}{\cal D}={\cal L}_{V^+}=\int
d^3x~\oint_0^Tdt~i\partial_+
\label{Q+sq}\eeq
The operator (\ref{Q+def}) generates the appropriate $N=1$ supersymmetry
transformations on the field multiplet and the supersymmetry algebra
(\ref{Q+sq}) coincides with the pertinent $N=1$ supersymmetry algebra
(\ref{Nsusyalg}). The above construction therefore provides a geometric
representation of the superalgebra (\ref{Nsusyalg}) on the space of fields of
the supersymetric field theory. A different representation of $\Sigma^\mu$
leads to a different choice of preferred $Q_\alpha$ in (\ref{Q+sq}) and a
different decomposition of the fields into loop space coordinates and 1-forms
in (\ref{dextsusy}) and (\ref{contrsusy}). It is now possible to examine how
various supersymmetric quantum field theories decompose with respect to the
above equivariant cohomological structure on the space of fields. The canonical
choice is the Wess-Zumino model which is defined by the sigma-model action
\beq
{\cal S}_{WZ}=\int d^3x~\oint_0^Tdt~\int
d\theta~d\bar\theta~\left(\frac{1}{2}\bar D\bar\Phi\cdot D\Phi+W[\Phi]\right)
\label{WZmodel}\eeq
where $W[\Phi]$ is some super-potential. With respect to the above
decompositions it is possible to show that the supersymmetric action decomposes
into a sum of a loop space scalar $\cal H$ and a loop space 2-form $\Omega$,
${\cal S}={\cal H}+\Omega$. Because of the boundary conditions on the fields of
the theory, the supersymmetry charges, which geometrically generate
translations in the chosen light-cone direction, are nilpotent on the spaces of
fields. By separating the loop space forms of different degrees, we find that
the supersymmetry of the action, $Q_+{\cal S}=0$, implies separately that
${\cal D}\Omega=0$ and ${\cal D}{\cal H}=-{\cal I}_{V^+}\Omega$. Thus the
supersymmetric model admits a loop space symplectic structure and the
corresponding path integral can be written as a super-loop space (i.e. phase
space) functional integration. Furthermore, because $Q_+^2=0$ on the space of
fields, the symplectic potential $\vartheta$ with ${\cal D}\vartheta=\Omega$
obeys ${\cal I}_{V^+}\vartheta={\cal H}$, so that the action is always locally
a supersymmetry variation, ${\cal S}=Q_+\vartheta=({\cal D}+{\cal
I}_{V^+})\vartheta$.

Thus, any generic quantum field theory with Poincar\'e supersymmetry group
admits a loop space symplectic structure and a corresponding $U(1)$ equivariant
cohomology responsible for localization of the supersymmetric path integral.
The key feature is an appropriate auxilliary field formalism which defines a
splitting of the fields into loop space ``coordinates" and their associated
``differentials" \footnote{\baselineskip=12pt Note that the problem of choosing
an appropriate set of auxilliary fields is the infinite-dimensional analog of
finding a preferred set of coordinates for an isometry generator (c.f.
Subsection 5.2).}. Notice that in general although the fields of the
supersymmetric theory always split up evenly into loop space coordinates and
1-forms, the coordinates and 1-forms involve both bosonic and fermionic fields.
It is only in the simplest cases (e.g. $N=\frac{1}{2}$ supersymmetric quantum
mechanics) that the pure bosonic fields are identified as coordinates and the
pure fermionic ones as 1-forms. In the auxilliary field formalism outlined
above, the supersymmetry of the model is encoded within the model independent
loop space equivariant cohomology defined by $Q_+$. In this way, one obtains a
geometric interpretation of general Poincar\'e supersymmetric quantum field
theories and an explicit localization of the supersymmetric path integral onto
the constant modes (zeroes of $\partial_+$).

We shall not present any explicit examples of the above general constructions
here. They have been verified in a number of cases. To check this formalism in
special instances one needs to impose certain additional constraints on the
multiplets \cite{palo} (e.g. the chirality condition mentioned above). The
above constructions have been in this way explictly carried out in
\cite{morniemi1,morniemi2} for $N=1$ supersymmetric quantum mechanics (i.e. the
(0 + 1)-dimensional Wess-Zumino model (\ref{WZmodel})), the Wess-Zumino model
(\ref{WZmodel}) in both 2 and 4 dimensions, and 4-dimensional $N=1$
supersymmetric $SU(N)$ Yang-Mills theory defined by the action
\beq
{\cal S}_{YM}=\int
d^4x~\left(-\frac{1}{4}F^a_{\mu\nu}F^{a,\mu\nu}+\frac{i}{2}\bar\psi
{\nabla\slash}_{\!\!A}\psi\right)
\label{susyYM}\eeq
where $\psi$ are Majorana fermion fields in the adjoint representation of the
gauge group. The model (\ref{susyYM}) can be reduced to a Wess-Zumino model by
eliminating the unphysical degrees of freedom and representing the theory
directly in terms of transverse (physical) degrees of freedom. The equivariant
localization framework has also been applied to the related supersymmetries of
Parisi-Sourlas stochastic quantization \cite{parisi} in \cite{morniemi2}. Palo
\cite{palo} applied these constructions to the 2-dimensional supersymmetric
non-linear sigma-model (i.e. the Wess-Zumino model (\ref{WZmodel}) with a
curved target space -- see the next Subsection).

\subsection{Supergeometry and the Batalin-Fradkin-Vilkovisky Formalism}

The role that the Cartan exterior derivative of equivariant cohomology plays in
localization resembles a Lagrangian BRST quantization in terms of the
gauge-fixing of a Lagrangian field theory over $\man$ with a gauge field
$\theta_\mu$ \cite{birm}. In \cite{nersess,nersesseq1} equivariant localization
was interpreted in terms of the Batalin-Vilkovisky Langrangian anti-field
formalism. In this formalism, a theory with first-stage reducible constraints
or with open gauge algebras is quantized by introducing an antibracket
\cite{batfrad}--\cite{batvil} which naturally introduces a new supersymmetry
element into the BRST quantization scheme. This formulation is especially
important for the construction of the complete quantum actions for topological
gauge theories (especially those of Schwarz-type) \cite{birm}. In this
Subsection we shall sketch some of the basic conceptual and computational ideas
of this formalism, which in the context of localization for dynamical systems
lead to more direct connections with supersymmetric and topological field
theories.

We return to the simpler situation of a generic Hamiltonian system
$(\man,\omega,H)$. We have seen that the localization prescription naturally
requires a formulation of objects defined over a super-manifold (i.e. one with
bosonic and Grassmann coordinates), namely the cotangent bundle
$M_S\equiv\man\otimes T^*\man$. In the case of the supersymmetric quantum field
theories that we considered in the previous Subsection, our fields where
defined on a superspace and the path integral localizations were carried out
over a superloop space. We would now like to try to exploit the mathematical
characteristics of a super-manifold and reformulate the localization concepts
in the more rigorous framework of supergeometry. This is particularly important
for some of the other localization features of topological field theories that
we shall discuss in the next 2 Subsections.

First, we shall incorporate the natural geometrical objects of the
Batalin-Vilkovisky formalism into the equivariant localization framework. The
local coordinates on the super-manifold $M_S$ are denoted as
$z^A=(x^\mu,\eta^\mu)$. We define a Grassmann-odd degree symplectic structure
on $M_S$ by the non-degenerate odd symplectic 2-form
\beq
\Omega^1=dz^A\wedge\Omega^1_{AB}dz^B=\omega_{\mu\nu}dx^\mu\wedge
d\eta^\nu+\frac{\partial\omega_{\mu\nu}}{\partial x^\lambda}\eta^\lambda
d\eta^\mu\wedge d\eta^\nu
\label{Om1odd}\eeq
The 2-form (\ref{Om1odd}) determines an odd Poisson bracket on $M_S$ called the
anti-bracket. It is defined by
\beq
\{\!\!\{{\cal A},{\cal B}\}\!\!\}_1=\frac{\partial{\cal A}}{\partial
z^A}\Omega_1^{AB}{\cal B}\frac{\overleftarrow\partial}{\partial
z^B}=\omega^{\mu\nu}\left(\frac{\partial{\cal A}}{\partial x^\mu}{\cal
B}\frac{\overleftarrow\partial}{\partial\eta^\nu}-\frac{\partial{\cal
B}}{\partial x^\mu}{\cal
A}\frac{\overleftarrow\partial}{\partial\eta^\nu}\right)+\frac{\partial
\omega^{\mu\nu}}{\partial x^\lambda}\eta^\lambda\frac{\partial{\cal
A}}{\partial\eta^\mu}{\cal B}\frac{\overleftarrow\partial}{\partial\eta^\nu}
\label{oddpoisson1}\eeq
where ${\cal A}(x,\eta)$ and ${\cal B}(x,\eta)$ are super-functions on $M_S$.
The grading and antisymmetry properties are opposite to those of the ordinary
graded Poisson bracket,
\beq\new{\begin{array}{c}
\{\!\!\{{\cal A},{\cal B}\}\!\!\}_1=-(-1)^{(p({\cal A})+1)(p({\cal
B})+1)}\{\!\!\{{\cal B},{\cal A}\}\!\!\}_1\\\{\!\!\{{\cal A},{\cal B}{\cal
C}\}\!\!\}_1=\{\!\!\{{\cal A},{\cal B}\}\!\!\}_1{\cal C}+(-1)^{p({\cal
B})(p({\cal A})+1)}{\cal B}\{\!\!\{{\cal A},{\cal C}\}\!\!\}_1\\\{\!\!\{{\cal
A},\{\!\!\{{\cal B},{\cal C}\}\!\!\}_1\}\!\!\}_1-(-1)^{(p({\cal A})+1)(p({\cal
C})+1)}\{\!\!\{{\cal B},\{\!\!\{{\cal A},{\cal
C}\}\!\!\}_1\}\!\!\}_1=\{\!\!\{\{\!\!\{{\cal A},{\cal B}\}\!\!\}_1,{\cal
C}\}\!\!\}_1\end{array}}
\label{antiids}\eeq
where $p({\cal A})$ is the Grassmann degree of the super-function ${\cal
A}(x,\eta)$ with the property $p(\{\!\!\{{\cal A},{\cal B}\}\!\!\})=p({\cal
A})+p({\cal B})+1$. In particular, the super-coordinate antibrackets are
\beq
\{\!\!\{x^\mu,x^\nu\}\!\!\}_1=0~~~,~~~\{\!\!\{x^\mu,\eta^\nu\}\!\!\}_1
=-\{\!\!\{
\eta^\nu,x^\mu\}\!\!\}_1=\omega^{\mu\nu}~~~,~~~\{\!\!\{\eta^\mu,\eta^\nu\}
\!\!\}_1=-\{\!\!\{\eta^\nu,\eta^\mu\}\!\!\}_1=\frac{\partial\omega^{\mu\nu}}
{\partial x^\lambda}\eta^\lambda
\label{supcoordanti}\eeq

We define a mapping on $C^\infty(\man)\to C^\infty(M_S)$ using the
super-function $\omega(z)=\frac{1}{2}\omega_{\mu\nu}(x)\eta^\mu\eta^\nu$ by
\beq
f(x)\to{\cal Q}_f(z)=\{\!\!\{f,\omega\}\!\!\}_1=\frac{\partial f}{\partial
x^\mu}\eta^\mu
\label{antifmap}\eeq
Then the antibracket coincides with the original Poisson bracket of the phase
space, $\{\!\!\{{\cal A},f\}\!\!\}_1=\{{\cal A},f\}_\omega$. In particular, the
dynamical systems $(\man,\omega,H)$ and $(M_S,\Omega^1,{\cal Q}_H)$ determine a
bi-Hamiltonian pair. The corresponding equations of motion are
\beq
\dot x^\mu=\{\!\!\{x^\mu,{\cal
Q}_H\}\!\!\}_1=\{x^\mu,H\}_\omega=V^\mu~~~,~~~
\dot\eta^\mu=\{\!\!\{\eta^\mu,{\cal Q}_H\}\!\!\}_1=\partial_\nu V^\mu\eta^\nu
\label{antieqsmotion}\eeq
Generally, it is readily seen that the operation
$\{\!\!\{\omega,\cdot\}\!\!\}_1$ acts as exterior differentiation $d$,
$\{\!\!\{H,\cdot\}\!\!\}_1$ acts like interior multiplication $i_V$ with
respect to the Hamiltonian vector field $V$, and $\{\!\!\{{\cal
Q}_H,\cdot\}\!\!\}_1$ acts as the Lie derivative $\lie_V$ along $V$. The
antibracket provides an equivalent supersymmetric generalization of the
ordinary Hamiltonian dynamics.

The key feature is that the supersymmetry of the odd Hamiltonian system
$(M_S,\Omega^1,{\cal Q}_H)$ is equivalent to the equivariant cohomology
determined by the equivariant exterior derviative $D_V=d+i_V$. If $\man$ admits
an invariant Riemannian metric tensor $g$ (the equivariant localization
constraints), then the super-function
\beq
{\cal I}_H=\frac{1}{2}g_{\mu\nu}V^\mu\eta^\nu
\label{antiactionvar}\eeq
is an integral of motion for the Hamiltonian system $(M_S,\Omega^1,{\cal
Q}_H)$, i.e. $\{\!\!\{{\cal Q}_H,{\cal I}_H\}\!\!\}_1=0$. Furthermore, ${\cal
I}_H$ determines the usual bi-Hamiltonian structure on $\man$ because
$\{\!\!\{\omega,{\cal
I}_H\}\!\!\}_1=\frac{1}{2}(\Omega_V)_{\mu\nu}\eta^\mu\eta^\nu$ and
$\{\!\!\{H,{\cal I}_H\}\!\!\}_1=K_V$. With these observations one can easily
now establish the (classical) equivariant localization principle. The usual
localization integral (\ref{zs}) can be written as an integral over the
super-manifold $M_S$,
\beq
{\cal Z}(s)=\frac{1}{(iT)^n}\int_{M_S}d^{4n}z~\e^{iT(H+\omega)
-s\{\!\!\{H-\omega,{\cal I}_H\}\!\!\}_1}
\label{Zsanti}\eeq
where as always the classical partition function is $Z(T)={\cal Z}(0)$. The
volume form $d^{4n}z=d^{2n}x~d^{2n}\eta$ is invariant under the equivariant
transformations of $D_V$ and $\lie_V$ determined by the anti-brackets.
Furthermore, we have
\beq\new{\begin{array}{c}
\{\!\!\{H-\omega,\e^{iT(H+\omega)-s\{\!\!\{H-\omega,{\cal
I}_H\}\!\!\}_1}\}\!\!\}_1=\{\!\!\{{\cal
Q}_H,\e^{iT(H+\omega)-s\{\!\!\{H-\omega,{\cal
I}_H\}\!\!\}_1}\}\!\!\}_1=0\\\{\!\!\{{\cal Q}_H,{\cal
I}_H\e^{iT(H+\omega)-s\{\!\!\{H-\omega,{\cal
I}_H\}\!\!\}_1}\}\!\!\}_1=0\end{array}}
\label{anti0s}\eeq
The first 2 vanishing conditions just represent the invariance of the integrand
in (\ref{Zsanti}) under the actions of the operators $D_V$ and $\lie_V$, so
that the usual equivariant cohomological structure for localization in
(\ref{Zsanti}) is manifested in the supersymmetry of the antibracket formalism.
With the identities (\ref{anti0s}), it is straightforward to establish that
$\frac{d}{ds}{\cal Z}(s)=0$, and hence the localization principle (i.e. the
Duistermaat-Heckman theorem).

Thus the lifting of the original Hamiltonian system to the odd one defined over
a supermanifold has provided another supersymmetric way to interpret the
localization, this time in terms of the presence of supersymmetric
bi-Hamiltonian dynamics with even and odd symplectic structures which is the
usual Batalin-Vilkovisky procedure for the evaluation of BRST gauge-fixed path
integrals. The representation (\ref{Zsanti}) of the canonical localization
integral formally coincides with the representation of differential forms in
the case where the original space $\man$ is a supermanifold. In
\cite{schwarzzab}, Schwarz and Zaboronsky derived some general localization
formulas for integrals over a finite-dimensional supermanifold $\man$ where the
integrand is invariant under the action of an odd vector field $W$. Using the
supergeometry of $\man$, they formulated sufficient conditions which generalize
those above under which the integral localizes onto the zero locus of the
c-number part $W(\eta=0)$ of $W$. Their theorems quite naturally generalize the
usual equivariant localization principles and could apply to physical models
such as those where the Batalin-Vilkovisky formalism is applicable
\cite{schwarz1} or in the dimensional reduction mechanism of the Parisi-Sourlas
model \cite{parisi}.

Nersessian has also demonstrated how to incorporate the anti-bracket structure
into the other models of equivariant cohomology (other than the Cartan model --
see Appendix B) in \cite{nersess1} and how the usual equivariant characteristic
class representations of the localization formulas appear in the
Batalin-Vilkovisky formalism in \cite{nersesseq1}. The superspace structure of
cohomological field theories in this context has been studied by Niemi and
Tirkkonen in \cite{niemitirk3}. They discussed the role of the BRST model of
equivariant cohomology in non-abelian localization (see Appendix B) and
topological field theories and showed how an appropriate superfield formulation
can be used to relate these equivariant cohomological structures to the
Batalin-Fradkin-Vilkovisky Hamiltonian quantization of constrained systems with
first stage reducible constraints. This suggests a geometric (superspace)
picture of the localization properties of some topological quantum field
theories such as 4-dimensional topological Yang-Mills theory (defined by the
action $\int\tr~F_A\wedge F_A$). This picture is similar to those of the
Poincar\'e supersymmetric theories described in the previous Subsection in that
the BRST charge of the cohomological field theory can be taken to generate
translations in the $\eta$-direction in superspace and then the connection
between equivariant cohomology and Batalin-Fradkin-Vilkovisky quantization of
4-dimensional topological Yang-Mills theory becomes transparent. These
superfield formalisms therefore describe both equivariant cohomology in the
symplectic setting relevant for localization and the BRST structure of
cohomological field theories. This seems to imply the existence of a unified
description of localization in the symplectic loop space, the supersymmetric
loop space of the last Subsection, and in cohomological field theory. Indeed,
it is conjectured that all lower dimensional integrable models are obtainable
as dimensional reductions of 4-dimensional self-dual Yang-Mills theory (i.e.
$F_A=\star F_A$) which is intimately connected to topological Yang-Mills
theory.

Finally, the incorporation of the Batalin-Vilkovisky formalism into loop space
localization has been discussed recently by Miettinen in \cite{miet1}. For path
integral quantization, one needs an even Hamiltonian and symplectic structure.
This can be done provided that $\man$ has on it a Riemannian structure. An even
symplectic structure on $M_S$ is given by the super-symplectic 2-form
\beq
\Omega^i=\frac{1}{2}\left(\omega^i_{\mu\nu}+R_{\mu\nu\lambda\rho}\eta^\lambda
\eta^\rho\right)dx^\mu\wedge dx^\nu+\frac{1}{2}g_{\mu\nu}D_g\eta^\mu\wedge
D_g\eta^\nu
\label{Omeven}\eeq
where
\beq
D_g\eta^\mu=d\eta^\mu+\Gamma^\mu_{\nu\lambda}\eta^\nu dx^\lambda
\label{Dgeta}\eeq
is the covariant derivative on $\man$, and the subscript $i=0,2$ labels the
Hamiltonian systems $(\man,\omega^0=\omega,H_0=H)$ and
$(\man,\omega^2=\Omega_V,H_2=K_V)$. The corresponding symplectic 1-forms, with
$\Omega^i=d\Theta^i$, are then
\beq
\Theta^0=\theta_\mu dx^\mu+g_{\mu\nu}\eta^\mu
D_g\eta^\nu~~~,~~~\Theta^2=g_{\mu\nu}V^\nu dx^\mu+g_{\mu\nu}\eta^\mu
D_g\eta^\nu
\label{Thetaevens}\eeq
The 2-forms $\Omega^i$ determine the even Poisson brackets on $M_S$
\beq
[\![{\cal A},{\cal B}]\!]_i=(\nabla_\mu{\cal
A})[\omega_{\mu\nu}^i+R_{\mu\nu\lambda\rho}\eta^\lambda
\eta^\rho]^{-1}\nabla_\nu{\cal B}+g^{\mu\nu}\frac{\partial{\cal
A}}{\partial\eta^\mu}{\cal B}\frac{\overleftarrow\partial}{\partial\eta^\nu}
\label{Poissoneven}\eeq
where
\beq
\nabla_\mu=\partial_\mu-\Gamma^\lambda_{\mu\nu}\eta^\nu\frac{\partial}
{\partial\eta^\lambda}
\label{supercovderiv}\eeq
Then the equations of motion for the odd and even Poisson brackets on $M_S$
coincide, $[\![z^A,{\cal H}_i]\!]_i=\{\!\!\{z^A,{\cal Q}_H\}\!\!\}_1$, where
${\cal H}_i=H_i+\Omega_V$. Thus the odd and even Poisson brackets provide
bi-Hamiltonian structures also on the super-symplectic manifold $M_S$ and
therefore the Hamiltonian system is also integrable on $M_S$.

Given this integrability feature on $M_S$, we can now examine the corresponding
localizations \cite{schwarzzab}. We write the partition functions ${\cal
Z}_i(T)={\rm str}\|\e^{-iT{\cal H}_i}\|$ as path integrals over a super-loop
space corresponding to $LM_S$ by absorbing the Liouville measure factors
associated with $\Omega^i$ into the argument of the action in the usual way
(c.f. Section 4). Then the Hamiltonian systems $(M_S,\Omega^i,{\cal H}_i)$ have
the quantum actions \cite{miet1}
\beq\new{\begin{array}{ll}
S_0=&\int_0^Tdt~\left(\theta_\mu\dot x^\mu-H+\frac{1}{2}\eta^\mu
g_{\mu\nu}\frac{D_g\eta^\nu}{dt}+\frac{1}{2}(\Omega_V)_{\mu\nu}\eta^\mu
\eta^\nu+\frac{1}{2}\omega_{\mu\nu}\lambda^\mu\lambda^\nu
\right.\\&~~~~~~~~~~\left.+\frac{1}{2}R_{\mu\nu\lambda\rho}\lambda^\mu
\lambda^\nu\eta^\lambda\eta^\rho+\frac{1}{2}g_{\mu\nu}F^\mu
F^\nu\right)\\S_2=&\int_0^Tdt~\left(g_{\mu\nu}V^\nu\dot
x^\mu-K_V+\frac{1}{2}\eta^\mu
g_{\mu\nu}\frac{D_g\eta^\nu}{dt}+\frac{1}{2}(\Omega_V)_{\mu\nu}\eta^\mu\eta^\nu
+\frac{1}{2}(\Omega_V)_{\mu\nu}\lambda^\mu\lambda^\nu\right.\\&~~~~~~~~~~
\left.+\frac{1}{2}R_{\mu\nu\lambda\rho}\lambda^\mu
\lambda^\nu\eta^\lambda\eta^\rho+\frac{1}{2}g_{\mu\nu}F^\mu
F^\nu\right)\end{array}}
\label{superquactions}\eeq
where the quantum partition functions are
\beq
{\cal Z}_i(T)=\int_{LM_S\otimes
L\Lambda^1M_S}[d^{2n}x]~[d^{2n}\eta]~[d^{2n}F]~[d^{2n}\lambda]~\e^{iS_i[x,
\eta;F,\lambda]}
\label{superquparts}\eeq
and we have introduced auxilliary anticommuting variables $\lambda^\mu\sim
dx^\mu$ and bosonic variables $F^\mu\sim d\eta^\mu$ to exponentiate the
determinant factors in the usual way. If $H=0$ (the topological limit), then
the action $S_0+\int_0^Tdt~\frac{1}{2}g_{\mu\nu}\dot x^\mu\dot x^\nu$ is that
of the (0 + 1)-dimensional $N=1$ supersymmetric non-linear sigma-model, i.e.
the action of $N=1$ DeRham supersymmetric quantum mechanics in background
gravitational and gauge fields\footnote{\baselineskip=12pt The action for $N=1$
supersymmetric quantum mechanics can be obtained from the Wess-Zumino model
action (\ref{WZmodel}) by integrating out the auxilliary field $F^\mu$
resulting from the chiral superfields (\ref{chiralsupfield}) and integrating
over the $\theta$ coordinates of the superspace. The action for $N=\frac{1}{2}$
supersymmetric quantum mechanics discussed in Subsection 4.2 can be obtained
from the $N=1$ model by setting $\lambda^\mu=\eta^\mu$ above. Notice that the
Riemann curvature term then drops out because of its symmetry properties.}.

One can now develop the standard machinery to evaluate these path integrals
using super-loop space equivariant cohomology. This has been done explicitly in
\cite{niemipalo3}. The super-loop space equivariant exterior derivative is
\beq
Q=\int_0^Tdt~\left(\lambda^\mu\frac{\delta}{\delta
x^\mu}+F^\mu\frac{\delta}{\delta\eta^\mu}
+(\dot x^\mu-V^\mu)\frac{\delta}{\delta\lambda^\mu}+(\dot\eta^\mu+\partial^\mu
V^\nu\eta_\nu)\frac{\delta}{\delta F^\mu}\right)
\label{superloopQ}\eeq
with $Q^2=\lie_S=\lie_{\dot x}-\lie_V$ where $S$ is the classical action
associated with the original Hamiltonian system $(\man,\omega,H)$. Notice that
a canonical conjugation $Q\to\e^{-\Phi}Q\e^\Phi\equiv\tilde Q$ does not alter
the cohomology groups of the derivative operator $Q$. Choosing the loop space
functional
\beq
\Phi=\int_0^Tdt~\Gamma^\lambda_{\mu\nu}\eta_\lambda\lambda^\nu\frac{\delta}
{\delta F^\mu}
\label{PHI}\eeq
the (topologically equivalent) operator $\tilde Q$ can be explicitly worked out
(see \cite{niemipalo3}). With regards to this supersymmetry charge, the
pertinent action $S_0$ in (\ref{superquactions}) can be obtained from the
2-dimensional $N=1$ supersymmetric sigma-model by partial localization of it to
a 1-dimensional model. This is done by breaking its (left-right) $(1,1)$
supersymmetry explicitly by the Hamiltonian flow. In this procedure the usual
boson kinetic term (in light-cone coordinates)
$g_{\mu\nu}\partial_+\phi^\mu\partial_-\phi^\nu$ drops out. The path integral
${\cal Z}_0(T)$ can be evaluated by adding an explicit gauge-fixing term
$\tilde Q\psi$ to the action for an appropriate gauge fermion $\psi$. Taking
$\psi=\int_0^Tdt~(g_{\mu\nu}F^\mu\eta^\nu+\frac{s}{2}g_{\mu\nu}(\dot
x^\mu-V^\mu)\lambda^\nu)$ localizes the path integral in the limit $s\to\infty$
onto the $T$-periodic classical trajectories of the original action $S$,
\beq
{\cal Z}_0(T)=\sum_{x(t)\in L\man_S}{\rm
sgn}[\det\|\delta^2S(x(t))\|]\e^{iS[x(t)]}
\label{PoinHopfloop}\eeq
On the other hand, selecting
$\psi=\int_0^Tdt~(g_{\mu\nu}F^\mu\eta^\nu+\frac{s}{2}g_{\mu\nu}\dot
x^\mu\lambda^\nu)$ we find that the path integral localizes onto an ordinary
integral over equivariant characteristic classes of the phase space $\man$,
\beq
{\cal Z}_0(T)=\int_\man\ch_V(-iT\omega)\wedge E_V(R)
\label{loopGaussBon}\eeq

The equality of these 2 expressions for the quantum partition function ${\cal
Z}_0(T)$ can be thought of as an equivariant, loop space generalization of the
relation (\ref{phgbc}) in Morse theory between the Gauss-Bonnet-Chern and
Poincar\'e-Hopf theorems for the representation of the Euler characteristic
$\chi(\man)$ of the manifold $\man$. Indeed, in the limit $H,\theta_\mu\to0$
the quantities (\ref{PoinHopfloop}) and (\ref{loopGaussBon}) reproduce exactly
the relation (\ref{phgbc}). These relations have been used to study the set of
Hamiltonian systems which satisfy the Arnold conjecture on the space of
$T$-periodic classical trajectories for (time-dependent) classical Hamiltonians
\cite{niemi1,niemipalo3}. Thus, the path integrals associated with the
supermanifolds defined by the Batalin-Vilkovisky formalism for dynamical
systems lead to loop space and equivariant generalizations of other familiar
topological invariants. Furthermore, these models are closely related to
supersymmetric non-linear sigma-models which ties together the ``hidden"
supersymmetry of the given Hamiltonian system with the Poincar\'e supersymmetry
of the localizable quantum field theories. These ideas lead us naturally into
the final topic of this Review which emphasizes these sorts of relations
between equivariant cohomology and topological quantum field theories. The
discussion of this Subsection then shows how this next topic is related to the
equivariant localization formalism for dynamical systems.

\subsection{Equivariant Euler Numbers, Thom Classes and the Mathai-Quillen
Formalism}

We have now almost completed the connections between the equivariant
localization formalism, cohomological field theories and their relations to
physical systems, thus uniting most of the ideas presented in this Review. The
last Subsection showed how the localization formalism connects phase space path
integrals of dynamical systems to some basic topological field theory models
(namely supersymmetric sigma-models). Conversely, in Subsection 8.2 we
demonstrated that arbitrary field theoretical models of these types could be
placed into the loop space equivariant localization framework so that there is
a sort of equivalence between dynamical systems and field theory models in this
geometric context. The discussion of Subsection 8.1 then illustrated certain
genuine, geometrical equivalences between physical and topological gauge
theories which demonstrates the power of the formalisms of both topological
field theory and equivariant localization of path integrals in describing the
quantum characteristics of physical systems.

There is one final step for this connection which is the Atiyah-Jeffrey
geometric interpretation of generic cohomological field theories
\cite{atiyahjef} which is based on the Mathai-Quillen construction of
Gaussian-shaped Thom forms \cite{mathai}. This construction is the natural
arena for the study of the localization properties of topological field
theories, and in its infinite dimensional versions it can be used to build up
topological gauge models. This approach, although based on equivariant
cohomology, is rather different in spirit than the equivariant localization
formalisms we have discussed thus far and we therefore only very briefly
highlight the details for the sake of completeness. At the end of this Section,
we will discuss a bit the connections with the other ideas of this Review. More
detailed reviews of the Mathai-Quillen formalism in topological field theory
can be found in \cite{blau2,blauthom,cordes}.

The basic idea behind the Mathai-Quillen formalism is the relation
(\ref{phgbc}) between the Poincar\'e-Hopf and Gauss-Bonnet-Chern
representations of the Euler characteristic. It represents the localization of
an explicit differential form representative of the Euler class of a vector
bundle onto the zero locus of some section of that bundle. The original idea
for the application and generalization of this relation to cohomological field
theories traces back to Witten's connection between supersymmetric quantum
mechanics and Morse theory \cite{witten1}. Let us start with a simple example
in this context. Given the local coordinates $(x,\eta)$ on the cotangent bundle
$\man\otimes T^*\man$ of some phase space $\man$, we denote
$p_\mu\equiv\frac{\partial}{\partial x^\mu}$ and
$\bar\eta_\mu\equiv\frac{\partial}{\partial\eta^\mu}$. In the spirit of the
previous Subsection, we then interpret $(x,\bar\eta)$ as local coordinates on a
supermanifold $S^*\man$ and $\eta^\mu\sim dx^\mu,p_\mu\sim d\bar\eta_\mu$ as
the local basis for the cotangent bundle of $S^*\man$.

The nilpotent exterior derivative operator on $S^*\man$ can be written as
\beq
d=\eta^\mu\frac{\partial}{\partial
x^\mu}+p_\mu\frac{\partial}{\partial\bar\eta_\mu}
\label{extderivSM}\eeq
The invertible conjugation $d\to\e^{-\Phi}d\e^\Phi$ produces another linear
derivation which generates the same cohomology as $d$. If $\man$ has metric
$g$, then we can select
$\Phi=-\Gamma^\lambda_{\mu\nu}\eta^\nu\bar\eta_\lambda\frac{\partial}{\partial
p_\mu}$ (as in (\ref{PHI})) so that (\ref{extderivSM}) conjugates to
\beq
d=\eta^\mu\frac{\partial}{\partial
x^\mu}+(p_\mu+\Gamma^\lambda_{\mu\nu}\eta^\nu\bar\eta_\lambda)\frac{\partial}
{\partial\bar\eta_\mu}+(\Gamma^\lambda_{\mu\nu}p_\lambda\eta^\nu-
\frac{1}{2}R^\lambda_{\mu\rho\nu}\eta^\nu\eta^\rho\bar\eta_\lambda)
\frac{\partial}{\partial p_\mu}
\label{dconj1}\eeq
The action of (\ref{dconj1}) on the local coordinates of the cotangent bundle
of the supermanifold $S^*\man$
\beq\new{\begin{array}{ll}
dx^\mu=\eta^\mu~~~&,~~~dp_\mu=\Gamma^\lambda_{\mu\nu}p_\lambda\eta^\nu-
\frac{1}{2}R^\lambda_{\mu\rho\nu}\eta^\nu\eta^\rho\bar\eta_\lambda\\d\eta^\mu=0
{}~~~&,~~~d\bar\eta_\mu=p_\mu+\Gamma^\lambda_{\mu\nu}\eta^\nu\bar\eta_\lambda
\end{array}}
\label{dconjcoords}\eeq
coincides with the standard infinitesimal transformation laws of $N=1$ DeRham
supersymmetric quantum mechanics.

We now consider the following integral
\beq
{\cal Z}^*=\int_{S^*\man\otimes
T^*(S^*\man)}d^{2n}x~d^{2n}p~d^{2n}\eta~d^{2n}\bar\eta~
\e^{d\psi[x,p;\eta,\bar\eta]}
\label{Z*def1}\eeq
Since $d^2=0$ the integral (\ref{Z*def1}) is formally independent of the
function $\psi$ on the super-manifold $S^*\man\otimes T^*(S^*\man)$, i.e.
$\frac{\delta{\cal Z}^*}{\delta\psi}=0$. We can therefore evaluate
(\ref{Z*def1}) in 2 equivalent ways. First, we introduce a Hamiltonian vector
field $V$ on $\man$ and take $\psi=\psi_V=\frac{1}{2}V^\mu\bar\eta_\mu$ so that
\beq
d\psi_V=p_\mu V^\mu+\eta^\mu\nabla_\mu V^\nu\bar\eta_\nu
\label{dpsiV}\eeq
The integration in (\ref{Z*def1}) can then be carried out explicitly. The
integration over $p_\mu$ produces a delta-function $\delta(V)$ localizing the
integral onto the zero locus $\man_V$ of the vector field $V$. The integration
over the Grassmann coordinates in (\ref{dpsiV}) yields a determinant of $\nabla
V$. Computing the relevant Jacobian for the transformation $x\to V(x)$ (c.f.
Subsection 2.6), we arrive finally at
\beq
{\cal Z}^*=\sum_{p\in\man_V}{\rm sgn}\det\hess(p)
\label{Z*hess}\eeq
Next, we take $\psi=\psi_g=g^{\mu\nu}p_\mu\bar\eta_\nu$ so that
\beq
d\psi_g=g^{\mu\nu}p_\mu
p_\nu-\frac{1}{2}R^\lambda_{\mu\rho\nu}\eta^\nu\eta^\rho\bar\eta_\lambda
g^{\mu\sigma}\bar\eta_\sigma
\label{dpsig}\eeq
Evaluating the Gaussian integrals over $p_\mu$ and $\bar\eta_\mu$ in
(\ref{Z*def1}) leads to
\beq
{\cal Z}^*=\int_{\man\otimes
T^*\man}d^{2n}x~d^{2n}\eta~\pfaff\left(\frac{1}{2}R^\mu_{\nu\lambda\rho}
\eta^\lambda\eta^\rho\right)
\label{Z*curv}\eeq
which we recognize as the Euler class of the tangent bundle $T\man$. Thus the
equality of (\ref{Z*hess}) and (\ref{Z*curv}) leads immediately to the relation
(\ref{phgbc}).

The exterior derivative operator (\ref{dconj1}) produces the Mathai-Quillen
representative of the Euler class of the tangent bundle of $\man$. The above
derivation is a special case of a more general construction of explicit
differential form representatives for the Euler numbers of vector bundles
$E\to\man$. These representatives are so-called Gaussian-shaped Thom forms
whose constructions are best understood within the framework of equivariant
cohomology. The idea is that one realizes the vector bundle $E\to\man$ as its
associated principal $G$-bundle $P\times W$ (with $W$ the standard fiber space
of $E$) and constructs a particular representative (the Thom class) of the
$G$-equivariant cohomology of $P\times W$. Given a section $V:\man\to E$, the
regularized Euler class $E_V(F_A)$ (with $F_A$ the curvature of a connection
$A$ of the bundle) is then the pullback of the Thom class to $\man$ under this
section. It can be expressed as
\beq
E_V(F_A)=\int d^m\eta~\e^{-\frac{1}{2}\|V\|^2+\frac{1}{2}\eta_\mu
F_A^{\mu\nu}\eta_\nu+i\nabla_AV^\mu\eta_\mu}
\label{eulerthomrep}\eeq
where $m=\dim\man$ and $\eta_\mu$ are Grassmann variables. The norm $\|V\|^2$
is with respect to a fixed fiber metric on $E$ and $\nabla_A$ is a compatible
connection. We shall not go into the details of the construction of Thom
classes using equivariant cohomology, but refer the reader to
\cite{blau2,cordes,kalkman1} for lucid accounts of this formalism. The
important features of the Mathai-Quillen representative (\ref{eulerthomrep})
are as follows. For general $V$, integrating out the Grassmann variables shows
that $E_V(F_A)$ is a $2m$-form, and the fact that it is closed follows from the
invariance of the exponent in (\ref{eulerthomrep}) under the supersymmetry
transformations
\beq
\delta V^\mu=\nabla_AV^\mu~~~~~,~~~~~\delta\eta^\mu=iV^\mu
\label{MQsusy}\eeq
with the additional condition $\delta x^\mu=\eta^\mu$ when (\ref{eulerthomrep})
is integrated over $x\in\man$. Note that setting $V=0$ in (\ref{eulerthomrep})
and integrating out the Grassmann coordinates we see that it coincides with the
usual Euler characteristic class $E(F_A)=\pfaff(F_A)$. Since the associated
Thom class is closed, (\ref{eulerthomrep}) is independent of the chosen section
$V$, i.e. $E_V(F_A)$ is cohomologous to $E(F_A)$ for any $V$. This means that
the Euler characteristic $\chi(E\to\man)=\int_\man E_V(F_A)$ can be evaluated
by rescaling $V$ by $s\in\IR$ and localizing the integral in the limit
$s\to\infty$ onto the zeroes of the section $V$ (note the curvature term in
this limit does not contribute), thus reproducing in this way the standard
relation between the Poincar\'e-Hopf and Gauss-Bonnet-Chern theorems for
generic vector bundles over $\man$. Thus the Thom class not only yields a
representative of the Euler class of a vector bundle, but it also produces the
Poincar\'e-dual form of the zero locus of a given section of the bundle. The
use of equivariant cohomology and localization techniques therefore also
reproduce some classical results from geometry and topology.

Niemi and Palo \cite{niemipalo2} have shown how to construct equivariant
generalizations of the Mathai-Quillen formalism. For this one considers the
usual Cartan equivariant exterior derivative $D_V$ and the associated Lie
derivative $\lie_V=D_V^2$ on the super-manifold $S^*\man$. If the Christoffel
connection satisfies $\lie_V\Gamma=0$, then the conjugation by $\Phi$
introduced above of these operators produces an action on the local coordinates
of the cotangent bundle of $S^*\man$ for $\lie_V$ which generates the usual
covariant coordinate transformation laws with respect to coordinate change
defined by the Hamiltonian
vector field $V$. Thus the integral
\beq
{\cal Z}_V^*=\int_{S^*\man\otimes
T^*(S^*\man)}d^{2n}x~d^{2n}p~d^{2n}\eta~d^{2n}\bar\eta~\e^{i\phi\otimes
(H+\omega)+D_V\psi}
\label{ZV*def}\eeq
is formally independent of any generally covariant function $\psi$ on the
cotangent bundle of $S^*\man$. This is again just the equivariant localization
principle. The measure in (\ref{ZV*def}) is the invariant Liouville measure on
the extended phase space. Evaluating (\ref{ZV*def}) using the 2 choices for
$\psi$ mentioned above, we arrive at the relation
\beq
\sum_{p\in\man_V}\e^{i\phi\otimes H(p)}~{\rm sgn}\det\hess(p)=\int_{\man\otimes
T^*\man}d^{2n}x~d^{2n}\eta~\e^{i\phi\otimes(H+\omega)}~\pfaff\left[\nabla_\nu
V^\mu+\frac{1}{2}R^\mu_{\nu\lambda\rho}\eta^\lambda\eta^\rho\right]
\label{eqphgbc}\eeq
which can be recognized as an equivariant generalization of (\ref{phgbc}). Thus
an appropriate equivariantization leads to a Mathai-Quillen representative for
the {\it equivariant} Euler number of an equivariant vector bundle. In the
limit $V,\phi\to0$, (\ref{eqphgbc}) reduces to the usual relation. The
non-degenerate cases are also possible to treat in this way \cite{niemipalo2}.

\subsection{The Mathai-Quillen Formalism for Infinite-dimensional Vector
Bundles}

In this final Subsection of this Review, we shall discuss briefly the explicit
connection to cohomological field theories. This will make explicit the
relations of localization quite generically to topological field theory that we
have mentioned through out this Paper. As originally pointed out by Atiyah and
Jeffrey \cite{atiyahjef}, although the Euler number itself does not make sense
for an infinite dimensional vector bundle, the Mathai-Quillen form $E_V(F_A)$
can be used to define regularized Euler numbers $\chi_V(E\to\man)=\int_\man
E_V(F_A)$ of such bundles for those choices of $V$ whose zero locus is
finite-dimensional so that the localization makes these quantities
well-defined. Although these numbers are not independent of $V$ as in the
finite-dimensional cases, they are naturally associated with $\man$ for certain
choices. The functional integrals which arise in this way are equivalent to
ordinary finite-dimensional integrals and represent the fundamental property of
topological field theories, i.e. that their path integrals represent
characteristic classes. This has been noted throughout this review as our
central theme, and indeed most topological field theories can be obtained or
interpreted in terms of the infinite-dimensional Mathai-Quillen formalism
\cite{cordes}.

The simplest example is the regularized Euler number of the loop space
$L\man\to\man$ over a manifold $\man$. The canonical vector field associated
with this bundle is $\dot x^\mu(t)$. Now the loop space version of the integral
(\ref{Z*def1}) is a path integral over an extended superloop space, and we
replace the exterior derivative $d$ there by the equivariant, loop space one
$Q_{\dot x}$ on $L(S^*\man)$. The Lie derivative $\lie_{\dot x}=Q_{\dot x}^2$
as before is the generator of time translations, and employing the standard
conjugation above the path integral can be localized using any single-valued
functional $\psi$ on $L(S^*\man)$. This follows from the equivariant
localization principle for the model independent $S^1$-action. Choosing the
natural loop space extensions of the functionals $\psi$ used in the
finite-dimensional calculations above, we arrive at precisely the same results
(\ref{Z*hess}) and (\ref{Z*curv}) for the Euler characteristic of $\man$. The
path integral analog of (\ref{eulerthomrep}) with $V=\dot x$ yields the action
of $N=1$ DeRham supersymmetric quantum mechanics. In Subsection 4.2 we saw that
the path integral for $N=\frac{1}{2}$ Dirac supersymmetric quantum mechanics
localized onto constant modes and yielded the index of the twisted spin complex
of $\man$. In the present case the localization of this Witten index onto
constant loops yields the index of the DeRham complex of $\man$ (i.e. the Euler
characteristic). This was the fundamental observation of Witten \cite{witten1}
and was one of the main ingredients in the birth of topological field theory.
If the target space manifold has a K\"ahler structure then the sigma-model
actually has 2 independent (holomorphic and anti-holomorphic) supersymmetries.
Restricting the computation of the supersymmetric quantum mechanics partition
function to the anti-holomorphic sector of the Hilbert space as described
earlier leads to the representation of the index of the Dolbeault complex in
terms of the Todd class.

Equivariant generalizations of this simple example are likewise possible. In
(\ref{ZV*def}) the path integration now involves the action $S$ rather than the
Hamiltonian, and $D_V$ gets replaced by $Q_S$ in the usual routine of Section
4. Now we conjugate the relevant operators and find that the localization
priniciple requires the localization functionals $\psi$ to be generally
covariant and single-valued. The resulting path integrations yield precisely
the computation at the end of Subsection 8.3 above. For Hamiltonians which
generate circle actions, the right-hand side of (\ref{PoinHopfloop}) coincides
with the left-hand side of (\ref{eqphgbc}) because of the structure of the set
$L\man_S$ discussed at the beginning of Subsection 4.6. Thus in this case we
again obtain the ordinary finite-dimensional relation (\ref{eqphgbc}). These
relations play a deeper role when the Hamiltonian depends explicitly on time
$t$. Then the right-hand side of (\ref{PoinHopfloop}) represents a regularized
measure of the number of $T$-periodic classical trajectories of the given
dynamical system \cite{niemi1,niemipalo3}. Thus the classical dynamics of a
physical system can be characterized in this way via the localization
properties of supersymmetric non-linear sigma-models. In \cite{niemipas1},
these relations were related to a functional Euler character in the quantum
cohomology defined by the topological non-linear sigma-model and also to a loop
space generalization of the Lefschetz fixed point theorem.

Besides supersymmetric quantum mechanics the localization features of more
complicated topological gauge theories can be studied by the computing the
Euler numbers of vector bundles over the infinite-dimensional space ${\cal
A}/G$ of gauge connections modulo gauge transformations of a principal
$G$-bundle. One can either start with a given topological field theory and
analyse its localization characteristics using the techniques of this Section,
or conversely by applying the Mathai-Quillen formalism to some vector bundle
over ${\cal A}/G$ and reconstructing the action of the corresponding
topological gauge theory from there. The resulting path integrals always
compute sorts of intersection numbers on moduli space. A discussion of these
models is beyond the scope of this review and we refer to \cite{cordes} for an
extensive discussion of the theories which can be viewed in this way. The basic
example is Donaldson theory \cite{birm} which is the prime example of a
cohomological field theory and is used to calculate intersection numbers of
moduli spaces of instantons for the study of 4-manifolds. Topological
Yang-Mills theory in 4-dimensions is another interesting application of this
formalism. The field theoretic generalization of supersymmetric quantum
mechanics, i.e. the topological sigma-model \cite{birm}, is the appropriate
setting for studying the quantum symmetries of string theory and more generally
super-conformal field theories. The Mathai-Quillen formalism applied to
2-dimensional topological gravity could presumably shed light on its
equivalence with physical gravity in 2 dimensions. The coupling of the
topological sigma-model to topological gravity can be interpreted as
topological string theory and studied using these methods. Finally, viewing
2-dimensional Yang-Mills theory as a topological field theory (see Subsection
8.1 above) leads in this way to a localization onto the rather complicated
Hurwitz space of branched covers of the Riemann surface. This construction has
been exploited recently as a candidate for a string theoretical realization of
2-dimensional Yang-Mills theory \cite{cordes}.

Thus, the Mathai-Quillen formalism serves as the natural arena for the
localization properties of cohomological field theories.
However, the connection between the localization formalisms of the earlier
Sections of this Review (i.e. the stationary-phase formula) and the
constructive Mathai-Quillen formalism above has yet to be completely clarified,
as the latter relies on quite different cohomological symmetries than the
ordinary BRST supersymmetries responsible for equivariant localization
\cite{niemipalo2}. Recall these models all possess a Grassmann-odd symmetry
$\delta$ that defines a supersymmetry transformation (\ref{MQsusy}) which
resembles the usual BRST supersymmetries of equivariant localization. It is
possible to argue \cite{blauthom,cordes} that the $\delta$-action is not free
and that the path integral receives contributions from some arbitrarily small
$\delta$-invariant tubular neighbourhood of the fixed point set of $\delta$.
The integration over the directions normal to this fixed point set can be
calculated in a stationary-phase approximation. One readily sees from
(\ref{MQsusy}) that the fixed point set of $\delta$ is the precisely the moduli
space $\man_V$ described by the zero locus of $V$ and its tangents $\psi$
satisfying the linearized equation $\nabla_A V(\psi)=0$. In this way the
topological field theory path integral reduces to an integration of
differential forms over $\man_V$. It remains though to still obtain a more
precise connection between these BRST fixed points, localization, and the
interpretation of the geometrical and topological features of path integrals in
terms of the Mathai-Quillen formalism which shows how such infinite-dimensional
integrations are {\it a priori} designed to represent finite-dimensional
integrals. The antibracket formalism developed in Subsection 8.3 above is a key
stepping stone between the Mathai-Quillen localization features of topological
field theory path integrals, and the path integral localizations of generic
Hamiltonian systems. The supersymmetric formulation of equivariant cohomology
developed in \cite{niemitirk3}, and its connections with 4-dimensional
topological Yang-Mills theory, could serve as another approach to this
connection. This might give a more direct connection between localization and
some of the more modern theories of quantum integrability \cite{das}, such as
$R$-matrix formulations and the Yang-Baxter equation. This has been discussed
somewhat in \cite{gorsky}. These connections are all important and should be
found in order to have full understandings of the structures of topological and
integrable quantum field theories, and hence generic physical models, from the
point of view of loop space equivariant localization.

\setcounter{section}{0}
\setcounter{subsection}{0}

\section*{Acknowledgements}

I would like to thank I. Kogan and G. Semenoff for advice and encouragement
during various stages of the writing of this Review and for helpful
discussions. I am grateful to L. Paniak for his participation in the various
calculational and conceptual aspects of Section 7, and to O. Tirkkonen for
clarifying discussions. I would also like to thank D. Austin and R. Douglas for
comments and suggestions on some of the more mathematical aspects of this
Paper. This work was supported in part by the Natural Sciences and Engineering
Research Council of Canada.

  \addtocounter{section}{1}
  \setcounter{equation}{0}
  \setcounter{equnum}{0}
  \renewcommand{\thesection}{\Alph{section}}
  \section*{Appendix A \ \ \  BRST Quantization}
  \addcontentsline{toc}{section}{Appendix \thesection\ \ \  BRST Quantization}

BRST quantization was first introduced in the quantization of Yang-Mills theory
as a useful device for proving the renormalizability of non-abelian gauge
theories in 4 dimensions. It was shown that a global fermionic symmetry was
present after Yang-Mills gauge fixing which incorporated the original gauge
invariance of the model and ultimately led to straightforward derivations of
the Ward identities associated with the gauge symmetry in both quantum
electrodynamics and quantum chromodynamics. New impetus came when the BRST
theory was applied to the quantization of Hamiltonian systems with first class
constraints \cite{henneaux}. For completeness, in this Appendix we shall
outline the essential features of the BRST quantization scheme of which the
loop space localization principle can be thought of as a special instance.

Consider any physical system with symmetry operators $K^a$ that (possibly
locally) generate a closed Lie algebra $\bf g$,
\beq
[K^a,K^b]=f^{abc}K^c
\label{symmalg}\eeq
Introduce Faddeev-Popov ghost and anti-ghost fields $\theta^a$,
$\bar\theta^a\sim\frac{\partial}{\partial\theta^a}$ which are anticommuting
Grassmann variables that transform in the adjoint representation of $\bf g$.
They have the canonical anticommutator
\beq
[\bar \theta^a,\theta^b]_+=\delta^{ab}
\label{cananticomm}\eeq
We define the ghost number operator as
\beq
U=\theta^a\bar \theta^a
\label{ghostnumop}\eeq
whose eigenvalues are integers running from 0 to $\dim{\bf g}$.

We now introduce the operator
\beq
Q=\theta^aK^a-\frac{1}{2}f^{abc}\theta^a\theta^b\bar\theta^c
\label{brstcharge}\eeq
In the physics literature the operator $Q$ is known as the BRST charge, while
in the mathematics literature it is the Lie algebra coboundary operator that
computes the cohomology of the Lie algebra $\bf g$ with values in the
representation defined by the operators $K^a$. The crucial property of $Q$ is
that it is nilpotent, $Q^2=0$, which can be seen from (\ref{symmalg}) and the
identity
\beq
f^{abc}f^{cde}+f^{bdc}f^{cae}+f^{dac}f^{cbe}=0
\eeq
which follows from (\ref{symmalg}) via the Jacobi identity for the Lie bracket.
Let ${\cal H}^k$ be the Hilbert space of states of ghost number $k$, i.e.
$U\Psi=k\cdot\Psi$ for $\Psi\in{\cal H}^k$. We say that a state $\Psi\in{\cal
H}^k$ is BRST invariant if it is annihilated by $Q$, $Q\Psi=0$, where in
general the action of $Q$ on any state raises the ghost number by 1. Any other
state $\Psi'=\Psi+Q\chi$ of ghost number $k$ is regarded as equivalent to
$\Psi\in{\cal H}^k$ for any other state $\chi\in{\cal H}^{k-1}$. The space of
$Q$-equivalence classes of ghost number $k$ is called the BRST-cohomology in
the physics literature. Mathematically, it forms the $k$-th cohomology group
$H^k({\bf g};R)$ of the Lie algebra $\bf g$ with values in the representation
$R$ carried by the symmetry operators $K^a$.

Of particular interest from a physical standpoint are the BRST-invariant states
of ghost number 0. From (\ref{ghostnumop}) it follows that a state $\Psi$ of
ghost number 0 must be annihilated by all of the anti-ghost fields
$\bar\theta^a$, so that the action of $Q$ on such a state is
\beq
Q\Psi=\theta^aK^a\Psi~~~~~,~~~~~\Psi\in{\cal H}^0
\label{ghostnum0}\eeq
The anticommutation relations (\ref{cananticomm}) imply that a state
annihilated by all $\bar\theta^a$ cannot be annihilated by any of the ghost
fields $\theta^a$, and so the condition $Q\Psi=0$ is equivalent to
\beq
K^a\Psi=0~~~,~~~a=1,\dots,\dim{\bf g}~~~~~;~~~~~\Psi\in H^0({\bf g};R)
\label{ghostQ0}\eeq
Therefore a state $\Psi$ of ghost number 0 is  BRST-invariant if and only if it
is ${\bf g}$-invariant, and thus the cohomology group $H^0({\bf g};R)$
coincides with the space of ${\bf g}$-invariant states that do not contain any
ghosts, i.e. the physical states.

In a gauge theory with gauge group $G$, the partition function must be
evaluated as always with gauge-fixing functions $g^a$, $a=1,\dots,\dim G$,
which specify representatives of the gauge equivalence classes of the theory
and restrict the functional integration to a subspace $U_0$ of the original
configuration space of the field theory defined by the zeroes of the functions
$g^a$. Then the path integral can be written symbolically as
\beq
\int_{U_0}\e^{iS}=~{\rm vol}(G)\int_{U_0}\prod_{a=1}^{\dim
G}\delta(g^a)\det\|V^b(g^c)\|\e^{iS}
\label{gaugeint}\eeq
where $V^a$ are as usual vector fields associated with an orthonormal basis
$\{X^a\}$ of $\bf g$ (i.e. $\tr(X^aX^b)=\delta^{ab}$). Here $S$ is the
classical $G$-invariant gauge field action and the volume factor ${\rm vol}(G)$
is infinite for a local gauge field theory. Modulo this infinite factor, the
right-hand side of (\ref{gaugeint}) is what is taken as the definition of the
quantum gauge theory partition function. Introducing Faddeev-Popov ghost fields
and additional auxilliary fields $\phi^a$, we can absorb the additional factors
on the right-hand side of (\ref{gaugeint}) into the exponential to write
\beq
\int_{U_0}\e^{iS}=~{\rm vol}(G)\int_{U_0}\e^{iS_q}
\eeq
where
\beq
S_q=S+\phi^ag^a+\bar\theta^aV^a(g^b)\theta^b
\label{gaugequpart}\eeq
is the gauge-fixed, quantum action.

The BRST-symmetry of this model is defined by the following differential,
\beq
s(\Phi)=V^a(\Phi)\theta^a~~~,~~~s(\phi^a)=0~~~;~~~
s(\theta^a)=-\frac{1}{2}f^{abc}
\theta^b\theta^c~~~,~~~s(\bar\theta^a)=-\phi^a
\label{gaugeBRST}\eeq
where $\Phi$ is any scalar-valued functional of the gauge fields of the theory.
With this definition we have $s^2=0$, $s(S)=0$ and the quantum action
(\ref{gaugequpart}) can be written as
\beq
S_q=S+s(-g^a\bar\theta^a)
\label{quactionBRST}\eeq
Thus $s$ is a BRST operator that determines an $N=\dim G$ supersymmetry of the
gauge-fixed field theory, and the statement that the partition function
(\ref{gaugeint}) is independent of the choice of gauge-fixing functions $g^a$
is equivalent to the fact that the path integral depends only on the
BRST-cohomology class of the action $S$, not on its particular representative.
Thus the BRST-supersymmetry here represents the local gauge symmetry of the
theory. The gauge variation of any functional $\cal O$ of the fields of the
theory is then represented as a graded commutator
\beq
\{s,{\cal O}\}=s{\cal O}-(-1)^p{\cal O}s
\eeq
with the fermionic charge $s$, where $p$ is the ghost-degree of $\cal O$. The
physical (i.e. gauge-invariant) Hilbert space of the gauge theory is the space
of BRST-cohomology classes of ghost number 0.

(\ref{gaugeint}) and (\ref{quactionBRST}) demonstrate the explicit relationship
between equivariant localization of path integrals and BRST quantization (see
the localization principle in Subsection 4.4). In the next Appendix we shall
make this connection a bit more explicit. As we have mentioned, BRST-cohomology
is the fundamental structure in topological field theories \cite{birm}. By
definition, a topological action is a Witten-type action if the the classical
action $S$ is BRST-exact, while it is a Schwarz-type action if the gauge-fixed,
quantum action $S_q$ is BRST-exact (but not the classical one). In the case of
the localization formalism for phase space path integrals, the BRST-operator is
identified with the loop space equivariant exterior derivative of the
underlying equivariant cohomological structure, and the ``Hilbert space" of
physical states consists of loop space functionals which are invariant under
the flows of the loop space Hamiltonian vector field. This BRST-supersymmetry
is always the symmetry that is responsible  for localization in these models.
The BRST formalism can also be applied to Hamiltonian systems with first-class
constraints, i.e. those whose constraint functions $K^a$ generate a Poisson
subalgebra representation of a Lie algebra (\ref{symmalg}). The supersymmetric
states then represent the observables which respect the constraints of the
dynamical system (as in a gauge theory).

  \addtocounter{section}{1}
  \setcounter{equation}{0}
  \setcounter{equnum}{0}
  \renewcommand{\thesection}{\Alph{section}}
  \section*{Appendix B \ \ \  Other Models of Equivariant Cohomology}
  \addcontentsline{toc}{section}{Appendix \thesection\ \ \  Other Models of
Equivariant Cohomology}

In this Appendix we shall briefly outline some of the other standard models for
the $G$-equivariant cohomology of a differentiable manifold $\man$ and compare
them with the Cartan model which was used extensively throughout this Review.
We shall also discuss how these other models apply to the derivation of some of
the more general localization formulas which were just briefly sketched in
Subsection 4.9, as well as their importance to other ideas in topological
quantum field theory.

\subsection{The Topological Definition}

As with ordinary cohomology, equivariant cohomology has a somewhat direct
interpretation in terms of topological characteristics of the manifold $\man$
(and in this case also the Lie group $G$) \cite{atiyahbott}. This can be used
to develop an axiomatic formulation of equivariant cohomology which in the
usual way provides properties that uniquely characterize the cohomology groups
\cite{cordes}. This topological definition resides heavily in the topology of
the Lie group $G$ through the notion of a classifying space \cite{hu}. A
classical theorem of topology tells us that to $G$ we can associate a very
special space $E_G$ which is characterized by the fact that it is contractible
and that $G$ acts on it without fixed points. The space $E_G$ is called the
universal $G$-bundle. The classifying space $B_G$ for $G$-bundles is then
defined as the base space of a universal bundle whose total space is $E_G$. The
space $B_G$ is unique up to homotopy and $E_G$ is unique up to equivariant
homotopy (i.e. smooth continuous equivariant deformations of the space).
The bundle $E_G\to B_G$ has 2 remarkable universal properties. The first one is
that any given principal $G$-bundle $E\to\man$ over a manifold $\man$ has an
isomorphic copy sitting inside $E_G\to B_G$. The isomorphism classes of
principal $G$-bundles are therefore in one-to-one correspondence with the
homotopy classes of maps $f:\man\to B_G$. The second property is that all
natural ways of measuring the topology of $E\to\man$ can be obtained from
$H^*(B_G)$.

For example, when $G=\bigoplus_{i=1}^n\IZ$, we have $E_{\IZ^n}=\IR^n$ and
$B_{\IZ^n}=(S^1)^n$,  while for $G=U(1)$ we get
$E_{U(1)}=S(H)=\bigcup_{n=0}^\infty S^{2n+1}$ (the Hilbert sphere) and
$B_{U(1)}=\IC P^\infty=\bigcup_{n=0}^\infty\IC P^n$. In gauge theories $G$ is
the group of local gauge transformations, so that $E_G$ is the space ${\cal A}$
of Yang-Mills potentials while $B_G={\cal A}/G$ is the space of gauge orbits.
In string theory $G$ is the semi-direct product of the diffeomorphism and Weyl
groups of a Riemann surface $\Sigma^h$ of genus $h$, $E_G$ is the space of
metrics on $\Sigma^h$, and $B_G$ is the moduli space $\man_{\Sigma^h}$ of
$\Sigma^h$. From this point of view, one can define topological field theory
and topological string theory as the study of $H^*(B_G)$ and related
cohomologies using the language of local quantum field theory.

Given a smooth $G$-action on a manifold $\man$, we thus have 2 spaces $E_G$ and
$\man$ on which $G$ acts. Thus $G$ also acts on the Cartesian product space
$\man\times E_G$ via the diagonal action
\beq\new{\begin{array}{c}
G\times(\man\times E_G)\to\man\times E_G\\(g,x,e)\to(g\cdot x,g\cdot
e)\end{array}}
\label{diagaction}\eeq
Like the $G$-action on $E_G$, this action is also free and thus the quotient
space
\beq
\man_G=(\man\times E_G)/G
\label{homquot}\eeq
is a smooth manifold called the homotopy quotient of $\man$ by $G$. Since $E_G$
is contractible, $\man\times E_G$ is homotopic to $\man$. Furthermore, if the
$G$-action on $\man$ is free then $\man_G$ is homotopic to $\man/G$, so that
both spaces have the same (ordinary) cohomology groups. In the general case we
can regard $\man_G$ as a bundle over $B_G$ with fiber $\man$. These
observations motivate the topological definition of equivariant cohomology as
\beq
H_{G,{\rm top}}^k(\man)=H^k(\man_G)
\label{topeqcoh}\eeq
Notice that if $\man$ is the space consisting of a single point, then
$\man_G\simeq E_G/G=B_G$. Thus
\beq
H_{G,{\rm top}}^k({\rm pt})=H^k(B_G)
\label{cohtoppt}\eeq
and so the $G$-equivariant cohomology of a point is the ordinary cohomology of
the classifying space $B_G$. This latter cohomology can be quite complicated
\cite{atiyahbott}, and the topological definition therefore shows that the
equivariant cohomology measures much more than simply the cohomology of a
manifold modulo a group action on it. It is this feature that makes the
non-abelian localization formalisms of topological field theories and
integrable models very powerful techniques.

\subsection{The Weil Model}

The topological definition of the $G$-equivariant cohomology above can be
reformulated in terms of nilpotent differential operators
\cite{atiyahbott,kalkman1}. In this formulation, the equivariant cohomology is
obtained in a more algebraic way by exploiting differential properties of the
Lie algebra $\bf g$ of the group $G$. More precisely, to describe the exterior
differential calculus of the symmetric algebra $S({\bf g}^*)$, we introduce the
Weil algebra
\beq
W({\bf g})=S({\bf g}^*)\otimes\Lambda{\bf g}^*
\label{weilalg}\eeq
As in Subsection 2.6, (\ref{weilalg}) algebraically describes the exterior
bundle of $S({\bf g}^*)$ where the exterior algebra $\Lambda{\bf g}^*$ consists
of multilinear antisymmetric forms on $\bf g$ and it is generated by an
anti-commuting basis of Grassmann numbers $\theta^a$, $a=1,\dots,\dim G$ (i.e.
the 1-forms $\theta^a\sim d\phi^a$). The Weil algebra has the usual
$\IZ$-grading (ghost number), i.e. the generators $\phi^a$ of $S({\bf
g}^*)$ have degree 2 while the generators $\theta^a$ of $\Lambda{\bf g}^*$ have
degree 1. Both of these sets of generators are dual to the same fixed basis
$\{X^a\}_{a=1}^{\dim G}$ of $\bf g$.

There are 2 differential operators of interest acting on $W({\bf g})$. The
first is the ``abelian" exterior derivative on $\Lambda{\bf g}^*$
\beq
d_0=\phi^a\otimes I_a
\label{d0}\eeq
where $I_a=\frac{\partial}{\partial\theta^a}$ is the interior multiplication of
degree $-1$ on $W({\bf g})$. (\ref{d0}) identifies $\phi^a$ as the superpartner
of $\theta^a$ and its non-vanishing actions are
\beq
d_0\theta^a=\phi^a~~~~~,~~~~~d_0\frac{\partial}{\partial\phi^a}=-I_a
\label{d0actioN}\eeq
Next, we introduce the linear derivations of degree 0
\beq
L_a=-f^{abc}\left(\phi^b\frac{\partial}{\partial\phi^c}+\theta^bI_c\right)
\label{La}\eeq
which generate the coadjoint action of $G$ on $W({\bf g})$ explicitly by
\beq
L_a\phi^b=f^{abc}\phi^c~~~~~,~~~~~L_a\theta^b=f^{abc}\theta^c
\label{Laaction}\eeq
and which yield a representation of $G$ on $W({\bf g})$,
\beq
[L_a,L_b]=f^{abc}L_c
\label{Laalg}\eeq
Using (\ref{La}), we define our second differential operator, the coboundary
operator
\beq
d_{\bf g}=\theta^aL_a+\frac{1}{2}f^{abc}\theta^a\theta^bI_c
\label{coop}\eeq
which computes the $W({\bf g})$-valued Lie algebra cohomology of $\bf g$, i.e.
it is the BRST operator associated with the constraint operators $L_a$ acting
on $W({\bf g})$ (see (\ref{brstcharge})).

The sum of (\ref{d0}) and (\ref{coop}) is known as the Weil differential,
\beq
d_{\cal W}=d_0+d_{\bf g}
\label{weildiff}\eeq
whose action on the generators of $W({\bf g})$ is
\beq
d_{\cal W}\theta^a=\phi^a-\frac{1}{2}f^{abc}\theta^b\theta^c~~~~~,~~~~~d_{\cal
W}\phi^a=-f^{abc}\theta^b\phi^c
\label{dWaction}\eeq
These 3 differential operators are all nilpotent derivations of degree 1,
\beq
d_{\cal W}^2=d_0^2=d_{\bf g}^2=0
\label{wilnilpot}\eeq
and they act as exterior derivatives on $W({\bf g})$. (\ref{weildiff}) makes
the Weil algebra into an exterior differential algebra. However, the cohomology
of the Weil differential $d_{\cal W}$ on $W({\bf g})$ is trivial. This can be
seen by redefining the basis of $W({\bf g})$ by the shift
$\phi^a\to\phi^a-\frac{1}{2}f^{abc}\theta^b\theta^c$. In the new basis,
$\phi^a=d_{\cal W}\theta^a$ are exact and the cohomology of $d_{\cal W}$
coincides with that of $d_0$ on $W({\bf g})$ so that
\beq
H^k(W({\bf g}),d_{\cal W})=\IR
\label{trivcoh}\eeq

The cohomology can be made non-trivial using the 2 derivations $I_a$ and $L_a$
on $W({\bf g})$ introduced above. We notice first of all the analogy between
(\ref{dWaction}) and the algebra of connections and curvatures. The first
relation in (\ref{dWaction}) is the definition of the curvature of a connection
1-form $A\sim\theta^a$ on a principal $G$-bundle $E\to\man$, while the second
one is the Bianchi identity for the curvature 2-form $F\sim\phi^a$, i.e.
\beq
dA=F-\frac{1}{2}[A~{\buildrel\wedge\over,}~A]~~~~~,~~~~~dF=-[A~{\buildrel
\wedge\over,}~F]
\label{curvconndef}\eeq
Here we recall that the characteristic classes of $E\to\man$ are constructed
from $A\in\Lambda^1(E,{\bf g})$, which can be regarded as a map $A:{\bf
g}^*\to\Lambda^1E$, and from the field strength $F\in\Lambda^2(E,{\bf g})$,
which can be regarded as a map $F:{\bf g}^*\to\Lambda^2E$. These maps generate
a differential algebra homomorphism $(W({\bf g}),d_{\cal W})\to(\Lambda E,d)$
which is called the Chern-Weil homomorphism. This homomorphism is unique and it
maps the algebraic connection and curvature $(\theta^a,\phi^a)$ to the
geometric ones $(A,F)$ \cite{mathai}. In this setting, a connection on a
principal $G$-bundle $E\to\man$ is just the same thing as a homomorphism
$W({\bf g})\to\Lambda E$. Thus the Weil algebra is an algebraic analog of the
universal $G$-bundle $E_G$. Like $E_G$, it possesses universal properties, and
therefore it provides a universal model of connections on $G$-bundles. In
particular, the contractibility of $E_G$ is the analog of the triviality
(\ref{trivcoh}) of the cohomology of the Weil algebra.

Pursuing this analogy between $E_G$ and $W({\bf g})$, we can find non-trivial
and universal cohomology classes by considering the so-called ``basic forms"
\cite{atiyahbott}. First, we note that the operators $I_a$ and $L_a$ above are
the algebraic analogues of the interior multiplication and Lie derivative of
differential forms with respect to the infinitesimal generators $\phi^a$ of the
$G$-action on $W({\bf g})$. Indeed, the operator (\ref{La}) can be expressed in
terms of the Weil differential as
\beq
L_a=I_ad_{\cal W}+d_{\cal W}I_a\equiv[d_{\bf g},I_a]_+
\label{Laliederiv}\eeq
and furthermore we have
\beq
[I_a,I_b]=0~~~~~,~~~~~[L_a,I_b]=f^{abc}I_c
\label{LaIarels}\eeq
Thus the derivation $L_a$ has the natural structure of a Lie derivative on
$W({\bf g})$ that commutes with all the derivatives above,
\beq
[d_{\cal W},L_a]=[d_{\bf g},L_a]=[d_0,L_a]=0
\label{dLarels}\eeq
In particular, the (anti-)commutation relations above among $d_{\cal W}$, $L_a$
and $I_a$ are all independent of the choice of basis of $\bf g$. As we saw in
Subsection 2.3, these relations also reflect the differential geometric
situation on a manifold $\man$ with a $G$-action on it, and the Chern-Weil
homomorphism above maps $(d_{\cal W},L_a,I_a)\to(d,\lie_{V^a},i_{V^a})$ between
the differential algebras $W({\bf g})\to\Lambda\man$.

We can finally define the Weil model for equivariant cohomology. For this we
consider the tensor product $W({\bf g})\otimes\Lambda\man$ of the Weil algebra
with the exterior algebra over the manifold $\man$. The replacement
$\Lambda\man\to W({\bf g})\otimes\Lambda\man$ for the description of the
equivariant cohomology is the algebraic equivalent of the replacement $\man\to
E_G\times\man$ in Subsection B.1 above. The basic subalgebra $(W({\bf
g})\otimes\Lambda\man)_{\rm basic}$ of $W({\bf g})\otimes\Lambda\man$ consists
of those forms which have no vertical component (i.e. the horizontal forms) and
which are $G$-invariant (i.e. have no vertical variation). These 2 conditions
mean, respectively, that the basic forms are those annihilated by all the
operators $I_a\otimes{\bf1}+{\bf1}\otimes i_{V^a}$ and
$L_a\otimes{\bf1}+{\bf1}\otimes\lie_{V^a}$ (recall that $i_V\alpha$ is the
component of $\alpha$ along the (vertical) vector field $V\in T\man$), so that
\beq
\left(W({\bf g})\otimes\Lambda\man\right)_{\rm basic}=\left(\bigcap_{a=1}^{\dim
G}{\rm ker}\left(I_a\otimes{\bf1}+{\bf1}\otimes
i_{V^a}\right)\right)\cap\left(\bigcap_{b=1}^{\dim G}{\rm
ker}\left(L_a\otimes{\bf1}+{\bf1}\otimes\lie_{V^a}\right)\right)
\label{basiccomplex}\eeq
This subalgebra is stable under the action of the extended DeRham exterior
derivative
\beq
d_T=d_{\cal W}\otimes{\bf1}+{\bf1}\otimes d
\label{derhamext}\eeq
and the cohomology of (\ref{derhamext}) on (\ref{basiccomplex}) is the
algebraic definition of the $G$-equivariant cohomology of $\man$,
\beq
H_{G,{\rm alg}}^k(\man)=H^k((W({\bf g})\otimes\Lambda\man)_{\rm basic},d_T)
\label{algeqcoh}\eeq
The Chern-Weil homomorphism $W({\bf g})\to\Lambda E$ with $E=E_G$ then reduces
to an isomorphism of cohomology groups if $G$ is compact and connected
\cite{atiyahbott} and we have
\beq
H_{G,{\rm alg}}^k(\man)\simeq H_{G,{\rm top}}^k(\man)
\label{algtopeq}\eeq
so that the algebraic and topological definitions of equivariant cohomology are
equivalent.

We close this Subsection with a remark concerning the basic subcomplex $B_{\bf
g}$ of $W({\bf g})$ ($V^a=0$ in (\ref{basiccomplex})). Since $d_{\cal W}$
vanishes on this basic subcomplex, we have $H^k(B_{\bf g},d_{\cal W})=B_{\bf
g}$. Horizontality in this case means $\theta^a$-independence, so that the
basic forms in $W({\bf g})$ lie in only $S({\bf g}^*)$. $G$-invariance in this
case translates into invariance under the coadjoint action of $G$ on ${\bf
g}^*$. Thus
\beq
H^k(B_{\bf g},d_{\cal W})=B_{\bf g}=S({\bf g}^*)^G
\label{basicweil}\eeq
and so the basic subalgebra of the Weil algebra coincides with the algebra of
invariant polynomial functions on the Lie algebra $\bf g$, i.e. $B_{\bf g}$ is
the algebra of corresponding Casimir invariants. It is known \cite{atiyahbott}
that if $G$ is a compact connected Lie group, then $H^k(B_G)=H^k(B_{\bf
g},d_{\cal W})$, and so comparing (\ref{basicweil}) with (\ref{cohtoppt}) we
find that the $G$-equivariant cohomology of a point is simply the algebra of
$G$-invariant polynomials on $\bf g$. This is in agreement with what we found
in
Subsection 2.3 from the Cartan model. In the next Subsection we shall indeed
find that this correspondence is no accident.

\subsection{The BRST Model}

The final model for the $G$-equivariant cohomology of a manifold $\man$
interpolates between the Cartan and Weil models, and it therefore relates the
Cartan model to the topological characteristics of this cohomology theory. It
also ties in with the BRST quantization ideas that are directly related to the
localization formalisms and it is the model of equivariant cohomology which
arises naturally in the physical context of topological field theories. The
(unrestricted) BRST algebra $B$ of topological models on quotient spaces is the
same as that for the Weil model as a vector space, $B=W({\bf
g})\otimes\Lambda\man$ \cite{kalkman1,osvb}. Now, however, the differential on
it is the BRST operator
\beq
\delta=d_{\cal W}\otimes{\bf1}+{\bf1}\otimes
d+\theta^a\otimes\lie_{V^a}-\phi^a\otimes i_{V^a}
\label{BRSTop}\eeq
which is a nilpotent graded derivation of degree 1 on $B$. The BRST operator is
the natural nilpotent extension of the Cartan equivariant exterior derivative
$D_{\bf g}$ defined in (\ref{donforms}). The Weil differential $d_{\cal W}$ in
(\ref{BRSTop}) takes care of the non-nilpotency (\ref{weilid}) of the Cartan
model derivative.

The Kalkman parametric model for the equivariant cohomology is defined by the
Kalkman differential \cite{kalkman1}
\beq
\delta_t=\e^{t\theta^a\otimes i_{V^a}}d_T\e^{-t\theta^a\otimes
i_{V^a}}=d_T+t\theta^a\otimes\lie_{V^a}-t\phi^a\otimes
i_{V^a}+\frac{1}{2}t(1-t)f^{abc}\theta^a\theta^b\otimes i_{V^c}
\label{kalkderiv}\eeq
where $t\in[0,1]$. Notice that $\delta_0=d_T$ and $\delta_1=\delta$, so that
(\ref{kalkderiv}) interpolates between the differentials of the Weil and BRST
models. Furthermore, it satisfies
\beq
[\delta_t,I_a\otimes{\bf1}+(1-t){\bf1}\otimes
i_{V^a}]=L_a\otimes{\bf1}+{\bf1}\otimes\lie_{V^a}
\label{kalkcomm}\eeq
so that we obtain in this way a family of Lie super-algebras acting on $W({\bf
g})\otimes\Lambda\man$. Notice also that
\beq
d_T=\e^{-\frac{1}{2}f^{abc}\theta^a\theta^b\frac{\partial}{\partial\phi^c}}
\left(d+
d_0\right)\e^{\frac{1}{2}f^{abc}\theta^a\theta^b\frac{\partial}
{\partial\phi^c}}
\label{dTconj}\eeq
and thus the cohomology of $\delta$ on $B$ coincides with the DeRham cohomology
of $d$ on $\Lambda\man$.

Thus the BRST operator $\delta$ does not capture the $G$-equivariant cohomology
of $\man$, because the $W({\bf g})$ part of (\ref{BRSTop}) can be conjugated to
the cohomologically trivial operator (\ref{d0}) (equivalently the cohomology of
$d_{\cal W}$ is trivial). We have to accompany $\delta$ with a restriction of
its domain in the same way that (\ref{donforms}) computes the equivariant
cohomology when restricted to the $G$-invariant subspace (\ref{eqdiff}).  The
appropriate restriction is to the $\theta^a$-independent and $G$-invariant
subalgebra of $B$. This reduction maps $B$ to $\Lambda_G\man$ and $\delta$ to
$D_{\bf g}$, so that the mapping $\theta^a\to0$, $B\to S({\bf
g}^*)\otimes\Lambda\man$ induces the isomorphism of complexes
\beq
(B_{\rm basic},d_{\cal W})\simeq(\Lambda_G\man,D_{\bf g})
\label{complexiso}\eeq
between the Weil and Cartan models. This restriction can be formulated by
introducing another nilpotent operator $\cal W$ whose kernel is the desired
$G$-invariant and $\theta^a$-independent subalgebra \cite{niemitirk3,osvb}.

For this, we introduce another copy $\bar W({\bf g})$ of the Weil algebra. It
is generated by $\bar\phi^a$ and $\bar\theta^a$ which are the ${\bf
g}^*$-valued coefficients corresponding, respectively, to
$\theta^a$-independence (generated by $I_a$) and $G$-invariance (generated by
$L_a\otimes{\bf1}+{\bf1}\otimes\lie_{V^a}$). A nilpotent operator with kernel
$\Lambda_G\man$ is
\beq
{\cal W}=d_{\bar{\bf
g}}\otimes{\bf1}\otimes{\bf1}+\bar\theta^a\otimes(L_a\otimes{\bf1}+{\bf1}
\otimes\lie_{V^a})-\bar\phi^a\otimes I_a\otimes{\bf1}
\label{calW}\eeq
where $d_{\bar{\bf g}}$ is the Lie algebra coboundary operator (\ref{coop}) on
$\bar W({\bf g})$ which makes the overall combination in (\ref{calW})
nilpotent.
The action of $\delta$ on $\bar W({\bf g})$ is taken as that of the abelian
differential (\ref{d0}) so that $\delta$ commutes with $\cal W$. As the
cohomology of $d_{\bar0}$ on $\bar W({\bf g})$ is trivial, this alteration of
$\delta$ does not affect its cohomology. The equivariant BRST operator is
therefore
\beq
\delta={\bf1}\otimes d_{\cal W}\otimes{\bf1}+{\bf1}\otimes{\bf1}\otimes
d+d_{\bar0}\otimes{\bf1}\otimes{\bf1}+{\bf1}\otimes\theta^a\otimes\lie_{V^a}
-{\bf1}\otimes\phi^a\otimes i_{V^a}
\label{eqBRSTop}\eeq
and $\delta$ and $\cal W$ satisfy the nilpotent algebra
\beq
\delta^2=[{\cal W},\delta]_+={\cal W}^2=0
\label{BRSTWalg}\eeq
The $G$-equivariant cohomology of $\man$ is isomorphic to the cohomology of
$\delta$ on the subalgebra of $\bar W({\bf g})\otimes W({\bf
g})\otimes\Lambda\man$ which is annihilated by $\cal W$. This defines the BRST
model of equivariant cohomology.

We remark that it is also possible to formulate the restriction onto the basic
subcomplex in the Weil model using the nilpotent operator
\beq
{\cal W}_W=d_{\bar{\bf
g}}\otimes{\bf1}\otimes{\bf1}+\bar\theta^a\otimes
(L_a\otimes{\bf1}+{\bf1}\otimes
\lie_{V^a})-\bar\phi^a\otimes(I_a\otimes{\bf1}+{\bf1}\otimes i_{V^a})
\label{calWweil}\eeq
and the corresponding extension of the Weil differential is
\beq
\bar d_{\cal W}={\bf1}\otimes d_T+d_{\bar0}\otimes{\bf1}\otimes{\bf1}
\label{dweilext}\eeq
The operators (\ref{calWweil}) and (\ref{dweilext}) are similarity transforms
of (\ref{calW}) and (\ref{eqBRSTop}), respectively, just as in (\ref{dTconj}).
They therefore obey the algebra (\ref{BRSTWalg}) as well, as required for the
appropriate restriction process above.

It is this model that is relevant for the construction of non-abelian
generalizations of the Duistermaat-Heckman integration formula, such as the
Witten localization formalism (see Subsection 3.8). Modelling the equivariant
cohomological structure as above is the correct way to incorporate the idea of
equivariant integration that we discussed earlier. In these models the
generators $\phi^a$ of the symmetric algebra $S({\bf g}^*)$ generate dynamics
of their own and only after they are fully incorporated as above can one define
properly the required equivariant localization. All 4 equivariant derivations
discussed in this Review -- the Cartan, Weil, Kalkman and BRST differentials,
have been related by Nersessian \cite{nersesseq2,nersess1} to the geometric
anti-bracket structure of the Batalin-Vilkovisky formalism (see Subsection
8.3). In these formalisms, one constructs antibrackets for $W({\bf g})$ in
addition to the usual ones for the cotangent bundle $\man\otimes\Lambda^1\man$.

\subsection{Loop Space Extensions}

Loop space generalizations of the constructions above have been presented by
Tirkkonen in \cite{tirk2} in the context of the dynamical localizations of
Subsection 4.9. For this, we introduce superpartners $\theta^a(t)$ for the
multipliers $\phi^a(t)$ and make them dynamical by adding a kinetic term for
the Grassmann coordinates $\theta^a$ to the action,
\beq
S\to S_T=S+\int_0^Tdt~\theta^a(t)\dot\theta^a(t)
\label{ST}\eeq
 The circle action generator in
(\ref{semiaction}) is also extended to include $\theta^a$, i.e.
\beq
V_{S^1}\to V_{S^1}+\int_0^Tdt~\dot\theta^a(t)\frac{\delta}{\delta\theta^a(t)}
\label{VS1ext}\eeq
Thus the path integral is now formulated over a superloop space with local
coordinates $(x,\theta)$ (recall from (\ref{d0actioN}) that $\theta^a$ are
coordinates and $\phi^a=d_0\theta^a$ are the corresponding 1-forms in the Weil
algebra). To exploit the loop space isometry generated by the semi-direct
product $LG\semi S^1$, we need to construct the corresponding equivariant
operators. Because the fields $\phi^a$ and $\theta^a$ are dynamical and are
therefore an important part of the path integration, this has to be done in
terms of the BRST model above.

In the corresponding equivariant BRST operator, the part corresponding to the
$LG$-action in (\ref{semiaction}) generated by $V_{LG}$ on $L\man$ is just
lifted from the corresponding equivariant BRST operator (\ref{eqBRSTop}) for
the $G$-action on $\man$. The part associated with the circle action in
(\ref{semiaction}) as generated by (\ref{VS1ext}) on the superloop space is
given by the Cartan model operator $Q_{\dot x}=d_L-i_{\dot x}$. Since $Q_{\dot
x}^2=\int_0^Tdt~\frac{d}{dt}$, the operator $Q_{\dot x}$ is nilpotent on the
loop space $L\man$. The total equivariant BRST operator for the semi-direct
product $LG\semi S^1$ then combines (\ref{eqBRSTop}) lifted to the loop space
with $Q_{\dot x}$,
\beq
Q_T=d_L+d_{\cal W}^{(L)}+\int_0^Tdt~\left(\theta^a\lie_{V^a}-\phi^ai_{V^a}+\dot
x^\mu\frac{\delta}{\delta\eta^\mu}+\dot\theta^a\frac{\delta}{\delta\phi^a}
\right)
\label{QTBRST}\eeq
where $d_{\cal W}^{(L)}$ is the Weil differential (\ref{weildiff}) lifted to
the superloop space in the standard way. Furthermore, to the restriction
operator (\ref{calW}) we add a piece corresponding to the $S^1$-action
\cite{tirk2}
\beq
{\cal W}_T={\cal
W}_L+\int_0^Tdt~{\dot{\bar{\theta}}}^a(t)\frac{\delta}{\delta\phi^a(t)}
\label{calWloop}\eeq
with ${\cal W}_L$ the lifting of the restriction operator (\ref{calW}) to the
superloop space. The algebra of $Q_T$ and ${\cal W}_T$ is then
\beq
Q_T^2=\int_0^Tdt~\frac{d}{dt}~~~~~,~~~~~{\cal W}_T^2=[Q_T,{\cal W}_T]_+=0
\label{QWalgloop}\eeq

The function $F(\phi)=\frac{1}{2}(\phi^a)^2$ which is effectively added to the
action as discussed in Subsection 4.9 can now be interpreted as a symplectic
2-form on the Weil algebra $W({\bf g})$ (as $\phi^a$ are interpreted as super
1-forms). The superloop space symplectic 2-form is then
\beq
\Omega_T=\Omega+\int_0^Tdt~\frac{1}{2}\left(\phi^a(t)\right)^2
\label{OmT}\eeq
and the equivariance properties of the phase space path integral are summarized
by the equations
\beq
Q_T(S_T+\Omega_T)={\cal W}_T(S_T+\Omega_T)=0
\label{patheqT}\eeq
The first vanishing condition in (\ref{patheqT}) identifies the superloop space
action $S_T$ as the moment map for the action of the semi-direct product
$LG\semi S^1$ on $L\man$, while the second one states that this group action is
symplectic. To carry out the Niemi-Tirkkonen localization procedure over the
superloop space with the BRST operator $Q_T$, we generalize the gauge fermion
field (\ref{ntirkpsi}) to the Weil algebra,
\beq
\psi_T=\int_0^Tdt~\left(g_{\mu\nu}\dot x^\mu\eta^\nu+\hat\phi^a\hat\theta^a
\right)
\label{psiT}\eeq
where we have left the zero mode part of $\phi^a\theta^a$ out of $\psi_T$. We
can now carry out the same localization procedure which led to the
Niemi-Tirkkonen formula, and thus arrive at the localization formula
(\ref{nonabpartloc}) \cite{tirk2}.

The above construction is a non-abelian generalization of the model-independent
auxilliary field formalism which we discussed in Subsection 4.9. In the abelian
case where we can {\it a priori} fix any function $F(\phi)$ for the
localization, instead of modifying the loop space symplectic structure as in
(\ref{OmT}), which in the model-independent formalism appears as the functional
determinant $\det\|\partial_t\|$, we shift the gauge fermion field (\ref{psiT})
as
\beq
\psi_T\to\psi_T-\int_0^Tdt~\left(\theta_\mu\eta^\mu-\left\{\xi(\phi)-
\frac{1}{2}\bar\phi\right\}\bar\theta\right)
\label{psiTab}\eeq
The standard Niemi-Tirkkonen localization procedure then leads to the
localization formula (\ref{partfhloc}) \cite{niemitirk2}. For the generic,
non-abelian group actions discussed above, the situation for loop space
localization is different, and even the second Weil algebra copy $\bar W({\bf
g})$ is made dynamical. For discussions of the relations between the BRST model
of equivariant cohomology and Witten-type topological field theories, see
\cite{blau2,cordes,niemitirk3}.

\end{document}